\documentclass[11pt,a4paper]{hiepadraft2}
\usepackage{besphysics}
\usepackage{authblk}
\usepackage{amsmath}
\usepackage{graphicx}
\usepackage{epsfig}
\usepackage{epstopdf}
\usepackage{multirow}
\usepackage{overpic}
\usepackage{colortbl}
\usepackage{tabularx}
\usepackage{url}
\usepackage{multirow}
\usepackage{setspace}
\usepackage{amssymb}
\usepackage{bm}

\usepackage{color}
\usepackage{amsfonts,amsmath,amssymb}
\usepackage{bm}
\usepackage{titletoc}
\usepackage{makecell}
\usepackage[colorlinks,linkcolor=blue,anchorcolor=blue,citecolor=blue]{hyperref}
\usepackage{listings}
\hypersetup{
colorlinks=true,
allcolors=blue
}
\usepackage{float}
\usepackage{subfig}
\usepackage[section]{placeins}
\usepackage{verbatim}
\usepackage{siunitx}
\usepackage[numbers,sort&compress]{natbib}
\usepackage{chapterbib} 
\usepackage{textcomp}

\title{Volume I - Physics \& Detector }

\author[3]{M. Achasov}
\author[82]{X. C. Ai}
\author[38]{R. Aliberti}
\author[54]{L. P. An}
\author[63, 72]{Q. An}
\author[63, 72]{X. Z. Bai}
\author[62]{Y. Bai}
\author[39]{O. Bakina}
\author[3, 50]{A. Barnyakov}
\author[3, 50, 51]{V. Blinov}

% ...
\affil[3]{\it Budker Institute of Nuclear Physics, Novosibirsk 630090, Russia}
% ...
\affil[38]{\it Johannes Gutenberg University of Mainz, Johann-Joachim-Becher-Weg 45, D-55099 Mainz, Germany}
\affil[39]{\it Joint Institute for Nuclear Research, 141980 Dubna, Moscow region, Russia}
% ...
\affil[50]{\it Novosibirsk State Technical University, Novosibirsk 630073, Russia}
\affil[51]{\it Novosibirsk State University, Novosibirsk 630090, Russia}
% ...
\affil[54]{\it Peking University, Beijing 100871, China}
% ...
\affil[62]{\it Southeast University, Nanjing 211189, China}
\affil[63]{\it State Key Laboratory of Particle Detection and Electronics, Beijing 100049, Hefei 230026, China}
% ...
\affil[72]{\it University of Science and Technology of China, Hefei 230026, China}
% ...
\affil[82]{\it Zhengzhou University, Zhengzhou 450001, China}

\abstracttext{
The Super $\tau$-Charm facility~(STCF) is an electron-positron collider proposed by the Chinese particle physics community. It is designed to operate in a center-of-mass energy range from 2 to 7 GeV with a peak luminosity of $\stcflum$  or higher. The STCF will produce a data sample about a factor of 100 larger than that of the present $\tau$-Charm factory --- the BEPCII, providing a unique platform for exploring the asymmetry of matter-antimatter (charge-parity violation), in-depth studies of the internal structure of hadrons and the nature of non-perturbative strong interactions, as well as searching for exotic hadrons and physics beyond the Standard Model. The STCF project in China is under development with an extensive R\&D program. This document presents the physics opportunities at the STCF, describes conceptual designs of the STCF detector system, and discusses future plans for detector R\&D and physics case studies. 
}

\begin{document}

%%%%%%%%%%%%%%%%%%%%%%%%%%%%%%%%%%%%%%%%%%%%%%%%%%%%%%%%%%%%%%%%%%%%%%%%%%%%%%
\setlength{\baselineskip}{0.5cm}
%===========================================================================
%================== Table of contents ======================================
%===========================================================================
\tableofcontents
\newpage

%\begin{comment}
\chapter{Introduction}
%\addcontentsline{toc}{chapter}{Introduction}

Starting with the discovery of the charmed quark and the $\tau$ lepton during the 1974 and 1975
%``November Revolution''
~\cite{Galison:1992us}, the results
from low-energy $e^+e^-$ collider experiments with a center-of-mass energy (CME) in the 2$\sim$6 GeV ($\tau$--charm threshold)
region have played a key role in elucidating the properties of these intriguing particles. Historically, there have
been several generations of $\tau$--charm facilities (TCFs) in the world, including
the Mark II and Mark III detectors~\cite{Abrams:1989cm,Bernstein:1983wk}, DM2~\cite{Augustin:1980ad}, CLEO-c~\cite{Asner:2004yu},
and BEPC/BES \cite{Bai:1994zm}, which have produced numerous critical contributions to the establishment of the
SM and to searches for new physics beyond the Standard Model~(SM). Of these, the BEPC/BES facility in Beijing, China, is no doubt one of the
most prolific TCFs.
This program, which started in the late 1980s, has produced many interesting physics results, such as precision
measurements of the $\tau$-lepton mass~\cite{Bai:1992bu,Ablikim:2020tau-mass} and the $e^+e^-$ annihilation cross section~\cite{Bai:2001ct}, the first observations of purely leptonic charmed
meson decays~\cite{Bai:1998cg}, the discovery of the $X(1835)$ as a baryonium state candidate~\cite{Bai:2003sw,Ablikim:2005um,BESIII:2011aa,besiii-ppetap}, and clear elucidation of the $\sigma$
($f_0(500)$)~\cite{Ablikim:2004qna} and $\kappa$ ($K_0(700)$)~\cite{Ablikim:2010ab}, the lowest-lying scalar mesons.

The currently operating BEPCII/BESIII~\cite{bepcii,Ablikim:2009aa} complex, which is a major upgrade of BEPC/BESII~\cite{Bai:1994zm} that includes separate electron
and positron magnet rings as part of the highest-ever-luminosity TCF and a completely new, state-of-the-art detector,
is the only facility in the world that can address the physics opportunities in this interesting energy range. BEPCII/BESIII's
unique capabilities and excellent performance have attracted a large collaboration of researchers from all over the
world that has been very successful in producing numerous high-quality, frequently cited physics results. After
ten years of operation, BEPCII is operating reliably at its designed luminosity of $10^{33}$~cm$^{-2}$s$^{-1}$ at
$\sqrt{s}= 3.77$~GeV. A continuous injection system has recently been implemented that increases its integrated luminosity
by 30\%, and its CME upper limit has been extended from 4.6 to 4.9~GeV, thereby providing access to charmed baryon thresholds.

In 2019, BESIII achieved one of its main data-taking goals with the successful accumulation of 10 billion $J/\psi$ events
for studies of light hadron physics. An early payoff from this unprecedentedly large data sample was the discovery of an
isospin-singlet $\eta\eta^{\prime}$ meson resonance with manifestly exotic $J^{PC}=1^{-+}$ quantum numbers~\cite{BESIII:2022riz, BESIII:2022iwi}. This is best
explained as a ``smoking-gun'' candidate for a Quantum Chromodynamics~(QCD) hybrid meson comprising a quark--antiquark pair plus a valence gluon,
a hadronic substructure that was predicted over forty years ago~\cite{Horn:1977rq} but has only recently started to
emerge experimentally thanks to the availability of enormous datasets such as the BESIII 10~billion $J/\psi$ event sample. Other notable light
hadron physics results from BESIII include the discoveries of an anomaly in the $X(1835)\to\pi^+\pi^-\eta^{\prime}$ line
shape at the $p\bar{p}$ mass threshold~\cite{besiii-ppetap, BESIIICollaboration:2022kwh}, an anomalously large partial width for the isospin-violating
$\eta(1405)\to f_0(980) \pi^0$ decay~\cite{BESIII:2012aa}, the first observation of $a_0(980)\leftrightarrow f_0(980)$
mixing~\cite{Ablikim:2018pik} (another forty-year-old prediction~\cite{Achasov:1979xc} that was eventually confirmed by
BESIII) and precise measurements of the hyperon decay parameters and tests of charge-conjugate and parity~($CP$) invariance in
$J/\psi\to\Lambda\bar{\Lambda}$ decays~\cite{BESIII:2018cnd, BESIII:2022qax} and $J/\psi\to\Xi^{-}\bar{\Xi}^{-}$~\cite{BESIII:2021ypr}.

For studies of charmed mesons and baryons, BESIII has accumulated samples of 1.7 million tagged
$D^+D^-$ events and 2.8~million tagged $D^0\bar{D}^0$ events produced via $\psi(3770)\to D\bar{D}$ decays, 0.30~million
tagged $D_s^+D_s^{*-}$ events from $\psi(4160)\to D_s^+D_s^{*-}$ decays and 90~thousand tagged $\Lambda_c^+\bar{\Lambda}_c^-$
events from $e^+e^-\to \Lambda_c^+\bar{\Lambda}_c^-$ with CMEs above 4.6~GeV. Measurements of purely leptonic and
semileptonic decays of $D$ and $D_s$ mesons produced the world's best measurements of the Cabibbo-Kobayashi-Maskawa~(CKM)
matrix elements $|V_{cs}|$ and $|V_{cd}|$~\cite{Ablikim:2017lks,Ablikim:2015ixa,Ablikim:2018junl,Ablikim:2016duz}. Absolute $\Lambda_c$ branching
fraction measurements based on the tagged $\Lambda_c$ baryon sample~\cite{Ablikim:2015flg}, including measurements for a number of
previously unseen modes, dominate the Particle Data Group (PDG)~\cite{ParticleDataGroup:2022pth} listings for this state. In addition, the large samples of $CP$-tagged
$D^0$-meson decays have been used to make precise measurements of final-state strong interaction phase decays, which are critical
inputs to the LHCb and Belle (II) measurements of the $CP$-violating angle $\gamma$ of the CKM unitary
triangle~\cite{Ablikim:2014gvw,Ablikim:2020lpk,Ablikim:2020cfp}.

Measurements of $e^+e^-$ annihilations for CME values between 2.0 and 3.67~GeV have provided $R$ measurements, defined as the ratio of cross section at lowest-order between the inclusive hadronic process $e^{+}e^{-}\to hadrons$ and the Quantum Electrodynamics~(QED) process $e^{+}e^{-}\to\mu^{+}\mu^{-}$, with an unprecedented precision of
$\sim$3\%~\cite{BESIII:2021wib}, which were critical inputs to SM calculations of $\alpha_{\rm QED}(m_Z^2)$~\cite{Davier:2017zfy}
that were used in fits to the electroweak sector of the model that provided accurate predictions of the Higgs
boson mass that were spectacularly confirmed in 2012 by LHC experiments. BESIII R measurements for $e^+e^-\to\pi^+\pi^-$
at CMEs below 1~GeV, extracted from $e^+e^-\to\gamma_{\rm ISR}\pi^+\pi^-$ events~\cite{Ablikim:2015orh}, where $\gamma_{\rm ISR}$ is an
initial-state radiation, offer significant improvements in accuracy over previous results and will enable future SM calculations
of $(g-2)_{\mu}$ matching the higher precision that is expected for imminent measurements from currently operating
experiments at Fermilab~\cite{Grange:2015fou,Muong-2:2021ojo} and JPARC~\cite{Mibe:2010zz}. In addition, this data sample is being used for numerous low-energy QCD
studies, including measurements of the proton, neutron and $\Lambda$ time-like form factors with improved
precision~\cite{Ablikim:2019eau, BESIII:2021tbq, Ablikim:2017pyl, Ablikim:2019vaj}; first measurements of the $\Sigma$ and $\Xi$ form
%Editor: Please ensure that the intended meaning has been maintained in the above edit.
factors~\cite{Ablikim:2020kqp,Ablikim:2020rwi};
and studies of the enigmatic $Y(2175)$ resonance~\cite{Ablikim:2020pgw,Ablikim:2020coo,Ablikim:2018iyx,Ablikim:2019tpp}.

A data sample of $\sim$20~fb$^{-1}$ integrated luminosity accumulated at a variety of CME values between 4.0 and 4.6~GeV
support detailed studies of charmonium-like $XYZ$ states, including some of BESIII's most remarkable results, such as
the discoveries of the charged charmonium-like states $Z_c(3900)$ and $Z_c(4020)$~\cite{Ablikim:2013mio,Ablikim:2013wzq};
the $Z_{cs}(3985)$, the first example of a charmonium-like state with a nonzero strangeness~\cite{Ablikim:2020Zcs-discovery};
an anomalous line shape for the $Y(4260)$ resonance\footnote{also known as $\psi(4230)$, and was $\psi(4260)$}~\cite{Ablikim:2016qzw}; and a large partial decay width for the
radiative process $Y(4260)\to\gamma X(3872)$~\cite{Ablikim:2013dyn}.

The measurement of the $\tau$-lepton mass by the original BES experiment~\cite{Bai:1992bu} in 1992 yielded a result that was 7~MeV ($\sim$2$\sigma$)
lower than the average of all previous measurements. It clarified what was the major discrepancy with the SM at that
time~\cite{Marciano:1991pr}. Since then, the BES program has made further improved $\tau$ mass measurements; the latest BESIII result
is in good agreement with the original BES measurement but with an order of magnitude better
precision~\cite{Ablikim:2020tau-mass}.

The BEPC/BES program followed by its upgrades has
significantly advanced our understanding of elementary
particle physics.
The primary task of the particle physics community during the next two decades will be to mount a comprehensive challenge
to the SM and to develop an understanding of the laws of nature at a more fundamental level. This will require a
coordinated multidimensional program including
precise predictions for measurable quantities in the framework of the SM.
These predictions, in turn, will need to be compared with experimental
measurements with state-of-the-art sensitivities and well-controlled systematic errors. The physics potential of the
current BEPCII/BESIII program is limited by its luminosity and CME range. Higher luminosities will be crucial for
investigating many of the key questions that can be uniquely addressed in the $\tau$--charm threshold
energy region, such as more precise measurements of the SM's free parameters, a better understanding of the internal
compositions of exotic hadron states such as the $XYZ$ and other charmed mesons and baryons as well as quark--gluon states
and their underlying dynamics, measurements of $CP$ violation~(CPV) in hyperon decays and other systems, $\tau$ physics and
probes for possible new physics beyond the SM. Next-generation studies of charmed baryons, especially the newly
discovered doubly charmed baryon states~\cite{Aaij:2017ueg}, will require an increased CME range. Because of strict spatial
constraints, there is insufficient space on the IHEP campus in Beijing to accommodate an upgrade of BEPCII that would meet
these luminosity and energy goals. As a result, after BEPCII/BESIII completes its mission in the near future, there
will be a need for a new collider with two orders of magnitude higher luminosity and a much broader (by a factor of 2) energy range in order
to continue pursuing and extending the scientific opportunities in the $\tau$--charm region.

The proposed STCF~\cite{Peng:2022loi} is an electron--positron collider with separated electron and positron rings and symmetric beam energy to be constructed in China.
It is designed to have a CME range spanning from 2~to~$\sim$7 GeV, with a peak luminosity of at least
$0.5\times 10^{35}$ cm$^{-2}$s$^{-1}$ optimized for a CME of 4~GeV. In addition to the boost in luminosity, the extended accessible energy
region will provide opportunities to study the recently discovered doubly charmed baryons~\cite{Aaij:2017ueg}. The
proposed design leaves space for higher-luminosity upgrades and for the implementation of a polarized $e^-$
beam in a future phase II~\cite{Luo:2019gri} of the project. To achieve such a high luminosity, several advanced technologies, such as
the introduction of a crabbed-waist beam-crossing scheme with a large-Piwinski-angle interaction region~\cite{crabbedwaist}, will be
implemented in the machine.

Some of the physics that such an STCF could access can also be investigated by the Belle II~\cite{Adachi:2018qme} and LHCb~\cite{Alves:2008zz}
experiments. Detailed descriptions of the physics programs of Belle II and LHCb can be found in Refs.~\cite{Kou:2018napp,Bediaga:2018lhg}, respectively. Both of these experiments can produce more $\tau$ leptons and charmed hadrons and mesons than the STCF. However, STCF
data samples will have distinctly lower backgrounds, near-100\% detection efficiencies, almost full detector-acceptance, better full-event reconstruction
rates, well-controlled systematic uncertainties, etc. The STCF will also have several unique features that are not available at
Belle II and LHCb, including the direct production of $1^{--}$ resonances such as charmonium ($J/\psi$, $\psi(3686)$
and $\psi(3770)$) and nonstandard charmonium-like mesons, such as $Y(4260)$, $Y(4320)$ and $Y(4660)$, as well as operation near
particle--antiparticle thresholds, thus providing the ability to fully reconstruct events with final-state neutrinos,
neutrons/antineutrons or $K_L$ mesons with high efficiency.

%The STCF project is still in its research and development (R\&D) stage.
To achieve these goals, a sophisticated, machine-compatible detector will be required to maximize the physics potential.
The detector is expected to exhibit considerably improved performance in each subsystem compared to the BESIII detector.
%Currently, the STCF detector design, shown in Fig.~\ref{stcf},
%as visualized using DD4hep~\cite{Petric:2017psf}, 
%is still under research and development.
%Several features must be considered for this detector.

%\begin{figure}[htbp]
%\begin{center}
%\includegraphics[width=0.7\textwidth]{Figs_01_Introduction/detector_stcf.jpg}
%\caption{
%    \label{stcf}
%Schematic view of the STCF detector.
%}
%\end{center}
%\end{figure}

%***{\color{red} modify the following to make sure statements are correct}*****

The STCF detector features large solid-angle coverage, low noise, high detection efficiency and resolution and excellent particle identification capabilities.
It is also required to have fast trigger response, high rate capability and high levels of radiation tolerance.
From the interaction point outward, the STCF detector consists of a tracking system, a particle identification (PID) system, an electromagnetic calorimeter (EMC), a superconducting solenoid~(SCS) and a muon detector (MUD), where the tracking system is composed of both inner and outer trackers. 

Among all the subdetectors, the inner tracker is the closest to the interaction point and hence is exposed to the highest level of radiation.
To withstand the high radiation background,
a novel micropattern gaseous detector, based on $\mu$RWELL technology and consisting of three cylindrical layers located 6, 11 and 16 cm away from the interaction point, is proposed as a baseline option for the inner tracker.
A low-mass silicon pixel detector based on the complementary metal-oxide-semiconductor~(CMOS) monolithic active pixel sensor~(MAPS) technology is also considered as an alternative option.
A large cylindrical drift chamber with ultralow material,
%Editor: Please consider replacing 'ultralow material' with 'ultralow-background material', 'ultralow-density material', or another description that better conveys your intended meaning.
spanning from 200 to 820~mm in radius and operating with a helium-based gas mixture, is proposed as the outer tracker. The momentum resolution in a 1 T magnetic field is expected to be better than 0.5\% for charged tracks with a momentum of 1~$\gevc$, and the dE/dx resolution should be better than 6\%, which can be exploited for particle identification for low-momentum charged particles.
The PID system uses two different Cherenkov detector technologies, one in the barrel region and one in the endcap region, to achieve a 3$\sigma$ separation between kaons and pions with a momentum up to 2~$\gevc$. 
%A separation capability of 3$\sigma$ between muons and pions with a momentum between 0.2 and 0.6~$\gevc$ is also available with the PID system.
A homogeneous electromagnetic calorimeter composed of trapezoid-shaped pure CsI crystal scintillators is proposed for the EMC to achieve an excellent energy resolution ($\sim$2.5\% at an energy of 1~$\gev$) and a good position resolution ($\sim$5~mm at an energy of 1~$\gev$) in a high radiation background. The EMC also has timing capability allowing the effective separation of photons from neutrons and $K_{L}^{0}$ in the energy region of interest. 
A SCS magnet surrounding the EMC provides the tracking system with a magnetic field of 1~T.
A hybrid of resistive plate chamber (RPC) with 3 inner layers and plastic scintillator (7 outer layers) detectors is proposed as the MUD,
which is expected to provide an excellent capability to efficiently separate muons from pions with an efficiency of 95\% and a misidentification rate of less than 3\% or even better.
Advanced data acquisition (DAQ) and trigger systems are required to handle a high event rate in the range from 60~kHz to 400~kHz.

%The baseline detector concept has been studied in depth through simulation, and the results demonstrate that the system can deliver the performance necessary to achieve the physics goals of the STCF. Critical R\&D tasks have already been under development, with preliminary and promising results achieved.

The rest of the document is organized as follows: The physics opportunities at the STCF are discussed in Chapter~\ref{CDR_phys}. In Chapter~\ref{CDR_det}, the conceptual designs of the STCF detector system are described. 
In Chapter~\ref{chap_phyper}, performances of several benchmark 
physics processes are introduced, and Chapter~\ref{chap_plan} is
the future plans and R\&D prospects.

%In Section V, several topics related to QCD, such as $R$-value and Collins effect measurements, the $Q^2$ behavior of
%baryon form factors, precision tests of rare/forbidden decays and CP violation in $\eta/\eta'$ and hyperon decays, and
%studies of glueballs and hybrids are discussed. In Section VI, the discovery potentials for new, beyond-the-SM light particles
%are presented. Section VII is a summary.

\begin{chapter}{Physics}
\label{CDR_phys}
\section{Motivation}
\subsection{Challenges in particle physics and the \texorpdfstring{$\tau$}{}-charm energy region}
The Standard Model (SM) of particle physics, comprising the unified electroweak (EW) and quantum
chromodynamics (QCD) theories, successfully explains almost all experimental results related to the
microscopic world. For example, it successfully predicted the existence of weak neutral current interactions
and the masses, widths and many other properties of the $W^{\pm}$ and $Z^0$ bosons; large particle--antiparticle
differences, so-called $CP$ violations, in specific $B$-meson decay channels, which were subsequently
confirmed by experiments; and the existence of the Higgs scalar boson, which was discovered at CERN in
2012~\cite{Aad:2012tfa,Chatrchyan:2012ufa} --- 50 years after its existence was predicted --- with properties that closely match the model's
expectations. As a result, the SM is currently universally accepted as the theory of elementary particles
and their interactions.

However, despite its considerable successes, the SM also has a number of shortcomings, including the following:

\noindent
{\bf Many free parameters} {The SM with the minimal particle contents (the gauge particle, the generations of left-handed quark doublets, the right-handed quark singlet, the left-handed lepton doublets, the right-handed charged leptons, and the Higgs doublet)} has 19 free parameters that must be extracted from experimental measurements. These include the quark, lepton and Higgs masses; the mixing angles of the Cabibbo--Kobayashi--Maskawa (CKM) quark-flavor mixing matrix; and the couplings of the electric, weak and QCD color forces (see Fig.~\ref{fig:SM-free-params}(a)). { Considering the neutrino mass, there are even more free parameters.}

\noindent
{\bf Baryon asymmetry of the universe}  The model's mechanism for $CP$ violation fails to explain our existence
in a matter-dominated universe by approximately ten orders of magnitude. There must be additional $CP$-violating
mechanisms in nature beyond those contained in the SM.

\noindent
{\bf Quark/gluon--hadron disconnect}  The strongly interacting particles of the SM are quarks and gluons,
whereas the strongly interacting particles that are measured in experiments are hadrons. In principle, QCD
accurately describes the transitions between quarks and hadrons. However, at the relevant distance scale
on the order of 1~fm (see Fig.~\ref{fig:SM-free-params}(b)), QCD is a strongly coupled theory, and perturbation theory is not directly applicable.
As a result, the spectrum and properties of
various particles that are seen in experiments haven't been well understood in theory.

\begin{figure}[htbp]
\begin{center}
\begin{overpic}[width=15cm, height=6.cm, angle=0]{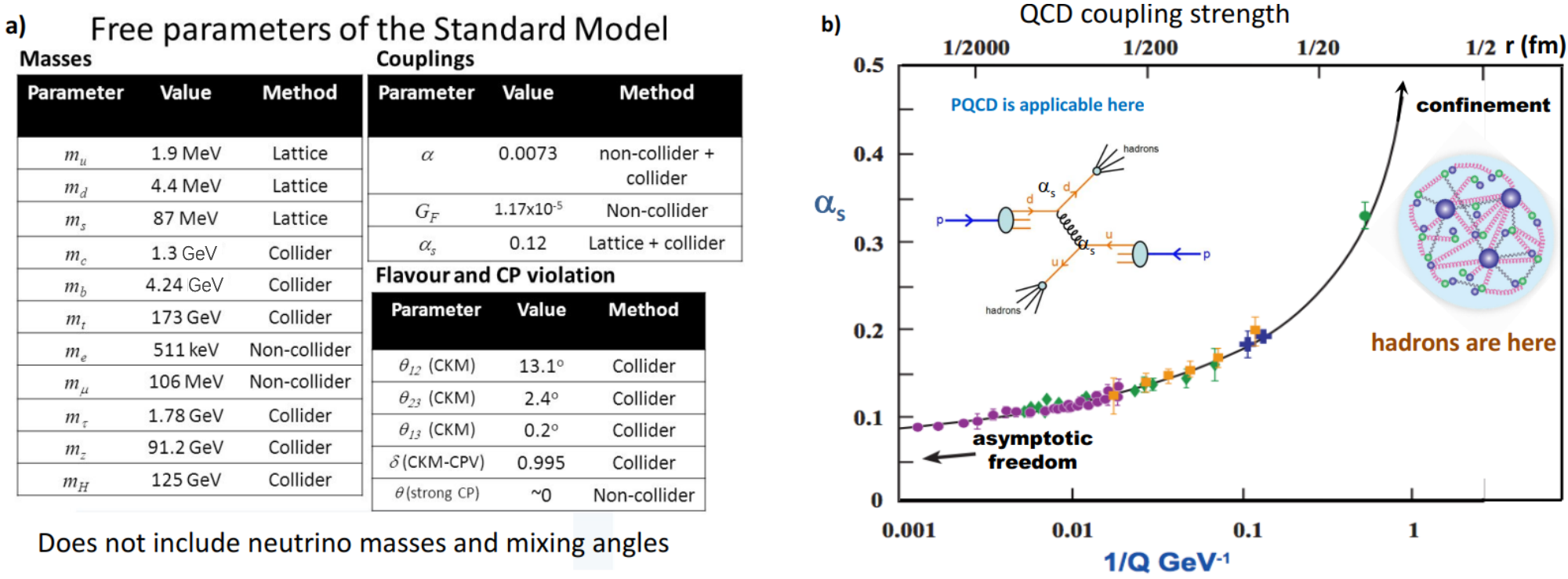}
\end{overpic}
\caption{\footnotesize
{\bf a)} The free parameters of the Standard Model. Note that the neutrino masses and mixing angles
are not included here. {\bf b)} The behavior of the QCD coupling strength $\alpha_s(Q)$ {\it vs.} $1/Q$ (bottom
axis) and distance (top axis).
}
\label{fig:SM-free-params}
\end{center}
\end{figure}

\noindent
{\bf Gravity, dark matter, neutrino masses, number of flavors, etc.} {The SM does not include a quantized theory for gravity. It does not explain the dark matter in the universe, the neutrino masses, the number of particle generations, etc. Because of these shortcomings, the SM cannot be taken as a perfect theory. }

There is considerable enthusiasm among particle physicists to search for evidence
of new, non-SM physics phenomena. In general, this requires evaluating and testing SM predictions with
ever-increasing energies and/or levels of precision. For example, at the {\em energy frontier}, the ATLAS and CMS
experiments at CERN search for new massive particles that are not accounted for in the SM and make precise
determinations of Higgs decay coupling strengths to search for deviations from SM predictions. At the {\em intensity
frontier}, the LHCb experiment at CERN, the recently commissioned Belle II experiment at KEK, and long-baseline
neutrino experiments are searching for evidence
%of non-SM sources of $CP$ violation in the $b$- and $c$-quark and neutrino sectors, and
of deviations from SM predictions for processes that are mediated by quantum loops containing
massive virtual particles, which could be signs of the influence of non-SM particles that are too heavy to be
accessed by experiments at LHC energies. In lower energy regions, the tests of the SM include precision measurements of
$(g - 2)_\mu$; tests of the unitarity of the CKM matrix, especially the $CP$ triangle; {searches for non-SM sources of $CP$ violation in neutrinos}; searches for lepton flavor
violations; investigation of the consistency among various approaches for determining the value of the Weinberg weak interaction angle $\theta_W$;
etc. These, in general, require precisely measured values of SM parameters, usually as inputs from other sources, and
independent determinations of the influence of long-distance hadron effects. Thus, a comprehensive challenge to the
SM requires a coordinated multidimensional program that includes careful refinement of theoretical predictions
coupled with experimental measurements with state-of-the-art sensitivities.

At low energies, some of the nonperturbative effects of QCD have an important influence on the determination of
fundamental parameters. An electron--positron ($e^+e^-$) collider operating at the transition interval between
nonperturbative QCD and perturbative QCD at the few-GeV level --- a $\tau$--charm facility ---
is uniquely well suited to play an important role in the determination of these parameters. Such a facility can
address a very broad physics program covering tests of QCD, investigations of hadron spectroscopy, precise tests of electroweak interactions,
and searches for new, beyond-the-SM physics. Currently, the only facility operating in this energy region is the Beijing
Electron--Positron Collider II (BEPCII) -- BEijing Spectrometer III (BESIII)~\cite{bepcii,Ablikim:2009aa}, which has significantly advanced progress
in elementary particle physics. A comprehensive description of the physics program and potential of BESIII can be found in Refs.~\cite{Asner:2008nq,beswhite}. BESIII has been in operation for more than 10 years and will complete its mission soon. An advanced facility that will continue investigating and extend
to additional research topics in the relevant energy region with significantly enhanced sensitivity is therefore necessary to address many of the remaining unsolved problems. A Super
$\tau$--Charm Facility (STCF), with a luminosity higher by two orders of magnitude,
%Editor: Please ensure that the intended meaning has been maintained in the above edit. Alternatively, you may mean 'with a luminosity higher by a factor of two'.
would be a natural extension and a viable option.
The successful construction and operation of the proposed STCF would play a crucial role in continuing China's leading worldwide role in
research at the high-intensity frontier of elementary particle physics.

%The SM has 19 free parameters that must be supplied by precision experiments. The discoveries of the
%charmed mesons and the $\tau$ lepton as well as the successful mapping of the charmonium meson spectrum in the 20th century
%opened a window to the rich physics program that is addressable in the $\tau$--charm region. The original BES
%experiment's determination of the $\tau$-lepton mass provided a stringent test of lepton flavor universality in combination with other measurements, such as couplings,
%and precision measurements of the cross section for $e^+e^-$ annihilation into hadrons by BESII were crucial inputs
%for fits to the electroweak sector of the model that provided an accurate prediction of the Higgs boson mass, prior
%to its discovery in 2012. BESIII has also made the world's most precise measurements of the $|V_{cs}|$ and $|V_{cd}|$
%elements of the Kobayashi--Maskawa quark-flavor mixing matrix. A deeper understanding of the underlying theory
%of the fundamental interactions will require even more precise measurements of all of these parameters.
%An STCF will contribute significantly to this goal.

%\input{01_01_TCF-slo_v2}
%\input{01_02_STCF}
% \input{01_03_DataSample}
%\input{01_04_Systematics-slo}
\subsection{Physics potential at the STCF}

An STCF operating at CMEs ranging from 2 to $\sim$7~GeV would be of great importance to the entire field of elementary particle physics. It would address a very broad range of physics topics,
including QCD tests, hadron spectroscopy, precise tests of the electroweak sector of the SM, and searches for new physics beyond the SM.
The proposed luminosity of the STCF is above $\stcflum$; at this level, it is expected to deliver more than 1~$\invab$ of data
samples each year. A possible data-taking plan for the STCF, along with the expected numbers of conventional events and/or particles, is shown in Table~\ref{tablelumi}, where the number of events at different CME is calculated based on 1~$\invab$ integrated luminosity. It should be noted that the final baseline logic for the data-taking plan needs to be fine-tuned depending on the scan results of the STCF CME range.

\par
\begin{table}[htbp]
\begin{center}
\caption{The expected numbers of events per year at different STCF energy points.}
\label{tablelumi}
\begin{tabular}{c|c|c|c|c| c}
     \hline \hline
     CME ({\rm GeV})   & Lumi ($\invab$)   &  ~~~~~~~Samples~~~~~~~ & $ ~~~\sigma ({\rm nb})~~~$  & ~No. of Events~ & Remarks \\ \hline
       3.097                  & 1                  &  $\jpsi$                &  3400 & $3.4\times 10^{12}$   \\ \hline
       3.670                  & 1                  &  $\tau^+\tau^-$         &  2.4  & $2.4\times 10^{9} $   \\ \hline
       \multirow{3}{*}{3.686} & \multirow{3}{*}{1} &  $\psip $               &  640  & $6.4\times 10^{11}$  \\
                              &                    &  $\tau^+\tau^-$         &  2.5  & $2.5\times 10^{9} $   \\
                              &                    &  $\psip\to \tau^+\tau^-$&       & $2.0\times 10^{9} $   \\ \hline
       \multirow{5}{*}{3.770} & \multirow{5}{*}{1} &  $\DDzbar $             &  3.6  & $3.6\times 10^{9} $   \\
                              &                    &  $\DpDm   $             &  2.8  & $2.8\times 10^{9} $   \\
                              &                    &  $\DDzbar $             &       & $7.9\times 10^{8} $ & Single tag \\
                              &                    &  $\DpDm   $             &       & $5.5\times 10^{8} $ & Single tag \\
                              &                    &  $\tau^+\tau^-$         & 2.9   & $2.9\times 10^{9} $\\  \hline
       \multirow{4}{*}{4.009} &  \multirow{4}{*}{1}&  $D^{*0}\bar{D}^{0}+c.c$  & 4.0  & $1.4\times 10^{9} $ & $\rm CP_{\DDzbar}=+$ \\
                              &                    & $D^{*0}\bar{D}^{0}+c.c$   & 4.0  & $2.6\times 10^{9} $ & $\rm CP_{\DDzbar}=-$ \\
                              &                    &  $\Dsp\Dsm$             & 0.20  & $2.0\times 10^{8} $ &       \\
                              &                    &  $\tau^+\tau^-$         & 3.5   & $3.5\times 10^{9} $ \\  \hline
       \multirow{3}{*}{4.180} &  \multirow{3}{*}{1}&  $\Dsps\Dsm$+c.c.       & 0.90  & $9.0\times 10^{8} $      \\
                              &                    &  $\Dsps\Dsm$+c.c.       &       & $1.3\times 10^{8} $ & Single tag          \\
                              &                    &  $\tau^+\tau^-$         & 3.6   & $3.6\times 10^{9} $  \\  \hline
       \multirow{3}{*}{4.230} &  \multirow{3}{*}{1}&  $\jpsi\pppm$           & 0.085 & $8.5\times 10^{7} $  \\
                              &                    &  $\tau^+\tau^-$         & 3.6   & $3.6\times 10^{9} $  \\
                              &                    &  $\gamma X(3872)$       &       &                      \\ \hline
       \multirow{2}{*}{4.360} &  \multirow{2}{*}{1}&  $\psip\pppm$           & 0.058 & $5.8\times 10^{7} $\\
                              &                    &  $\tau^+\tau^-$         & 3.5   & $3.5\times 10^{9} $  \\  \hline
       \multirow{2}{*}{4.420} &  \multirow{2}{*}{1}&  $\psip\pppm$           & 0.040 & $4.0\times 10^{7} $ \\
                              &                    &  $\tau^+\tau^-$         & 3.5   & $3.5\times 10^{9} $   \\  \hline
       \multirow{2}{*}{4.630} & \multirow{4}{*}{1} &  $\psip\pppm$            & 0.033& $3.3\times 10^{7} $     \\
                              &                    &  $\Lambda_c\bar\Lambda_c$& 0.56 & $5.6\times 10^{8} $  \\
                              &                    &  $\Lambda_c\bar\Lambda_c$&      & $6.4\times 10^{7} $ & Single tag \\
                              &                    &  $\tau^+\tau^-$          & 3.4  & $3.4\times 10^{9} $   \\  \hline
       4.0--7.0                & 3                  &  \multicolumn{4}{c}{300-point scan with 10~MeV steps, 1 $\invfb$/point}   \\
        $>5$                  & 2--7                &  \multicolumn{4}{c}{Several $\invab$ of high-energy data, details dependent on scan results} \\

   \hline\end{tabular}
\end{center}
\end{table}

\par

B-factory experiments and BESIII have found a striking failure of the charmonium model to provide an explanation
of the spectrum of hidden charm states with masses above $2m_D = 3.73$~GeV, which is a threshold for open charm meson
production. In addition to some conventional $c\bar{c}$ charmonium states, a larger number of unexplained
charmonium-like meson states, the so-called $XYZ$ states, with masses in the 3.8$\sim$5 GeV mass region, have been discovered.
These discoveries underline a glaring weakness of the SM: the lack of understanding of how QCD, the strong interaction sector of the theory that deals only with quarks and gluons, explains experimental data that involve only hadrons. In addition,
after a decade of searches, strong candidates for light non-$q\bar q$ hadrons such as glueballs and $q\bar q$--gluon QCD
hybrids with exotic spin-parity quantum numbers $J^{PC} = 1^{-+}$ have been found in large samples of radiative
$J/\psi$ decays. Both the $XYZ$ and exotic light hadrons point to entirely new hadron spectra
%Editor: Please ensure that the intended meaning has been maintained in the above edit.
that must be explored
and understood. At the moment, many of the properties of the $XYZ$ particles are unknown, and there is no clearly
identifiable pattern to the $XYZ$ particle spectrum. In certain circumstances, it is even unclear whether the
$XYZ$ resonance signals are partially or totally produced by kinematic singularities. These uncertainties prevent
us from obtaining an unambiguous mass spectrum and obscure insight into the inner structure of the $XYZ$ particles.
At the STCF, not only can large data samples of conventional particles be collected, as summarized in Table~\ref{tablelumi}, but
copious $XYZ$-particle event samples will also be produced; the expected event numbers for some of the $XYZ$ states are given
in Table~\ref{tableXYZ}. These large data samples will enable detailed studies of the properties of the $XYZ$ states through precisely studying Argand plots, searching for rare decays, and precisely measuring masses and widths, which will lead
to more conclusive results.

\begin{table}
\begin{center}
\caption{The expected numbers of produced $XYZ$-particle events before reconstruction per year at the STCF.}
\label{tableXYZ}
\begin{tabular}{c c|c|c|c|cl}
     \hline \hline
     & XYZ  &  $Y(4260)$  &  $Z_c(3900)$ & $ Z_c(4020) $   & $X(3872)$  \\ \hline
     & No. of events  & $10^{9}$                  &  $ 10^8$                &  $10^8$ & $5\times10^6$   \\ \hline
   \hline\end{tabular}
\end{center}
\end{table}

In addition to mesons containing a charmed--anticharmed quark pair, new heavy baryons containing a charmed quark and
doubly charmed baryons have been discovered, opening the way to new territories for QCD spectroscopic studies. A comprehensive
portfolio of high-precision and comprehensive measurements of these spectra could challenge and calibrate predictions from LQCD,
which is rapidly emerging as a powerful theoretical tool for performing precision first-principles QCD calculations for
long-distance phenomena. The STCF's high luminosity will help us complete the task of constructing a comprehensive and
precise spectrum of these hadrons. The extension of the STCF's high-energy coverage to approximately 7 GeV is motivated by the need
to understand the dynamics of these doubly charmed heavy baryons.

%%%

\par
With the ability to produce the large data samples indicated in Table~\ref{tablelumi}, the STCF will serve as an ideal facility
for studies of the physics of charmed hadron decays. A large $D$-meson production rate will support rigorous tests
of the SM. For example, purely leptonic decays of tagged $D^\pm$ and $D_s^{\pm}$ mesons produced in large numbers at the $\psi(3770)$
and $\psi(4040)$ (or $\psi(4160)$) resonances would enable precise measurements of the $|V_{cd}|$ and $|V_{cs}|$ matrix elements to test
the second-row unitarity of the CKM matrix and uniquely address the Cabibbo angle anomaly, i.e., the $\sim 4\sigma$ discrepancy
in the $\theta_c$ values measured in different processes~\cite{Grossman:2019bzp}. In addition, the $D^0-\bar D^0$ mixing parameters could be measured with
significantly improved precision. Measurements of and searches for rare and forbidden decays with improvements of up to two orders of magnitude
in sensitivity could be realized as part of a search for new physics.

The $\tau$, as the heaviest charged lepton, occupies a unique place in the SM. It has more decay channels than the muon and thus can provide unique access to new physics beyond the SM. At the STCF, the number of accumulated $\tau^+\tau^-$ pair events
will be approximately three orders of magnitude higher than the currently accumulated number of such events at BESIII. As many as a few billion $\tau$ pairs could be obtained in a one-year run at the CME$=2m_{\tau}$ threshold. Operation near the threshold would provide
the STCF with unique advantages over Belle II~\cite{Kou:2018napp} and LHCb~\cite{Bediaga:2018lhg}, even though the latter would have larger $\tau$-pair event samples.
For example, these events, together with well-controlled background studies using data accumulated just below the threshold, would be uniquely well suited for a high-sensitivity
study of the anomalous ($\sim$3$\sigma$) sign of $CP$ violation in $\tau \to K_S \pi \nu_\tau$ decays that was reported by
BaBar~\cite{BABAR:2011aa}.
%@article{BABAR:2011aa,
%    author = "Lees, J.P. and others",
%    collaboration = "BaBar",
%    title = "{Search for CP Violation in the Decay $\tau^- -> \pi^- K^0_S (>= 0 \pi^0) \nu_tau$}",
%    eprint = "1109.1527",
%    archivePrefix = "arXiv",
%    primaryClass = "hep-ex",
%    reportNumber = "SLAC-PUB-14556, BABAR-PUB-11-009",
%    doi = "10.1103/PhysRevD.85.031102",
%    journal = "Phys. Rev. D",
%    volume = "85",
%    pages = "031102",
%    year = "2012",
%    note = "[Erratum: Phys.Rev.D 85, 099904 (2012)]"
%}
Another unique advantage of $\tau$ pairs that are produced near the threshold is that they are primarily
produced in an $S$-wave, and thus, if the electron beam is polarized, this polarization translates nearly 100\% into
a well-understood polarization of the two final-state $\tau$ leptons~\cite{Tsai:1994rc}.
Therefore, operation of the STCF with a polarized electron beam just above the $\tau$-pair threshold would enable a high-sensitivity
search for $CP$-violating asymmetries in $\tau^{\mp}\to\pi^{\mp}\pi^0\nu$ decays~\cite{Tsai:1994rc}.
%@article{Tsai:1994rc,
%    author = "Tsai, Yung Su",
%    title = "{Production of polarized tau pairs and tests of CP violation using polarized e+- colliders near threshold}",
%    eprint = "hep-ph/9410265",
%    archivePrefix = "arXiv",
%    reportNumber = "SLAC-PUB-6685",
%    doi = "10.1103/PhysRevD.51.3172",
%    journal = "Phys. Rev. D",
%    volume = "51",
%    pages = "3172--3181",
%    year = "1995"
%}
The same data sample would also enable better determinations of the SM $\tau$-lepton parameters and stringent tests of
the lepton-flavor universality of weak interactions and might reveal possible clues toward the understanding and
study of $g-2$ for the $\tau$, which may shed light on the anomaly in $g-2$ for the muon.

The large matter--antimatter asymmetries in the $b$-quark sector observed by $B$-factory experiments confirmed
the CKM ansatz as the SM mechanism for $CP$ violation. This model can also explain the $CP$ violations that were first observed
in neutral kaon mixing and kaon decays into two and three pions. However, this mechanism fails to explain the baryon asymmetry
of the universe by approximately ten orders of magnitude, which strongly suggests the presence of additional, non-SM $CP$-violating
interactions. Promising channels for searching for new sources of $CP$ violation include the weak decays of the $\Lambda$ and $\Xi$
hyperons, where SM-CPV effects are small but effects of new, beyond-the-SM interactions could be large~\cite{Adlarson:2019jtw}. These measurements can be elegantly
done with high-statistics samples of quantum-entangled hyperon--antihyperon pair events produced via
$J/\psi \to \Lambda \bar \Lambda$ and $\Xi \bar \Xi$ decays. A one-year STCF run at the $J/\psi$ resonance
would produce data samples of 160M (60M) fully reconstructed $J/\psi \to \Lambda \bar \Lambda$ ($\Xi \bar \Xi $) events
and more than an order-of-magnitude improvement over the BESIII $CP$ sensitivity. With $\sim$80$\%$ electron beam polarization,
this sensitivity would be improved by an additional factor of four.

BESIII measurements demonstrate that the hadronic final states produced in radiative $J/\psi$ decays are replete with
QCD hybrids and glueballs and are ideally well suited for studying the spectra of these mostly unexplored systems.
Searches for anomalous weak decays of the $J/\psi$ at the STCF would have sensitivities extending all the way down to the
level of SM expectations. With STCF data running at a variety of energies, interesting $Q^2$-dependent quantities could
be studied with high precision. These include time-like nucleon form factors, which
could be measured for $Q^2$ values as high as 50~GeV$^2$ with the best precision matching those of the existing measurements in the space-like
region. The puzzling threshold behavior and peculiar oscillation patterns observed in recent low-statistics experiments could
be studied in precise detail. Moreover, unlike space-like form-factor measurements, which are possible only for the proton
and neutron, such time-like form-factor studies could be repeated for the $\Lambda$, $\Sigma$, $\Xi$ and $\Omega^-$ strange
hyperons, offering a unique new window on the baryon structure.
With this very high luminosity, high-sensitivity searches for new light particles and new interactions that are predicted by
a number of beyond-the-SM theories could be performed using decays of all of the weakly and electromagnetically decaying
particle systems that are accessible in the STCF energy range.

In short, the STCF will undoubtedly cover a very broad physics program %covering QCD tests, hadron spectroscopy, precisely tests of electroweak interactions of the SM and hunt for new physics beyond.
and could support a multidimensional program of experimental
measurements with state-of-the-art sensitivities, allowing many of the challenges of the SM to be addressed.
In the following chapters, more details on some of the highlighted physics topics that could be addressed at the STCF
are provided. These materials include discussions of research opportunities regarding particles ranging from the high-mass $XYZ$
states to low-mass systems such as hyperons, glueball/hybrid states, and possible new, beyond-the-SM light particles. Some
potential studies that could extract important SM information for nonresonance energies will also be presented. Studies of decays and
interactions can provide essential information to both flesh out the SM and search for clues toward new physics beyond the SM.
In addition to spectroscopic issues, the precision of the determination of strong- and weak-interaction parameters, the sensitivities
of measurements of and searches for rare and forbidden decays and $CP$-violating asymmetries, and how new particles
and new, beyond-the-SM interactions might manifest are discussed.
%It is hoped that such a coordinated multi-dimensional program at STCF will enable us to have a much
%more in-depth understanding challenges facing the SM and hopefully to provide some solutions to them.

%In Section II, the charmonium and $XYZ$-meson systems are discussed, with an emphasis on opportunities for solving the
%$XYZ$ puzzle and the discovery of other higher charmonium states. In Sections III and IV, charmed meson and baryon physics
%and tau physics are discussed, including the determination of SM parameters as well as searches for rare and forbidden decays and
%$CP$ violations.
%In Section V, several topics related to QCD, such as $R$-value and Collins effect measurements, the $Q^2$ behavior of
%baryon form factors, precision tests of rare/forbidden decays and CP violation in $\eta/\eta'$ and hyperon decays, and
%studies of glueballs and hybrids are discussed. In Section VI, the discovery potentials for new, beyond-the-SM light particles
%are presented. Section VII is a summary.

%{\color{red} References need more work!!!!}

\newpage
\section{Charmonium and XYZ Physics}
\label{sec:charmonium}
%At STCF it is an ideal place to study charmonium states and the exotic states containing a $c\bar c$ pair.
%For these states with the quantum number $J^{PC} = 1^{--}$ they can be copiously produced in their threshold
%energy regions. For states with other quantum numbers they can be searched in certain decay products.
%With the design luminosity, large event numbers of the states with the quantum numbers other than $1^{--}$
%can be expected. With the energy regions of STCF, a systematic study of highly excited charmonium states
%and $XYZ$-states can be performed with the statistics never reached before. The mass spectrum of charmonia below the $D\bar D$ threshold has been successfully described with the quark model with a confining potential between a charm- and an anti-charm quark. However, many excited charmonium
%states predicted by the model are still missing, or their properties are poolly known.
%The $XYZ$ states discovered in the last decade can not be predicted by the quark model, or it has difficulties
%to explain the existence of $XYZ$ states. It is unclear if they are hadronic molecule, tetraquark states, hadro-charmonia and threshold effects. Experimentally, the answers of these questions can be found at STCF
%combined with studies of theory.

\subsection{The \texorpdfstring{$XYZ$}{} puzzles}

The STCF is an ideal place to study charmonium states and exotic states containing a $c\bar c$ pair.
Before 2003, it was thought that charmonium states, being bound states of a charm and an anticharm quark, should be well described by nonrelativistic potential quark models.
However, since the discovery of the $X(3872)$ by Belle in 2003, a large number of new resonance(-like) structures have been observed in the charmonium mass region by various experiments, including BESIII, BaBar, Belle, CDF, D0, ATLAS, CMS and LHCb (see, e.g., Refs.~\cite{Swanson:2006st,Voloshin:2007dx,Chen:2016qju,Hosaka:2016pey,Lebed:2016hpi,Esposito:2016noz,Guo:2017jvc,Ali:2017jda,Olsen:2017bmm,Karliner:2017qhf,Yuan:2018inv,Kou:2018napp,Cerri:2018ypt,ParticleDataGroup:2022pth,Liu:2019zoy,Brambilla:2019esw,Guo:2019twa,Chen:2022asf} for recent reviews), as shown in Fig.~\ref{fig:ccbarspec} in comparison with the predictions of the Godfrey--Isgur quark model~\cite{Godfrey:1985xjl}. Most of them have peculiar features that deviate from quark model expectations:
\begin{itemize}
\item The masses are a few tens of MeV away from the quark model predictions for charmonia with the same quark numbers and cannot be easily accommodated in quark model spectra. Examples include the $X(3872)$, $Y(4260)$, and $Y(4360)$ masses; see Fig.~\ref{fig:ccbarspec}.
\item All of the $XYZ$ states are above or at least in the vicinity of the open-charm thresholds. For those above the thresholds, one would expect them to predominantly decay into open-charm channels because of the OZI rule. However, many of them have only been seen as peaks in final states of a charmonium and light mesons/photon. For instance, four resonant structures have been observed in the $J/\psi\phi$ final states, namely, $X(4140)$, $X(4274)$, $X(4500)$ and $X(4700)$, while no $XYZ$ signal was reported in open-charm channels.
\item Charged structures have been observed, including the $Z_c(3900)$, $Z_c(4020)$, $Z_c(4050)$, $Z_c(4250)$, $Z_c(4200)$ and $Z_c(4430)$. More recently, charged $Z_{cs}$ structures with explicit strangeness have also been reported~\cite{Ablikim:2020hsk,Aaij:2021ivw}. If they are hadron resonances, they must contain at least four quarks, making them explicitly exotic multiquark states beyond the conventional quark model.
\end{itemize}
Because of these features, $XYZ$ particles are excellent candidates for exotic hadrons, which include hadronic molecules, tetraquarks, hadro-charmonia and hybrids in this context and have been sought for decades.

%%%%%%%%%%%%%%%%%%% Fig 3 %%%%%%%%%%%%%%%%%%%%%%%%
\begin{figure*}[t]
	\centering
	\includegraphics[width=\textwidth]{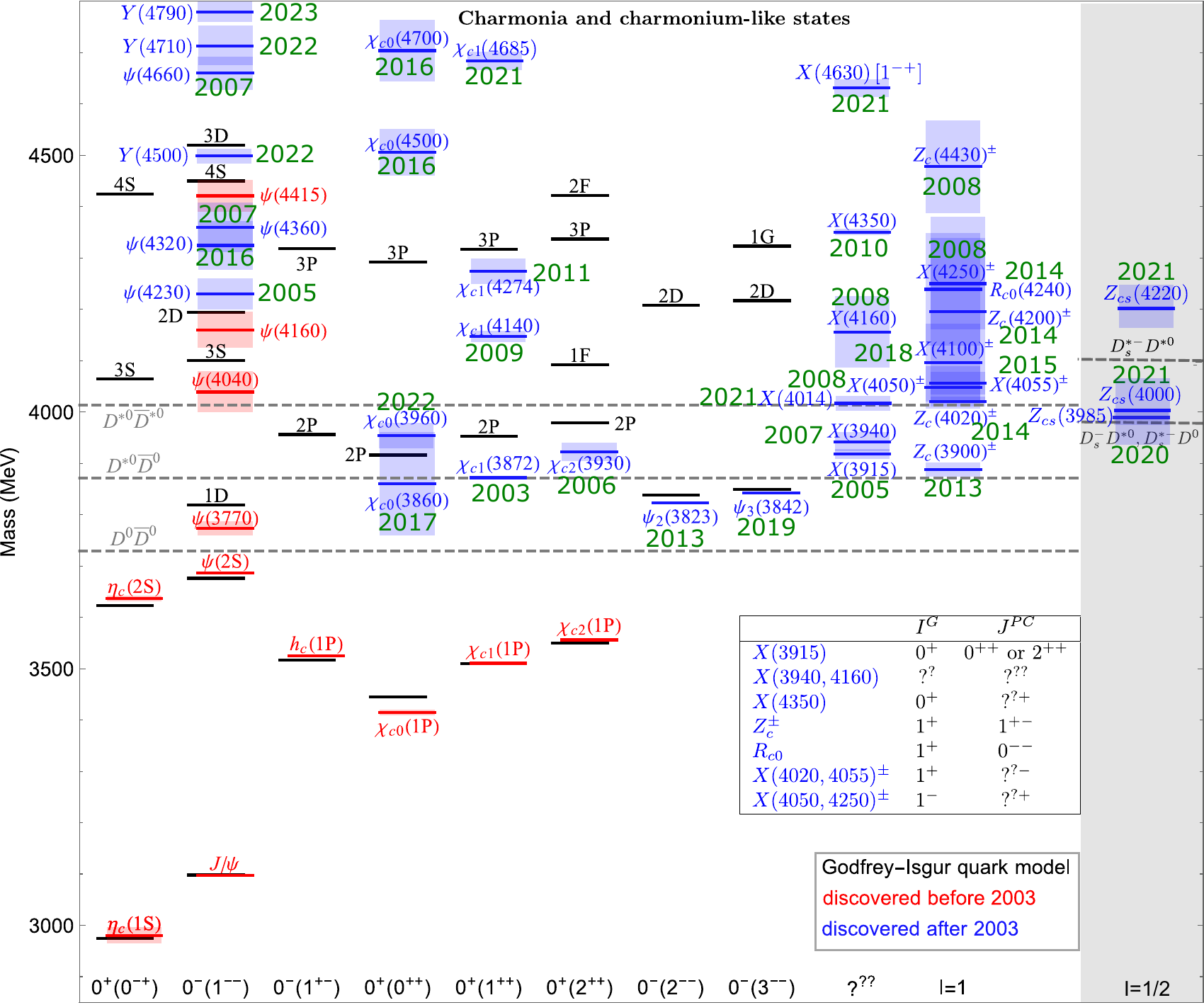}
\vspace{0cm}
\caption{The mass spectrum of charmonia and $XYZ$ states in comparison with the predictions from the Godfrey--Isgur quark model~\cite{Godfrey:1985xjl}.
		\label{fig:ccbarspec}}
\end{figure*}
%%%%%%%%%%%%%%%%%%%%%%%%%%%%%%%%%%%%%%%%%%%%%%%%%%

Within the energy region of the STCF at the designed luminosity, a large number of states with the quantum numbers $J^{PC} = 1^{--}$ can be produced. States with other quantum numbers can also be searched for in certain decay products.
A systematic study of the intriguing $XYZ$ states and related highly excited charmonium states can thus be performed with unprecedented statistics. With collaborative inputs from theory and lattice QCD~(LQCD), answers to the $XYZ$ puzzles and a deeper understanding of how color confinement organizes the QCD spectrum are foreseen.

\subsection{Limitations of current experiments}

\begin{table}[tb]
%\small
  \caption{\label{tab:XYZ} Some of the $XYZ$ states in the charmonium mass region as well as the observed production processes and decay modes. For the complete list and more detailed information, we refer to the latest version of the Review of Particle Physics (RPP)~\cite{ParticleDataGroup:2022pth}. }
\centering
% \begin{tabularx}{\textwidth}{l c c c }
%\begin{ruledtabular}
\begin{tabular}{l c c c}
% \hline
     $XYZ$ & $I^G(J^{PC})$ & Production processes
            & Decay modes % & Experiments
                 \\\hline
           \multirow{2}{*}{$X(3872)$} & \multirow{2}{*}{$0^+(1^{++})$} & $B \to K X/ K\pi X$, $e^+e^-\to\gamma X$,  & $\pi^+\pi^-J/\psi$, $\omega J/\psi$, $D^{*0}\bar{D}^0$, \\
           & &  $pp/p\bar p$ inclusive, PbPb, $\gamma\gamma^*$ &  $\gamma J/\psi$, $\gamma \psi(3686)$
                %\multirow{4}{*}{$X(3872)$}
                    \\ \hline
    $X(3915)$ & $0^+(0~\text{or}~2^{++})$ & $B \to K X$, $\gamma\gamma\to X$ & $\omega J/\psi$ \\ \hline
    $X(4140)$ & $0^+(1^{++})$ & $B \to K X$, $p\bar p$ inclusive & \multirow{4}{*}{$\phi J/\psi$} \\
    $X(4274)$ & $0^+(1^{++})$ & \multirow{3}{*}{$B \to K X$} & \\
    $X(4500)$ & $0^+(0^{++})$ & & \\
    $X(4700)$ & $0^+(0^{++})$ & & \\ \hline
   $X(3940)$ & $?^?(?^{??})$ &  \multirow{2}{*}{$e^+e^- \to J/\psi + X$} & $D {\bar D}^*$\\
  $X(4160)$ & $?^?(?^{??})$ &      & $D^* {\bar D}^*$\\ \hline
  $X(4350)$ & $0^+(?^{?+})$ & $\gamma\gamma \to X$ & $\phi J/\psi$\\ \hline\hline
    $Y(4008)$ & $0^-(1^{--})$  &  $e^+e^- \to Y$  & $\pi\pi J/\psi$ \\ %\hline
  $Y(4260)$ & $0^-(1^{--})$ & $e^+e^- \to Y$ & $\pi\pi J/\psi$, $D\bar D^*\pi$,$\chi_{c0}\omega$, $h_c\pi\pi$ \\ \hline
   $Y(4360)$ & $0^-(1^{--})$ &  \multirow{2}{*}{$e^+e^- \to Y$} & $\pi \pi \psi(3686)$ \\
  $Y(4660)$ & $0^-(1^{--})$ & & $\pi \pi \psi(3686)$, $\Lambda_c\bar \Lambda_c$ \\ \hline\hline
  $Z_c(3900)$ & $1^+(1^{+-})$ & $e^+e^-\to \pi Z_c$, inclusive $b$-hadron decays & $\pi J/\psi$, $D\bar D^*$ \\
  $Z_c(4020)$ & $1^+(?^{?-})$ & $e^+e^-\to \pi Z_c$  & $\pi h_c$, $D^*\bar D^*$ \\\hline
    $Z_1(4050)$ & $1^-(?^{?+})$ & \multirow{2}{*}{$B \to K  Z_c$} & \multirow{2}{*}{$\pi^\pm \chi_{c1}$}\\
   $Z_2(4250)$ & $1^-(?^{?+})$ & & \\ \hline
   $Z_c(4200)$ & $1^+(1^{+-})$ & \multirow{2}{*}{$B \to K Z_c$} & $\pi^\pm J/\psi$ \\
    $Z_c(4430)$ & $1^+(1^{+-})$ & & $\pi^\pm J/\psi$, $\pi^\pm \psi(3686)$ \\ \hline
    $Z_{cs}(3985)$ & $\frac12(?^?)$ & $e^+e^-\to K Z_{cs}$ & $\bar D_s D^*$, $\bar D_s^* D$\\
    $Z_{cs}(4000)$ & $\frac12(1^+)$ & $B^+\to \phi Z_{cs}$ & $J/\psi K$\\
    $Z_{cs}(4220)$ & $\frac12(1^+)$ & $B^+\to \phi Z_{cs}$ & $J/\psi K$
\end{tabular}
%\end{ruledtabular}
\end{table}
Most of the $XYZ$ states reported thus far, together with their observed production processes and decay modes, are listed in Table~\ref{tab:XYZ}. There are several basic types of production processes: $e^+e^-$ collisions, including the direct production and initial-state radiation (ISR) processes; $B$ decays with a kaon in the final state; $pp$ or $p\bar p$ collisions; photon--photon fusion and heavy-ion production. The first two are the main types of interest because they have cleaner backgrounds than hadron collisions and higher rates than photon--photon fusion processes. However, they need to be improved upon in the following aspects:
\begin{itemize}
\item $B\to K X$: The maximum mass of the $X$ or $Z_c$ states that can be found via this type of reaction is approximately 4.8~GeV, the mass difference between the $B$ meson and the kaon. To date, the heaviest charmonium-like state that has been observed is the $X(4700)$. Similarly, the $P_c$ states are difficult to be effectively study in $B$ decays. Additional complexity comes from the fact that these charmonium-like states were all observed as invariant mass distribution peaks in final states with two or more hadrons. Consequently, further complications arise in analyzing data coming from 1) resonances from cross channels and 2) possible triangle singularities (see Ref.~\cite{Guo:2019twa} for a review). Thus, the structures observed in the $B$ decays need to be further confirmed in other reactions, such as $e^+e^-$ collisions.
\item $e^+e^-$ collisions: Charmonia and charmonium-like states with vector quantum numbers can be easily produced either directly or via ISR processes. As a result, the $Y(4260)$ has been studied with unprecedented precision at BESIII. The heaviest vector $Y$ states is the $Y(4660)$, which is above the $\Lambda_c\bar \Lambda_c$ threshold. %which is beyond the current energy range of BES-III.
Charmonium-like states with other quantum numbers can only be produced from the decays of heavier vector states, along with the emission of pions or a photon. Thus, BESIII has observed only the $X(3872)$, $Z_c(3900)$, $Z_c(4020)$ and $Z_{cs}(3985)$ among the many nonvector states.
\end{itemize}

\subsection{Opportunities for solving the $XYZ$ puzzles}

To date, no clear pattern has emerged for the complicated spectrum of the $XYZ$ states. To establish a pattern such that the $XYZ$ states can be classified, more measurements will be necessary, including searches for new charmonium-like structures. There are a few guidelines for possible measurements: 1) No matter what kind of internal structure the states have, there should be partners in the same heavy quark spin multiplet~\cite{Cleven:2015era}, which need to be searched for.
Complications arise from the mixing of them and their partners with spin multiplets of other structures (such as $c\bar c$) with the same quantum numbers. These can only be sorted out through observations, and this can only be done with sufficient measurements.
For instance, the $(0^{++}, 1^{++}, 2^{++}, 1^{+-})$ states have the $J^{PC}$ quantum numbers of $P$-wave $c\bar c$. Thus, the states with these quantum numbers having masses of approximately 3.9~GeV need to be systematically studied in decays to as many final states as possible. 2) It will be important to disentangle the contributions of kinematic singularities from those of resonances in order to establish the correct mass spectrum, and thus, the energy dependence of structures such as the $Z_c$ and $Z_{cs}$ will need to be measured. 3) Some of the structures that have been reported have similar masses and might have the same origin. To check this, it will be important to search for them in other channels and to measure their properties more precisely. 4)
It is expected to be worthwhile to pay special attention to energies around the $S$-wave open-charm thresholds.

Below, we list the opportunities at the STCF regarding the physics of hidden-charm $XYZ$ states:
\begin{itemize}
\item At the STCF luminosity of $0.5\times10^{35}$\,cm$^{-2}$s$^{-1}$ optimized at $\sqrt{s}=4$~GeV, two orders of magnitude higher than that of BEPCII, the vector charmonium-like states that are being investigated at BESIII can be studied in much more detail. Greatly improved knowledge of the intriguing $Z_c(3900)$ and $Z_c(4020)$ through $e^+e^-\to \pi^\pm Z_c^\mp$, as well as that of the $Z_{cs}(3985)$ and its possible spin partner in $e^+e^-\to K^\pm Z_{cs}^\mp$, will be obtained at various CMEs. The dependence of the $Z_c$ and $Z_{cs}$ line shapes and production rates on the CME will be crucial to keep kinematic effects from triangle singularities~\cite{Wang:2013cya,Pilloni:2016obd,Yang:2020nrt} under control.
\item Among all of the $PC=++$ $XYZ$ states, only the $X(3872)$ has been observed in $e^+e^-$ collisions, associated with a photon, and all others have only been seen in $B$ decays. This is because of the low production rates of the radiative processes and the fact that $X(3872)$ production receives an enhancement due to its large coupling to the $D\bar D^*$ pair. At the STCF with $E_\text{cm}\gtrsim4.7$~GeV, it will be possible to produce the $J^{++}$ states, $X(3915)$, $\chi_{c0}(3860)$ and $\chi_{c2}(3930)$, via $e^+e^-\to \omega X$ transitions, which should have much higher rates than the radiative processes.
\item  At the STCF with $E_\text{cm}\gtrsim5$~GeV, the $J^{++}$ states observed in the $\phi J/\psi$ invariant mass distributions can be investigated via $e^+e^-\to \phi X$. Searching for these states and others mentioned in the above item will be crucial for establishing the spectrum in the highly excited charmonium mass region and thus important in understanding the effects of the hadron thresholds on the spectrum and confinement. In addition to the abovementioned transitions, processes such as $e^+e^-\to \eta X$ should also be studied.

\item Energies higher than 5~GeV at the STCF will be useful for searches for the hadronic transitions to the spin partners of the $Z_c(3900)$ and $Z_c(4020)$ exotic states, named the $W_{c}$, as well as for conventional but not-yet-observed charmonium states.
The spin partners of the $Z_c$ are similar to those of the $Z_b$ proposed in Ref.~\cite{Bondar:2011ev}. They are isospin vector states with $J^{PC}=0^{++}$, $1^{++}$ and $2^{++}$, where the $C$ parity is for the charge-neutral state. The neutral ones can decay into $J/\psi \pi^+\pi^-$.
The $W_c$ can be studied in $e^+e^-\to \rho X$ transitions.

%     %%%%%%%%%%%%%%%%%%% Fig 4 %%%%%%%%%%%%%%%%%%%%%%%%
% \begin{figure*}[tp]
% 	\centering
% 	\includegraphics[width=0.92\textwidth]{Figs_02_CharmoniumXYZ/BB00}
% 	\caption{The spectrum of hadronic molecules consisting of a pair of charmed-anticharmed hadrons with negative parity and $(\text{isospin}, \text{strangeness})=(0,0)$ predicted in Ref.~\cite{Dong:2021juy}.
%     The colored rectangle, green for a bound state and orange for a virtual state, covers the uncertainty of the predicted mass. Thresholds are marked by dotted horizontal lines.
%     The rectangle closest to, but below, the threshold corresponds to the hadronic molecule in that system. When the masses of two hadronic molecules overlap, small rectangles are used with the left (right) one for the system with the higher (lower) threshold. The blue line (band) represents the center value (error) of the mass of the $\psi(4230)$~\cite{Zyla:2020zbs}.}\label{fig:specBB00}
% \end{figure*}
% %%%%%%%%%%%%%%%%%%%%%%%%%%%%%%%%%%%%%%%%%%%%%%%%%%
\item The lowest charmed baryon--antibaryon threshold, $\Lambda_c\bar \Lambda_c$, is at 4.57~GeV.
The BESIII measurement of the $e^+e^-\to \Lambda_c\bar \Lambda_c$ near-threshold production cross section indicates a state below the $\Lambda_c\bar \Lambda_c$ threshold~\cite{Cao:2019wwt,Dong:2021juy},
which is the lowest among a wealth of charmed baryon--antibaryon molecules recently predicted~\cite{Dong:2021juy}.
With $E_\text{cm}\gtrsim5$~GeV, the STCF will be able to reveal the expected rich phenomena due to the charmed baryon--antibaryon channels as well as those of excited charmed mesons.

\item With $E_\text{cm}\gtrsim 5$~GeV, it will also be possible to study hidden-charm pentaquark states in processes such as $e^+e^-\to J/\psi p\bar p$ and $e^+e^-\to \Lambda_c \bar D\bar p$. Similar to the $XYZ$ states above the $D\bar D$ threshold, there should be rich phenomena above the $\Lambda_c\bar D$ threshold. The cross section for $e^+e^-\to J/\psi p\bar p$ between 5 and 7~GeV may be estimated as
% \begin{equation}
$\sigma(e^+e^-\to J/\psi p\bar p) =\mathcal{O}(4~\text{fb})$~\cite{Chao:2023} based on the result for the $e^+e^-\to J/\psi gg$ cross section estimated using nonrelativistic QCD (NRQCD)~\cite{Ma:2008gq}. With an integrated luminosity of 2~ab$^{-1}$/year, $\mathcal{O}(8\times10^3)$ $J/\psi p\bar p$ events can be produced per year. A similar amount is expected for $J/\psi n\bar n$, and it will be possible to study this process at the STCF, whereas this is impossible for LHCb. The open-charm final states are expected to have larger cross sections. Furthermore, the hidden-charm pentaquarks are expected to decay much more easily into $\Lambda_c\bar D^{(*)}$ than into $J/\psi N$~\cite{Shen:2016tzq}, and the $\Sigma_c^{(*)}\bar D^{(*)}$ hadronic molecules, proposed by many authors to explain the LHCb $P_c$ states, couple strongly to $\Sigma_c^{(*)}\bar D^{(*)}$. Therefore, promising channels for the search for hidden-charm pentaquarks at the STCF include $e^+e^-\to  \Lambda_c \bar D^{(*)} \bar p$ and $\Sigma_c^{(*)} \bar D^{(*)} \bar p$. The STCF will provide good opportunities to search for hidden-charm $P_c$ and anomalous $P_{cs}$ pentaquarks.

\item For the interpretation of the nature of well-established highly excited charmonium states, detailed measurements of the production rates of open-charm final states such as $D^{(*)}\bar D^{(*)}$, $D_s^{(*)} \bar D_s^{(*)}$, and $D^{(*)}\bar D^{(*)} \pi(\pi)$ throughout the whole energy range of the STCF will be necessary.
To measure the cross sections of the three independent $D^*\bar D^*$ processes, namely, the $P$-wave with total spin $S=0$, the $P$-wave with $S=2$, and the $F$-wave with $S=2$, studies of the angular correlations of the $D^*$ decay products will need to be performed.

%%%%%%%%%%%%%%%%%%% Fig 2 %%%%%%%%%%%%%%%%%%%%%%%%
% \begin{figure*}[t]
% 	\centering
% 	\includegraphics[width=0.45\textwidth]{Figs_02_CharmoniumXYZ/ee2psietac}~~
% 	\includegraphics[width=0.45\textwidth]{Figs_02_CharmoniumXYZ/ee2psicc}
% 	\vspace{0cm}
% 	\caption{The cross sections for $e^+e^-\to J/\psi\eta_c$ (left) and $e^+e^-\to J/\psi c\bar c$ (right) calculated using NRQCD with the charm quark mass fixed at 1.5~GeV. The solid and dashed curves represent the results from the next-to-leading order and leading order calculations, respectively.
% 		\label{fig:xsec_NRQCD}}
% \end{figure*}
%%%%%%%%%%%%%%%%%%%%%%%%%%%%%%%%%%%%%%%%%%%%%%%%%%
\item There is a unique physics opportunity with $E_\text{cm}\in [6,7]$~GeV: this energy range offers an opportunity to study physics related to the production of two $c\bar c$ pairs. The production cross sections for  $e^+e^-\to J/\psi c\bar c$ based on the NRQCD calculations in Refs.~\cite{Zhang:2005cha,Zhang:2006ay} are on the order of tens of fb; see also Section~\ref{subsec:qcd}.
% are shown in Fig.~\ref{fig:xsec_NRQCD}.
In addition to double-charmonium production, which is also of interest, this energy range is ideal for the search for fully charmed tetraquark states, which are expected to have a mass of above 6~GeV~(see Refs.~\cite{Chao:1980dv,Karliner:2016zzc,Debastiani:2016xgg,Wang:2019rdo}). While it is uncertain whether the ground state $cc\bar c\bar c$ is below the double-$J/\psi$ or double-$\eta_c$ threshold, the low-lying $cc\bar c\bar c$ states are expected to decay predominantly into final states containing a pair of charm and anticharm hadrons via the annihilation of a $c\bar c$ pair into a gluon, with widths on the order of 100~MeV~\cite{Chao:1980dv,Anwar:2017toa}. Excited states with a mass well above the 6.2~GeV threshold can also easily decay into $J/\psi J/\psi$.
The LHCb measurement of the double-$J/\psi$ invariant mass spectrum in semi-inclusive processes of $pp$ collisions shows clear evidence for the existence of such states~\cite{Aaij:2020fnh}.
Searching for fully charmed tetraquarks in final states other than charged leptons is difficult at hadron colliders due to the high background; hence, the STCF is rather unique in its ability to support such a search.

% ******{\color{red} figure 5 should be figure 4 if appear here. It appear again in 5.1.5. May be should combined materials for these two small sections!}******

\item Charmonium-like hybrid candidates are also important targets to be searched for at the STCF, among which the most intriguing one is the lowest $1^{-+}$ state since the quantum numbers are prohibited for quark--antiquark states and from extensive lattice studies, it is expected to be the lowest charmonium-like hybrid. The mass is approximately $4.1\sim4.3$~GeV. In addition, one expects a hybrid supermultiplet including $(0,1,2)^{-+}$ and $1^{--}$ states with nearly degenerate masses of approximately 4.4 GeV~\cite{Liu:2012ze}. At the STCF with $E_\text{cm}\gtrsim4.5$~GeV, the
$(0,1,2)^{-+}$ states can be produced either from the hadronic and radiative transitions from highly excited charmonia, such as $\psi(4S)$ and higher excitations, or from the final-state radiations in $e^+e^-$ annihilations. The $Y(4260)$ has a possible assignment of a $1^{--}$ charmonium-like hybrid~\cite{Zhu:2005hp}, but further experimental and theoretical efforts should be made to unravel its nature. At the STCF, with its much higher luminosity than BEPCII/BESIII, it will be possible to measure the decay properties of the $Y(4260)$ more precisely and to search for other open-charm decay modes, along with the possible connections between $Y(4260)$, $X(3872)$ and $Z_c(3900)$. It is expected that the STCF will enable the final determination of the status of the $Y(4260)$.

\end{itemize}

\subsection{Opportunities in higher charmonium states}

Closely related to the $XYZ$ puzzles, there are also predictions from the quark model~\cite{Godfrey:1985xjl} and from LQCD~\cite{Liu:2012ze} of states that have not been identified. Some of them, such as the $2P$ states, are certainly intertwined with the $X$ states with the same quantum numbers. However, there are also still missing states that are believed to be relatively clean, such as the $1D$ state $\eta_{c2}$ and other higher-$L$ excitations.

% \subsubsection*{2.1.1 $2P$ charmonia}

% $2P$ charmonium states are called $\chi_{cJ}'$ and $h_c'$ according to the naming rules of hadrons. Quark model
% studies predict their masses to be around 4.0 GeV. The $\chi_{c2}'$ state (named as $\chi_{c2}(3930)$ by PDG2018~\cite{Tanabashi:2018oca}) is almost established with $M=3927.2\pm 2.6$ MeV and $\Gamma=24\pm 6$. Belle observes a wide resonance structure of $D\bar{D}$ with $M=3862^{+26+40}_{-32-13}$ MeV and $\Gamma=201^{+154+88}_{-67-82}$ MeV in the process $e^+e^-\rightarrow J/\psi D\bar{D}$ using the full amplitude analysis~\cite{Chilikin:2017evr}, which is now tentatively assigned to be $\chi_{c0}(3860)$ by PDG2018. PDG2018 now names $X(3872)$ to be $\chi_{c1}(3872)$, but admit that its properties are different from a conventional $q\bar{q}$ state
% and can be a candidate for an exotic structure. There is no clear evidence of $h_c'$ yet. Obviously, the
% masses of these $2P$ states or candidates are a little lower than the quark model prediction. This raises natural questions what on earth their inner dynamics are. On future STCF, $2P$ states can be produced by the radiative transitions from higher vector charmonia. This requires a considerable statistics accumulated at the $\psi(4040)$, $\psi(4160)$ and $\psi(4415)$ energy scales.

% \subsubsection*{2.1.2  $1D$ charmonia $\eta_{c2}$ and $\psi_3$}

The supermultiplet of $1D$ states includes $1^3 D_{1,2,3}$ and $1^1 D_2$ (named $\eta_{c_2}$), with the quantum numbers $(1,2,3)^{--}$ and $2^{-+}$, respectively. Apart from the well-known $\psi(3770)$, the $2^{--}$ state has likely been observed by Belle~\cite{Bhardwaj:2013rmw} and BESIII~\cite{Ablikim:2015dlj} and is labeled $\psi_2(3823)$ in the 2018 PDG. Very recently, LHCb reported a candidate for the $1^3 D_3$ state (named $\psi_3$). However, the spin singlet $1D$ state $\eta_{c_2}$ continues to evade experimental searches. LQCD studies predict that the mass of $\eta_{c2}$ is approximately $3.8$ GeV~\cite{Liu:2012ze,Yang:2012mya}, nearly degenerate with other $1D$ states. Experimentally, $\eta_{c2}$ can be produced directly from $\psi(4040)$ through the M1 transition. If the partial width of $\psi(4040)\to \gamma \eta_{c2}$ is a few keV, then the
corresponding branching fraction is $\mathcal{O}(10^{-5})$. Therefore, it is difficult for BESIII to observe $\eta_{c2}$ in this process (the number of $\psi(4040)$ events at BESIII is $\mathcal{O}(10^{6})$~\cite{Asner:2008nq}). However, the STCF, with a luminosity 100 times higher, will have the possibility to search for $\eta_{c2}$. Since it has no open-charm decay modes, the hadronic transitions, such as the decay modes $\chi_{c1}\pi\pi$ and $J/\psi \pi^0 \pi^- \pi^+$, and the E1 radiative transition $\eta_{c2}\to \gamma h_c$ can be important.

% $\psi_3$ can be search in the processes $e^+e^-\to \pi\pi \psi_3$ with $\psi_3\to \gamma \chi_{c2}$, similar to the case of $\psi_2(3823)\to \gamma \chi_{c1}$, and  also $\psi_3 \to D\bar{D}$. BESIII does not observe $\psi_3$ in the $e^+e^-\to \pi\pi D\bar{D}$ process~\cite{Ablikim:2019faj}. This is understandable since the partial decay width of $\psi_3\to D\bar{D}(L=3)$ is highly suppressed by the centrifugal potential barrier. Non-relativistic models predict that that both the decay widths of $\psi_3\to \gamma \chi_{c2}$ and $\psi_2\to \gamma \chi_{c1}$ are around 280 keV~\cite{Barnes:2005pb}

% \subsubsection*{2.1.4  Radiative tansitions and decays of charmonia}

Apart from spectroscopy, the understanding of the known charmonium states can be greatly improved through more precise measurements of their radiative and hadronic states~\cite{Asner:2008nq}. In the following, the two types of decays will be discussed.

\par
For the radiative transitions, at the STCF, it will be possible to measure the rare electric-dipole transitions $\eta_c(2S)\rightarrow h_c\gamma$ and $\psi(3770)\rightarrow \chi_{c0}\gamma$ and the magnetic-dipole transitions $\psi(3686)\rightarrow \eta_c(2S)\gamma$, $\eta_c(2S)\rightarrow J/\psi\gamma$, and $h_c\rightarrow \chi_{c0}\gamma $. It will also be possible to measure the total and leptonic or two-photon widths with high precision.
These transitions and decay widths can be calculated in both the quark model and LQCD. Comparisons between experimental data and these theoretical predictions will help us more clearly understand the inner structure of charmonia.

A more systematic and comprehensive study of the decays of low-lying charmonia can also be performed at the STCF. These states are below the threshold for $D$-meson production and decay predominantly into hadrons consisting of light $u$, $d$ and $s$ quarks. However, information about their decays is incomplete at present. For the well-known $J/\psi$ meson, only approximately 40\% of its hadronic decays have been measured. For other states, the situation is even worse. The high luminosity of the STCF will facilitate more precise measurements of the properties of light hadrons from low-lying charmonium decays and the subsequent acquisition of a more complete understanding of the scenario of low-energy strong interactions.

%\input{02_ref_CharmoniumXYZ_v2}
%\bibliography{stcf-xyz}

\newpage
\section{Charmed Hadron Physics}
\label{sec:charmedhadron}
The discovery of the charm quark in 1974 was a great milestone in
the development of particle physics and the establishment of the
Standard Model (SM).
%A high-luminosity Super $\tau$-Charm Factory  (STCF),
The high-luminosity STCF,
which will be capable of producing approximately $10^9\sim 10^{10}$ quantum-coherent $D^0 \bar{D}^0$ meson pairs,
$D^+$ or $D^+_s$ mesons, and $\Lambda_c^+$ baryons, will be an important
low-background playground for testing the SM and probing new
physics beyond relative to the experience at BESIII~\cite{Li:2021iwf}. In particular, it will serve as a unique tool
for determining the Cabbibo--Kobayashi--Maskawa (CKM) matrix elements
$V_{cd}$ and $V_{cs}$, measuring the $D^0$--$\bar{D}^0$ mixing
parameters, probing $CP$ violation in the charm sector, searching for
SM rare and forbidden charmed hadron decays, and studying other fundamental
problems associated with the charmed hadrons. Many of the golden measurements at the STCF will be dominated by systematic uncertainties; therefore, a state-of-the-art detector with excellent performance will be required, especially for identifying the different types of charged particles, detecting low-momentum charged particles and measuring photons.

\subsection{Charmed mesons}

\subsubsection{$D^+_{(s)}$ leptonic decays}

Direct determination of the CKM matrix elements $|V_{cd}|$ and
$|V_{cs}|$ is one of the most important targets in charm physics.
These two quark-flavor mixing quantities not only govern the rates
of leptonic $D^+$ and $D^+_s$ decays but also play a crucial role in
testing the unitarity of the CKM matrix.
Hence, the precise measurement of $|V_{cd}|$ and $|V_{cs}|$
is a priority of the STCF experiment.
%A determination of $|V_{cd}|$ and $|V_{cs}|$ to
%a much better degree of accuracy is therefore desirable at STCF.

%%%%%%%%%%%%%%%%%%%%%%%%%% Table 1 %%%%%%%%%%%%%%%%%
\begin{table*}[htbp]
\centering
\caption{\label{tab:pure_LP}\small
For studies on $D^+_{(s)}\to \ell^+\nu_\ell$, the precisions achieved at BESIII and the projected precisions at the STCF
and Belle II.
Considering that the LQCD uncertainty of $f_{D_{(s)}^+}$
has been updated to be approximately 0.2\%~\cite{Bazavov:2017lyh},
the $|V_{cd}|$ value measured at BESIII has been recalculated; this recalculated value is marked with $^*$.
For Belle II, we assume that the systematic uncertainties can be reduced by
a factor of 2 compared to the Belle results.}
\begin{tabular}{lcccc} \hline\hline
\multicolumn{1}{c}{} & BESIII & STCF
& Belle II \\ \hline
\multicolumn{1}{c}{Luminosity} &2.93 fb$^{-1}$ at 3.773 GeV & 1 ab$^{-1}$ at 3.773 GeV
& 50 ab$^{-1}$ at $\Upsilon(nS)$ \\ \hline
${\mathcal B}(D^+\to \mu^+\nu_\mu)$ & $5.1\%_{\rm stat}\,1.6\%_{\rm syst}$~\cite{bes3_muv} &$0.28\%_{\rm stat}$ & $2.8\%_{\rm stat}$~\cite{Kou:2018nap}  \\
$f^{\mu}_{D^+}$ (MeV) &$2.6\%_{\rm stat}\,0.9\%_{\rm syst}$~\cite{bes3_muv} &$0.15\%_{\rm stat}$ &--   \\
$|V_{cd}|$ &$2.6\%_{\rm stat}\,1.0\%_{\rm syst}^*$~\cite{bes3_muv} &$0.15\%_{\rm stat}$  & --   \\
${\mathcal B}(D^+\to \tau^+\nu_\tau)$ &$20\%_{\rm stat}\,{10\%_{\rm syst}}$~\cite{Ablikim:2019rpl}  &  $0.41\%_{\rm stat}$ &-- \\
$\displaystyle \frac{{\mathcal B}(D^+\to \tau^+\nu_\tau)}{{\mathcal B}(D^+\to \mu^+\nu_\mu)}$&$21\%_{\rm stat}\,13\%_{\rm syst}$~\cite{Ablikim:2019rpl}  & $0.50\%_{\rm stat}$ &   --
\\ \hline \hline
\multicolumn{1}{c}{Luminosity} &6.3 fb$^{-1}$ at (4.178, 4.226) GeV &1 ab$^{-1}$ at 4.009 GeV
& 50 ab$^{-1}$ at $\Upsilon(nS)$ \\ \hline

${\mathcal B}(D^+_s\to \mu^+\nu_\mu)$ &$2.4\%_{\rm stat}\,3.0\%_{\rm syst}$~\cite{BESIII:2021anh}&$0.30\%_{\rm stat}$ & $0.8\%_{\rm stat}\,1.8\%_{\rm syst}$\\
$f^{\mu}_{D^+_s}$ (MeV) &$1.2\%_{\rm stat}\,1.5\%_{\rm syst}$~\cite{BESIII:2021anh}&$0.15\%_{\rm stat}$ &-- \\
$|V_{cs}|$ &$1.2\%_{\rm stat}\,1.5\%_{\rm syst}$~\cite{BESIII:2021anh}
&$0.15\%_{\rm stat}$ &-- \\\hline

${\mathcal B}(D^+_s\to \tau^+\nu_\tau)$ &$1.7\%_{\rm stat}\,2.1\%_{\rm syst}$~\cite{HFLAV:2022pwe} &$0.24\%_{\rm stat}$ &$0.6\%_{\rm stat}\,2.7\%_{\rm syst}$ \\
$f^{\tau}_{D^+_s}$ (MeV) &$0.8\%_{\rm stat}\,1.1\%_{\rm syst}$~\cite{HFLAV:2022pwe} &$0.11\%_{\rm stat}$ & -- \\
$|V_{cs}|$ &$0.8\%_{\rm stat}\,1.1\%_{\rm syst}$~\cite{HFLAV:2022pwe} &$0.11\%_{\rm stat}$ &-- \\ \hline

$\overline f^{\mu \&\tau}_{D^+_s}$ (MeV) &$0.7\%_{\rm stat}\,0.9\%_{\rm syst}$&$0.09\%_{\rm stat}$&$0.3\%_{\rm stat}\,1.0\%_{\rm syst}$\\
$|\overline V_{cs}^{\mu \&\tau}|$ &$0.7\%_{\rm stat}\,{0.9\%_{\rm syst}}$&$0.09\%_{\rm stat}$ &-- \\ \hline \hline
$f_{D^+_s}/f_{D^+}$ &$1.4\%_{\rm stat}\,1.7\%_{\rm syst}$~\cite{BESIII:2021bdp} &$0.21\%_{\rm stat}$ &-- \\ 
$\displaystyle \frac{{\mathcal B}(D^+_s\to \tau^+\nu_\tau)}{{\mathcal B}(D^+_s\to \mu^+\nu_\mu)}$&$2.9\%_{\rm stat}\,3.5\%_{\rm syst}$& $0.38\%_{\rm stat}$&$0.9\%_{\rm stat}\,3.2\%_{\rm syst}$
\\ \hline \hline
\end{tabular}
\end{table*}
%%%%%%%%%%%%%%%%%%%%%%%%%%

The most precise way to determine $|V_{cd}|$ and $|V_{cs}|$ at the STCF will be via pure leptonic decays of the form $D_{(s)}^+ \to \ell^+ \nu^{}_\ell$ (where $\ell = e, \mu, \tau$), as the semileptonic decays suffer from large uncertainties in the LQCD calculations of the form factors.
The product of the decay constant $f_{D_{(s)}^+}$ and $|V_{cd(s)}|$ can be directly accessed by measuring the widths
of $D_{(s)}^+ \to \ell^+ \nu^{}_\ell$. Then, with $f_{D_{(s)}^+}$ from LQCD as input, the values of $|V_{cd(s)}|$ or $f_{D_{(s)}^+}$ can be obtained.
Listed in Table~\ref{tab:pure_LP} are the most precise determinations to date of $|V_{cs(d)}|$ and $f_{D^+_{(s)}}$~\cite{bes3_muv,Ablikim:2019rpl,Ablikim:2018jun} at BESIII and the projected precisions at the STCF.
Note that for ${\mathcal B}(D^+\to \tau^+\nu_\tau)$,
several $\tau^+$ decay channels,
such as $\tau^+\to \pi^+\overline{\nu}_\tau$, $e^+\overline{\nu}_\tau\nu_e$,
$\mu^+\overline{\nu}_\tau\nu_\mu$, and $\rho^+\overline{\nu}_\tau$,
are combined to improve the statistical sensitivity.

The systematic uncertainties at the STCF are to be optimized to a subleading level, as the statistical uncertainties are expected to be less than 0.5\%. To reduce the systematic uncertainties due to background and fitting, it will be optimal for the STCF to study $D_s^+\to \ell^+\nu_\ell$ using $\ee\to D_s^+ D_s^{-}$ at 4.009 GeV.
Thus far, the $f_{D_{(s)}^+}$ values have been calculated via LQCD with precisions of approximately 0.2\%~\cite{Bazavov:2017lyh}; specifically,
$f_D^+=212.7\pm0.6$~\mev, $f_{D_s}^+=249.9\pm0.4~\mev$ and $f_{D_s}^+/f_D^+=1.1749\pm0.0016$. At the time when the STCF comes online,
%Editor: Please ensure that the intended meaning has been maintained in the above edit.
their precisions are expected to be below 0.1\%. This means that the sizes of the systematic uncertainties at the STCF will be crucial and must be improved to the similar level. The feasibility studies of $D^{+}\to\mu^{+}\nu_{\mu}$ and $D_{s}^{+}\to \tau^{+}\nu_{\tau}$ are presented in Ref.~\cite{Liu:2021qio,Li:2021ala}.  
In particular, the efficiencies of muon and electron identification will be critical and must be optimized to constrain the total uncertainty to reach the expected level.

On the other hand, precise measurements of the semileptonic branching fractions for $D_{(s)}\to h \ell^+\nu_\ell$, where $h$ is a charmless hadron, will be used to calibrate the LQCD calculations of the form factors involved by introducing the $|V_{cd(s)}|$ values from global CKM fits (such as those of CKMfitter~\cite{Charles:2004jd,CKMfitter_web} and UTfit~\cite{Bona:2005vz,utfit_web}).
For the case of $D_{(s)}\to V(h_1h_2) \ell^+\nu_\ell$ (where $V$ denotes a vector meson, decaying into hadrons $h_1$ and $h_2$), time reversal ($T$) invariance can be tested with high precision by constructing triple-product $T$-odd observables~\cite{Belanger:1991vx}. This will serve as a sensitive probe of $CP$ violation mechanisms beyond the Standard Model and of new physics~\cite{Grossman:2003qi}, such as models with multi-Higgs doublets or leptoquarks. Ref.~\cite{Wang:2019wee} proposes combined measurements of $D\to K_1 \ell^+\nu_\ell$ and $B\to K_1 \gamma$ to unambiguously determine the photon polarization in $b\to s\gamma$ in a clean way to probe right-handed couplings in new physics.

%%%%%%Test on the lepton universality %%%%%%%%%%

Lepton flavor universality (LFU) can also be tested in charmed meson leptonic decays. LFU violation may occur in $c\to s$ transitions due to an amplitude that includes a charged Higgs boson, which arises in a two-Higgs-doublet model, interfering with the SM amplitude involving a $W^{\pm}$ boson~\cite{prd91_094009}. In the SM, the ratio of the partial widths of
$D^+_{(s)}\to \tau^+\nu_\tau$ and $D^+_{(s)}\to \mu^+\nu_\mu$
is predicted to be
\begin{eqnarray}
R_{D^+_{(s)}}=\frac{\Gamma(D^+_{(s)}\to \tau^+\nu_\tau)}{\Gamma(D^+_{(s)}\to \mu^+\nu_\mu)}=
\frac{m^2_{\tau^+}\left(1-\frac{m^2_{\tau^+}}{m^2_{D^+_{(s)}}} \right )^2}{m^2_{\mu^+} \left(1-\frac{m^2_{\mu^+}}{m^2_{D^+_{(s)}}} \right )^2}.
\end{eqnarray}
Using the world average values of the masses of the leptons and $D^+_{(s)}$~\cite{ParticleDataGroup:2022pth}, one obtains $R_{D^+}=2.67\pm 0.01$ and $R_{D^+_s}=9.75\pm0.01$.
The measured values of $R_{D^+}$ and $R_{D^+_s}$ reported by BESIII are $3.21\pm 0.64_{\rm stat}\pm 0.43_{\rm syst}$~\cite{Ablikim:2019rpl} and $9.72\pm 0.37$~\cite{BESIII:2021bdp}, respectively, which agree with the SM predictions.
%Editor: Please ensure that the intended meaning has been maintained in the above edit.
However, these measurements are currently statistically limited.
At the STCF, as listed in Table~\ref{tab:pure_LP}, the statistical precision for $R_{D_{(s)}^{+}}$ will be comparable to the uncertainties of the predictions in the SM. Hence, it will provide a meaningful test of LFU via these channels.

Another LFU test could be performed via the semileptonic decay modes, of which the semitauonic decay is
kinematically forbidden or suppressed. Measurements of the ratios of the partial widths of $D^{0(+)}\to h \mu^+\nu_\mu$
over those of $D^{0(+)}\to h e^+\nu_e$ in different $q^2$ intervals
would constitute a test of LFU complementary to those using tauonic decays.
BESIII has reported precise measurements of the ratios ${\mathcal B}(D^0\to\pi^-\mu^+\nu_\mu)/{\mathcal B}(D^0\to\pi^-e^+\nu_e)=0.922\pm0.030\pm0.022$
and ${\mathcal B}(D^+\to\pi^0\mu^+\nu_\mu)/{\mathcal B}(D^+\to\pi^0e^+\nu_e)=0.964\pm0.037\pm0.026$~\cite{bes3_pimuv}.
These results are consistent with the SM predictions within $1.7\sigma$ and $0.5\sigma$~\cite{bes3_pimuv}, respectively.
These measurements are currently statistically limited~\cite{bes3_kmuv,bes3_pimuv}, and they could be significantly improved with 1 ab$^{-1}$ of data taken at the center-of-mass energy of 3.773 GeV at the STCF.

For the above LFU tests at STCF, control of systematic uncertainties will be an essential issue to enhance the sensitivity. Hence, a double ratio of two different types of leptonic decay modes could provide significant cancellation of detection systematics in the further measurement. 

\subsubsection{$\dzero$--$\dzerobar$ mixing and $CP$ violation}

The phenomenon of meson--antimeson mixing has been of great interest
throughout the long history of particle physics.
In contrast to the $B$-meson and kaon systems, $CP$ violation in the mixing of $D$ mesons has not been observed. The STCF will be an
ideal place for the study of $D^0$--$\bar{D}^0$ mixing and $CP$ violation. By convention,
the mass states of the two neutral $D$ mesons are written as
\begin{eqnarray}
|D^{}_1\rangle \hspace{-0.2cm} & = & \hspace{-0.2cm} p |D^0\rangle +
q |\bar{D}^0\rangle \; ,
\nonumber \\
|D^{}_2\rangle \hspace{-0.2cm} & = & \hspace{-0.2cm} p |D^0\rangle -
q |\bar{D}^0\rangle \; ,
%       (4)
\end{eqnarray}
where $|p|^2 + |q|^2 = 1$. The $D^0$--$\bar{D}^0$ mixing
parameters are defined as $x \equiv (M^{}_2 - M^{}_1)/\Gamma$ and $y
\equiv (\Gamma^{}_2 - \Gamma^{}_1)/(2\Gamma)$, where $M^{}_{1,2}$
and $\Gamma^{}_{1,2}$ are the masses and widths, respectively, of $D^{}_{1,2}$.
Additionally, $\Gamma \equiv (\Gamma^{}_1 + \Gamma^{}_2)/2$, and $M \equiv (M^{}_1
+ M^{}_2)/2$. This system is unique because it is the only
meson--antimeson system whose mixing (or oscillation) takes place via
the intermediate states with down-type quarks. It is also the only
meson--antimeson system whose mixing parameters $x$ and $y$ are
notoriously difficult to calculate in the SM, as they involve large long-distance uncertainties in this nonperturbative regime.
%
%The mixing between $D^0$ and $\bar{D}^0$ mesons arises from the fact
%that they couple to a subset of virtual or real intermediate states.
%While $x$ may be sensitive to new physics via the $\Delta C =2$
%contribution, $y$ is dominated by the SM (i.e., $\Delta C= 1$)
%contribution. They can be estimated at either the quark level or the
%hadron level, but both of them involve large long-distance
%uncertainties.
One expects $x\sim y\sim \sin^2\theta^{}_{\rm C}
\times [{\rm SU(3) ~ breaking}]^2$ as a second-order effect of the
flavor SU(3) symmetry breaking. A more careful analysis yields the
order-of-magnitude estimates $x\lesssim y$ and $10^{-3} < |x| <
10^{-2}$ \cite{Falk2004}.
A global fit to the world measurements of $x$ and $y$, carried out by the Heavy Flavor Averaging Group \cite{Amhis:2019ckw,hflav_web}, gives
intervals of $1.6 \times 10^{-3} \lesssim x \lesssim 6.1 \times 10^{-3}$
and $5.2 \times 10^{-3} \lesssim y \lesssim 7.9 \times 10^{-3}$
at the $95\%$ confidence level~\cite{Amhis:2019ckw,hflav_web}. We see that the allowed regions for $x$ and $y$ are
essentially consistent with the theoretical estimates (i.e.,
$x\lesssim y \sim 7 \times 10^{-3}$).
Much more precise measurements of
these two $D^0$--$\bar{D}^0$ mixing parameters can be achieved at the
STCF. Although their accurate values might not help much to
clarify the long-distance effects in $D^0$--$\bar{D}^0$ mixing, they
will be of great help in probing the presumably small effects of $CP$ violation
in neutral $D$-meson decays and mixing \cite{SuperB}.

%The following decay modes have been measured in different experiments
%to globally determine the values
%of $x$ and $y$ \cite{PDG}: $D^0 \to K^{(*)+}\ell^-
%\overline{\nu}^{}_\ell$, $K^+K^-$, $\pi^+\pi^-$, $K^+\pi^-$,
%$K^+\pi^-\pi^0$, $K^+\pi^-\pi^+\pi^-$, $K^0_{\rm S}\pi^+\pi^-$,
%$K^0_{\rm S}K^+K^-$ {\it etc} and (or) their $CP$-conjugate processes,
%together with the coherent $\psi(3770) \to D^0\bar{D}^0 \to f^{}_1 f^{}_2$
%decays.

The charm sector is a precision laboratory for exploring possible
$CP$-violating new physics because the SM-induced $CP$-violating asymmetries
in $D$-meson decays are typically in the range of $10^{-4}$ to $10^{-3}$~\cite{Xing:2007zz}
and are very challenging to detect in experiments. The $CP$-violating asymmetries
in the singly Cabibbo-suppressed $D$-meson decays are now expected to be much larger than
those in the Cabibbo-favored and doubly Cabibbo-suppressed decays~\cite{SuperB},
where such asymmetries vanish.
%One may easily understand this point by considering the
%{\it charmed} unitarity triangle of the CKM matrix as defined by the
%relation $V^*_{ud} V_{cd} + V^*_{us} V_{cs} + V^*_{ub}
%V_{cb} = 0$ in the complex plane, in which two sides are
%comparable in magnitude and much longer than the third one. In other
%words, the shape of this unitarity triangle is too sharp. Given
%${\rm Im}\left(V^*_{ud} V_{cd} + V^*_{us} V_{cs}\right) =
%-{\rm Im}\left(V^*_{ub} V_{cb}\right) \sim \lambda^6\sin\gamma$
%with $\lambda \simeq 0.22$ and $\gamma \simeq 74^\circ$ \cite{PDG},
%the ratio of the CP-violating part to the CP-conserving part in many
%$D$-meson decays is characterized by ${\rm Im}\left(V^*_{ud}
%V_{cd} + V^*_{us} V_{cs}\right)/ \left(|V^*_{ud} V_{cd}| +
%|V^*_{us} V_{cs}|\right) \sim \lambda^5 \sin\gamma \sim 5\times
%10^{-4}$.
There are, in general, three different types of $CP$-violating effects in
neutral and charged $D$-meson decays \cite{Xing97}: 1) $CP$ violation
in $D^0$--$\bar{D}^0$ mixing, 2) $CP$ violation in the direct decay, and 3)
$CP$ violation from the interplay of decay and mixing.
% 4) CP violation in the $CP$-forbidden decay of coherent $D^0$ and $\bar{D}^0$ mesons.
In addition to these three types of $CP$-violating effects in $D$-meson
decays, one may expect an effect of $CP$ violation induced by
$K^0$--$\bar{K}^0$ mixing in some decay modes with $K^{}_{\rm S}$ or
$K^{}_{\rm L}$ in their final states. The magnitude of this effect is typically $2
{\rm Re}(\epsilon^{}_K) \simeq 3.3 \times 10^{-3}$, which may be
comparable to or even larger than the {\it charmed} $CP$-violating
effects~\cite{Xing:1995jg,Yu:2017oky}.
To date, much effort has been put into searching for $CP$ violation
in $D$-meson decays. The LHCb collaboration has discovered
$CP$ violation in combined $D^0\to \pip\pim$ and $D^0\to K^+K^-$ decays with a
significance of 5.3$\sigma$. The time-integrated $CP$-violating asymmetry
is given as
\begin{eqnarray}
\Delta a_{CP}&=&\frac{\Gamma(D\to \kk)-\Gamma(\bar{D}\to\kk)}{\Gamma(D\to \kk)+\Gamma(\bar{D}\to\kk)}-\frac{\Gamma(D\to\pip\pim)-\Gamma(\bar D\to\pip\pim)}{\Gamma(D\to\pip\pim)+\Gamma(\bar D\to\pip\pim)}  \nonumber \\
&=& (-0.154\pm0.029)\%,
%     (5)
\end{eqnarray}
where $D$($\bar{D}$) is a $D^0$($\bar{D}{}^0$) at time $t$=0~\cite{Aaij:2019kcg},
and it mainly arises from direct $CP$ violation in the charm-quark decay~\cite{Saur:2020rgd}.
This result is consistent with some theoretical estimates within the SM
(see, e.g., Refs. \cite{Cheng,Li,Grinstein,Gronau,Italy,Italy2,Li:2019hho,Grossman:2019xcj});
however, the latter involve quite large uncertainties.
The STCF will have a sensitivity at the level of $10^{-4}$ in systematically searching
for $CP$ violation in different types of charmed-meson decays.
In particular, the advantage of kinematical constraints on the initial four-momenta of the $\ee$ collisions will make the STCF competitive in studies
of $CP$-violating asymmetries in multibody $D$ decays~\cite{Bigi:2011em},
such as 4-body hadronic decays and the $CP$ asymmetries therein in the local Dalitz region.
Considering that the CKM mechanism of $CP$ violation in the SM fails to explain the puzzle of
the observed matter--antimatter asymmetry in the Universe
by more than 10 orders of magnitude~\cite{Morrissey:2012db}, there is strong motivation to search for
new (heretofore undiscovered) sources of $CP$ violation associated with both quark and lepton
flavors. In this context, the charm-quark sector is certainly a promising playground.

Note that the STCF will be a unique place for the study of $D^0$--$\bar{D}^0$
mixing and $CP$ violation by means of the quantum coherence of $D^0$ and
$\bar{D}^0$ mesons produced at energy points near the threshold. In fact, a $D^0\bar{D}^0$ pair can be coherently produced
through the reactions $\ee \to (D^0\bar{D}^0)^{}_{\rm CP=-}$ at 3.773 GeV and
$\ee\to D^0\bar{D}^{*0} \to \pi^0 (D^0\bar{D}^0)^{}_{\rm
CP=-}$ or $\gamma (D^0\bar{D}^0)^{}_{\rm CP=+}$ at 4.009 GeV. One may
therefore obtain useful constraints on $D^0$--$\bar{D}^0$ mixing and $CP$-violating
parameters in the respective decays of correlated $D^0$ and
$\bar{D}^0$ events \cite{Xing97}.
For example, the $D^0$--$\bar{D}^0$ mixing rate $R_M=(x^2+y^2)/2$ can
be accessed via the same charged final states $(K^\pm\pi^\mp)(K^\pm\pi^\mp)$
or $(K^\pm\ell^\mp\nu)(K^\pm\ell^\mp\nu)$ with a sensitivity of $10^{-5}$
with 1 ab$^{-1}$ of data collected at 3.773 GeV.
Considering $\ee\to \gamma D^0\bar{D}^0$ at 4.009 GeV, the $D^0\bar{D}^0$ pairs
are in $C$-even states, and the charm mixing contribution is doubled compared with
the time-dependent (uncorrelated) case. With 1 ab$^{-1}$ of data at 4.009 GeV,
it is expected that the measurement sensitivities for the mixing parameters
($x$ and $y$) will reach a level of 0.05\%, and those for $|q/p|$ and $\arg(q/p)$ will be 1.5\% and $1.4^\circ$, respectively~\cite{Bondar:2010qs}. Another possible case is that the decay mode
$\left( D^0\bar{D}^0 \right)^{}_{\rm CP = \pm} \to \left( f^{}_1
f^{}_2 \right)^{}_{\rm CP = \mp}$, where $f^{}_1$ and $f^{}_2$ are
proper $CP$ eigenstates (e.g., $\pi^+\pi^-$, $K^+K^-$ and $K^{}_{\rm
S} \pi^0$), is a $CP$-forbidden process and can only occur due to $CP$ violation.
The rate for a pair of $CP$-even final states $f_+$ (such as $f_+=\pi^+\pi^-$)
can be expressed as
\begin{equation}
\Gamma^{++}_{D^0\bar{D}^0 } = \left[
\left(x^2+y^2\right)\left(\cosh^2 a_m - \cos^2 \phi\right) \right] \Gamma^2(D \to f_+),
%     (6)
\end{equation}
where $\phi = \arg(p/q)$, $R_M=|p/q|$, and $a_m=\log R_M$~\cite{Atwood:2002ak}.

$CPT$ is conserved in all locally Lorentz-invariant theories,
including the SM and all of its commonly discussed extensions.
When $CPT$ is conserved, $CP$ violation implies the violation of time reversal symmetry ($T$).
However, $CPT$ violation might also arise in string theory or some extradimensional
models with Lorentz-symmetry violation in four dimensions.
Hence, direct observation of $T$ violation without the presumption of $CPT$
conservation is very important~\cite{Shi:2016bvo}.
Experimental studies of the time evolution of $CP$-correlated $D^0$--$\bar{D}^0$
states at the STCF could be complementary to the $CPT$-violation studies at
super $B$ factories and the LHCb experiments~\cite{Kostelecky:2001ff}.
However, this becomes very challenging with symmetric $\ee$ collisions, as the produced $D$ mesons have very low momentum in the laboratory frame and hence have flight distances that are too short to be detected. Only an asymmetric $\ee$ collision mode can be feasible for such investigations.

The quantum correlation of a $D^0\bar D^0$ meson pair offers a
unique feature for probing the amplitudes of the $D^0$ decays and
determining the strong-phase difference between their Cabibbo-favored and
doubly Cabibbo-suppressed amplitudes. Measurements of the strong-phase difference
are well motivated from several perspectives: understanding the nonperturbative
QCD effects in the charm sector, serving as essential inputs for extracting
the angle $\gamma$ of the CKM unitarity triangle (UT), and relating the
measured mixing parameters $(x', y')$ in hadronic decay to the mass and width
difference parameters $(x, y)$~\cite{Amhis:2019ckw}.

Measurements of the CKM UT angles $\alpha$,
$\beta$, and $\gamma$ in $B$ decays are important tests of CKM unitarity
and provide another avenue to search for possible sources of $CP$ violation beyond the SM. Any discrepancy in
measurements of the UT involving tree- and loop-dominated processes would
indicate the existence of new heavy degrees of freedom contributing to
the loops. Among the three CKM angles, $\gamma$ is of particular importance
because it is the only $CP$-violating observable that can be determined
using tree-level decays. Currently, the world-best single measurement of $\gamma$
is from LHCb: $\gamma = (63.8^{+3.5}_{-3.7})^\circ$~\cite{LHCb:2022awq}.
The precision measurement of $\gamma$ will be one of the top priorities for the
LHCb upgrade(s) and the Belle II experiment.

The most precise method of measuring $\gamma$ is
based on the interference between the $B^{+}\to\bar{D}^{0}K^{+}$
and $B^{+}\to D^{0}K^{+}$ decays~\cite{GLW1, GLW2, ADS1, ADS2, GGSZ}.
In the future, the statistical uncertainties of these measurements
will be greatly reduced by using the large $B$ meson samples collected
by LHCb and Belle II. Hence, limited knowledge of the strong phases of
the $D$ decays will systematically restrict the overall sensitivity.
A 20 fb$^{-1}$ dataset collected at 3.773 GeV at BESIII would lead to a
systematic uncertainty of $\sim$0.4$^\circ$ for the $\gamma$ measurement~\cite{Ablikim:2019hff}.
Hence, to match the anticipated future statistical uncertainty of less than $0.4^\circ$
in the future LHCb upgrade II,
the STCF could provide important constraints to reduce the systematic
uncertainty from the $D$ strong phase to less than 0.1$^\circ$ and enable
detailed comparisons of the $\gamma$ results from different decay modes.

\subsubsection{Rare and forbidden decays}

With its high luminosity, clean collision environment and excellent
detector performance, the STCF has great potential to
perform searches for rare and forbidden $D$-meson decays,
which may serve as a useful tool for
probing new physics beyond the SM. Such decays can be classified into three
categories: (1) decays via a flavor-changing neutral current
(FCNC), such as the $D^{0(+)} \to \gamma V^{0(+)}$, $D^0 \to \gamma\gamma$,
$D^0 \to \ell^+\ell^-$, and
$D \to \ell^+\ell^- X$ channels (where $\ell = e, \mu$)
and $D \to \nu \overline{\nu} X$, which provide SM-allowed transitions between
$c$ and $u$ quarks; (2)
decays with lepton flavor violation (LFV), such as the $D^0 \to \ell^+
\ell^{\prime -}$ and $D \to \ell^+\ell^{\prime -} X$ channels (for
$\ell \neq \ell^\prime$), which are forbidden in the SM; and (3) decays
with lepton number violation (LNV), such as the $D^+ \to \ell^+\ell^{\prime +}
X^-$ and $D^+_s \to \ell^+\ell^{\prime +} X^-$ channels (for either
$\ell = \ell^\prime$ or $\ell \neq \ell^\prime$), which are also
forbidden in the SM. The discovery of neutrino oscillations has
confirmed the occurrence of LFV in the lepton sector, and LNV is also possible if the massive
neutrinos are Majorana particles. It is therefore needed to
search for the LFV and LNV phenomena in the charm-quark sector.

Although FCNC decays of $D$ mesons are allowed in the SM, they
can only occur via loop diagrams and hence are strongly
suppressed. The long-distance dynamics are expected to dominate the
SM contributions to such decays, but their branching fractions are
still tiny. For instance, ${\cal B}(D^0 \to \gamma\gamma) \sim 1
\times 10^{-8}$ and ${\cal B}(D^0 \to \mu^+\mu^-) \sim 3 \times
10^{-13}$ in the SM \cite{Burdman}, but they can be significantly
enhanced by new physics \cite{Golowich}. The current experimental bounds
on these two typical FCNC channels are ${\cal B}(D^0 \to
\gamma\gamma) < 8.5 \times 10^{-7}$ and ${\cal B}(D^0 \to \mu^+\mu^-)
< 6.2 \times 10^{-9}$ \cite{ParticleDataGroup:2022pth}. However, the semileptonic decays
$D^0\to\pp\MM$, $\kk\MM$ and $K^-\pi^+ \MM$
have been observed at LHCb with BFs at a level of $10^{-7}$~\cite{ParticleDataGroup:2022pth}.
In addition to the removal of helicity suppression dominating the highly suppressed BF for $D^0 \to \mu^+\mu^-$, the observed BFs for the semileptonic decays indicate nontrivial contributions from complicated long-distance effects.
At the STCF, it will be better to study di-electron modes of the form $D\to \ee X$~\cite{TheBESIIICollaboration2018a},
which will provide sensitivities of $10^{-8}\sim 10^{-9}$ for $m_{\ee}$
in the range less polluted by long-range resonance contributions.
Compared to Belle II and LHCb, the STCF has competitive sensitivities in channels that contain neutral final states, such as photons and $\pi^0$, benefit from the almost full acceptance and quasi background-free advantages.
Furthermore, BESIII carried out world-first search for charmed meson decays into di-nutrinos $D^0 \to \pi^0 \nu \overline{\nu}$ and set the upper limit of ${\cal B}(D^0 \to \pi^0 \nu \overline{\nu})$ to be $2.1\times 10^{-4}$~\cite{BESIII:2021slf}. The STCF has the advantage of best constraining the upper limits on the
BFs for $D$ rare decays with neutrinos, such as
$D^0 \to \pi^0 \nu \overline{\nu}$ and $D^0 \to \gamma \nu \overline{\nu}$.

No evidence has been found for forbidden $D_{(s)}$-meson decays with
either LFV or LNV or both of them. The present experimental bounds
on the LFV decays are generally at the level of $10^{-6}$ to $10^{-5}$
(with the exception of ${\cal B}(D^0 \to \mu^\pm e^\mp)<1.3\times 10^{-8}$)~\cite{ParticleDataGroup:2022pth}.
The STCF will provide more stringent limits on
such interesting LFV and LNV decay modes, with a sensitivity of
$10^{-8}$ to $10^{-9}$ or smaller, taking advantage of its clean environment
and accurate charge discrimination.

\subsubsection{Charmed-meson production and spectroscopy}

The STCF will also act as a good playground for studying the production of charmed mesons and exploring charmed-meson spectroscopy.
To date, all of the 1$S$ and 1$P$ $D_{(s)}$ states have been found in experiments~\cite{Chen:2016spr,HFLAV:2022pwe}. However, almost all of the other predicted excited states in QCD-derived effective models are missing.
Furthermore, many excited open-charm states have been reported in experiments, and attempts to formulate an understanding of their nature remain controversial. Some of them are candidates for exotic mesons. For instance, the narrow $D^*_{sJ}(2632)$ state was observed by SELEX, but CLEO, BaBar and FOCUS all reported negative search results.
The unexpectedly low masses of the $D_{s0}^{*}(2317)$ and $D_{s1}(2460)$ have given rise to various exotic explanations, such as the $D^{(*)}K$ molecule state~\cite{Guo:2017jvc}. It has been claimed that strong $S$-wave $D^{(*)}K$ scattering contributes to the mass drop.
Thus, further systematic research on the open-charm meson spectra is highly desired.

At the STCF, it will be possible to produce excited charmed-meson states $D^{**}$ via direct
$\ee$ production processes, such as $\ee\to D^{**}\bar{D}^{(*)}(\pi)$, in the energy range from 4.1 to 7.0 GeV.
This will allow higher excited open-charm states to be studied through their hadronic or radiative decays~\cite{Kato:2018ijx} to lower open-charm states.
Systematic studies at the STCF on the open-charm meson spectra will provide important data for exploring nonperturbative QCD in the charm regime and testing various theoretical models.

%\documentclass[aps,eqsecnum,preprint,floats,epsf,epsfig,nofootinbib,letter]{revtex4}
%%\hoffset -.52in
%%\voffset 0.0in %% AS
%%\voffset 0.6in  States
%%\voffset -0.9in  %% pr, phys10
%%\voffset -0.5in    %% archive
%\textwidth 6.5in \textheight 9.0in
%\usepackage{epsfig,color}
%\usepackage{multirow}
%\usepackage{subfigure}
%
%\renewcommand{\baselinestretch}{1.00}
%
%\begin{document}
%\def\B{{\cal B}}
%\def\ov{\overline}
%\def\pr{{\sl Phys. Rev.}~}
%\def\prl{{\sl Phys. Rev. Lett.}~}
%\def\pl{{\sl Phys. Lett.}~}
%\def\np{{\sl Nucl. Phys.}~}
%\def\zp{{\sl Z. Phys.}~}
%\def\lsim{ {\ \lower-1.2pt\vbox{\hbox{\rlap{$<$}\lower5pt\vbox{\hbox{$\sim$}
%}}}\ } }
%\def\gsim{ {\ \lower-1.2pt\vbox{\hbox{\rlap{$>$}\lower5pt\vbox{\hbox{$\sim$}
%}}}\ } }
%
%\font\el=cmbx10 scaled \magstep2{\obeylines\hfill December, 2019}
%
%\vskip 1.5 cm
%
%\centerline{\large\bf Physics of Charmed Baryons}
%
%\vskip 1.5 cm
%
%\bigskip
%\bigskip
%\centerline{\bf Hai-Yang Cheng}
%\medskip
%\centerline{Institute of Physics, Academia Sinica}
%\centerline{Taipei, Taiwan 115, Republic of China}
%\medskip
%
%\bigskip
%\bigskip
%%\centerline{\bf Abstract}
%\bigskip
%\small
%
%
%
%
%\pagebreak
%
%\tableofcontents
%
%\newpage
%
%%\section{Introduction}

\subsection{Charmed baryons}

Theoretical interest in hadronic weak decays of charmed baryons peaked around the early 1990s and then faded away.
Nevertheless, there have been many progress in recent charmed baryon experiments in regard to hadronic weak decays of $\Lambda_c^+$. BESIII has played an essential role in these new developments. Motivated by the experimental progress, theoretical activity is growing in the study of hadronic weak decays of singly charmed baryons.

Charmed baryon spectroscopy provides an excellent basis for studying the dynamics of light quarks in the environment of a heavy quark. In the past decade,
many new excited charmed baryon states have been
discovered by BaBar, Belle, CLEO and LHCb. $B$ decays and the $e^+e^-\to c\bar c$ continuum are both very rich sources of charmed baryons. Many efforts have been made to identify the quantum numbers of these new states and to understand their properties.

\subsubsection{Hadronic weak decays}

Hadronic weak decays of singly charmed baryons, especially the two-body decay modes, provide essential inputs to understand the dynamics of strong interaction in the charm sector.

\begin{itemize}
\item{Nonleptonic decays of singly charmed baryons}

\noindent \underline{$\Lambda_c$ decays}
\vskip 0.2 cm
The branching fractions of the Cabibbo-allowed two-body decays of $\Lambda_c^+$ are listed in Table \ref{tab:BRs}. Many of these decays, such as $\Sigma^+\phi$, $\Xi^{(*)}K^{(*)+}$ and $\Delta ^{++}K^-$, can proceed only through $W$ exchange. Their experimental measurements imply the importance of $W$ exchange, which is not subject to color suppression in charmed baryon decays. Both Belle \cite{Zupanc} and BESIII \cite{BES:pKpi} have measured
the absolute branching fraction of the decay $\Lambda_c^+\to pK^-\pi^+$.
An average of $(6.28\pm0.32)\%$ for this benchmark mode is quoted by the PDG \cite{ParticleDataGroup:2022pth}. Furthermore,
the doubly Cabibbo-suppressed decay $\Lambda_c^+\to pK^+\pi^-$ has been observed by Belle \cite{Belle:DCS} and LHCb \cite{LHCb:DCS}. To complete the knowledge on the two-body decays, it is important to search for $pK^{0(*)}$ and $nK^{+(*)}$, which are doubly Cabibbo suppressed.

\begin{table}[hbtp]
\caption{The measured branching fractions of the Cabibbo-allowed two-body decays of $\Lambda_c^+$ (in units of \%) taken from the PDG~\cite{ParticleDataGroup:2022pth}. We have included the new BESIII measurements of $\Lambda_c^+\to \Lambda \rho^+$, $\Sigma^{*+}\pi^0$ and $\Sigma^{*0}\pi^+$~\cite{BESIII:2022udq}.} \label{tab:BRs}
\begin{center}
\begin{tabular}{lc | lc|lc}
\hline\hline ~~~Decay & $\B$ & ~~~Decay & $\B$ & ~~~Decay & $\B$  \\
\hline
~~$\Lambda^+_c\to \Lambda \pi^+$~~ & 1.30$\pm$0.07 & ~~$\Lambda^+_c\to \Lambda \rho^+$~~ & $ 4.06\pm 0.52$ & ~~$\Lambda^+_c\to \Delta^{++}K^-$  &  $1.08\pm0.25$\\
\hline
~~$\Lambda^+_c\to \Sigma^0 \pi^+$~~ & 1.29$\pm$0.07 & ~~$\Lambda^+_c\to \Sigma^0 \rho^+$  & & ~~$\Lambda^+_c\to \Sigma^{*0} \pi^+$ & $0.65\pm 0.10 $ \\
\hline
~~$\Lambda^+_c\to \Sigma^+ \pi^0$~~ & 1.25$\pm$0.10 & ~~$\Lambda^+_c\to \Sigma^+ \rho^0$~~  & $<1.7$ & ~~$\Lambda^+_c\to \Sigma^{*+} \pi^0$& $0.59\pm 0.08$\\
\hline
~~$\Lambda^+_c\to \Sigma^+ \eta$~~ & 0.44$\pm$0.20 & ~~$\Lambda^+_c\to \Sigma^+ \omega$~~  &  1.70$\pm$0.21 & ~~$\Lambda^+_c\to \Sigma^{*+}\eta$~~ & $1.05\pm0.23$ \\
\hline
~~$\Lambda^+_c\to \Sigma^+ \eta^\prime$~~ & 1.5$\pm$0.6 & ~~$\Lambda^+_c\to \Sigma^+ \phi$~~ & 0.38$\pm$0.06 & ~~$\Lambda^+_c\to \Sigma^{*+} \eta^\prime$ &\\
\hline
~~$\Lambda^+_c\to \Xi^0 K^+$~~ & 0.55$\pm$0.07 & ~~$\Lambda^+_c\to \Xi^0 K^{*+}$~~ &  & ~~$\Lambda^+_c\to \Xi^{*0}K^+$~~  & 0.43$\pm$0.09 \\
\hline
~~$\Lambda^+_c\to p K_S$~~ &  1.59$\pm$0.08 & ~~$\Lambda^+_c\to p \bar K^{*0}$~~  & 1.96$\pm$0.27 & ~~$\Lambda^+_c\to \Delta^+\bar K^0$~~ &  \\
\hline \hline
\end{tabular}
\end{center}
\end{table}

Various theoretical approaches to weak decays of heavy baryons have been investigated, including the current algebraic approach, the factorization scheme, the pole model, the relativistic quark model, the quark diagram scheme and the SU(3) flavor symmetry.
In general, the decay rates predicted by most models except the current algebraic scheme are below the experimental measurements.
Moreover, the decay asymmetries of the two-body hadronic weak decays of charmed baryons, defined as
$\alpha\equiv\frac{2{\rm Re}(s^* p)}{|s|^2+|p|^2}$, can be investigated. Here,
$s$ and $p$ represent the parity-violating $s$-wave and
parity-conserving $p$-wave amplitudes in the decay, respectively.
The pole model as well as the covariant quark model and its variants all predict a positive decay asymmetry $\alpha$ for both $\Lambda_c^+\to \Sigma^+\pi^0$ and $\Sigma^0\pi^+$; however, it was measured to be $-0.45\pm0.31\pm0.06$ for $\Sigma^+\pi^0$ by CLEO \cite{CLEO:alpha}. In contrast, the current algebraic approach always leads to a negative decay asymmetry for the two aforementioned modes: $-0.49$ in \cite{CT93}, $-0.31$ in \cite{Verma98}, $-0.76$ in \cite{Zen:1993} and $-0.47$ in \cite{Datta}. The issue with the sign of $\alpha_{\Sigma^+\pi^0}$ has finally been resolved by BESIII and Belle. The decay asymmetry parameters of $\Lambda_c^+\to \Lambda\pi^+$, $\Sigma^0\pi^+$, $\Sigma^+\pi^0$ and $pK_S$ were all recently measured by BESIII~\cite{BES:deasy}; for example, $\alpha_{\Sigma^+\pi^0}=-0.57\pm0.12$ was obtained. Hence, the negative sign of $\alpha_{\Sigma^+\pi^0}$ measured by CLEO has been confirmed by BESIII. Later, Belle confirmed the negative sign with the result of   $\alpha_{\Sigma^+\pi^0}=-0.463\pm0.018$~\cite{Belle:2022uod}.

\vskip 0.15 cm
\noindent \underline{$\Xi_c$ and $\Omega_c$ decays}
\vskip 0.2 cm
The absolute branching fractions of $\Xi_c^0\to \Xi^-\pi^+$ and $\Xi_c^+\to \Xi^-\pi^+\pi^+$
were measured by Belle \cite{Belle:Xic0,Belle:Xic+} to be
\begin{eqnarray}
\B(\Xi_c^0\to \Xi^-\pi^+)=(1.80\pm0.50\pm0.14)\%, \quad \B(\Xi_c^+\to \Xi^-\pi^+\pi^+)=(2.86\pm1.21\pm0.38)\%.
\end{eqnarray}
From these measurements, the branching fractions of other $\Xi_c^0$ and $\Xi_c^+$ decays can be inferred. No absolute branching fractions have been measured for the $\Omega_c^0$. The hadronic weak decays of the $\Omega_c^0$ have been theoretically studied in great detail in \cite{Dhir}, where most of the decay channels of $\Omega_c^0$ decays were found to proceed only through the
$W$-exchange diagram.

It is conceivable that nonleptonic decay modes of $\Lambda_c^+$ and $\Xi_c^{+,0}$ could be measured at the STCF with significantly improved precision. Priority will be given to the decay asymmetries $\alpha$ in various charmed baryon decays and the absolute branching fractions of $\Omega_c^0$ decays.

\item{Charm-flavor-conserving nonleptonic decays}

There is a special class of weak decays of charmed baryons that
can be studied reliably, namely, heavy-flavor-conserving
nonleptonic decays. Some examples are the singly
Cabibbo-suppressed decays $\Xi_c\to\Lambda_c\pi$ and
$\Omega_c\to\Xi_c\pi$. In these decays, not only
the light quarks inside the heavy baryon will participate in weak
interactions, but the charm quark also contributes to the $W$-exchange diagram through the transition $cs\to dc$.
The synthesis of the heavy quark and chiral
symmetries provides a natural setting for investigating these
reactions \cite{ChengHFC}. The predicted branching
fractions for the charm-flavor-conserving decays
$\Xi_c^0\to\Lambda_c^+\pi^-$ and $\Xi_c^+\to\Lambda_c^+\pi^0$ in early days were
of the order of $10^{-3}\sim 10^{-4}$ \cite{ChengHFC}. Recently, the first measurement of
the charm-flavor-conserving decay $\Xi_c^0\to\Lambda_c^+\pi^-$ was achieved by LHCb, with a branching fraction of $(0.55\pm0.02\pm0.18)\%$ \cite{Aaij:2020wtg}, which is confirmed later by Belle~\cite{Belle:2022kqi}. The theoretical estimates of $\B(\Xi_c\to\Lambda_c\pi)$ have been improved recently in \cite{Niu:2021qcc,Cheng:2022kea,Cheng:2022jbr}. The STCF should be able to cross-check the current measurements and search for another $c$-flavor-conserving weak decay, namely, $\Xi_c^+\to\Lambda_c^+\pi^0$.

\end{itemize}

\subsubsection{Semileptonic decays}

Exclusive semileptonic decays of charmed baryons, namely,
$\Lambda_c^+\to\Lambda e^+(\mu^+)\nu_{e(\mu)}$, $\Xi_c^+\to \Xi^0
e^+\nu_e$ and $\Xi_c^0\to \Xi^-e^+\nu_e$, have been observed
experimentally. Their rates depend on the ${\cal B}_c\to{\cal B}$
form factors $f_i(q^2)$ and $g_i(q^2)$ ($i=1,2,3$), defined as
\begin{eqnarray} \label{eq:FF}
 \langle {\cal B}_f(p_f)|V_\mu|{\cal B}_c(p_i)\rangle &=& \bar{u}_f(p_f)
[f_1(q^2)\gamma_\mu+if_2(q^2)\sigma_{\mu\nu}q^\nu+f_3(q^2)q_\mu] u_i(p_i),  \nonumber \\
 \langle {\cal B}_f(p_f)|A_\mu|{\cal B}_c(p_i)\rangle &=& \bar{u}_f(p_f)
[g_1(q^2)\gamma_\mu+ig_2(q^2)\sigma_{\mu\nu}q^\nu+g_3(q^2)q_\mu]\gamma_5
u_i(p_i).
\end{eqnarray}
These form factors have been evaluated using the
nonrelativistic quark model \cite{Marcial,Singleton,CT96,Pervin},
the MIT bag model \cite{Marcial}, the relativistic quark model \cite{Ivanov96,Gutsche,Faustov:semi}, the light-front quark model \cite{Luo}, QCD
sum rules \cite{Carvalho,Huang,Azizi} and LQCD \cite{Meinel:LamcLam,Meinel:Lamcn}. Many of the early predictions of
$\B(\Lambda_c^+\to\Lambda e^+\nu_e)$ are smaller than the first measurement of the absolute branching fraction of $(3.6\pm0.4)\%$ reported by BESIII \cite{BESIII:Lambdaenu}. However, the LQCD calculations in \cite{Meinel:LamcLam} show good agreement with the experimental results for both $\Lambda_c^+\to\Lambda e^+\nu_e$ and $\Lambda_c^+\to\Lambda \mu^+\nu_\mu$. Needless to say,
the semileptonic decays of $\Lambda_c^+$ (including the yet-to-be-observed $\Lambda_c^+\to ne^+\nu_e$), $\Xi_c^{+,0}$ and $\Omega_c^0$, which can be used to discriminate between different form-factor models, will be thoroughly studied at the STCF.

\subsubsection{Electromagnetic and weak radiative decays}

The electromagnetic decays of interest in the charmed baryon sector are the following:
(i) $\Sigma_c \rightarrow
\Lambda_c + \gamma$ and $\Xi^\prime_c \rightarrow \Xi_c + \gamma$; (ii)
$\Sigma^\ast_c \rightarrow \Lambda_c + \gamma$ and $\Xi^{
\ast}_c \rightarrow
\Xi_c + \gamma$; (iii) $\Sigma^\ast_c \rightarrow
\Sigma_c + \gamma$, $\Xi^{ \ast}_c \rightarrow
\Xi^\prime_c + \gamma$, and $\Omega^\ast_c \rightarrow \Omega_c
+ \gamma$; and (iv) $\Lambda_c(2595, 2625)\to\Lambda_c+\gamma$ and $\Xi_c(2790,2815)\to \Xi_c+\gamma$.
Among them, the decay modes $\Xi'^0_c\to\Xi_c^0\gamma$, $\Xi'^+_c\to
\Xi^+_c\gamma$ and $\Omega_c^{*0}\to\Omega_c^0\gamma$ have been experimentally observed.

{
%\squeezetable
\begin{table}[tp]
\centering
\footnotesize
\caption{Electromagnetic decay rates (in units of keV) of $s$-wave charmed
baryons in heavy hadron chiral perturbation theory to LO \cite{Cheng97,Cheng93}, NLO \cite{Jiang} and NNLO  \cite{Wang:2018}. } \label{tab:em}
\begin{center}
\begin{tabular}{c c c c c c c c c c}
\hline\hline  & $\Sigma^+_c\to \Lambda_c^+\gamma$ & $\Sigma_c^{*+}\to\Lambda_c^{+}\gamma$ & $\Sigma_c^{*++}\to\Lambda_c^{++}\gamma$ & $\Sigma_c^{*0}\to\Sigma_c^0\gamma$ &  $\Xi'^+_c\to\Xi_c^+\gamma$  &  $\Xi^{*+}_c\to\Xi_c^+\gamma$ & $\Xi^{*0}_c\to\Xi_c^0\gamma$ & $\Xi'^0_c\to\Xi_c^0\gamma$ & $\Omega_c^{*0}\to\Omega_c^0\gamma$ \\ \hline
 LO & 91.5 & 150.3 & 1.3 & 1.2 & 19.7 & 63.5 & 0.4 & 1.0 & 0.9 \\
 NLO & 164.2 & 893.0 & 11.6 & 2.9 & 54.3 & 502.1 & 0.02 & 3.8 & 4.8 \\
 NNLO & 65.6 & 161.8 & 1.2 & 0.49 & 5.4 & 21.6 & 0.46 & 0.42 & 0.32 \\
 \hline \hline
\end{tabular}
\end{center}
\end{table}
}

The calculated results of \cite{Cheng97,Cheng93}, \cite{Jiang} and \cite{Wang:2018}, denoted by (i), (ii) and (iii), respectively, in Table \ref{tab:em}, can be regarded as the predictions of heavy hadron chiral perturbation theory (HHChPT) to the leading order (LO), next-to-leading order (NLO) and next-to-next-to-leading order (NNLO), respectively.
It is not clear why the predictions of HHChPT to NLO are quite different from those to LO and NNLO for the following three modes: $\Sigma_c^{*+}\to\Lambda_c^+\gamma$, $\Sigma_c^{*++}\to \Sigma_c^{++}\gamma$ and $\Xi^{*+}_c\to\Xi_c^+\gamma$.
It is naively expected that all HHChPT approaches should agree with each other to the lowest order of chiral expansion provided that the coefficients are inferred from the nonrelativistic quark model. This issue can be clarified by the STCF through the measurement of these decay rates.

Very recently, Belle observed the electromagnetic decays of the orbitally excited charmed baryons $\Xi_c(2790)$ and $\Xi_c(2815)$ for the first time \cite{Belle:charme.m.}. The partial widths of $\Xi_c(2815)^0\to\Xi_c^0\gamma$ and $\Xi_c(2790)^0\to\Xi_c^0\gamma$ were measured to be $320\pm45^{+45}_{-80}$ keV and $\sim 800$ keV, respectively. However, no signal was found for the analogous decays of $\Xi_c(2815)^+$ and $\Xi_c(2790)^+$.

Weak radiative decays such as $\Lambda_c^+\to\Sigma^+\gamma$ and $\Lambda_c^+\to p\gamma$ can proceed through the bremsstrahlung processes $cd\to us\gamma$ (Cabibbo allowed) and $cd\to ud\gamma$ (Cabibbo suppressed), respectively. Upper limits on the branching fraction of the former have been set to be $2.6\times 10^{-4}$ and $4.4\times 10^{-4}$ by Belle \cite{Belle:2022raw} and BESIII \cite{BESIII:2022rox}, respectively, which are in agreement with standard model expectations.

\subsubsection{$CP$ violation}

The CKM matrix contains a single phase that implies the
existence of $CP$ violation. This means that $CP$ violation can be studied in baryons as well. However, the predicted $CP$-violating asymmetries are small for charmed
baryons.
The search for $CP$ violation in charmed baryon decays has gained new momentum
with the large samples of $\Lambda_c$ obtained by BESIII and LHCb. For two-body
decays of the $\Lambda_c^+$, $CP$ violation can be explored through the measurement of the $CP$-violating asymmetry ${\cal A}=(\alpha+\bar\alpha)/(\alpha-\bar\alpha)$, which corresponds to the asymmetries $\alpha$ for the $\Lambda_c^+$ decays and $\bar\alpha$ for the $\bar{\Lambda}_c^-$ decays. For example, the most precise single measurement of 
${\cal A}$ in $\Lambda_c^+\to \Lambda K^+$ and $\bar\Lambda_c^-\to\bar\Lambda K^-$ is reported by BELLE to be $(-58.5\pm4.9\pm1.8)\%$ \cite{Belle:2022uod}. At the STCF, much more sensitive searches for $CP$ violation will be carried out by combining single-tag $\Lambda_c^+$ data~\cite{BES:deasy} with double-tag $\Lambda_c^+\bar{\Lambda}{}_c^-$ data, where the $\Lambda_c^+\bar{\Lambda}{}_c^-$ pairs are quantum correlated in regard to the alignment of their spins with the initial spins of the virtual photons.
In particular, with polarized beams~\cite{Bondar:2019zgm}, the unique advantage of enhanced sensitivities to the decay asymmetries and $CP$ violation can be achieved with prior knowledge of the spin direction of the produced $\Lambda_c^+$.
Regarding three-body decays,
LHCb has measured $\Delta A_{CP}$ as the difference between the $CP$ asymmetries in
the $\Lambda_c^+ \to p K^+ K^-$ and $\Lambda_c^+ \to p\pi^+ \pi^-$ decay channels.
The result is $\Delta A_{CP} = (0.30 \pm 0.91 \pm 0.61)\%$ \cite{Aaij:2017xva},
to be compared with the generic SM prediction of a fraction of 0.1\%~\cite{Bigi:2012ev}.
To probe the SM contribution to such asymmetries, it will be necessary to increase the available statistics by at least a factor of 100.

For $\Lambda_c^+$ decays with multiple hadrons in the final state, such as $\Lambda_c^+\to pK^-\pi^+\pi^0$, $\Lambda_c^+\to\Lambda\pi^+\pi^+\pi^-$ and $\Lambda_c^+\to pK_S\pi^+\pi^-$, $CP$ violation can be exploited through several $T$-odd observables. By virtue of its characteristics of high luminosity, broad center-of-mass energy acceptance, abundant production and a clean environment, the STCF will serve as an excellent platform for this kind of study.
A fast Monte Carlo simulation~\cite{Shi:2019vus} of 1 ab$^{-1}$ $e^+e^-$ annihilation data at $\sqrt{s}=4.64$ GeV, which is expected to be available at the future STCF, indicates that a sensitivity at the level of (0.25--0.5)\% is accessible for the three abovementioned decay modes. This will be sufficient to measure nonzero $CP$-violating asymmetries as large as 1\%.

\subsubsection{Spectroscopy}

\begin{table}[tp]
\caption{Antitriplet and sextet states of charmed baryons.
The mass differences $\Delta m_{\Xi_c\Lambda_c}\equiv m_{\Xi_c}-m_{\Lambda_c}$, $\Delta m_{\Xi'_c\Sigma_c}\equiv m_{\Xi'_c}-m_{\Sigma_c}$, and $\Delta m_{\Omega_c\Xi'_c}\equiv m_{\Omega_c}-m_{\Xi'_c}$ are all in units of MeV. } \label{tab:3and6}
\begin{center}
\begin{tabular}{c| ccc } \hline\hline
  & $J^P(nL)$ & States & Mass difference(s)  \\
 \hline
 ~~${\bf \bar 3}$~~ & ~~${1\over 2}^+(1S)$~~ &  $\Lambda_c(2287)^+$, $\Xi_c(2470)^+,\Xi_c(2470)^0$ & ~~$\Delta m_{\Xi_c\Lambda_c}=183$ ~~  \\
 & ~~${1\over 2}^-(1P)$~~ &  $\Lambda_c(2595)^+$, $\Xi_c(2790)^+,\Xi_c(2790)^0$ & $\Delta m_{\Xi_c\Lambda_c}=198$  \\
 & ~~${3\over 2}^-(1P)$~~ &  $\Lambda_c(2625)^+$, $\Xi_c(2815)^+,\Xi_c(2815)^0$ & $\Delta m_{\Xi_c\Lambda_c}=190$  \\
 & ~~${1\over 2}^+(2S)$~~ &  $\Lambda_c(2765)^+$, $\Xi_c(2970)^+,\Xi_c(2970)^0$ & $\Delta m_{\Xi_c\Lambda_c}=200$  \\
 & ~~${3\over 2}^+(1D)$~~ &  $\Lambda_c(2860)^+$, $\Xi_c(3055)^+,\Xi_c(3055)^0$ & $\Delta m_{\Xi_c\Lambda_c}=201$  \\
 & ~~${5\over 2}^+(1D)$~~ &  $\Lambda_c(2880)^+$, $\Xi_c(3080)^+,\Xi_c(3080)^0$ & $\Delta m_{\Xi_c\Lambda_c}=196$  \\
 \hline
 ~~${\bf 6}$~~ & ~~${1\over 2}^+(1S)$~~ &  $\Omega_c(2695)^0$, $\Xi'_c(2575)^{+,0},\Sigma_c(2455)^{++,+,0}$ & ~~~~$\Delta  m_{\Omega_c\Xi'_c}=119$, $\Delta m_{\Xi'_c\Sigma_c}=124$~~  \\
 & ~~~${3\over 2}^+(1S)$~~~ &  $\Omega_c(2770)^0$, $\Xi'_c(2645)^{+,0},\Sigma_c(2520)^{++,+,0}$ & ~~~~ $\Delta m_{\Omega_c\Xi'_c}=120$, $\Delta m_{\Xi'_c\Sigma_c}=128$~~ \\
 \hline\hline
\end{tabular}
\end{center}
\end{table}

The observed antitriplet and sextet states of charmed baryons are listed in Table \ref{tab:3and6}. At present, the $J^P={1\over 2}^+$, $\frac12^-$, $\frac32^+$, $\frac32^-$ and $\frac52^+$ antitriplet states of $\Lambda_c$ and $\Xi_c$ and the
$J^P={1\over 2}^+$ and ${3\over 2}^+$ sextet states of $\Omega_c$, $\Xi'_c$, and $\Sigma_c$
have been established. The highest state $\Lambda_c(2940)^+$ in the $\Lambda_c$ family was first discovered by BaBar in the $D^0p$ decay mode~\cite{BaBar:Lamc2940}, but its spin-parity assignment is quite diverse (see Refs.~\cite{Cheng:2015,HFLAV:2022pwe}for  review). The constraints on its spin and parity were recently found to be $J^P=\frac32^-$ by LHCb~\cite{LHCb:Lambdac2880}. It was suggested in Ref.~\cite{Cheng:Omegac} that the quantum numbers of $\Lambda_c(2940)^+$ are likely to be $\frac12^-(2P)$ based on the Regge analysis. However, it was argued in Ref.~\cite{Luo:2019qkm} that $\Lambda_c(2940)^+$ is a $\frac32^-(2P)$ state and that there exists a state $\frac12^-(2P)$ higher than the $\Lambda_c(2P, 3/2^-)$. 
This issue can be clarified by the STCF.

In 2017, LHCb explored the charmed baryon sector of the $\Omega_c$ and observed five narrow excited $\Omega_c$ states decaying into $\Xi_c^+K^-$: $\Omega_c(3000)$, $\Omega_c(3050)$, $\Omega_c(3066)$, $\Omega_c(3090)$ and $\Omega_c(3119)$ \cite{LHCb:Omegac}. With the exception of the $\Omega_c(3119)$ state, the first four states were also later confirmed by Belle \cite{Belle:Omegac}. This has triggered considerable interest in the possible identification of their spin-parity quantum numbers. In addition to the five previously observed excited $\Omega_c^0$ states, LHCb has recently reported two new excited states, $\Omega_c(3185)$ and $\Omega_c(3227)$, observed in the $\Xi_c^+ K^-$ spectrum \cite{LHCb:2023rtu}.

Within the energy region of the STCF up to $7$~GeV, it will be feasible to study the spectra of the singly charmed baryon states $\Lambda_c$, $\Sigma_c$, $\Xi_c^{(\prime)}$, $\Omega_c$ and their excited states in the energy range of $5 \sim 7$ GeV. It will be important for the STCF to explore their possible structure and spin-parity quantum number assignments, especially for the five new narrow $\Omega_c$ resonances.
If the energy region were to be extended to above $7.4$~GeV, the production of the doubly charmed baryon $\Xi^{++}_{cc}$ would also be allowed. This would enable a more detailed study of the recently discovered doubly charmed baryons.
%It is a very tempting future project for upgrade.

%\input{03_ref_CharmHadron_v2}

\newpage
\section{Tau Physics}
\label{sec:tau}
At the STCF, as many as $3.5\times10^{9}$ $\tau^+\tau^-$ pairs can be produced per year at $\sqrt{s} = 4.26~\GeV$, which is approximately 3 orders of magnitude higher than the currently accumulated number of $\tau^+\tau^-$ events at BESIII. At the production threshold, there could be as many as $10^{8}$ $\tau^+\tau^-$ events per year. Under near-threshold conditions, data from just below the threshold can be used to understand the background to achieve better control over systematic errors compared with BESIII~\cite{YB}. In this regard, the STCF has advantages over both LHCb and Belle II for $\tau$ physics studies. In the energy range covered by the STCF, good control can also be exerted over the polarizations of the $e^+$ and $e^-$ beams to extract new information about $\tau$ physics. The STCF will tremendously increase the statistical significance for $\tau$-related physics studies and will reach a level of precision that has never been achieved before.

The $\tau$ lepton occupies a unique place in the SM. Being the heaviest charged lepton, it has many more decay channels than the next lighter charged lepton, the muon ($\mu$). With an unprecedented number of $\tau$s produced not far from the threshold and possible polarization information at the STCF, one can gain more precise knowledge of not only the properties of the $\tau$ itself but also how it interacts with other particles; thus, one can more precisely determine the SM parameters, probe possible new interactions and possibly also shed light on some of the related anomalies in particle physics. In the following, we describe some of the interesting subjects in $\tau$ physics that can be addressed at the STCF.

\subsection{Precision measurement of the $\tau$ properties}

To test the SM and search for new physics in the $\tau$ sector, it is important for the properties of the $\tau$ to be known with great precision. Here, we list a few measurements at the STCF that can improve our understanding of the $\tau$ properties.

\subsubsection{$\tau$ mass and lifetime}
Many of the tests for the SM and beyond involve precise measurements of the $\tau$ mass ($m_\tau$) and lifetime. While at the threshold for $\tau^+\tau^-$ pair production, measurement of the $\tau$ lifetime is difficult, at the high-energy ($5\sim 7$~GeV) end of the STCF range, it could be possible to measure it by reconstructing the $\tau^+\tau^-$ vertex. With sufficiently high statistics, there is a chance to improve the measurement of the $\tau$ lifetime, for which a more dedicated study would be needed. On the other hand, the mass measurement can also be improved. The mass has been measured at the 70~ppm level, with a world average of~\cite{ParticleDataGroup:2022pth} $m_\tau = 1776.86\pm0.12~\MeV$. In charged-current induced leptonic decays, $\tau \to \nu_\tau l \bar \nu_l$ $(l= e, \mu)$, the decay widths are proportional to the fifth power of $m_\tau$. Consequently, a small error in the mass can cause significant deviations in tests of the universality of the SM and in the search for new physics. At the STCF, the number of $\tau$s produced may be one to three orders of magnitude greater than at BESIII, which will greatly enhance the statistical significance achieved. With further improvements in particle ID and energy measurement capabilities, the improved sensitivity can increase the accuracy by a factor of 7 to reach a level better than 10~ppm. This improved $\tau$ mass measurement will consolidate the basis for any further $\tau$ physics studies. It should be noted that with high precision foreseen, the formation of the ditauonium state cannot be 
ignored in the $\tau$ mass measurement~\cite{dEnterria:2023yao}.

\subsubsection{Measurement of $a_\tau = (g_\tau-2)/2$}
The anomalous magnetic dipole moment of the $\tau$ lepton, $a_\tau$, is another property of fundamental importance. The corresponding $a_l$ values for the electron and muon have been measured to high precision. For the electron, there is a $2\sigma$ deviation between the measurement and the SM prediction, $\Delta a_e = a^\textrm{exp}_{e} - a^\textrm{SM}_e =-78(36)\times 10^{-14}$~\cite{electron-mdm}. On the other hand, there is a longstanding and larger discrepancy for the muon moment $a_\mu$, which is currently being measured at Fermilab and J-PARC. Very recently, Fermilab reported their new result from the Run 1 measurement~\cite{Fermilab}. Upon combining it with previous data from BNL, the discrepancy is now $\Delta a_\mu = a^\textrm{exp}_\mu - a^\textrm{SM}_\mu = (251\pm 59)\times 10^{-11}$, and its significance level has been enhanced from $3.7\sigma$ to $4.2\sigma$. As this may be an indication of new physics, it has motivated extensive theoretical studies within the SM and beyond to understand possible causes. It is therefore important to test whether there is also a deviation in $a_\tau$. This is especially important for testing models of new physics that include states whose couplings are proportional to mass.

However, the measurement of $a_\tau$ is drastically different from that of $a_{e,\mu}$ due to the short lifetime of the $\tau$. The SM prediction for $a_\tau$ is $1177.21(5)\times 10^{-6}$~\cite{tau-mdm}. Currently, $a_\tau$ has been measured from the production cross section for $\tau$ pairs together with the spin or angular distributions of the $\tau$ decays; for instance, the current bounds of $-0.052 \leq a_\tau \leq 0.013$ (95\% C.L.) were obtained by the DELPHI collaboration~\cite{Abdallah:2003xd} from the cross section for the process $e^+e^-\to e^+e^-\tau^+\tau^-$ under the assumption that the SM tree-level result is modified only by the anomalous magnetic moment. These measurements are still far from constituting a precision test for the SM, and conventional measurements through similar processes may never reach the necessary level of precision. To overcome this bottleneck, a new method has recently been proposed in Ref.~\cite{Chen:2018cxt}, in which it was found to be feasible to reach a precision level of $1.75\times 10^{-5}$ at Belle II before considering systematics. In addition, it was shown some time ago that in $e^+e^- \to \tau^+\tau^-$ with a polarized electron beam, it would be plausible to achieve this precision goal at the STCF by measuring the transverse and longitudinal polarizations of the $\tau$ lepton~\cite{bernabeu-mdm}. It has been argued that if $\tau^+\tau^-$ pairs are produced on top of the narrow $\Upsilon(1S,2S,3S)$ resonances, with a very well-controlled background near the threshold, a precision even better than that of Belle II can be expected. Nevertheless, it has also been pointed out in Ref.~\cite{Eidelman:2016aih} that an energy spread with $e^+e^-$ beams on the order of a few MeV, which is likely to occur, would make such a measurement impractical because the resonant contributions would be contaminated by nonresonant ones of at least similar size, which would need to be subtracted to extract the dipole moment. In addition, the momentum transfer is too large to be directly related to dipole moments. The authors of Ref.~\cite{Eidelman:2016aih} proposed another method of measuring dipole moments, i.e., by means of radiative decays of the form $\tau^-\to l^-\nu_\tau\bar\nu_l\gamma$. However, they estimated that the sensitivity to $a_\tau$ would be approximately $0.085$ ($0.012$) using the full data of Belle (Belle II), which offers no meaningful improvement compared to the sensitivity of $0.017$ at DELPHI,
%Editor: Please ensure that the intended meaning has been maintained in the above edit.
and the sensitivity to $d_\tau^\gamma$ cannot be improved either. Therefore, more critical studies are needed.

\subsection{Determination of the SM parameters}
\label{subsec:tausm}
The $\tau$ lepton has well-defined interactions with other particles in the SM. The experimental measurements are consistent with the SM predictions~\cite{TauR}. With a large sample of $\tau$s, many of the interaction parameters in the SM can be determined with great precision. Here, we discuss some of the most important of these tests: the universality properties, the Michel parameters, the strong coupling constant $\alpha_s$, and the element $V_{us}$ in the Cabibbo--Kobayashi--Maskawa (CKM) mixing matrix.

\subsubsection{The universality test}

The charged-current interaction of the left-handed leptons with the $W$ boson is described by
\begin{eqnarray}
{\cal L} = -{g_i\over \sqrt{2}} \bar l_i \gamma^\mu P_L \nu_i W^-_\mu + \textrm{H.C.},
\end{eqnarray}
where $P_L = (1-\gamma_5)/2$. The term `charged lepton universality' refers to the fact that $g_e=g_\mu = g_\tau$. This is indeed the case in the SM but is not necessarily so in models beyond the SM. Therefore, these quantities can be measured to test the SM. One can obtain the following~\cite{ExpTau} using the very good approximation $B(\mu\to e\bar\nu_e\nu_\mu(\gamma))\approx 1$:
\begin{eqnarray}
{g_\tau\over g_e} &=& \sqrt{ B(\tau^- \to \mu^- \bar \nu_\mu \nu_\tau(\gamma)) {\tau_\mu\over \tau_\tau}
{m^5_\mu\over m^5_\tau} {F_\textrm{corr}(m_\mu, m_e)\over F_\textrm{corr}(m_\tau, m_\mu)}} \;,
\nonumber\\
{g_\tau\over g_\mu} &=& \sqrt{ B(\tau^- \to e^- \bar \nu_e \nu_\tau(\gamma)) {\tau_\mu\over \tau_\tau}
{m^5_\mu\over m^5_\tau} {F_\textrm{corr}(m_\mu, m_e)\over F_\textrm{corr}(m_\tau, m_e)}} \;,
\end{eqnarray}
where $F_\textrm{corr}(m_i, m_j)$ includes radiative corrections and corrections due to the different charged lepton masses. The current data $g_\tau/g_e=1.0029\pm 0.0015$, $g_\mu /g_e = 1.0019\pm 0.0014$, and $g_\tau/g_\mu = 1.0010\pm 0.0015$~\cite{ExpTau} are consistent with the prediction of universality. As discussed earlier, by improving the measurement of the value of $m_\tau$ to a level better than 10~ppm, the universality prediction could be tested at a level more than 3 times better to constrain the allowed room for new physics.

Universality tests could also be carried out by combining the decays $\tau \to P\nu_\tau$ and $P\to l \bar \nu_l$ (with $P=\pi$ and $K$, $l = \mu$ and $e$, and $\nu_l = \nu_\mu$ and $\nu_e$), as the ratio of their decay widths is proportional to $g_\tau^2/g_l^2$:
\begin{eqnarray}
R_l = {\Gamma(\tau \to P\nu_\tau) \over \Gamma(P\to l \bar \nu_l)} {m_\tau/(m^2_\tau - m^2_P)^2\over m_P/(m^2_P - m^2_l)^2} = {g^2_\tau\over g^2_l} \;.
\end{eqnarray}
All these decays have been measured experimentally, with $B(\tau^- \to \pi^-\nu_\tau) = (10.82\pm 0.05)\% $,
$B(\tau^- \to K^-\nu_\tau) = (6.96\pm0.10)\%$,
$B(\pi^- \to \mu^- \bar \nu_\mu) = (99.98770\pm 0.00004)\%$,
$B(\pi^- \to e^- \bar \nu_e) = ( 1.230\pm0.0004)\%$,
$B(K^- \to \mu^- \bar \nu_\mu) = (63.56\pm 0.11)\%$,
and
$B(K^-\to e^-\bar\nu_e) = ( 1.582\pm0.007)10^{-5}$~\cite{ParticleDataGroup:2022pth}.
The error bars for the $\tau \to \pi(K)\nu_\tau$ decays are presently not as good as those for the pure leptonic $\tau \to \nu_\tau l \bar \nu_l$ decays and yield a weaker constraint. However, with improved sensitivity for $\tau \to\pi(K)\nu_\tau$ (and especially with more monochromatic $\pi(K)$ data near the $\tau^+\tau^-$ production threshold) at the STCF together with improved higher-order theoretical corrections, these decays will provide complementary universality tests.

\subsubsection{The Michel parameters}

Decays of the form $\tau \to l \bar \nu_l \nu_\tau$ provide sensitive constraints on other forms of interactions due to new physics. Barring exotic interactions such as tensor couplings, the most general form of new physics can be parameterized in terms of the Michel parameters $\rho$, $\eta$, $\xi$, and $\delta$~\cite{ParticleDataGroup:2022pth}:
\begin{eqnarray}
&&{d^2\Gamma(\tau \to l \bar \nu_{l} \nu_\tau) \over x^2 dx d\cos\theta}
{96\pi^3\over G^2_F m^5_\tau}
\nonumber\\
&=&3(1-x) + \rho_l \bigg({8\over 3}x -2\bigg) +
6\eta_l {m_l\over m_\tau}{(1-x)\over x}
-P_\tau\xi_l\cos\theta \bigg[(1-x) + \delta_l\bigg({8\over 3}x
-2\bigg)\bigg]\;,
\end{eqnarray}
where $P_\tau$ is the degree of $\tau$ polarization, $x= E_{l}/ E^\textrm{max}_{l}$, and $\theta$ is the angle between the $\tau$ spin and the $l$ momentum direction. In the SM, the Michel parameters are
\begin{eqnarray}
\rho_l = {3\over 4}\;,\;\;\eta_l = 0\;,\;\;\xi_l = 1\;,\;\;\xi_l\delta_l = {3\over 4}\;.
\end{eqnarray}
Experimentally, the values are~\cite{ParticleDataGroup:2022pth}
\begin{eqnarray}
&&\rho_e = 0.747\pm 0.010,\;\rho_\mu = 0.763\pm 0.020,\;\;\xi_e = 0.994\pm 0.040,\;\;\xi_\mu = 1.030\pm 0.059,
\\
&&\eta_e = 0.013\pm 0.020,\;\;\eta_\mu = 0.094\pm 0.073,\;\;
(\xi\delta)_e = 0.734\pm 0.028,\;\;(\xi\delta)_\mu = 0.778\pm 0.037.
\nonumber
\end{eqnarray}
Again, the experimental measurements are consistent with the SM predictions.

With the production of a larger number of $\tau$s and improved sensitivities, the STCF will be capable of reducing the error bars by at least a factor of 2. In addition, rare decays such as radiative leptonic decays~\cite{Arbuzov:2016ywn,Lees:2015gea,Shimizu:2017dpq} and multi-charged-lepton decays~\cite{Eidelman:2016aih,Flores-Tlalpa:2015vga} can also be studied at the STCF. This will help to examine the SM electroweak interactions and place limits on new physics contributions.

\subsubsection{Extraction of the strong coupling $\alpha_s$}

It is well known that the strong coupling constant $\alpha_s$ can be extracted from the following ratio~\cite{Tau0}:
\begin{equation}
R_\tau = \frac{\Gamma (\tau^- \to \nu_\tau {\rm hadrons} )}{\Gamma (\tau^-\to\nu_\tau e^- \bar\nu_e)}.
\end{equation}
The theoretical predictions of this ratio have been carefully examined in \cite{Tau1,Tau2}. In accordance with the structure of the weak interactions and the classification of the final states, the ratio can be decomposed as follows:
\begin{equation}
R_\tau = R_{V,ud} + R_{A,ud} + R_{\tau,s}.
\end{equation}
Here, $R_{\tau,s}$ is the contribution from final states containing an $s$ quark, while $R_{V,ud}$ ($R_{A,ud}$) comes from nonstrange final states involving an even (odd) number of pions. Each term contains perturbative and nonperturbative contributions. The perturbative contributions are currently determined at the 5-loop level, while the nonperturbative contributions are estimated via QCD sum rules. Because of the large quark mass $m_s$, a large power correction exists in $R_{\tau,s}$, whose theoretical estimate therefore cannot reach the level of precision of $R_{V,ud}$ and $R_{A,ud}$. The analysis presented in~\cite{TauR} gives the value
\begin{equation}
  \alpha_s (m_\tau) = 0.331\pm 0.013,
\end{equation}
with one set of parameterizations of nonperturbative contributions. To improve the determination, an experimental study at the STCF will be important. Specifically, a precise measurement of $R_{\tau,s}$ and the spectral function containing the strange quark will help to understand the nonperturbative contributions and to precisely extract the CKM matrix element $V_{us}$.

\subsubsection{Extraction of the CKM matrix element $V_{us}$}

The experimental study of hadronic decays of $\tau$ has yielded one of the most precise measurements of $V_{us}$ to date. There are two main methods of determining this parameter. One is by measuring the ratio of the decay widths for $\tau^-\to\pi^- \nu_\tau$ and $\tau^- \to K^-\nu_\tau$, and the other is by measuring the ratio $R_\tau = R_{V,ud} + R_{A,ud} + R_{\tau,s}$ as discussed earlier. Theoretically,
\begin{eqnarray}
&&{B(\tau \to K^- \nu_\tau)\over B(\tau^- \to \pi^- \nu_\tau)} = {f^2_K\over f_\pi^2} {\vert V_{us}\vert^2\over \vert V_{ud}\vert^2}
{(m^2_\tau - m^2_K)^2\over (m^2_\tau - m^2_\pi)^2} {1+\delta R_{\tau/K}\over 1+\delta R_{\tau/\pi}} (1+\delta R_{K/\pi})\;,
\nonumber
\\
&&\vert V_{us}\vert ^2 =
{R_{\tau,s}\over [(R_{V,ud}+R_{A,us})/\vert V_{ud}\vert^2 - \delta R_\textrm{theory}]}\;.
\end{eqnarray}
With the known values from theoretical calculations and experimental measurements~\cite{ExpTau}, namely, $f_K/f_\pi=1.1930\pm 0.0030$,
$V_{ud} = 0.97417\pm 0.00021$, $1+\delta R_{\tau/K} = 1+(0.90\pm 0.22)\%$, $1+\delta R_{\tau/\pi} = 1+(0.16\pm 0.14)\%$, $1+\delta R_{K/\pi} = 1+(-1.13\pm 0.23)\%$, and $\delta R_\textrm{theory} = 0.242\pm 0.032$, one respectively obtains the following results from the above two methods:
\begin{eqnarray}
\vert V_{us} \vert_{\tau K/\pi} = 0.2236\pm 0.0018\;,\;\;\vert V_{us}\vert_{\tau s} = 0.2186\pm 0.0021\;.
\end{eqnarray}
The first value is $1.1~\sigma$ away from the value determined by the unitarity relation, $\vert V_{us}\vert_\textrm{uni} \approx \sqrt{1-\vert V_{ud}\vert^2}= 0.2258\pm0.0009 $, and the second is $3.1~\sigma$ away from $\vert V_{us}\vert_\textrm{uni}$. These deviations need to be further understood with better precision before evidence of new physics beyond the SM can be claimed.

The STCF can measure the values of $R_i$ and may therefore confirm or refute these deviations.

\subsection{$CP$ symmetry tests}

How $CP$ symmetry is broken may hold the key to why our universe contains more matter than antimatter. The violation of $CP$ symmetry is one of the required conditions to understand this. There is insufficient $CP$ violation in the SM to explain this fundamental question affecting our very existence in the Universe, and therefore, new sources of $CP$ violation are demanded. The search for new $CP$-violating effects is one of the most active areas in particle physics. Physical processes involving the $\tau$ lepton are potential sectors in which new $CP$-violating effects may appear.

\subsubsection{$CP$ violation in $\tau^- \to K^0_S \pi^- \nu_\tau$}

In the SM, because of the $CP$ violation in $K^0$--$\bar K^0$ mixing, a detectable $CP$-violating effect is predicted for this process~\cite{Bigi:2005ts1, Bigi:2005ts2}:
\begin{eqnarray}
A_Q = {B(\tau^+ \to K^0_S \pi^+ \bar \nu_\tau) - B(\tau^- \to K^0_S \pi^- \nu_\tau) \over
B(\tau^+ \to K^0_S \pi^+ \bar \nu_\tau) + B(\tau^- \to K^0_S \pi^- \nu_\tau)} = (+0.36\pm 0.01)\%\;.
\end{eqnarray}
While Belle observed no $CP$ violation in the angular distributions for the exclusive decays~\cite{Bischofberger:2011pw},
BaBar yielded a value of $A_Q=(-0.36\pm0.23\pm0.11)\%$ for the inclusive decays with $\ge 0\pi^0$ in the final states~\cite{BABAR:2011aa}, which is $2.8\sigma$ away from the SM prediction.

The above deviation represents a challenge to the SM. Theoretical efforts have been made to reconcile this deviation. However, even with beyond-the-SM effects included, it is not easy to obtain the central value of the BaBar data. The STCF can provide a crucial check with a large number of $\tau^+\tau^-$ pairs produced not far from the threshold, where the background can be well controlled. At the STCF, the expected luminosity of 1~ab$^{-1}$/year at an energy of 4.26~GeV can allow a statistical sensitivity of $9.7\times 10^{-4}$ to $CP$ violation to be reached. With 10 years of operation, the sensitivity can reach $3.1\times 10^{-4}$~\cite{Sang:2020ksa}, which will be comparable to the sensitivity of $10^{-4}$ projected for Belle II with a luminosity of 50~ab$^{-1}$~\cite{Chen:2020uxi}. The STCF can thus provide crucial information for resolving the $A_Q$ discrepancy.

\subsubsection{Measurement of the electric dipole moment of the $\tau$}

The initial state of an $e^+e^-$ pair in the center-of-mass system is a $CP$ eigenstate. Therefore, $CP$ tests can be conveniently performed at any $e^+e^-$ collider. By measuring the decay products from $\tau$ decays, a $CP$ test can be conducted based on the $e^+e^-\to  \tau^+\tau^-$ process, as suggested in~\cite{CPTau1, CPTau2}. By measuring $CP$-odd observables, one can determine the electric and weak dipole moments of the $\tau$. In the SM, these moments are predicted to be extremely small (for example, the electric dipole moment is expected to be on the order of $10^{-34}$~e~cm). If either of the two moments is nonzero at a level much larger than the SM predictions, it will be a clear signal of new physics beyond the SM. These two moments have been studied at LEP and $B$ factories. While the weak dipole moment is suppressed at low energy by the large masses of the weak gauge bosons, the electric dipole moment $d^\gamma_\tau$ can be probed at $B$ and $\tau$--charm factories. The newest result for the electric dipole moment obtained from the Belle experiment~\cite{TauBelle}, in units of $10^{-16}$~e~cm, is
\begin{equation}
-0.22 < {\rm Re} (d^\gamma_\tau ) < 0.45, \ \ \ \ -0.25 < {\rm Im} ( d^\gamma_\tau) < 0.08.
\end{equation}
These bounds can be tightened by 2 or 3 orders of magnitude through experiments at the STCF.

\subsubsection{CP violation with polarized beams}

With polarized $e^+$ and/or $e^-$ beams, highly polarized $\tau^\pm$s can be produced. $\tau$ polarizations normal ($N$) to their production plane can be measured by studying semileptonic decays of the form $\tau^\pm\to\pi^\pm/\rho^\pm\bar\nu_\tau(\nu_\tau)$. One can then construct asymmetry observables with respect to the left-hand ($L$) and right-hand ($R$) sides of the plane, which are directly related to the electric dipole moment of the $\tau^\pm$~\cite{bernabeu-cp}:
\begin{eqnarray}
A^\pm_N = {\sigma_L^\pm - \sigma_R^\pm\over \sigma} = \alpha_\pm {3\pi \beta\over 8 (3-\beta^2)}{2m_\tau \over e} \textrm{Re}(d^\gamma_\tau)\;,
\end{eqnarray}
where $\sigma$ is the cross section for $e^+e^- \to \tau^+\tau^- \to (\pi^+/\rho^+)\bar\nu_\tau (\pi^- /\rho^-)\nu_\tau$, $\beta = \sqrt{1-4m^2_\tau/s}$, and $\alpha_\pm$ is the polarization analyzer in the $\tau^\pm\to\pi^\pm/\rho^\pm\bar\nu_\tau(\nu_\tau)$ decays. Belle II can reach a sensitivity of $3\times 10^{-19}$~e~cm with an integrated luminosity of $50~\textrm{ab}^{-1}$. At the STCF, the sensitivity can be improved by a factor of approximately 30, reaching $10^{-20}$~e~cm.

With polarized $e^+$ and $e^-$ beams, one can also construct new $T$-odd observables to measure $CP$-violating effects. An interesting observable is the triple product $P^{\tau^{\pm}}_z\hat z\cdot(\vec p_{\pi^\pm}\times\vec p_{\pi^0})$ from measuring the two pion momenta in the $\tau^\pm\to\pi^\pm\pi^0\bar\nu_\tau(\nu_\tau)$ decays~\cite{paultsai}.
Here, $P^\tau_z = [(w_{e^-} + w_{e^+})/(1+w_{e^+}w_{e^-})][(1+2a)/(2+a^2)]$ is the component of the polarization vector of the $\tau$ obtained upon averaging over its momentum direction, with $w_{e^\pm}$ being the components of the polarization vectors of the $e^\pm$ in the $e^-$ beam direction $\hat z$ and $a = 2m_\tau/\sqrt{s}$. If the difference in the triple products for $\tau^+$ and $\tau^-$ is nonzero, this will be a signal of $CP$ violation. Since the SM predicts very small values for the triple products, the measurement of a nonzero difference would already signal new physics beyond the SM. This measurement can be done at the STCF to provide new information about sources of $CP$ violation. Similar measurements can be carried out by replacing $\pi^\pm$ with $K^\pm$.

\subsection{Flavor-violating $\tau$ decays}

FCNC interactions of the $\tau$ are suppressed in the SM when the neutrino masses and mixing are incorporated. In new physics models beyond the SM, larger FCNC effects may appear in some decays, such as $\tau$ decays into $3l$, $l \gamma$, and one or more hadrons plus charged leptons. With the increased statistics for $\tau$ events at the STCF, these decays can be searched for to test the SM and beyond.

\subsubsection{The $\tau^- \to 3l$ decay}

The $\tau^- \to 3l$ decay is one of the most sensitive probes of FCNC interactions. The current upper bound is on the order of $10^{-8}$. At Belle II, upon the accumulation of $50~\textrm{ab}^{-1}$ of integrated luminosity, the sensitivity can reach $4\times 10^{-10}$. When running the STCF at its peak energy ($\sqrt{s} = 4.26$ GeV), it will be possible to produce $3.5\times 10^{9}$ $\tau$ pairs each year, which could be used to push the branching fraction down to a level of $1.9\times 10^{-10}$ with 10~ab$^{-1}$ of luminosity~\cite{tau23l}.

\subsubsection{The $\tau^-\to l\gamma$ decays}

Equally interesting are the $\tau\to l\gamma$ decays, where $l=e$ and $\mu$. The current limits for these decays are also on the order of $10^{-8}$. Since initial-state radiation effects are strongly suppressed near the $\tau^+\tau^-$ production threshold, the STCF has an advantage over $B$ factories in a search for these decays~\cite{Bobrov:2012kg}. At the STCF, the sensitivity for the  branching fraction of $\tau\to\mu\gamma$ will be able to reach around $5.7\times 10^{-9}$ with 10~ab$^{-1}$ of luminosity~\cite{Xiang:2023mkc}.

\subsubsection{The $\tau^- \to lP_1P_2$ decays}

The $\tau^\pm \to l^\pm P_1P_2$ decays, where $P_i = \pi$ and $K$, have been previously searched for with a sensitivity on the order of $10^{-8}$. Similar to these decays are the lepton-number-violating $\tau^\pm\to l^\mp P_1^\pm P_2^\pm$ decays, for which the current bounds are also on the order of $10^{-8}$. At the STCF, the sensitivity for these decays can be increased by two orders of magnitude to a few times $10^{-10}$.

As mentioned earlier, FCNC interactions are highly suppressed in the SM. In some new physics models, however, FCNC interactions can be generated at the tree level and may therefore induce some of the above processes at a level close to their current bounds. In this circumstance, the STCF will be capable of providing very useful information on those models.

\newpage
\section{Topics in QCD Studies and Light Hadron Physics}
\label{sec:qcd}

The formation of the observed hadrons from QCD partons is still not understood. Experimentally, an $e^+e-$ collider is a suitable place to
study hadronization because its initial states are leptons, whereas such studies at a hadron collider will suffer from uncertainties due to the presence of initial hadrons.
At the STCF, such a study can be performed by measuring the $R$ value for a totally inclusive cross section and by measuring the inclusive production
of one or two hadrons. The latter will provide important information about various parton fragmentation functions.
In addition to inclusive processes, exclusive processes will also be studied at the STCF. There are interesting near-threshold phenomena
in $e^+e^-\to B\bar B$, where $B$ is a baryon. Because the STCF will run at center-of-mass energies of up to 7 GeV, it will be possible to exclusively produce two charmonia. The study of the exclusive and inclusive
production of quarkonia will provide important tests of theoretical predictions of nonrelativistic QCD.

In addition to the abovementioned processes, which can be theoretically studied via perturbative QCD at a certain level, many phenomena at the STCF are totally nonperturbative.
These nonperturbative processes can also be well studied at the STCF to provide more insights
into nonperturbative QCD and even new physics beyond the SM.

\subsection{QCD Physics}
\label{subsec:qcd}
\subsubsection{$R$ value}
The $R$ value is defined as
\begin{equation}
   R(s) = \frac{\sigma_{\rm tot} ( e^+ e^- \to \gamma^* \to  {\rm hadrons)}}{\sigma (e^+e^-\to \gamma^* \to \mu^+ \mu^-)},
\end{equation}
which is a function of $s$. An early measurement of $R$ was made at BES \cite{BES-R-2000, BES-R-2002}. Recently, it has also been measured by the KEDR and BESIII~\cite{KEDR, BESIII:2021wib}.

From experimental measurements of $R$, one can determine the running of the electroweak coupling and conduct precision
tests of the SM, as demonstrated in a recent study of the global SM fit \cite{Gfit}.
Precise measurements of $R$ enable the determination of the coupling constants in the SM.
Currently, the possible deviation of $(g-2)_\mu$ of the $\mu$ lepton has motivated many efforts to improve
the precision of the theoretical predictions and to explain this deviation as an effect of new physics beyond the SM. The newest result indicates that there are 4.2 standard deviations between the experimentally measured and theoretically predicted $(g-2)_\mu$ values \cite{NewM}.
An important contribution to the uncertainty of $(g-2)_\mu$ is the contribution from hadronic vacuum polarization. This contribution can be extracted from $R$ as measured in experiments.
Therefore, a precise measurement of $R$ can play an important role in precision tests of the SM.
It is clear that more precise results for the $R$ value will be obtained at the STCF.

\subsubsection{Inclusive production of a single hadron}

For a sufficiently large $\sqrt{s}$, the inclusive production of a single hadron in $e^+e^- \to h +X$ can be predicted from QCD
via the QCD factorization theorem \cite{Book}:
\begin{eqnarray}
\frac{d\sigma (e^+ e^- \rightarrow h +X)} {d z}  &=& \sum_{a=q,\bar q ,g} \int \frac{d\xi}{\xi} H_a (\frac{z}{\xi},Q^2,\mu^2) D_{a\rightarrow h} (\xi,\mu^2)
\nonumber\\
 &=& \sum_q \sigma (e^+ e^- \rightarrow q\bar q) \biggr ( D_{q\rightarrow h} (z) +   D_{\bar q\rightarrow h} (z) \biggr ) + {\mathcal O}(\alpha_s),
\label{FF}
\end{eqnarray}
where $z$ is the fraction of the energy carried by the observed hadron $h$, the functions $H_a$ ($a=q$, $\bar q$, and $g$) can be calculated
via perturbation theory, and $D_{a\rightarrow h}$ denotes parton fragmentation functions describing the hadronization of a parton $a$ to $h$.
Eq. (\ref{FF}) is the expression from QCD for collinear factorization.
The fragmentation functions are universal for any process in which QCD factorization is applicable. Extracting fragmentation functions
at rather low energy, such as the energy region of the STCF near 4--5~GeV, is especially important because with these extracted fragmentation
functions, it is possible to test their energy evolution from a rather low energy scale to high energy scales.

\subsubsection{The Collins effect in the inclusive production of two hadrons}

If two hadrons in the final state are observed in the kinematic region such that the two hadrons are almost back to back, collinear factorization cannot be used.
However, there is another type of factorization, called transverse-momentum-dependent~(TMD) factorization, that holds in this region~\cite{TMDEP}. The angular distributions in this kinematic region are determined by TMD quark fragmentation functions. These functions describe the fragmentation of an initial parton into the observed hadron, where the hadron has a small transverse momentum with respect to the momentum of the initial parton.
The general form of these angular distributions can be found in Ref.~\cite{TMDFF}. Studies of the production
in this region are expected to yield many interesting results regarding TMD parton fragmentation functions. Among them, one, called the Collins function,
is of particular interest. This function describes how a transversely polarized quark fragments into a hadron~\cite{Collins}. Its value is zero
if there is no $T$-odd effect. Belle, operating at $\sqrt{s}=10.6$~GeV, has performed a study of the Collins function~\cite{ColBelle}. It will be interesting to see whether the
Collins function can be measured at the STCF. Theoretical predictions concerning the Collins effect in the energy region of $\sqrt{s}\sim 4$~GeV have been presented in Ref.~\cite{SY}.
In general, by studying the angular correlations of the two produced hadrons in the kinematic region, one can extract various TMD quark fragmentation functions. These functions contain information on how quarks are hadronized into a hadron. Studies of TMD parton fragmentation
functions will be important not only for understanding hadronization but also for exploring the inner structure of hadrons
in semi-inclusive deep in-elastic scattering~(DIS), for which one needs to know the TMD parton fragmentation functions in order to extract the TMD parton distribution functions.

\subsubsection{Form factors of hadrons}

Measuring the electromagnetic~(EM) form factors of nucleons has played an important role in exploring the inner structure
of nucleons. At present, these form factors are still the simplest structure observables for testing the nonperturbative QCD and the related phenomenological models.
In the past, these form factors have been studied mostly in the space-like region. Recently, however,
such studies have been extended to the time-like region. In the time-like region, it is also possible to measure
EM form factors for baryons other than nucleons.
Currently available
time-like experiments demonstrate several puzzling features of EM form factors. Enhancement has been observed by BES \cite{BES1} near the threshold of the $p\bar p$ system. BaBar has also reported enhancement in the
$e^+ e^- \to p\bar p$, $\Lambda \bar \Lambda$, and $\Sigma^0 \bar \Sigma^0$ processes \cite{BaB}.
In the space-like region, the EM form factors of the proton and neutron have been measured with precision at the $1\sim 2\%$ level. In the time-like region, the EM form factors of the proton have a precision o $3.4\%$ \cite{Ablikim:2019eau}, while those of the neutron have an error at the $20\%$ level\cite{BESIII:2021tbq}.
At the STCF with the suggested luminosity, it will be possible to measure the EM form factors in the time-like region with a precision
%Editor: Please ensure that the intended meaning has been maintained in the above edit.
of $0.4\%$ for the
proton and $2\%$ for the neutron, comparable to that in the space-like region. Moreover,
it will be possible to study the enhancement in the production of baryon--antibaryon pairs near the threshold more precisely
to extract information about the interaction between a baryon and an antibaryon,
and it will also be possible to extend this study to the $\Lambda_c \bar \Lambda_c$ system to see whether enhancement occurs in the heavy baryon--heavy antibaryon system.
%Editor: Please ensure that the intended meaning has been maintained in the above edit.

Processes such as $\gamma\gamma\to {\rm hadrons}$ can be studied at the STCF. In addition to the general interest in photon--photon physics, particular quantities of interest are the transition form factors of mesons.
These form factors determine the leading contributions to hadronic light--light scattering, which is closely related
to the precise prediction of the interesting quantity $(g-2)_\mu$. Precise results for these form factors will greatly help to reduce the uncertainties in the contribution to $(g-2)_\mu$ from hadronic light--light scattering.

\subsubsection{Production of charmonia}

The inclusive production of a doubly charmonium has been observed near the $\Upsilon(4S)$ resonance at Belle. The ratio has been measured as \cite{Belle09}
\begin{equation}
   R_{c\bar c} = \frac{\sigma (e^+ e^- \to J/\psi + c +\bar c)}
       {\sigma(e^+ e^- \to J/\psi  +X_{non. c\bar c} ) }   \approx 1.72.
\end{equation}
This is in conflict with theoretical expectations. Some progress in theoretically explaining
this result has been made by including various higher-order corrections. Although the experimental result can be explained by adding one-loop corrections \cite{MZC1,MZC2,BGJW1,BGJW2}, the outcome may be not consistent. If one includes the so-called color-octet contributions estimated
from the hadroproduction of $J/\psi$, there is still conflict between experiment and theory (see also Ref.~\cite{REVW}).
Belle has also observed the exclusive production of double charmonia, $e^+ + e^-
\to J/\psi + \eta_c$~\cite{Belle02}. Theoretically, the measured cross section is still not well explained,
even with the inclusion of two-loop predictions in the theory \cite{FJS}.
\par
With the STCF running at $\sqrt{s}$ larger than 6~GeV, it will be possible to experimentally study these production processes more precisely. This will be helpful for gaining a better understanding of production. This energy range offers a unique opportunity to study physics related to the production of two $c\bar c$ pairs. The production cross sections for  $e^+e^-\to J/\psi c\bar c$ based on the NRQCD calculations in Refs.~\cite{ZGC1,ZGC2} are shown in Fig.~\ref{fig:xsec_NRQCD}. These cross sections can be tested at the STCF.

%%%%%%%%%%%%%%%%%%% Fig 2 %%%%%%%%%%%%%%%%%%%%%%%%
\begin{figure*}[t]
	\centering
	\includegraphics[width=0.45\textwidth]{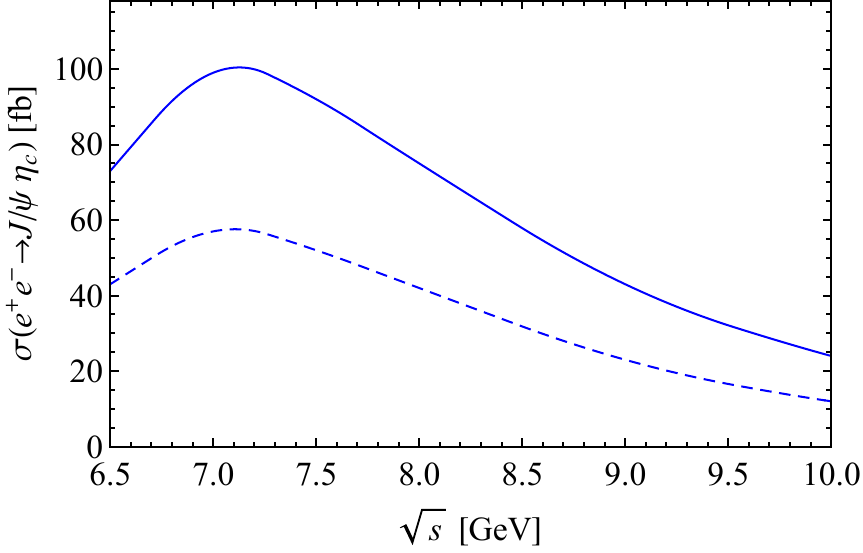}~~
	\includegraphics[width=0.45\textwidth]{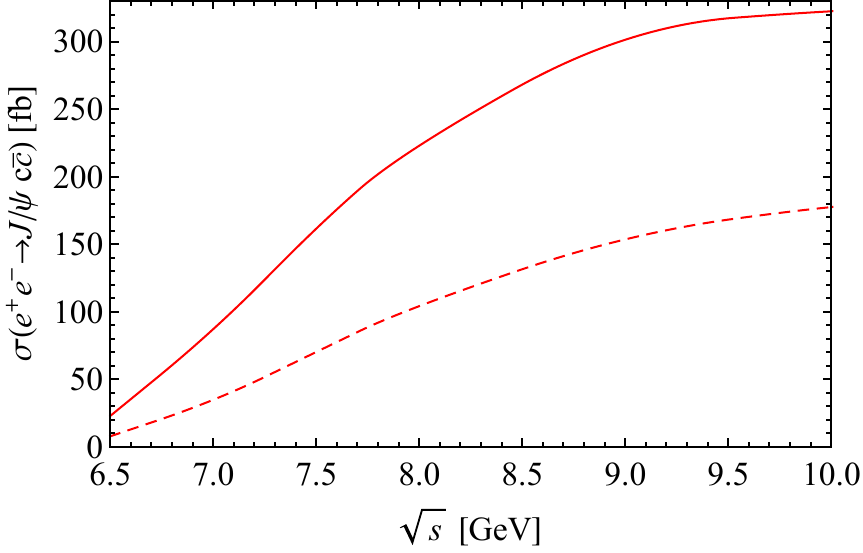}
\vspace{0cm}
\caption{Cross sections for $e^+e^-\to J/\psi\eta_c$ (left) and $e^+e^-\to J/\psi c\bar c$ (right) as calculated using NRQCD with the charm quark mass fixed at 1.5~GeV. The solid and dashed curves represent the results from the next-to-leading-order and leading-order calculations, respectively.
		\label{fig:xsec_NRQCD}}
\end{figure*}
%%%%%%%%%%%%%%%%%%%%%%%%%%%%%%%%%%%%%%%%%%%%%%%%%%

% *****{\color{red} Figure 5 already appear early in section 2.3. Need to make coherent display. May be combine the materials for this two small sections?}*****

%\input{06_ref_QCD}

%\input{05_01_Spectroscopy}
\subsection{Spectroscopy}

The spectrum of the light hadrons serves as an excellent probe of nonperturbative QCD~\cite{Brambilla:2014jmp,Meyer:2010ku,Crede:2008vw,Klempt:2007cp,Amsler2004,Godfrey:1998pd}. The complexity of strong QCD manifests itself in hadrons, their properties and their internal structures. The quark model suggests that mesons are formed from a constituent quark and antiquark and that baryons consist of three such quarks. QCD, however, allows a richer spectrum of color singlets that takes into account not only the quark degrees of freedom but also the gluonic degrees of freedom. Additionally, excited and exotic hadronic states are sensitive to the details of quark confinement, which is only poorly understood within QCD.

%Lattice-QCD calculations of both the baryon and the meson spectra have made tremendous progress
%and have now reached a maturity so that they can provide some guidance in the experimental efforts.

%The mass spectrum of hadrons is clearly organized according to flavor content, spin, and parity.
In intermediate- and long-distance phenomena such as hadron properties, the full complexity of QCD emerges, which makes it difficult to understand hadronic phenomena at a fundamental level.
%However, many states are not well established and evidence remains vague,
%particularly in the baryon sector.
Based on quark model expectations, the experimental meson spectrum appears to be overpopulated, which has inspired speculation about states beyond the $q\bar{q}$ picture, whereas fewer states have been observed in the baryon spectrum, which has led to the problem of the so-called missing baryon resonances. Even for several well-established baryons, their spins and parities have never been measured and are based merely on quark model assignments, particularly for resonances involving strange quarks. Whether glueballs made of multiple gluons and hybrids made of gluons and quarks, as predicted by LQCD~\cite{Lee:1999kv,Bali:1993fb, Morningstar:1997ff, chen:2005mg, Lacock:1996ny, MILC:1997usn, Dudek:2011tt, Dudek:2013yja}, truly exist is still an open question. These are some of the important issues limiting the current understanding of hadronic physics.
Another critical and poorly studied sector is the light vector mesons, especially strangeonium states, which can provide critical information on the connection between the light quark and heavy quark sectors. Hadron production via $e^+e^-$ collisions with ISR~\cite{Druzhinin:2011qd} plays an important role. Current and future experiments present a real opportunity for a dramatic improvement in our knowledge of the spectrum.

At present, BESIII remains unique in its ability to study and search for QCD exotics and new excited baryons~\cite{bes3yellowbook}, as its high-statistics data sets of charmonia provide a gluon-rich environment with clearly defined initial- and final-state properties~\cite{beswhite}.
%Recent progress and future plan of light hadron physics at BESIII has been reviewed in ~\cite{bes3whitepaper}.
At the STCF, many more data sets of charmonia will be obtained. The expected high-statistics data samples for $J/\psi$ and $\psi(3686)$ decays, including both hadronic and radiative decay channels,
%Editor: Please ensure that the intended meaning has been maintained in the above edit.
will provide an unprecedented opportunity to obtain a better understanding of the spectrum of light hadrons, their properties and their couplings to all the channels in which they appear and, from these, to learn about the composition of these states, including glueballs and hybrid states.
An interesting example is the study of the glueball nature of some states using data from the STCF. The production properties suggest a prominent glueball nature of $f_0(1710)$ and a flavor octet structure of $f_0(1500)$~\cite{beswhite}. However, the scalar meson sector is the most complex one, and the interpretation of the
states' natures and their nonet assignments are still very controversial. There is no question that more states than can be
accommodated by a single meson nonet have been found. However, the nature of all of these states is still open
for discussion. At the STCF, a year of operation will provide $\sim$3~T $J/\psi$ and $\sim$500~B $\psi(3686)$ events at their peak cross sections for
exploring light hadron physics. Traces of glueballs and hybrid states may be found in some more confirmed ways. Measurements of electromagnetic couplings to glueball candidates would be extremely useful for the clarification of the nature of these states. The radiative transition rates of a relatively pure glueball would be anomalous relative to the expectations for a conventional $q\bar{q}$ state. The dilepton decay modes of the light unflavored mesons are expected to provide deeper insight into the meson structure, allowing the transition form factors to be measured in the time-like region. A glueball should have suppressed couplings to $\gamma\gamma$, which can be measured at the STCF. There has been a long history of experimental searches for the spin-exotic states $J^{PC}$ quantum numbers that can not be formed by a simple quark-antiquark pair. Recently, an isoscalar resonance with exotic $1^{-+}$ quantum numbers, $\eta_1(1855)$, has been observed by BESIII experiment~\cite{BESIII:2022riz}. At the STCF, further studies with more production mechanisms and decay modes will help clarify the nature of the
$\eta_1(1855)$. In addition, more precise study on the isoscalar $1^{-+}$ $\eta_1(1855)$, combined
with previous and also future measurements of the isovector $\pi$ states, will provide critical clue of searching for other partners of exotic supermulitplets.

Nevertheless, the extraction of resonance properties from experimental data is far from straightforward; the resonances tend to be broad and plentiful, leading to intricate interference patterns, or buried under a background in the same and other waves. The key to success lies in high statistical precision complemented by sophisticated analysis methods. Partial
wave or amplitude analysis (PWA) techniques~\cite{Battaglieri:2014gca} are the state-of-the-art way to disentangle the contributions from individual, even small, resonances and to determine
their quantum numbers. Nevertheless, the extremely high statistics at the STCF will present new challenges for
data handling and processing. High-performance computing harnessing heterogeneous acceleration (e.g., Ref.~\cite{Berger:2010zza}) will be a key requirement. The correct analytical properties of the amplitude will be essential for extrapolation from the experimental data to the complex plane to determine the pole positions. A key component of the necessary PWA will be close cooperation between experimentalists and theorists.

\subsection{Precision tests with light hadrons}
\subsubsection{Light meson decays}

At the STCF, it is expected that approximately $3.4\times10^{12}$ $J/\psi$ events will be collected per year; thus, the STCF will be a factory for light mesons due to their high production rates in $J/\psi$ decays. Taking $\eta/\eta'$ mesons as an example,
Table~\ref{tab:eta} indicates that more than $10^9$ $\eta/\eta'$ events could be produced through $J/\psi$ radiative or hadronic decays. Accordingly, the STCF will offer an unprecedented opportunity to explore
light meson decays for a variety of physics at low energy scales, including precision tests
of effective field theories, investigations of the quark structure of the light mesons, tests of
fundamental symmetries, and searches for new particles.

At low energies,
nonperturbative QCD calculations are usually performed using an effective field theory called chiral
perturbation theory (ChPT). High-quality and precise measurements of low-energy hadronic processes
are necessary to verify the systematic expansion of ChPT. Thus, studies of light meson
decays can provide important guidance for our understanding of how QCD works in the nonperturbative
regime. In particular, the $\eta^\prime$ meson, which is much heavier than the
Goldstone bosons of chiral symmetry breaking, plays a special role
as the predominant singlet state arising from
the strong axial $U(1)$ anomaly. The decays of light mesons, such as the $\eta/\eta^\prime$ and $\omega$, as well as their excited states can provide useful information about chiral perturbation theory through hadronic decays~\cite{Gasser:1983yg, Kaiser:2000gs} and
anomalous Wess--Zumino--Witten (WZW) processes~\cite{Wess:1971yu,Witten:1983tw, Bijnens:1989jb}, such as ${\eta}^{\prime} \to \rho^0 \gamma$ and ${\eta}^{\prime} \to \pip\pim\pip\pim$. $\eta/\eta'$ decays can also provide model-independent information about low-energy meson interactions, such as vector meson dominance (VMD)~\cite{Sakurai:1960ju,Landsberg:1986fd}.
The $\eta/\eta^\prime\to\gamma\gamma\pi^0$ decays are of particular interest for tests of ChPT at the two-loop level. Since light vector mesons play a critical role in these models, the dynamical role of the vector mesons must be systematically included in the context of either VMD or the Nambu--Jona--Lasinio model~\cite{Nambu:1961tp,Nambu:1961fr} to reach a deeper understanding of these decays.

%In this case, the expected $\eta/\eta^\prime$ decays could reach about $10^9$. Therefore STCF is also a of light mesons. In Table.~\ref{tab:eta}, $\eta/\eta'$ events per year through several $J/\psi$ decay channels are %listed.
%Both $\eta$ and $\eta^\prime$ mesons are very well suited for tests of the SM.

The $\eta/\eta^\prime\to\gamma l^+l^-$ ($l=e,\mu$) Dalitz decays, where the lepton pair is formed through the internal conversion of an intermediate virtual photon and
the decay rates are modified by the electromagnetic structure arising at the vertex of the transition, are of special interest. Deviations of measured quantities from their QED predictions are usually described in terms of a
time-like transition form factor, which, in addition to being an important probe into the meson's structure~\cite{Landsberg:1986fd}, has an important role in the evaluation of the hadronic light-by-light contribution to the muon anomalous magnetic moment (see the nice review in Ref.~\cite{Aoyama:2020ynm} for details).
In addition, using the expected large data sample to be collected at a center-of-mass energy above the $J/\psi$ peak at the STCF, measurements of the space-like transition form factors in the decay
$e^+e^-\rightarrow e^+e^- \pi^0(\eta,\eta^\prime)$ via $\gamma\gamma$ interactions in the transfer momentum ($Q^2$) range of $[0.3, 10]$ GeV/c$^2$ will be feasible. The measured space-like transition form factors will uniquely cover the $Q^2$ range that is relevant to the hadronic
light-by-light correction for the evaluation of the muon anomalous moment.

%Although $\eta/\eta^\prime$ cannot be produced directly from $e^+e^-$ collisions, their high production rate in $J/\psi$ decays provide an efficiency source of a great number of $\eta/\eta^\prime$ mesons. The STCF is %designed to have a luminosity of $~10^{35}$
%$ cm^{-2}s^{-1}$ and the goal is to have at least $10^{12}$ $J/\psi$ events produced per year.
%In this case, the expected $\eta/\eta^\prime$ decays could reach about $10^9$. Therefore STCF is also a of light mesons. In Table.~\ref{tab:eta}, $\eta/\eta'$ events per year through several $J/\psi$ decay channels are %listed.
%Both $\eta$ and $\eta^\prime$ mesons are very well suited for tests of the SM.

%The neutral members of the ground state pseudoscalar nonet, both $\eta$ and $\eta^\prime$
%play an important role in understanding low energy QCD. Decays of the $\eta/\eta^\prime$ probe a wide variety of
%physics issues {\it e.g.} $\pi^0-\eta$ mixing, light quark masses
%and pion-pion  scattering.
%In addition,
%being the eigenstates of the C, P and CP operators, the decays of $\eta/\eta^\prime$ offer a unique opportunity for testing these  fundamental
%discrete symmetries. Their rare and forbidden decays can also provide information about new physics beyond SM.
%\cite{UA1,TH01}.

%{\color{red}

The $\eta$ and $\eta^\prime$ mesons are eigenstates of $P$, $C$ and $CP$ whose strong and electromagnetic decays are either anomalous or forbidden at the lowest order by $P$, $C$, $CP$ and angular momentum conservation. Therefore, their decays provide a unique laboratory for testing the fundamental symmetries in flavor-conserving processes, as extensively reviewed in Ref.~\cite{Gan:2020aco}.

A straightforward way to test these symmetries is to search for $P$- and $CP$-violating
$\eta/\eta^\prime$ decays into two pions. 
In the SM, the branching fractions for these modes are at a level of $10^{-28}$~\cite{Jarlskog:2002zz}, but they may be enhanced by about ten orders of magnitudes due to $CP$ violation in the extended Higgs sector of the electroweak theory~\cite{Jarlskog:1995gz}.
Therefore, an observation of the $\eta\rightarrow 2\pi$ decay, with a rate considerably higher than that quoted above, would imply new sources of $CP$ violation beyond the SM.
Experimentally, $\eta/\eta^\prime\rightarrow l^+l^- \pi^0$ decays could be used to test charge-conjugation invariance. In the SM, this process can proceed via a two-virtual-photon exchange, whereas an one-photon exchange would violate $C$-parity. Within the framework of the VMD model, the most recent predictions~\cite{Escribano:2020rfs} for the branching fractions
are on the order of $10^{-9}$ for $\eta\rightarrow l^+l^- \pi^0$ and $10^{-10}$ for $\eta^\prime\rightarrow l^+l^- \pi^0 (\eta)$. Thus,
a significant enhancement of the branching fractions exceeding the predictions of the two-photon model may be indicative of $C$ violation.
With the expected $3.4\times 10^{12}$ $J/\psi$ events at the STCF, the branching fractions can reach a new high precision on the order of $10^{-9}$, making the investigation of these rare decays very promising.
Many other decays of the $\eta/\eta'$ mesons, as summarized in Table~\ref{tab:rareetap} for $\eta^\prime$ decays, are also useful for tests of the SM.
For example, the $\eta\rightarrow \mu^+\mu^-$ and $\eta\rightarrow e^+e^-$ decays are of interest when
searching for physics beyond SM.
Within the framework of the SM, the decays are dominated by a two-photon intermediate state, which suppresses the
branching fractions. However, beyond-the-SM interactions, such as leptoquark exchange, can enhance the branching fractions. Therefore,
larger-than-expected measurements will provide information about non-SM interactions, and the same can be concluded for their flavor-violating counterparts, $\eta\rightarrow e^\pm \mu^\mp$.

\begin{table}[htbp]
\begin{center}
%\centering
 \caption{\label{tab:eta} The expected numbers of $\eta/\eta^\prime$ events as calculated from the $3.4\times 10^{12}$ $J/\psi$ events anticipated to be produced at the STCF per year.}
 \begin{tabular}{l c c c }\hline\hline
        Decay mode    &       $\mathcal{B}$ ($\times 10^{-4}$) ~\cite{ParticleDataGroup:2022pth}   & $\eta/\eta^\prime$ events \\ \hline
      $J/\psi\rightarrow\gamma\eta^\prime$ &$52.1\pm1.7$ &    $1.8\times 10^{10}$ \\ \hline
          $J/\psi\rightarrow\gamma\eta$ &$11.08\pm0.27$ &$3.7\times 10^9$\\  \hline
       $J/\psi\rightarrow\phi\eta^\prime$ & $7.4\pm0.8$ &    $2.5\times 10^9$ \\ \hline
          $J/\psi\rightarrow\phi\eta$ & $4.6\pm0.5$&   $1.6\times 10^9$ \\  \hline
                  \end{tabular}
                  \end{center}
\end{table}

\begin{table*}
\centering
  \caption{The statistical sensitivities to rare and forbidden $\eta^\prime$ decays. The expected sensitivities are estimated by considering
the detector efficiencies for different decay modes at the STCF. We
assume that there is no background dilution and that the observed number of signal
events is zero. The STCF limits are given at the 90\% confidence
level. } {\begin{tabular}{llccc}
\hline
Decay mode &
% Best upper limits (measurements$^*$) & BES-III limit \\
Best upper limit  &STCF limit & Theoretical  & Physics \\
&90\% CL  &$(3.4\times 10^{12}$ $J/\psi$ events)  & prediction  & \\
\hline
$\eta^\prime \rightarrow e^+e^- $ &  $ 5.6\times 10^{-9} $  & 1.5 $\times 10^{-10}$  & $1.1\times 10^{-10}$ & leptoquark\\
$\eta^\prime \rightarrow \mu^+\mu^-$ &  $-$  & 1.5 $\times 10^{-10}$ & $1.1\times 10^{-7}$ & leptoquark\\
$\eta^\prime \rightarrow e^+ e^- e^+ e^- $ &  $-$ & 2.4$\times 10^{-10}$  & $1\times 10^{-4}$& $\gamma^*\gamma^*$\\
$\eta^\prime \rightarrow \mu^+ \mu^- \mu^+ \mu^- $ &  $-$  &2.4$\times 10^{-10}$ &  $4\times 10^{-7}$ &$\gamma^*\gamma^*$\\
%$\eta^\prime \rightarrow \pi^+\pi^- e^+ e^- $ &  $6.0\times 10^{-3}$  & 1.4$\times 10^{-7}$  & VMD, TFF\\
%$\eta^\prime \rightarrow \pi^+\pi^- e^+ e^-$ &  $(24^{+13}_{-10})\times 10^{-4}$  & 1.4$\times 10^{-7}$  \\
%$\eta^\prime \rightarrow \pi^+ \pi^- \mu^+ \mu^- $ &  $2.9\times 10^{-5}$  & 8$\times 10^{-10}$  & $2.2\times 10^{-5}$ & VMD, TFF\\
$\eta^\prime \rightarrow \pi^0 \mu^+ \mu^- $ &  $6.0 \times 10^{-5}$  & 2.4$\times 10^{-10}$  & &$C$ violation\\
$\eta^\prime \rightarrow \pi^0 e^+ e^- $ &  $1.4 \times 10^{-3}$  & 2.4$\times 10^{-10}$  & &$C$ violation\\
%$\eta^\prime \rightarrow \pi^0\gamma $ &  $-$  &7$\times 10^{-10}$  & &angular momentum \\
$\eta^\prime \rightarrow \pi^0\pi^0$ &  $9.0 \times 10^{-4}$  & 2.9$\times 10^{-9}$ & &$CP$ violation \\
$\eta^\prime \rightarrow \pi^+\pi^-  $ &  $2.9 \times 10^{-3}$  & 1.5$\times 10^{-10}$  & &$CP$ violation\\
$\eta^\prime \rightarrow \mu^+ e^- + \mu^- e^+ $ &  $4.7 \times 10^{-4}$  &1.5$\times 10^{-10}$ && LPV\\
$\eta^\prime \rightarrow $ invisible  &  $5.3\times10^{-4}$  & 3.3$\times 10^{-8}$  & &Dark matter\\
$\eta^\prime \rightarrow \eta e^+e^- $ &  $2.4 \times 10^{-3}$     & 5.9$\times 10^{-10}$ & &$C$ violation\\
$\eta^\prime \rightarrow \eta \mu^+\mu^- $ &  $1.5 \times 10^{-5} $     &  5.9$\times 10^{-10}$ & &$C$ violation\\
\hline
\end{tabular}\label{tab:rareetap}}
\end{table*}

In addition to the $\eta/\eta^\prime$ decays, the high production of other light mesons, $\omega$, $a_0(980)$, $f_0(980)$, and $\eta(1405)$, as well as other excited states, is also an important source for exploring many aspects of particle physics at low energy. The $\omega\to \pi^+\pi^-\pi^0$ decay could be employed to investigate the $\omega$ decay mechanism by comparing a high-statistics Dalitz plot density distribution with the predictions within the dispersive theoretical framework~\cite{Niecknig:2012sj,Danilkin:2014cra}; moreover, the $a_0(980)$--$f_0(980)$ mixing is sensitive to the quark structure of the light scalars, and the $\eta(1405)\to 3\pi$ process may help reveal 
the well-known triangle singularity mechanism~\cite{triangle}.

In general, despite the impressive progress that has been achieved in recent years, many light meson decays are still unobserved and need to be explored.
With the advantages of high production rates and excellent performance at the STCF, the highly abundant and clean samples of $e^+e^-$ annihilations will bring the study of light meson decays into a precision era, will certainly play an important role in the further development of chiral effective field theory and LQCD, and will make significant contributions to the understanding of hadron physics in the nonperturbative regime.

\subsubsection{Hyperon decays}
The ongoing experimental studies of $CP$-symmetry violation in particle decays aim to find effects
that are not expected in the SM such that new
dynamics is revealed. The existence of $CP$ violation in kaon and beauty
meson decays is well established \cite{Christenson:1964fg,
Aubert:2001nu,Abe:2001xe}. The first observation of $CP$ violation in charm mesons was
reported in 2019 by the LHCb experiment \cite{Aaij:2019kcg}, but thus far, there is no evidence in the baryon sector. All the observations are consistent with the SM expectation. Baryons with strange quark (hyperon) decays offer promising possibilities for searches for new $CP$-violating effects since they involve $p$-wave decay amplitudes, which neutral kaon decays do not \cite{Donoghue:1985ww}.
%Editor: Please ensure that the intended meaning has been maintained in the above edit.
A possible signal of $CP$ violation would be a difference
in the decay distributions between charge-conjugate decay modes. The
main decay modes of the ground-state hyperons are weak hadronic transitions
into a baryon and a pseudoscalar meson, such as $\Lambda\to p\pi^-$
(${\cal B}\approx64\ \%$) and $\Xi^-\to\Lambda\pi^-$
(${\cal B}\approx100\ \%$) \cite{ParticleDataGroup:2022pth}. They involve two amplitudes: one for a parity-conserving decay to the relative $p$ state and one for a parity-violating decay to the $s$
state. The angular distribution and polarization of the daughter
baryon are described by two decay parameters:
$\alpha=2{\rm Re}(s^*p)/(|p|^2+|s|^2)$ and $\phi={\rm
arg}((s-p)/(s+p))$.  Here, we denote the decay parameters $\alpha$ for $\Lambda\to p\pi^-$ and
$\Xi^-\to\Lambda\pi^-$ by $\alpha_\Lambda=0.7519(43)$~\cite{BESIII:2018cnd,BESIII:2022qax} and $\alpha_\Xi=-0.376(8)$~\cite{BESIII:2021ypr},
respectively. In the $CP$-symmetry-conserving limit, the parameters
$\alpha_D/\alpha_{\bar D}$ and $\phi_D/\phi_{\overline{D}}$ for the charge-conjugated $D/\bar D$ decay modes have the same absolute values but opposite signs.
The $CP$ asymmetry $A_D$ is defined as follows:
\begin{align}
  A_{D}&\equiv\frac{\alpha_D+\alpha_{\bar D}}{\alpha_D-\alpha_{\bar D}}\approx -\tan(\delta_p-\delta_s)\sin(\zeta_p-\zeta_s),\label{eq:AD}
%  &\approx -\frac{\sqrt{1-\alpha_D^2}}{\alpha_D}\sin\phi_D\sin(\zeta_p-\zeta_s),
\end{align}
where $\delta_p-\delta_s$ is the strong $(p\!-\!s)$-wave phase difference in the decay due to final-state interaction and $\zeta_p-\zeta_s$ is the weak $CP$-violating phase difference.

The best limit for $CP$ violation in the strange baryon sector was obtained by comparing the complete
$\Xi^-\to\Lambda\pi^-\to p\pi^-\pi^-$ and c.c. decay chains of unpolarized $\Xi$ baryons at
the dedicated HyperCP (E871) experiment~\cite{Holmstrom:2004ar} by determining the
asymmetry
$A_{\Xi\Lambda}=(\alpha_\Lambda\alpha_\Xi-\alpha_{\bar\Lambda}\alpha_{\bar\Xi})/(\alpha_\Lambda\alpha_\Xi+\alpha_{\bar\Lambda}\alpha_{\bar\Xi})\approx A_{\Xi}+A_{\Lambda}$.
The result, $A_{\Xi\Lambda}=(0.0\pm5.1\pm4.7)\times10^{-4}$, is consistent with the SM prediction: $\left|A_{\Xi\Lambda}\right|\le
5\times10^{-5}$~\cite{Tandean:2002vy}. Moreover, an improved
preliminary HyperCP result presented at the BEACH
2008 Conference suggests a large asymmetry value of $A_{\Xi\Lambda}=(-6.0\pm2.1\pm2.0)\times10^{-4}$
\cite{Materniak:2009zz}. However, it is difficult to interpret this result in terms of the weak $CP$-violating phase difference. The $A_D$ asymmetries are, in general, not sensitive probes of the weak $CP$-violating phase
difference since the $\tan(\delta_p-\delta_s)$ term is very small and not well known. The values are
$-0.097(53)$ for  $\Lambda\to p\pi^-$ and $0.087(33)$ for $\Xi^-\to\Lambda\pi^-$,
as determined from the values of the $\phi_D$ decay parameters using the following relation:
\begin{equation}
   \tan(\delta_p-\delta_s)\approx -\frac{\sqrt{1-\alpha_D^2}}{\alpha_D}\sin\phi_D\ .
\end{equation}
A much more sensitive, independent determination of $\zeta_p-\zeta_s$ is obtained
by comparing the $\phi_D$ and  $\phi_{\bar D}$
parameters:
\begin{equation}
  \Delta\phi_D\equiv\frac{\phi_{D} + {\phi}_{\bar D}}{2}\approx \frac{\alpha_D}{\sqrt{1-\alpha_D^2}}\sin(\zeta_p-\zeta_s)\ . \label{eq:Dphi}
\end{equation}

\begin{table*}
\begin{tabular}{p{3cm}llrr}
\hline\hline
Decay mode&${\cal B}$ (units of $10^{-4}$)&Angular distribution&Detection&\multicolumn{1}{l}{No. of events }\\
          &&parameter $\alpha_\psi$ &efficiency& expected at the STCF\\
\hline
$J/\psi\to\Lambda\bar\Lambda$ \vphantom{$\int\limits^M$} &${19.43\pm0.03\pm0.33}$&$\phantom{-}0.469\pm0.026$&40\%&$1100\times10^6$\\
$\psi(3686)\to\Lambda\bar\Lambda$&$\phantom{0}{3.97\pm0.02\pm0.12} $&$\phantom{-}0.824\pm0.074$&40\%&$130\times10^6$\\
$J/\psi\to\Xi^{0}\bar\Xi^{0}$&$11.65\pm 0.04 $&$\phantom{-}0.66\pm 0.03$&14\%&$230\times10^6$\\
$\psi(3686)\to\Xi^{0}\bar\Xi^{0}$&$\phantom{0}2.73\pm 0.03 $&$\phantom{-}0.65\pm0.09$&14\%&$32\times10^6$\\
$J/\psi\to\Xi^{-}\bar\Xi^{+}$&$10.40\pm 0.06 $&$\phantom{-}0.58\pm0.04$&19\%&$270\times10^6$\\
$\psi(3686)\to\Xi^{-}\bar\Xi^{+}$&$\phantom{0}2.78\pm 0.05 $&$\phantom{-}0.91\pm0.13$&19\%&$42\times10^6$\\
\hline\hline
\end{tabular}
\caption[]{Branching fractions for some $J/\psi,\psi(3686)\to B\bar B$
  decays and the estimated sizes of the data samples from the full
  data set of $3.4\times 10^{12}\ J/\psi$ and $3.2\times 10^{9}\ \psi(3686)$
  to be collected
  by the
  STCF.  The approximate detection
  efficiencies for the final states reconstructed using the $\Lambda\to
  p\pi^-$ and $\Xi\to\Lambda\pi$ decay modes are based on the
  published BESIII analyses using partial data sets
  \cite{Ablikim:2017tys,Ablikim:2016sjb,Ablikim:2016iym}.
\label{tab:data}}
\end{table*}

With a well-defined initial state, the charmonium decay into a strange
baryon--antibaryon pair offers an ideal system for testing fundamental
symmetries. The vector charmonia $J/\psi$ and $\psi(3686)$ can be directly
produced in an electron--positron collider with high yields and
have relatively large branching fractions into hyperon--antihyperon
pairs; see Table~\ref{tab:data}.
The potential power of such measurements was shown in a recent BESIII analysis using a
data set of $4.2\times10^{5}$  $e^+e^-\to
J/\psi\to\Lambda\bar\Lambda$ events reconstructed via the $\Lambda\to p\pi^-$ $+$
c.c. decay chain~\cite{BESIII:2018cnd}. The determination of the asymmetry parameters was possible
due to the
transverse polarization and spin correlations of the
$\Lambda$ and $\bar\Lambda$. In the analysis, the complete multidimensional
information of
the final-state particles was used in an unbinned maximum log likelihood fit
to the fully differential angular expressions from Ref.~\cite{Faldt:2017kgy}.
This
method allows direct comparison of the decay parameters of the
charge-conjugated decay modes and enables a test of the $CP$ symmetry.

In Ref.~\cite{Perotti:2018wxm}, the formalism was extended to
describe processes that include decay chains of multiple strange
hyperons, such as the $e^+e^-\to\Xi\bar\Xi$ reaction with the
$\Xi\to\Lambda\pi$ and $\Lambda\to p\pi^-$ $+$ c.c. decay sequences. The
expressions are much more complicated than the single-step weak decays
in $e^+e^-\to\Lambda\bar\Lambda$. The joint
distributions for $e^+e^-\to\Xi\bar\Xi$ allow all decay parameters to be determined
simultaneously, and the statistical uncertainties are independent of
the size of the transverse polarization in the production process.
The uncertainties of the $CP$-odd asymmetries that can be extracted from the exclusive
analysis were estimated in Ref.~\cite{Adlarson:2019jtw}.
\begin{table*}
\centering
  \begin{tabular}{lrrrrrrr}
\hline\hline
&$A_\Xi$&$A_\Lambda$&$A_{\Xi\Lambda}$&$(\zeta_p-\zeta_s)_\Xi$&$(\zeta_p-\zeta_s)_\Xi$\\
&&&&Eq.~\eqref{eq:AD}&Eq.~\eqref{eq:Dphi}\\\hline
$J/\psi\to\Lambda\bar\Lambda$ \vphantom{$\int\limits^M$}&$-$&$1.7\times10^{-4}$&$-$&$-$&$-$\\
    $J/\psi\to\Xi^-\bar\Xi^+$ ($\Delta\Phi=0$)&$2.2\times10^{-4}$&$2.1\times10^{-4}$&$2.5\times10^{-4}$&$2.4\times10^{-3}$&$6.5\times10^{-4}$\\
%$J/\psi\to\Xi^0\bar\Xi^0$ ($\Delta\Phi=\pi/2$)&2.0&3.0&5.2&2.9&15
%    &1.4&4.0&1.5&3.4&0.77&4.4&3.7&10\\
    \hline\hline
  \end{tabular}
  \caption[]{Statistical sensitivity for asymmetry parameters extracted using STCF data samples.
    The input values of the parameters
    are taken from Table~\ref{tab:data} and Ref.~\cite{Adlarson:2019jtw}.\label{tab:sigpar}}
  \end{table*}
To study the angular distribution for the
$e^+e^-\to \Xi^-\bar\Xi^+$ reaction, we fix the decay
parameters of the $\Lambda$ and $\Xi^-$ to the central values
from the PDG~\cite{ParticleDataGroup:2022pth}. For the production process,
the
phase $\Delta\Phi$ is an unknown parameter, but the result shows almost no dependence
on this parameter, and we set $\Delta\Phi=0$. In
Table~\ref{tab:sigpar}, we report the statistical uncertainties in
the $J/\psi\to\Xi^-\bar\Xi^+$ decay.
By exploiting the spin entanglement between the $\Xi^{-}$ baryon and its antiparticle $\bar{Xi}^{+}$, 
BESIII has enabled a direct determination of the weak-phase difference, $(\zeta_p-\zeta_s)=(1.2\pm3.4\pm0.8)\times10^{-2}$rad~\cite{BESIII:2021ypr}.

The sensitivities for the $A_\Xi$, $A_\Lambda$ and $A_{\Xi\Lambda}$ asymmetries
with a data sample of $3.4\times 10^{12}$ $J/\psi$ events at the STCF (see Table~\ref{tab:data}) are given in Table~\ref{tab:sigpar}. The statistical uncertainty for the $A_{\Xi\Lambda}$ asymmetry from the dedicated HyperCP experiment will be surpassed at the STCF.
The SM predictions for the $A_{\Xi}$ and $A_{\Lambda}$ asymmetries
are $-3\times10^{-5}\le A_\Lambda\le
4\times10^{-5}$ and $-2\times10^{-5}\le A_\Xi\le 1\times10^{-5}$~\cite{Tandean:2002vy}.

Under the assumption of a value of $0.037$ for the
$\phi_\Xi$ parameter, five-sigma significance would require $3.1\times 10^5$
exclusive $\Xi^-\bar\Xi^+$ events. Reaching a statistical uncertainty
of 0.011, as in the HyperCP
experiment~\cite{Huang:2004jp}, would require $1.4\times10^{5}$
$J/\psi\to\Xi^-\bar\Xi^+$ events, while the single-cascade HyperCP result is based on
$114\times10^6$ events. In contrast, the present PDG precision of $\phi_{\Xi^{0}}$ could be achieved with only
$3\times 10^{2}$ $\Xi^{0}\bar\Xi^{0}$ events.

The sensitivities for the weak phase difference $(\zeta_p-\zeta_s)_\Xi$~\cite{Donoghue:1985ww}
using the two independent methods are also given in Table~\ref{tab:sigpar}. The SM estimate for $(\zeta_p-\zeta_s)_\Xi$ is $8\times10^{-5}$. However, it should be stressed that the SM predictions
for all asymmetries need to be updated in view of the recent and forthcoming
BESIII results on hyperon decay parameters using the collected $10^{10}$ $J/\psi$ events.
A wide range of precision $CP$ tests can be conducted
based on a single measurement. Thus, the spin-entangled cascade--anticascade system is a promising probe for testing fundamental symmetries in the strange baryon sector.

%\subsubsection{$\Lambda - \bar{\Lambda}$ oscillation in $J/\psi \to \Lambda \bar{\Lambda}$ decay}

A large data sample of $\Lambda\bar\Lambda$ from $J/\psi$ decays can also be used to study $\Lambda$--$\bar\Lambda$ oscillations.
The seesaw mechanism, as an explanation for the small neutrino masses \cite{Dutta:2005af}, predicts the existence of $\Delta(B-L)=2$
interactions and baryon--antibaryon oscillations.
To date, searches for processes that violate the baryon number by two units have been performed only in neutron--antineutron oscillation experiments~\cite{BaldoCeolin:1994jz}. Searches for $\Lambda$--$\bar\Lambda$ oscillations in the $J/\psi \to \Lambda \bar{\Lambda}$ decay at BESIII have been proposed \cite{Kang:2009xt}.
With 10 billion $J/\psi$ decay events at BESIII, the expected sensitivity of the measurement of
$\Lambda$--$\bar{\Lambda}$ oscillation is $\delta m_{\Lambda\bar\Lambda}<10^{-15}$ MeV
at the 90\% confidence level. This corresponds to a lower limit of $10^{-7}$ s on the oscillation time.
At the STCF, the expected constraint on $\delta m_{\Lambda\bar\Lambda}$ can be improved to the $10^{-17}$ MeV level or even better. This upper limit is already much larger than the lifetime of the $\Lambda$, and further significant improvements would require other approaches, such as the use of certain long-lived hypernuclei~\cite{Dalitz:1962eb1,Dalitz:1962eb2,Dalitz:1962eb3}.

\subsection{Tests of the $CPT$ invariance with $\jpsi$ decays}
While violations of $C$,~$P$,~$T$,~and~$CP$~symmetries have been well established and characterized, the validity
of $CPT$~symmetry remains intact at a high level of sensitivity. The $CPT$~theorem ~\cite{Schwinger:1951xk}
states that any quantum field theory  that is {\it Lorentz invariant}, has {\it local point-like interaction vertices},
and is {\it Hermitian} ({\it i.e.}, conserves probability) is invariant under the combined operations of
$C$, $P$~and~$T$. Since the three quantum field theories that make up the Standard Model---QED, QCD, and
Electroweak theory---all satisfy these criteria, $CPT$~symmetry has been elevated to  some kind mystical status
in particle physics.  However, there is good reason to believe that $CPT$, like all of the other discrete symmetries,
is violated at least a mass scale of ${\mathcal O}(10^{-35} m)$ {\it i.e.}, at the so-called Planck scale.

One of the requirements for a $CPT$-invariant theory is that it is {\it local}, which means that the
couplings at each vertex occurs at a single point in space-time. But theoretical physics has always had
troubles with point-like quantities. For example, the classical self-energy of the electron is
\begin{equation}
  W_{e}=\frac{e^2}{4\pi\eps_{0} r_{e}},
\end{equation}
which diverges for $r_e\rt 0$.  The {\it classical radius of the electron}, {\it i.e.}, the value of $r_e$
that makes $W_e=m_ec^2$, is $r^{\rm c.r.e.}_e= 2.8\times 10^{-13}$~cm (2.8 fermis),  which is three times the
radius of the proton, and $\sim$300 times larger than experimental upper limits on the electron radius, which
are $<10^{-16}$~cm~\cite{ZEUS:2003eqd}.  Infinities associated with point-like objects persist in quantum field
theories, where they are especially troublesome. In second-order and higher perturbation theory, all of
the diagrams that have virtual particle loops involve integrals over all possible configurations of the virtual
particles in the loops that conserve energy and momentum.  Whenever two of the point-like vertices coincide,
the integrands become infinite and cause the integrals to diverge.

In the QED, QCD and Electroweak quantum field theories that make up the Standard Model, these infinities are
removed by the well established methods of renormalization~\cite{Bethe:1947id,Dyson:1949bp,Wilson:1973jj}. In all
three of these theories, the perturbation expansions are in increasing powers of a dimensionless coupling
strength, $\alpha_{\rm QED}, \alpha_s$ and, $\alpha_{\rm EW}=\sqrt{2}M^2_WG_F/\pi$.\footnote{Specifically
  not just $G_F$, which has the dimension of mass$^{-2}$.}
As a result of this, in the renormalization procedure, relations that exist between different orders of the
perturbation expansion reduce the number of observed quantities that are needed to subtract off divergences.
In QED, for example, there are only two, the electron's mass, and charge (three, if the diagram includes
muons).   However, in quantum theories of gravity, where a massless spin=2 {\it graviton} plays the role of
the photon in QED, the  expansion constant is Newton's gravitational constant $G=\hbar c/M^2_{\rm P}$, where
$M_{\rm P}\equiv\sqrt{\hbar c/G}=1.2\times 10^{19}$~GeV is the {\it Planck mass}. Because of this, every
order in the perturbation expansion has different dimensions and, thus, a distinct observed quantity
is needed to carry out the subtraction at each step, which means that complete renormalization would
infinite number of observed quantities to complete the renormalization.  This means that a
$CPT$-conserving quantum theory of gravity would, in principle, be {\it nonrenormalizable}\cite{Weinberg:1980kq}.
Although difficulties associated with non-renormalizability ({\it i.e.}, higher-order perturbative effects)
will never show up at mass scales below the Planck mass, this problem demonstrates that there is nothing
especially sacred about $CPT$-invariance that prevents from being violated at a lower mass scale.  Because of
its close connection with the fundamental assumptions of the Standard Model, stringent experimental tests of
$CPT$~invariance should have high priority.

\subsubsection{Kaon mixing and tests of the $CPT$~theorem}

Among the consequences of $CPT$~symmetry are that particle and antiparticle masses and lifetimes are equal.
Since lifetime differences can only come from on-mass-shell intermediate states and do not probe short distance,
high mass physics, these are unlikely to exhibit any $CPT$-violating asymmetry. Instead, the focus here is on the
possibility that particle and antiparticle masses may be different.

The particles with the best measured masses are the stable electron and proton, and, according to the
PDG~\cite{ParticleDataGroup:2022pth}:
\begin{eqnarray}
  |m_{e^+}-m_{e^-}|&<& 4\times 10^{-9}~{\rm MeV}\\
  \nonumber
  |m_{\bar{p}}-m_{p}|&<& 7\times 10^{-7}~{\rm MeV}.
\end{eqnarray}
However, these limits do not provide the best tests of $CPT$; the most stringent experimental restriction on
$CPT$ violation comes from the difference between the $\Kzbar$ and $\Kz$ masses:
\begin{equation}
  |M_{\Kzbar}-M_{\Kz}|\leq 5 \times 10^{-16}~{\rm MeV},
\end{equation}
which is $7$-$9$~orders of magnitude more strict than those from the electron and proton mass measurements
even though the value of $M_{\Kz}$ itself is only known to~$\pm 13$~keV. This is because the
Fig.~\ref{fig:k-mix_c-quark-KM} diagrams, taken together with the quantum mechanics of $\Kz$-$\Kzbar$ mixing,
maps the $M_{\Kzbar}-M_{\Kz}$ difference into the quantity $\Delta M=M_{\KL}-M_{\KS}\approx 3.5\times 10^{-12}$~MeV,
which is 14 orders of magnitude lower than $M_{\Kzbar}$ or $M_{\Kz}$ and the independent quantity 
$\Delta\Gamma=\Gamma_{\KS}-\Gamma_{\KL}\approx 7.4\times 10^{-12}$~MeV (which is, coincidently,
$\approx 2\times \Delta M$).

\begin{figure}[!bp]
\centering
\includegraphics[width=0.99\textwidth]{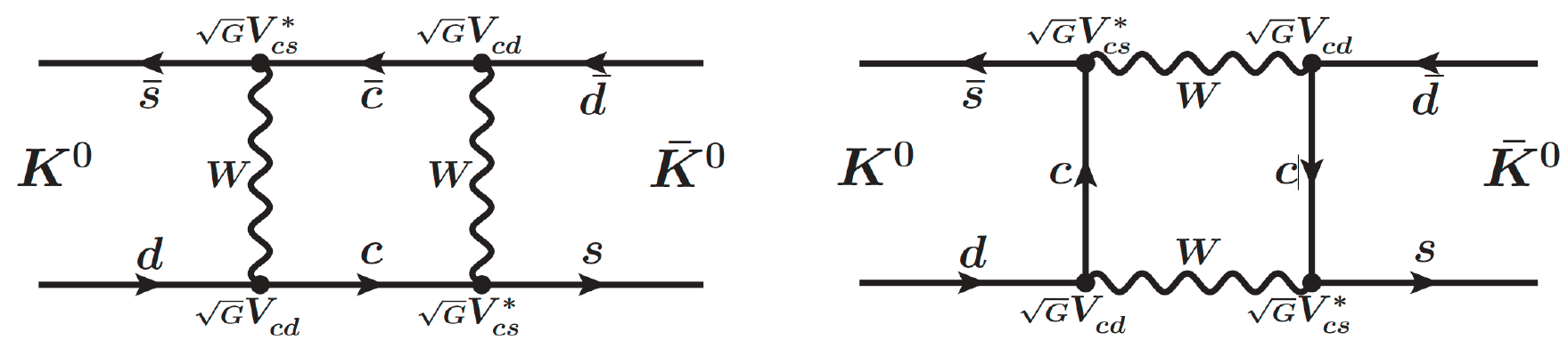}
\caption{\footnotesize The box diagrams for the short-distance contributions to  $\Kz$-$\Kzbar$
mixing.}
 \label{fig:k-mix_c-quark-KM}
  \end{figure}

The mapping is done by computing the difference between the proper-time-dependence of the $K\rt\pipi$
decay rates for neutral $K$ mesons that are tagged as strangeness \Str$=+1$ and \Str$=-1$ at their time of
production ({\it i.e.}, $\tau=0$), and are denoted here as $\Kz(\tau)$ and $\Kzbar(\tau)$, respectively.
At STCF, this tagging is automatically done in $\jpsi\rt\Km\pip\Kz$ and $\jpsi\rt\Kp\pim\Kzbar$ decays,
by the sign of the charged-kaon's electric charge: a $\Km$ tags a $\Kz(\tau)$ and a $\Kp$ tags a $\Kzbar(\tau)$.
These events are quite distinct at a center-of-mass system $\ee\rt\jpsi$ collider as shown in Fig.~\ref{fig:k0kpi-STCF}a, and occur
with a branching fraction $BF[\jpsi\rt K^\mp\pi^\pm\Kz(\Kzbar)]=(0.56\pm 0.05)\%$, which, for $\jpsi$ decays,
is substantial. Moreover, at a c.m. collider these events are pretty much background free, the only significant
backgrounds are misidentified  $K\rt\pi^\pm\ell^\mp\nu$ decays that also depend on $\Delta M$.

\begin{figure}[!]
\centering
\includegraphics[width=0.99\textwidth]{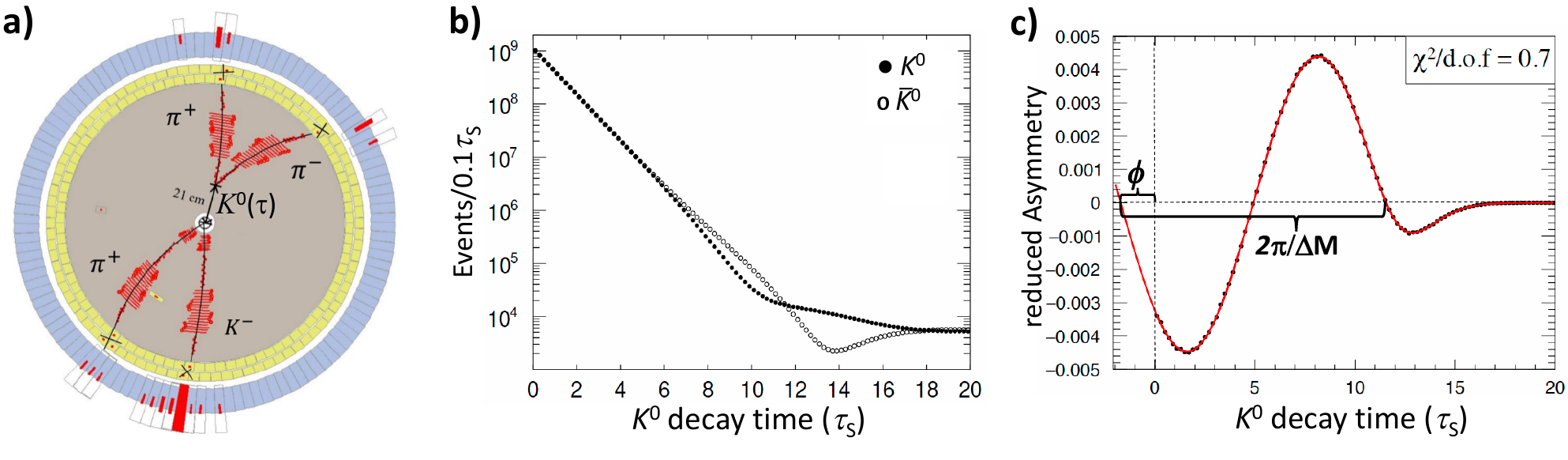}
\caption{\footnotesize {\bf a)}  A simulated $\jpsi\rt K^-\pip\Kz(\tau)$; $\Kz(\tau)\rt\pipi$ event in
  the BESIII detector.
  {\bf b)} The solid circles show the proper time distribution for simulated strangeness-tagged
  $\Kz(\tau)\rt\pipi$ decays (the open circles are $\Kzbar(\tau)\rt\pipi$ decays). 
  {\bf c)} The reduced asymmetry,
  ${\mathcal A}^{\rm reduced}_{\pipi}={\mathcal A}_{\pipi}\times e^{-{\scriptstyle \frac{1}{2}}\Delta\Gamma\tau}$,
  which weights the asymmetries according to their relative statistical significance, is plotted
  for the events shown in panel~b.  Here $\phi=\phi_{\rm SW}+\delta\phi^{CPT}_{\eta}$.) 
  }
 \label{fig:k0kpi-STCF}
 \end{figure}

Although the event shown in  Fig.~\ref{fig:k0kpi-STCF}a looks superficially like $K^\mp\pi^\pm\KS$, here
the neutral kaon is not a $\KS$ mass eigenstate with a simple exponential lifetime.  Instead, separate plots
of the proper-decay-times for simulated $K^-$-tagged $\Kz(\tau)\rt\pipi$ and $K^+$-tagged $\Kzbar(\tau)\rt\pipi$
events, shown in Fig.~\ref{fig:k0kpi-STCF}b, tell a different, and much more interesting story. Instead of
single exponential curves that decrease indefinitely with the $\KS$ lifetime, the decay curves contain
interfering $\KS$ and $\KL$ components, with interference patterns for $\Kz(\tau)$ and $\Kzbar(\tau)$ decays
that quite different. There is a pronounced asymmetry between the two modes that, according to numerous
available descriptions of the quantum mechanics of the neutral kaon system (see,
{\it e.g.}, ref.~~\cite{Schubert:2014ska}),  can be expressed as
\begin{eqnarray}
  \label{eqn:KKbar-asymm-CPT} 
   {\mathcal A}_{\pipi}&\equiv&\frac{\bar{N}(\Kzbar(\tau))-N(\Kz(\tau))}{\bar{N}(\Kzbar(\tau))+N(\Kz(\tau))}\\
   \nonumber
   &~&~~~~~~=2\Re(\eps-\delta)-2\frac{|\eta_{+-}|e^{{\scriptstyle \frac{1}{2}}\Delta\Gamma\tau}\cos(\Delta M\tau-\phi_{\rm SW}
                 +\delta\phi^{CPT}_{\eta})}{1+|\eta_{+-}|^2e^{\Delta\Gamma\tau}}.
\end{eqnarray}
Here the first term on the r.h.s. of the equation is a  small ($\sim$0.3\%) constant offset involving $\eps$, the
familiar mass-matrix parameter that characterizes the $CP$ impurities in the $\KS$ and $\KL$ mass eigenstates, and
$\delta$, which is an even smaller $CPT$-violating parameter (that is explained below). Inherent differences in the $\Kp$
and $\Km$ detection efficiencies make it impossible to measure this term with any significant precision. On the
other hand the phase of the  oscillation described by the second term, $\phi_{\rm SW} +\delta\phi^{CPT}_{\eta}$, is
not sensitive to experimental efficiencies and can, in principle, be accurately measured. Here $\phi_{\rm SW}$ is
the ``superweak phase'' and is equal to $\tan^{-1}(2\Delta M/\Delta \Gamma)$, where $\Delta M$ can be
measured from the wavelength of the oscillation, and $\Delta\Gamma$ is determined from the $\KS$ and $\KL$
decay curves.\footnote{There  is a small, ${\mathcal O}(0.03^\circ)$ theoretical correction to the
         $\phi_{\rm SW}=\tan^{-1}(2\Delta M/\Delta \Gamma)$ that is well understood~\cite{Bell:1965mn}.}
The difference between the measured phase of the eqn.~\ref{eqn:KKbar-asymm-CPT} oscillation term and 
$\phi_{\rm SW}$ is $\phi^{CPT}_{\eta}=\delta_{\perp}/\eps$, where
\begin{equation}
  \delta_{\perp}=\frac{M_{\Kzbar}-M_{\Kz}}{2\sqrt{2}\Delta M};
\end{equation}
a non-zero value would indicate the existence of a $\Kz$-$\Kzbar$ mass difference, and a violation of $CPT$..   

This translation of $M_{\Kzbar}-M_{\Kz}$ into a relation involving the $M_{\KL}$-$M_{\KS}$ is a trick that is
unique to the kaon system and is not applicable to other particles. Although the neutral $D^0$, $B^0$ and $B^0_s$
mesons mix with their antiparticles, they do not have a measurable equivalent of the kaon's $\eps$ mass-matrix
mixing parameter and, moreover, have a
large number of decay modes that are common to both of their mass eigenstates that make $CPT$-related analyses
that are easily used for neutral kaons impractical~\cite{Bell:1965mn}. As a result, tests of $CPT$~with neutral
$B$ mesons by BaBar~\cite{BaBar:2006zzh} and Belle~\cite{Higuchi:2012kx} did not place any limits on 
$|M_{\bar{B}^0}-M_{B^0}|$.

Although it is apparent from eqn.~\ref{eqn:KKbar-asymm-CPT} that the ${\mathcal A}_{\pipi}$ asymmetry increases
with the proper-time $\tau$, the statistical precision of measurements rapidly decreases with $\tau$. Because
of this, the CLEAR group, which measured this asymmetry in $\bar{p}_{\rm stop}p\rt K^\mp\pi^\pm\Kz(\Kzbar)$ events
produced in the annihilation of stopped antiprotons~\cite{Apostolakis:1999zw}, suggested that an alternative
{\it reduced asymmetry}, defined as
\begin{equation}
  {\mathcal A}^{\rm reduced}_{\pipi}={\mathcal A}_{\pipi}\times e^{-{\scriptstyle \frac{1}{2}}\Delta\Gamma\tau},
\end{equation} 
be used for display since it emphasizes the importance of the higher-statistics, low and intermediate
decay-time measurements that provide the bulk of the measurement sensitivity. The reduced asymmetry between
the decay curves for $\Kz(\tau)\rt\pipi$ and $\Kzbar(\tau)\rt\pipi$ events shown in Fig.~\ref{fig:k0kpi-STCF}b,
is plotted in Fig.~\ref{fig:k0kpi-STCF}c.

The  simulated data shown in Figs.~~\ref{fig:k0kpi-STCF}b~\&~c correspond to 3.8B~tagged $K\rt\pipi$
decays that are almost equally split between $\jpsi\rt\Km\pip\Kz$ and $\jpsi\rt\Kp\pim\Kzbar$ that were generated 
with $\phi^{CPT}_{\eta}=0$ and $\phi_{\rm SW}=43.4^\circ$. This corresponds to what one would expect for a
total of $10^{12}$~$\jpsi$ decays in a detector that covered a $|\cos\theta|\le 0.85$ solid angle and was
otherwise almost perfect.  The red curve in the figure is the result of a fit to the data that determined
\begin{equation}
  \phi_{\rm SW}+\phi^{CPT}_{\eta} = 43.51^\circ\pm 0.05^\circ~~~~
          \Rightarrow~~~~\phi^{CPT}_{\eta} < 0.15^\circ~~(90\%~{\rm C.L.}),
\end{equation}
where the errors are statistical only. To date, the best existing measurements are a 1999 result from the
CPLEAR~$\bar{p}_{\rm stop}p$ experiment at CERN~\cite{Apostolakis:1999zw} that used a sample of $\sim$70M tagged
$\Kz(\tau)\rt\pipi
i$ and $\Kzbar(\tau)\rt\pipi$ events and a 1995 result from Fermilab experiment
E773~that used $\sim$2M~$K\rt\pipi$ and $\sim$0.5M~$K\rt\piz\piz$ decays produced downstream of a regeneration
  located in a high-energy $\KL$ beam~\cite{Schwingenheuer:1995uf}:
\begin{eqnarray}
  {\rm CPLEAR:}~~~\phi_{\rm SW}+\phi^{CPT}_{\eta}&=&42.91^\circ\pm 0.53^\circ {\rm (stat)}\pm 0.28^\circ  {\rm (syst)},\\
  \nonumber
  {\rm E773:~~}~~~\phi_{\rm SW}+\phi^{CPT}_{\eta}&=&42.94^\circ\pm 0.58^\circ {\rm (stat)}\pm 0.49^\circ  {\rm (syst)},
\end{eqnarray}
where the systematic errors include $0.19^\circ$ (CPLEAR)~\&~$0.35^\circ$ (E733) contributions from uncertainties in
the regeneration phases.   The potential statistical error for a 1~trillion $\jpsi$~event data sample at STCF is
factor of ten better than previous experiments. The issue will be whether or not the systematic errors can be controlled
accordingly.  In the STCF detector the production and a fraction of the $K\rt\pipi$ decays occur in a high vacuum, and
the material traversed by rest of the decaying neutral kaons is very small, This will substantially reduce the effects of
regeneration, which was the largest source of systematic error in the previous experiments. 

As the results in the previous paragraph indicate, constraints on the validity of the $CPT$ theorem have not changed
in the twenty-five years.  This is not because these are not important or interesting, instead it is because that
no tests of $CPT$ invariance for particle systems other than the neutral kaons have anything like the same sensitivity,
and none of the world's active particle physics facilities have been capable of producing enough kaons to match, much less
improve on, the CPLEAR and E773 results. (The BESIII 10B $\jpsi$ event sample only has $\sim$20M tagged $\Kz(\tau)\rt\pipi$
and $\Kzbar(\tau)\rt\pipi$ decays, less than a third of the CPLEAR data sample.) The STCF project will provide a unique 
opportunity to make an order of magnitude sensitivity improvement over earlier measurements.

\newpage
\section{New light particles beyond the SM}
\label{sec:newphys}
In this section, we briefly describe the beyond the SM~(BSM) motivations for the STCF. Since the Higgs boson was discovered, for the first time, a complete theory is available to describe the electroweak and strong interactions. A drawback
to the success of the SM is that one loses guidance for future directions of research. Under such circumstances, one must scrutinize all possibilities, such as the STCF, super B factories, the LHC and other facilities, to find clues on how to proceed. %{\color{red}
The STCF, with its high luminosity, will be sensitive to new light particles and their interactions with SM particles.
%}
We have listed
three categories of motivations in terms of BSM physics: (1) forbidden and rare decays, (2) $CP$ violation, and (3) searches for new weakly interacting light particles. Nevertheless, we should also point out that the BSM possibilities are still more extensive than those listed here,
and other new topics can also be investigated.

Here, we mainly focus on the possibility of new light particles in a hidden sector
that has weak coupling with the SM sector.
The candidates for such new light particles include dark photons,
new light scalars,
and millicharged particles.

\subsection{Particles in the dark sector}

The existence of a dark sector that weakly couples to the SM sector is well motivated by many BSM theories. Some new physics particles may exist at the TeV scale or above and can only be probed at high-energy colliders. However, the messengers connecting the dark sector to the SM sector may exist at lower energies, such as the GeV scale. These messengers can be scalars, pseudoscalars, or gauge bosons that interact with SM particles through certain ``portals'' \cite{Essig:2013lka}. Because this new light sector would need to interact with SM particles very weakly to evade the constraints from current experiments, it is generally dubbed the ``dark sector''.

A particular motivation for such a scenario follows from the observations of anomalous cosmic-ray positrons. In 2008, the PAMELA collaboration reported excess positrons above $\sim$ 10~GeV \cite{Adriani:2008zr}, and this observation has since been confirmed by many other experiments, such as ATIC \cite{Chang:2008aa}, Fermi-LAT \cite{Abdo:2009zk} and AMS02 \cite{Aguilar:2013qda}. In one class of dark matter models, dark matter particles with masses of $\sim \mathcal{O}(\mathrm{TeV})$ annihilate into a pair of light bosons with masses of $\sim \mathcal{O}(\mathrm{GeV})$, which then decay into charged leptons \cite{ArkaniHamed:2008qn,Pospelov:2008jd}.

These light bosons may be massive dark photons in models with an extra U(1) gauge symmetry. These dark photons couple to photons through the kinetic mixing $\frac{\epsilon}{2} F^{\mu \nu} F'_{\mu \nu}$. Since QED is a well-tested model, the mixing strength $\epsilon$ should be small. In theory, $\epsilon$ can be generated by high-order effects \cite{ArkaniHamed:2008qp}. Therefore, $\epsilon$ is naturally $\sim 10^{-2}$--$10^{-3}$ or smaller. The dark photon could acquire a mass through the Higgs mechanism or the Stueckelberg mechanism. Some models could predict that the mass of the dark photon should be at the $\sim \mathcal{O}(\mathrm{MeV})$--$\mathcal{O}(\mathrm{GeV})$ scale \cite{ArkaniHamed:2008qp,Cheung:2009qd}. This suggests that the structure of the dark sector may be complicated. There could be a broad class of light particles, including scalars, pseudoscalars, gauge bosons and fermions, at the GeV scale.

Since the interaction between the dark sector and the SM sector must be very weak, a search for light dark photons (or other light particles) at the intensity frontier is well motivated. In the phenomenology of the dark photon, the most important parameters are the dark photon mass $m_{A'}$ and the mixing strength $\epsilon$. Fig.~\ref{fig:limits} shows the constraints on $\epsilon$ and $m_A'$ from the measurements of the electron and muon anomalous magnetic moments, low-energy $e^+e^-$ colliders, beam dump experiments and fixed-target experiments \cite{Essig:2013lka}. Due to their high luminosity and low center-of-mass energy, which should be close to the mass of the dark photon, electron--positron colliders are also suitable for probing dark photons through either direct production or rare decays of mesons.

\begin{figure}[!htbp]
\centering
\includegraphics[width=0.5\textwidth]{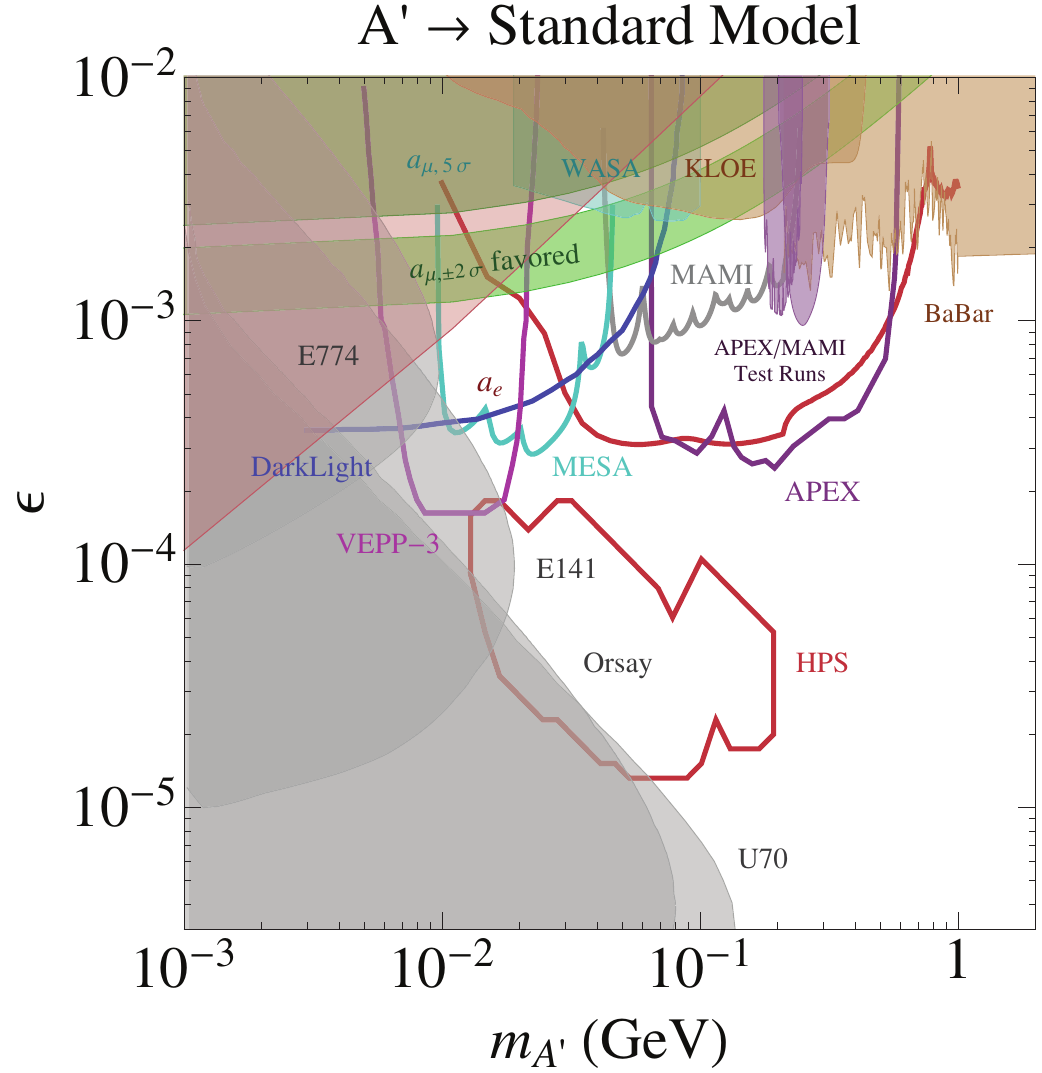}
\caption{Constraints on the mixing strength $\epsilon$ versus the dark photon mass $m_{A'}>1$ MeV from the measurements of the electron and muon anomalous magnetic moments, low-energy $e^+e^-$ colliders, beam dump experiments and fixed-target experiments. For details, see Ref. \cite{Essig:2013lka}. Reproduced from Ref.~\cite{Essig:2013lka}.}
\label{fig:limits}
\end{figure}

Electron--positron collisions could directly produce dark photons, which would subsequently decay into charged leptons, via $e^+e^-\rightarrow \gamma+A' (\rightarrow l^+l^-)$ \cite{Zhu:2007zt,Fayet:2007ua,Reece:2009un,Essig:2009nc,Yin:2009mc}. In comparison with the irreducible QED background $e^+e^-\rightarrow \gamma l^+l^-$, dark photon production is suppressed by a factor of $\epsilon^2$. To reduce the influence of this background, precise reconstruction of the dark photon mass and a high luminosity are important. Such studies for $\Upsilon \rightarrow \gamma+A' (\rightarrow \mu^+\mu^-)$ have been performed by interpreting results from the BaBar experiment \cite{Aubert:2009cp,Bjorken:2009mm,Reece:2009un}. 
Since no new peak has been found in the data, the mixing strength $\epsilon$ is constrained to be smaller than $\sim 2 \times 10^{-3}$ for a dark photon with a mass of $\sim$ 1~GeV.
Later, Belle reported similar restriction on the coupling strength of $A' (\rightarrow \mu^+\mu^-)$ via $\Upsilon \rightarrow \gamma+A'$~\cite{Belle:2021rcl}.
Preliminary studies show that this limit could be reduced to $5\times 10^{-4}$ at SuperB \cite{O'Leary:2010af}.
At Belle II, under the hypothesis of the coupling strength about few $\times 10^{-4}$, one can expect to observe an excess of
events due to a dark photon decays to charged leptons, taking advantage of about two orders of magnitude larger data statistics than those available at BaBar and Belle~\cite{Kou:2018nap}.
With 20~fb$^{-1}$ of data collected at $\psi(3770)$, sensitivity to an $\epsilon$ as low as $2\times10^{-3}$ could be achieved for $e^+e^-\rightarrow \gamma+A' (\rightarrow l^+l^-)$
with $m_{A'}\sim 1$~GeV~\cite{Li:2009wz}. A similar estimate can be performed for the STCF, indicating that the sensitivity for $\epsilon$ will be
$\mathcal{O}(10^{-4})$ for $m_{A'} \sim 0.6-3.7$ GeV with 1~ab$^{-1}$ of data, as shown in Fig.~\ref{fig:BESIII}.

\begin{figure}[!htbp]
\centering
\includegraphics[width=0.6\textwidth]{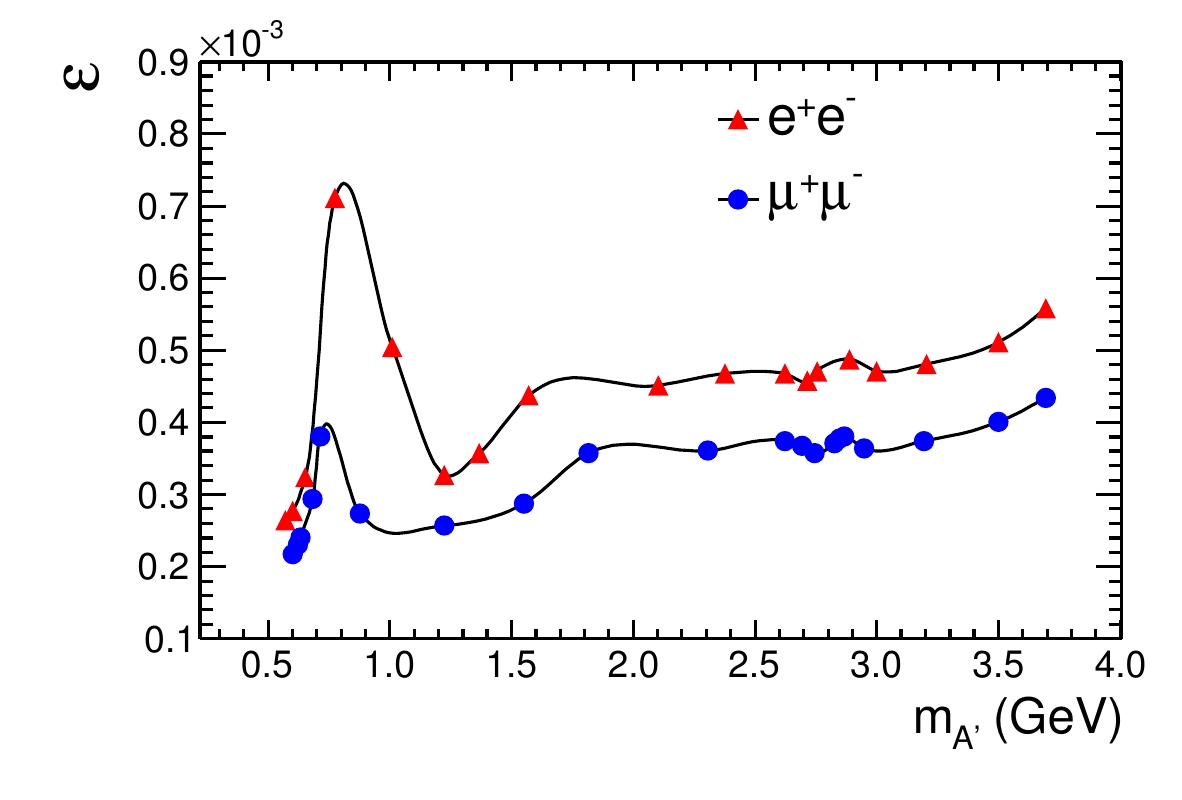}
\caption{The sensitivity to the mixing strength $\epsilon$ at the STCF for $e^+e^-\rightarrow \gamma+A' (\rightarrow l^+l^-)$ with 1~ab$^{-1}$ of data. Reproduced from Ref.~\cite{Li:2009wz}.}
\label{fig:BESIII}
\end{figure}

If there is also a light Higgs $h'$ that provides the mass of the dark photon, with a mass of $\sim \mathcal{O}(\mathrm{MeV})$--$\mathcal{O}(\mathrm{GeV})$, in the dark sector, then some new processes can be used to investigate the dark sector at electron--positron colliders \cite{Baumgart:2009tn,Batell:2009yf}. If $m_{h'}>2 m_{A'}$, then the signal process $e^+e^-\rightarrow A'+h' (\rightarrow 2 A')\rightarrow 3 l^+l^-$ will be very clean for such research due to the presence of several resonances in the lepton pairs. If $m_{h'}< m_{A'}$, then $h'$ can only decay into lepton pairs via loop processes. In this case, the lifetime of the $h'$ will be long; possible signals are displaced vertices or even missing energy in the detector. Note that other light bosons may also exist, such as gauge bosons under an extra non-Abelian symmetry, in the dark sector \cite{Baumgart:2009tn}. The final states of direct production may contain more lepton pairs. In this case, it will be easier to extract the signals from large QED backgrounds via the reconstruction of resonances.

In general, if mesons have decay channels into photons, they could also decay into dark photons with branching ratios of approximately $\epsilon^2 \times \mathrm{BF}(\mathrm{meson}\rightarrow \gamma)$ \cite{Reece:2009un,Li:2009wz}. Since low-energy electron--positron colliders produce numerous mesons, such as $\pi$, $\rho$, $K$, $\phi$, and $J/\psi$, it is possible to search for dark photons in the rare decays of mesons. For instance, one can search for a resonance in the $\phi \rightarrow \eta+A'$ and $\pi/\eta\rightarrow \gamma+A'$ processes with $ A'\rightarrow l^+l^-$. At the STCF, where a large sample of charm mesons will be produced, charmonium decay channels, such as $J/\psi \rightarrow e^+e^-+A'$ \cite{Zhu:2007zt} and $\psi(3686)\rightarrow \chi_{c1,2}+A'$, can be used to probe dark photons.

Light $Z'$ bosons that decay predominantly into invisible states can be probed by looking for missing energy in the final states at electron colliders; see, e.g.\, ref.\ \cite{Adachi:2019otg} for the recent Belle II results. Other light particles, such as light axion-like particles, can also be searched for at electron colliders; see, e.g.\, ref.\ \cite{BelleII:2020fag} for the recent Belle II results.

\subsection{Millicharged particles}

Particles with an electric charge that is significantly
smaller than that of an electron are often referred to as millicharged
particles.
A variety of BSM models predict millicharged
particles;
for example, millicharged fermions in the hidden sector
can naturally arise via kinetic mixing
\cite{Holdom:1985ag, Holdom:1986eq, Foot:1991kb}
or Stueckelberg mass mixing
\cite{Kors:2004dx, Cheung:2007ut, Feldman:2007wj}.
Millicharged
particles have previously been searched for
at various mass scales both at terrestrial laboratories
and in cosmological/astrophysical processes
(see, e.g.\,  \cite{Jaeckel:2010ni} for a review).
Electron colliders operating at the GeV scale can probe
the previously allowed millicharged
particles parameter space for masses
in the MeV--GeV range \cite{Liu:2018jdi, Liang:2019zkb}.
At the MeV--GeV energy scale,
the existing laboratory constraints
on millicharged
particles include constraints from colliders \cite{Davidson:1991si},
the SLAC electron beam dump experiment \cite{Prinz:1998ua},
and neutrino experiments \cite{Magill:2018tbb}.

\begin{figure}[htbp]
\begin{center}
\includegraphics[width=0.45 \columnwidth]{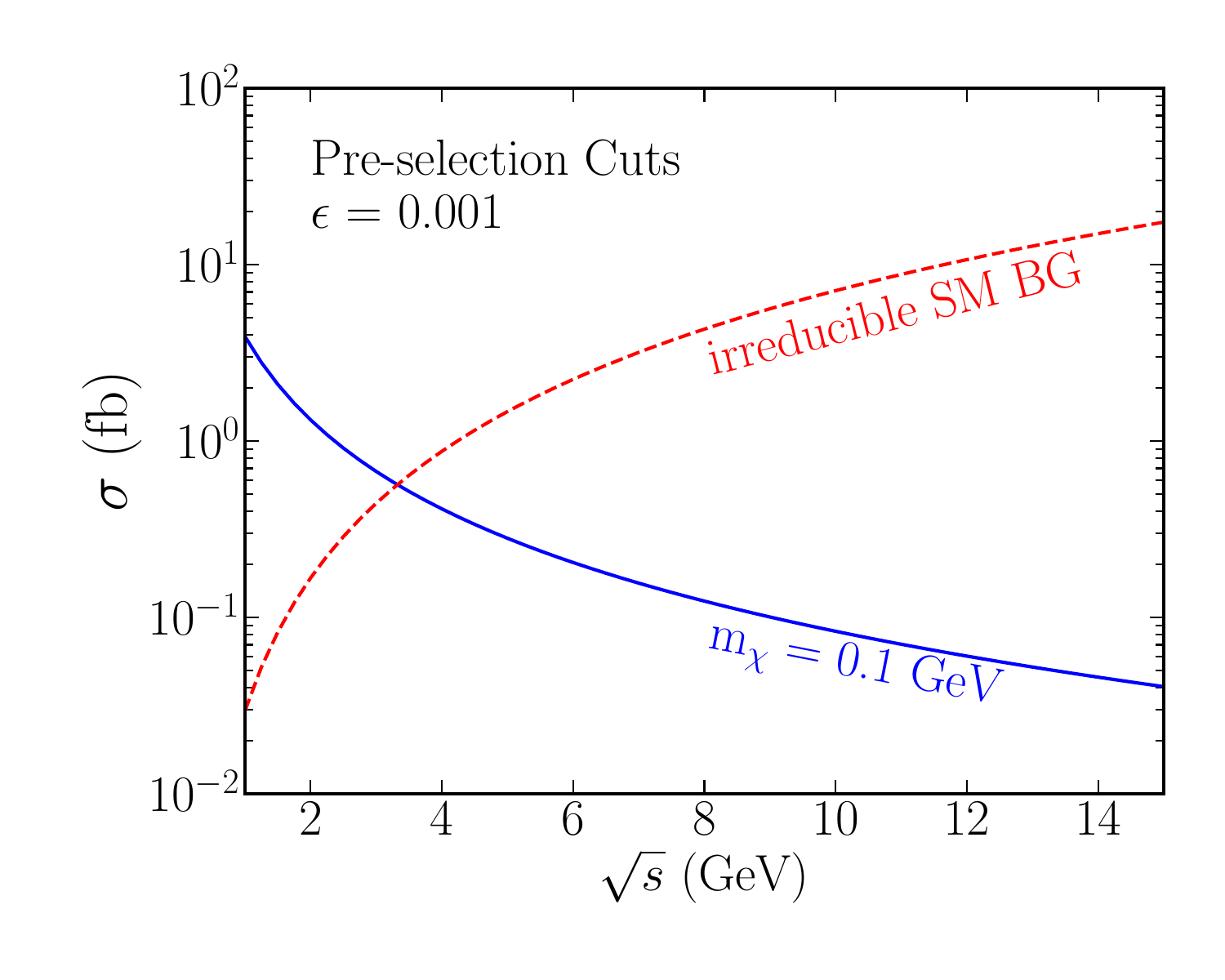}
\caption{
Monophoton cross sections
for millicharged
particles (solid) and for the SM irreducible BG (dashed)
versus the collision energy $\sqrt{s}$.
These cross sections are computed with the
following detector preselection cuts:
$E_\gamma > 25$ MeV for $\cos\theta_\gamma<0.8$
and
$E_\gamma > 50$ MeV for $0.86 < \cos\theta_\gamma< 0.92$.
The model parameters
$\epsilon=0.001$ and $m_\chi=0.1$ GeV
are used for the millicharged
particles model.
Taken from Ref. \cite{Liang:2019zkb}.
}
\label{Fig-SB-compare}
\end{center}
\end{figure}

A small fraction of the dark matter (DM) can be
millicharged in nature.
Recently, the EDGES experiment
detected an anomalous absorption signal in the global 21 cm background
near a redshift of $z=17$ \cite{Bowman:2018yin}.
Millicharged dark matter models have been invoked to
provide sufficient cooling of cosmic hydrogens
\cite{Munoz:2018pzp, Berlin:2018sjs, Barkana:2018qrx};
because the interaction cross section between
millicharged DM and baryons increases
as the universe cools, constraints from the early
universe can be somewhat alleviated.

Because the ionization signals from millicharged
particles are
so weak that typical detectors in particle colliders
are unable to detect millicharged
particles directly,
millicharged
particles can be searched for at electron colliders via
the monophoton final state
\cite{Liu:2018jdi, Liang:2019zkb}.
%
%MCPs can be searched for at the electron colliders via
%the monophoton final state
%\cite{Liu:2018jdi, Liang:2019zkb}.
%This is because the ionization signals from MCPs is
%so weak that typical detectors in particle colliders
%are unable to detect MCPs directly.
%Searches
The analysis of searching for millicharged
particles via the monophoton state at the STCF can be
easily extended to a variety of invisible
particles in the hidden sector.
In millicharged
particle models, monophoton events can be produced via
$e^+ e^- \to \bar \chi \chi \gamma$,
where $\chi$ is the millicharged
particle.
The irreducible monophoton background
processes have the form $e^+ e^- \to \bar \nu \nu \gamma$,
where $\nu$ is a neutrino. There are also
reducible monophoton backgrounds
due to the limited coverage of the detectors.
There are two types of reducible backgrounds:
the ``bBG'' background, which occurs when
all other visible final-state particles are emitted along the beam directions,
and the ``gBG'' background, which
is due to visible particles escaping the detectors via gaps
\cite{Liang:2019zkb}.

Fig.~\ref{Fig-SB-compare} shows the monophoton cross sections
for millicharged
particles and for the SM irreducible background,
where the analytical differential cross sections for these
processes are taken from Ref.\ \cite{Liu:2018jdi}.
The monophoton cross section for millicharged
particles
increases as the collision energy decreases, as shown in Fig.~\ref{Fig-SB-compare}.
In contrast, the monophoton irreducible
background grows with increasing collision energy.
Thus, an electron collider with a lower collision energy
has a better sensitivity to kinematically accessible millicharged
particles.

\begin{figure}[htbp]
\begin{center}
\includegraphics[width=0.45 \columnwidth]{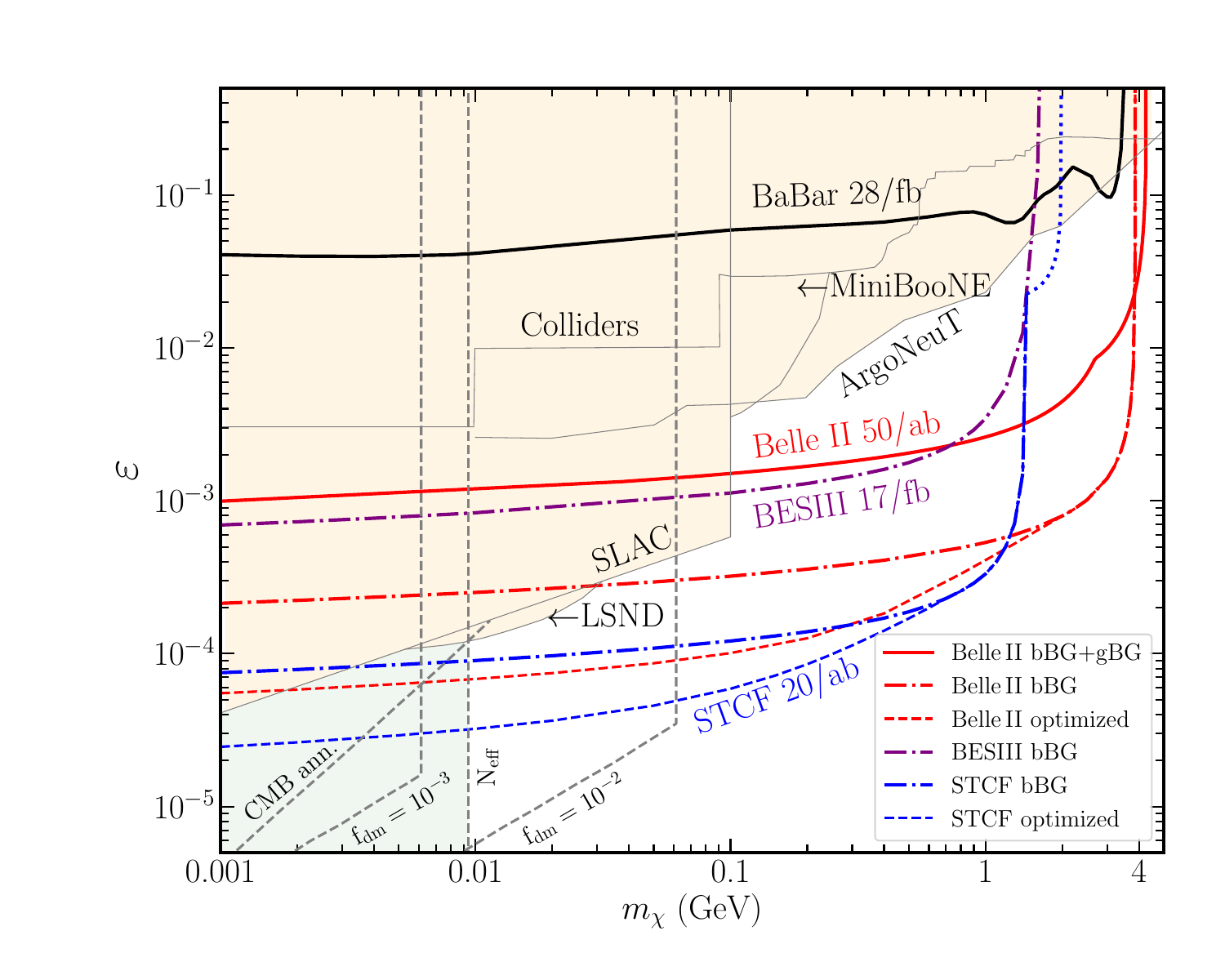}
\caption{
The expected 95\% C.L.\ upper bounds on millicharged
particles
from the STCF as well as from Belle II, BESIII, and BaBar.
The dot-dashed curves are obtained with
the bBG cut for the STCF, BESIII, and Belle II,
while gBG is neglected \cite{Liang:2019zkb}.
Taken from Ref. \cite{Liang:2019zkb}.
}
\label{fig:exclusion}
\end{center}
\end{figure}

To analyze the sensitivity of the proposed STCF experiment to millicharged particles,
the STCF detector is assumed to have the same
acceptance as the BESIII detector.
The STCF sensitivity to millicharged
particles in the MeV--GeV mass range is shown
in Fig.~\ref{fig:exclusion},
under the assumption of 20 ab$^{-1}$ of data collected at $\sqrt{s}= 4$ GeV.
The STCF can probe a large parameter space below that of the SLAC
electron beam dump experiment for millicharged
particles,
from $\sim$4 MeV to 0.1 GeV.
Millicharged
particles with $\epsilon \lesssim (0.8-3) \times 10^{-4}$
and masses from $\sim$4 MeV to 1 GeV
can be probed by the STCF with
20 ab$^{-1}$ of data at $\sqrt{s}= 4$ GeV.
This also eliminates a significant portion of the
parameter space in which the 21 cm anomaly
observed by the EDGES experiment can be
explained \cite{Munoz:2018pzp}.
The expected constraints on millicharged
particles from the STCF
analyzed on the basis of 20 ab$^{-1}$ of data collected at $\sqrt{s}= 4$ GeV
are better than those from Belle II with 50 ab$^{-1}$ of data
for millicharged
particles from 1 MeV to 1 GeV.
The increase in sensitivity is largely due to the
fact that the collision energy of the STCF
is lower than that of Belle II, which is $\sim 10.6$ GeV.
Thus,
the STCF has unprecedented sensitivity to the millicharged parameter space
for the MeV--GeV mass scale that has not been explored by current experiments.

\begin{figure}[htbp]
\begin{center}
\includegraphics[width=0.45 \columnwidth]{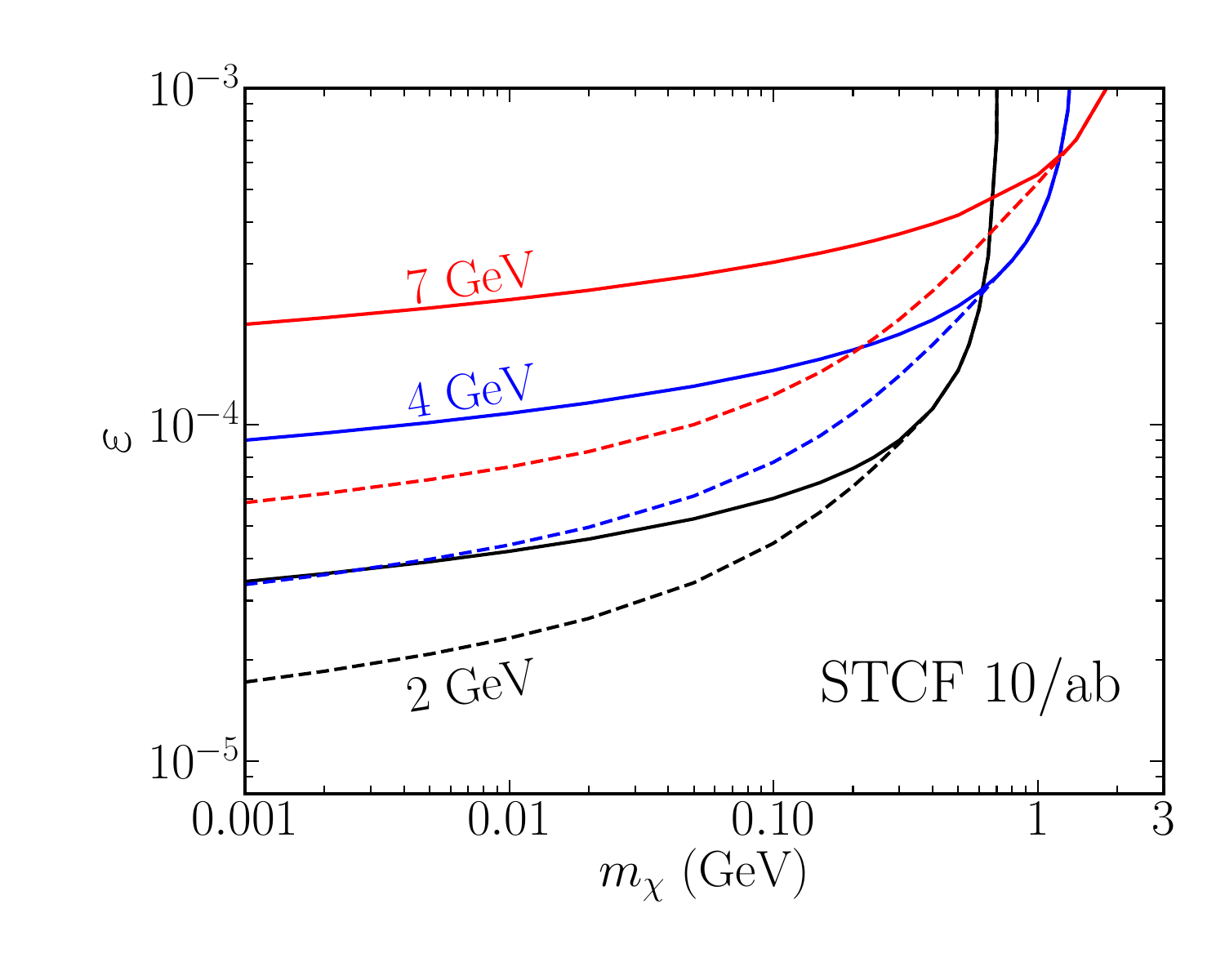}
\caption{
The expected 95\% C.L.\ upper bounds on millicharged
particles
with 10 ab$^{-1}$ of data assumed for each of the three
STCF $\sqrt{s}$ values.
The solid curves are analyzed with the bBG cut.
Taken from Ref. \cite{Liang:2019zkb}.
}
\label{F-STCF-limit}
\end{center}
\end{figure}

For simplicity, a single collision energy of $\sqrt{s}= 4$ GeV
with 20 ab$^{-1}$ is assumed to obtain the limits in Fig.~\ref{fig:exclusion}.
However, because the STCF will be operated at various energy points,
as shown in Table~\ref{tablelumi},
the actual limit should be analyzed by considering
various collision energies and detailed detector simulations.
The STCF sensitivities to millicharged
particles at three different collision
energies are compared in Fig.~\ref{F-STCF-limit},
where 10 ab$^{-1}$ of data is assumed for each collision
energy.
Although the low-energy mode loses sensitivity to heavy millicharged
particles,
it has better sensitivity than the high-energy mode for
probing light millicharged
particles.
For example, 10 ab$^{-1}$ of data with $\sqrt{s}=2$ GeV can be used to probe millicharged
particles
down to $\sim 4 \times 10^{-5}$ for a 10 MeV mass, as shown
in Fig.~\ref{F-STCF-limit}, outperforming the
$\sqrt{s}=7$ GeV mode by a factor of $\sim 5$.
The constraints on millicharged
particles at various collision energies are also shown in
Table~\ref{tab:mcp} for several benchmark points.
\begin{table}[htbp]
\begin{center}
\begin{tabular}{|c|c|c|c|c|c|c|}
\hline 
$\sqrt{s}$ (GeV)  & 2  & 2 & 4 & 4 & 7 & 7 \\ \hline
$m$ (MeV) & 1 & 100 & 1 & 100 & 1 & 100 \\ \hline
$\epsilon \lesssim$ 
& $3 \times 10^{-5}$ &  $7\times 10^{-5}$
& 9 $\times 10^{-5}$ &  $1\times 10^{-4}$
& 2 $\times 10^{-4}$ &  $3\times 10^{-4}$ \\ \hline
\end{tabular}
\caption{
The expected 95\% C.L.\ upper bounds on $\epsilon$ for millicharged
particles 
with 10 ab$^{-1}$ of data for three
STCF $\sqrt{s}$ values, namely, 2 GeV, 4 GeV, and 7 GeV, as 
analyzed with the bBG cut \cite{Liang:2019zkb}.}
\label{tab:mcp}
\end{center}
\end{table}

\newpage
\section{Summary}

The proposed STCF is a $4\pi$-solid-angle particle detector operating at an $e^{+}e^{-}$ collider with a high luminosity
($>0.5\times10^{35}$~cm$^{-2}$s$^{-1}$), center-of-mass energies spanning the range of $2\sim7$~GeV, and the future option
for a polarized $e^-$ beam. We have presented some of the interesting physics potential at the STCF, mainly on the basis of various particle systems ranging from higher masses, starting from the $XYZ$ states, to lower-mass systems, such as hyperon and glueball/hybrid states, as well as possible new light particles beyond the SM along with information can be extracted from their decays and interactions. Some studies that are important for extracting SM information not associated with energies on/near resonances have also been discussed. With unprecedentedly high luminosity, studies of the spectra in the relevant energy ranges can provide much more precise knowledge about known states and possibly other new and exotic states, and studies of decays and how they interact with other SM particles can yield new insights along with the most precise information about parameters in SM electroweak interactions, the perturbative and nonperturbative nature of strong QCD interactions, and possible new particles and interactions beyond the SM. The topics presented are not all inclusive; instead, we have focused on measurements that are unique to the STCF, with emphasis on reactions that challenge the SM, are sensitive to new physics and address poorly understood features in existing data. We summarize the highlights in the following.

\noindent
{\bf QCD dynamics and hadron physics:} The STCF will run in the energy range of 2 $\sim 7$ GeV, which is in the transition interval between nonperturbative and perturbative QCD based on the $SU(3)_C$ gauge interaction. Experimental data from the STCF will provide much more information to study the QCD dynamics of confinement
through the study of hadron spectra and interactions (see Sections~\ref{sec:charmonium}, \ref{sec:charmedhadron} and \ref{sec:qcd}). The energy region covers the pair production thresholds for the recently discovered doubly charmed baryon states, $XYZ$ states, charmed baryons, charmed mesons, $\tau$ leptons, and all of the strange hyperons. With the high luminosity of the STCF, firmer traces of the QCD-predicted glueballs/hybrids may finally be observed, and with operation above 7 GeV, higher-mass doubly charmed baryons or other possible new states may be studied in more detail.

More detailed discussions of hadron spectroscopy have been provided in the relevant sections. Two remarkable recent developments in hadron spectroscopy are worth emphasizing again:
1) the failure of hadron models to anticipate the rich charmonium spectrum of
hidden-charm states with masses above the open-charm pair threshold and
2) the emergence of clear experimental evidence for new, light-hadron spectra of QCD hybrids
and glueballs. These developments boost confidence that
more detailed spectroscopic studies are needed and will be fruitful. The STCF will be an ideal place to study related issues.
The mapping out of the {\it XYZ}, glueball and hybrid
spectra will require comprehensive measurements of as many decay modes as possible and
more sophisticated analysis techniques to extract and interpret exotic mesons that
overlap with conventional $q\bar{q}$ states.
Year-long STCF runs will produce data samples containing $\sim$3~T $J/\psi$ events and $\sim$500~B $\psi(3686)$ events
for in-depth explorations of light hadron physics. At CME~$\approx4230$~MeV, the STCF will function as an
``$XYZ$-meson factory'', producing $\sim$1B $Y(4260)$ events, $\sim$100M each of $Z_{c}(3900)$ and $Z_{c}(4020)$ events, and
$\sim$5M $X(3872)$ events
per year, thereby enabling precision Argand plots, studies of rare (including nonhidden-charm)
decays, precise mass and width measurements, etc.
With close cooperation
between high-precision experiments at the STCF and the LQCD community, a robust,
first-principles understanding of the confinement of quarks and gluons will be produced in the foreseeable future.

Many perturbative QCD properties can also be tested at the STCF, such as the determination of the $R$ values at various energies and therefore also the strong interaction coupling $\alpha_s$, especially at the $\tau$ threshold, as discussed in Section~\ref{subsec:tausm}. More data from the STCF can also be used to test NRQCD predictions for charmonium production with high precision.

The STCF can also measure strong interactions at the hadron level with great precision. Some of the interesting possible measurements are the time-like nucleon and hyperon form factors, as discussed. This will greatly help to improve the currently poor understanding of the structure of nucleons.
%After 100 years of experimental studies, the structure of nucleons is still poorly understood.\cite{Nayak,Yang}
In this context, time-like measurements are expected to play an increasingly important role.
Moreover, time-like pair production measurements are not restricted to nucleons;
the form factors of all of the weakly decaying hyperons can be measured and compared, thereby opening a new, previously
unexplored dimension of investigation. Currently available
(statistically limited) time-like experiments exhibit puzzling features in their threshold cross sections
and electric and magnetic form factors. At the STCF, the time-like nucleon and hyperon
form factors will be measured for $Q^{2}$ values as high as 40~GeV$^{2}$ with precisions that match
those of existing results for the proton and neutron in the space-like region. Moreover, hyperon polarizations will enable new determinations
of their parity-violating decay asymmetries and can be used to extract the complex phases between their electric and
magnetic form factors. At the STCF, the Collins effect will be able to be measured in the inclusive production
of two hadrons at the percent level, thereby providing valuable input for the interpretation of nucleon spin-structure
measurements at high-energy electron--ion colliders that are currently under construction in China
and the U.S.

\noindent
{\bf  Electroweak interaction, flavor physics and $CP$ violation:}
The electroweak interaction based on $SU(2)_L\times U(1)_Y$ is an integral part of the SM.
There are many free parameters in the theory. It is fair to say that the underlying structure of the SM is flavor physics;
most of the 19 fundamental parameters of the SM are the masses of the quarks and leptons and their flavor mixing angles.
As discussed in Section~\ref{sec:tau}, the large number of $\tau$ pairs produced at the STCF will provide much more accurate measurements of the electroweak couplings and test the universality of the weak interaction. Many other SM parameters can also be measured to further test the SM.

One of the most important tests of the SM is to see how well the CKM mechanism works. The most general tests of the SM that involve the CKM matrix are to confirm its unitarity and
the internal consistency of its elements. The SM coupling strengths for the $u\leftrightarrow s$
and $c\leftrightarrow d$ transitions are both equal to $G_{F}|\sin\theta_{c}|$, with a small, well-understood
$\mathcal{O}(10^{-4})$ correction. Here, $G_{F}$ is the Fermi constant, and $\theta_{C}$ is the Cabibbo angle.
Any significant difference in $|\sin\theta_{c}|$ extracted from different quark transitions
would be an unambiguous sign of new physics.
%Figure~\ref{fig}(b) summarizes the current status of
%$|\sin\theta_{c}|$ derived from different transitions,\cite{CKM1,CKM2,CKM3,CKM4,CKM5} where the mutual agreement
%is poor, the so-called $Cabbibo$ $angle$ $anomaly$.

The $\sim$0.2\% precision from nuclear $\beta$
($|V_{ud}|$) and kaon ($|V_{us}|$) decays is more than an order of magnitude better than the precision from
$D_s$ ($|V_{cs}|$) and $D$ ($|V_{cd}|$) decays, which is
$\sim$3\%, based on statistics-limited BESIII measurements of the $D_s^{+}\to\mu^+\nu$,
$D^+\to\mu^+\nu$ and $D^0\to K^-(\pi^-)\ell^+\nu$ decays. The clean environments for $D$ and $D_s$
mesons produced by $\psi(3770)\to D\bar{D}$ and $\psi(4160)\to D_s^*\bar{D}_s$, respectively, which are
unique to an STCF-like facility, are especially well suited for low-systematic-error $c$-quark transition
measurements. Year-long STCF runs at 3.773~GeV and 4.160~GeV would reduce the errors on $c$-quark-related
determinations of $|\sin\theta_c|$ to the 0.1$\sim$0.2\% level and match those from $\beta$ and kaon decays. The STCF will produce a large number of $\tau$ pairs, allowing more precise measurements of $\tau \to K^- \nu_\tau$ and $\tau \to \pi^- \nu_\tau$ to be carried out. This will also enable the determination of the value of $\theta_c$.

Searching for a non-SM source of $CPV$ is a promising strategy for uncovering signs of physics beyond
the SM. To date, intensive investigations of $CPV$ with beauty and charmed mesons and in the neutral kaon
system have not revealed any deviations from expectations based on the Kobayashi--Maskawa mechanism.
The good agreement between the SM calculation of $\epsilon '/\epsilon$ and its measured value
restricts the level of non-SM $CPV$ for non-SM parity-changing decays involving $s$ quarks to
$<6\times 10^{-5}$ but allows for asymmetries at $\mathcal{O}(10^{-3})$ in hyperon
parity-conserving decay processes such as $\Lambda\to p \pi^-$ and $\Xi^-\to \Lambda \pi^-$.
At the STCF, using quantum-entangled, coherent $\Lambda\bar{\Lambda}$ and $\Xi^-\bar{\Xi}^+$ pairs produced via
$J/\psi$ decays, a comprehensive search for non-SM CPV asymmetries could probe the sensitivity level between
$10^{-3}$ and the SM level of $\sim6\times10^{-5}$. Notably, the sensitivities for
CPV in hyperon decays depend linearly on the hyperon polarization, and thus, a future option for an $\sim$80\% polarized
$e^-$ beam at the STCF would boost the discovery potential for hyperon CPV by more than an order of magnitude.

Various CPV processes involving $\tau$ leptons have been discussed.
A particularly interesting one is CPV in $\tau\to K_{S}\pi\nu$. Until now, the sensitivity for CPV has been at only the
$\mathcal{O}(1\%)$ level when studying  $\tau\to K_{S}\pi\nu$ decays using unpolarized $\tau$ leptons.
The corresponding CPV sensitivity for one year of STCF data at $E_{\rm c.m.}=4.26$~GeV will be
$\mathcal{O}(10^{-4})$, which is the level expected for the well-understood influence of SM CPV effects in the
neutral kaon meson system. The future polarized $e^-$ beam option would enable unambiguous probes
for new-physics sources of CPV in $\tau$-lepton decays to final states that do not contain neutral kaons, such as
$\tau^-\to \pi^-\pi^0\nu$.
Searches for $CP$ violation in heavy hadron decays and $\eta/\eta' \to \pi \pi$ decays could also be carried out at the STCF.

%a different approach,
%i.e., one based on high-statistics studies of decays to final states
%that do not contain neutral kaons and with strict controls on systematic errors, is needed.
%Such a method for sensitive searches for non-SM CP violations in semileptonic $\tau\to\pi\pi^{0}\nu$
%decays of polarized $\tau$ leptons produced near threshold in an $e^{+}e^{-}$ collider was presented
%in detail by Tsai in ref.~\cite{tau4}. Here he established that a threshold facility like STCF with a
%polarized $e^{-}$ beam would be ideally well suited for high sensitivity searches fo CP-violating
%asymmetries.

\noindent
{\bf Other searches for new physics beyond the SM:}
With high luminosity, a clean collision environment and excellent detector performance, the STCF will have great potential
to search for rare and forbidden decays and will serve as a powerful instrument for other investigations of physics beyond the SM. Such searches can be classified into three categories:
(1) decays via a FCNC process,
(2) decays with LFV and
(3) decays with LNV.
The STCF will support searches for $\tau$-lepton LFV and LNV decays with sensitivities of $10^{-8}$ to
$10^{-9}$. In addition, as discussed in Section~\ref{sec:newphys}, it will serve as a platform to search for proposed new low-mass particles such as dark photons, light scalars and millicharged particles.

The physics program at the STCF is a multidimensional program. We emphasize that the unprecedentedly high luminosity in the energy region of $2\sim 7$ GeV offers great physics potential, enabling us to develop a much more in-depth understanding of the challenges facing the SM and hopefully providing some clues or solutions for overcoming them. It will play a crucial role in leading the high-intensity frontier of elementary particle physics worldwide.

\end{chapter}

%\begin{comment}
    
\chapter{Detector}
\label{CDR_det}
\section{Physics Requirements}
\label{sec:phys_requirements}

To access the physics potential and scientific merits of the STCF as discussed in Chapter~\ref{CDR_phys} and to understand the demands on the
machine and detector performance from a physics perspective,
the physics requirements of the detector design are proposed from several benchmark processes,
as listed in Table~\ref{phyreq} along with the corresponding parameters to be optimized.
The processes cover a wide range of physics programs
including $CPV$ tests in the $\tau$ lepton, baryon and charm meson sectors, searches for charged lepton flavor violations~(cLFVs), tests of the unitarity of the CKM matrix, nucleon structure researches via electromagnetic form factors and fragmentation functions {\it, etc. }
In Table~\ref{phyreq}, and in the text overall, $chrg1/chrg2$ denotes the sepatation of the signal particle
$chrg1$ from the background particle $chrg2$. Both the misidentification rate and the
suppression power are used to evaluate the power of the particle identification.

\begin{table}[htbp]
\caption{Summary of the physics processes and the corresponding responses to be optimized.}
\label{phyreq}
\begin{center}
\begin{tabular}{c|c}
\hline
\hline
  Physics Process  &  Optimized Parameter   \\
\hline
 $\tau\to K_{s}\pi\nu_{\tau}$; $J/\psi\to\Lambda\bar{\Lambda}$    & Vertex reconstruction; Tracking (efficiency, momentum resolution) \\
 $\tau\to\gamma\mu$; $\tau\to lll$; $D_{s}\to\mu\nu$; $D\to\pi\mu\nu$ 	 & PID (range, $\mu/\pi$ suppression power, efficiency) \\
 $e^{+}e^{-}\to\pipi+X$, $KK+X$; $D_{s}\to\tau \nu_{\tau}$        & PID (range, $\pi/K$ and $K/\pi$ misidentification, efficiency)  \\
 $\tau\to\gamma\mu$; $J/\psi\to\Lambda\bar{\Lambda}$              &  Photon  (position/energy resolution)\\
 $e^{+}e^{-}\to n\bar{n}$, $e^{+}e^{-}\to\gamma n\bar{n}$         & $n$  (position/time resolution)\\
 $D^{0}\to K_{L}\pi^{+}\pi^{-}$                                   & $K_{L}$  (position/time resolution) \\
\hline
\hline
\end{tabular}
\end{center}
\end{table}

\subsection{Charged Particles}

\begin{figure}[htbp]
\begin{center}
\begin{overpic}[width=9cm, height=7cm, angle=0]{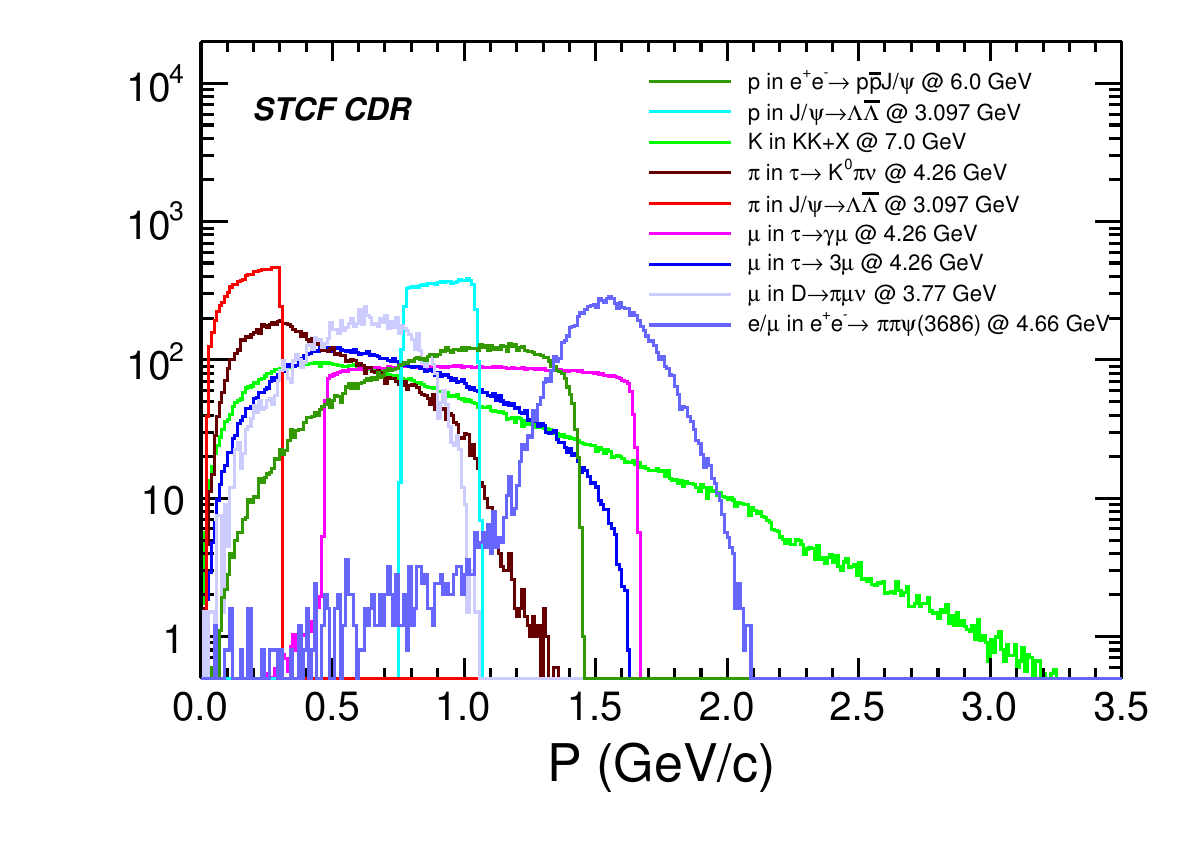}
\end{overpic}
\caption{ Momentum distributions of charged particles from various processes at the truth level, normalized to 10k entries.}
\label{pr:chgmom}
\end{center}
\end{figure}

The main goal of the STCF detector is to precisely measure the production and decay properties of various particles, {\it i.e.}, charmonium states, $XYZ$ particles,
$\tau$-leptons, charm hadrons and all hyperons.
 It is crucial to be able to detect charged final-state particles with an excellent tracking
efficiency and momentum resolution.
Figure~\ref{pr:chgmom} shows the momentum distribution for charged particles ($e$, $\mu$, $\pi$, $K$ and $p$)
from several physics processes.
The momenta spectra cover a large range, spanning as high as 3.5~GeV/c, while most particles have momenta less than 2.0~GeV/c. In addition, there are a considerable number of particles with momenta values lower than 0.4 GeV/c.
This requires the detector to be able to cover a large momentum range with high reconstruction efficiency.

\subsubsection{Vertex Resolution}
The STCF has unique features of rich resonances and the threshold production of hadron pairs,
which makes it a clean place to study their properties with high detection
precision and a low background.
Due to the threshold characteristics, the decay lengths of most particles with finite
lifetimes, {\it i.e.}, $D^{0}$, $D^{\pm}$, end immediately after the production points.
Therefore, the time-dependent analysis of $D^{0}$ decay is not applicable at STCF for
the study of $D^{0}-\bar{D}^{0}$ mixing.
Instead, with the charm meson pair produced near the threshold, {\it i.e.}, $e^{+}e^{-}\to D^{*0}\bar{D}^{0}+c.c$ at $\sqrt{s}=4.009$~GeV, the mixing parameters of $D^{0}-\bar{D}^{0}$ can be estimated with much better
sensitivity
by means of a quantum coherence approach.
As a consequence, the vertex measurement of the tracks with a precision of hundreds of $\mu$m is sufficient to meet the needs of
most physics programs at the STCF. For some special physics such as $\tau$ lifetime measurement, vertex resolution requirements should be re-evaluated.

\subsubsection {Tracking Efficiency}
In the benchmark processes of testing $CPV$s in hyperon decays, $J/\psi\to\Lambda\bar{\Lambda}$ and $J/\psi\to\Xi^{-}\bar{\Xi}^{+}$,
the charged pions in the final states have low momentum within the region of (0, 0.3)~GeV/c
while the momentum of protons is relatively high within the region of (0.6, 1.0)~GeV/c.
In the process of testing $CPV$s in $\tau$-lepton decay, $\tau^{-}\to K_{s}\pi^{-}\nu_{\tau}$, the charged pions mostly accumulate in the low-momentum region.
For most charm meson $D$ or $D_{s}$ decays, the momenta of the final states are low due to large multiplicities.

\begin{figure}[hbtp]
\begin{center}
\begin{overpic}[width=7.3 cm, height=5.0 cm, angle=0]{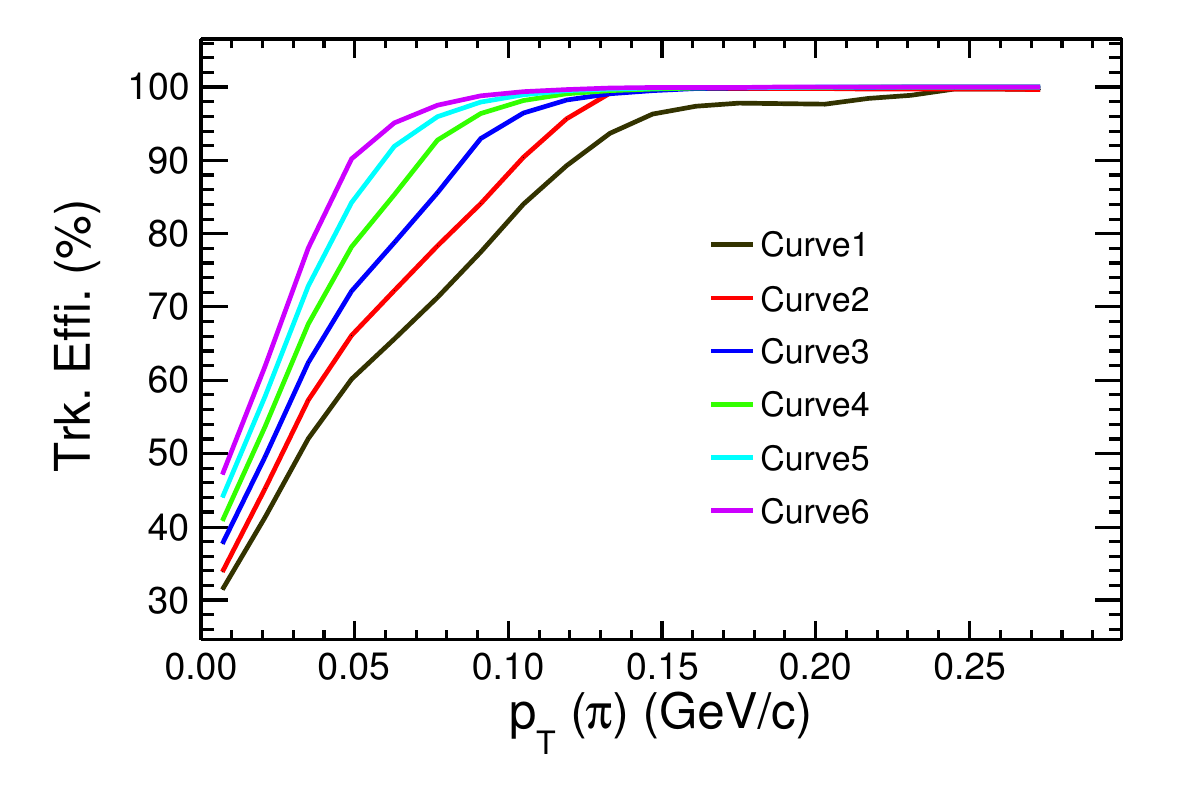}
\put(23,55){\textbf{(a)}}
\end{overpic}
\begin{overpic}[width=7.3 cm, height=5.0 cm, angle=0]{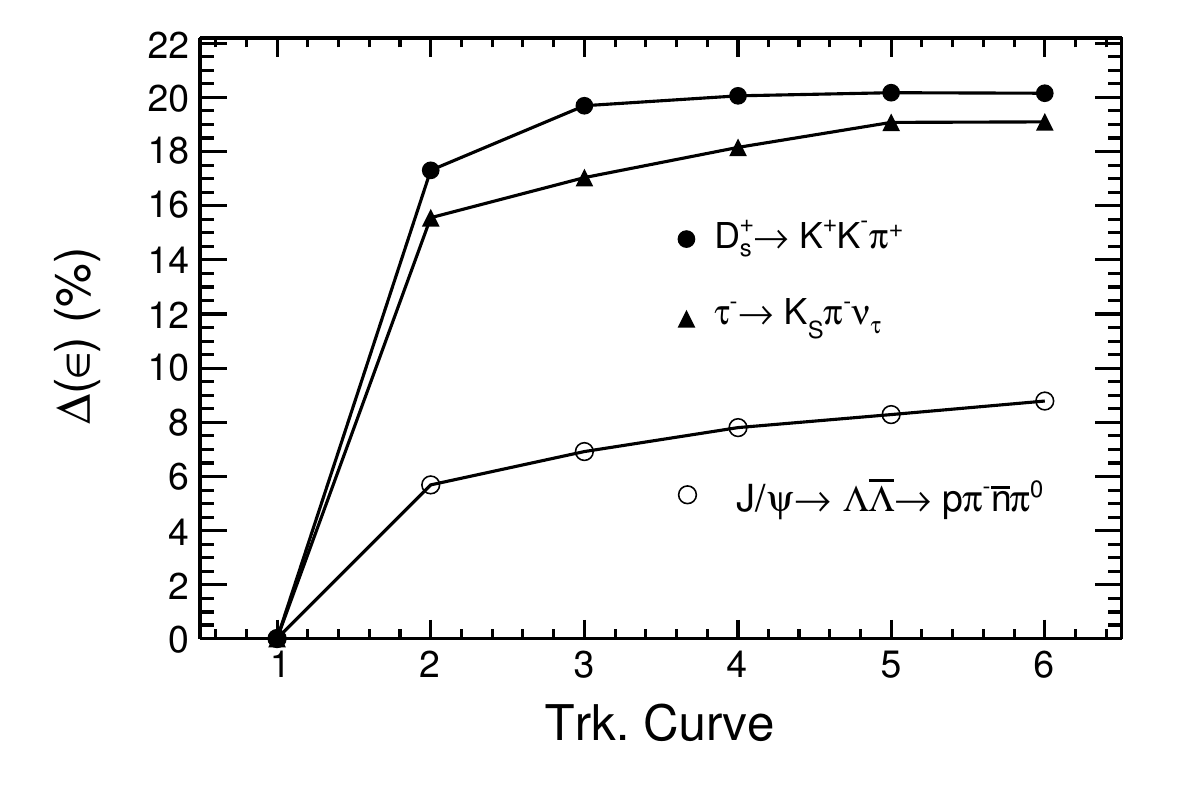}
\put(23,55){\textbf{(b)}}
\end{overpic}
\end{center}
\caption{ (a) Six tracking efficiency curves of charged pions used for optimization of benchmark processes.
(b) The relative improvement in the detection efficiency $(\Delta(\epsilon)$ for the processes for the six efficiencies, where solid circles represent $D_{s}^{+}\to K^{+}K^{-}\pi^{+}$, triangles represent $\tau^{-}\to K_{S}\pi^{-}\nu_{\tau}$ and open circles represent $J/\psi\to\Lambda\bar{\Lambda}\to p\pi^{-} \bar{n}\pi^{0}$.}
\label{trackeff1}
\end{figure}

The optimization of tracking efficiency is studied for charged pions.
Six different efficiency curves of charged pions are applied, as shown in Fig.~\ref{trackeff1}(a),
where curve1 is the real detection efficiency curve obtained from BESIII detector, while
curve 2-6 are obtained by multiplying the detection efficiency of curve 1 by a factor of 1.1, 1.2, 1.3, 1.4 and 1.5, respectively.
These efficiency curves are applied to various benchmark processes, and the
improvement in the detection efficiency $\Delta(\epsilon)$ to that of the unoptimized state is depicted in Fig.~\ref{trackeff1}(b).
The value of $\Delta(\epsilon)$ is related to the number of final charged particles with low momentum; thus,
$\Delta(\epsilon)$ is larger in
$D_{s}^{+}\to K^{+}K^{-}\pi^{+}$ than in $J/\psi\to\Lambda\bar{\Lambda}\to p\pi^{-} \bar{n}\pi^{0}$.
The efficiency is most optimized for curve 2 in all the processes,
which corresponds to a pion tracking efficiency of 90\% at $p_{T}=0.1$~GeV/c
within the detector acceptance.
Under this optimized efficiency curve, the physics requirements can be achieved:
sensitivities for $CPV$ in $\Lambda$ hyperon decay
on the order of $\Delta A_{CP}^{\Lambda}< \mathcal{O}(10^{-4})$ and in
$\tau$ lepton decay on the order of $\mathcal{O}(10^{-3})$ and plenty of
reconstructed $D_{s}$ mesons for further (semi)leptonic decays.

The polar angles of the particles from various processes in Fig.~\ref{pr:chgmom} are usually uniformly distributed,
and therefore, a large acceptance, {\it i.e.} $|\cos\theta|<0.93$, is needed.

\subsubsection {Momentum Resolution}

A good momentum resolution for charged particles, especially for low-momentum tracks, is important in
distinguishing signals from background events. For example, in the rare semileptonic decays of hyperons,
$\Xi^{-}\to\Lambda e^{-}\nu_{e}$, the dominant background comes from pionic decay $\Xi^{-}\to \Lambda \pi^{-}$.
A good momentum resolution for low-momentum pions and electrons is essential to separate the signal from the background
in this process.
%It is essential to improve the momentum resolution of low-momentum $e$ and $\pi$ to have a good separation
%in the reconstruction of events.

The momenta of charged particles are usually measured by their flight trajectories in a magnetic field.
With more points measured on the trajectory, more accurate track position is obtained, and better momentum resolution
can be obtained. In addition, for tracks associated with low momentum, the main effect on the momentum resolution
comes from multiple Coulomb scattering on the material in the detector. Therefore, a material with a low atomic number Z is
required in the detector.

\begin{figure}[htbp]
\begin{center}
\begin{overpic}[width=7.5cm, height=5cm, angle=0]{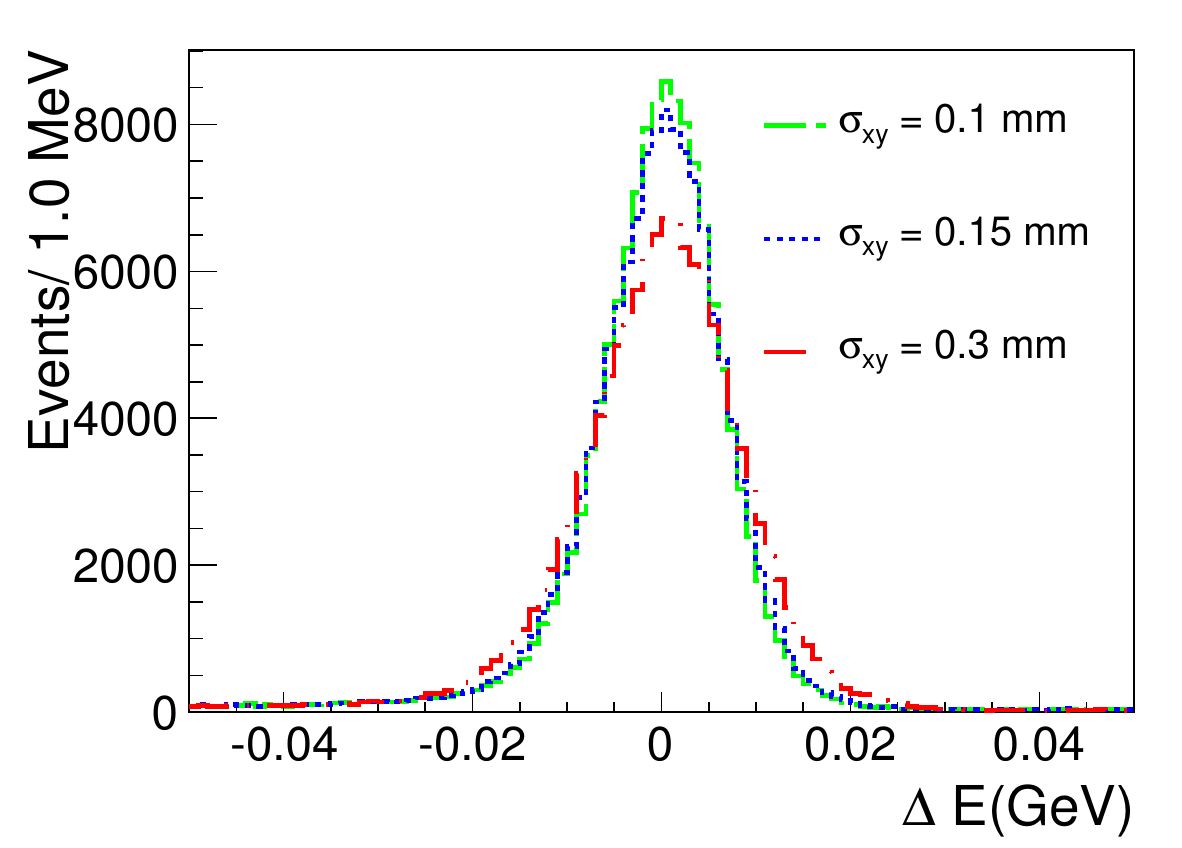}
\put(20,50){\small{(a)}}
\end{overpic}
\begin{overpic}[width=7.5cm, height=5cm, angle=0]{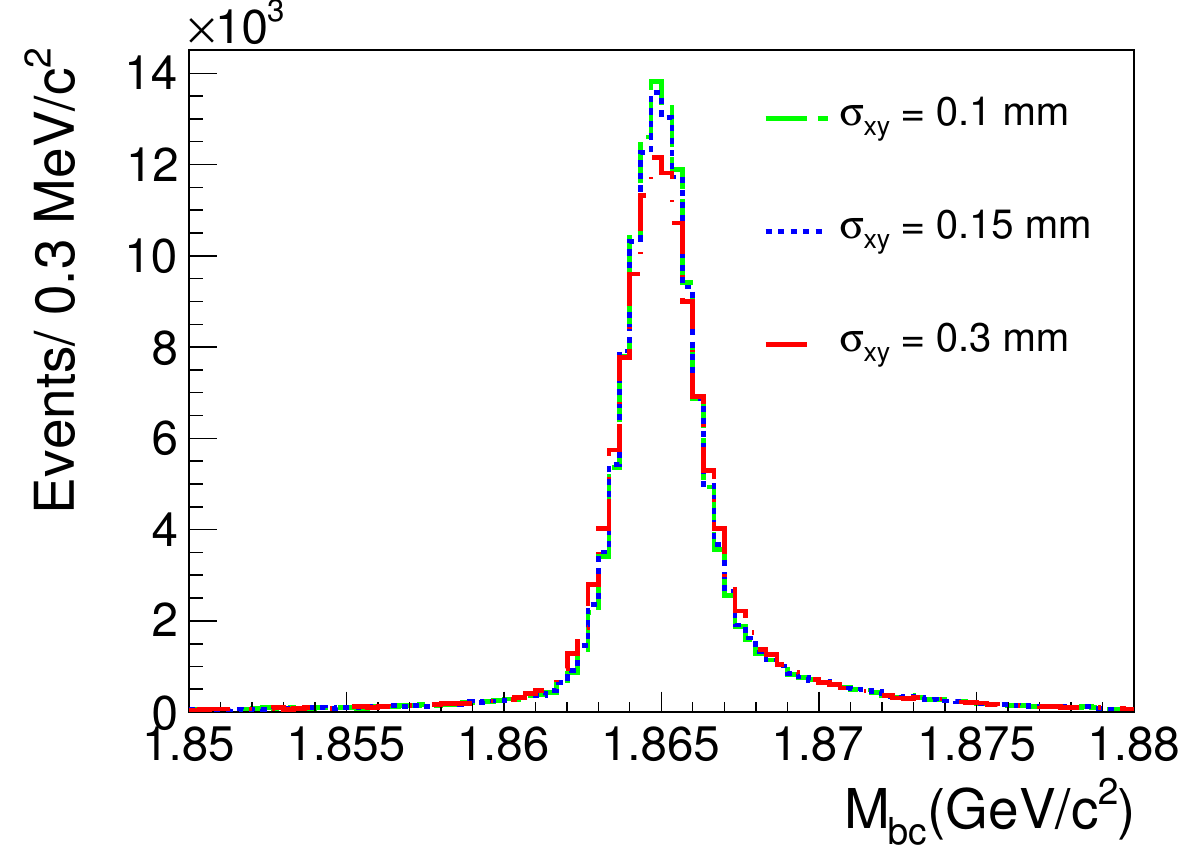}
\put(20,50){\small{(b)}}
\end{overpic}
\begin{overpic}[width=7.cm, height=5cm, angle=0]{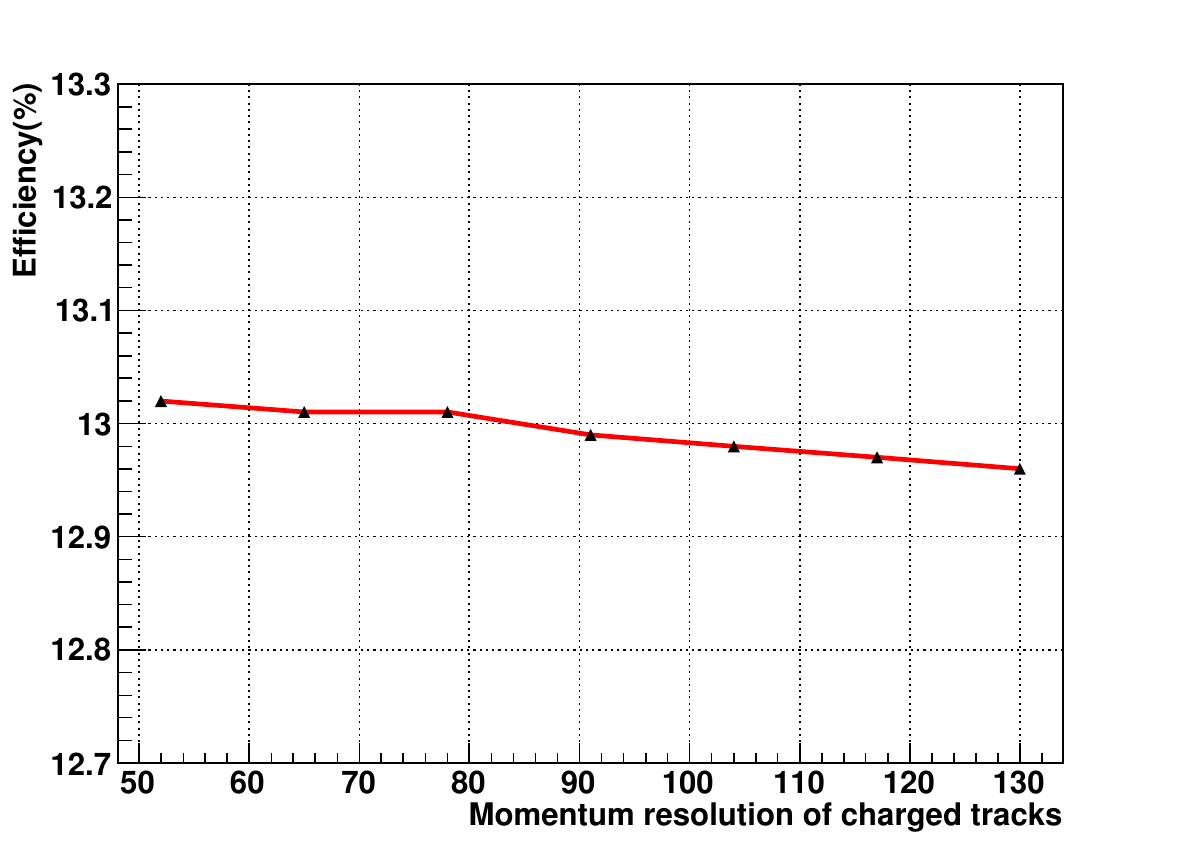}
\put(20,50){\small{(c)}}
\end{overpic}
\begin{overpic}[width=7.cm, height=5cm, angle=0]{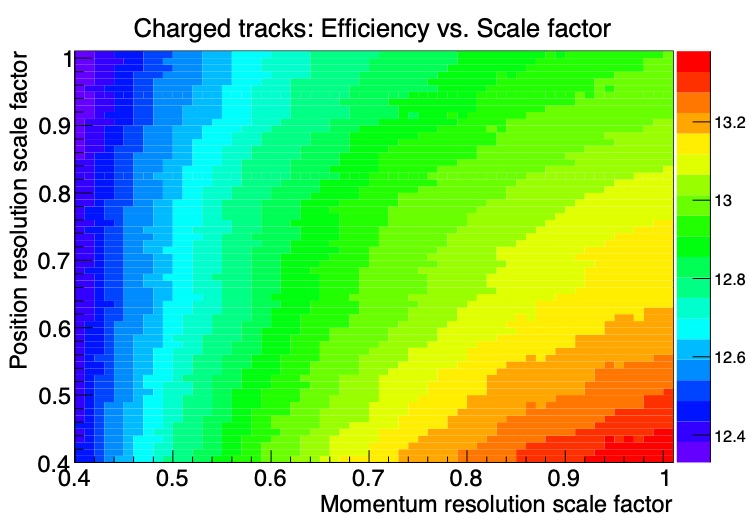}
\put(20,50){\small{(d)}}
\end{overpic}
\end{center}
\caption{
The distribution of (a) $\Delta E$ and (b) $M_{BC}$ associated with the spatial resolution of the track system
in process $e^{+}e^{-}\to D^{0}\bar{D}^{0}$ at $\sqrt{s}=3.77$~GeV with $D^0\to K^-\pi^+$.
The different colored lines represent different spatial resolutions. (c) Detection efficiency with different
charged track position resolutions in the $J/\psi\to\Lambda\bar{\Lambda}$ process. (d) 2D scattering plot of position resolution versus momentum resolution in the $J/\psi\to\Lambda\bar{\Lambda}$ process. }
    \label{D0}
\end{figure}

Next, we discuss the effect of position resolution on momentum measurement from $D$ meson decay processes
whose final states are in a moderate momentum region and $J/\psi\to\Lambda\bar{\Lambda}$ with low-momentum pions.
The reconstruction of $e^{+}e^{-}\to D^{0}\bar{D}^{0}$ produced near the threshold exploits two key variables that
discriminate the signal from the background, the energy difference $\Delta E=E-E_{\rm beam}$ and the beam constrained mass $M_{BC} = \sqrt{E_{\rm beam}^2/c^4-p^2/c^2}$,
where $E_{\rm beam}$ is the beam energy and $E$ and $p$ are the total measured energy and three-momentum of the charm meson, respectively.
Three sets of position resolutions are applied, {\it, i.e.}, $\sigma_{xy}=$100, 150 and 300~$\mu$m.
The results indicate that a better position resolution for the charged track
would improve the resolution of $\Delta E$ and $M_{BC}$, as shown in Fig.~\ref{D0}(a) and (b), but the improvement is not significant.
In addition, the variation in detection efficiency in $J/\psi\to\Lambda\bar{\Lambda}$ is examined, yielding similar
results, as shown in Fig.~\ref{D0}(c) and (d).

Considering that an acceptable position resolution that trivially affects the momentum measurement
compared with multiple scattering is sufficient for physics,
a charged particle position resolution
better than 130~$\mu$m and a corresponding charged
track momentum resolution of $\sigma_{p}/p=0.5\%$ at $p=1$~GeV/c are needed.
Possible contribution from multiple scattering in the momentum measurement will be discussed in Sec.~\ref{sec:mdc}.

\subsection{Photons}

\begin{figure}[htbp]
\begin{center}
\begin{overpic}[width=9cm, height=7cm, angle=0]{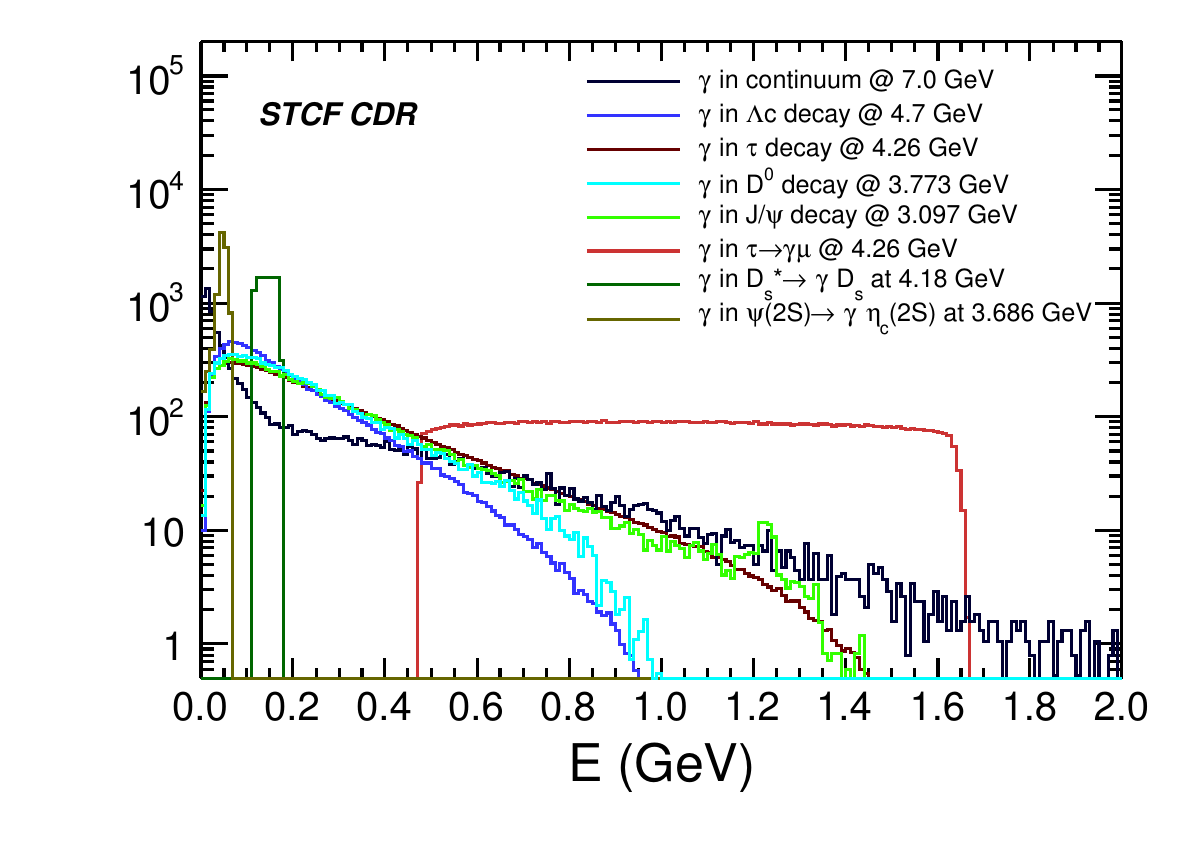}
\end{overpic}
\caption{ Energy distribution of photons at the truth level, normalized to 10k entries.}
\label{pr:phoene}
\end{center}
\end{figure}

Photons are one of the most important particles in the final states at the STCF, and they are involved in many physics programs,
as shown in Fig.~\ref{pr:phoene}.
The energy~($E$) of photons can be as high as
3.5~GeV, although multiplicity is rare when the energy is $E>2$~GeV.
For example, the $e^{+}e^{-}\to\gamma\gamma$ process with energy of photon equal to beam energy 
is essential for the luminosity measurement.
In addition, the photon energy can be as low as dozens of MeV, such as $\psi(3686)\to\gamma\eta_{c}(2S)$ and $D_{s}^{*}\to\gamma D_{s}$. The requires the detector
to be able to cover a large energy range with high efficiency.
The energy coverage of photon detection is therefore required to be from 25~MeV to 3.5~GeV.
The energy and position resolution
are two key parameters for photon detection.
Besides, to distinguish the neutral tracks and suppress the noise photons, a time resolution of hundred ps is required for the 
calorimeter system, which will be further discussed in Sec.~\ref{neutralhadrons}.

\begin{figure}[htbp]
\begin{center}
\begin{overpic}[width=7.5cm, height=5cm, angle=0]{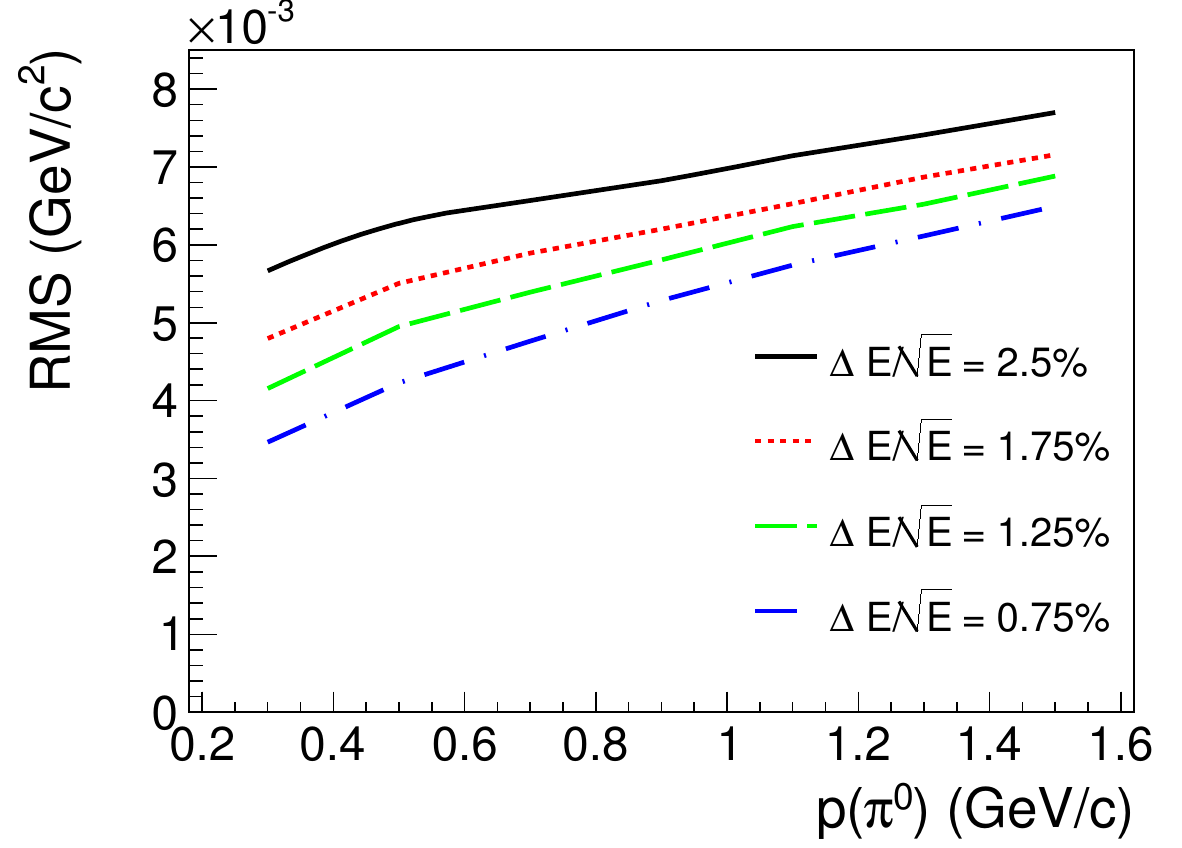}
\put(26,58){\small{(a)}}
\end{overpic}
\begin{overpic}[width=7.5cm, height=5cm, angle=0]{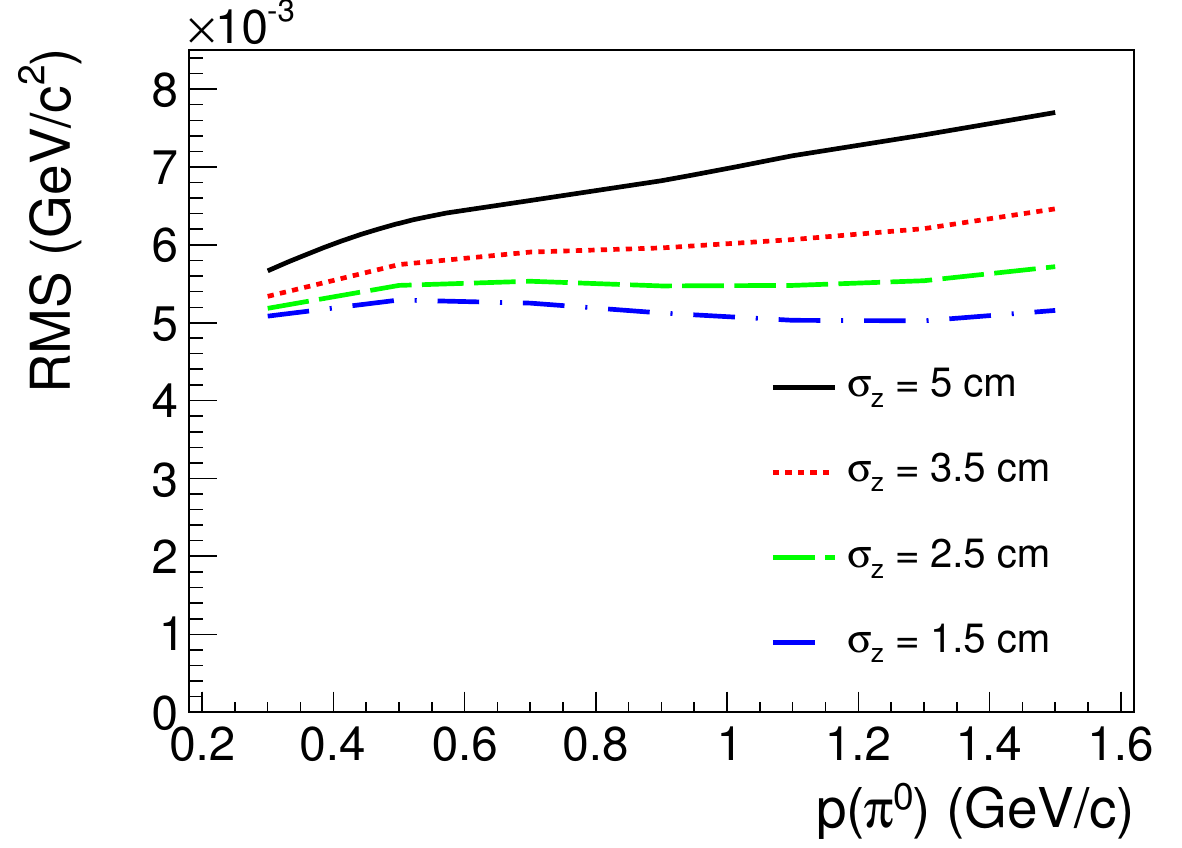}
\put(26,58){\small{(b)}}
\end{overpic}
\end{center}
\vskip -0.6 cm
\caption{ The RMS of $M_{\pi^0}$ versus the momentum of $\pi^{0}$ under different (a) energy resolution $\Delta E/\sqrt{E}$ and
(b) spatial resolution $\sigma_z$ when $E=1$~GeV. }
    \label{mpi0}
\end{figure}

The resolution of photon detection is crucial for the reconstructed mass spectra of various particles containing photons,
such as $\pi^{0}$, where the energy of most $\pi^{0}$ is located below 1.5~GeV at the STCF.
The evolution of the $\pi^{0}$ mass spectra is studied with various
photon energy/position resolutions.
The root mean square (RMS) value of $M_{\pi^0}$ shows that for low momentum $\pi^0$, the invariant mass resolution is significantly improved with
a finer photon energy resolution,
while for high momentum $\pi^0$, the mass resolution of $\pi^{0}$ is significantly improved with
finer photon position resolution, as shown in Fig.~\ref{mpi0}.
Similar conclusions can be drawn in the study of other mass spectra.

In the charged lepton-flavor-violation~(cLFV) process $\tau\to\gamma \mu$, the signals are distinguished from
the background by constructing the mass and energy distribution of the $\gamma\mu$ system, where the reconstruction of photons
is the main source of the resolution of these spectra.
%photon is the dominant source on the resolutions.
Moreover, one of the dominant background signals comes from the misidentification of $\gamma$ from $\pi^{0}$.
A good energy and position resolution can help to improve the resolution of signals and hence the signal-to-background ratio.
According to Fig.~\ref{pr:phoene}, the energy of photons in $\tau\to\gamma\mu$ ranges from 0.5 to 1.7 GeV, a range in which the position resolution of photons matters more.
Different sets of detector responses are applied in the estimation of experimental sensitivity in $\tau\to\gamma\mu$,
with the energy and position resolutions of photons varying from 2.0\% to 2.5\% and from 3~mm to 6~mm, respectively.
It is found that to obtain the ability to probe
the cLFV process $\tau\to\gamma\mu$ with an upper limit of better than $10^{-8}$ at the 90\% confidence level~(C.L. ),
the energy and position resolutions of photons with 1~GeV energy are required to be better than 2.5\% and
5~mm, respectively.
It should be noted that, though the detector is required to reconstruct photons with energy up to 3.5~GeV, the energy resolution is not that strict since the $e^{+}e^{-}\to\gamma\gamma$ process can be easily distinguished
from hadronic backgrounds
with a proper energy window and back-to-back angle requirements.

A good position resolution is also important in the detection of other neutral particles.
In process $J/\psi\to\Lambda\bar{\Lambda}$ with $\Lambda\to p\pi^{-}$
and $\bar{\Lambda}\to\bar{n}\pi^{0}$, the dominant background comes from $J/\psi\to\Lambda\Sigma^{0}$ with an additional final state photon.
The various position resolutions can affect the mass window for the reconstructed
$\bar{\Lambda}$. The simulation result shows that with a better position resolution, the resolution
of $M_{\bar{\Lambda}}$ can be improved significantly.
The efficiency for reconstructing the process is increased by 9.3\% when the position resolution is
improved by a factor of 20\%, {\it i.e.}, from 6~mm to 5~mm in the photon case.

\subsection{Neutral Hadrons}
\label{neutralhadrons}
\begin{figure}[htbp]
\begin{center}
\begin{overpic}[width=9cm, height=7cm, angle=0]{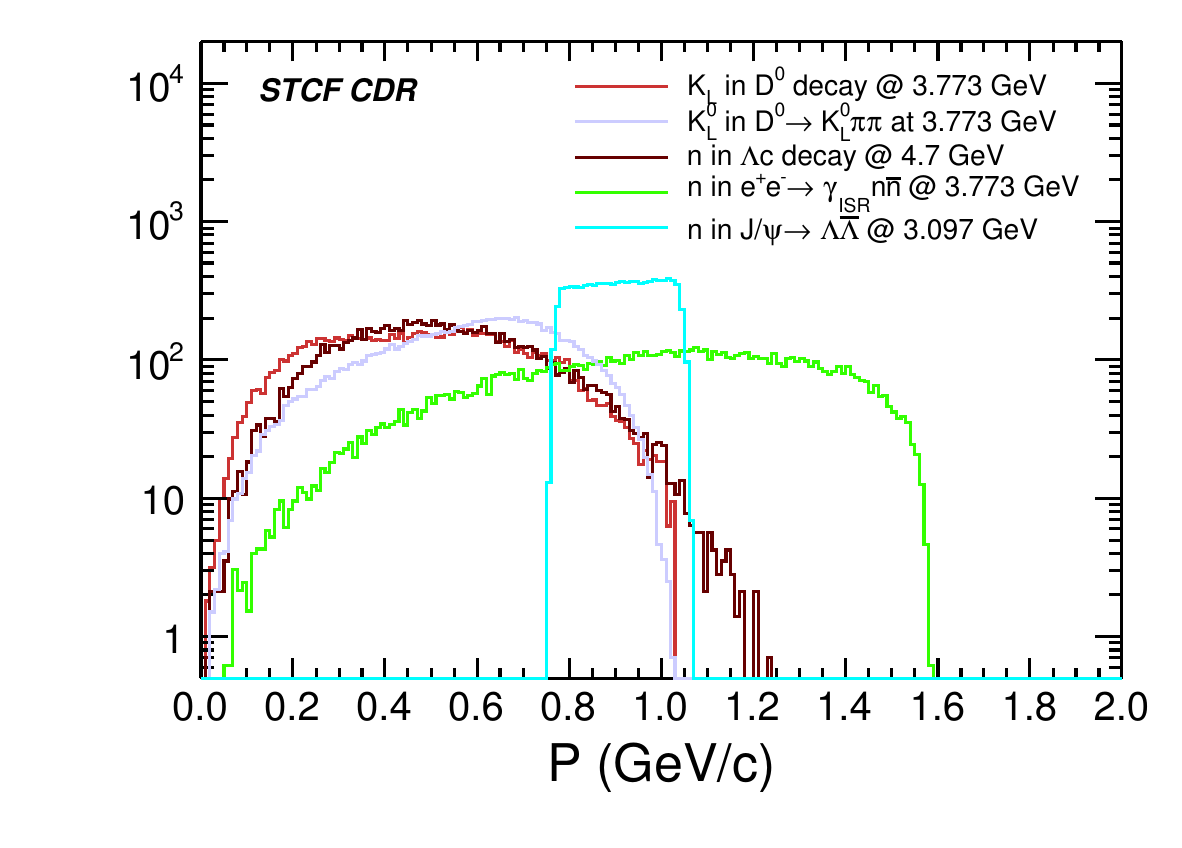}
\end{overpic}
\caption{ Momentum distribution of neutral particles
($K_{L}/n$) from various physics processes, which is depicted at the truth level and normalized to 10k entries.}
\label{pr:neuhad}
\end{center}
\end{figure}

Apart from photons, other neutral particles, such as neutrons and $K_{L}$, are also involved in many interesting physics programs,
as shown in Fig.~\ref{pr:neuhad}:
the studies of neutrons are important to understanding their internal structures and improving
knowledge for hyperon decays containing neutrons; $D^{0}\to K_{L}\pi^{+}\pi^{-}$
is one of the benchmark processes for the study of $D^{0}-\bar{D}^{0}$ mixing
and $CPV$ by means of the quantum coherence of $D^{0}$ and $\bar{D}^{0}$ production.

\begin{figure}[hbtp]
\begin{center}
\begin{overpic}[width=9cm, height=6cm, angle=0]{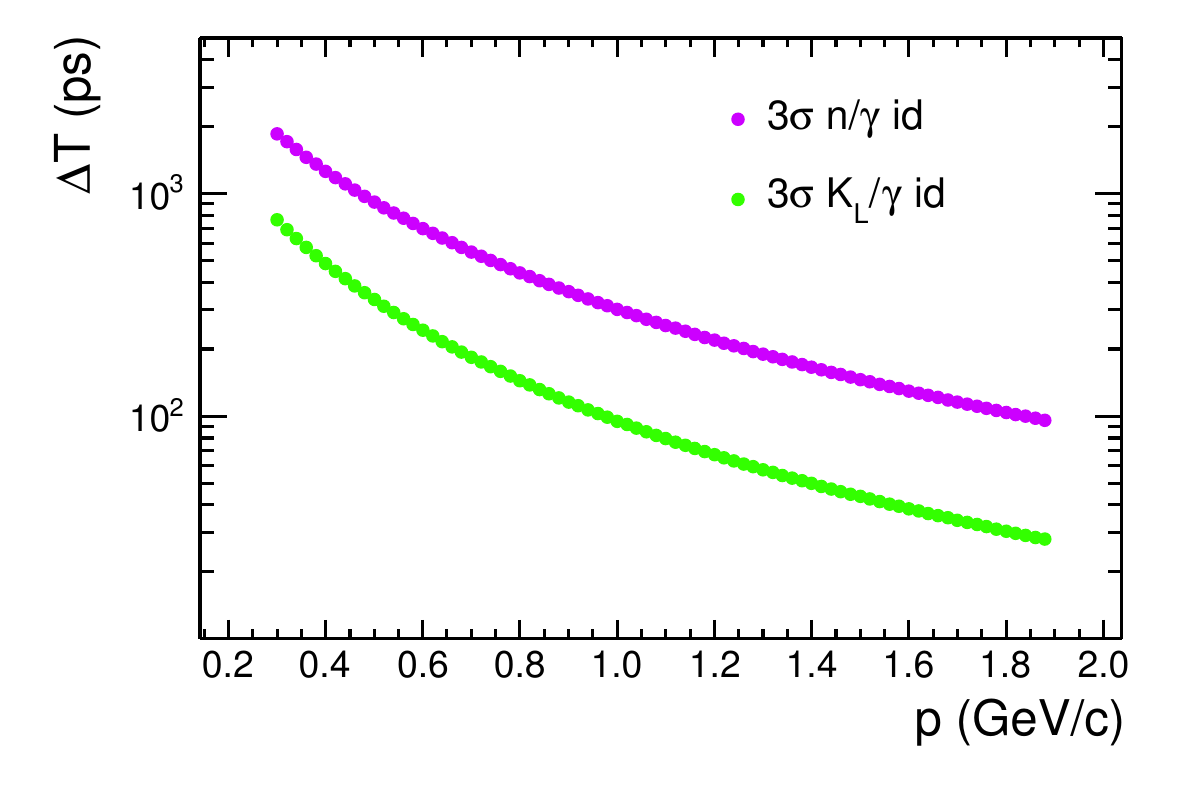}
\end{overpic}
\end{center}
\caption{Expected time resolution ($\Delta T$) for distinguishing neutral particles, neutrons/$K_{L}$ from photons with
a separation power of $3\sigma$ versus their incident momenta for a flight length of 1.5~m. }
\label{time}
\end{figure}

It is essential to have a good ability to separate neutral hadrons and photons
since the latter are the dominant background sources for the former.
For example, in processes $e^{+}e^{-}\to n\bar{n}$ and $D_{0}\to K_{L}\pi^{+}\pi^{-}$, the dominant backgrounds come from the
photon contamination in
$e^{+}e^{-}\to\gamma\gamma$ and $D_{0}\to K_{S}\pi^{+}\pi^{-}$ with $K_{S}\to\pi^{0}\pi^{0}$, respectively.
Due to the electrically neutral nature of these hadrons and that they can deposit only part of their energy into the calorimeter, distinguishing neutral hadrons from photons is difficult.
According to the demand, more information from the detector is needed for the identification of neutral hadrons.
In fact, a time-of-flight difference can serve as an effective way to help distinguish neutrons and $K_{L}$ from
photons.
Figure~\ref{time} shows the expected time resolution for the separation of neutrons/$K_{L}$ from photons
with a power of $3\sigma$ with respect to their momentum for a flight length of 1.5~m.
We therefore propose a momentum-dependent time resolution for neutral hadrons,
{\it i.e.}, $\sigma_{T} = 300/\sqrt{p^{3}\rm (GeV^{3})}~\rm{ps}$, that fulfills $3\sigma$ $\gamma/n$ separation
and an approximately $2\sigma$ $\gamma/K_{L}$ separation.
Under these requirements, the physics goals of measuring mixing parameters of $D^{0}-\bar{D}^{0}$ with a precision better than 0.05\%,
and electric-magnetic neutron form factors with a precision better than 1\% can be achieved.

\subsection{Particle Identification}

\subsubsection{ $\pi/K$ Identification}

The identification of $\pi$ and $K$ at the STCF is essential for charm physics, $\tau$ physics 
and fragmentation function~(FF) studies etc.
The momentum region of $\pi/K$ produced from charm hadrons or $\tau$ leptons are within 1.5~GeV
and that from FF studies can be as high as 3.5~GeV. 
The quark-hadron FF is essential for understanding the formation of observed
hadrons from QCD partons, while an $e^{+}e^{-}$ collider
provides a clean place for the study of hadronization.
Both the unpolarized and polarized FFs can be measured at the STCF by inclusive production
of one or two hadrons. Since a wide momentum range of FFs with high precision
are necessary at the STCF, the STCF must have the ability to identify
all kinds of final state hadrons with excellent separation power.

Taken the study of FFs as an example,
in the study of the Collins FF via $e^{+}e^{-}\to KK+X$ at $\sqrt{s}=7$~GeV, the dominant background
comes from $\pi/K$ contamination.
The momentum spectra of $\pi$ and $K$ can be up to 3.5~GeV/c, with most events accumulating less than 2.0~GeV/c according to Fig.~\ref{pr:chgmom},
which requires good $\pi/K$ identification abilities with momentum up to at least 2.0~GeV.
In addition, the yields of $K$ are at least one magnitude lower than that of $\pi$ in the production, leading to a worse situation in the study of $e^{+}e^{-}\to KK+X$ due to $\pi$ contamination.
It is found that background contamination affects the asymmetry distributions of
final hadrons, which leads to an underestimation of the asymmetry.
The background effects are studied under two situations:
first with $\pi/K$ misidentification of 10\% at $p=2$~GeV/c, where the background level from
$\pi$ contamination in $e^{+}e^{-}\to KK+X$ is over 50\%, and second
with $\pi/K$ misidentification of $<2$\% at $p=2$~GeV/c, where
the background level from $\pi$ contamination is $4\%$.
Because the precision of the Collins effect must be better than
7\% for spin structure measurements in electron-ion colliders,
the $\pi/K$ misidentification must
be less than $2\%$ at $p=2.0$~GeV/c and the corresponding identification efficiencies for the hadrons must be over 97\% at the STCF.\\

\subsubsection{ $\mu/\pi$ Identification}
In electron-positron collider experiments, the identification of muons is of great importance
for a wide range of physics program involving XYZ physics, $\tau$ physics and (semi)leptonic decays of
charm mesons, rare decays {\it etc. }
According to Fig.~\ref{pr:chgmom}, muon identification with $p<2$~GeV is essential.
A good $\mu$ detection efficiency and suppression power of $\mu/\pi$ are required at the STCF.

Two benchmark processes are applied to study the requirement for $\mu/\pi$ separation at high energy, {\it e.g.}, $p>0.5$~GeV/c,
the $CPV$ of $\tau$ decay $\tau^{-}\to K_{s}\pi^{-}\nu_{\tau}$
with another $\tau^{+}\to \mu^{+}\nu_{\mu}\bar{\nu}_{\tau}$, and the pure leptonic decay of $D_{s}$ decay $D_{s}\to\mu\nu_{\mu}$.
Three values for the misidentification rate of $\mu/\pi$, 1\%, 1.6\%, and 3\% at $p=1$~GeV/c, are tested;
these values correspond to muon identification efficiencies of 85\%, 92\% and 97\%, respectively.
It is found that higher $\mu$-ID efficiency is favored under the low background level.
Therefore, a $\mu/\pi$ suppression power of 30 up to a momentum of 2~GeV/c with
a high identification efficiency for muon to be 95\% with momentum $p>1$~GeV/c are needed,
which can meet the physics goal of achieving a sensitivity of 0.15\% for the measurement of CKM matrix element $|V_{cs}|$ in $D_{s}\to\mu\nu_{\mu}$.
In addition, in the semileptonic decays of $D_{(s)}$, such as the $D_{(s)}\to\pi^{-}\mu^{+}\nu_{\mu}$ process,
low-momentum $\mu/\pi$ separation is essential in the background
estimation for the purity of the selected sample; 
thus, good $\mu/\pi$ separation at low momentum is needed.

\subsection{Summary of the Physics Requirements}
From the discussions illustrated above, the quantified requirements from the physics perspective are listed below and summarized in Table~\ref{phyreqv2}.
\begin{itemize}
\item Tracking: An excellent tracking efficiency better than 99\% is required for charged tracks
in the high $p_{T}$ region, {\it i.e.}, $p_{T}>0.3$~GeV/c, and good tracking efficiency at low momentum is needed, {\it i.e.}, larger than $90\%$
at $p_{T}=0.1$~GeV/c.
Good momentum resolution for charged tracks is needed, {\it, e.g.}, $\sigma_{p}/p=0.5\%$ at $p=1$~GeV/c,
where the position resolution provided by the tracking system should be better than 130~$\mu$m.
The magnetic field for the current machine running in the tau-charm region is set at 1~T.
However, to obtain a better detection efficiency for low-momentum tracks and
good momentum resolution for all charged tracks, the mean value
and uniformity of the magnetic field should be optimized further.

\item Particle identification:
The $\pi/K$ or $K/\pi$ misidentification rate must be less than 2\% at $p=2$~GeV/c with the corresponding PID efficiency for hadrons to be over 97\%.
A $\mu/\pi$ suppression power of 30 up to a momentum of 2~GeV/c is proposed, and a
high PID efficiency for muons is needed, {\it i.e.}, larger than 95\% at $p=1$~GeV/c.

\item Photons: The photons need to be detected in the EMC within a wide energy range, from $E=25$~MeV to 3.5~GeV.
A good energy resolution is needed, {\it i.e.}, $\sigma_{E}\approx2.5\%$ at $E=1$~GeV, and the position
resolution is required to be $\sigma_{\rm pos}\approx5$~mm at $E=1$~GeV.
Moreover, the granularity of the detector affects the identification of
high energy $\pi^{0}$, where the opening angle of two photons
from $\pi^{0}$ decay is small, which should be considered in the design of the EMC.

\item Other neutral particles: a momentum-dependent time resolution is required for $\gamma/n/K_{L}$ identification, {\it i.e.}, $\sigma_{T} = 300/\sqrt{p^{3}\rm (GeV^{3})}~\rm{ps}$.

\end{itemize}

\begin{table}[htbp]
\caption{Benchmark physics processes used to determine the physics requirements of the STCF detector.}
\label{phyreqv2}
\footnotesize
\begin{center}
\begin{spacing}{1.3}
\begin{tabular}{cccc}
\hline
\hline
\vspace{0.2cm}
\multirow{2}{*}{Process}  &   \multirow{2}{*}{Physics Interest} & Optimized  & \multirow{2}{*}{Requirements}  \\
                                  &                                      & Subdetector & \\
\hline
 $\tau\to K_{s}\pi\nu_{\tau}$,    &    CPV in the $\tau$ sector,  &   \multirow{3}*{ITK+MDC}  &  acceptance: 93\% of $4\pi$; trk. effi.:  \\
 $J/\psi\to\Lambda\bar{\Lambda}$, &    CPV in the hyperon sector,  &                          &   $>99\%$ at $p_{T}>0.3$~GeV/c; $>90$\% at $p_{T}=0.1$~GeV/c\\
 $D_{(s)}$ tag                          &    Charm physics     &                          &   $\sigma_{p}/p=0.5\%$, $\sigma_{\gamma\phi}=130~\mu$m at 1~GeV/c     \\
\hline
$e^{+}e^{-}\to KK+X$,            &    Fragmentation function,  & \multirow{2}*{PID}   &      $\pi/K$ and $K/\pi$ misidentification rate $<2\%$ \\
$D_{(s)}$ decays                 &    CKM matrix, LQCD etc. &                      &          ~~~~~~~~~~~~~~~~~PID efficiency of hadrons $>97\%$ at $p<2$~GeV/c   \\
%$D_{0}\to K_{1} e\mu$			  &    photon polarization   &                      &        \\
\hline
 $\tau\to \mu\mu\mu$, $\tau\to\gamma\mu$,             &   cLFV decay of $\tau$,  & \multirow{2}*{PID+MUD}  &   $\mu/\pi$ suppression power over 30 at $p<2$~GeV/c, \\
 $D_{s}\to\mu\nu$                  &   CKM matrix, LQCD etc. &                          &   $\mu$ efficiency over 95\% at $p=1$~GeV/c    \\
\hline
$\tau\to\gamma\mu$,                 &   cLFV decay of $\tau$,  & \multirow{2}*{EMC}     &    $\sigma_{E}/E\approx2.5\%$ at $E=1$~GeV    \\
$\psi(3686)\to\gamma\eta(2S)$     &   Charmonium transition    &                        &    $\sigma_{\rm pos}\approx 5$~mm at $E=1$~GeV\\
\hline
$e^{+}e^{-}\to n\bar{n}$,           &    Nucleon structure     & \multirow{2}*{EMC+MUD}  &   \multirow{2}*{ $\sigma_{T} = \frac{300}{\sqrt{p^{3}\rm (GeV^{3}})}~\rm{ps}$ }   \\
$D_{0}\to K_{L}\pi^{+}\pi^{-}$     &     Unity of CKM triangle      \\
\hline 
\hline
\end{tabular}
\end{spacing}
\end{center}
\end{table}

\newpage

\newpage
\clearpage
\section{Experimental Conditions}
\label{sec:expcon}

\subsection{Machine Parameters}
The STCF is an electron-positron collider with one interaction point~(IP) and two symmetrical storage rings with a circumference of approximately 600 meters.
For the current design, the CMEy ranges from $2$ to $7\gev$, and the target luminosity is over \stcflum at the optimized CME of $4\gev$. Motivated by the wide range of physics programs at a CME approximately 4~GeV, {\it e.g.}, charm physics and $XYZ$ physics, experimental conditions with peak luminosity at a CME of 4~GeV are considered in the following discussion. The luminosity changes moderately within a few hundred MeVs around a CME of 4~GeV.

The STCF will carry out collision with flat beams, in which case the luminosity can be calculated by:
\begin{eqnarray}
  \mathcal{L}&=&\frac{\gamma f_0 N_b}{2r_e\beta^*_y}\xi_y,
  \label{eq:Luminosity_MachineParameters}
\end{eqnarray}
where $\gamma$ is the relativistic factor, $f_0$ is the collision frequency, $r_e$ is the classical radius of an electron, $N_b$ is the number of particles per bunch, $\beta^*_y$ is the vertical betatron function at the IP and $\xi_y$ is the vertical beam-beam parameter.

To inhibit the hourglass effect with a small $\beta_y*$, a large Piwinski angle collision is adopted, and the crossing angle is 60~mrad to obtain a large Piwinski angle, which results in a larger boost than that of the BEPCII and slightly limits the coverage of the detector. Additionally, the synchro-betatron coupling resonances are suppressed with the crab waist scheme.
The machine parameters are listed in Table~\ref{tab:MachineParameters}. 
More details on the design of the STCF accelerator can be found in Chapter~\ref{CDR_phys}.

\begin{table*}[htb]
    \caption{The designed machine parameters for the STCF.}
    \label{tab:MachineParameters}
    \centering
    \begin{tabular}{c|c}
      \hline
      Parameter                                                              &  Value \\ \hline
      Circumference (m)                                                   &  600       \\
      Beam energy range (GeV)                          &  $1\sim3.5     $        \\
      Optimized beam energy (GeV)                                   &  2                       \\
      Current (A)                                                              &  2                        \\
      Crossing angle $2\theta$ (mrad)                                & 60                    \\
      Natural energy spread                                               & $4.0\times10^{-4}$  \\
      Bunch length (mm)                                                   & 12                          \\
      Luminosity ($\times 10^{35}$~cm$^{-2}$s$^{-1}$)                &  $>0.5$                 \\
      \hline
    \end{tabular}
\end{table*}

\subsection{Machine Detector Interface}
\begin{figure}[hbtp]
  \begin{center}
    \begin{overpic}[width=15.cm,angle=0]{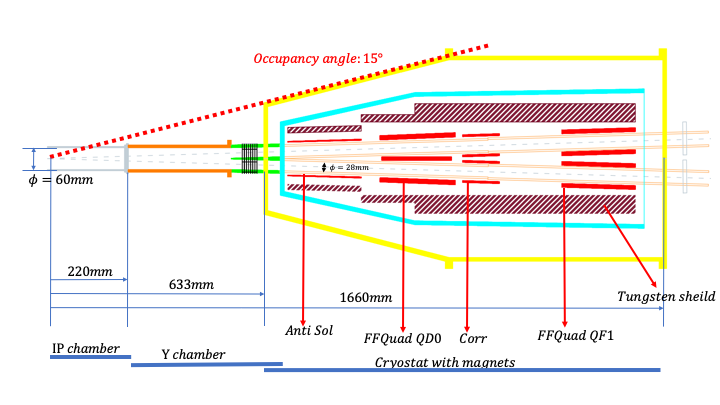}
    \end{overpic}
    \caption{The MDI structure layout includes an inner pipe (white), Y-shaped pipe (orange and green), separated pipe (pink), stainless shield (yellow), copper shield (blue), tungsten shield (dark red) and magnets (red).}
    \label{fig:MDI2D}
  \end{center}
\end{figure}

The final focus magnet QD0 is a crucial element in the machine detector interface~(MDI) and is currently set 0.9~m away from the IP to balance the requirements of the accelerator and detector. QD0 is a double aperture magnet because the distance of electron and positron beams 0.9~m from the IP exceeds the aperture limit, with a considerable requirement for the magnetic field gradient. The electron and positron beams are 0.5~m from the IP when their distance is large enough to smoothly transition to two beam pipes.
The layout of the MDI structure considering the crucial issues above is shown in Fig.~\ref{fig:MDI2D}. The layout includes a beam pipe and a cryostat that contains the magnets near the IP and a helium channel to cool the magnet. In order of distance from the IP, the magnets consist of an anti-solenoid, a final defocus magnet (QD0), a correcting magnet and a final focus magnet (QF0). From innermost to outermost, the beam pipe is composed of an inner pipe, a Y-shaped pipe and a separated pipe.

Considering the reconstruction precision and the background level, the inner beam pipe is designed to have an inner radius of 30~mm and thickness of 1~mm. The precision and background level define the boundary conditions of the radius of the ITK.
With a narrower radius of 10~mm, the expected precision of the reconstructed position would be improved by $15\%$ only for low-momentum tracks ($p=0.1\gev$/c), with little difference for high-momentum tracks. However, the background from lost particles would increase by approximately nine-fold.
The angle between the marginal line of the MDI structure and the midline of the two beams is $15^\circ$, which limits the corresponding angular acceptance of the detector.

\subsection{Beam Background}
\label{sec:mdi_bkg}

With high luminosity and narrow beam design, the estimation and inhibition of background is a crucial issue for the STCF. The main sources of background are the luminosity-related background, such as radiative Bhabha scattering and two-photon processes, and beam-related background, including the Touschek and beam-gas effects. All of these background sources are simulated by different source generators and then transmitted to {\sc Geant4} for full simulations.
The background level at the beam energy of 2.0~GeV is estimated from simulation with a full luminosity of $1\times10^{35}$~cm$^{-2}$s$^{-1}$ since the luminosity is the largest at this beam energy.
However, the background level at other beam energies does not exceed that at the optimized energy with a decrease in luminosity.

\subsubsection{Background Sources}
\paragraph{Luminosity-related Background}
\begin{itemize}
\item Radiative Bhabha scattering: In the process of $e^{+}e^{-} \to e^{+}e^{-}{(n)\gamma}$, both electron-positron pairs and photons are potential background signals and are considered in the simulation. The {\sc Geant4} source particles come from the {\sc BBBrem} generator for $|\cos\theta|<0.9$ and {\sc BabaYaga} for $|\cos\theta|>0.9$.

%\subparagraph{Two Photon processes}
%\quad\\
\item Two-photon processes: In the reaction of $e^{+}e^{-} \to \gamma^{*} \gamma^{*} \to e^{+}e^{-}e^{+}e^{-}$, the original electron-positron pair cannot be considered due to the extreme forward focus of the momentum after the reaction. Thus, only the low-energy electron-positron pair is considered and simulated by {\sc DIAG 36}, while the calculated particle rate is listed in Table~\ref{tab:5.2.01}.
\end{itemize}

\begin{figure*}[bt]
  \begin{center}
    \begin{overpic}[width=12.cm,angle=0]{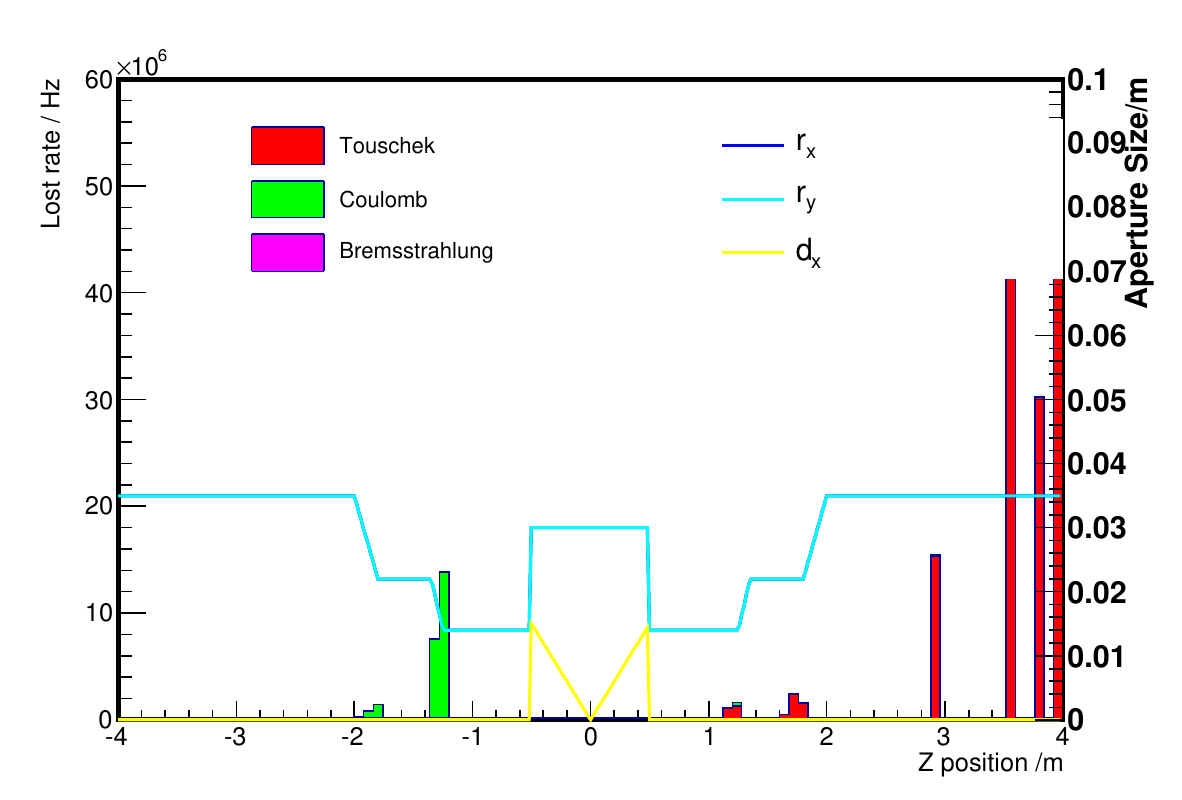}
    \end{overpic}
    \caption{Particle distribution near the IP if it hits the vacuum chamber for the electron beam, which is supposed to transfer from left to right. The positron beam is considered lost at the symmetrical position from the IP.The dark blue line and light blue line means the radius of the beamline in X and Y direction, respectively. The yellow line means the bias distance between the center of the electron/positron cluster and the center of the beam pipe in the shared pipe section.}
    \label{fig:distribution_SAD}
  \end{center}
\end{figure*}

\paragraph{Beam-related Background}
\begin{itemize}
%\subparagraph{Thouschek Effect}
%\quad\\
\item Thouschek effect: The Touschek effect is caused by the collision of particles in a beam bunch, which transforms the transverse momentum to longitudinal momentum, causing bunch spread and particle loss. The Touschek scattering rate can be calculated by the Touschek lifetime from the $\rm{Br\ddot{u}ck}$ model~\cite{BKG:bkg_bruckmodel}. The Touschek effect mostly occurs at the IP, where the beam is most compressed. However, the background is more affected by the Touschek effect from upstream because the original particles are transported for a distance in the ring instead of being lost immediately. %as elaborated in Sec.\ref{section:SAD}.

%\subparagraph{Beam-gas Effect}
%\quad\\
\item Beam-gas effect: The beam-gas effect mainly includes Coulomb scattering and bremsstrahlung, both of which are caused by the reaction between particles and residue gas in the ring and are highly influenced by the gas pressure in the vacuum chamber. The Coulomb scattering rate is approximately proportional to the $\beta$ function. Therefore, Coulomb scattering from upstream of the IP is very dangerous to the detector.

%\subparagraph{Transport in the ring}
%\label{section:SAD}
%\quad\\
\item Transport in the ring: The particles lost from the beam due to the Touschek effect and the beam-gas effect transport a distance in the ring, which is simulated by the {\sc SAD} program developed by KEK~\cite{bkg_SAD}. The aperture in the {\sc SAD} simulation is described according to the MDI structure.
The lost particles near the IP are input to {\sc Geant4} for further simulation, and the corresponding distribution is shown in Fig.~\ref{fig:distribution_SAD}.
\end{itemize}

\paragraph{Other Sources of Background}
\begin{itemize}
\item Synchrotron radiation (SR): The synchrotron radiation generated in the upstream pipe may influence the inner part of the detector system and should be carefully considered in the accelerator design to prevent it from directly entering the detector.

%\subparagraph{Injection}
%\quad\\
\item Injection: During injection, the background maybe increases by one or two orders of magnitude higher than the normal level, which is determined by the method of injection and accelerator status. In Belle II and BESIII, the injection background is not the major background source; thus, the detector response is not simulated in this step.
\end{itemize}

%%%%%%%%%%%%%%%%%  TABLE  %%%%%%%%%%%%%%%%%%%%%%%%
\begin{table*}[htb]
    \caption{Calculated particle rates for various background sources. The thresholds for the scattering angle and radiated photon energy for radiative Bhabha scattering are set to 4.47~mrad and 1~MeV, respectively. The average number of radiated photons in a radiative Bhabha event is denoted by $\bar{n}_\gamma$. No threshold is set for the two photon process. The particle rate of the beam-related background is calculated by the theoretical lifetime.}
    \label{tab:5.2.01}
    \vspace{0pt}
    \centering
    \begin{tabular}{cccc}
        \hline
        \thead[c]{Luminosity-related} & \thead[c]{RBB $e^{+}e^{-}$}& \thead[c]{RBB photon} &\thead[c]{Two photon process} \\
        \hline
        Cross-section (mb) &2.99 &2.99, $\bar{n}_\gamma$=1.3573 &5.15 \\
        Luminosity (/cm$^{2}$/s)& & $1\times10^{35}$  & \\						
        Particle rate (Hz) &$5.98\times10^{8}$ &$1.07\times10^{8}$ &$1.03\times10^{9}$ \\
        \hline
        \end{tabular}
        \begin{tabular}{cccc}
        \hline
        \thead[c]{Beam-related} &\thead[c]{Touschek effect}  &\thead[c]{Coulomb scattering}  &\thead[c]{Bremsstrahlung}\\
        \hline				
        Particle rate (Hz) &$1.12\times10^{9}$ &$2.09\times10^{8}$ &$2.1\times10^{6}$ \\
        \hline
    \end{tabular}
\end{table*}
%%%%%%%%%%%%%%%%%%%%%%%%%%%%%%%%%%%%%%%%%%%%%%%%%%

\subsubsection{Background Simulation Results}
\label{sec:bkg_sim}
%\paragraph{Simulation results}
\quad\\
As mentioned before, the luminosity-related background, Touschek background and beam-gas background are the main background components, which are fully simulated with {\sc Geant4}, and the physics list QGSP$\textunderscore$BERT$\textunderscore$HP is chosen. In the full simulation, 6 kinds of background signals are set as the primary particles in {\sc Geant4}, the electron-positron pairs generated by radiative Bhabha scattering, two photon processes, the Touschek effect, Coulomb scattering and bremsstrahlung and the photons generated by radiative Bhabha scattering. Each is simulated by $1\times10^{6}$ particles to obtain accurate background estimation. After the simulation, the detector responses to the 6 kinds of background signals are weighted and summed according to the particle generation rate shown in Table~\ref{tab:5.2.01}. In this case, the background influences on the detector system can be estimated well. In the simulation, there are three main parameters that we are most concerned with: the total ionizing dose (TID), nonionizing energy loss (NIEL) damage and background count rate of the STCF detector system.

\begin{figure*}[htb]
	\centering
	\includegraphics[width=50mm]{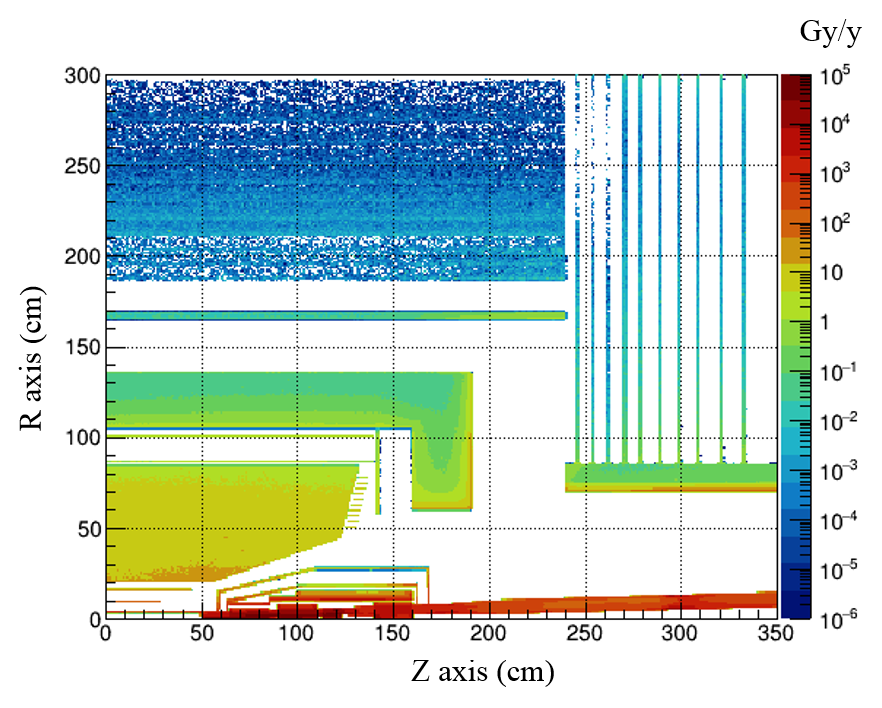}
\hspace{3 mm}
    \includegraphics[width=50mm]{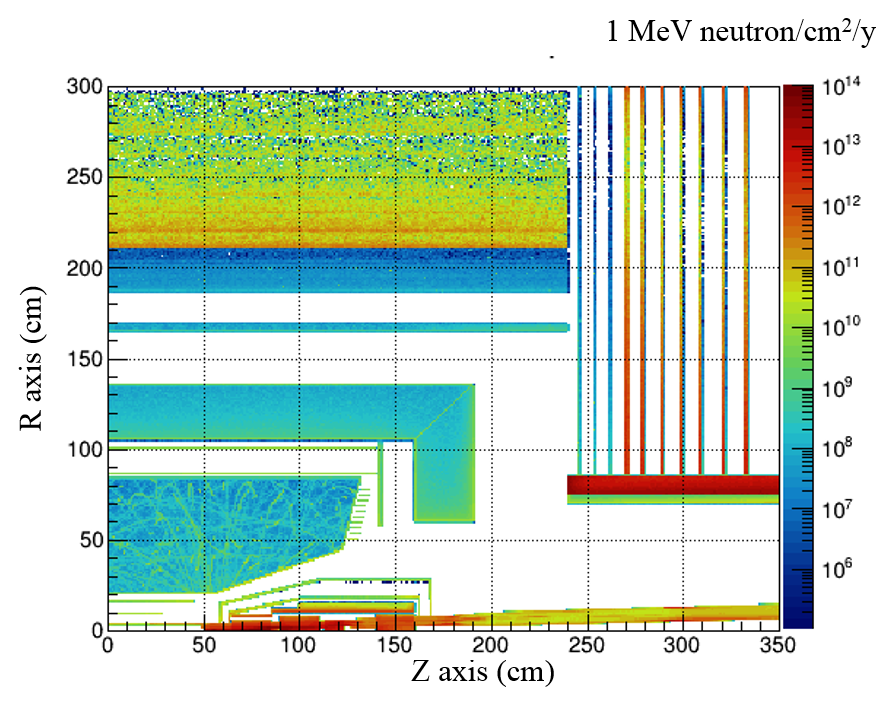}
\hspace{3 mm}
    \includegraphics[width=50mm]{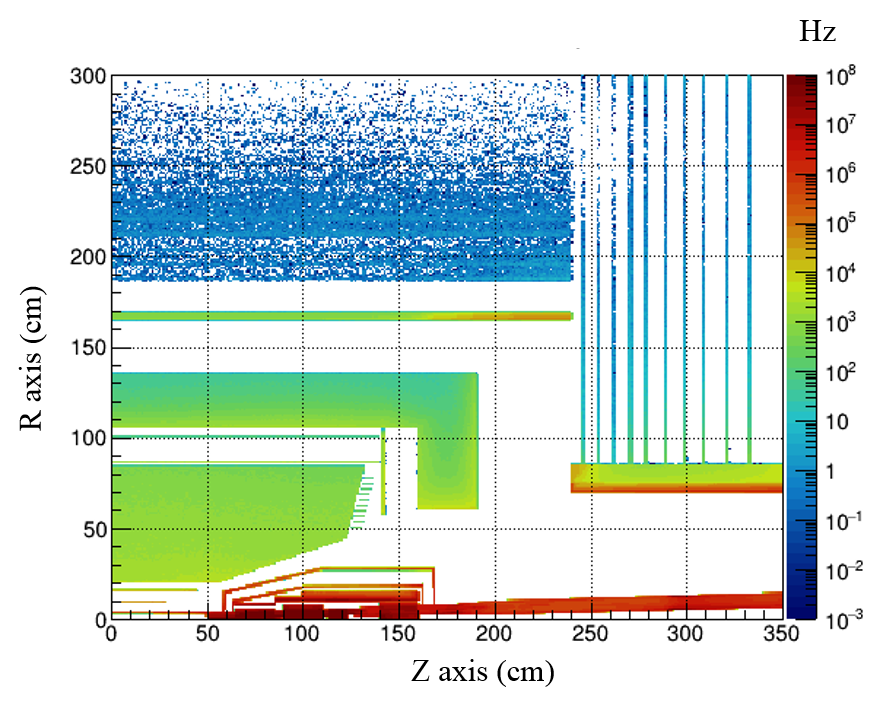}
\vspace{0cm}
\caption{The TID value (left), NIEL damage (middle) and background count rate (right) distributions of the STCF detector system.}
    \label{fig:5.3.01}
\end{figure*}

Fig.~\ref{fig:5.3.01} (left) shows the TID value distribution in the RZ plane, which is divided into 1 cubic cm pixels. The average value of TID in various STCF detector subsystems and electronic systems, as well as the maximum TID value, are calculated. These results indicate that the first layer of the silicon-based ITK has the highest TID, with a value of 1170~Gy/y. For detectors and electronics other than the ITK, the TID is less than 20 Gy/y, which is tolerable for the current technologies. Fig.~\ref{fig:5.3.01}~(middle) displays the NIEL damage distribution in the RZ plane. The simulated result shows that the NIEL damage of all of the important detector and electronic systems is below $10^{11}$ 1~MeV neutron/cm$^{2}$/y (for silicon). For the plastic scintillator detector in the MUD, the NIEL damage is on the order of $10^{11}$ 1~MeV neutron/cm$^{2}$/y because many more low-energy protons are produced via the neutron elastic reaction, which has an extremely high equivalent neutron coefficient.
The background count rate distribution is shown in Fig.~\ref{fig:5.3.01}~(right). For different detectors, suitable thresholds are used. For the gaseous detector, the deposition energy threshold of the background count is set to 100~eV. For scintillator detectors, the deposition energy threshold of background hits is 0.15-0.5~MeV. For Cherenkov detectors, the kinetic energy of incident electrons is 0.185~MeV, corresponding to the Cherenkov light generation demand. It is shown that the MDC has the highest background count level because this detector is composed of 48 layer wires. The maximum background rate per cm$^{2}$ occurs in the first layer of the ITK, necessitating a good detector design to realize an acceptable occupancy level.

%%%%%%%%%%%%%%%%%%% Fig %%%%%%%%%%%%%%%%%%%%%%%%%%
%%%%%%%%%%%%%%%%%%%%%%%%%%%%%%%%%%%%%%%%%%%%%%%%%%

\begin{figure*}[htbp!]
	\centering
	\includegraphics[width=50mm]{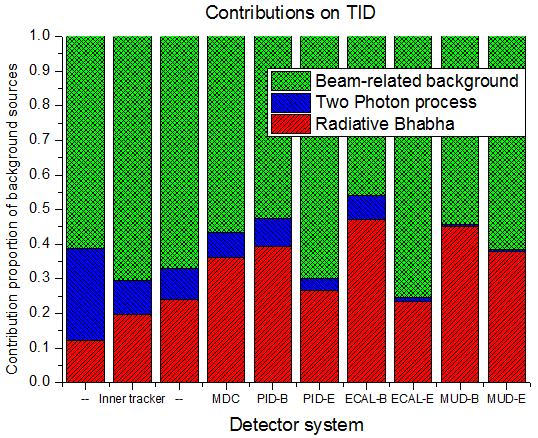}
\hspace{3 mm}
    \includegraphics[width=50mm]{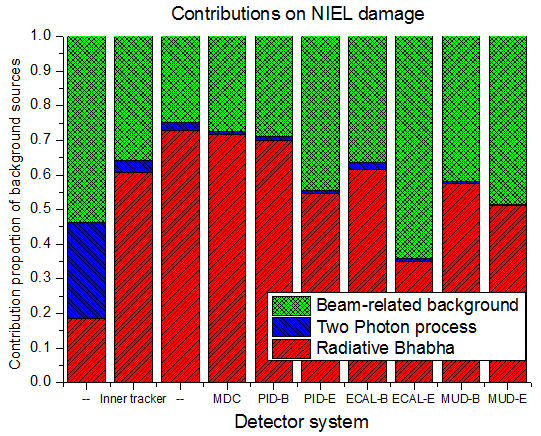}
\hspace{3 mm}
    \includegraphics[width=50mm]{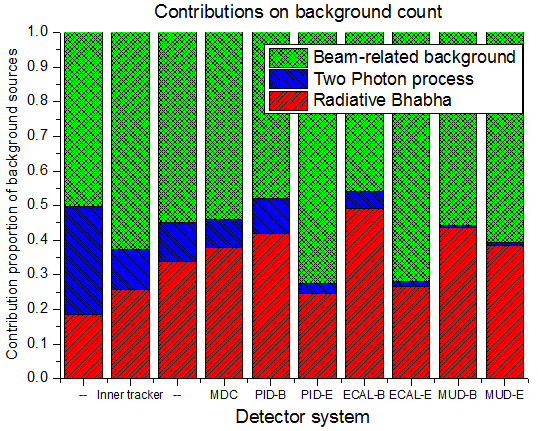}
\vspace{0cm}
\caption{The contribution of the background to the TID, NIEL damage and count.}
    \label{fig:5.3.04}
\end{figure*}

%\quad\\
Fig.~\ref{fig:5.3.04} shows the contributions of the luminosity-related background and beam-related background to the TID, NIEL damage and background count. Tables~\ref{tab:TIDNIEL_mean}-\ref{tab:TIDNIEL_eletronic} list the TID, NIEL and count rate of the detector subsystems and electronic subsystems, respectively.
These simulated results indicate that the beam-related background sources, especially Coulomb scattering, are the major contributors under full luminosity conditions in the inner detector subsystems. Radiative Bhabha scattering is the main influence on the outer detector subsystems, while the two-photon process has less influence. Coulomb scattering is the major component of the beam-related background because the Touschek- and Bremsstrahlung-produced electron-positron pairs have large probabilities of appearing downstream of the beam pipe, while Coulomb scattering-generated particle loss appears around the IP.

%%%%%%%%%%%%%%%%%%% Fig %%%%%%%%%%%%%%%%%%%%%%%%%%
%%%%%%%%%%%%%%%%%%%%%%%%%%%%%%%%%%%%%%%%%%%%%%%%%%

%%%%%%%%%%%%%%%%%%% Table %%%%%%%%%%%%%%%%%%%%%%%%%%
\begin{table*}[htbp!]
    \caption{{\sc Geant4} simulated TID and NIEL in the STCF subdetectors. The numbers are given as the mean values along the beam direction for each subdetector. For the ITK, the results are given for two different design options, the silicon pixel-based and the $\mu$RWELL-based designs.}
    \label{tab:TIDNIEL_mean}
    \vspace{0pt}
    \centering
    \begin{tabular}{llll}
        \hline
        \thead[l]{Detector} & \thead[l]{TID \\value (Gy/y)}& \thead[l]{NIEL damage\\ (1~MeV neutron/cm$^{2}$/y)}& \thead[l]{Total count\\ rate (Hz)}\\
        \hline
        Silicon-inner-1	&1170	&$2.71\times10^{10}$ & $3.90\times10^{8}$  \\
        Silicon-inner-2	&243	&$1.02\times10^{10}$ & $3.59\times10^{8}$  \\
        Silicon-inner-3	&64.9	&$1.71\times10^{10}$ & $2.92\times10^{8}$  \\
        \hline
        $\mu$RWELL-inner-1 &10.9    &$9.95\times10^{9}$ & $5.35\times10^{8}$  \\
        $\mu$RWELL-inner-2 &4.55    &$1.15\times10^{10}$ & $4.75\times10^{8}$  \\
        $\mu$RWELL-inner-3 &4.66    &$1.44\times10^{10}$ & $6.81\times10^{8}$  \\
        \hline
        \end{tabular}
    \begin{tabular}{llll}
        \hline
        \thead[l]{Detector} & \thead[l]{TID \\value (Gy/y)}& \thead[l]{NIEL damage\\ (1~MeV neutron/cm$^{2}$/y)}& \thead[l]{Total count\\ rate (Hz)}\\
        \hline
        MDC	&11.0	&$4.27\times10^{10}$ & $7.27\times10^{8}$ \\
        PID-Barrel~(RICH)	&2.96	&$8.67\times10^{9}$ & $4.50\times10^{8}$  \\
        PID-Endcap~(DTOF)	&1.34	&$4.65\times10^{9}$ & $8.30\times10^{8}$  \\
        EMC-Barrel	&0.35	&$1.41\times10^{10}$ & $2.64\times10^{9}$  \\
        EMC-Endcap	&0.32	&$7.26\times10^{9}$ & $9.38\times10^{8}$  \\
        MUD-Barrel-RPC	&0.028	&$3.23\times10^{8}$ & $5.58\times10^{6}$  \\
        MUD-Barrel-Scintillator	&0.040	&$3.89\times10^{11}$ & $1.06\times10^{7}$  \\
        MUD-Endcap-RPC	&0.017	&$7.03\times10^{7}$ & $3.53\times10^{6}$  \\
        MUD-Endcap-Scintillator	&0.027	&$1.86\times10^{11}$ & $1.22\times10^{7}$  \\
        \hline
        \end{tabular}
\end{table*}

\begin{table*}[htbp!]
    \caption{{\sc Geant4} simulated TID and NIEL in the STCF subdetectors. The numbers are given as the maximum values along the beam direction for each subdetector. For the inner tracker, the results are given for two different design options, the silicon pixel-based and the $\mu$RWELL-based designs.}
    \label{tab:TIDNIEL_max}
    \vspace{0pt}
    \centering
        \begin{tabular}{llll}
        \hline
        \thead[l]{Detector} & \thead[l]{Highest TID value\\ per pixel (Gy/y)}& \thead[l]{Highest NIEL \\damage per pixel \\(1~MeV neutron/cm$^{2}$/y)} & \thead[l]{Highest count rate\\ per channel (Hz/channel)} \\
        \hline
        Silicon-inner-1	&3490	&$1.75\times10^{11}$ &  $2.61\times10^{2}$ \\
        Silicon-inner-2	&320	&$3.72\times10^{10}$ &  $2.74\times10^{1}$ \\
        Silicon-inner-3	&150	&$2.68\times10^{10}$ &  $8.51\times10^{0}$ \\
        \hline
        $\mu$RWELL-inner-1 &118    &$1.12\times10^{10}$ & $3.35\times10^{5}$  \\
        $\mu$RWELL-inner-2 &61.8    &$1.46\times10^{10}$ & $1.63\times10^{5}$  \\
        $\mu$RWELL-inner-3 &38.6    &$5.67\times10^{10}$ & $1.61\times10^{5}$  \\
        \end{tabular}
        \begin{tabular}{llll}
        \hline
        \thead[l]{Detector} & \thead[l]{Highest TID value\\ per pixel (Gy/y)}& \thead[l]{Highest NIEL \\damage per pixel \\(1~MeV neutron/cm$^{2}$/y)} & \thead[l]{Highest count rate\\ per channel (Hz/channel)} \\
        \hline
        MDC	 &60.5	&$4.87\times10^{10}$ &  $4.00\times10^{5}$ \\
        PID-Barrel~(RICH)	&4.25	&$1.07\times10^{10}$ &  $3.3\times10^{3}$ \\
        PID-Endcap~(DTOF)	 &44.3	&$1.98\times10^{10}$ &  $1.20\times10^{5}$ \\
        EMC-Barrel	&21.1	&$1.76\times10^{10}$ &  $9.00\times10^{5}$ \\
        EMC-Endcap	&45.1	&$1.88\times10^{10}$ &  $1.50\times10^{6}$ \\
        MUD-Barrel-RPC	&0.093	&$3.74\times10^{11}$ &  $1.76\times10^{3}$ \\
        MUD-Barrel-Scintillator &0.047	&$4.88\times10^{11}$ & $1.15\times10^{3}$ \\
        MUD-Endcap-RPC	&0.37	&$1.22\times10^{10}$ &  $2.83\times10^{4}$ \\
        MUD-Endcap-Scintillator	&0.24	&$2.79\times10^{12}$ &  $9.8\times10^{4}$ \\
        \hline
        \end{tabular}
\end{table*}

\begin{table*}[htbp!]
    \caption{{\sc Geant4} simulated TID and NIEL values in the STCF electronic subsystems.}
    \label{tab:TIDNIEL_eletronic}
    \vspace{0pt}
    \centering
        \begin{tabular}{lllllll}
        \hline
        \thead[l]{Electronic component} & \thead[l]{TID \\value (Gy/y)}& \thead[l]{NIEL damage\\ (1~MeV neutron/cm$^{2}$/y)}& & \thead[l]{Highest TID value\\ per pixel (Gy/y)}& \thead[l]{Highest NIEL \\damage per pixel \\(1~MeV neutron/cm$^{2}$/y)} \\
        \hline
        Inner-1-electronic	&1420	&$5.09\times10^{10}$ & &1460	&$5.94\times10^{10}$ & \\
        Inner-2-electronic	&238	&$2.22\times10^{10}$ & &250	&$2.35\times10^{10}$ & \\
        Inner-3-electronic	&95.9	&$2.95\times10^{10}$ & &97.2	&$3.24\times10^{10}$ & \\
        MDC-electronic	&5.2	&$6.44\times10^{9}$ & &7.4	&$2.20\times10^{10}$ & \\
        PID-Barrel-electronic	&2.45	&$6.87\times10^{9}$ & &2.95	&$8.37\times10^{9}$ & \\
        PID-Endcap-electronic	&1.02	&$2.70\times10^{9}$ & &6.81	&$3.96\times10^{9}$ & \\
        EMC-Barrel-electronic	&0.046	&$1.51\times10^{9}$ & &1.03	&$3.88\times10^{9}$ & \\
        EMC-Endcap-electronic	&0.67	&$9.44\times10^{8}$ & &60.5	&$1.78\times10^{10}$ & \\
        MUD-Barrel-electronic	&0.020	&$1.45\times10^{8}$ & &0.065	&$3.42\times10^{11}$ & \\
        MUD-Endcap-electronic	&0.28	&$1.87\times10^{8}$ & &3.56	&$1.79\times10^{9}$ & \\
        \hline
    \end{tabular}
\end{table*}
%%%%%%%%%%%%%%%%%%%%%%%%%%%%%%%%%%%%%%%%%%%%%%%%%%

\subsubsection{Comparison and Validation}
Since the beam energy range of the STCF is similar to that of BESIII, it is helpful to compare the background results with those of BESIII. The count rate of the BESIII MDC layer at radius = 20 cm is approximately 100 Hz/cm$^2$ with a beam current of 0.45 A and a luminosity of 0.3$\times$10$^{33}$ /cm$^2$/s at BESIII. The luminosity-related background is proportional to the luminosity. The beam-related background dominated by Touschek is roughly proportional to the beam current (4.4) and inversely proportional to the beam size (35). Thus, the expected count rate under ideal experimental conditions is approximately 1.35 MHz/channel for the innermost STCF MDC layer. The simulated data is 450 kHz/channel, indicating a 3 times of difference. Considering the uncertainty of the extrapolation, the 3 times of difference is acceptable and the simulation result is reasonable. Also, the high luminosity in STCF requires for a much better MDI design compared with that in BESIII, which would further suppress the background interference. Thus, the predicted background level is close to the simulation result, and the simulated data would be used to evaluate the detector spectrometer performance in this step.

\subsection{Conclusion}
The basic MDI geometry is designed with limits on the radius of the ITK larger than 33~mm and acceptance larger than $15^{\circ}$.
Several kinds of luminosity-related and beam-related background sources are generated by optimal generators and fully simulated with {\sc Geant4}. The simulation shows that the TIDs of the detectors and electronics are below 1500~Gy/y. Additionally, the NIEL damage in all of the detector and electronic volumes is below $10^{11}$ 1~MeV neutron/cm$^{2}$/y.
The background count simulation result indicates that in most detector subsystems, the occupancy and s/n ratio problems caused by the background count are moderate. With a well-designed detector and electronics, the STCF detector is expected to operate safely under these conditions. Although the current background level is acceptable, MDI upgrades, such as adding shieldings and collimators, will be the next stage of study to ensure safety in a real machine.

\quad\\
%\input{Chapters/Chapter_02_Experimentalconditions/02_Reference}
%\newpage
%\input{02_01_Charmonium}
%\input{02_02_XYZ}
%\input{02_ref_CharmoniumXYZ}

\newpage
\clearpage
\section{Detector Design Overview}

\subsection{General Considerations}
The STCF detector system has to detect and identify particles  in a large kinematic phase space and a high radiation and counting-rate environment
induced by the high luminosity,
as described in Sec.~\ref{sec:mdi_bkg}.
Adequate radiation resistance and fast response is required for the STCF detector, especially for the inner part of the detector. Data would be taken at a rate 2-3 orders of magnitude higher than that of the BEPCII. This places stringent demands on the performance of the trigger and data acquisition system.
 
For many benchmark physics studies at the STCF, as described in Chapter~\ref{chap_phyper}, systematic errors will be the dominant factor limiting the measurement precision and may come from
\begin{itemize}
\item uncertainty of detector acceptance and response, including uncertainties of the geometrical acceptance and detector efficiencies and non-linearity of detector response;
\item mismeasurements by detectors, such as misreconstructed tracks and photons, and particle misidentification;  
\item uncertainty in the luminosity measurement, energy calibration, trigger etc.
\end{itemize}

To achieve optimal detector performance for precise reconstruction of exclusive final states produced at the STCF, the general requirements for the STCF detector system include the following
\begin{itemize}
\item (nearly) $4\pi$ solid angle coverage for both charged and neutral particles and a uniform response for all particles;
\item excellent momentum and angular resolution for charged particles and high energy and position resolution for photons;
\item high reconstruction efficiency for low momentum/energy particles;
\item superior PID capability ($e/\mu/\pi/K/p/\gamma$ and other neutral particles);
\item precise luminosity measurement;
\item radiation hardness and high rate capability in the high radiation and counting-rate environment expected at the STCF.
\end{itemize}

Since the momenta of most final-state particles are below 1~GeV/c, a low-mass design is required for the tracking system, especially for the very low momentum region, where the multiple scattering effect dominates. A separate inner tracker is used in the tracking system to enhance tracking performance for low momentum particles. 
A low-mass tracking system would also greatly benefit the energy measurement of low-energy photons by the EMC.  Fast response is required for the crystal-based EMC to preserve its excellent intrinsic energy resolution. 
The extra high radiation level in the inner and forward regions of the STCF experiment demands detector and electronics technologies with significantly high radiation resistance and rate capability. 
Powerful trigger and DAQ systems are required to handle the very high physics event rate up to 400 kHz and the large data flow expected at the STCF.

\subsection{Overall Detector Concept}

\begin{figure}[htbp]
\begin{center}
\includegraphics[width=0.8\textwidth]{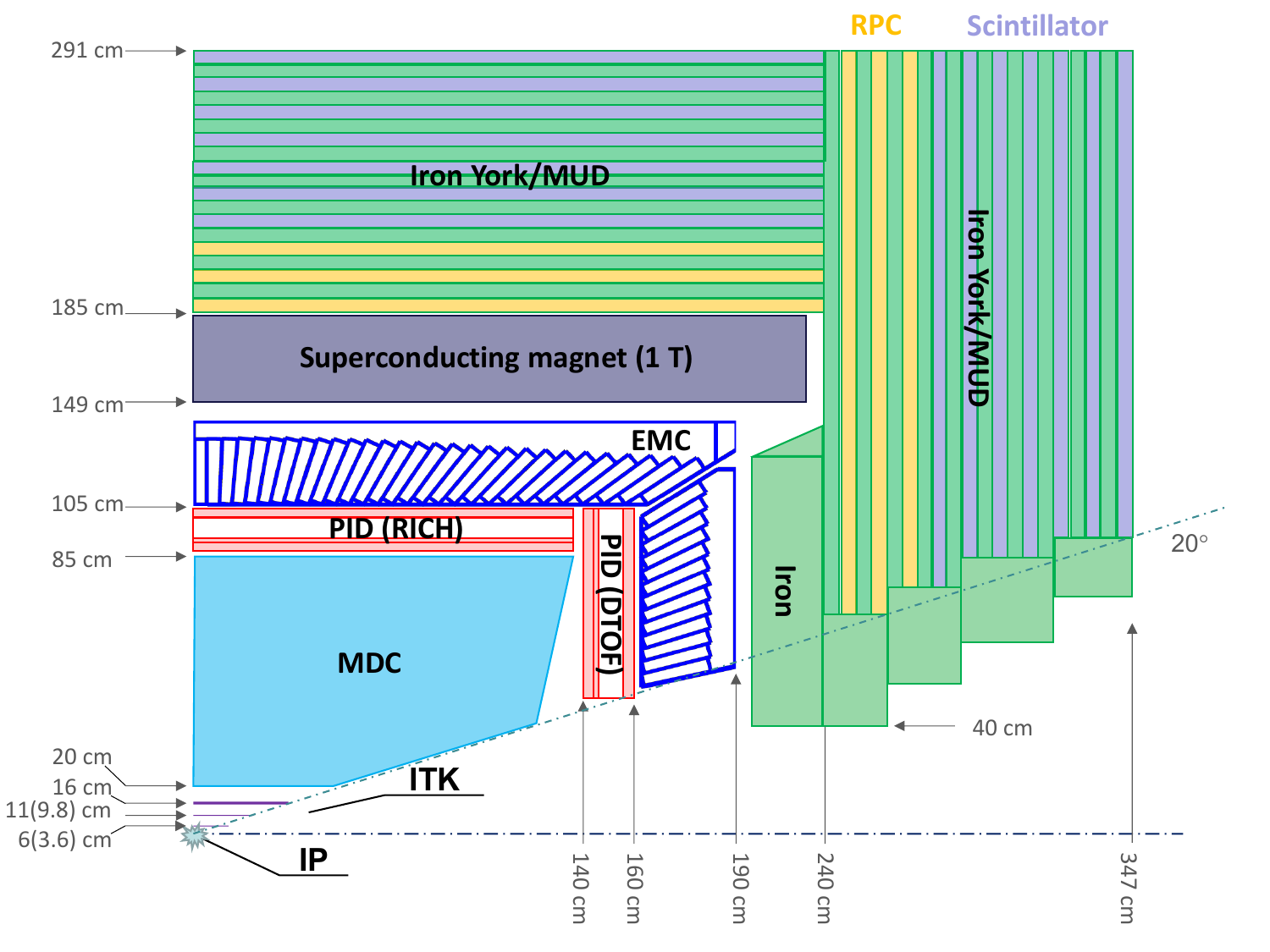}
\caption{Schematic layout of the STCF detector concept}
\label{fig:fulldetector_overview}
\end{center}
\end{figure}

\begin{figure}[htbp]
\begin{center}
\subfloat[][]{\includegraphics[width=0.8\textwidth]{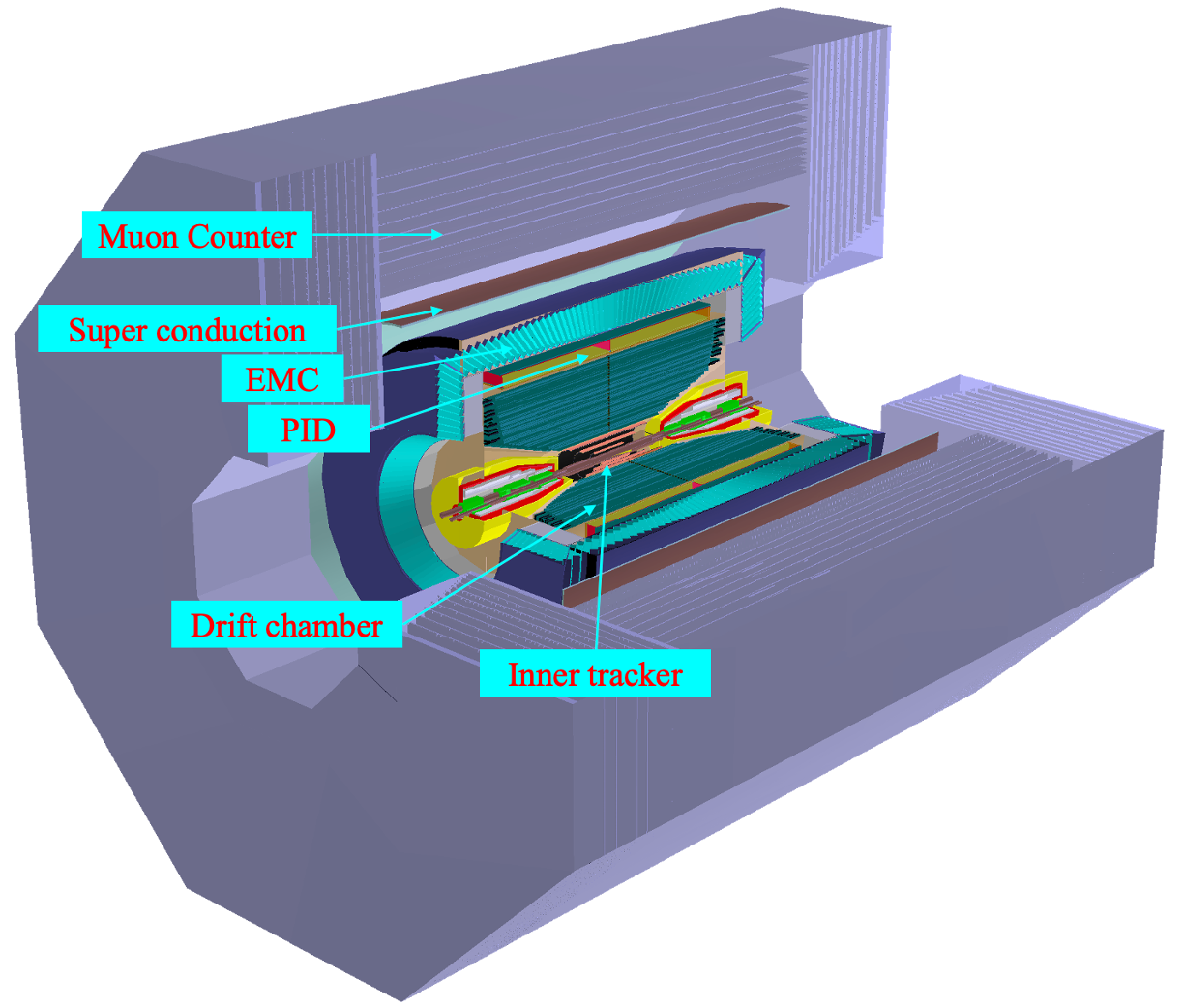}} \\
\subfloat[][]{\includegraphics[width=0.42\textwidth]{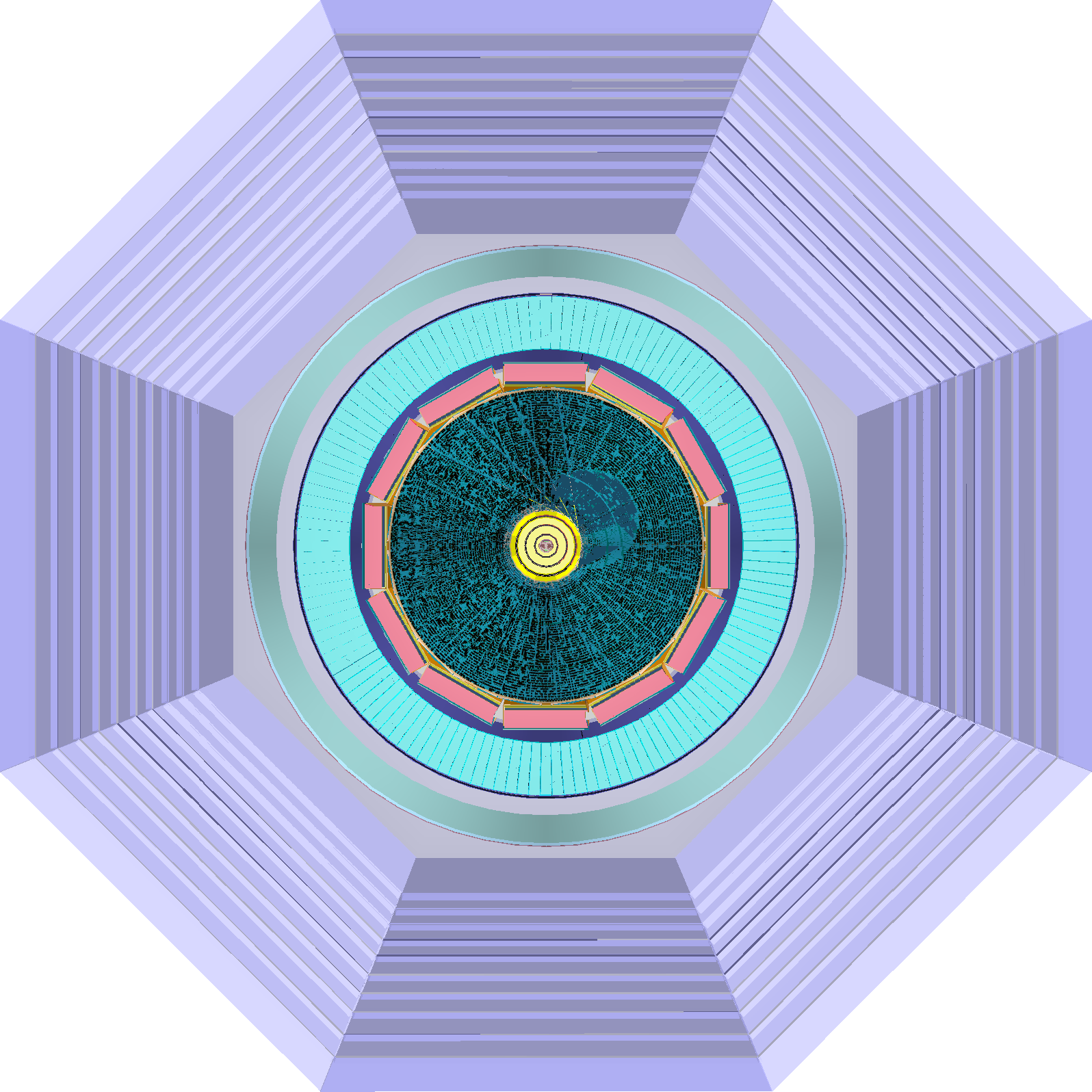}}
\hspace{5 mm}
\subfloat[][]{\includegraphics[width=0.42\textwidth]{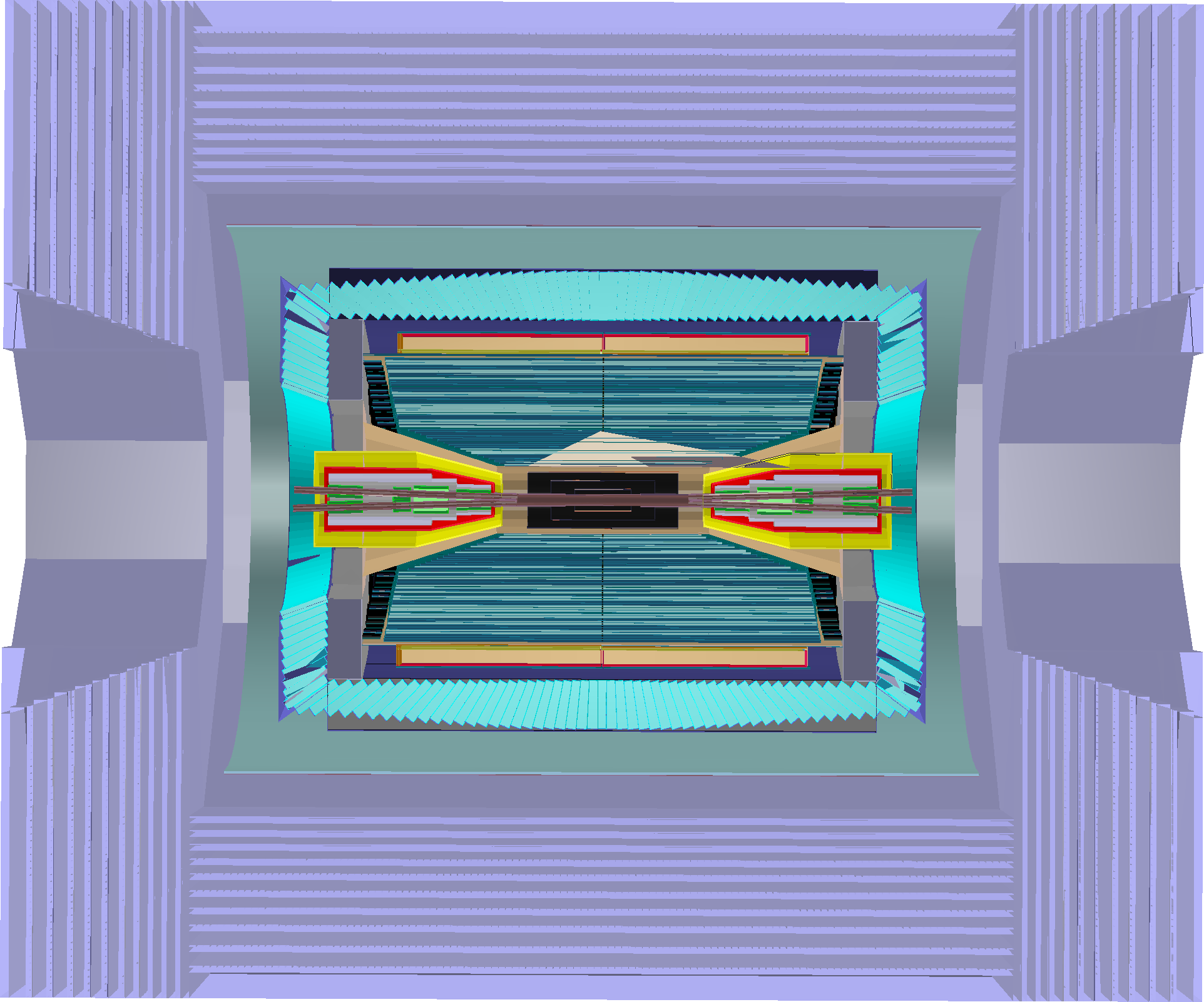}}
\caption{Geometry of the STCF detector: (a) 3D cutaway view, (b) cross-section view in the $x-y$ plane, and (c) cross-section view in the $z-r$ plane. It consists of ITK, PID system, EMC, SCS and MUD from inner to outmost. }
\label{fig:fulldetector_geo}
\end{center}
\end{figure}

The conceptual layout of the STCF detector system is shown in Fig.~\ref{fig:fulldetector_overview}.
The 2D and 3D detector geometries are shown in Fig.~\ref{fig:fulldetector_geo}.
Along the radial direction from the interaction region, the major detector components are as follows:
\begin{itemize}
\item an ITK consisting of three layers of low-material budget silicon or gaseous detectors, closest to the beam pipe and covering a polar angle range of 20 degrees to 160 degrees, to achieve high tracking efficiency for very low-momentum charged particles;
\item an MDC tracking detector based on He-gas to provide efficient and precise trajectory measurements for charged particles;
\item a RICH detector for PID in the barrel to distinguish charged hadrons at high momentum;
\item a time-of-flight detector based on the detection of the internal total-reflected Cherenkov light, DTOF, for PID in the endcap;
\item a homogeneous EMC composed of trapezoid-shaped pure CsI crystal scintillators to precisely determine the photon energy;
\item a superconducting solenoid outside the EMC to produce a uniform and stable magnetic field of 1~T;
\item a multilayer flux return yoke instrumented with plastic scintillator strips and resistive plate chambers (RPCs) to serve as a MUD to provide sufficient $\mu/\pi$ suppression power.
\end{itemize}

The tracking system consists of two components, the ITK and MDC, to cope with the high radiation level of the tracking layer closest to the beam pipe and to reduce the material budget as much as possible. The PID system is also split into two parts, the RICH detector in the barrel and the DTOF detector in the endcap, to take into account the different times of flight of particles with different polar angles.

The primary performance requirements for the STCF detector have been presented in Table~\ref{phyreqv2} and are listed below:
\begin{itemize}
\item Low-momentum tracking efficiency $>90\% @ 100$~MeV/c, low material budget ($\lt0.01X_{0}$) for ITK;
\item Momentum resolution $\lt0.5\% @1~$GeV/c, $dE/dx$ resolution $\lt6\%$, and low material budget ($\lt0.05X_{0}$) for MDC.
\item PID $\pi/K$ misidentification rate $<2\%$ and PID efficiency $>97\%$ up to 2~GeV/c with a modest material budget ($\lt0.3X_{0}$)
\item EMC energy resolution $\sim 2.5\% @ 1$~GeV, position resolution $\sim 5$~mm @ 1~GeV;
\item MUD $\mu/\pi$ suppression power $>30$, with $\mu$ detection efficiency $\gt$ 70\% @ $0.5<p<0.7$~GeV/c, and $\mu$ detection efficiency $\gt$ 95\% @ p $\gt$ 0.7 GeV/c.
\end{itemize}

\begin{comment}
\begin{table}[htbp]
\caption{Main physics requirements on the STCF detector.}
\label{phyreq_det}
\footnotesize
\begin{center}
\begin{spacing}{1.3}
\begin{tabular}{lc}
\hline
\hline
\vspace{0.2cm}

Subdetector  & Requirements \\
\hline
\multirow{3}*{ITK+MDC}        &  acceptance: 93\% of $4\pi$; trk. effi.:  \\
                              &   $>99\%$ at $p_{T}>0.3$~GeV/c; $>90$\% at $p_{T}=0.1$~GeV/c\\
                              &   $\sigma_{p}/p=0.5\%$, $\sigma_{\gamma\phi}=130~\mu$m at $p=1$~GeV/c     \\
\hline
\multirow{2}*{RICH+DTOF}      &  $\pi/K$ and $K/\pi$ misidentification rate $<2\%$ \\
                              &   PID efficiency for hadrons $>97\%$ at $p<2$~GeV/c   \\
\hline
\multirow{4}*{EMC}            &  $\sigma_{E}/E\approx2.5\%$ at $E=1$~GeV     \\
                              &   $\sigma_{\rm pos}\approx 5$~mm at $E=1$~GeV \\
                              &   \multirow{2}*{ $\sigma_{T} = \frac{300}{\sqrt{p^{3}\rm (GeV^{3}})}~\rm{ps}$ }   \\
                              & \\
\hline
\multirow{2}*{MUD}            &  $\mu$ efficiency over 95\% at $p=1$~GeV/c    \\
                              &  $\mu/\pi$ suppression power over 30 at $p<2$~GeV/c \\
                              & \\

\hline 
\hline
\end{tabular}
\end{spacing}
\end{center}
\end{table}
\end{comment}
\newpage

\newpage
\clearpage
\section{Inner Tracker~(ITK)}
\label{sec:itk}

\subsection{Introduction}
The STCF physics programs demand high detection efficiency and good spatial resolution for charged particle tracks while imposing no explicit requirements on vertex reconstruction. As a consequence, the main tasks of the STCF inner tracker are to detect particle hits of charged particles, especially those with very low momentum, below 100~MeV/c, and to facilitate the reconstruction of charged particle tracks with the MDC. In this chapter, the performance requirements of the inner tracker and the conceptual baseline designs are described. The resulting expected performance is also discussed through a simulation study.

\subsection{Performance Requirements and Technology Choices}
The major performance requirements for the inner tracker for realizing the expected detection capabilities are listed below:
\begin{itemize}
\item Low material budget: about 0.25\% X$_{0}$ for each detector layer.
\item Spatial resolution: single-hit spatial resolution better than 100~$\mu$m in the $r-\phi$ direction. 
\item Detector occupancy: not exceeding a few percent.
\item Radiation tolerance requirements as described in Sec.~\ref{sec:mdi_bkg}.
\end{itemize}

Several detector technologies can fulfill the requirements above, such as micropattern gaseous detectors (MPGDs) and silicon pixel detectors. 
A silicon pixel detector~ \cite{itk1} was used as the vertex detector at the STAR~\cite{itk2} and Belle II~\cite{belle2} experiments, showing good performance in spatial resolution and rate capabilities \cite{itk4}. MPGD-based detectors, 
a cylindrical gas electron multiplier (CGEM) detector~\cite{itk5}, was used as the inner tracker for the KLOE~\cite{itk6} detector and the BESIII upgrade~\cite{itk7}. 
In the CDR of the STCF, an MPGD-based design is the baseline choice for the inner track, while a silicon pixel detector is considered to be an alternative design. The sensitive lengths in each of the layers are determined by the required angular acceptance, {\it i.e.}, a polar angle range of 20 degrees to 160 degrees.

%% muRWell

\subsection{$\mu$RWELL-based Inner Tracker}

\subsubsection{Detector Design}

\begin{figure*}[htb]
	\centering
\subfloat[][]{\includegraphics[width=0.7\textwidth]{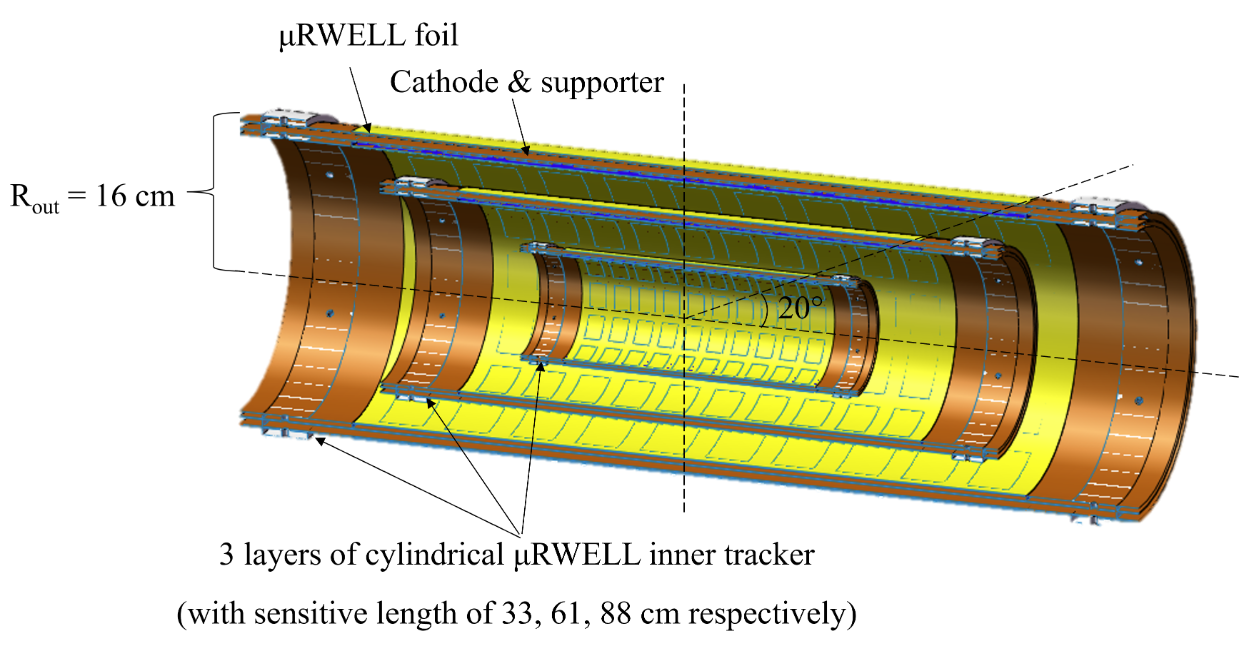}} \\
\subfloat[][]{\includegraphics[width=0.7\textwidth]{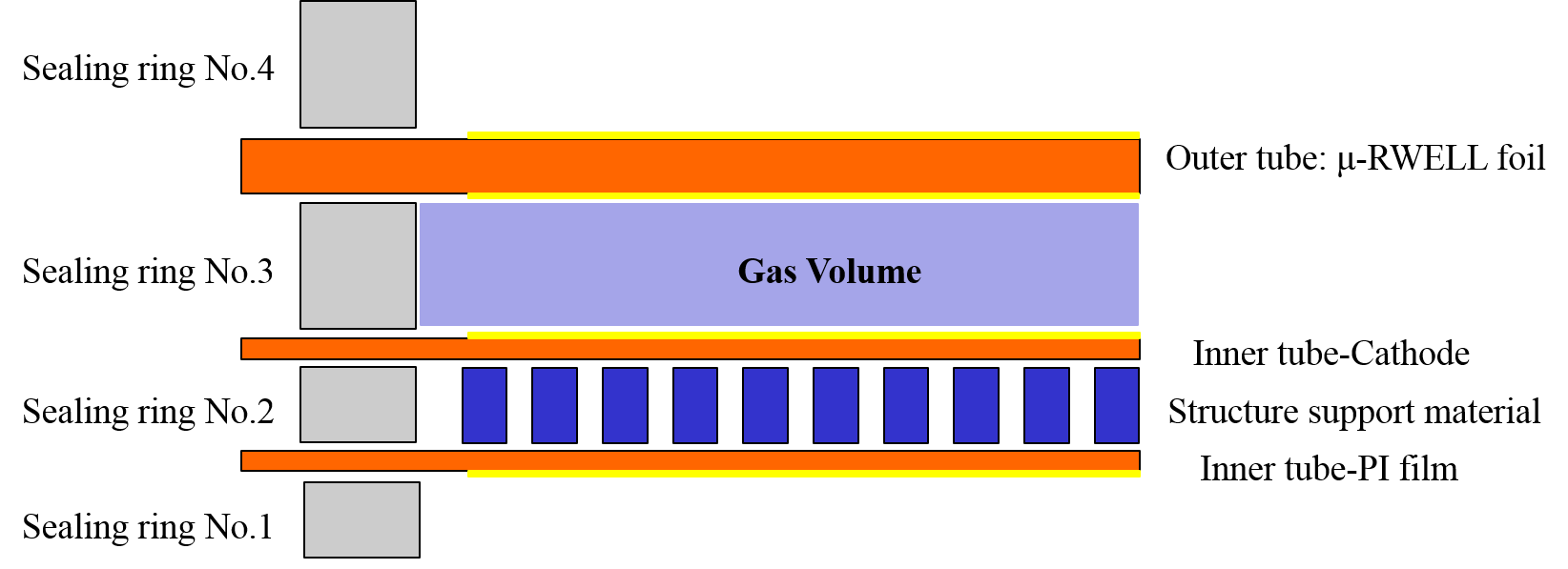}}
\vspace{0cm}
\caption{(a) The schematic structure of the $\mu$RWELL-based inner tracker. (b) The structure of each layer of the $\mu$RWELL detector.}
    \label{fig:4.1.02}
\end{figure*}

$\mu$RWELL~\cite{itk8,itk9,itk10} is a single-amplification stage resistive micropattern gaseous detector.
It has been introduced as a thin, simple and robust detector for very large area applications requiring operation in harsh radiation environments.
Figure~\ref{fig:4.1.02} shows the schematic structure of the inner tracker system based on the $\mu$RWELL detector, consisting of 3 cylindrical detector layers  with inner radii of 60~mm, 110~mm and 160~mm, and full-length of 330~mm, 610~mm, and 880~mm.
The radii are determined based on the radiation tolerance of the $\mu$RWELL detector and the MDC range according to the background simulation in Sec.~\ref{sec:mdi_bkg}. The innermost wire layer of the MDC (Sec.~\ref{sec:mdc}) has a radius of approximately 200~mm. The radii of the three layers of the $\mu$RWELL-based inner tracker ensure uniform gaps between the inner tracker hits and the first MDC hit.
The 3 layers are functionally and structurally independent of each other, thus simplifying the manufacturing and maintenance processes. As illustrated in Fig.~\ref{fig:4.1.02} (b), each detector layer consists of 2 cylinders and 4 pairs of sealing rings. The inner cylinder  has a sandwich-like structure made of polyimide film and supporting material, providing a detector frame with sufficient mechanical strength. The outer surface of the inner cylinder is the cylindrical $\mu$RWELL foil, which acts as both an electron multiplier and signal readout unit.  The outer cylinder is coated with a thin aluminum foil, acting as the drift cathode of the $\mu$RWELL detector. Four pairs of sealing rings are located at both ends of the cylindrical detector, sealing the gap between the polyimide films and the $\mu$RWELL foil. The space between the inner cylinder  and the outer cylinder  is the gas volume. To ensure the gas tightness of the detector, each seam of the gas volume is sealed with epoxy resin.

A very low material budget is crucial for the measurement of low-momentum charged particles. It is essential to reduce the material contributions of the inner tracker.
In this baseline design, the mechanical strength of the detector is mainly provided by the sandwich-like inner cylinder , and the material budget in the detection area can be limited to a very low level, as shown in Table~\ref{tab:4.1.01}.

%%%%%%%%%%%%%%%%%%% Fig %%%%%%%%%%%%%%%%%%%%%%%%%%
%%%%%%%%%%%%%%%%%%%%%%%%%%%%%%%%%%%%%%%%%%%%%%%%%%

%%%%%%%%%%%%%%%%%  TABLE  %%%%%%%%%%%%%%%%%%%%%%%%
\begin{table*}[htb]
\small
    \caption{The material budget of the $\mu$RWELL-based inner tracker design.}
    \label{tab:4.1.01}
    \vspace{0pt}
    \centering
    \begin{tabular}{llll}
        \hline
        \thead[l]{Structure} & \thead[l]{Material}& \thead[l]{Thickness\\(cm)} & \thead[l]{Material budget\\(X/X$_0$)}\\
        \hline
        Inner cylinder  &Aluminum (X$_0$=8.897 cm)           &0.001	 &0.011\%  \\
        	       &Polyimide (X$_0$=28.57 cm)	       &0.01	 &0.035\%  \\
	               &Aramid honeycomb/Rohacell (X$_0$$\simeq$267 cm)	&0.2	&0.075\% \\
        Gas volume  &Argon-based gas mixture (X$_0$=11760 cm)	&0.5	&0.00425\% \\
        Outer cylinder  ($\mu$RWELL foil)	&Alumium (X$_0$=8.897 cm)	&0.0015	&0.017\% \\
	               &Polyimide (X$_0$=28.57 cm)	&0.03	&0.106\% \\
	               &DLC (X$_0$=12.13 cm)	&0.0001	&0.00082\% \\
        Total	   &	                     &	                  &0.249\% \\
        \hline
    \end{tabular}
\end{table*}

\subsubsection{Detector Simulation and Optimization}

\begin{figure*}[htb]
	\centering
    \includegraphics[width=140mm]{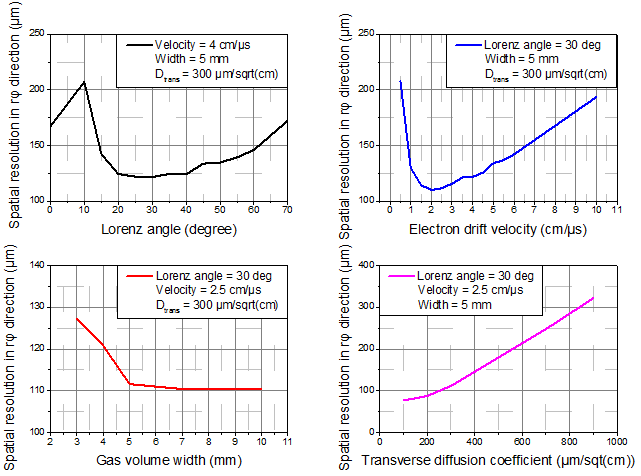}
\vspace{0cm}
\caption{The $r-\phi$ spatial resolution as a function of various parameters based on the {\sc Geant4} simulation.}
    \label{fig:4.1.ideal gas component}
\end{figure*}

%\paragraph{Working point and gas component optimization}
%\quad\\
To obtain optimal detection performance from the $\mu$RWELL-based inner tracker, it is crucial to determine the working point and gas component.
Based on a simulation study with {\sc Garfield-9}~\cite{garfield} and {\sc Geant4}~\cite{Geant4_ref}, the impact of these parameters on the detector spatial resolution is investigated, and the detector design is optimized.
\quad\\
The spatial resolution is dependent on various detector parameters, such as the drift electric field strength, working gas component and width of the gas volume. The optimization of these parameters must take into account the effect of the 1~T magnetic field in the $z$-direction. To simplify the optimization, the full simulation starts from an ideal gas component. Figure~\ref{fig:4.1.ideal gas component} shows the {\sc Geant4} simulated spatial resolution as a function of various parameters being studied. The Lorentz angle and the electron drift velocity have optimal values of approximately 30 degrees and 2~cm/$\mu$s, respectively. Additionally, a larger gas volume width and a smaller transverse diffusion coefficient enhance the spatial resolution. The optimal Lorentz angle, electron drift velocity and transverse diffusion coefficient can be realized by choosing a suitable gas component and electric field. However, a larger gas volume width leads to a longer signal duration time, decreasing the counting rate capabilities and increasing the occupancy ratio of the detector. As a compromise, the gap width is set to 5~mm.

%%%%%%%%%%%%%%%%%%% Fig %%%%%%%%%%%%%%%%%%%%%%%%%%
%%%%%%%%%%%%%%%%%%%%%%%%%%%%%%%%%%%%%%%%%%%%%%%%%%

\begin{figure*}[htb]
	\centering
\subfloat[][]{\includegraphics[width=50.5mm,height=40mm]{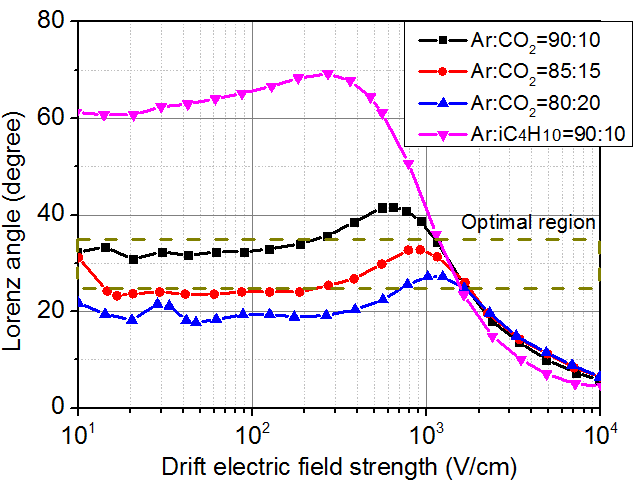}}
\hspace{3 mm}
\subfloat[][]{\includegraphics[width=50.5mm, height=40mm]{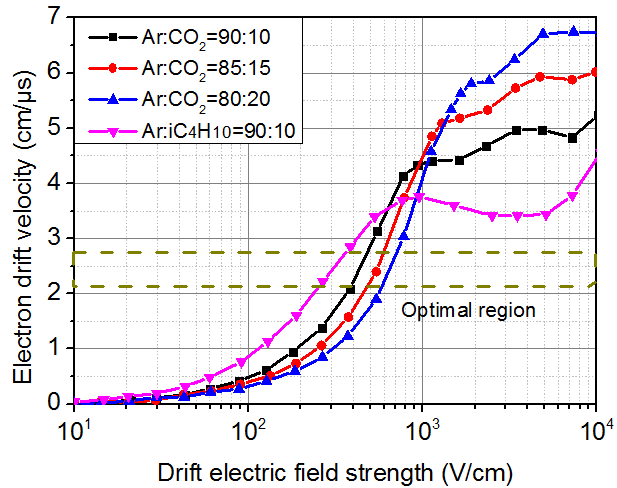}}
\hspace{3 mm}
\subfloat[][]{\includegraphics[width=50.5mm, height=40mm]{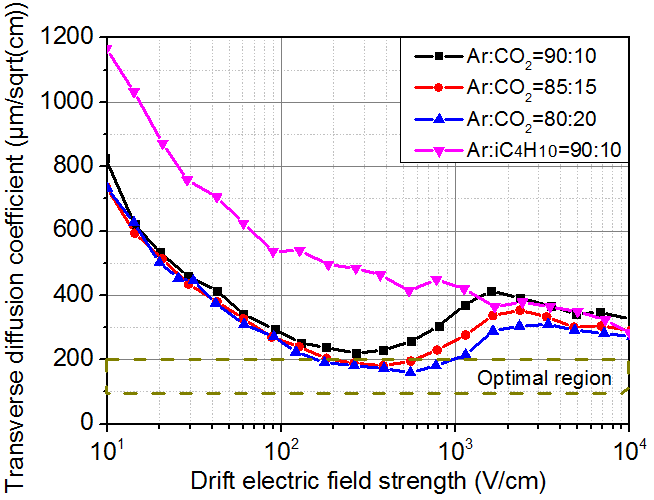}}
\vspace{0cm}
\caption{Dependence of the (a) Lorentz angle, (b) electron drift velocity and (c) transverse diffusion coefficient on the drift electric field strength with the 1 atm gas mixtures, simulated by {\sc Garfield-9}.}
    \label{fig:4.2.real gas component}
\end{figure*}

It is feasible to realize the expected performance of the gas mixture by adding an appropriate type and proportion of doping gas to the noble gas. With the {\sc Garfield-9} simulation and the analysis of tens of different argon-based gas mixtures, a suitable working gas component is found, as shown in Fig. \ref{fig:4.2.real gas component}. The {\sc Garfield-9} simulation shows that the gas mixture of Ar:CO$_{2}$=85:15 gives a Lorentz angle of 29.2 degrees, an electron drift velocity of 2.34~cm/$\mu$s, and a transverse diffusion coefficient of 191~$\mu$m/$\sqrt{\mathrm{cm}}$, with a drift electric field strength of 500~V/cm. All the parameters are within the preferred ranges, indicating that this gas mixture could be a suitable working gas for the $\mu$RWELL-based inner tracker.

%%%%%%%%%%%%%%%%%%% Fig %%%%%%%%%%%%%%%%%%%%%%%%%%
%%%%%%%%%%%%%%%%%%%%%%%%%%%%%%%%%%%%%%%%%%%%%%%%%%

\subsubsection{Detector Performance}

\begin{figure*}[htb]
	\centering
    \subfloat[][]{\includegraphics[height=50mm]{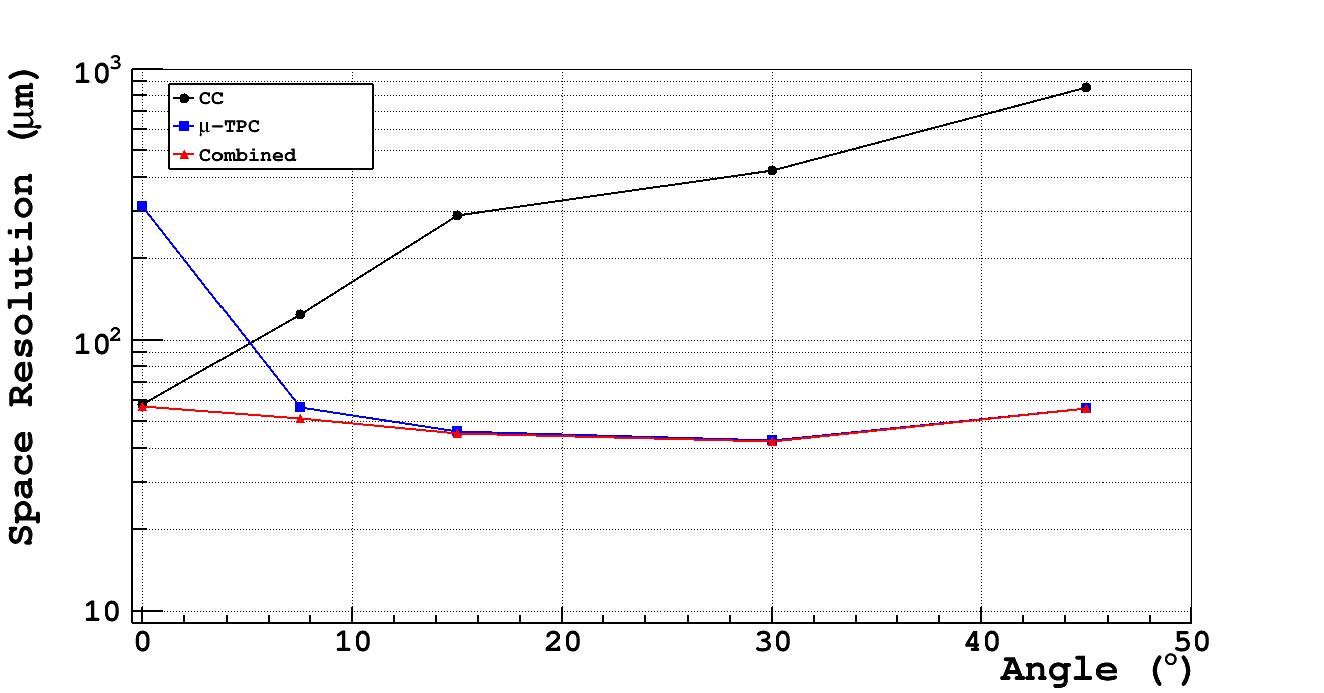}} \\
    \subfloat[][]{\includegraphics[height=50mm]{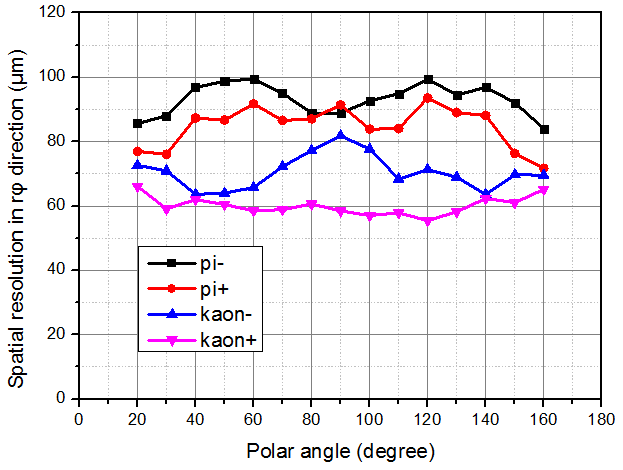}}
    \hspace {5 mm}
    \subfloat[][]{\includegraphics[height=50mm]{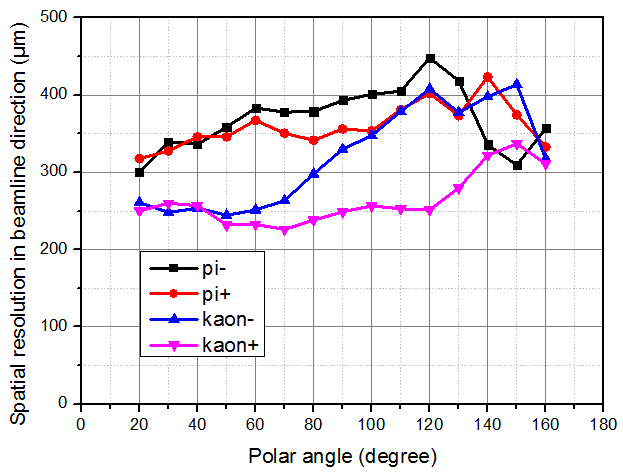}}
	\vspace{0cm}
	\caption{
        (a) The spatial resolution $\sigma$$_r$$_\phi$ as a function of the incidence angle with the $\mu$RWELL detector operated in the $\mu$-TPC mode.
        The spatial resolution in the (b) $r-\phi$ and (c) beamline direction of various particles with the same transverse momentum of 100 MeV/c and the $\mu$RWELL detector operated in the $\mu$-TPC mode.
        }
    \label{fig:4.3.spatial resolution simulation result}
\end{figure*}

%\paragraph{Spatial resolution of single point}
%\quad\\
A previous study~\cite{itk12} on $\mu$RWELL indicates that, with the $\mu$-time projection chamber (TPC) %Editor: Please ensure that the intended meaning has been maintained in this edit.
mode, the spatial resolution of the detector is almost flat over a wide range of incidence angles, as shown in Fig.~\ref{fig:4.3.spatial resolution simulation result} (a). In the $\mu$-TPC mode, each ionization cluster is projected into a 2-D spatial and 1-D time distribution inside the conversion gap. With the drift time measurement of the primary ionized electrons, the track segment in the gas volume can be reconstructed, and a precise spatial resolution can be obtained. In this mode, the $\mu$RWELL detector can achieve a spatial resolution of approximately 100~$\mu$m.
The spatial resolution of different types of particles with the same transverse momentum of 100 MeV/c is determined by using {\sc Geant4} simulation, as shown in Fig.~\ref{fig:4.3.spatial resolution simulation result}. The spatial resolution in the $r-\phi$ and beamline directions can be limited below 100~$\mu$m and 450~$\mu$m, respectively. Additionally, positively charged particles have a better spatial resolution than negatively charged particles due to the influence of the magnetic field. Both charged particles and ionized electrons are deflected during migration in the $\mu$RWELL gap since the Lorenz angle is not 0 in the magnetic field. For positively charged particles, the deflection directions of primary particles and ionized electrons are opposite, leading to a larger track projection range. For negative particles, the deflection directions of primary particles and ionized electrons are the same, resulting in a smaller tracker projection range. In the $\mu$-TPC mode, a larger track range is beneficial to the tracking performance. Thus, positively charged particles have a better spatial resolution than negatively charged particles.

%%%%%%%%%%%%%%%%%%% Fig %%%%%%%%%%%%%%%%%%%%%%%%%%
%%%%%%%%%%%%%%%%%%%%%%%%%%%%%%%%%%%%%%%%%%%%%%%%%%

\begin{figure*}[htb]
	\centering
\subfloat[][]{\includegraphics[height=50 mm]{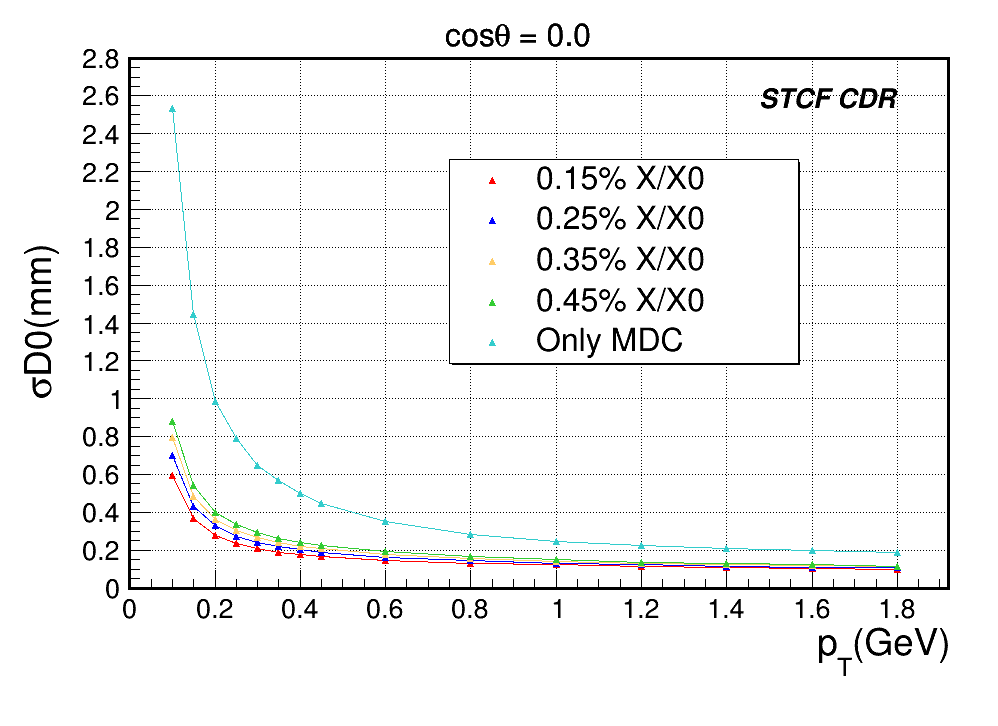}}
\subfloat[][]{\includegraphics[height=50 mm]{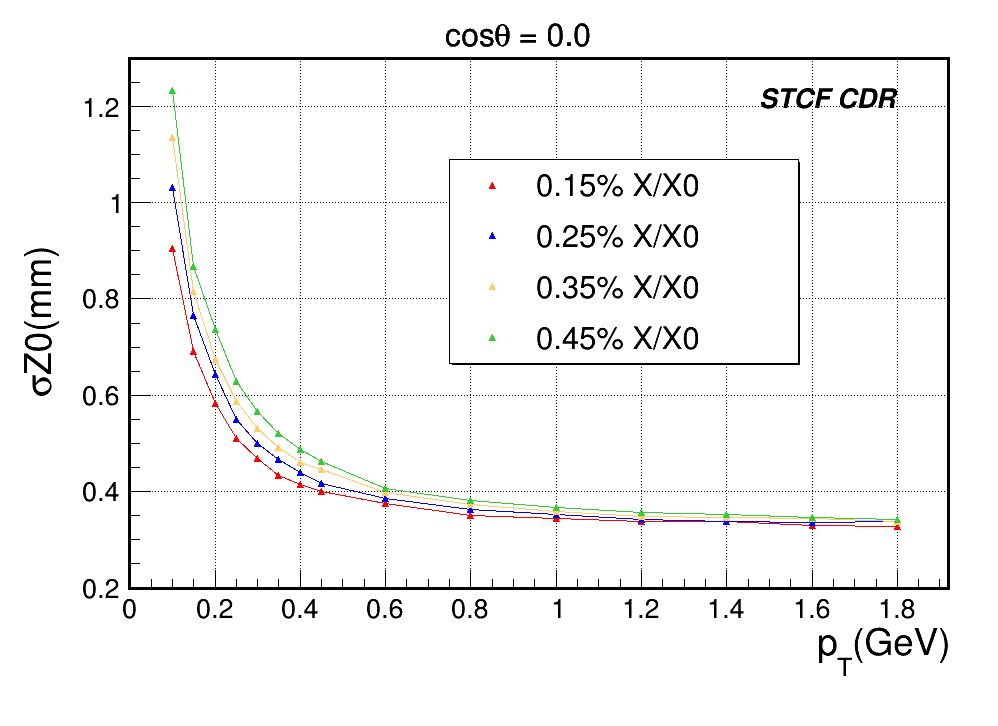}} \\
\subfloat[][]{\includegraphics[height=50 mm]{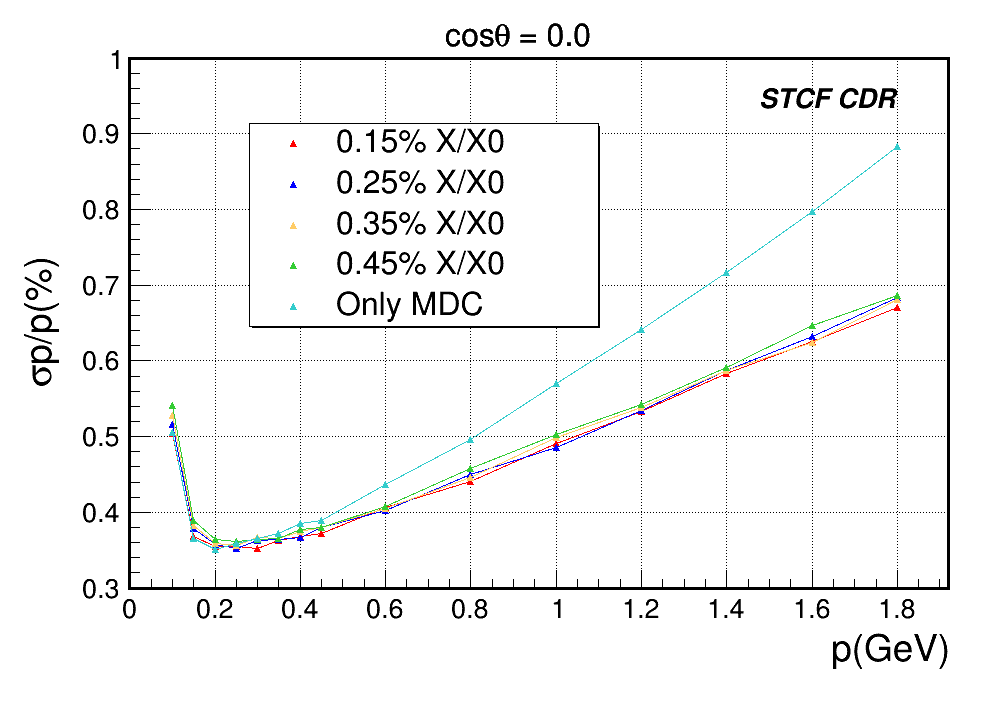}}
\subfloat[][]{\includegraphics[height=50 mm]{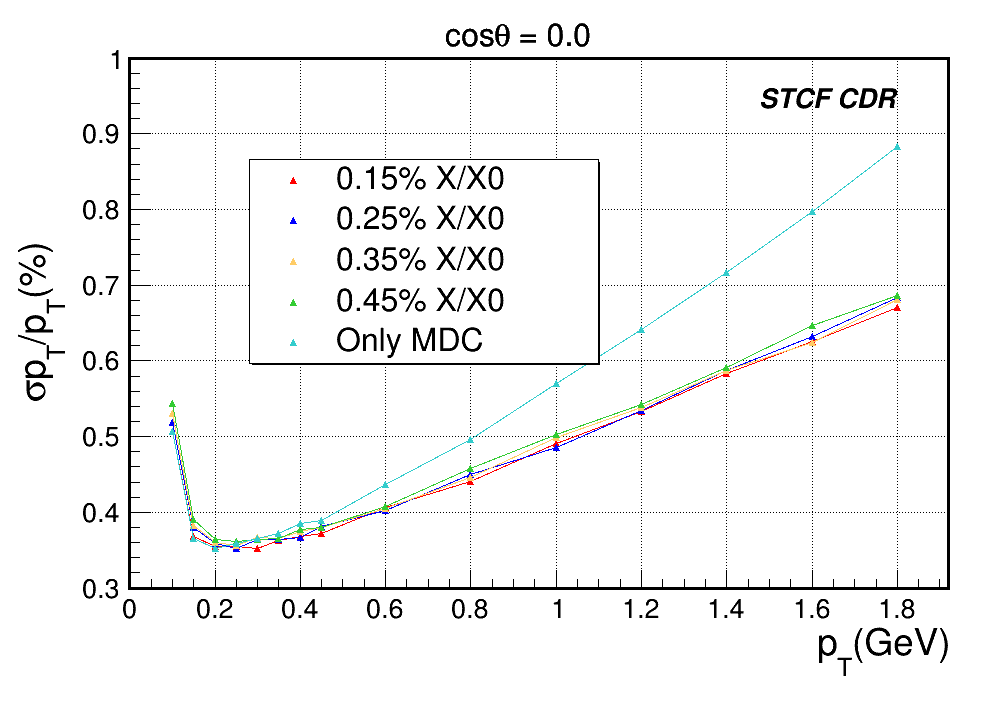}}
\vspace{0cm}
\caption{The simulated resolution of the impact parameters (a) $d_0$ and (b) $z_0$ and (c) the momentum $p$ and (d) transverse momentum $p_T$ as a function of the $p_T$ of the incident particle. The results with different material budgets, expressed in terms of the radiation length, are compared.}
    \label{fig:4.3.11}
\end{figure*}

%%%% material budget
To investigate the impact of the material budget on the performance of the ITK, the expected momentum resolution and position resolution with different material budgets, ranging from 0.15\%$X_0$ to 0.45\%$X_0$, are compared.
A single hit position resolution of $100\times400$~$\mu$m is assumed.
The results are shown in Fig.~\ref{fig:4.3.11} and are obtained from {\sc Geant4} simulation with a combined tracking fitting of the MDC + ITK tracking system. The incident particles ($\pi$) are assumed to have a polar angle of cos$\theta=0$.
In the low momentum range, the momentum resolution degrades with a higher material budget.
%Another important factor that needs to be considered is the track finding and reconstruction efficiency.

%%%%%%%%%%%%%%%%%%% Fig %%%%%%%%%%%%%%%%%%%%%%%%%%
%%%%%%%%%%%%%%%%%%%%%%%%%%%%%%%%%%%%%%%%%%%%%%%%%%

\subsubsection{Counting rate and layout of readout strips}
%\paragraph{Readout channel, electronics, counting rate and data size estimation}
%\quad\\
From the background simulation study, the highest background count rate in the inner tracker appears in the first layer and is approximately 26.8~kHz/cm$^{2}$ (as discussed in Sec.~\ref{sec:mdi_bkg} at the peak luminosity. This is well within the rate capability of the $\mu$RWELL detector technology. For example, it has been demonstrated that a $\mu$RWELL detector with 400~$\mu$m pitch strips has a rate capability ranging from a few tens of
kHz/cm$^2$ up to a few MHz/cm$^2$~\cite{itk10}.

\begin{figure*}[htb]
    \centering
{
        \includegraphics[width=0.4\textwidth]{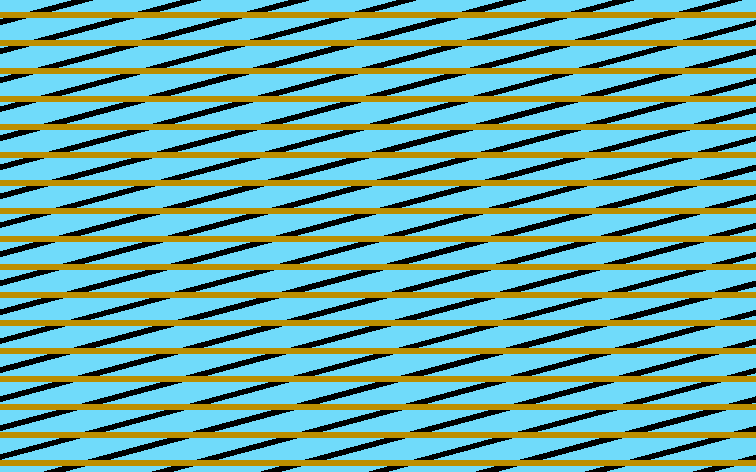}
}
\caption{The $X/V$ readout strips of the $\mu$RWELL detector.
}
    \label{fig:muRWELL_XV}
\end{figure*}

In the $\mu$RWELL-ITK conceptual design, two-dimensional readout strips ($X$ and $V$) are used with a crossing angle of 15 degrees, as shown in Fig.~\ref{fig:muRWELL_XV}. The $X$ readout strip is along the beamline direction. Both the X and V strips have a pitch of 400~$\mu$m in all 3 layers of the $\mu$RWELL film and cover a polar angle acceptance of $20^{\circ}$ $<$ $\theta$ $<$ $160^{\circ}$. Considering that the radii of the 3 $\mu$RWELL layers are 60~mm, 110~mm and 160~mm, the numbers of readout channels are 1919, 3517, and 5116, respectively, with 10552 in total. 
With this readout strips layout, each hit generates 10 signals in the X strips and 10 signals in the V strips on average, resulting in a highest count rate per channel as 367~kHz. This background level corresponds to an occupancy of 11.2\%, 5.4\%, and 5.32\% with a time window of 400~ns for the three layers of inner tracker, respectively. Obviously, it is necessary to optimize the detector design, particularly the layout of readout strips  to reduce the occupancy to a acceptable level. Splitting the X and V strips at Z = 0 into two parts would be one of the effective ways to decrease the occupancy. The average number of hits produced by the passage of a charged particle on the detector can be reduced by optimizing the gas gap width and the working gas. This would be another way to reduced the occupancy. 
%%%%%%%%%%%%%%%%%%% Fig %%%%%%%%%%%%%%%%%%%%%%%%%%
%%%%%%%%%%%%%%%%%%%%%%%%%%%%%%%%%%%%%%%%%%%%%%%%%%

%% muRWell electronics writen by Prof. Lei Zhao

\begin{figure*}[htb]
    \centering
{
        \includegraphics[width=160mm]{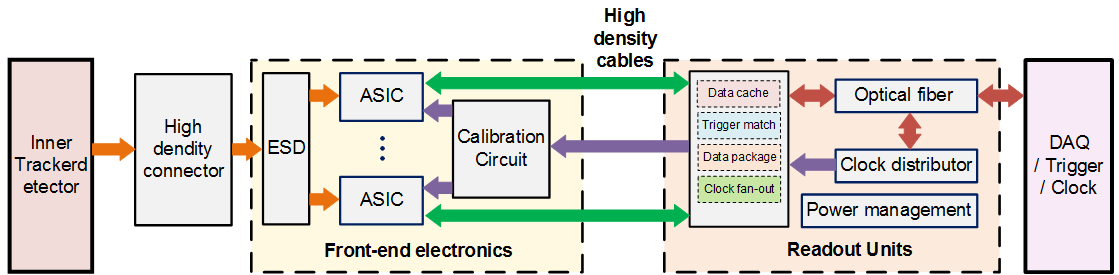}
}
\vspace{0cm}
\caption{Block diagram of the readout electronics for the $\mu$RWELL-based ITK detector.}
    \label{fig:4.1.01_electronic}
\end{figure*}

\paragraph{Readout Electronics}
\quad\\
The structure of the readout electronics of the $\mu$RWELL-based ITK detector is illustrated in Fig.~\ref{fig:4.1.01_electronic}. The front-end electronics are linked to detectors through high-density connectors, and a protection circuit is added at the input to protect the readout electronics from unexpected high-voltage discharge of the detector. Due to the small amplitude of the signal from the detector, the front-end readout electronics are placed close to the detector, and a high-density design is required for the front-end electronics. The front-end electronics are set near the endcaps of the inner tracker. The connector for $\mu$RWELL-based ITK is Hirose connector FH26W-71S-0.3SHW(60). It has a width of 23 mm and 0.3 mm channel pitch. The X and V strips of $\mu$RWELL are designed on two films, so that their readout electronics can be arranged in two complete circles. In this case, all the readout channels can arranged well by the FPCB connector. Additionally, high-precision signal measurements are required. For the reasons above, it is planned to design an application-specific integrated circuit (ASIC) chip that integrates front-end analog circuits, analog-to-digital conversion (ADC), and a charge $\&$ time calculation circuit within the chip. In addition, the calibration circuit is added to correct the mismatch among channels.
%%%%%%%%%%%%%%%%%%% Fig %%%%%%%%%%%%%%%%%%%%%%%%%%
%%%%%%%%%%%%%%%%%%%%%%%%%%%%%%%%%%%%%%%%%%%%%%%%%%

The output data of the front-end ASICs are transferred to a digital ASIC or field-programmable gate array (FPGA) for data packaging and finally transferred to the DAQ through high-speed serial data interfaces. Since the ITK readout electronics need to be synchronized with the global clock, the FPGA is also responsible for fanning out the clock to each front-end ASIC.

ITK electronics do not take part in generating the global trigger signal and only receive the trigger signal for trigger matching. The data from the front-end ASICs are first stored in RAMs, and when the FPGA receives the global trigger signal, it picks out the valid data through trigger match logic and transfers the data to the DAQ.

The hardware system is composed of front-end electronics ~(FEE), readout units ~(RUs), and clock, trigger submodules. Multiple front-end ASICs, which complete analog signal processing, A/D conversion, and charge $\&$ time calculation, are integrated into one FEE module. The output data of these ASICs are transferred to the RUs through a high-speed serial data interface.

 Both the charge/amplitude and time information of the hits are necessary for track reconstruction. The recorded data of each hit signal in one channel represent a 96-bit word, including 8 bits for the header, 16 bits for the trigger number, 34 bits for the timing information, 6 bits for the amplitude, 16 bits for the FEE number, 8 bits for the check and 8 bits for the tail. As a consequence, the total data stream sizes are approximately 6.89~GB/s for all 3 layers.

\begin{figure*}[htb]
    \centering
{
        \includegraphics[width=160mm]{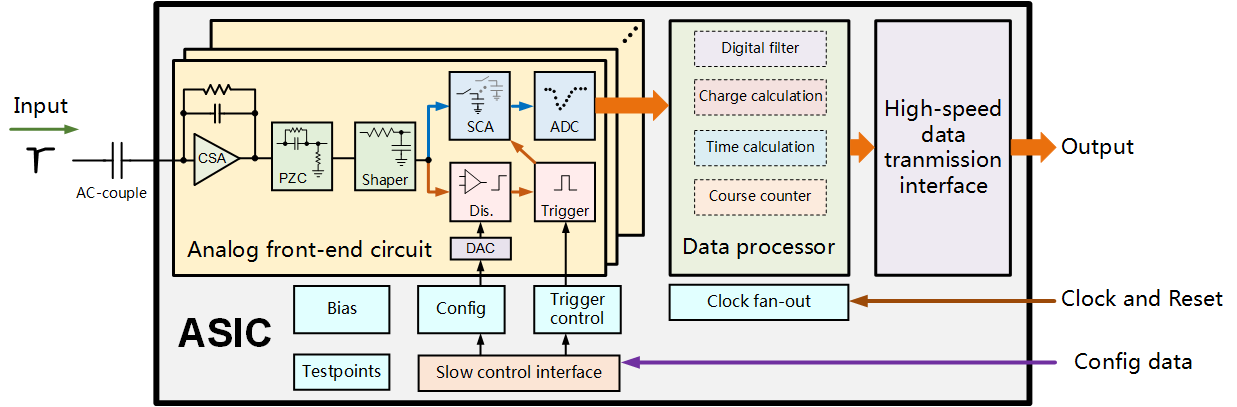}
}
\vspace{0cm}
\caption{Block diagram of the front-end ASIC.}
    \label{fig:4.1.02_electronic}
\end{figure*}

\paragraph{Front-End ASIC}
\quad\\
The high channel density of the $\mu$RWELL-ITK requires customized ASICs for its front-end electronics that integrate functions of analog signal processing, ADC and charge \& time calculation. As shown in Fig.~\ref{fig:4.1.02_electronic}, each channel comprises a charge sensitive amplifier (CSA) for low noise amplification, a semi-Gaussian shaper network, a discriminator for generating a self-trigger signal, a switched capacitor array (SCA) for waveform sampling, a Wilkinson ADC for digitization, a digital circuit for charge \& time calculation, and a high-speed data transfer interface.

Due to the high event rate, full waveform data transfer would place high pressure on the data transfer interface and corresponding power consumption. Therefore, it is preferable to integrate the charge \& time calculation circuit into the chip.

%%%%%%%%%%%%%%%%%%% Fig %%%%%%%%%%%%%%%%%%%%%%%%%%
%%%%%%%%%%%%%%%%%%%%%%%%%%%%%%%%%%%%%%%%%%%%%%%%%%

\begin{figure*}[htb]
    \centering
{
        \includegraphics[width=160mm]{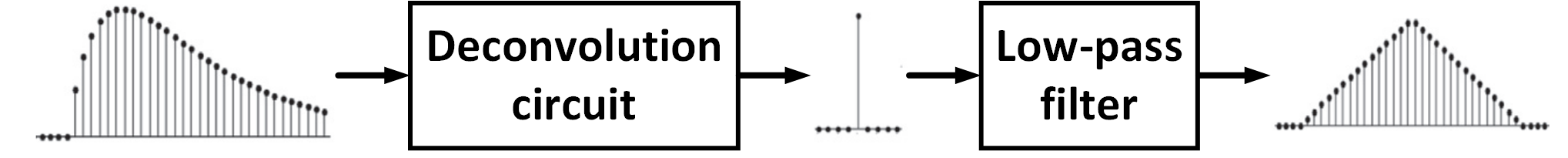}
}
\vspace{0cm}
\caption{Block diagram of the digital deconvolution and low-pass filter circuit.}
    \label{fig:4.1.03_electronic}
\end{figure*}

To further enhance the signal-to-noise ratio~(SNR) and to improve the accuracy of charge and time measurements, a digital deconvolution and filtering circuit is also integrated into the ASIC. The exponential signal is first unfolded into the unit impulse, and then the trapezoidal output pulse is processed by the moving average method (corresponding to the low-pass filter in Fig.~\ref{fig:4.1.03_electronic}) to filter out the high-frequency noise. The specific parameters of the deconvolution and moving average circuit need to be optimized according to the shaping time and knee frequency of the signal to achieve the best filtering result.

%%%%%%%%%%%%%%%%%%% Fig %%%%%%%%%%%%%%%%%%%%%%%%%%
%%%%%%%%%%%%%%%%%%%%%%%%%%%%%%%%%%%%%%%%%%%%%%%%%%

\subsubsection{R\&D on Cylindrical $\mu$RWELL}
\begin{figure*}[htb]
    \centering
{
        \includegraphics[height=30mm]{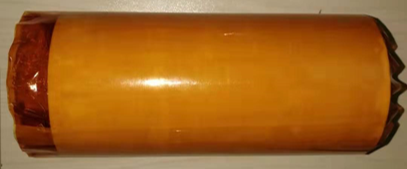}
}
\hspace{5 mm}
{
        \includegraphics[height=30mm]{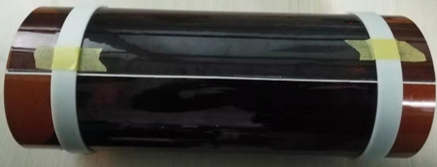}
}
\vspace{0cm}
\caption{The Rohacell foam-based inner cylinder  (left) and aramid honeycomb-based detector model (right) produced in the preresearch stage.}
    \label{fig:4.1.06}
\end{figure*}

%\quad\\
Several kinds of structure supporting materials have been tested, of which aramid honeycomb~\cite{itk13} and Rohacell foam~\cite{itk14} showed good performance in terms of both material budget and mechanical strength. Additionally, two bonding methods have been developed for the aramid honeycomb and Rohacell foam. A solid sandwich structure with a very low material budget of adhesive is feasible. Fig.~\ref{fig:4.1.06} shows a sandwich-like inner cylinder  made of the two kinds of material, and Table~\ref{tab:4.1.03} presents the performance of these cylinder . The mechanical strength of 2~mm aramid honeycomb is sufficient, while that of 1~mm Rohacell foam is lower. Thus, in the next step, thinner aramid honeycomb and thicker Rohacell foam will be manufactured and tested. Additionally, the study of a new X/V strip readout is still ongoing.

%%%%%%%%%%%%%%%%%%% Fig %%%%%%%%%%%%%%%%%%%%%%%%%%
%%%%%%%%%%%%%%%%%%%%%%%%%%%%%%%%%%%%%%%%%%%%%%%%%%

%%%%%%%%%%%%%%%%%  TABLE  %%%%%%%%%%%%%%%%%%%%%%%%
\begin{table*}[htb]
\small
    \caption{The material budgets of the inner cylinders  manufactured in the preresearch stage.}
    \label{tab:4.1.03}
    \vspace{0pt}
    \centering
    \begin{tabular}{lllllll}
        \hline
        \thead[l]{ } & \thead[l]{Inner PI film}& \thead[l]{Inner adhesive} &\thead[l]{Structure support } & \thead[l]{Outer adhesive}& \thead[l]{Outer PI film} & \thead[l]{Total}\\
        & & &material & & & \\
        \hline
        Honeycomb-based	&0.028\%	&0.009\%	&0.033\%	&0.009\%	&0.030\%	&0.105\% \\
        Rohacell-based	&0.028\%	&0.009\%	&0.010\%	&0.008\%	&0.029\%	&0.084\% \\
        \hline
    \end{tabular}
\end{table*}
%%%%%%%%%%%%%%%%%%%%%%%%%%%%%%%%%%%%%%%%%%%%%%%%%%

%% Silicon

\subsection{MAPS-based Inner Tracker}
The MAPS-based ITK is composed of three layers of silicon pixel detectors~(PXDs) and is located inside the MDC (see Sec.~\ref{sec:mdc}) at radii of 36~mm, 98~mm and 160~mm.
The radii of the two inner layers are smaller than those of the design of the $\mu$RWELL-based ITK to take into account the higher rate-capabilities of PXDs and to achieve better ITK spatial resolution.
A pixel size of $100~\mu\mathrm{m} \times 250~\mu\mathrm{m}$ is sufficient to meet the spatial resolution requirement of the STCF ITK.
To reduce the multiscattering effect for charged particles with low momentum, especially for $p < 200$~MeV/c, it is crucial to reduce the material budget as much as possible.
For the baseline design of the STCF PXDs, a radiation length of 0.25\% $X_0$ per layer is assumed, including the material budget from the sensor, readout electronics and supporting material.

The monolithic active pixel sensor (MAPS) technology has the potential to satisfy the low-material and high-rate requirements for the STCF inner tracker. 
This technology has an attractive advantage of having both the sensor and readout electronics in the same pixel, thus reducing the material budget,
and it has been developing rapidly in the particle physics community.
The first-generation MAPS-based vertex detector for the STAR upgrade successfully completed a 3-year physics run~\cite{starmaps1,starmaps2},
The new generation complementary metal-oxide-semiconductor (CMOS) pixel sensor (CPS) for the ALICE-ITS upgrade~\cite{aliceits,alpide} is in mass production.
The CMOS MAPS sensor is chosen as the pixel sensor technology for the silicon-based ITK. The ITK is called MAPS-based ITK in this case. 

The high luminosity of the STCF places additional stringent requirements on the design of the ITK detector, and the challenges include the high hit rate and pileup effects.
From Table~\ref{tab:TIDNIEL_max}, the highest expected hit rate is approximately $1.04\times 10^6$~Hz/cm$^{2}$ at the innermost layer of the MAPS-based~($\mu$RWELL-based) ITK.
The state of the art MAPS technology can easily handle such a hit rate . For example, The STAR ULTIMATE MAPS can cope with a hit rate of
approximately 1 MHz/cm$^{2}$s$^{-1}$~\cite{starmaps1}, and the ALICE-ITS ALPIDE sensor can
operate with a hit rate of 3~MHz/cm$^{2}$~\cite{aliceits}

\subsubsection{Expected Tracking Performance}
\begin{figure*}[htb]
	\centering
\subfloat[][]{\includegraphics[height=50 mm]{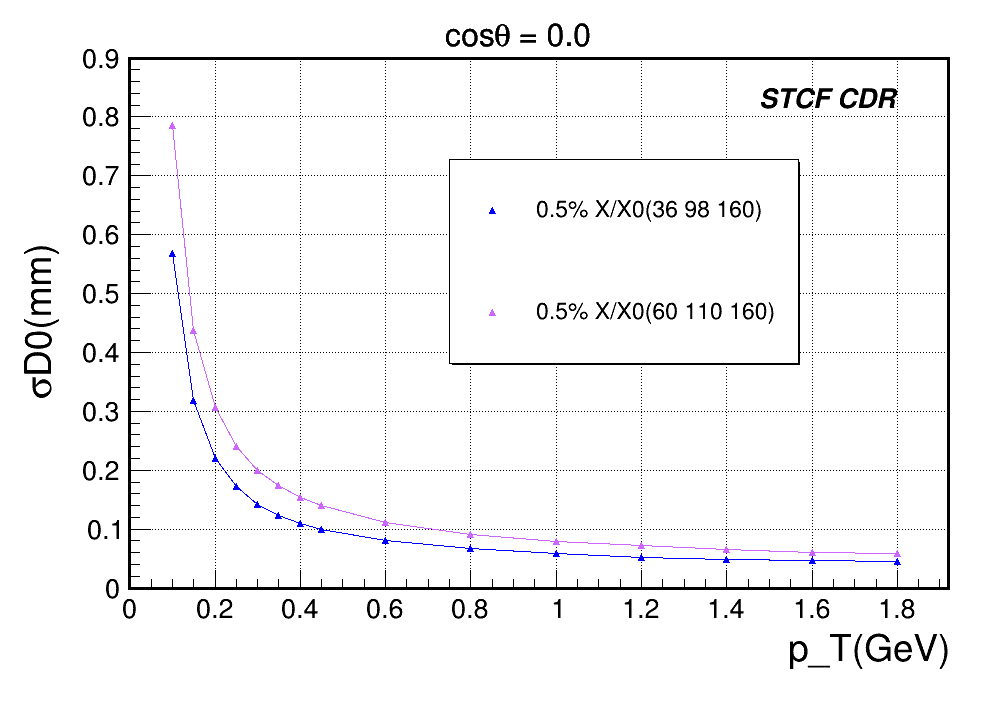}}
\subfloat[][]{\includegraphics[height=50 mm]{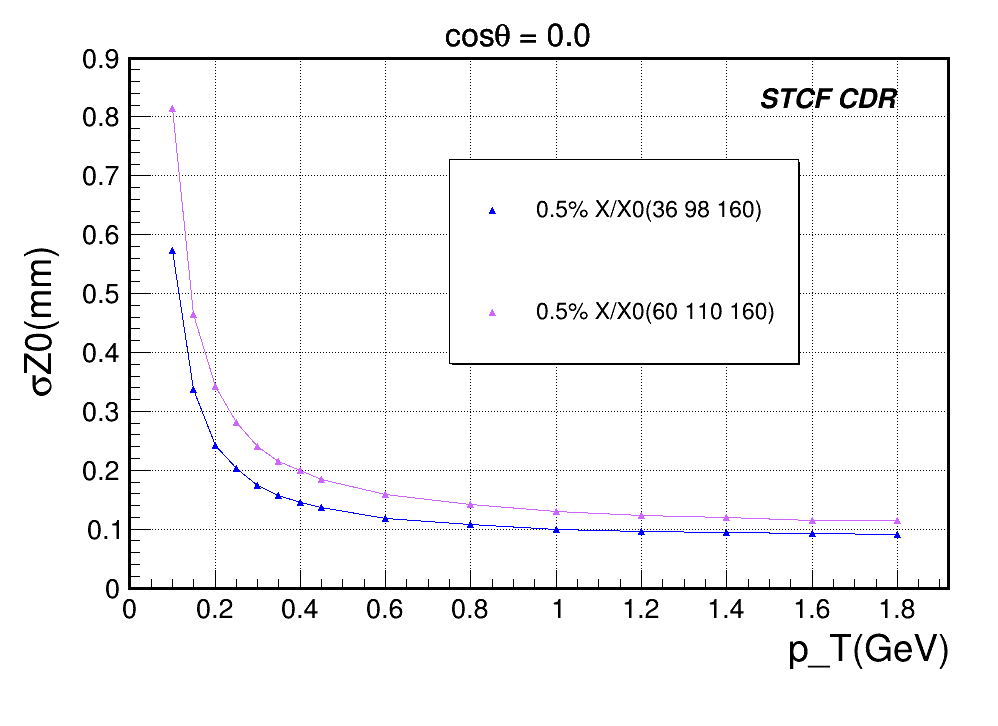}} \\
\subfloat[][]{\includegraphics[height=50 mm]{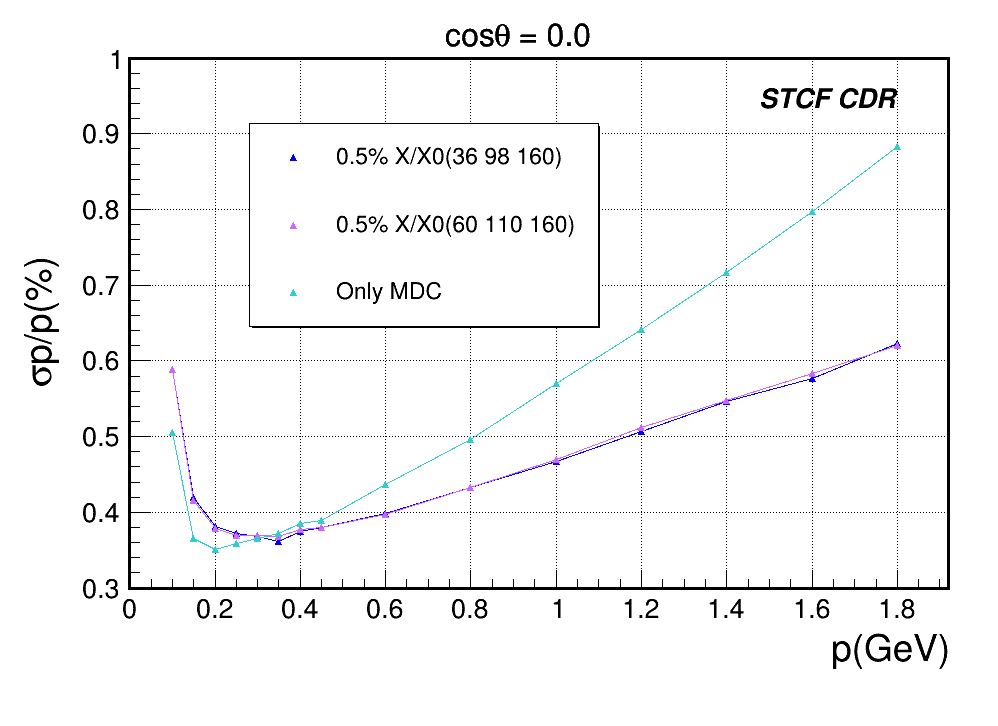}}
\subfloat[][]{\includegraphics[height=50 mm]{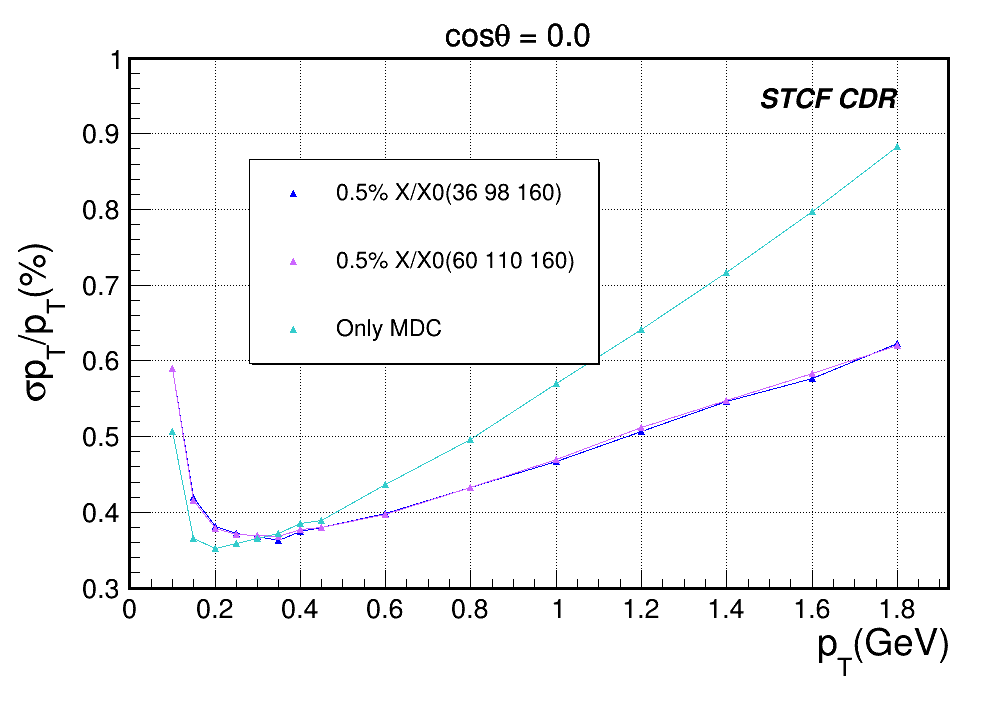}}
\vspace{0cm}
\caption{The simulated resolution of the impact parameters (a) $d_0$ and (b) $z_0$ and (c) the momentum $p$ and (d) transverse momentum $p_T$ as a function of the $p_T$ of the incident particle. The results with different layout configurations, the default with radii of 36 mm, 98 mm and 160 mm and alternative radii of 60 mm, 110 mm and 160 mm, are compared. }
    \label{fig:4.1.09.z}
\end{figure*}

%{Performance of the Silicon-based Inner Tracker}
Figure~\ref{fig:4.1.09.z} shows the expected performance of the momentum resolution and position resolution of the tracking system, comparing the two configurations MDC only and MDC + PXD with different settings for the radius of the PXD layers.
A single hit position resolution of 30~$\mu$~$\times$~75~$\mu$m is assumed for the PXD.
The results are obtained from simulation with combined track fitting of the MDC + PXD tracking system, and incident particles are assumed to have a polar angle of cos$\theta=0$.
With MDC + PXD, the momentum resolution is improved by a factor of approximately 1.5 at 1.8~GeV/c compared with MDC tracking only.

As expected, while the impact parameter resolution can be improved when the innermost layer is closer to the beam pipe, the resolution of the momentum and transverse momentum has little dependence on the radius of the innermost layer.

%%%%%%%%%%%%%%%%%%% Fig %%%%%%%%%%%%%%%%%%%%%%%%%%
%%%%%%%%%%%%%%%%%%%%%%%%%%%%%%%%%%%%%%%%%%%%%%%%%%

\begin{figure*}[htb]
	\centering
\subfloat[][]{\includegraphics[height=50 mm]{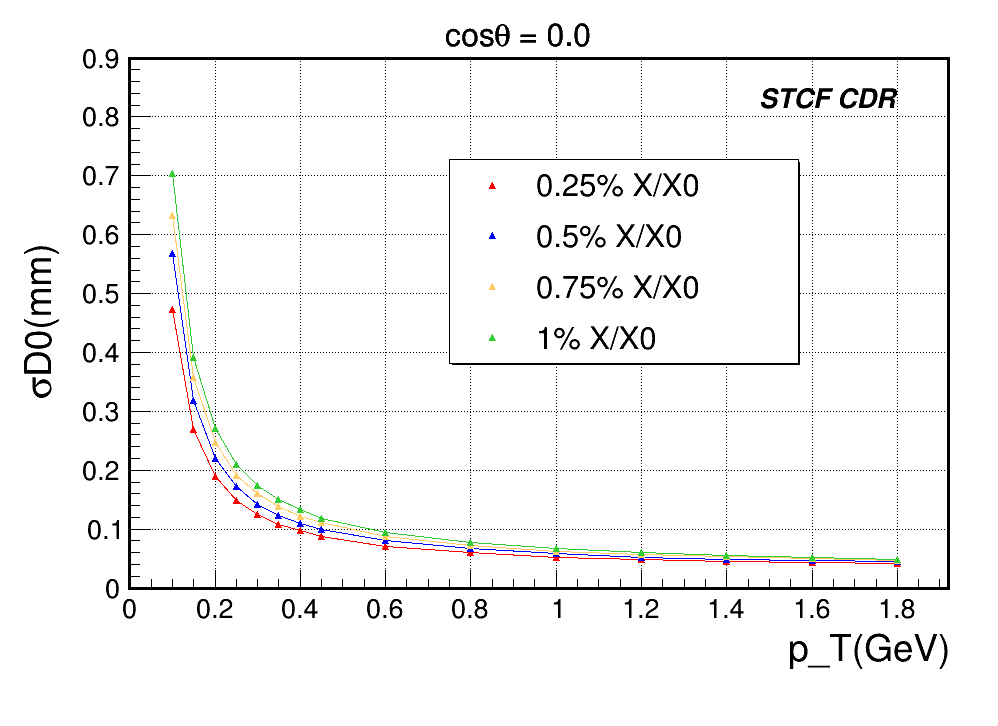}}
\subfloat[][]{\includegraphics[height=50 mm]{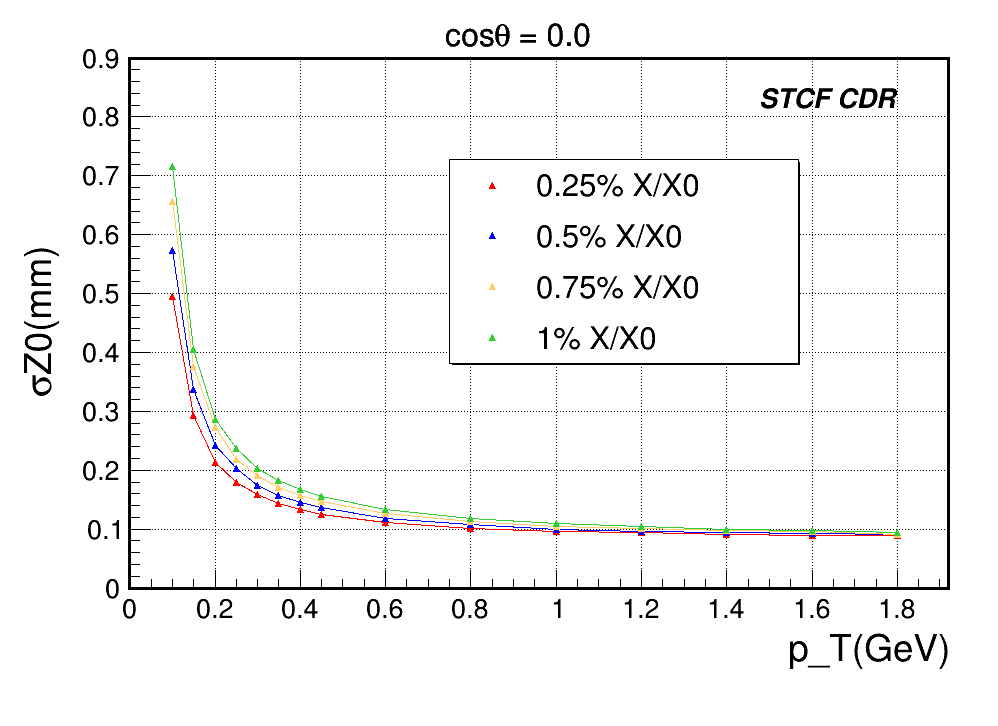}} \\
\subfloat[][]{\includegraphics[height=50 mm]{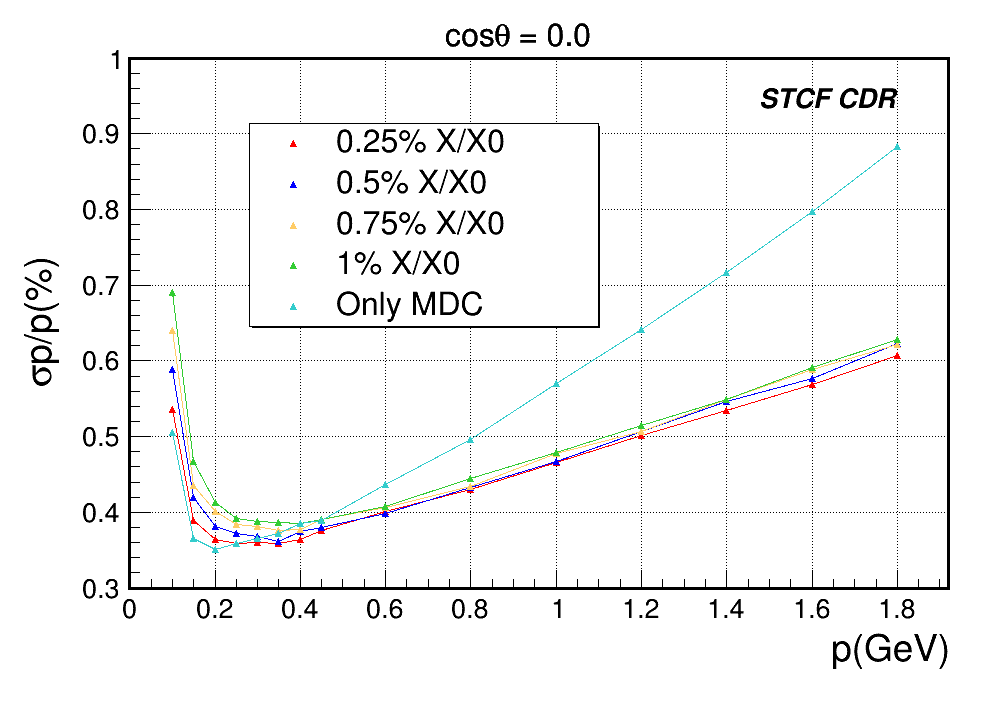}}
\subfloat[][]{\includegraphics[height=50 mm]{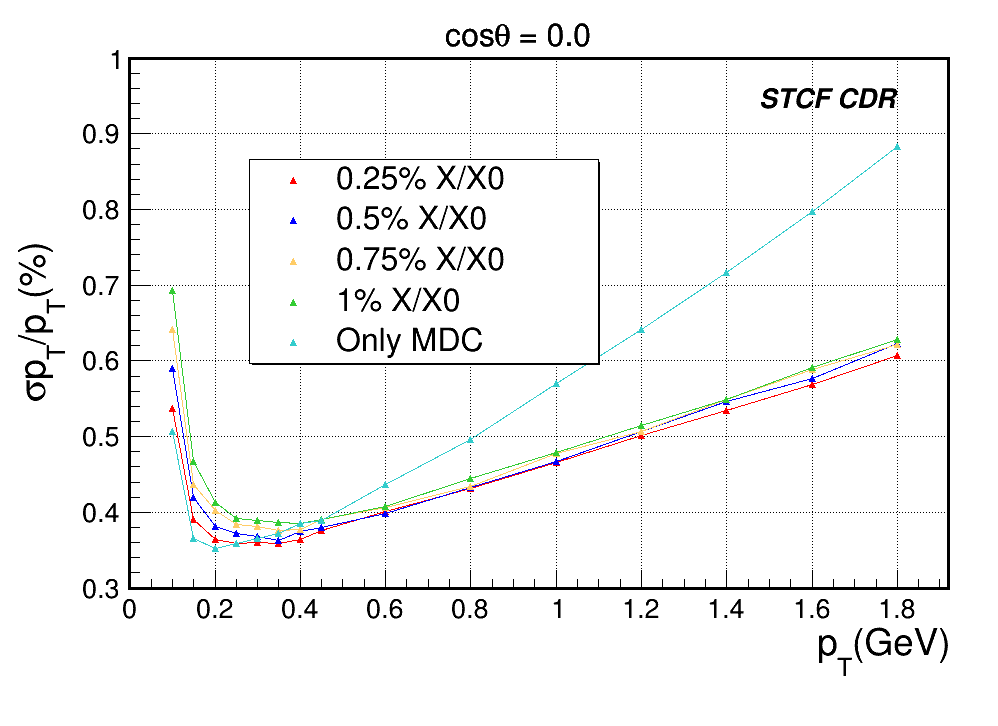}}
\vspace{0cm}
\caption{The simulated resolution of the impact parameters (a) $d_0$ and (b) $z_0$ and (c) the momentum $p$ and (d) transverse momentum $p_T$ as a function of the $p_T$ of the incident particle. The results with different material budgets, expressed in terms of the radiation length, are compared.}
    \label{fig:4.1.10.z}
\end{figure*}

To investigate the impact of the material budget on the performance of the ITK, the expected momentum resolution and position resolution with different material budgets, ranging from 0.25\%$X_0$ to 1.0\%$X_0$, are compared.
The results from the Geant4 simulation are shown in Fig.~\ref{fig:4.1.10.z}.
In the low momentum range, a degradation of the momentum resolution with a higher material budget is seen.
%Another important factor that needs to be considered is the track finding and reconstruction efficiency.
Further investigation is needed to understand the impact of the material budget on the track finding and reconstruction efficiency, which have a significant impact on the physics potential in the low momentum range at the STCF.

%%%%%%%%%%%%%%%%%%% Fig %%%%%%%%%%%%%%%%%%%%%%%%%%
%%%%%%%%%%%%%%%%%%%%%%%%%%%%%%%%%%%%%%%%%%%%%%%%%%

\subsubsection{CMOS MAPS for the Inner Tracker}
%{Prospect of MAPS for PXD}
A good starting point for the STCF ITK is the ALPIDE design, which was developed for the aforementioned ALICE-ITS upgrade.
The ALICE-ITS has achieved a material budget of approximately 0.3\%$X_0$ with a sensor thickness of 50~$\mu$m.
CMOS pixel sensors, for instance, JadePix~\cite{jadepix}, are also being proposed as the vertex detector for the conceptual design of the Circular Electron Positron Collider (CEPC)~\cite{cepc}. JadePix-1 features a small pixel size ($16\times 16~\mu\mathrm{m}^2$), with the main goal being achieving low power consumption and material budget. For the STCF, the pixel size requirement can be relaxed, and the main challenges are the low power consumption and fast readout necessary to cope with the high event rate (see Sec.~\ref{sec:tdaq}).
In contrast to an ordinary MAPS, which collects ionization charge mainly by diffusion, high-voltage~(HV)/high-resistivity~(HR)-CMOS designs collect ionization charge mainly via drift, as the sensor can be fully depleted.
The fast collection of signal charges and the low noise HV/HR-CMOSs also allow them to be more radiation tolerant.
HV-MAPSs have been prototyped for several experiments, such Mu3e~\cite{mu3e}, ATLAS~\cite{atlaspix} and CLIC~\cite{clicpix}.
Another promising option for the STCF ITK is the MuPix sensor~\cite{mupix} to be used for the Mu3e experiment, which was designed to detect extremely low-momentum tracks ($<50$~MeV/c) with very high tracking efficiency and momentum resolution.

\subsubsection{Readout Circuitry}
\begin{figure*}[htb]
	\centering
    \includegraphics[width=0.7\linewidth]{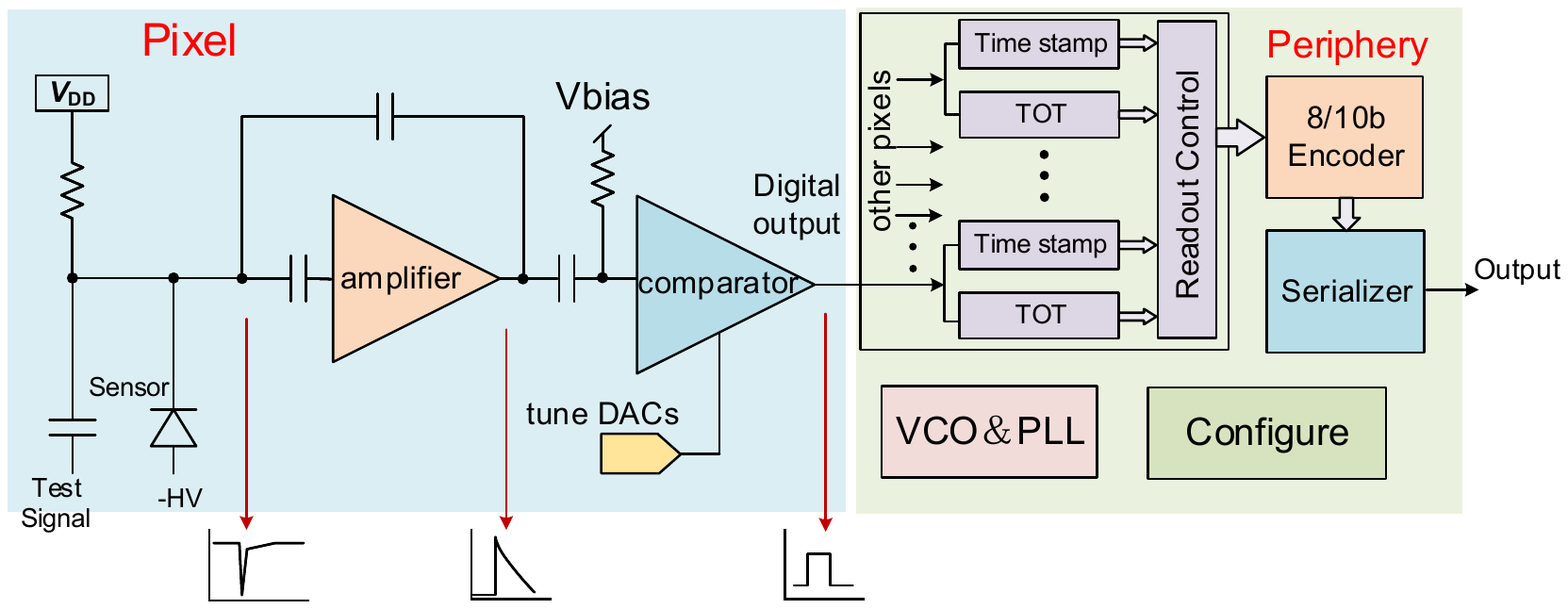}
\vspace{0cm}
\caption{Block diagram of the pixel sensor readout circuit.}
    \label{fig:pix_readout}
\end{figure*}

The STCF CMOS pixel sensor is expected to be a $2\times2$~cm$^2$ chip that contains 16k pixels of $100\times 250~\mu\mathrm{m}^{2}$ size.
The readout circuitry provides time stamping and charge measurement based on the time-over-threshold (TOT).
The structure of the readout circuitry is shown in Fig.~\ref{fig:pix_readout}.

The readout circuitry mainly consists of an in-pixel part and a periphery part.
There is a low-power charge-sensitive amplifier (CSA) and a voltage comparator in each pixel.
The CSA integrates the charge collected from the sensor and outputs a voltage signal to the following comparator,
and the voltage threshold of the comparator can be tuned by a local DAC.
All the pulse signals from the comparators are driven to the periphery, and the arrival time of each pulse is recorded as a timestamp.
Meanwhile, the pulse width is measured based on the TOT method.
The time stamps and TOT messages are read out through the readout control block to an 8/10-bit encoder
and, finally, to a serializer with an output data rate of several Gbits/s.
Additionally, the configuration block, voltage-controlled oscillator (VCO), and phase-locked loop (PLL) are integrated in the periphery.

\subsection{Pileup and Radiation Effects}
% pileup
\subsubsection{Pileup Effects}

The STCF detector is expected to operate at an event rate as high as 400 kHz (Sec.~\ref{sec:tdaq}). With such a high event rate, the probability of events overlapping is approximately 8 (18)\% within a time window of 200 (500) ns. This indicates that fast response or good timing performance is required for detectors to cope with such a high event rate. Recent studies ~\cite{aplide-improved,malta2} have shown that depleted MAPS (DMPAS) with improved sensor designs and processes could provide a timing resolution of about 2~ns or better. Such excellent timing capability would be very helpful in solving the overlapping events by providing each charged track with precise time measurement. 

% radition and aging effects
\subsubsection{Radiation Effects}
As the innermost subdetector of the STCF detector, the ITK receives the highest level of background radiation.  Radiation hardness is a crucial factor in the design of the ITK.
A detailed background radiation simulation is described in Sec.~\ref{sec:mdi_bkg}, and the expected radiation levels in individual subdetectors are given in Table~\ref{tab:TIDNIEL_mean} and \ref{tab:TIDNIEL_max}.
However, note that the radiation simulation in Sec.~\ref{sec:mdi_bkg} 
 is subject to large uncertainties since many technical details for the designs of the accelerator and the machine detector interface are not available yet at the conceptual design stage. In addition, beam background usually can't be simulated precisely given too many factors in the simulation that can't be well determined. For example, significant discrepancies between simulation and actual measurement of beam background have been observed in the Belle II experiment~\cite{belle2bkg}. Therefore, sufficient safety margins have to be taken into account when assessing beam background levels. 

 From Table~\ref{tab:TIDNIEL_max}, the TID and NIEL of the uRWELL-based ITK are 120 Gy/y and $1.1\times 10^{10}$~n/cm$^{2}$/y~(1 MeV neutron equivalent), respectively. Those of the MAPS-based ITK are 3.5~kGy/y and a NIEL of $1.8\times 10^{11}$~n/cm$^{2}$/y, respectively. These levels of radiation are well below the limits of the state of the art MPGD and MAPS detectors. For example, The ALPIDE pixel sensor for the ALICE-ITS has been demonstrated to have a radiation hardness of up to $10^{13}$~n/cm$^{2}$ (NIEL) and 27~kGy~\cite{alpide2}.

For the $\mu$RWELL-based ITK, potential degradation of performance due to aging effects, such as a
specific gas gain reduction or high voltage instabilities during operation caused by irradiation, should be considered.
In general, aging is induced by plasma-chemical processes during gas amplification processes.
A complete overview and description of the aging phenomena in gaseous detectors can be found in Ref.~\cite{gaseousaging}.
A dedicated aging study is planned for the $\mu$RWELL-based ITK. During the long-term operation of the STCF, it can be replaced if it malfunctions due to radiation damage.

\subsection{Conclusion and Outlook}
The cylindrical $\mu$RWELL-based detector, with the advantages of low material budget, robustness, scalability, and simplified manufacturing and maintenance processes, is the baseline design of the STCF inner tracker and is expected to provide promising performance. Additional studies are needed in the future to realize the baseline design and to further optimize the detector performance, including material budget limitation, optimization of the cylindrical $\mu$RWELL detector structure, manufacturing of the detector prototype and performance testing. As an alternative design choice, CMOS silicon pixel detectors are being considered. They are expected to provide better vertex resolution and greater radiation hardness. It will be crucial to continue the pixel sensor R\&D program and to develop CMOS pixel sensors with radiation tolerance, lower power consumption and fast readout electronics.
The expected performance of the different ITK designs are discussed in Sec.~\ref{sec:mdc}, taking into account both the inner tracker and outer tracker.
\quad\\

\clearpage
\newpage
\section{Main Drift Chamber (MDC)}
\label{sec:mdc}

\subsection{Introduction}
The MDC is the main part of the tracking system of the STCF detector, providing important functions including the following:
\begin{itemize}
\item Reconstructing charged tracks together with the ITK.
\item Measuring the momentum (~$p$~)/transverse momentum ($p_T$) of charged particles (with a resolution of $\sigma_{p_T}/p_{T} < 0.5$\%@1~GeV/c).
\item Measuring the energy loss of charged particles in each cell ($dE/dx$) and providing the $dE/dx$ information (with $dE/dx$ resolution $\sim$ 6\%) to facilitate particle identification, especially for low-momentum charged particles.
\item Providing critical input to the trigger decision of the STCF detector system.
\end{itemize}
With the advantages of robustness, low cost and low material budget, drift chambers have been used as the main tracking system in many particle physics experiments, for example, BESIII~\cite{besiii}, Belle II~\cite{belle2} and GlueX~\cite{gluex}.
The inner radius of the MDC is determined to be 200~mm, a compromise between the count rate capabilities and tracking performance, and the outer radius is 850~mm.

\subsection{Conceptual Design of the MDC}
\begin{figure*}[htb]
    \centering
      \includegraphics[width=0.5\textwidth]{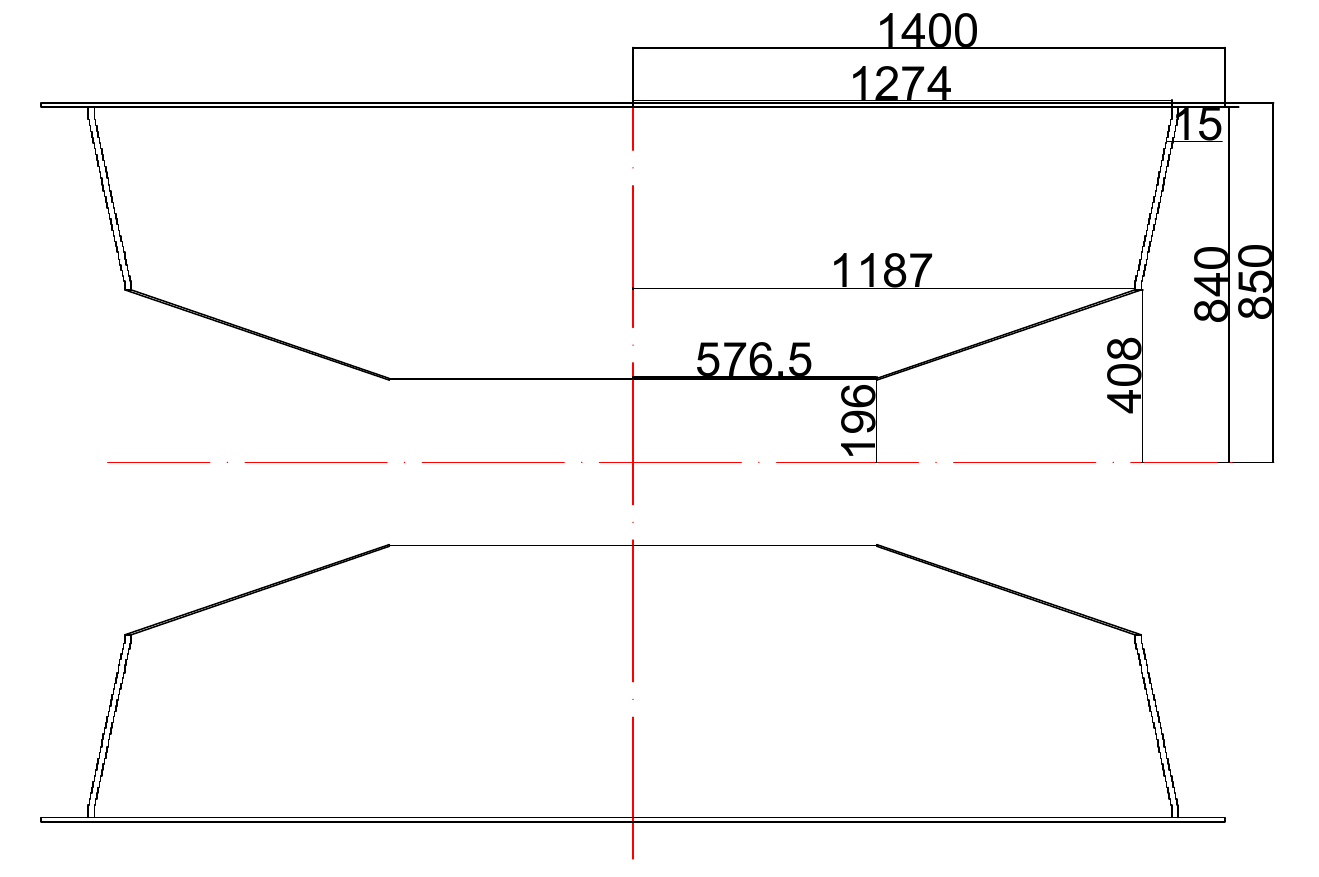}
\vspace{0cm}
\caption{The schematic structure of the MDC.}
    \label{fig:4.2.01.a}
\end{figure*}

\begin{figure*}[htb]
    \centering
\subfloat[][]{\includegraphics[width=0.5\textwidth]{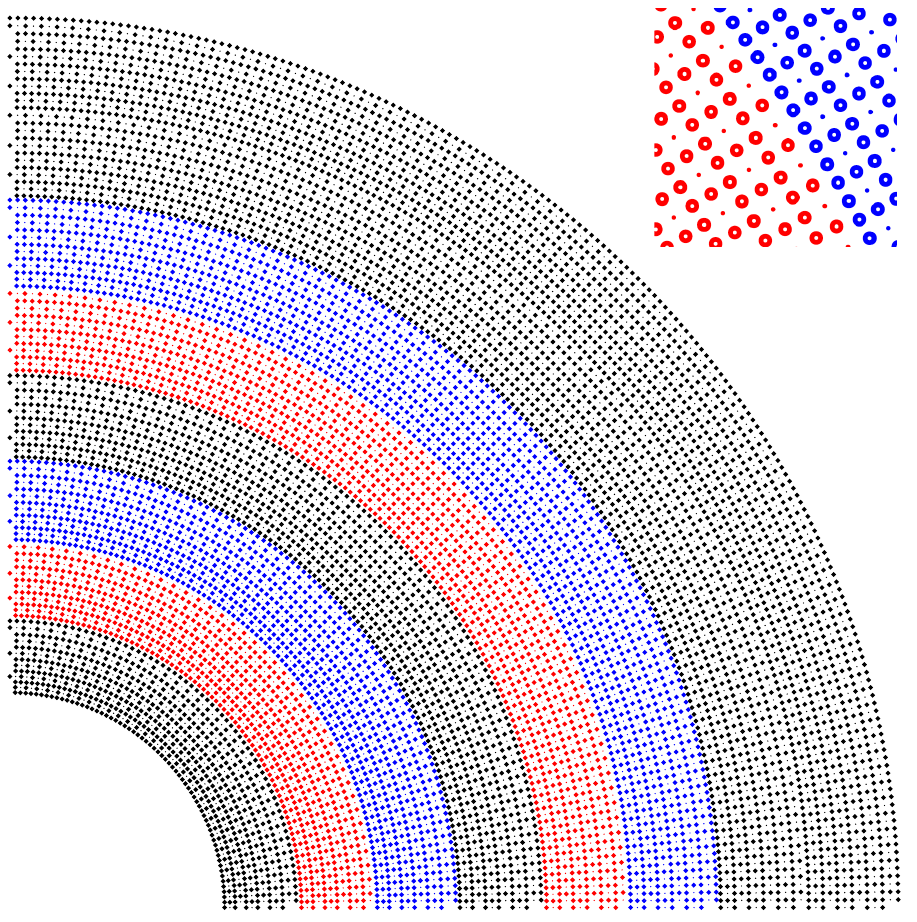}}
\hspace {5 mm}
\subfloat[][]{\includegraphics[width=0.15\textwidth]{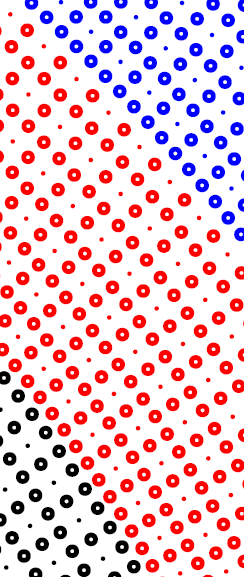}}
\vspace{0cm}
\caption{(a) Cross-section view of the layout of wire layers and superlayers. (b) Enlarged cross-section view of the wire layout. Open circles represent field wires, and dots represent sense wires. The square-shaped drift cell structure can be seen, with a sense wire in the center and field wires forming a square.}
    \label{fig:4.2.01.b}
\end{figure*}

Figure~\ref{fig:4.2.01.a} shows the main geometric parameters of the MDC.
The STCF MDC adopts a square cell and a superlayer wire configuration similar to those used at BESIII and Belle II.
There are six layers of drift cells in each superlayer. Each wire layer contains a field wire layer (with all field wires) and a sense wire layer (with alternating sense and field wires).
The sense wire is 20~$\mu$m-diameter and 0.5~$\mu$m-thick gold-coated tungsten wire, and the field wire is 100~$\mu$m-diameter and 0.5~$\mu$m-thick gold-coated aluminum wire.
The cell size increases gradually from the innermost layer to the outermost layer, approximately $9.8\times9.8$~mm$^2$ to $12.5\times 12.5$~mm$^2$ in the first superlayer and $13.3\times13.3$~mm$^2$ to $14.5\times 14.5$~mm$^2$ in the outermost superlayer, as shown in Table~\ref{tab:4.2.01}.
The innermost and outermost superlayers contain axial (``A'') layers to match the shape of the inner and outer cylinders. The intervening superlayers alternate between stereo (``U'' or ``V'') and axial layers. The stereo angles are listed in Table~\ref{tab:4.2.01}.
In total, there are 8 superlayers (AUVAUVAA) and 48 layers.
The working gas is He/C$_{3}$H$_{8}$~(60/40), together with the use of other low-mass materials, to minimize the effect of multiple scattering.
The main parameters of the MDC are summarized in Table~\ref{tab:4.2.01}. Figure~\ref{fig:4.2.01.b} shows a schematic of the MDC wire structure.

%%%%%%%%%%%%%%%%%  TABLE  %%%%%%%%%%%%%%%%%%%%%%%%
\begin{table*}[htb]
\small
    \caption{The main parameters of the STCF MDC conceptual design.}
    \label{tab:4.2.01}
    \vspace{0pt}
    \centering
    \begin{tabular}{llllll}
        \hline
        \thead[l]{Superlayer} & \thead[l]{Radius (mm)}&\thead[l]{Num. of Layers} & \thead[l]{Stereo angle (mrad)}&\thead[l]{Num. of Cells}&\thead[l]{Cell size (mm)}\\
        \hline
            A & 200.0 & 6 & 0 & 128 & 9.8 to 12.5 \\
            U & 271.6 & 6 & 39.3 to 47.6 & 160 & 10.7 to 12.9 \\
            V & 342.2 & 6 & -41.2 to -48.4 & 192 & 11.2 to 13.2 \\
            A & 419.2 & 6 & 0 & 224 & 11.7 to 13.5 \\
            U & 499.8 & 6 & 50.0 to 56.4 & 256 & 12.3 to 13.8 \\
            V & 578.1 & 6 & -51.3 to -57.2 & 288 & 12.6 to 14.0 \\
            A & 662.0 & 6 & 0 & 320 & 13.0 to 14.3 \\
            A & 744.0 & 6 & 0 & 352 & 13.3 to 14.5 \\
            total & 200 to 827.3 & 48 & & 11520 & \\
        \hline
    \end{tabular}
\end{table*}
%%%%%%%%%%%%%%%%%%%%%%%%%%%%%%%%%%%%%%%%%%%%%%%%%%

%%%%%%%%%%%%%%%%%%% Fig %%%%%%%%%%%%%%%%%%%%%%%%%%
%%%%%%%%%%%%%%%%%%%%%%%%%%%%%%%%%%%%%%%%%%%%%%%%%%

%%%%%%%%%%%%%%%%%%% Fig %%%%%%%%%%%%%%%%%%%%%%%%%%
%%%%%%%%%%%%%%%%%%%%%%%%%%%%%%%%%%%%%%%%%%%%%%%%%%

On each side of the MDC, there is a 15~mm-thick aluminum endcap flange. All the sense wires and field wires are fixed between the two flanges and fastened. The inner and outer surfaces of the MDC are made of cylindrical carbon fiber composite material, with thicknesses of 1~mm and 10~mm, respectively. The drift chambers in BESIII and Belle II demonstrated that this type of material has enough mechanical strength and a low material budget, which meets the requirements of the STCF MDC design.

\FloatBarrier

\subsection{MDC Simulation and Optimization}
An extensive simulation and optimization study is performed for the conceptual design of the MDC, including the wire material and diameter, cell structure and layout, and working gas choice.
The detector simulation is based on {\sc Geant4} with $\pi$ incident particles, and track fitting is then applied to evaluate the MDC performance.

\subsubsection{Drift wires}
%\quad\\
Aluminum wire and tungsten wire are widely used in the manufacturing of MWDCs due to their low material budget and high robustness, respectively. It is demonstrated that adding a coating of gold or silver on the surface can mitigate aging effects and enhance the conductivity of wires. Thus, gold-coated aluminum wires are used as sense wires, and gold/silver-coated tungsten wires are chosen as field wires, with a coating thickness of 0.5~$\mu$m.

\begin{figure*}[htb]
    \centering
{
\subfloat[][]{\includegraphics[width=70 mm]{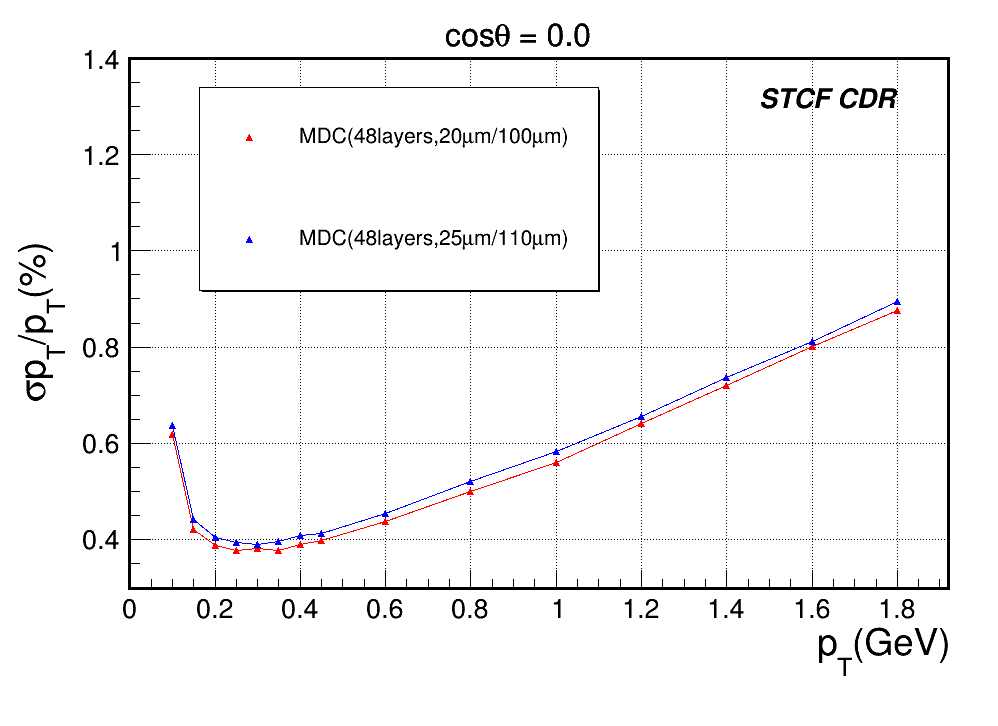}}
}
\hspace{2 mm}  
{
\subfloat[][]{\includegraphics[width=70 mm]{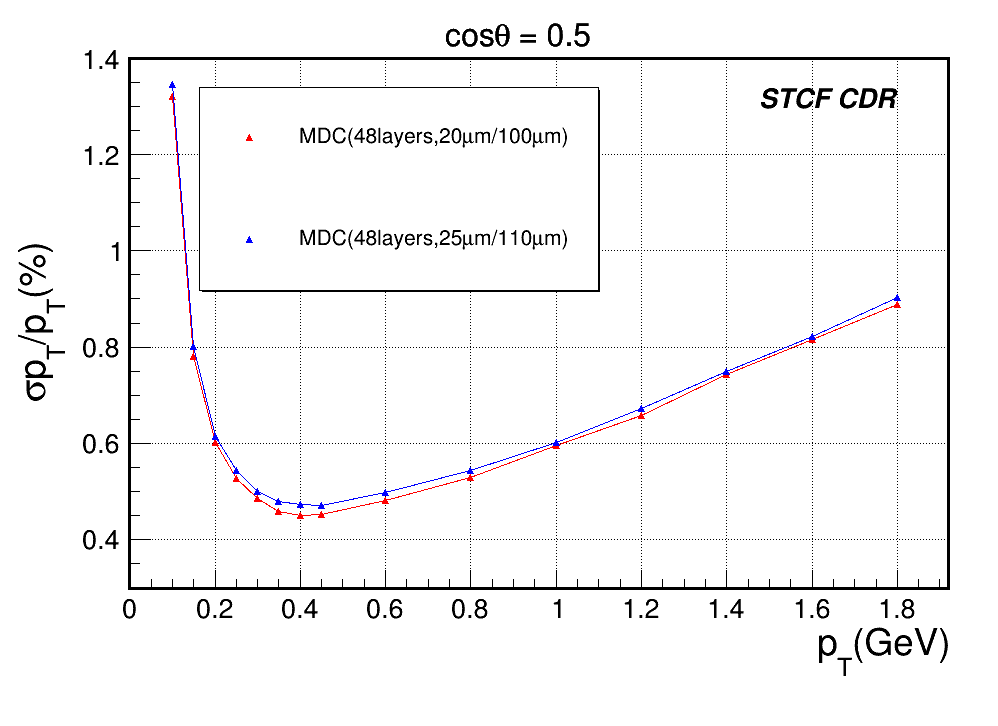}}
}
\vspace{0cm}
\caption{The simulated resolution of the transverse momentum of the MDC-only tracking system with different wire diameter settings, with polar angles of (a) cos$\theta$= 0 and (b) cos$\theta$ = 0.5.}
    \label{fig:4.2.04}
\end{figure*}

The wire diameter can influence the performance of the MDC. A thicker wire has better robustness but with the disadvantages of a larger material budget and a higher required working voltage.
In BESIII, the sense wire diameter is 25~$\mu$m and the field wire diameter is 110~$\mu$m with a working gas of He/C$_3$H$_8$~(60/40).
The drift chamber used in MEG-II~\cite{mdc1} indicates that the diameter of the field wire can be decreased to 50~$\mu$m with a gas mixture He/iC$_4$H$_{10}$~(90/10), resulting in a lower material budget and better momentum resolution.
A 20~$\mu$m sense wire diameter was proposed for a drift chamber for a future $e^{+}e^{-}$ collider experiment~\cite{idea}.
A compromise needs to be made between the low material budget and robustness.
Fig.~\ref{fig:4.2.04} shows the dependence of the transverse momentum resolution on the wire diameter from the simulation.
The simulation is performed with the combined inner tracker (Sec.~\ref{sec:itk}) and MDC tracking system.
Different configurations of the sense wire diameter (20~$\mu$m and 25~$\mu$m) and field wire diameter (100~$\mu$m and 110~$\mu$m) are investigated.
It can be seen that a thinner wire setting benefits the tracker system. It is decided to use a diameter of 20~$\mu$m for the sense wires and 100~$\mu$m for the field wires.

%%%%%%%%%%%%%%%%%%% Fig %%%%%%%%%%%%%%%%%%%%%%%%%%
%%%%%%%%%%%%%%%%%%%%%%%%%%%%%%%%%%%%%%%%%%%%%%%%%%

\subsubsection{Drift Cells}
%\quad\\
Square cells have been used for many small-cell drift chambers in particle physics experiments. The cell aspect ratio, {\it i.e.} ratio of cell width to height is around 1 in this case. 
The electric field inside a square drift cell is more symmetric and homogeneous than that in those cells with an aspect ratio other than 1, and square cells in the same layer can be easily arranged at the same radius. The square shape is adopted for drift cells in the MDC baseline design. 

%, and this is the main reason why it is selected as the baseline design.

\begin{figure*}[htb]
    \centering
{
        \includegraphics[width=70mm]{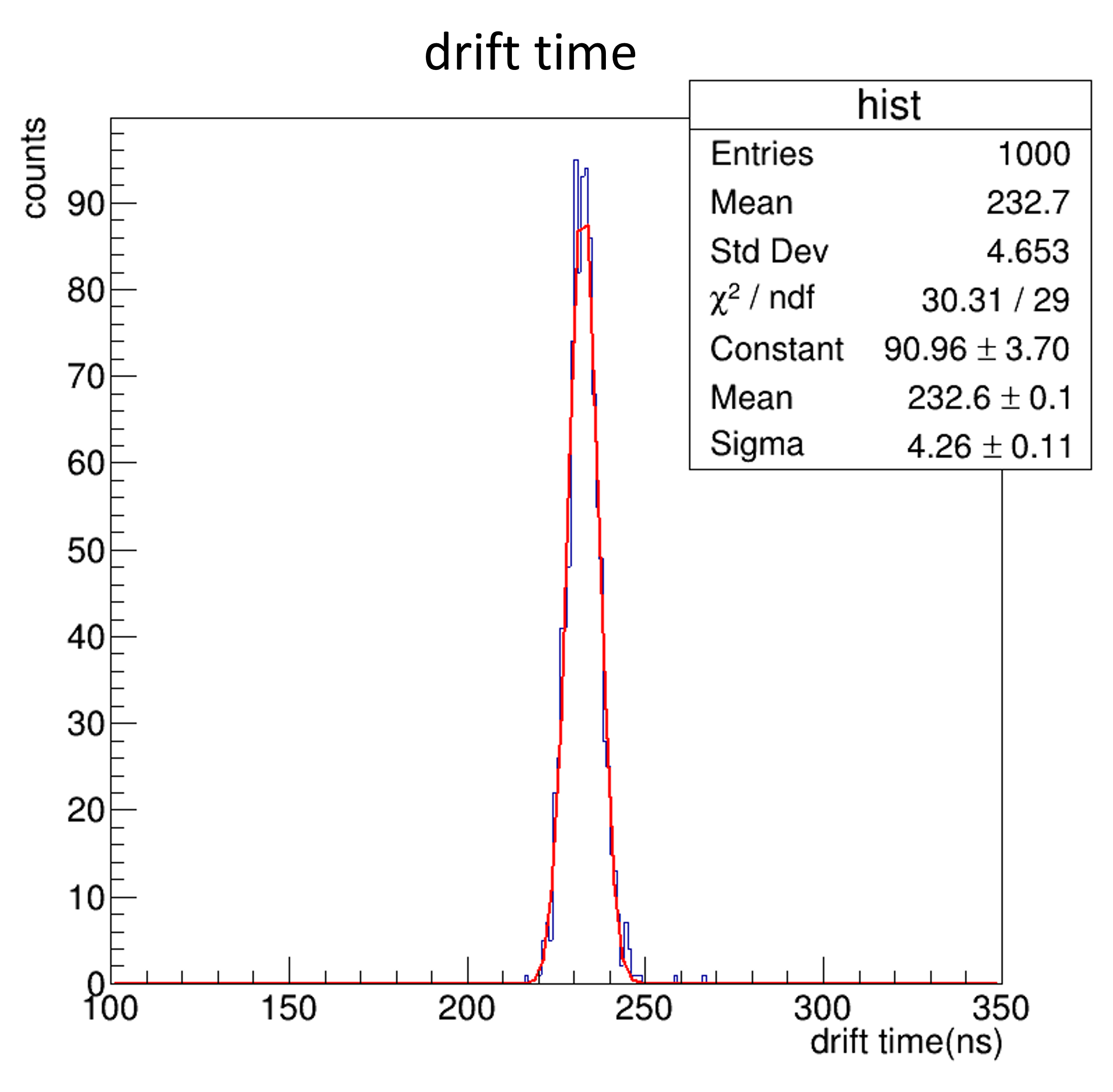}
}
\hspace{2 mm}    
{
        \includegraphics[width=70mm]{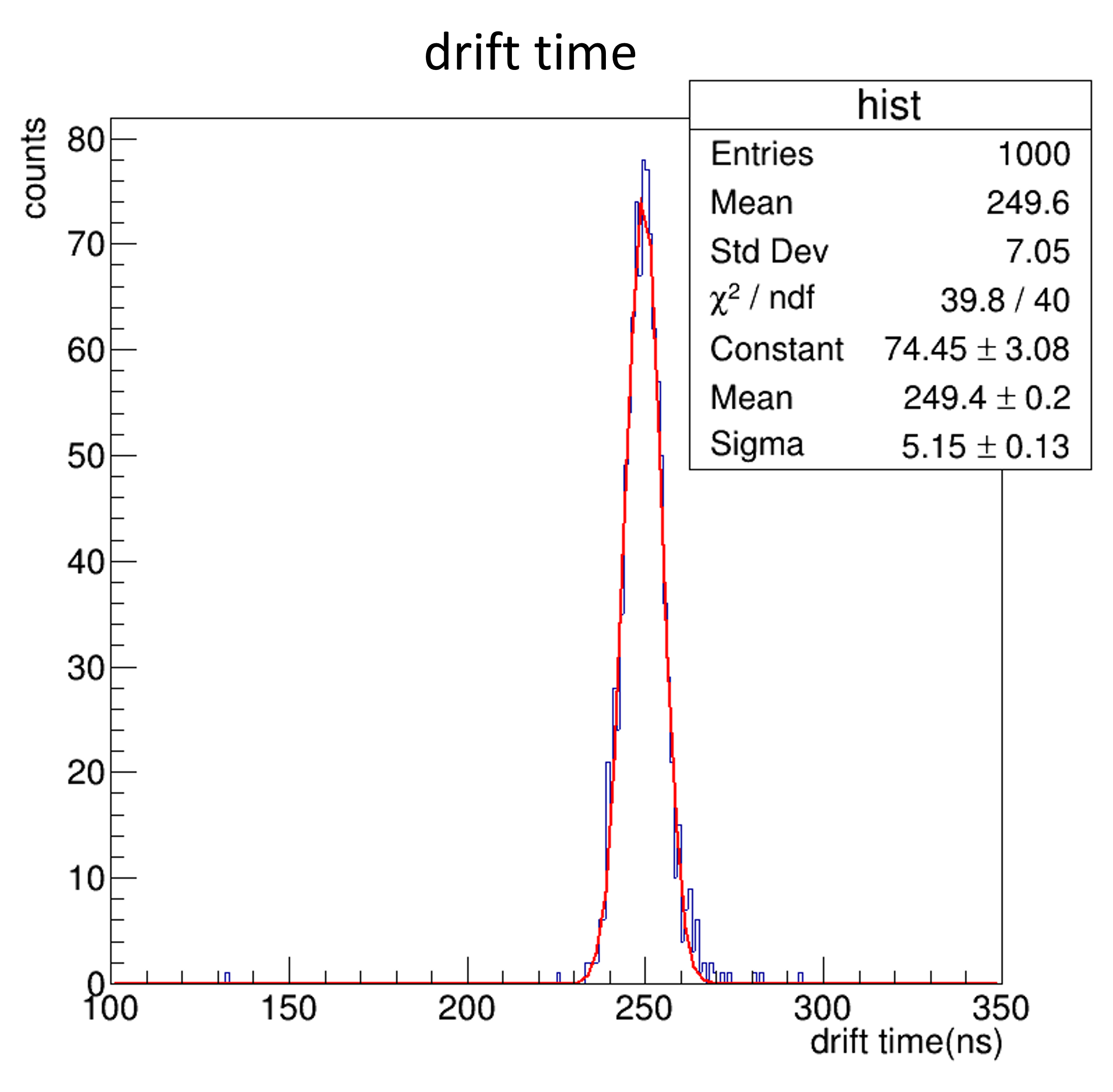}
}
\vspace{0cm}
\caption{ The simulated drift time for  particles entering the drift cell at a distance of half cell width from the sense wire and at a polar angle of 45 degree, with the  cell aspect ratio of 1 (left) and 1.1 (right), respectively. }
    \label{fig:4.2.05}
\end{figure*}

The cell aspect ratio affects the MDC drift time distribution directly. Fig.~\ref{fig:4.2.05}  compares the drift time simulated for two different cell aspect ratios. In the simulation, the incident particle passed a drift cell at 45 degree polar angle and a distance of half of the cell width from the sense wire. The drift time for the cell aspect ratio of 1 has sizably smaller spread compared to that for a cell aspect ratio of 1.1, suggesting a better spatial resolution.

%%%%%%%%%%%%%%%%%%% Fig %%%%%%%%%%%%%%%%%%%%%%%%%%
%%%%%%%%%%%%%%%%%%%%%%%%%%%%%%%%%%%%%%%%%%%%%%%%%%

\subsubsection{Layer Arrangement}
\begin{figure*}[htb]
    \centering
{
        \includegraphics[width=70mm]{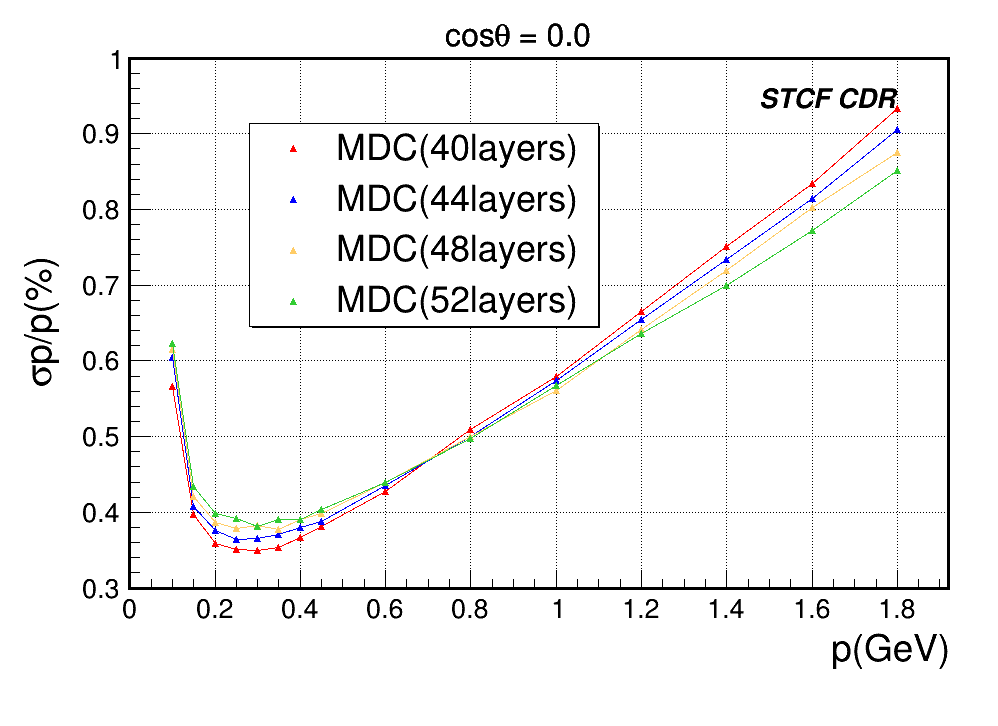}
}
\hspace{2 mm}   
{
        \includegraphics[width=70mm]{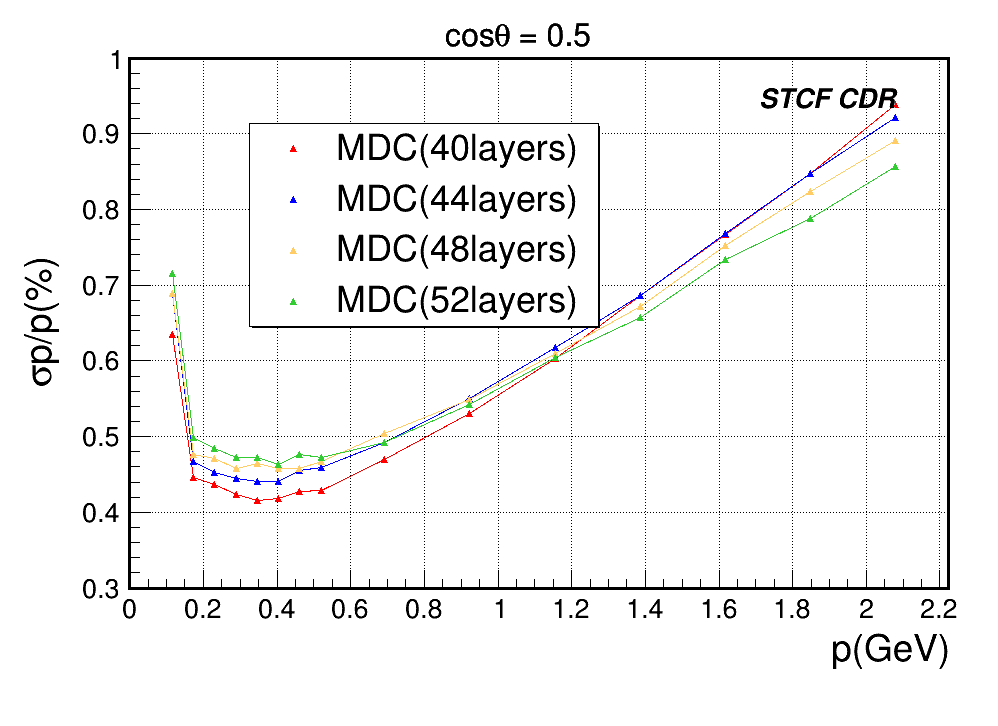}
}
\vspace{0cm}
\caption{The simulated transverse momentum resolution with different numbers of layer, with polar angles of cos$\theta$=0 (left) and cos$\theta$=0.5 (right).}
    \label{fig:4.2.06}
\end{figure*}

%{Optimization of the Layers Layout}
%\quad\\
In the design of the layer layout, the primary parameter is the number of layers. At the STCF, the inner radius of the MDC is approximately 200~mm, and the outer radius is approximately 850~mm, which allows 40 to 52 layers of cells to achieve the required spatial resolution and momentum resolution with an acceptable detector complexity.
Fig. \ref{fig:4.2.06} illustrates the transverse momentum resolution with various layer layouts at different incident polar angles, indicating that the range of 40 to 52 layers in the MDC exhibits tiny differences. Considering the radial/azimuthal cell width ratio of 1, a larger layer number results in a smaller cell size, which is beneficial, with higher count rate tolerance and decreased drift time. A greater number of layers could also improve the $dE/dx$ performance, with a larger number of hits. However, the cost will also be higher, and the increase in the material budget would degrade the momentum resolution of charged particles, especially for low-momentum tracks.
As a compromise, it is decided to use 48 layers.

%%%%%%%%%%%%%%%%%%% Fig %%%%%%%%%%%%%%%%%%%%%%%%%%
%%%%%%%%%%%%%%%%%%%%%%%%%%%%%%%%%%%%%%%%%%%%%%%%%%

\subsubsection{Working Gas Choices}
\begin{figure*}[htb]
    \centering
{
        \includegraphics[width=0.6\textwidth]{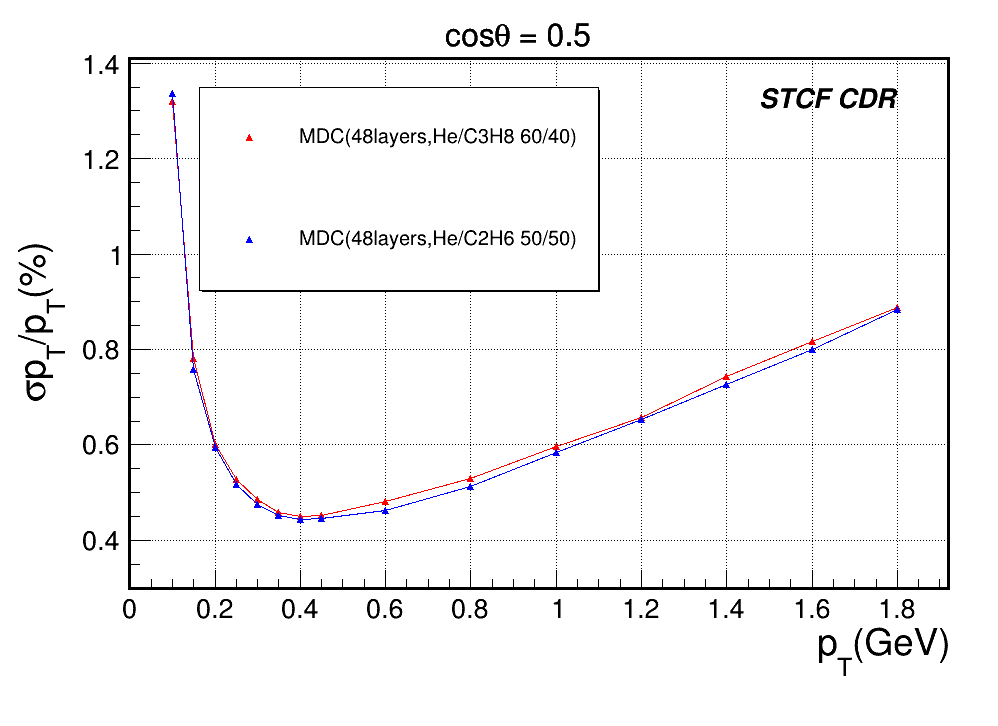}
}
\vspace{0cm}
\caption{The simulated resolution of the transverse momentum of the MDC with different working gases, with the polar angle of cos$\theta=0.5$.}
    \label{fig:4.2.03}
\end{figure*}

%\quad\\
The choice of working gas is essential for the performance of the MDC, affecting the time resolution, maximum count rate and other detector performance aspects.
The ideal working gas of the MDC should have a low material budget, fast drift velocity of ionized electrons and strong primary ionization for penetrating charged particles.
Mixtures of He/C$_{2}$H$_{6}$ (50/50) and He/C$_{3}$H$_{8}$ (60/40) were used in the drift chambers of BESIII and Belle II, respectively.
To find the optimal working gas choice, the characteristic parameters of different gas mixtures are calculated and compared, as summarized in Table~\ref{tab:4.2.02}.
It can be seen that the argon-based gas mixture has a small radiation length and large primary ionization power due to the high density, which leads to a larger material budget. The helium-based working gas has more balanced parameters, while the components of the gas mixture have a significant influence on the performance.
Fig. \ref{fig:4.2.03} illustrates the simulated transverse momentum resolution with different choices of working gas. The simulation is performed with the combined inner tracker-MDC tracking system at a polar angle of cos$\theta$=0.5. The result indicates that with He/C$_{2}$H$_{6}$ (50/50), the MDC has slightly better momentum resolution. However, He/C$_{3}$H$_{8}$ (60/40) has a significantly better particle stopping ability, resulting in a higher signal amplitude and better SNR. Additionally, other simulations demonstrate that the MDC with He/C$_{3}$H$_{8}$ (60/40) has a better position resolution and $dE/dx$ resolution than that with He/C$_{2}$H$_{6}$ (50/50).
As a compromise, the He/C$_{3}$H$_{8}$ (60/40) gas mixture is chosen as the working gas for the conceptual design of the STCF MDC.

%%%%%%%%%%%%%%%%%  TABLE  %%%%%%%%%%%%%%%%%%%%%%%%
\begin{table*}[htb]
\small
    \caption{The main parameters of several kinds of gas mixtures, pressure = 1 atm, temperature = 20 Celsius, magnetic field strength = 1~T .}
    \label{tab:4.2.02}
    \vspace{0pt}
    \centering
    \begin{tabular}{llllll}
        \hline
        \thead[l]{Gas Mixture} & \thead[l]{Ar/CO$_{2}$/CH$_{4}$\\(89/10/1)}&\thead[l]{He/CH$_{4}$\\(60/40)} & \thead[l]{He/C$_{2}$H$_{6}$\\(50/50)}&\thead[l]{He/C$_{3}$H$_{8}$\\(60/40)}&\thead[l]{He/iC$_{4}$H$_{10}$\\(80/20)}\\
        \hline
            Drift velocity of an electron & 5.0 & 3.7 & 4.0 & 3.8 & 3.4 \\
            v$_{d}$ (cm/$\mu$s) &&&&&\\
            Transverse diffusion coefficient & 233 & 191 & 170 &154 & 159 \\
            $\sigma$$_{L}$ ($\mu$m/$\sqrt {cm}$) @E=760 V/cm &&&&&\\
            Lorentz angle &41 &28 &29 &24 &21 \\
            $\theta$$_{L}$ (degree) @E=760 V/cm &&&&&\\
            Primary ionizing power & 30 & 10 & 23 & 30 & 21 \\
            (i.p./cm) &&&&&\\
            Radiation length& 124 & 808 & 640 & 550 & 807 \\
            (m) &&&&&\\
        \hline
    \end{tabular}
\end{table*}
%%%%%%%%%%%%%%%%%%%%%%%%%%%%%%%%%%%%%%%%%%%%%%%%%%

%%%%%%%%%%%%%%%%%%% Fig %%%%%%%%%%%%%%%%%%%%%%%%%%
%%%%%%%%%%%%%%%%%%%%%%%%%%%%%%%%%%%%%%%%%%%%%%%%%%

\FloatBarrier

\subsection{Expected Performance}
\subsubsection{Momentum and Spatial Resolution}
\begin{figure*}[htb]
	\centering
\subfloat[][]{\includegraphics[height=50 mm]{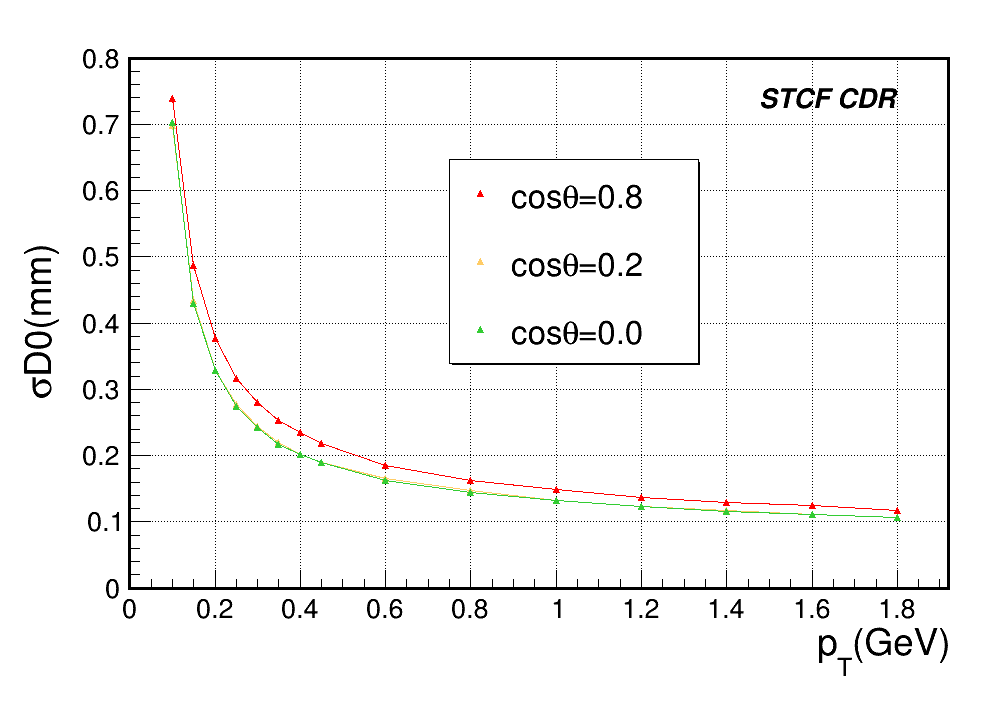}}
\subfloat[][]{\includegraphics[height=50 mm]{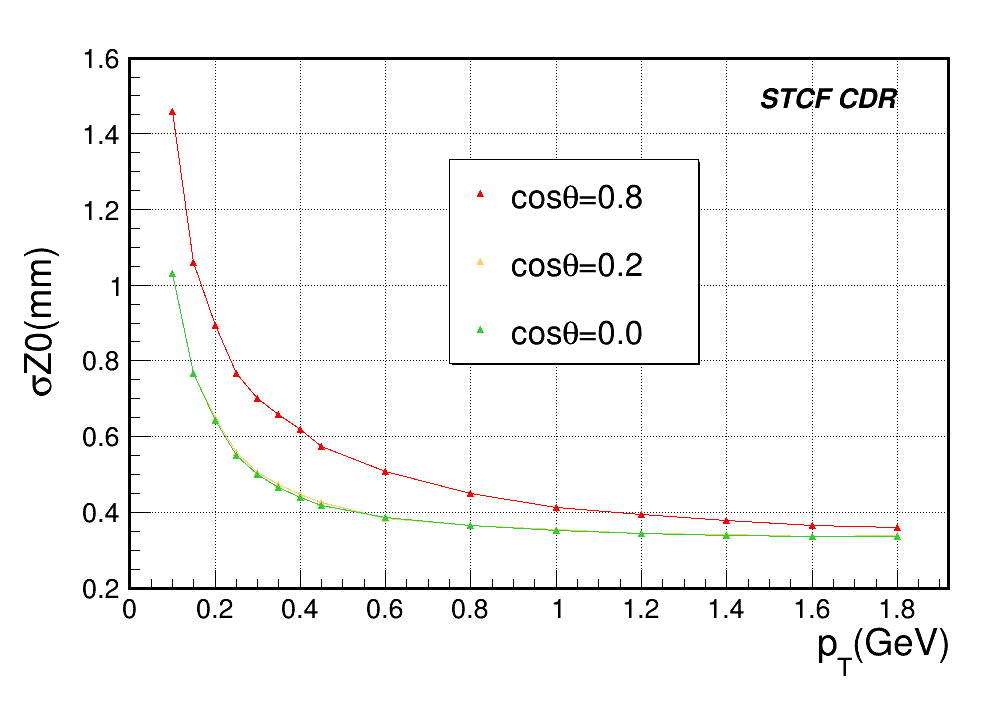}} \\
\subfloat[][]{\includegraphics[height=50 mm]{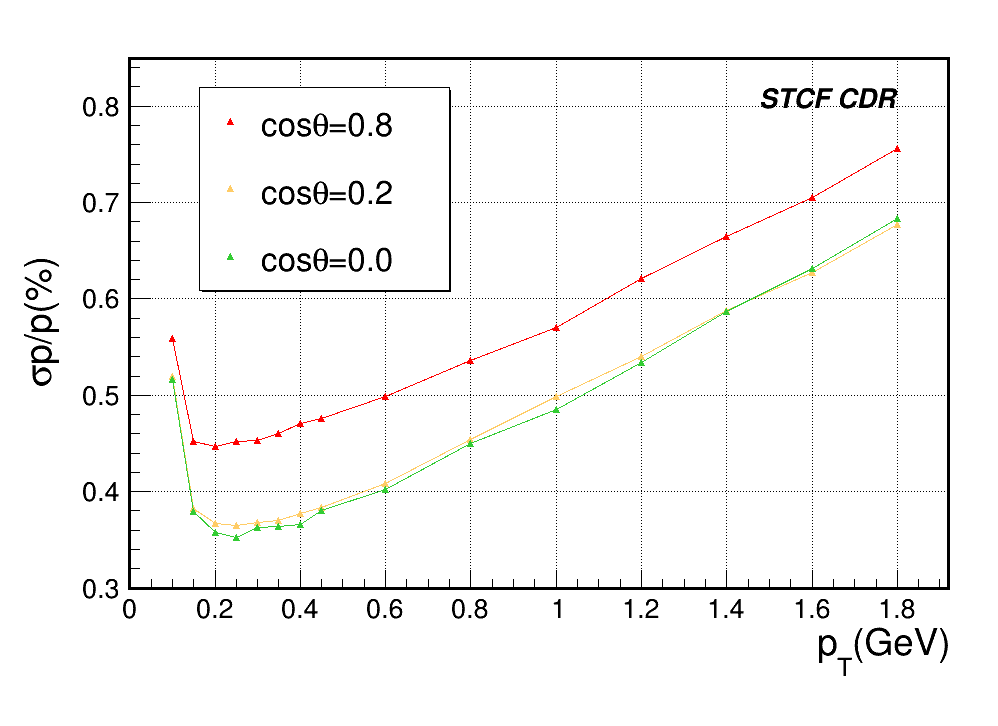}}
\subfloat[][]{\includegraphics[height=50 mm]{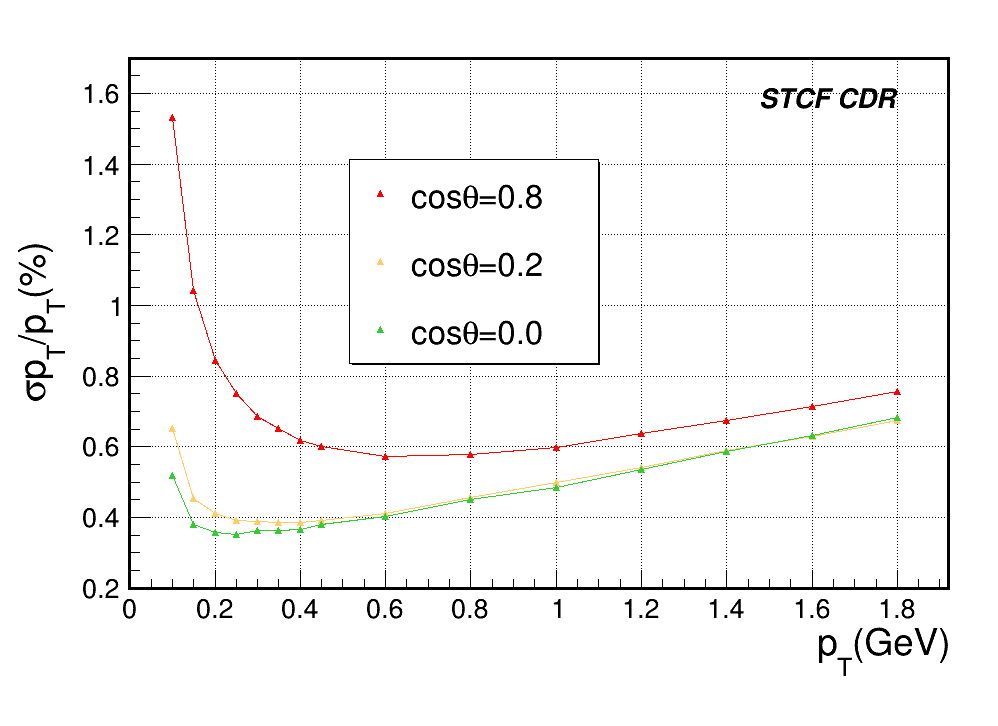}}
\vspace{0cm}
\caption{The simulated resolution of the impact parameters (a) $d_0$ and (b) $z_0$ and (c) momentum $p$ and (d) transverse momentum $p_T$ as a function of $p_T$. The results with different polar angles of incident particles, with cos$\theta$=0, 0.2 and 0.8, are compared.
}
    \label{fig:4.2.07}
\end{figure*}

%\subsubsection{Expected performance of baseline tracking system}
%\quad\\
The expected performance of the combined tracking system, with the 3-layer ITK and the 48-layer MDC, is evaluated, especially the momentum resolution at different polar angles.
The expected tracking performance results are obtained using a Geant4 simulation, without considering the background contribution, detector signal readout digitization or track reconstruction, and track fitting is then performed.
%A comparison of the MDC only and ITK-MDC tracking system is also performed.
Fig.~\ref{fig:4.2.07} shows the simulated results on the resolution of impact parameters and momentum resolution.
A minimum $\sigma$$_{p}/p$ can reach approximately 0.35\% for $\pi$ particles at $p = 0.2$~GeV/c with a polar angle of cos$\theta=0$.

%%%%%%%%%%%%%%%%%%% Fig %%%%%%%%%%%%%%%%%%%%%%%%%%
%%%%%%%%%%%%%%%%%%%%%%%%%%%%%%%%%%%%%%%%%%%%%%%%%%

\begin{figure*}[htb]
	\centering
\subfloat[][]{\includegraphics[height=50 mm]{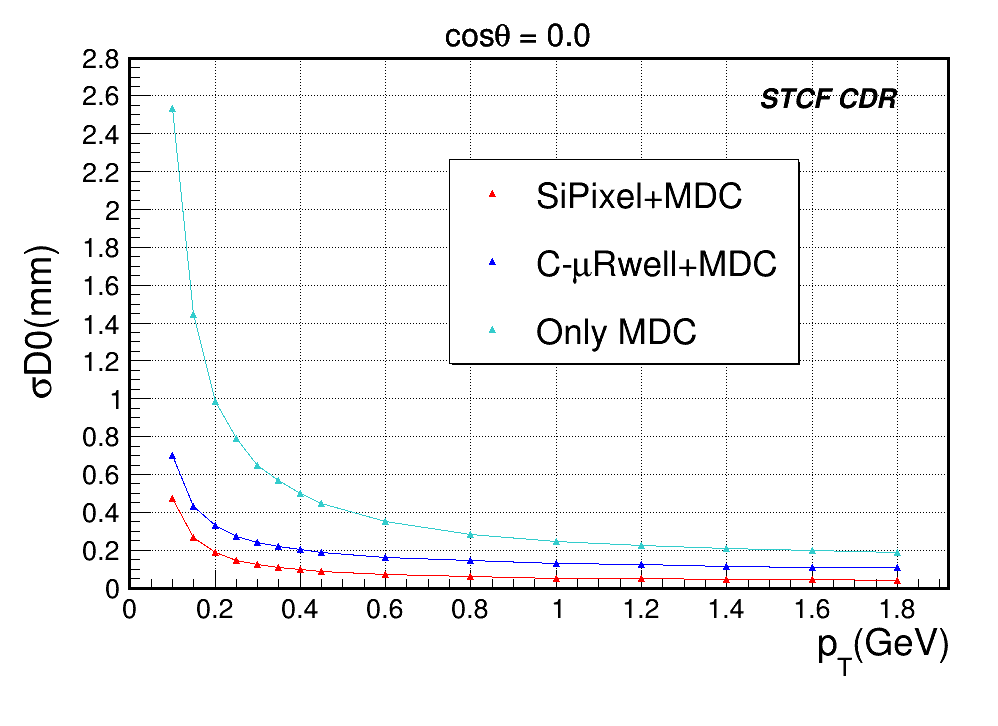}}
\subfloat[][]{\includegraphics[height=50 mm]{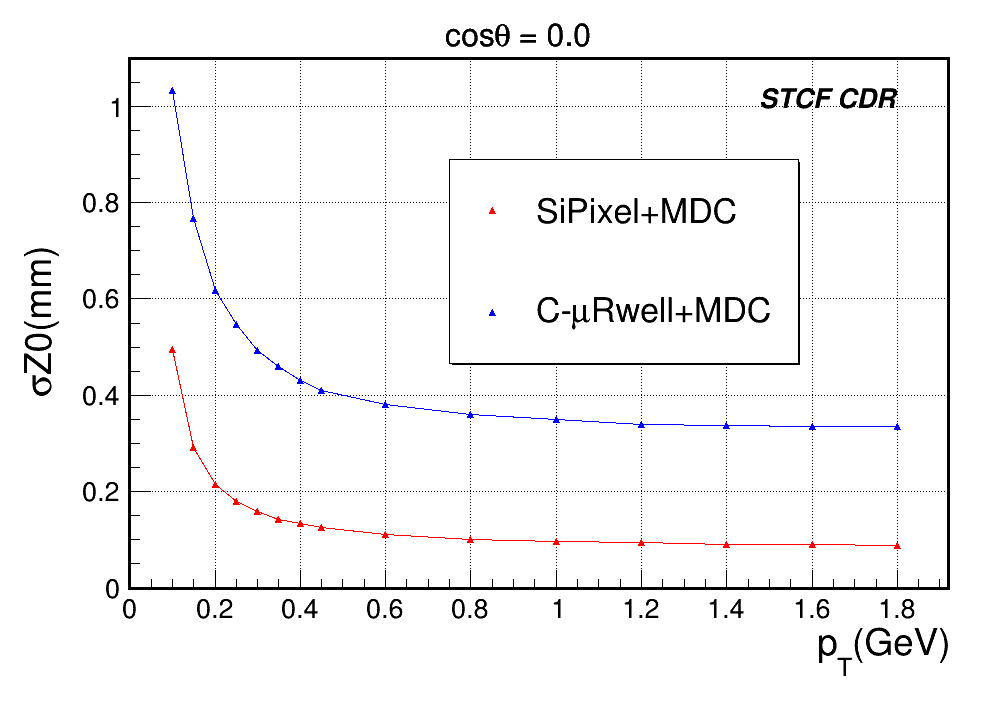}} \\
\subfloat[][]{\includegraphics[height=50 mm]{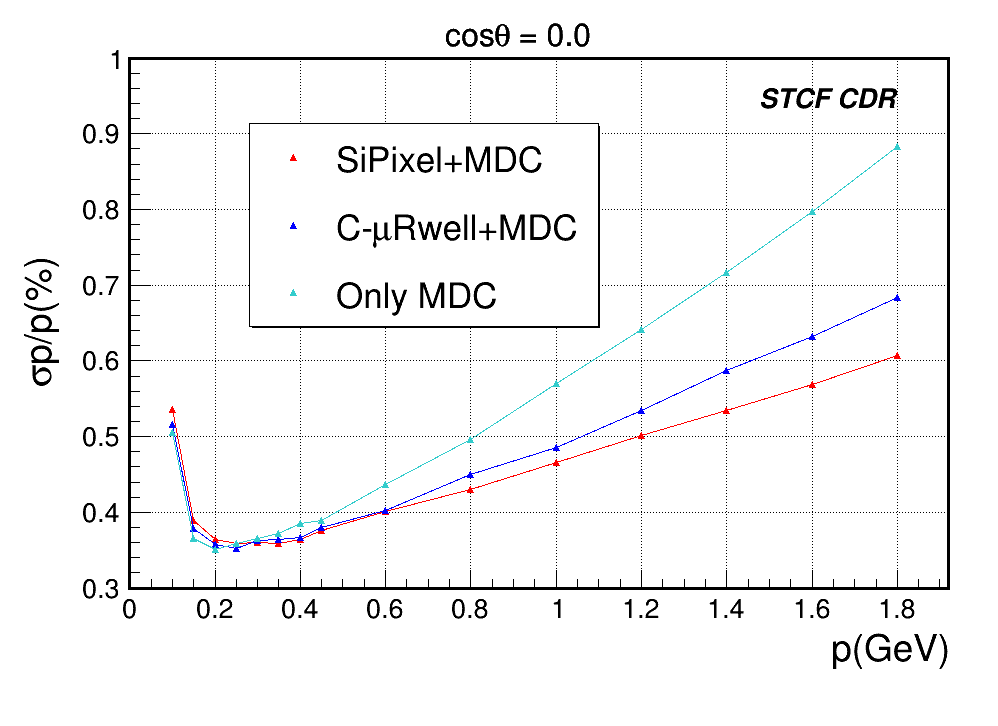}}
\subfloat[][]{\includegraphics[height=50 mm]{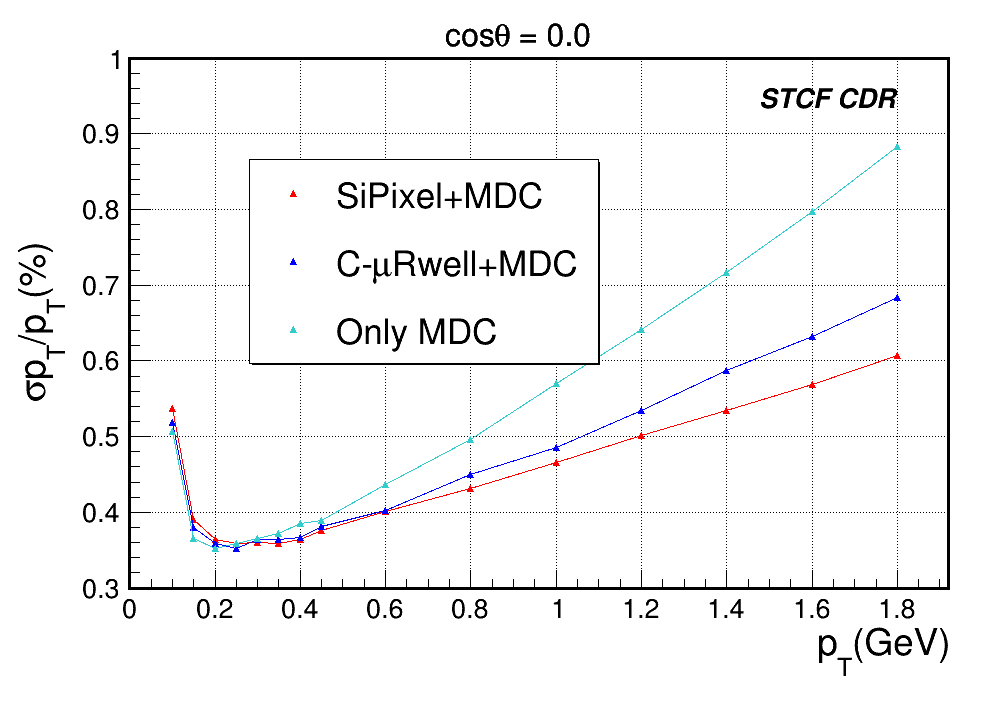}}
\vspace{0cm}
\caption{The simulated resolution of the impact parameters (a) $d_0$ and (b) $z_0$ and (c) momentum $p$ and (d) transverse momentum $p_T$ as a function of the $p_T$ of an incident particle with a polar angle of cos$\theta$=0. The results with different ITK designs are compared.
For the comparison of the $z_0$ resolution, the results for the MDC-only option are not shown since the design of the MDC alone cannot provide precise $z_0$ measurements.
}
    \label{fig:4.2.08}
\end{figure*}

%\subsubsection{Comparison of the performance of the ITK designs}
As described in Sec.~\ref{sec:itk}, the baseline design of the STCF ITK is a cylindrical $\mu$RWELL-based detector, and an alternative design using CMOS pixel sensors is also considered.
To evaluate the performance of different designs, the entire tracking system, with the MDC and the ITK combined, should be considered in the {\sc Geant4} simulation and track fitting.
Single hit position resolutions of $100\times400$~$\mu$m and $30\times75$~$\mu$m are assumed for the $\mu$RWELL detector and PXD, respectively.
A material budget of 0.25\%$X_0$ is assumed for both ITK designs.
Fig.~\ref{fig:4.2.08} shows the comparison of the expected performance of the two different tracking system designs.
As expected, for the spatial resolution, the PXD ITK gives much better performance than the $\mu$RWELL-based ITK, while for the momentum resolution, the two different ITK designs give similar detector performance in the low momentum range since the material budget is the limiting factor.
The results also indicate that the inclusion of the inner tracker improves the momentum resolution at p$_{T}$ $>$ 0.3~GeV/c. When the momentum of the incident particle is below 0.2~GeV/c and the polar angle is close to the zenith direction, the inner tracker exerts a negative influence due to the impact of the additional material budget. In general, the addition of the inner tracker can benefit the momentum resolution of the tracker system.

%%%%%%%%%%%%%%%%%%% Fig %%%%%%%%%%%%%%%%%%%%%%%%%%
%%%%%%%%%%%%%%%%%%%%%%%%%%%%%%%%%%%%%%%%%%%%%%%%%%

\subsubsection{$dE/dx$ Resolution and PID Performance}
The $dE/dx$ measurement provided by the MDC can be used for particle identification, especially for low momentum charged particles.
To explore the PID potential of the MDC, the $dE/dx$ measurement of the MDC is simulated for various types of particles.
In the simulation, the {\sc GEANT4 PAI} model is used to model the primary ionization process of high-energy charged particles with the gas medium in the MDC.
The {\sc HEED} module of the {\sc GARFIELD++} package~\cite{heed} is then invoked to produce the secondary ionization caused by the very energetic electrons produced in the primary ionization process. All ionization electrons are fed into the {\sc GARFIELD++} for simulation of the electron avalanche multiplication process.
Given the computationally intensive nature of this simulation, a fast approach is adopted here instead of simulating electron avalanche multiplication using {\sc GARFIELD++} for every ionization electron.
In this approach, the distribution of the total charge induced on a sense wire due to the avalanche multiplication process of a single electron was first obtained by performing the full {\sc GARFIELD++} simulation for many single electrons.
The charge of the induced signal due to each of the ionization electrons is then generated by sampling this distribution.
The simulated $dE/dx$ measurement by a drift cell is finally taken as the sum of the induced charge on the sense wire for all ionization electrons produced in the cell and normalized by the track length in the cell.
The fluctuations of the avalanche multiplication process are fully included in the simulation by this approach.

\begin{figure*}[htb]
    \centering
{
\subfloat[][]{\includegraphics[width=0.45\textwidth]{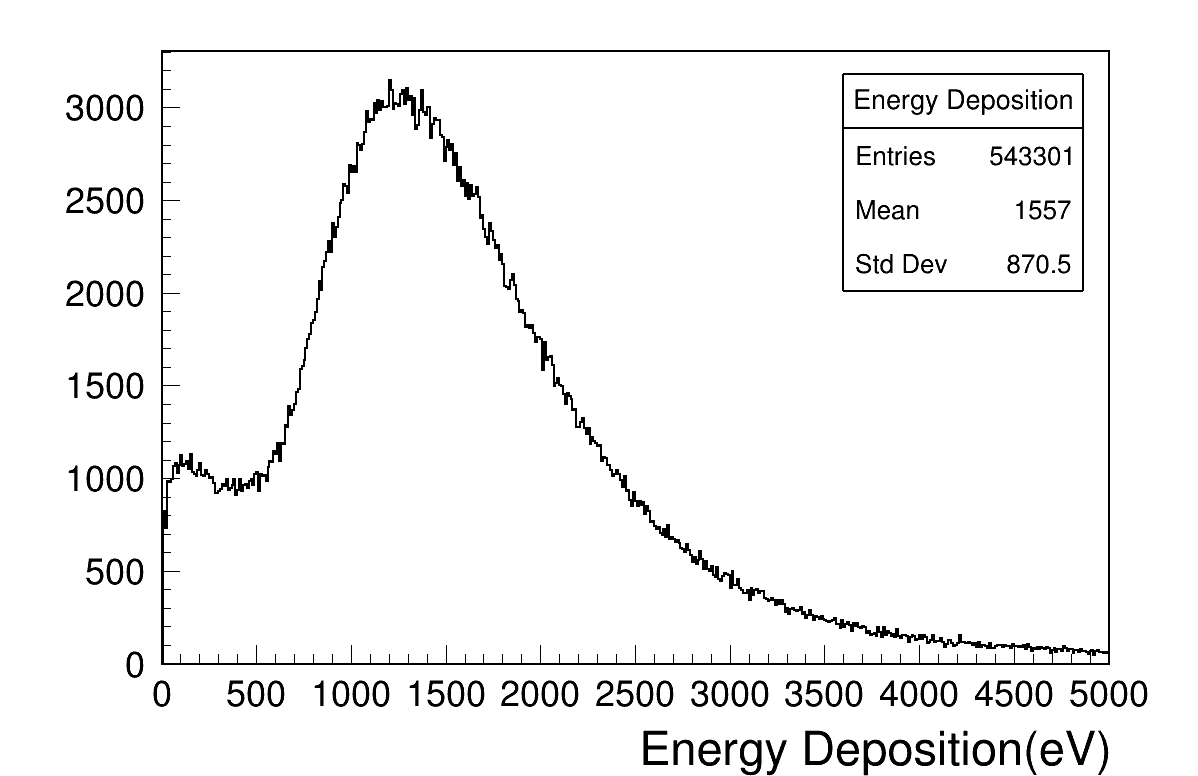}}
\subfloat[][]{\includegraphics[width=0.45\textwidth]{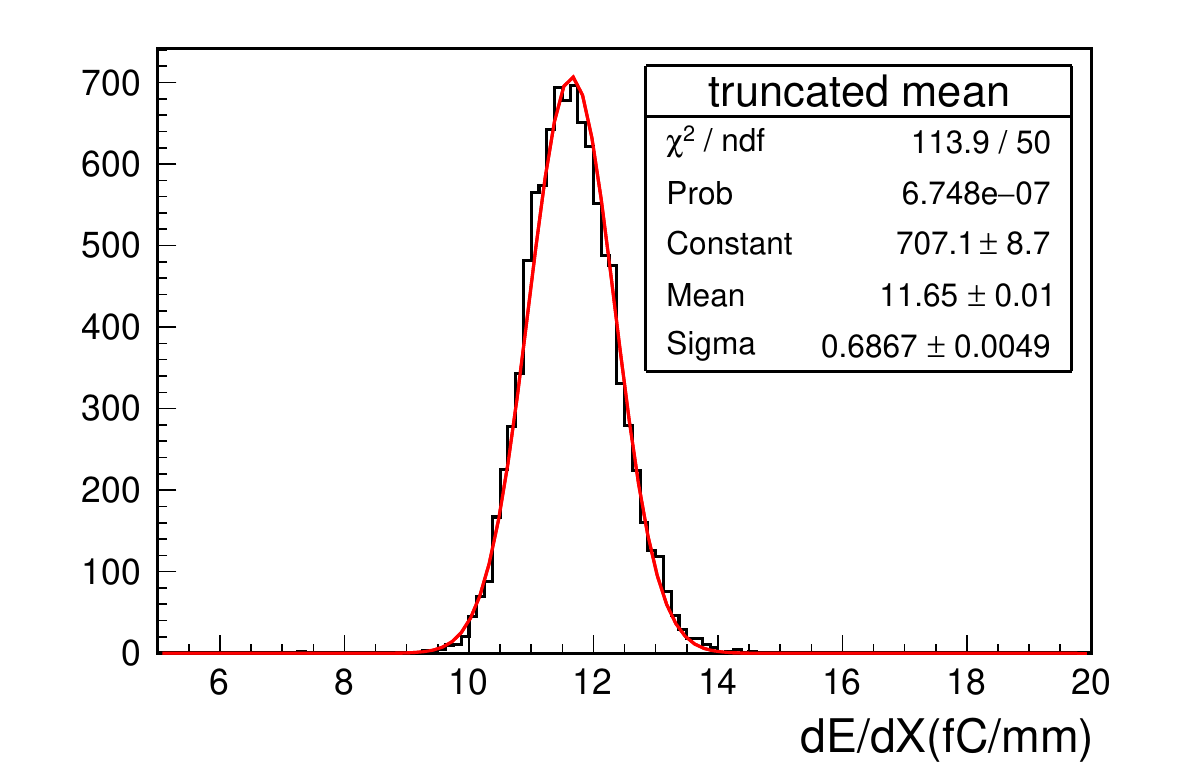}}
}
\vspace{0cm}
\caption{(a) The distribution of the original $\dedx$ in one MDC cell, with penetrating $\pi$ particles with $p = 0.5$~GeV/c.
(b) The calculated $dE/dx$ resolution of $p = 0.5$~GeV/c $\pi$ with the truncated average method.}
    \label{fig:4.2.14}
\end{figure*}

Fig.~\ref{fig:4.2.14}~(a) shows the $\dedx$ distribution within a single MDC cell by $\pi$ particles at $p = 0.5$~GeV/c.
A truncated-average method~\cite{dedx} is used in $dE/dx$ estimation to reduce the impact of the Landau tail of the ionization energy loss on the $dE/dx$ estimation, hence improving the $dE/dx$ resolution.
In this method, 25\% of the cell hits with the highest energy deposition are removed, and only the remaining 75\% are used to calculate the average $dE/dx$ of the track.
%Figure~\ref{fig:4.2.13} shows the $dE/dx$ distribution for $\pi$ particles at $p = 0.5$~GeV/c, with and without the truncated average method.
The simulation results indicate that the truncated average method can effectively improve the $dE/dx$ resolution.
The simulated $dE/dx$ resolution is approximately 5.89\% and smaller than the requirement of 6\%, as shown in Fig.~\ref{fig:4.2.14}.

%%%%%%%%%%%%%%%%%%% Fig %%%%%%%%%%%%%%%%%%%%%%%%%%
%%%%%%%%%%%%%%%%%%%%%%%%%%%%%%%%%%%%%%%%%%%%%%%%%%

\begin{figure*}[htb]
    \centering
\subfloat[][]{\includegraphics[width=0.45\textwidth]{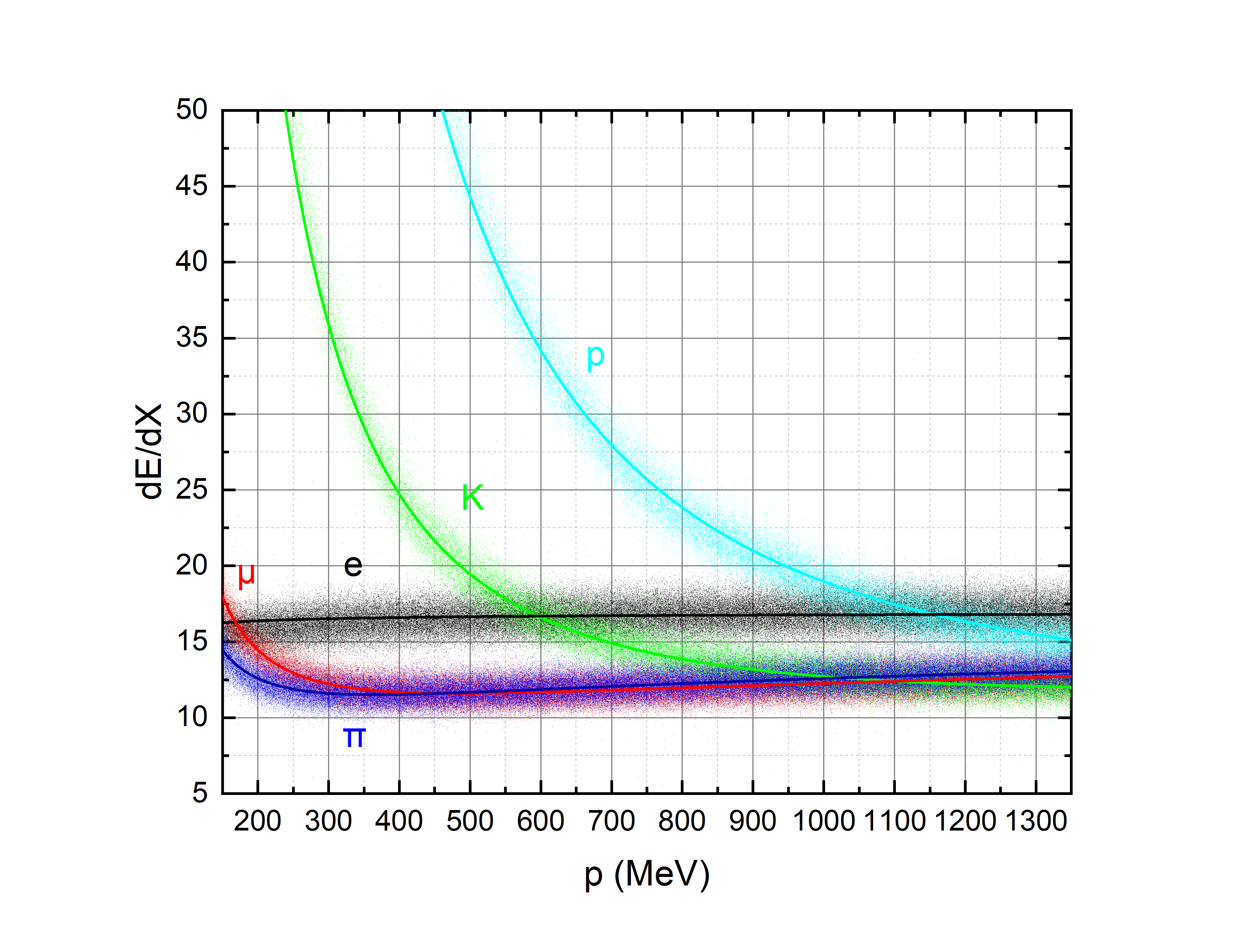}}
\subfloat[][]{\includegraphics[width=0.45\textwidth]{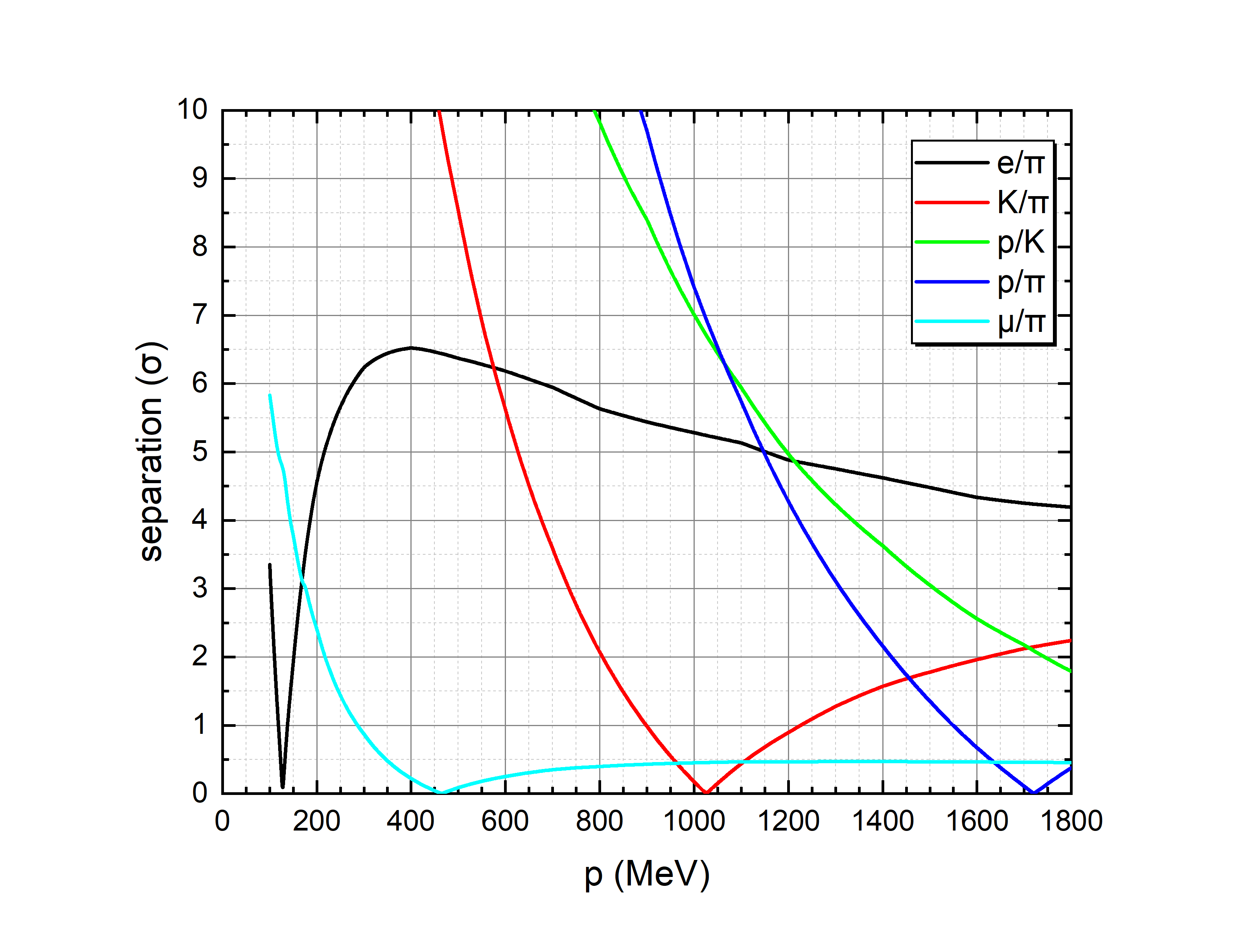}}
\vspace{0cm}
\caption{The simulated relationship between $dE/dx$ and momentum with various particles (left) and the simulated PID performance of the MDC (right).}
    \label{fig:4.2.16}
\end{figure*}

\begin{figure*}[htb]
    \centering
    {
        \includegraphics[height=60mm]{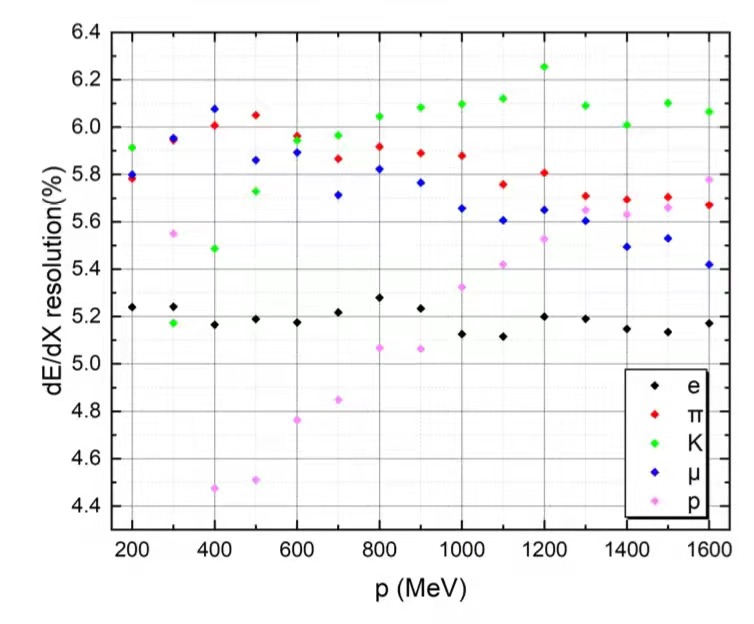}
    }
    \vspace{0cm}
	\caption{The simulated relationship between momentum and $dE/dx$ resolution for various particles.}
    \label{fig:dedxresolution_p}
\end{figure*}

The truncated mean of $dE/dx$ is simulated as a function of the momentum for different particle species, as shown in Fig.~\ref{fig:4.2.16}~(left), while the $dE/dx$ PID separation power is also extracted from the simulation, as shown in Fig.~\ref{fig:4.2.16}~(right).
The $dE/dx$ PID separation power for two particles A and B is defined as follows:
\begin{equation}
S_{AB} = \frac{\dedx_{A}-\dedx_{B}}{\sigma_{\dedx(AB)}},
\end{equation}
where $\dedx_{A}$($\dedx_{B}$) is the $\dedx$ for particle A(B) and $\sigma_{\dedx(AB)}$ is the average resolution (defined as the RMS of the $\dedx$ distribution for a given type of particle) of the two particles.
Fig.~\ref{fig:dedxresolution_p} presents that the $dE/dx$ resolution of most of the particles is below 6\%, except for MIP particles.
Fig.~\ref{fig:4.2.16} shows the simulated PID performance of the MDC, with five scenarios of hypothesized signals and background particles.
It can be seen that in the low momentum region, the MDC can achieve good particle separation power.
For example, for $K/\pi$ ($p/\pi$), the MDC PID separation power is over 3 $\sigma$ up to 700 MeV/c (1300 MeV/c).

%%%%%%%%%%%%%%%%%%%% Fig %%%%%%%%%%%%%%%%%%%%%%%%%%

%%%%%%%%%%%%%%%%%%%%%%%%%%%%%%%%%%%%%%%%%%%%%%%%%%%

%%%%%%%%%%%%%%%%%%% Fig %%%%%%%%%%%%%%%%%%%%%%%%%%
%%%%%%%%%%%%%%%%%%%%%%%%%%%%%%%%%%%%%%%%%%%%%%%%%%

\FloatBarrier
\subsection{Pileup and Radiation Effects}
\label{sec:mdc_pileup_effect}
Given that the STCF detector operates with very high luminosity, additional challenges in the design of the MDC detector must be taken into account,
such as pileup and radiation effects.
% pileup
\subsubsection{Pileup Effects}
\begin{figure*}[htb]
    \centering
{
        \includegraphics[height=60mm]{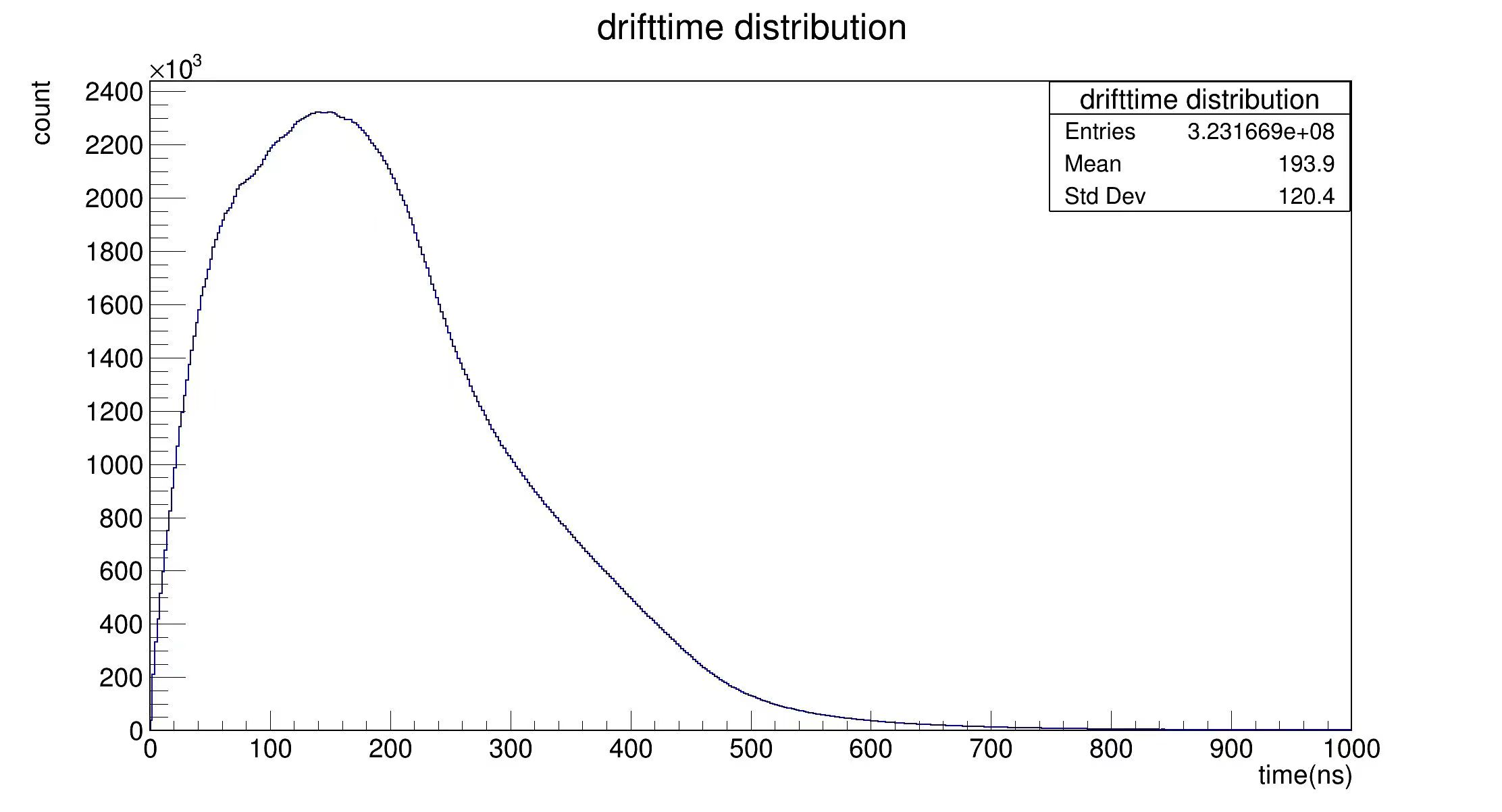}
}
\vspace{0cm}
\caption{Drift time distribution in the STCF MDC.}
    \label{fig:4.2.18}
\end{figure*}

From Table~\ref{tab:TIDNIEL_max}, the expected hit rate of the MDC is approximately $4\times 10^5$~Hz/channel for the innermost layer. This is an extremely high hit rate for the MDC given the maximum drift time in its drift cell being about 250 ns and the induced signal spreading over 500 ns (as shown in Fig.~\ref{fig:4.2.18}). There is a high probability of signals overlapping in one channel. This poses big challenges to readout electronics. The MDC counting rate may have to be reduced by modifying the MDC design depending on further studies.  

%%%%%%%%%%%%%%%%%%% Fig %%%%%%%%%%%%%%%%%%%%%%%%%%
%%%%%%%%%%%%%%%%%%%%%%%%%%%%%%%%%%%%%%%%%%%%%%%%%%

The STCF detector is expected to operate at an event rate up to 400~kHz (Sec.~\ref{sec:tdaq}). At such a high event rate, the probability of events piling up is approximately 8 (18)\% within a time window of 200 (500) ns. This poses a severe problem for the MDC which could have drift time spread over a few hundred ns (as shown in Fig.~\ref{fig:4.2.05}). Such a time spread implies a rather large integration time window for the MDC and hence a significantly high probability of tracks from different events overlapping in the MDC. One way to resolve the overlapping tracks is to exploit the timing capability of the STCF detector. In the current STCF detector design, both the CMOS-ITK and EMC have timing capability at different levels of precision, with the former being able to reach a few ns time resolution for charged particles~\cite{aplide-improved} ~\cite{malta2}  and the latter a level of a few hundred ps for energy deposits of ~100 MeV (see discussions in Sec.~\ref{sec:emc_perf}). The time measurements by the two subdetectors can be associated to each track recorded by the MDC by spatial matching of hits between the MDC and either the MAPS-ITK or EMC or both depending on availability of time measurements on the two subdetectors. The MDC tracks can then be assigned to different events according to their associated time measurements. As a result, the overlapping events recorded by the MDC are resolved. Further studies are needed to investigate the event overlapping or pileup problem with the MDC.

% radition and aging effects
\subsubsection{Radiation Effects}
The detailed background radiation simulation is described in Sec.~\ref{sec:mdi_bkg}, and the expected radiation levels in each subdetector are given in
Table~\ref{tab:TIDNIEL_mean} and \ref{tab:TIDNIEL_max}.
From Table~\ref{tab:TIDNIEL_max}, the MDC needs to withstand a TID of approximately 60~Gy/y and a NIEL of $4.9\times 10^{10}$ n/cm$^{2}$/y~(1 MeV neutron equivalent). Also, the highest accumulated charge can be calculated as approximately 24.2 mC/cm/y in the innermost layer. 
Therefore, aging effects in wire chambers, a permanent degradation of the operating characteristics under sustained irradiation, must be considered.
The classical aging effects are the results of chemical reactions occurring in avalanche plasma near anodes in wire chambers, leading to the formation of deposits on electrode surfaces~\cite{gaseousaging}.
For the MDC, the aging effect includes those from both anode aging and cathode aging, expected a further study in the future.

\FloatBarrier
\subsection{Readout Electronics}
%\quad\\
In the conceptual design, the MDC contains 11520 cells, each with a sense wire in the center. The total number of MDC readout electronic channels is 11520 as well. The readout electronics system is arranged in the endcap that is made of a 15 mm-thick aluminum plate for mechanical support.
The background simulation (Sec.~\ref{sec:bkg_sim}) indicates that the first (innermost) layer of the MDC is subjected to the highest background count rate, approximately 200 to 400 kHz, leading to severe interference in the measurement of charged tracks from signal events. The {\sc Garfield}~\cite{garfield} simulation demonstrates that the signal in each cell lasts for 200 to 500~ns due to the electrons distributed by the uncertainty of the distance from the particle track to the wire. As a consequence, the readout electronics must have a fast shaping time, and sufficient background shielding in the MDC endcap is necessary.

\begin{figure*}[htb]
    \centering
{
        \includegraphics[width=120mm]{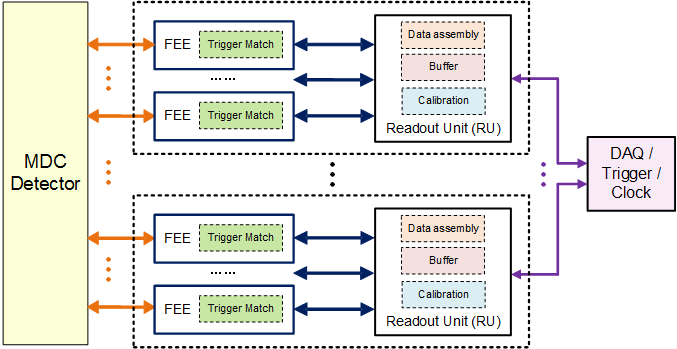}
}
\vspace{0cm}
\caption{Block diagram of the MDC electronics.}
    \label{fig:4.2.09}
\end{figure*}

High precision time (approximately 0.5~ns RMS) and charge measurement are required for the MDC readout electronics, with an input signal amplitude up to 1.8~pC.
The MDC readout electronics are composed of FEE modules and RUs, with the schematic structure shown in Fig. \ref{fig:4.2.09}.
%%%%%%%%%%%%%%%%%%% Fig %%%%%%%%%%%%%%%%%%%%%%%%%%
%%%%%%%%%%%%%%%%%%%%%%%%%%%%%%%%%%%%%%%%%%%%%%%%%%

\begin{figure*}[htb]
    \centering
{
        \includegraphics[width=120mm]{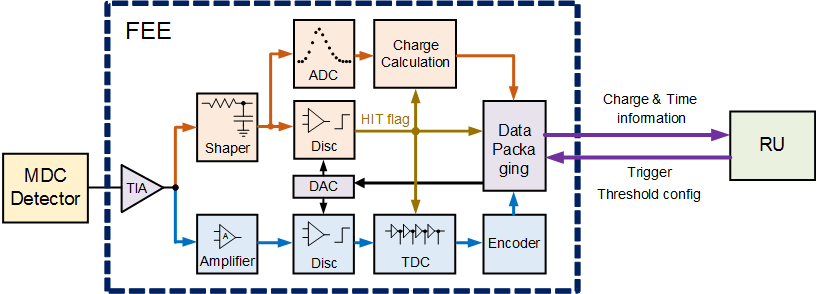}
}
\vspace{0cm}
\caption{Time and charge measurement circuit for the MDC.}
    \label{fig:4.2.10}
\end{figure*}

The FEE are responsible for analog signal manipulation, A/D conversion, and time \& charge measurement.
The circuit structure of the FEE is shown in Fig.~\ref{fig:4.2.10}.
The signals from the MDC detector are first amplified by fast trans-impedance amplifiers (TIAs) to ensure a fast response for time measurement and to provide charge measurement. The output signal from the TIA is then split into two paths: one connects to the charge measurement circuits and the other connects to the time measurement circuits.
%\quad\\
For the time measurement, considering the characteristic shape of the signal, the threshold of the discriminator is set to a low value. To achieve a high time precision, an amplifier located before the discriminator is used to enhance the signal slew rate. To filter out the situations when the noise crosses this low threshold, another high-threshold discriminator is used (after shaping), as shown in Fig.~\ref{fig:4.2.10}. Then, the time-to-digital converter ~(TDC) is used to digitize the time of the leading edge of the low-threshold discriminator.
%\quad\\
The charge measurement circuitry uses a shaping circuit to enhance the SNR, the output waveforms of which are digitized and sent to an FPGA chip for charge calculation. As mentioned above, the output signal of the shaper is fed to a discriminator with a high threshold. The output from this discriminator is used as the flag signal to start the charge calculation process and as a ``valid'' condition for time measurement from the output from the low-threshold discriminator. To suppress the effect of pileup, baseline restoration and digital processing on the signal waveform after digitization will be applied in the electronics design.

%%%%%%%%%%%%%%%%%%% Fig %%%%%%%%%%%%%%%%%%%%%%%%%%
%%%%%%%%%%%%%%%%%%%%%%%%%%%%%%%%%%%%%%%%%%%%%%%%%%

Outputs from the FEE are collected by the RUs, which further assemble these data and transfer them to the DAQ through optical fibers. The MDC readout electronics are required to be synchronized with a system clock signal, which is received by the RUs and then fanned out to the FEE. The RUs also receive the global trigger signal and fan it to the FEE to perform trigger matching to read out valid data.

\subsection{Conclusion}
In summary, an MWDC-based MDC is the baseline for the conceptual design of the main tracking system for the STCF, ensuring the robustness and stability of the whole tracker system.
The MDC consists of 48 layers of drift cells with an inner radius of 200~mm and an outer radius of 850~mm.
To improve its performance, several detector design parameters, including the working gas component, wire parameters, cell structure and layer layout, are optimized via simulation.
The study indicates that the baseline MDC can satisfy the physics requirements of the STCF, with a transverse momentum resolution of $\sigma_{p_T}/p_{T} < 0.5$\%@1~GeV/c. According to the experience of the BESIII experiment, a $dE/dx$ resolution of $\sim$ 6\% can be achieved for this MDC design.

\clearpage
\newpage
\section{Particle Identification in the Barrel (RICH)}
\label{sec:rich}

\subsection{Introduction}
\label{sec:rich_intro}

The particle identification for the full momentum range is essential for charm physics studies and fragmentation function studies. In particular, the precise measurement of the collision fragmentation function requires a $\pi/K$ misidentification rate less than $2\%$ up to $p=2.0\gevc$, with the corresponding identification efficiency being larger than 97\%. In addition, studies of $XYZ$ physics, $\tau$ physics and (semi)leptonic decays of charmed mesons require a good suppression power for $\mu/\pi$.

The identification of hadrons in the low momentum range is achieved through measurements of the specific energy loss rate ($dE/dx$) by the MDC. The identification of leptons and neutral particles is provided by the EMC and the MUD. The PID system of the STCF focuses on charged hadrons with a high momentum, from approximately 0.7\,$\gevc$ up to 2\,$\gevc$. To cover this range, the Cherenkov detector is one of the technologies that can fulfill those requirements.

The PID system of the STCF is placed between the EMC and MDC. The solid angle coverage $\cos\theta$ of the barrel PID is 0.83 and that of the end-cap PID ranges from $0.81$ to $0.93$.

Cherenkov radiation can be used for PID in a wide momentum range in modern high-energy physics experiments through the measurement of the characteristic radiation angle, which depends on the refractive index of the medium and the particle velocity. According to the PID requirements of various particle species and momentum ranges, different kinds of media (commonly called Cherenkov radiators) with different refractive indexes can be chosen to identify particles with momenta ranging from $\gevc$ to several hundred $\gevc$. The methods for realizing the measurement of Cherenkov light, ({\it e.g.}, the radiation angle or spatial-time hit pattern) are numerous. Two main types, namely, the RICH and the detection of internal total-reflected Cherenkov light (DIRC), are commonly employed in high luminosity experiments. Due to the space limit, the RICH detector is chosen as the baseline candidate for the PID barrel region and is described below.

Time of flight (TOF) is also a common PID technology. For the barrel region, the minimum flight distance is $\sim85$~cm. An overall TOF time resolution of $\sim 20$~ps is needed to effectively separate $\pi/K$ at $p \ge 2$\,$\gevc$. It is thus very challenging to apply the TOF technique to barrel PID detectors. However, for the end-cap region, where the flight distance is approximately $>1.5$~m, a TOF resolution of $\sim 35$~ps is adequate. The details is discussed in Sec.~\ref{sec:dtof}.

\subsection{RICH Detector Concept}

\subsubsection{Conceptual Design}
The barrel PID system is placed between the MDC and EMC, with a solid coverage of $0.83$ and a distance of approximately $85$~cm to the collision point. This requires the system to be thin enough to leave space for the calorimeter, and the material budget needs to be low to reduce the energy distortion. Hence, the thickness must be less than $20$~cm, and the material budget must be less than $30\%X{_0}$. The detector also needs to have a fast time response to operate under a high luminosity environment, with $\mathcal{L} =10^{35}$~cm$^{-2}$s$^{-1}$.	

\begin{figure}
\centering
\includegraphics[width=.8\textwidth]{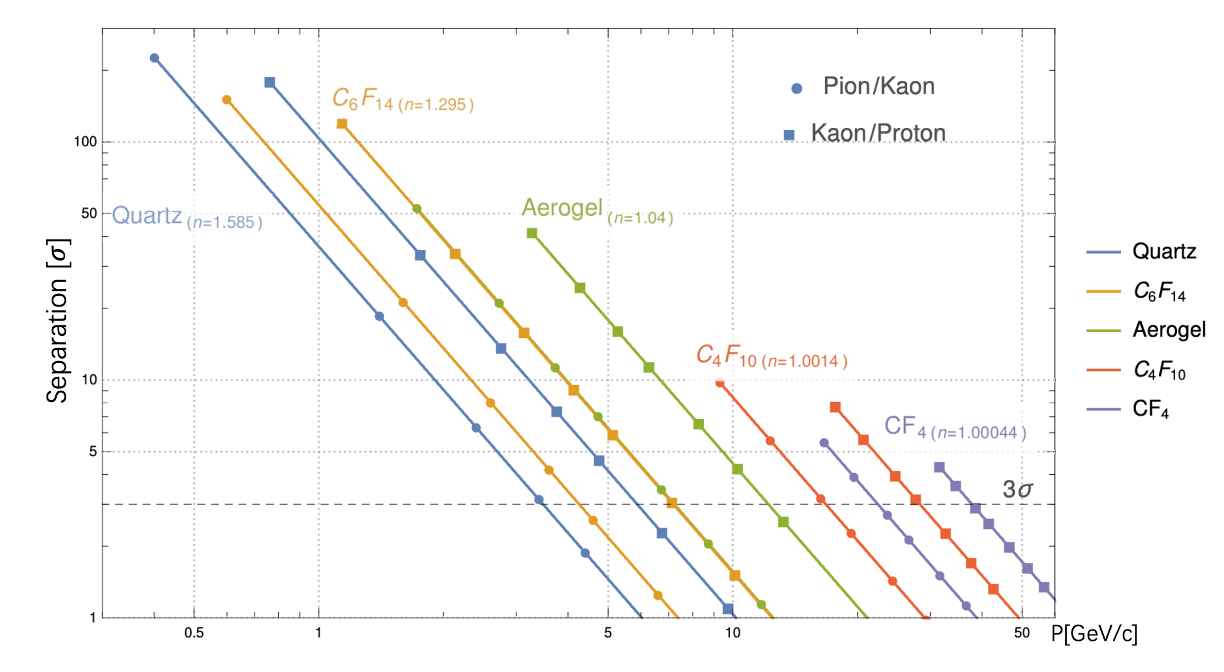}
\caption{The $\pi/K/p$ PID separation abilities of different radiators assuming $2.5\,$mrad angular resolution. The results for quartz and C$_6$F$_{14}$ with a refractive index of $180\,$nm are depicted.}
\label{fig:pidvsrad}
\end{figure}

To satisfy these requirements, a micropattern gas detector (MPGD)-based RICH detector is chosen as the baseline design for the barrel part of the PID detector. The RICH particle separation power can be estimated from Eq.\,\ref{eq:nsigma},
\begin{equation}
N_\sigma\approx\frac{|m_1^2 - m_2^2|}{2p^2\sigma_\theta\sqrt{n^2-1}} \label{eq:nsigma},
\end{equation}
where $m_1$ and $m_2$ are the masses from different particle hypotheses, $\sigma_\theta$ is the angular resolution, $p$ is the momentum and $n$ is the refractive index. Fig.~\ref{fig:pidvsrad} shows the $\pi/K$ and $K/p$ separation capability in terms of standard deviation vs. momentum for different types of radiators, where the angular resolution of the RICH detector is assumed to be $2.5\,$mrad. For the momentum range of interest of the STCF ($<2$\,$\gevc$), liquid and quartz have adequate refractive indexes.

\begin{figure}
\centering
\includegraphics[width=.8\textwidth]{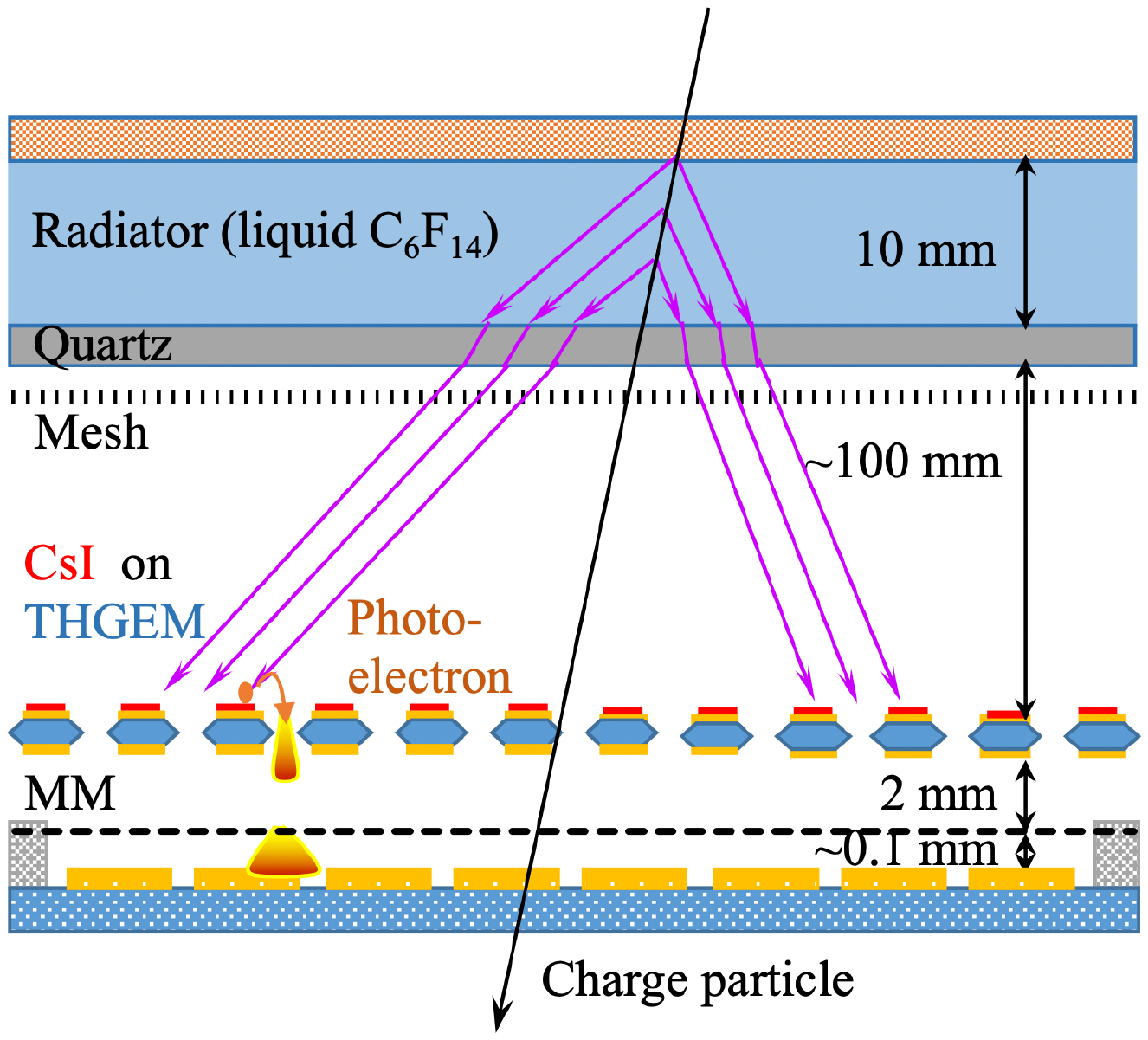}
\caption{The RICH detector structure.}
\label{fig:RICHstructure}
\end{figure}

A sketch of the approximately focused RICH detector is shown in Fig.~\ref{fig:RICHstructure}. It includes a liquid radiator layer, optical transparent quartz, a working gas, and photon detectors. A charged particle moving outward can emit photons in the radiator with a Cherenkov angle of $\theta_c=\cos^{-1}(1/\beta n)$ with respect to the particle direction, where $\beta$ is the velocity of the particle and $n$ is the refractive index of the radiator medium. These photons are approximately focused on the anode layer of the photon detectors to produce a circular ring image. For each detected photon, a value of $\theta_c$ can be calculated for comparison with different particle hypotheses.

The present baseline design of the RICH detector considering the above discussion is described as follows. Each module contains a 10.0\,mm-thick liquid radiator with a 3.0\,mm quartz window. A light propagation region of 100.0\,mm separates the quartz window and the cathode wires (mesh) from the CsI-coated thick gaseous electron multiplier (THGEM) placed under the cathode wires.

The radiator container, the chamber gas and the remaining components must be UV transparent or at least exhibit UV transparency comparable to that of the radiator. However, the gas may be contaminated by water vapor and oxygen, which reduces the transparency range. Hence, this contamination must be controlled to less than $10$\,ppm.

The RICH detector consists of a CsI-coated THGEM as the photocathode and a Micromegas (MM) layer to amplify the photoelectrons. The distance between the THGEM layer and the MM layer is 2~mm. A small reverse electric field is applied between the cathode and the top layer of the THGEM to increase the quantum efficiency. The converted photoelectrons drift to the bottom layer of the Micromegas plane with a gain of approximately 10$^5$. The induced signal is then picked up by the anode. The ions generated during the amplification drift along the electrical lines and bombard the photocathode. This induces a second signal and results in photocathode aging. With the hybrid combination of the THGEM and MM, the ion backflow is demonstrated to be less than $4$\%~\cite{compass}.
Considering the goal of ten years of operation, the total accumulated charge for the STCF RICH detector would be approximately $\approx 2\,\mu$C/cm$^2$.
The usage of large-area CsI photocathodes has been demonstrated in the ALICE~\cite{alice} and COMPASS~\cite{compass} experiments, and no severe aging effects have been seen under $10\,\mu$C/cm$^2$.

The overall material budget with aluminum plating as the container is listed in Table \,\ref{TAB:richbudget}. The total material budget is approximately 24\% $X_0$.

\begin{table}[h]
  \centering
  \begin{tabular}{|l|c|c|}\hline
    &thickness [mm]&$X/X_0$\\\hline
    Top ceramic plate &3&0.03\\
    Quartz window&3&0.03\\
    Radiator C$_6$F$_{14}$&10&0.05\\
    THGEM+Micromegas&0.4&0.01\\
    Anode+FEE&8&0.02\\
    Aluminum plate&5&0.05\\
    FEE cooling&5&0.05\\
    Total&&0.24\\\hline
  \end{tabular}
   \caption{Material budget of the RICH detector.}
  \label{TAB:richbudget}
\end{table}

\subsubsection{Systematic error for RICH reconstruction with S.P.E}

\begin{figure}
\centering
\includegraphics[width=.8\textwidth]{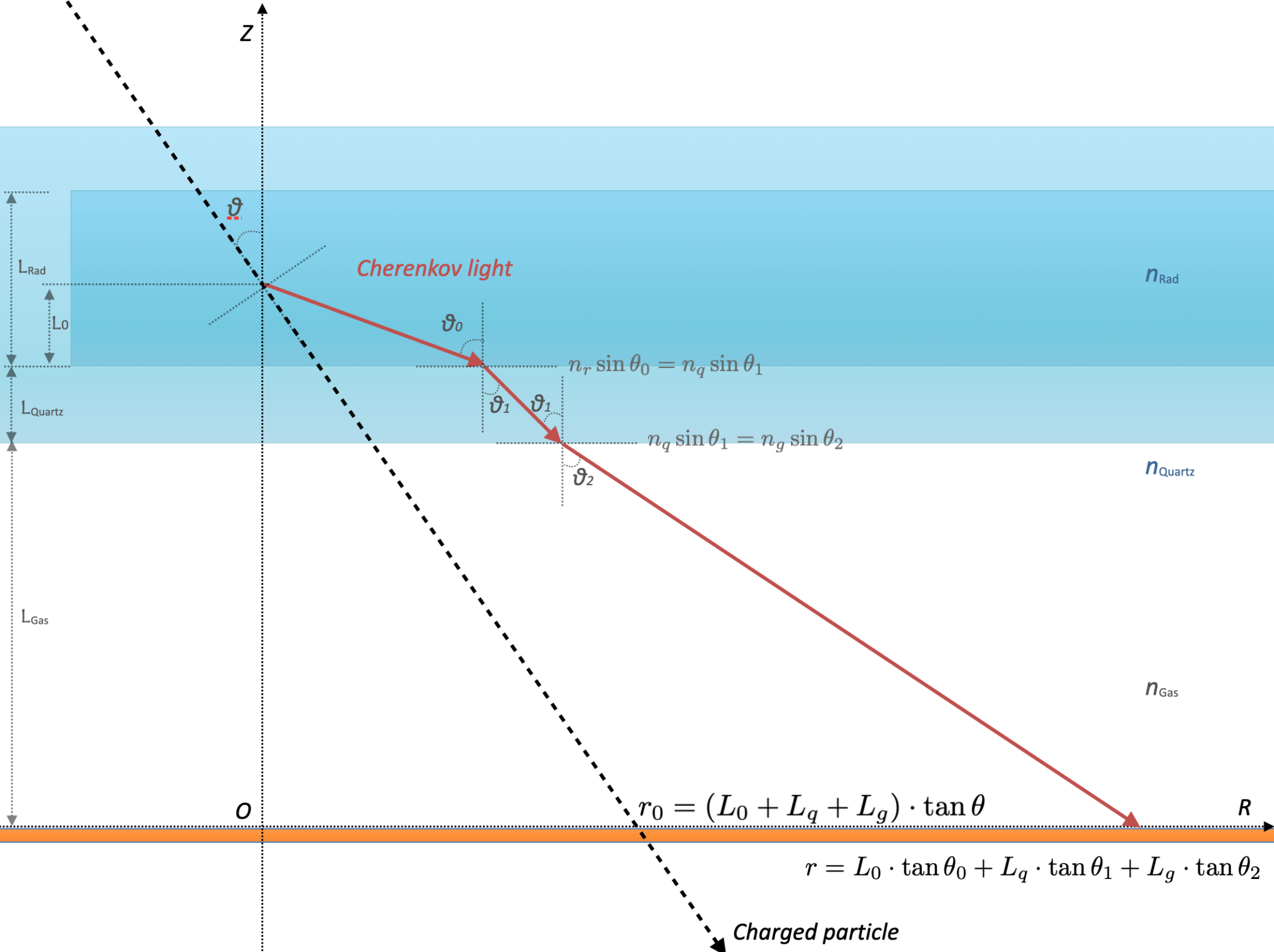}
\caption{The Cherenkov light propagation in the RICH detector.}
\label{fig:richrec}
\end{figure}

RICH signals can be simulated analytically based on the propagation of Cherenkov light, as shown in Fig.\,\ref{fig:richrec}, where $L$ is the thickness and $n$ is the refractive index. The subscript indexes $r$, $q$, and $g$ represent the radiator, the quartz box, and the transport gas medium, respectively. $L_0$ is the distance between the emission point where Cherenkov light is generated along a charged particle and the bottom of the radiator. $\theta_0$ can be calculated from the charged particle incident angle $\theta$, Cherenkov emission angle $\theta_c$ and azimuth angle $\phi$. Thus, the expected hit position on the anode can be expressed as $r = L_0\cdot\tan\theta_0+L_q\cdot\tan\theta_1+L_g\cdot\tan\theta_2$.

To reconstruct the Cherenkov angle $\theta_c$ from each hit position $r$, two assumptions are made: 1) the Cherenkov radiation is emitted from the center of the charged track trajectory inside the radiator because the absorption effects are neglected; 2) although Cherenkov radiation is a spectrum, the refractive indexes of each component at 180\,nm are taken because this is the most likely value considering the absorption of each medium and the quantum effect of the cathode.

For illustration purposes, it is useful to consider the case of a vertical incident particle ($\theta=0$). By defining $R_0 = L_g\frac{n\sin\theta_c}{\sqrt{1-n^2\sin^2\theta_c}}$, the Cherenkov angle can be expressed as:
\begin{equation}
\theta_c=acos(\frac{1}{\sqrt{\frac{\beta^2}{1+\frac{L_g^2}{R_0^2}}+1}})
\end{equation}

The angular resolution of this type of RICH detector includes contributions from the following components:
\begin{equation}
\sigma^2_\theta=\frac{\sigma^2_E + \sigma^2_{L_{rad}} + \sigma^2_{xy} + \sigma^2_{ms} + \sigma^2_{\theta_0}}{N_{pe}} \label{eq:rich_sigma_theta}
\end{equation}

The term $\sigma^2_E$ is the contribution from chromatic dispersion. Cherenkov radiation consists of a continuous spectrum of wavelengths extending from the ultraviolet region into the visible spectrum. Liquid perfluorohexane C$_6$F$_{14}$ and high purity quartz are both candidates for Cherenkov radiators for the STCF. The UV threshold is mainly set by quartz, at approximately 170~nm. Cesium iodide (CsI) provides a relatively high quantum efficiency ($\sim22\%@180$\,nm) in this region and decreases to zero at approximately 210\,nm. However, the refractive indexes of quartz and C$_6$F$_{14}$ vary by approximately $\sim 8$\% and $\sim 3$\% in this region, respectively. The chromatic uncertainty is intrinsic and the main source of the systematic error.

The term $\sigma^2_{L_{rad}}$ comes from the uncertainty of the emission point of the Cherenkov radiation. This is estimated by taking $L/\sqrt{12}$, where $L$ is the charged track length inside the radiator, and the expected emission point is near the center of the radiator. This contribution is proportional to the fraction of the radiator length $L_{rad}$ and the light propagation length.

The term $\sigma^2_{xy}$ comes from the anode spatial resolution. The granularity of the detector, i.e., the anode pad size for the readout, is driven by the required angular resolution, the light propagation distance and the number of photoelectrons. The gaseous detector provides a negligible material budget as well as possibly a long distance for light propagation. From our previous study, $5\,$mm anode pads with a $10\,$cm light propagation length would be sufficient.

The term $\sigma^2_{ms}$ represents the multiple scattering of charged particles inside the radiator medium. This dispersion can be estimated by taking $\sigma_{ms}\sim\Delta\theta_{ms}$, where $\Delta\theta_{ms}\propto (1/p)\sqrt{L/X_0}$. This term contributes mainly to the low momentum range and decreases rapidly when momentum increases.

The term $\sigma^2_{\theta_0}$ comes from the incident angle uncertainty. This can be estimated by extrapolating the reconstructed charged track from the MDC to the PID detector and is not taken into account for current optimization.

\begin{figure}
\centering
\includegraphics[width=.8\textwidth]{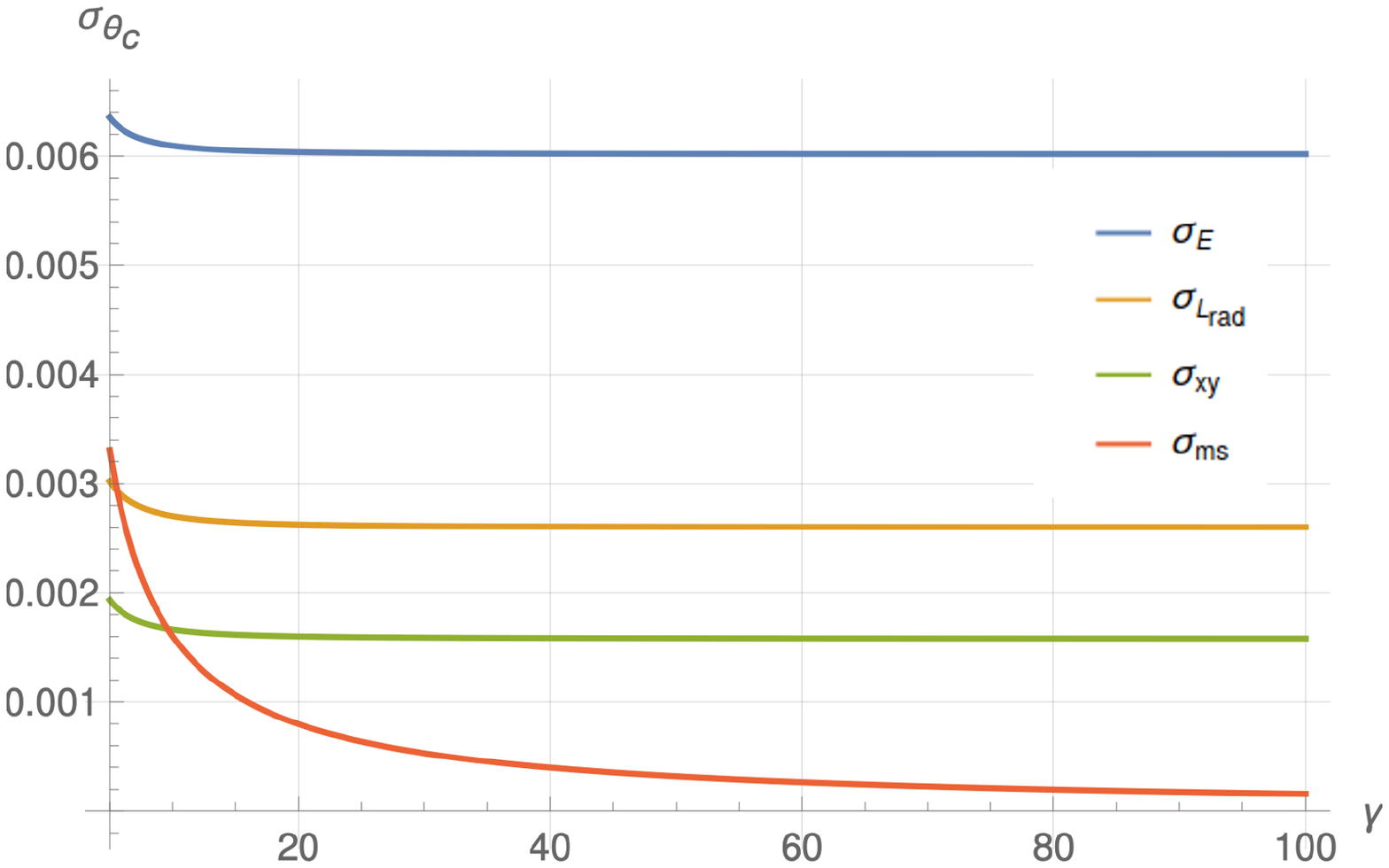}
\caption{RICH reconstruction system error versus the Lorentz factor $\gamma$. }
\label{fig:richsyserr}
\end{figure}

To estimate the RICH reconstruction accuracy, we can deduce the chromatic error related to the refractive index $n$ of the radiator, the geometric error related to the emission point of Cherenkov radiation $L_0$, the localization error related to the spatial resolution of the detector, and the multiple scattering error of the charged incident particle. By taking the thicknesses for each part as $L_r=10$\,mm, $L_q=3$\,mm and $L_g=100$\,mm, which are the default designs for the RICH detector, we can obtain the contribution of each systematic error on the angular resolution $\theta_c$. The results are summarized in Table~\ref{TAB:richsys}. Fig.\,\ref{fig:richsyserr} shows these four contributions to $\theta_c$ resolution as a function of $\gamma$ in this configuration, and it can be seen that the dominant contribution is the chromatic error and the geometric error. A comparison to the Geant4 simulation for $2\,$GeV/c $\pi$ is made, showing consistency.

\begin{table}[h]
  \centering
  \begin{tabular}{|l|c|c|}\hline
    Source&Error (mrad)&Simulation (mrad)\\\hline
    Chromatic&6.0&5.0\\
    Geometric&2.6&3.1\\
    Localization&1.6&1.8\\
    Multiple scattering&1.1&1.1\\
    Total&6.8&6.2\\\hline
  \end{tabular}
   \caption{Systematic error for RICH reconstruction.}
  \label{TAB:richsys}
\end{table}

\subsubsection{Number of Photoelectrons}
\begin{figure*}[htbp]
 \centering
  \subfloat[][]{\includegraphics[width=0.45\textwidth]{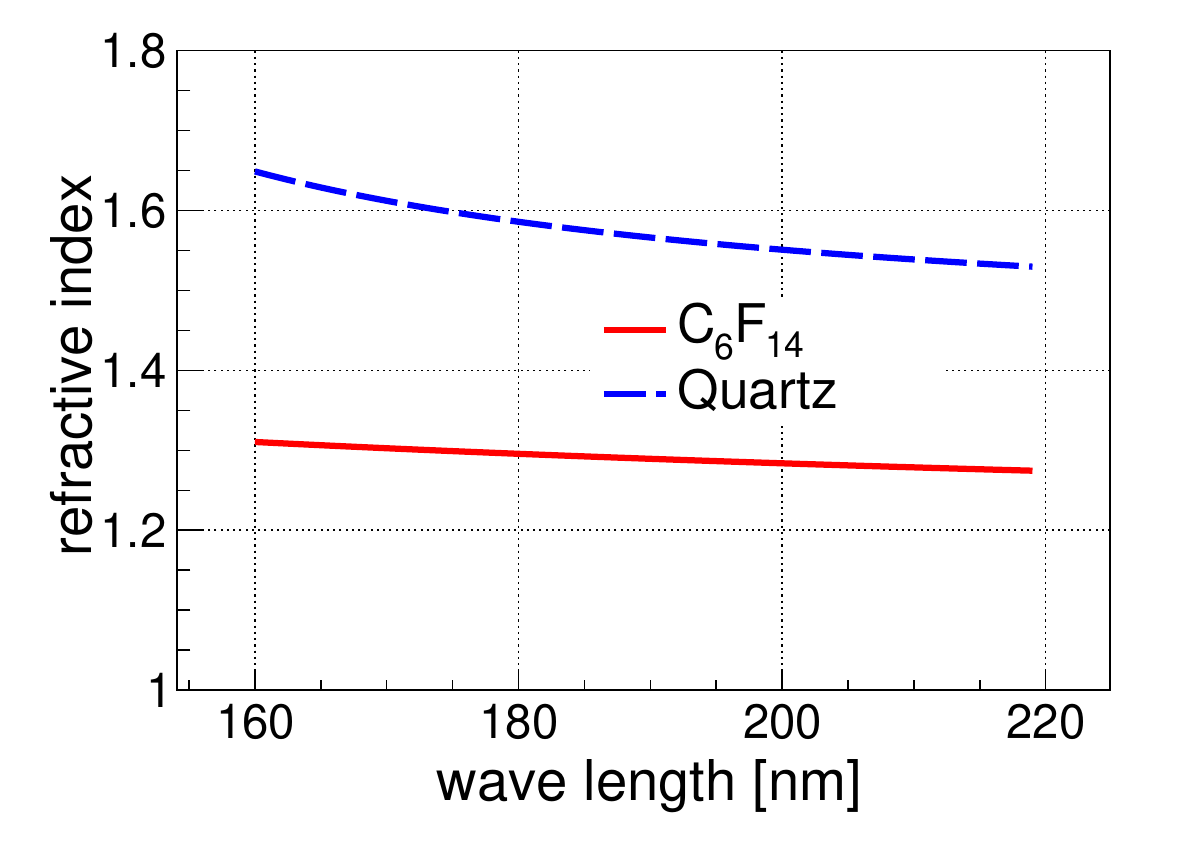}}\quad
\subfloat[][]{\includegraphics[width=0.45\textwidth]{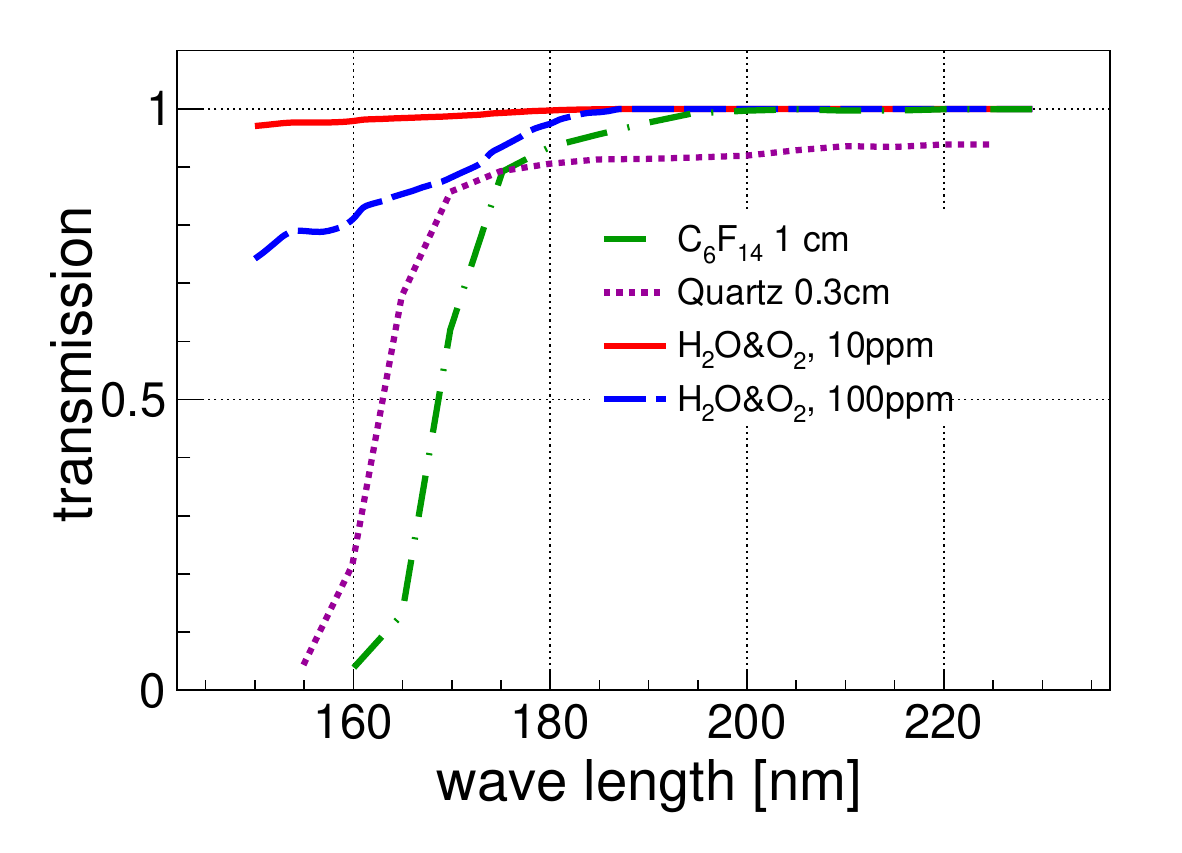}} \quad
  \subfloat[][]{\includegraphics[width=0.45\textwidth]{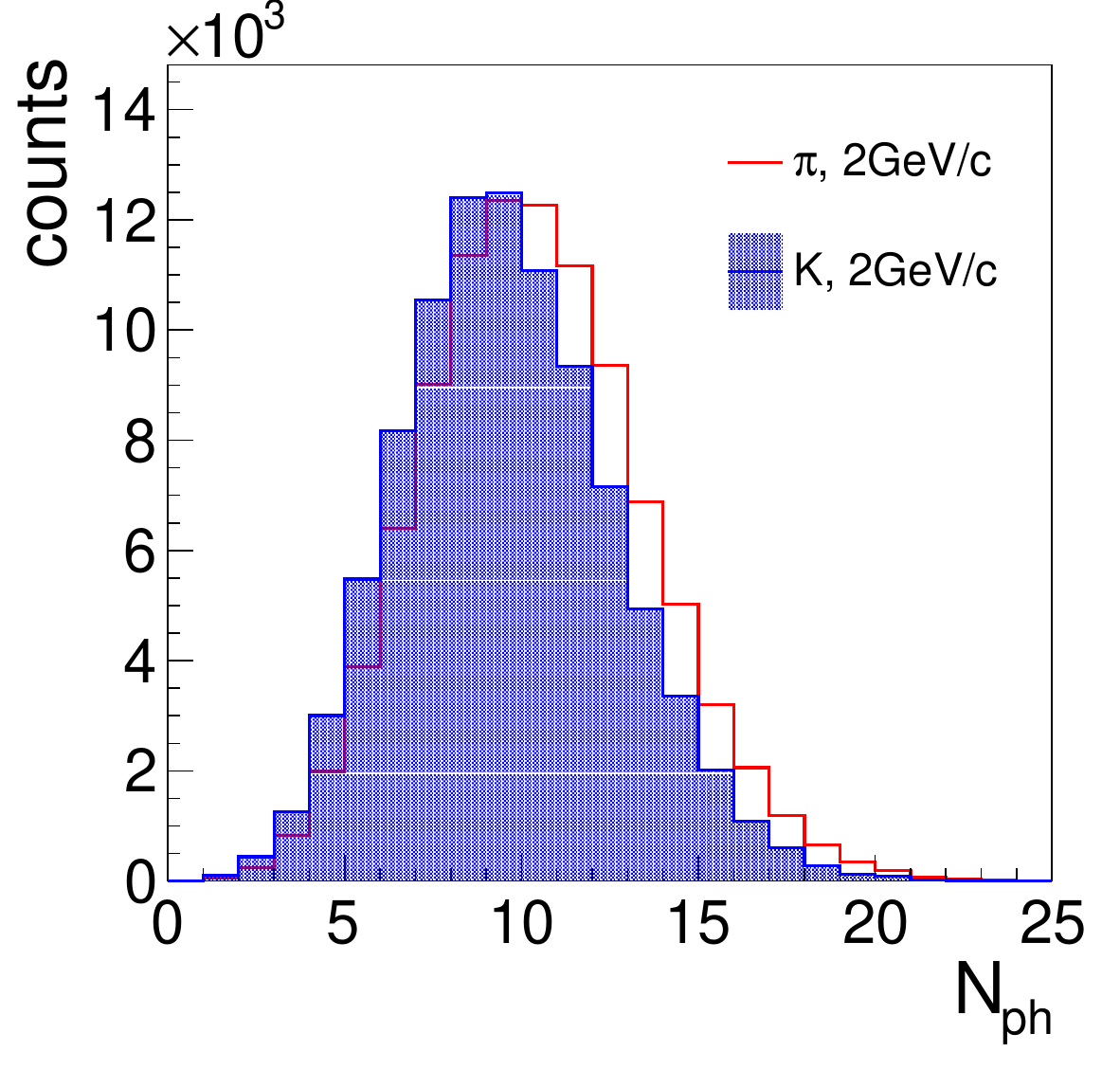}}\quad
  \subfloat[][]{\includegraphics[width=0.45\textwidth]{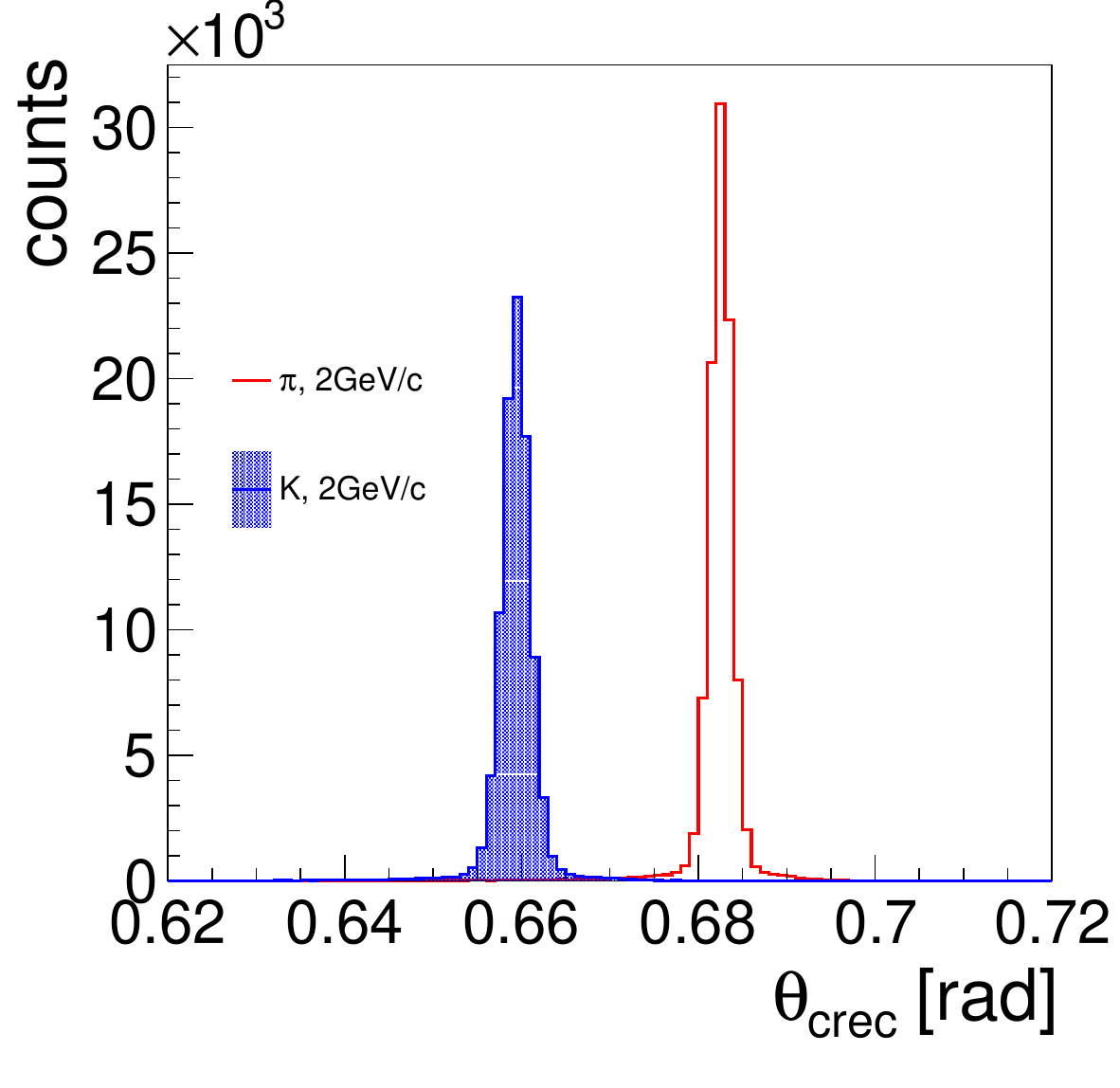}}
\caption{(a) Refractive indexes for liquid C$_6$F$_{14}$ and quartz, (b) transmission rate of each optical component, (c) photoelectron distribution, and (d) reconstructed Cherenkov angle distribution.}
\label{fig:RICHCompDist}
\end{figure*}

The number of photon electrons $N_{pe}$ depends on the thickness of the radiator, the attenuation length of the radiator and the light propagation distance, and the quantum efficiency of the CsI photocathode. This number is given by:
\begin{equation}
N_{pe}=\int{N_0 \cdot L_t \cdot \sin^2\theta_c \prod \limits_{i=0}^n e^{-l_i/L_{ai}}\epsilon(\lambda)d\lambda}
\end{equation}
where $N_0 \cdot L_t \cdot \sin^2\theta_c$ is the Cherenkov light output per unit thickness, which is related to the particle velocity $\beta$ and the refractive index $n$ of the radiator; $\theta_c$ is the mean Cherenkov angle over the detected photon energy spectrum; $L_t$ is the thickness of the particle passing through the radiator; $l_i$ is the distance traveled by the Cherenkov light passing through the $i$-th optical component; $L_{a}$ is the attenuation length of the $i$-th optical component; and $\epsilon(\lambda)$ is the CsI quantum efficiency, which is taken from Ref.~\cite{alice}.
Fig.~\ref{fig:RICHCompDist}(a) shows the refractive indexes of the liquid C$_6$F$_{14}$ and quartz used as the RICH radiators, (b) shows the transmission rate of each optical component, including the working gas with different humidity and oxygen contamination values, (c) shows the number of photoelectrons for 2~$\gevc$ $\pi$ and $K$, and (d) shows the reconstructed Cherenkov angle distribution for 2~$\gevc$ $\pi$ and $K$.
An average of $\sim10$\,p.e. is expected. In total, the RICH detector is expected to have $6.8$\,mrad from S.P.E. reconstruction, and better than $2.5$\,mrad angular resolution can be achieved.

\subsection{RICH Detector Performance Simulation}
\subsubsection{Expected PID Capabilities}

\begin{figure}[!htb]
  \centering
  \includegraphics[width=1.0\textwidth]{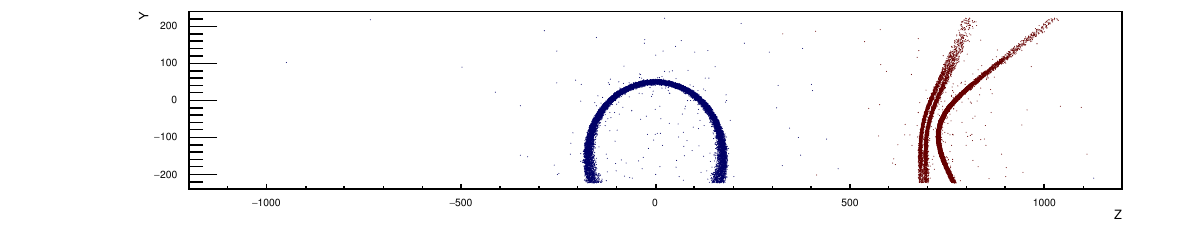}
  \caption{Examples of Cherenkov images in a RICH module. The blue image depicts the distribution of hits for $2\,\rm{GeV}/c$ pion with incident angle $\theta = 0^\circ$, perpendicular to RICH, while the red image depicts $\theta=40^\circ$.}
  \label{FIG:RICH_Cherenkov_Example}
\end{figure}

{\sc Geant4} simulations are performed to study the expected performance of the RICH detector. Incident particles $\pi/K/p$ are emitted from the IP inside a 1~T magnetic field. The optical properties of the radiator (quartz and C$_6$F$_{14}$) are defined according to the measurement results~\cite{alice}, and the absorption of gas is calculated from the H$_2$O and O$_2$ absorption cross-section, with an assumed contamination of $10$\,ppm for each. During the simulation, the detector response is considered using the CsI converter quantum efficiency. A momentum and azimuth angle scan is then performed, and a typical ring of the Cherenkov hit pattern is shown in Fig.\,\ref{FIG:RICH_Cherenkov_Example}. The blue and red curves represent $2\,\rm{GeV}/c$ pions with polar angles $\theta = 0^\circ$ and $40^\circ$, respectively.

To further evaluate the PID capabilities of the RICH, a likelihood-based PID method is studied. The number of Cherenkov photons collected by the $5\times 5\rm\,mm^2$ anode pads follows a Poisson distribution. Therefore, the probability density function for the signal of an anode pad can be constructed as:

\begin{equation}
  pdf_{i, h} = Poisson(N_i + 10^{-3}, mean_{i, h} + 10^{-3}),
  \label{PIDlikelihood}
\end{equation}
\noindent

where $i$ represents the pad index, $h$ denotes hadron species (in our case, $\pi,~K, ~p$), $N$ is the photon number collected by this pad, $mean_{i, h}$ represents the expected average number of photons of each anode pad, which is simulated in Geant4, and the constant $10^{-3}$ is a conservative estimation of the background level for each anode pad. Under each particle type hypothesis, the log-likelihood of the RICH detector is calculated by summarizing the log-likelihood of all pads:

\begin{equation}
  \ln\mathcal{L}_{h} = \sum_{i}^{\rm {npads}}\ln pdf_{i, h}
\end{equation}

\begin{figure*}[htbp]
 \centering
   \subfloat[][]{\includegraphics[width=0.45\textwidth]{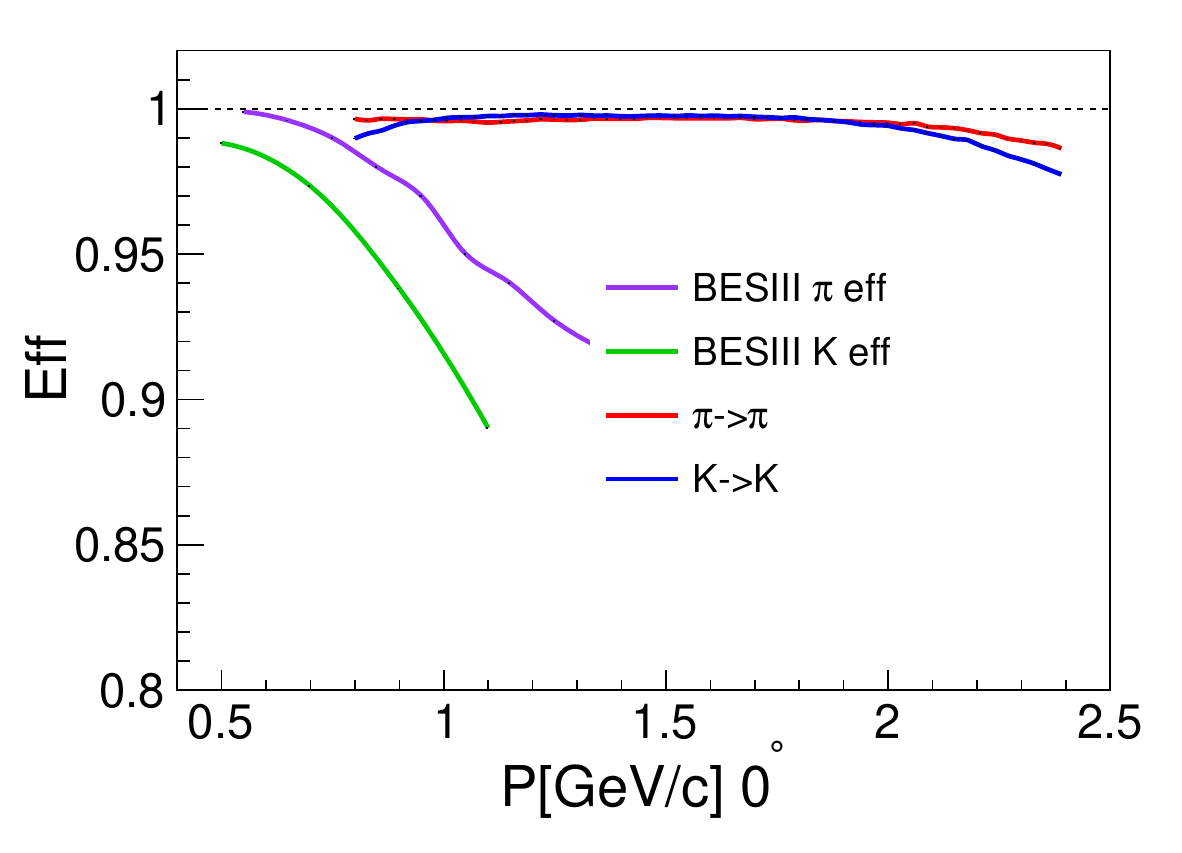}}
  \subfloat[][]{\includegraphics[width=0.45\textwidth]{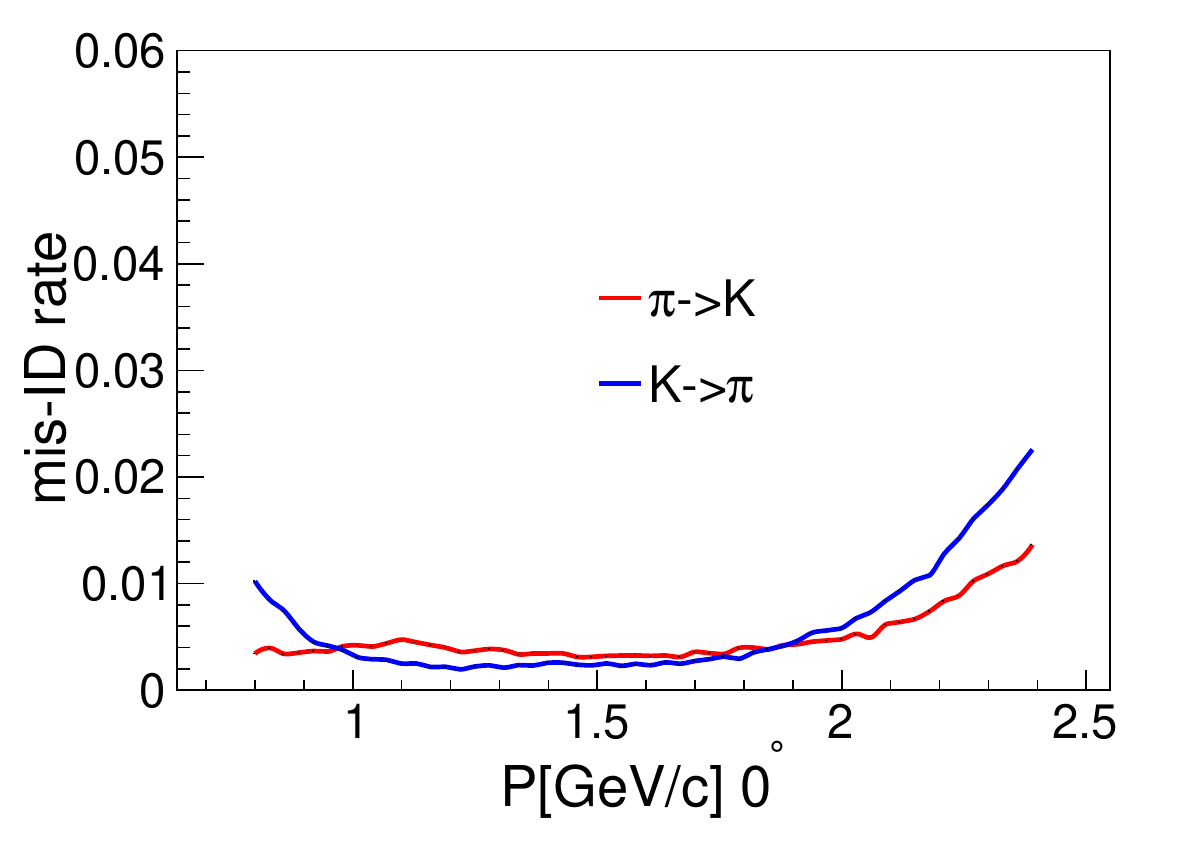}}
\caption{RICH PID capabilities in terms of (a) $\pi$/K efficiency and (b) misidentification efficiency.}

\label{fig:richpideffsim}
\end{figure*}

However, instead of calculating the absolute log-likelihood, in separating particle types, the difference in log-likelihood (DLL) between two hypotheses is calculated. For instance, in $\pi,~ K$ separation, DLL is defined as:

\begin{equation}
  {\rm {DLL}} = \sum_{i}^{\rm {npads}} \ln\frac{pdf_{i,\pi}}{pdf_{i,K}}
\end{equation}
\noindent
If $\rm DLL>0$, the $\pi$ hypothesis is accepted; otherwise, the $K$ hypothesis is accepted. The PID efficiency vs. momentum obtained by applying the reconstruction algorithm is shown in Fig.\,\ref{fig:richpideffsim}.

\begin{figure}[h!]
  \centering
   \subfloat[][]{\includegraphics[width=0.45\textwidth]{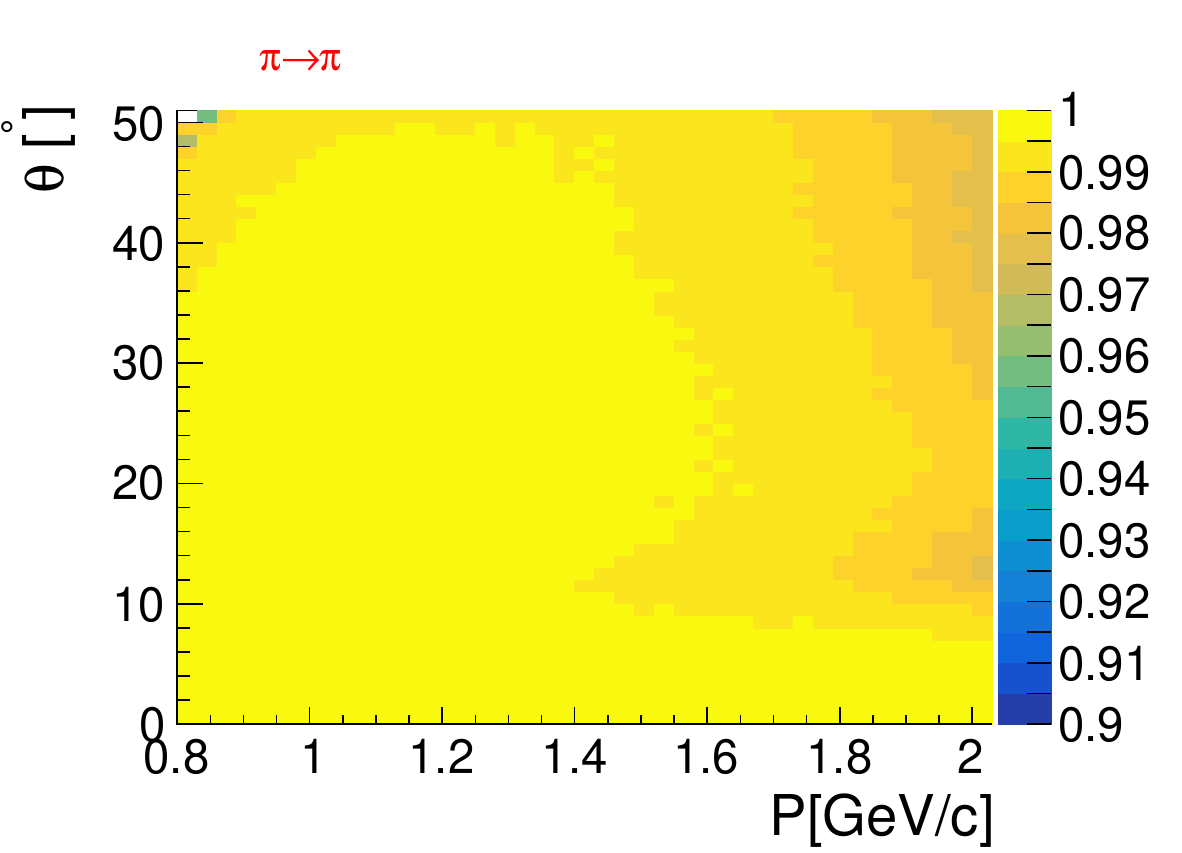}}
   \subfloat[][]{\includegraphics[width=0.45\textwidth]{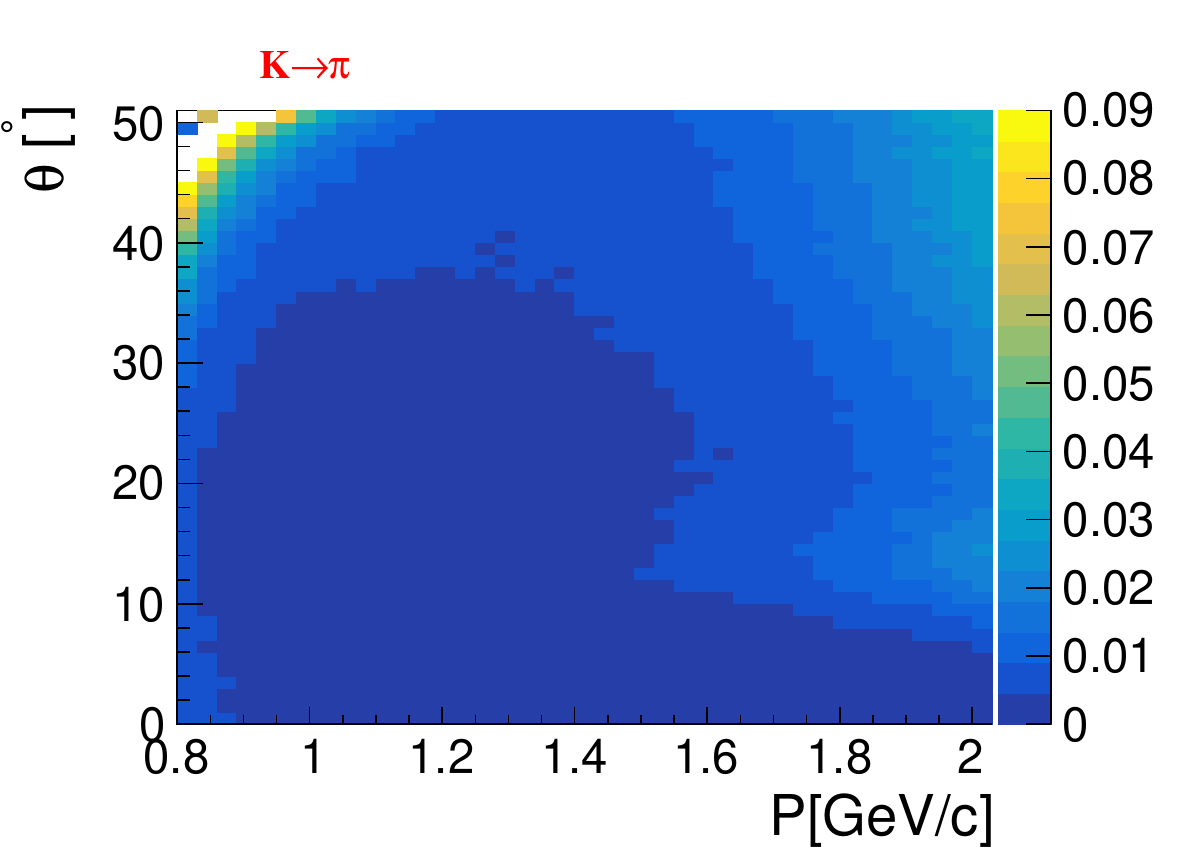}}
  \caption{PID capability scans for (a) $\pi$ efficiency, (b) $\pi/K$ mis-ID rate and (c) $\pi/p$ mis-ID rate.}\label{fig_RICH_pidetail}
\end{figure}

A detailed scan for the $\pi/K/p$ hadron hypothesis is also performed. The momentum of the tracks ranges from $0.8\,\rm {GeV}/c$ to $2.4\,\rm{GeV}/c$, and the incident polar angle ranges from $0^\circ $ to $50^\circ$. The momentum range is divided in $30\,{\rm MeV}/c$ step, and the polar angle step is $1^\circ$. The $\pi$ PID efficiency and $K/p$ mis-ID rate are shown in Fig.~\ref{fig_RICH_pidetail}. The PID efficiency increases with momentum and varies along the incident angle. Especially for a very large incident angle, such as an angle larger than 50 degrees, the Cherenkov light mainly comes from the quartz box instead of the radiator; thus, the PID efficiency decreases dramatically. Additionally, there are some small zones where the PID efficiency drops due to some other misidentifications. This is because of the difficulties in distinguishing the light from the quartz box and that from the radiator. In general, the RICH system can fulfill the hadron PID requirements of the STCF.

\subsubsection{Background simulation and occupancy}

As shown in Table~\ref{tab:5.2.01}, the major component of the background is expected to be luminosity-related background. To estimate the occupancy from the background, each charged track is assumed to produce a signal in the detector. Each Cherenkov photon that hits the anode plane and is sampled according to the quantum efficiency of the CsI photocathode is assumed to produce a signal. The charged tracks, including Cherenkov photon-electron signals, are summed as the RICH rate and listed in Table~\ref{tab:pidbackground}. Radiative Bhabha scattering is the dominant source of background. On average, the occupancy is approximately $4.3\times10^{-4}$~Hz for a $5\times5$~mm$^2$ anode pad with a 500~ns time window, which is smaller than the previous $10^{-3}$ background estimation from Eq.~\ref{PIDlikelihood}.

%%%%%%%%%%%%%%%%%  TABLE  %%%%%%%%%%%%%%%%%%%%%%%%
\begin{table*}[hptb]
\small
    \caption{The background simulation for the RICH detector.}
    \label{tab:pidbackground}
    \vspace{0pt}
    \centering
        \begin{tabular}{lrrr}
        \hline
        \hline
          &  \multicolumn{1}{l}{Generated rate (Hz)} & \multicolumn{1}{l}{RICH rate (Hz)} & \multicolumn{1}{l}{Counting rate (Hz/mm$^2$)} \\
        \hline
    RBB e$^{\pm}$   &  5.98$\times 10^8$   & 1.25$\times 10^8$ & 50.7 \\
    RBB $\gamma$  &  1.07$\times 10^8$    & 3.71$\times 10^6$ & 1.47 \\
    Two photon &  1.03$\times 10^9$          & 2.44$\times 10^7$ & 9.65 \\
    Touschek &  1.12$\times 10^9$               & 5.04$\times 10^6$  & 1.99 \\
    Coulomb  &  2.09$\times 10^8$            & 2.90$\times 10^8$ & 115 \\
    Brems & 2.10$\times 10^6$                  & 2.10$\times 10^2$      & negligible \\
    \hline
    \end{tabular}
\end{table*}
%%%%%%%%%%%%%%%%%%%%%%%%%%%%%%%%%%%%%%%%%%%%%%%%%%

\subsection{Detector Layout}

\begin{figure}[!htb]
  \centering
  \subfloat[][]{\includegraphics[width=0.45\textwidth]{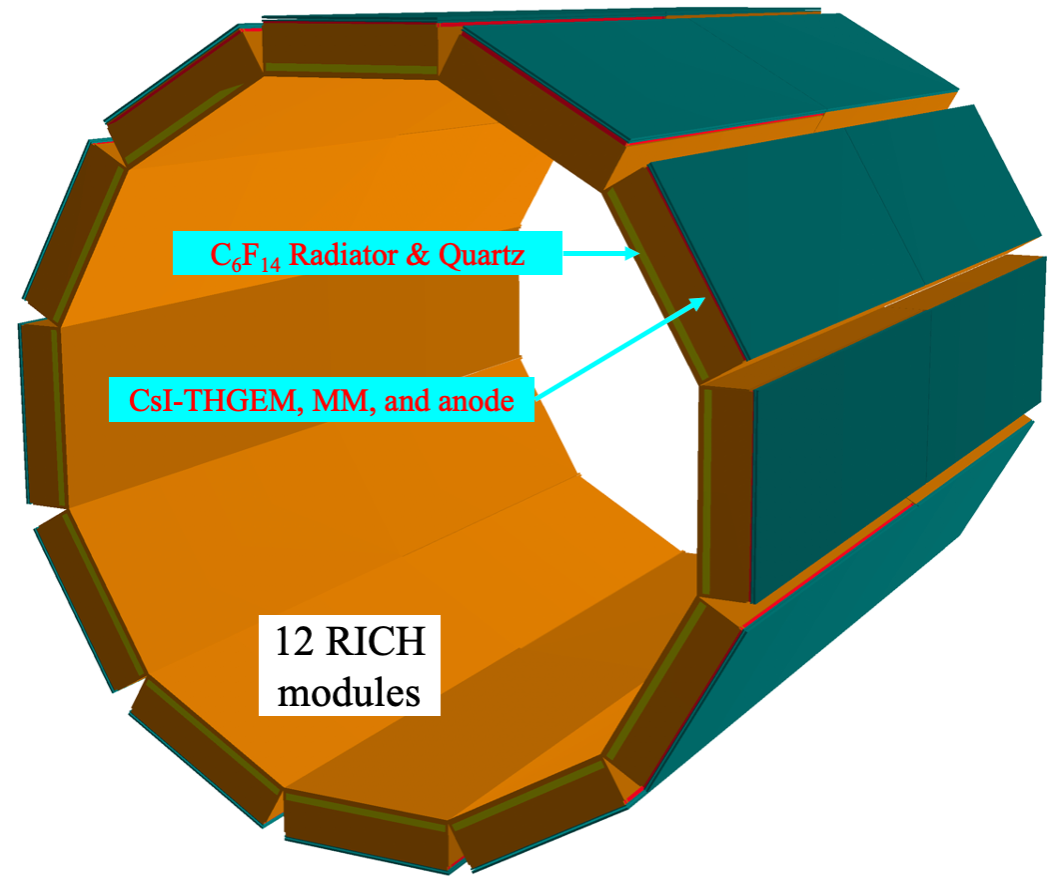}}
  \subfloat[][]{\includegraphics[width=0.45\textwidth]{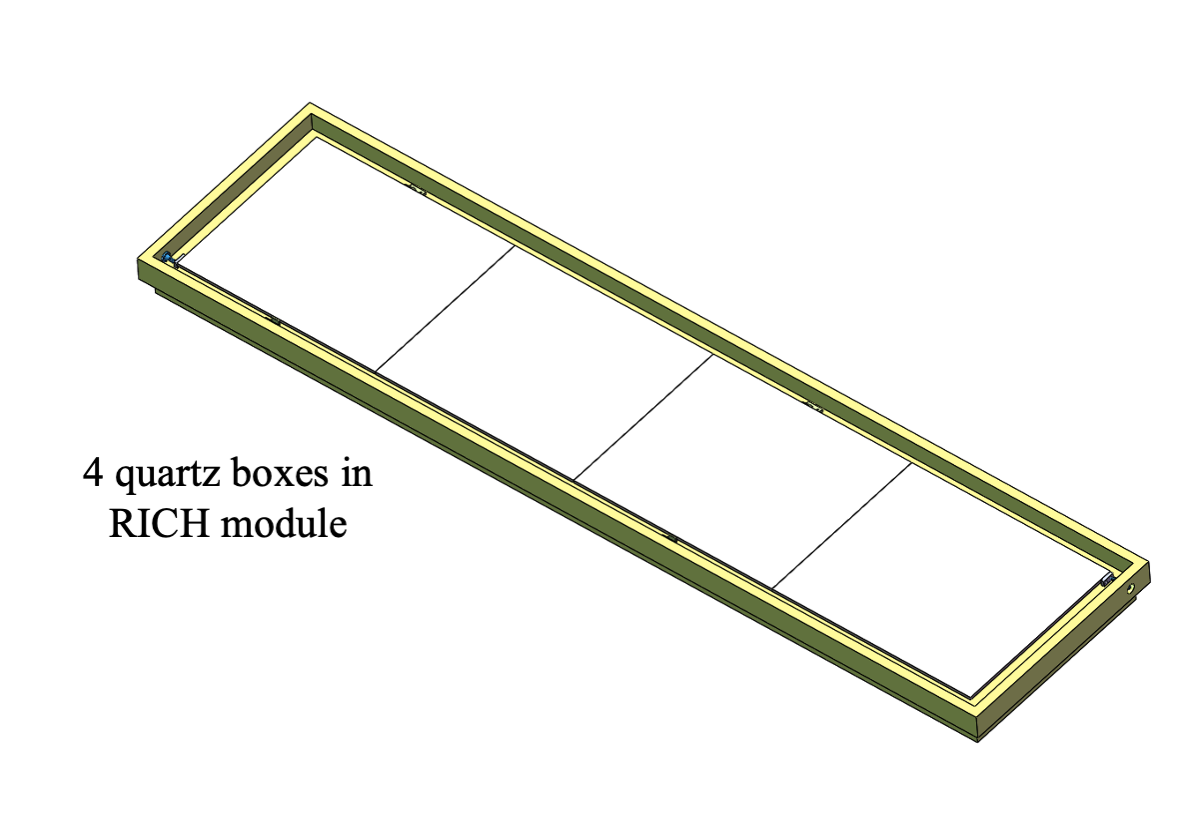}}
  \caption{The conceptual design of the RICH detector: (a) the overall layout and (b) a schematic view of the module box.}
  \label{FIG:RICH_Conceptual_Design}
\end{figure}

The RICH detector is placed between the MDC and EMC, with 0.83 solid coverage of the barrel part. The inner radius is $850$\,mm, and the outer radius is $~950$\,mm. As shown in Fig.\,\ref{FIG:RICH_Conceptual_Design}(a), the RICH detector consists of $12$ identical block modules. Each RICH module is $2400$\,mm long, $450$\,mm wide, and $130$\, mm in height. The whole module is enclosed in a light-tight aluminum box with support from each side.

Each RICH module consists of radiators, a light propagation zone, a CsI-coated THGEM layer, a MicroMegas layer, and anode pads.
A schematic view of the RICH module is shown in Fig.~\ref{FIG:RICH_Conceptual_Design}~(b).
The radiator contains $4$ quartz boxes that are $\sim600$\,mm long and $\sim450$\,mm wide. The quartz boxes are glued and sealed, the bottom layer of which is a UV transparent quartz plate. These liquid C$_6$F$_{14}$ radiators are sealed inside quartz boxes, with pipes connected one to another. A purification system is employed to continuously purify the liquid and pump it to the highest module. Modules are connected by pipes as well. The liquid is driven by gravity and flows back to the tank of the purification system. Note that due to the aluminum box and radiator quartz container, the insensitive area of barrel PID is less than $5\%$.
The light propagation zone is filled with an argon-based gas, which acts as the working gas for the photon detector. Since humidity and oxygen contamination result in the absorption of UV light, a purification system for the gas system is required.
The THGEM and Micromegas are the same size as the radiator boxes for convenience.
The readout pads are $5\times 5$\,mm$^2$. In total, there are $43200$ channels for each RICH module. The electronics are described in the next section.

\subsection{Readout Electronics}
\subsubsection{Design Overview}
High precision time ($~1$\,ns RMS @ $48$\,fC) and charge ($~1$\,fC RMS) measurements are required for the RICH readout electronics, with an input capacitance of approximately $20$\,pF. The total number of channels is estimated to be approximately $518,400$.

\begin{figure}
\centering
\includegraphics[width=1.0\textwidth]{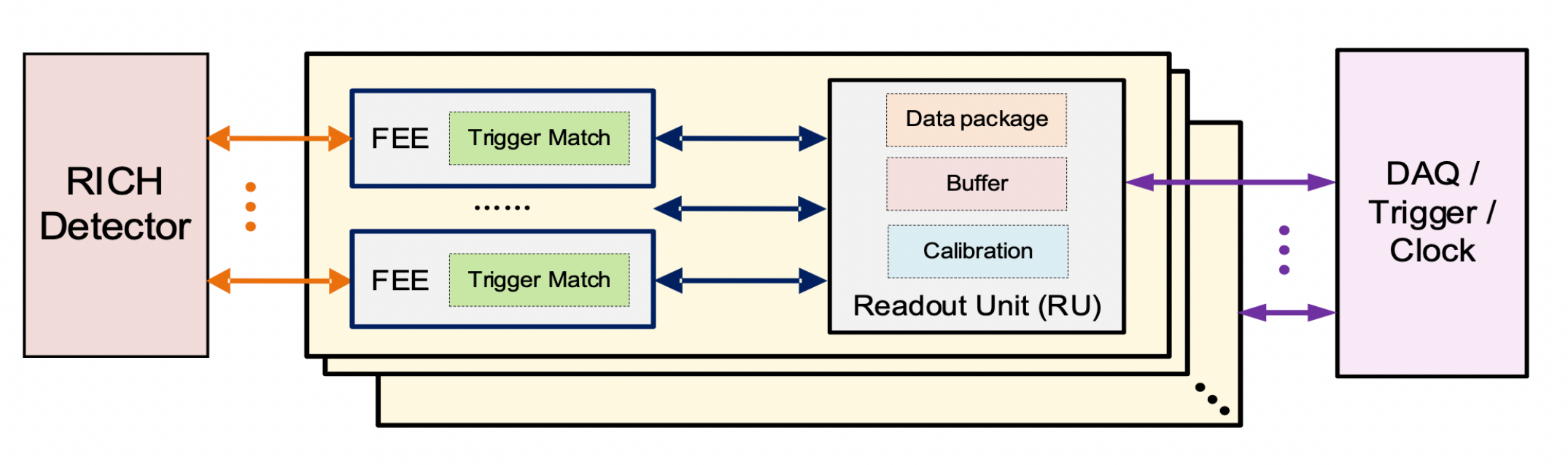}
\caption{Block diagram of the RICH electronics.}
\label{fig:rich-elec}
\end{figure}

The structure of the readout electronics of the PID RICH detector is illustrated in Fig.\,\ref{fig:rich-elec} and is composed of FEE and RUs. Multiple front-end ASICs, which perform analog signal manipulation and A/D conversion, are integrated into one FEE module, and the output data of these ASICs are transferred to a digital ASIC or FPGA, which is responsible for data packaging and transferring the data to the RU through a high-speed serial data interface. The RU receives the data from multiple FEE modules and finally transfers them to the DAQ.

RICH electronics do not participate in the generation of the global trigger signal. The RICH readout electronics must receive the global trigger signal and implement trigger matching in the FEE. The RICH detector readout electronics must also be synchronized with a global clock signal, and this clock is fanned out to the ASICs in the FEE through RUs.

\subsubsection{Front-End ASIC}
Given the large number of readout channels and the requirements for high precision (time and charge) measurements,
the RICH electronics must feature high density, low noise, and low power consumption, and thus, it is necessary to employ a suitable ASIC that satisfies these requirements.

\begin{figure}[!htb]
\centering
\includegraphics[width=1.0\textwidth]{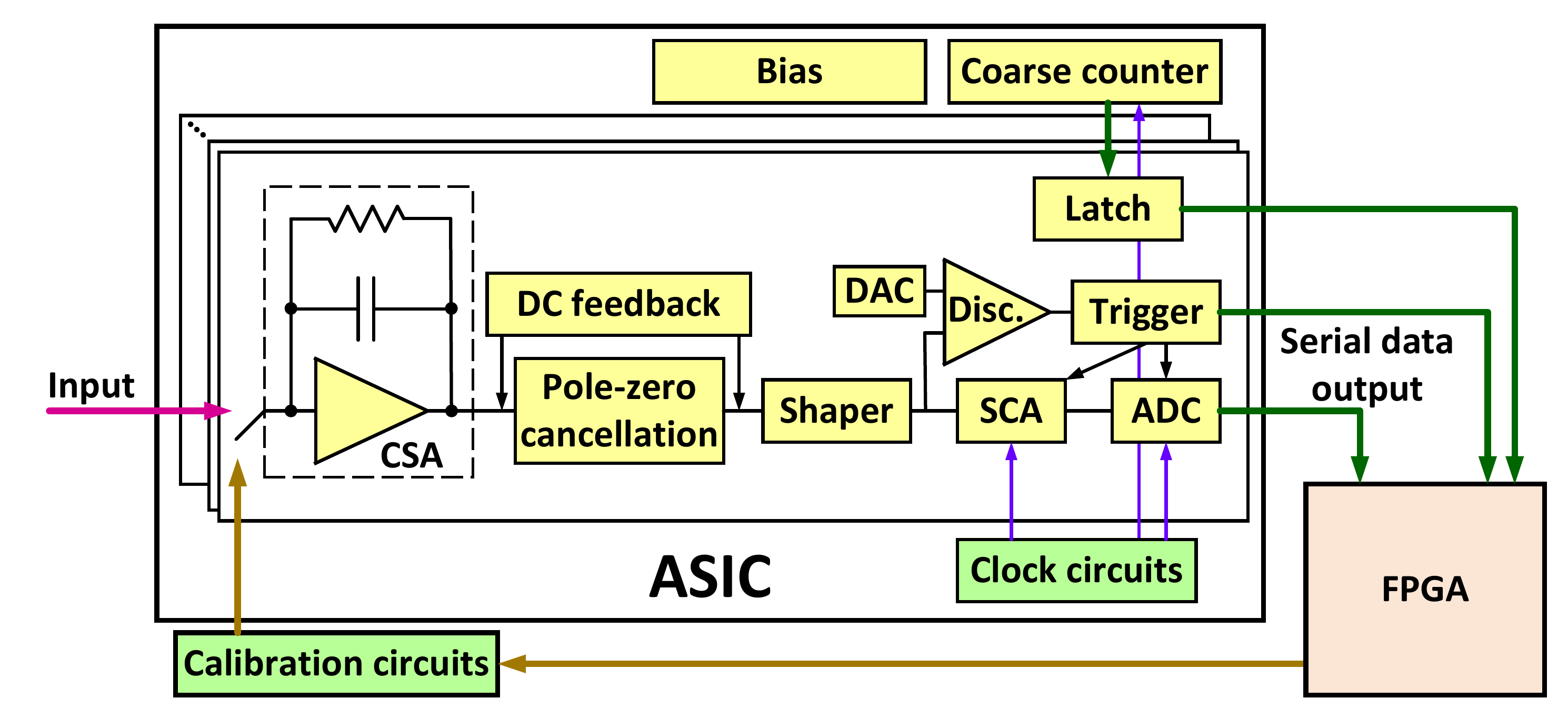}
\caption{Block diagram of the front-end ASIC.}
\label{fig:rich-asic}
\end{figure}

The block diagram of the front-end ASIC is shown in Fig.\,\ref{fig:rich-asic}. The CSA integrates the input signal and generates a signal at its output with an amplitude proportional to the input charge. This signal is then fed to a shaping circuit that outputs a semi-Gaussian pulse while enhancing the SNR. To digitize the waveform, SCAs followed by Wilkinson ADCs are integrated into the ASIC. The digitized signal is sent to the FPGA, and then we can obtain the charge information through peak detection or area calculation of the digitized waveform. Additionally, time information can be obtained using a leading-edge discrimination method. In addition, a coarse counter is designed to expand the time measurement range.

\newpage
\subsection{Summary and Outlook}

The baseline technology for the barrel PID detector at the STCF is a RICH detector with a C$_6$F$_{14}$ radiator.
The conceptual designs are described, and the expected performance is studied through MC simulation. After optimization, the RICH design is expected to meet the PID requirements at the STCF. Extensive R$\&$D works are underway to verify the designs and key technical aspects.

\newpage

\section{Particle Identification in the Endcap (DTOF)}
\label{sec:dtof}

\subsection{Introduction}

As discussed in Sec.~\ref{sec:rich_intro}, for the PID detector in the endcap region, a technology based on the detection of internal total-reflected Cherenkov light (DIRC) is adopted, and the conceptual design is described in this section.

The concept of DIRC was first introduced by the BaBar experiment~\cite{BabarTDR}. Cherenkov lights generated in long-fused silica bars are propagated to the ends through total internal reflections and then projected to an
array of photo sensors via a water expansion volume. The fused silica is taken as both a Cherenkov radiator and light guide. The angles of Cherenkov photons are maintained through hundreds of reflections, and the spatial pattern of the Cherenkov ring can be recognized for PID purposes.
An excellent $\pi/K$ separation can be achieved up to $p=4~\gevc$.
It is worth noting that with BaBar DIRC, the time resolution of a single photon is approximately $1$~ns, which is mainly used to suppress uncorrelated background through the setting of a proper time window.

With improved timing resolution, better PID capability is expected for new generation DIRC detectors, such as those proposed in the future PANDA experiment~\cite{PandaTDR} at FAIR or EID at EIC~\cite{EiC}. The 3-D measurements of $(x, y, t)$ are achieved by a multianode PMT with high-precision timing performance, namely, a microchannel plate photomultiplier tube (MCP-PMT). The volume of such a detector is rather compact, with typical thickness $\le 5$~cm (excluding the optical focusing part). The radiation resistance and mechanical robustness are good, as well as the rate capacity, all of which make it a suitable detector in high luminosity experiments.
The excellent timing capabilities of the new DIRC detectors has also led to the direct
application of the DIRC technology to high time resolution and high rate TOF measurement.
For example, assuming that the time resolution of a single photoelectron is $\sim100$~ps, then for $N_{PE} = 10$, the total timing resolution is $\sim30$~ps, where $N_{PE}$ is the number of photoelectrons.
Such a DIRC-like TOF detector was first proposed for the superB project~\cite{SuperB} and recently adopted
for the LHCb upgrade, namely, the Time Of internally Reflected CHerenkov light (TORCH) detector~\cite{torch}.
For TORCH, a single photon timing error of $\le 70$~ps and a light yield of $N_{PE} \sim 25 $ are predicted, which is expected to achieve a high precision with a time resolution of $\le 15$~ps.

According to the above discussion, a DIRC-like time-of-flight~(DTOF) detector
is supposed to meet the PID requirement for the endcap region of the STCF
owing to the
extended distance between the interaction point and the endcap PID detectors.
From physics requirements, a $\pi/K$ misidentification rate less than 2\% with corresponding identification efficiency larger than 97\% at $p=2$~GeV/c is needed. This is equivalent to a $4\sigma$ deviation
of two probability distributions. As a consequence,
with a flight length of 1.5~m for the hadrons,
a total time resolution of 35~ps is needed for the
TOF measurements.
However, the time measurement with DTOF benefits from both the time of flight of a
hadron and the time of propagation~(TOP) of a photon, which has the potential to provide a 40~ps time resolution, for the DTOF at the STCF.
In the following, the conceptual
design of the DTOF detector at the STCF and its geometry optimization are presented.

\subsection{DTOF Conceptual Design}
%{Conceptual Design of DTOF}

\begin{figure}[!htb]
	\centering
	\includegraphics[width=0.45\textwidth]{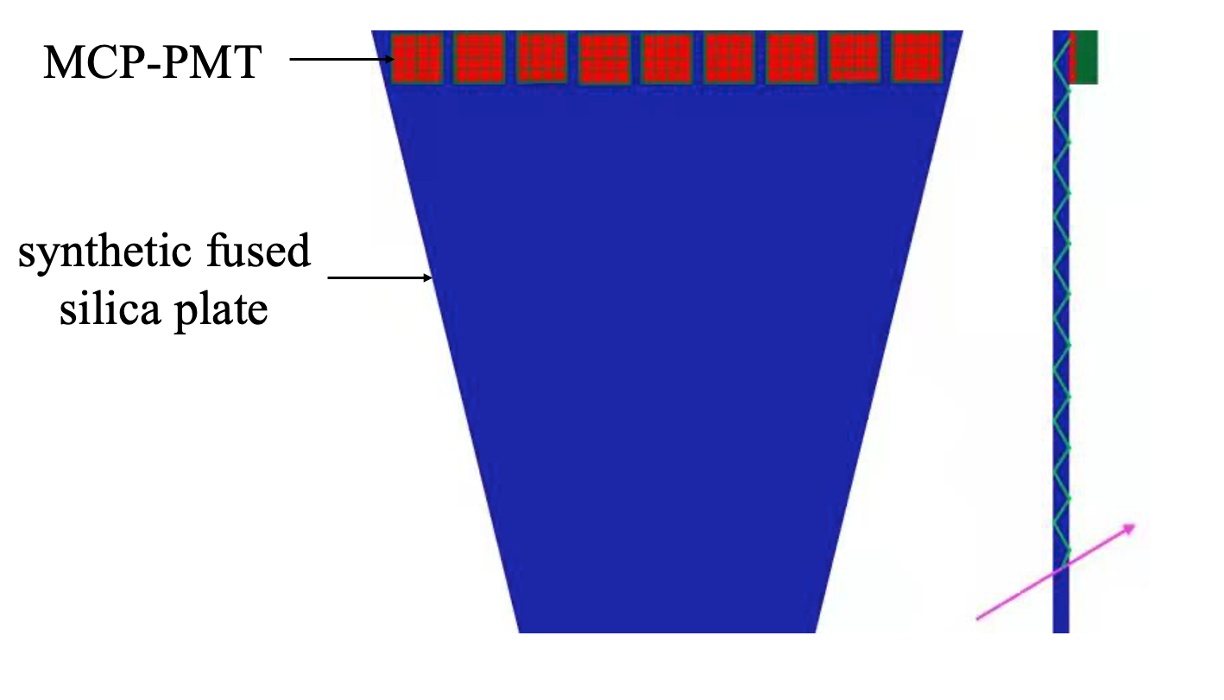}
\caption{An example of the radiator sector for the DTOF detector and the light path of the radiator.}
	\label{layout}
\end{figure}

The proposed DTOF detector consists of two identical endcap discs positioned at $\sim \pm 1400$~mm away from the collision point along the beam direction. Each disc is made up of several quadrantal sectors, as shown in Fig.~\ref{layout}, with an inner radius of $\sim 560$~mm and an outer radius of $\sim 1050$~mm, covering in polar angles of $\sim 22^\circ - 36^\circ$. The sensitive regions of the the DTOF and RICH detectors overlap, leaving no dead areas between the barrel and the endcap. In each sector, a synthetic fused silica plate is used as a radiator to generate Cherenkov photons. The supporting structure between sectors occupies $\sim 15~mm$ of space, approximately $1.25 \%$ of the total sensitive area. Considering the effect of the magnetic field on the photon sensors, an array of
multianode MCP-PMTs are optically coupled to the radiator along the outer side. Fig.~\ref{layout} also shows an example of the light path from a photon directly hitting the MCP-PMT. Note that there
are also alternative paths for photons reflecting off the lateral-side mirror.

\subsubsection{DTOF Time Resolution}

Time resolution is a key indicator of the performance of a DTOF detector.
It is necessary to analyze the factors that affect the timing uncertainty,
and the relative importance of the various factors must be investigated to optimize the time performance.
The main sources contributing to the timing uncertainty of the DTOF detector can be expressed by
\begin{equation}
\sigma^{2}_{\rm tot} \approx \sigma^{2}_{\rm trk} +\sigma^{2}_{\rm T_{0}} + (\frac{\sigma_{\rm elec}}{\sqrt{N_{PE}}})^{2} + (\frac{\sigma_{\rm TTS}}{\sqrt{N_{PE}}})^{2}  + (\frac{\sigma_{\rm det}}{\sqrt{N_{PE}}})^{2} 
\label{uncertainty}
\end{equation}
where $\sigma_{\rm trk}$ is the uncertainty caused by track reconstruction, which is $\sim 10$~ps;
$\sigma_{\rm T_{0}}$ is the event reference time (i.e. the time of $e^{+}e^{-}$ collision, $T_{0}$) uncertainty, which is $\sim 40$~ps (it is closely related to the bunch length, which is $\sim 12 mm$ in the current design, as depicted in the STCF accelerator conceptual design report);
$\sigma_{\rm elec}$ is the electronic timing accuracy;
$\sigma_{\rm TTS}$ is the single-photon transit time spread~(TTS) of the MCP-PMT;
and $\sigma_{\rm det}$ is the time reconstruction uncertainty of the DTOF detector.

\begin{figure}[!htb]
	\centering
	\includegraphics[width=0.6\textwidth]{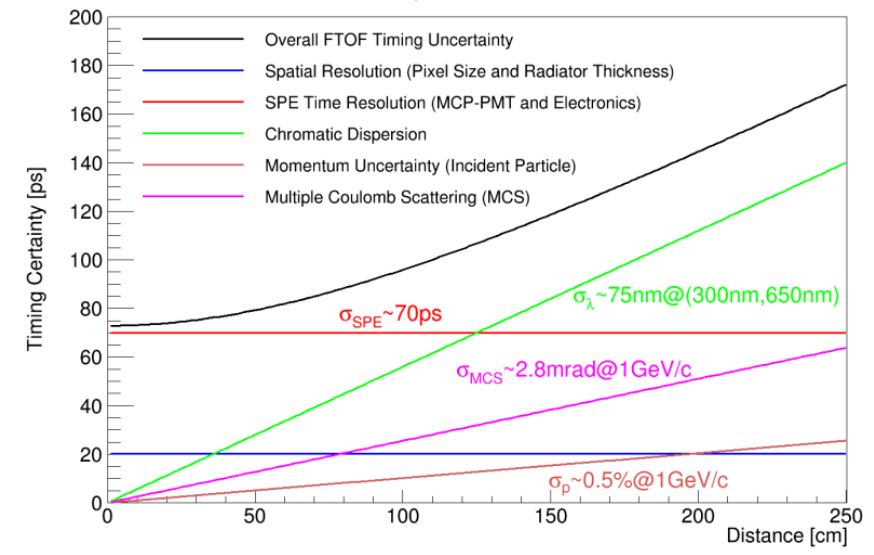}
\caption{Main DTOF timing error factors and their dependences on the distance from the incident point of the particle (kaon
at $p=1$~GeV/c) to the photon detector}
	\label{FIG:SPE_Resolution}
\end{figure}

The contribution from $\sigma_{\rm elec}$, $\sigma_{\rm TTS}$ and $\sigma_{\rm det}$
decreases with increasing $N_{PE}$, and the timing uncertainty of these three effects
can be estimated from the single photoelectron~(SPE) time resolution of
the photon sensor and the electronics, called $\sigma_{SPE}$.
The calculation results for the SPE parameter are shown in Fig.~\ref{FIG:SPE_Resolution} as a function of the photon transmission distance $D$. The timing jitter of the photon sensor plays a major role when $D$ is relatively short ($<1$~m), whereas the dispersion effect gradually becomes the
dominant factor when the distance from the incident point to the
photon sensor $D$ is large ($>1.5$~m). A proper optical design can be used; then, the dispersion effect can be corrected by position information if a very precise timing is maintained.
For the DTOF detector, a typical $D$ value is approximately $0.5-1$~m; hence, the timing uncertainty due to the dispersion effect is
smaller than the timing jitter of the photon sensor. This means that a compact design for the
DTOF with no focusing component is desirable. In addition, the spatial
resolution, including the thickness of fused silica and the pixel size of the photon sensor, has little
effect on the time uncertainty of the DTOF. Therefore, a photon sensor with a large pixel size can be
used to reduce the number of electronic readout channels. Notably, $N_{PE}$ may also
increase with the thickness of the radiator. However, this can cause an increase in the material
budget, although it has little influence on the SPE time resolution.

From the simulation, the average number of photons detected by the MCP-PMT arrays is $N_{PE}\approx 17$.
By applying a TOP-position calibration, where the average track length collected by each
sensor pixel and the average velocity of photons are used to calculate the TOP
(no dispersion effects are accounted for), a timing resolution of $\sim 20$~ps or better
for the latter three effects expressed in Eq.~\ref{uncertainty} can be obtained.
To further study the time resolution of the DTOF detector, a reconstruction algorithm is
required, which is presented in Sec.~\ref{DTOF_performance}.

\subsubsection{DTOF Detector Layout}
\begin{figure}[!htb]
	\centering
	\includegraphics[width=0.75\textwidth]{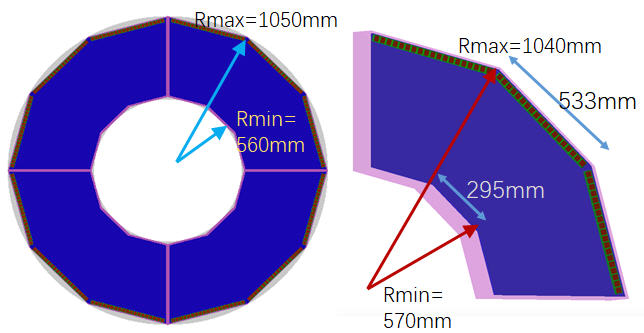}
	\includegraphics[width=0.20\textwidth]{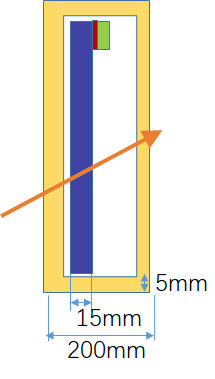}
\caption{The conceptual design of the DTOF detector.}
	\label{FIG:DTOF_Conceptual_Design}
\end{figure}

%{Layout of the DTOF}
\label{sec:dtof_layout}
Based on the results from both the simulation study and the experimental test, we developed the conceptual design of the DTOF at the STCF, as shown in Fig. \ref{FIG:DTOF_Conceptual_Design}. The detailed structure inside a sector is also depicted in Fig.~\ref{FIG:DTOF_Conceptual_Design}. The planar synthetic fused silica radiator is fan shaped and can be viewed as a composite structure of 3 trapezoidal units, each $\sim 295$~mm (inner side) $/$ $\sim 533$~mm (outer side) wide, $\sim 470$~mm high and $15$~mm thick. An array of $3 \times \sim 14-16$ multi-anode MCP-PMTs are directly coupled to the radiator along the outer side. The whole sector is enclosed in a light-tight black box made of $5$~mm thick carbon fiber, occupying $\sim 200$~mm space along the beam (Z) direction.

\subsection{DTOF Performance Simulation}
%{DTOF Performance}
\label{DTOF_performance}

\subsubsection{Reconstruction Algorithm}
\begin{figure}[!htb]
	\centering
	\includegraphics[width=0.6\textwidth]{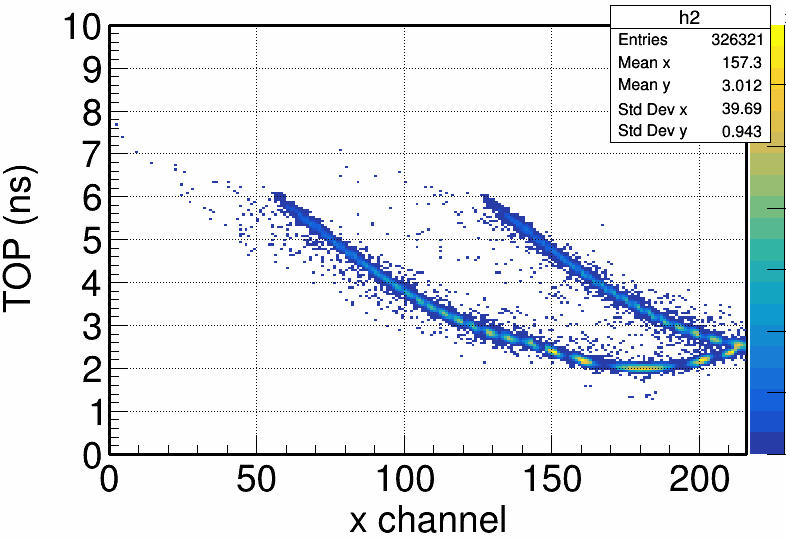}
\caption{The simulated TOP vs. hit position pattern of the DTOF detector.}
	\label{FIG:DTOF_TOP_Resolution}
\end{figure}

{\sc Geant4} simulations are performed to predict the performance of the DTOF detector.
A $20$~mm thick aluminum plate is added at a distance of $100$~mm in front of the DTOF detector to simulate the material budget of the MDC endcap. When tracking photon propagation, the inner and outer sides of the DTOF radiator are set to be absorptive, while the two lateral sides are set to be reflective (reflection factor $\sim 92\%$, typical for reflective coatings). The surface roughness of the radiator is simulated by randomizing the normal direction of the facet by $\sigma = 0.1^\circ$ (corresponding to a conservative average reflection factor of $\sim 97\%$). The wavelength dependence of the refractive index, absorption length for the radiator, and quantum efficiency of MCP-PMTs are carefully taken into account. Pion and kaon particles are emitted from the IP at different momenta and directions. Different polar angles, azimuth angles and particle momenta are tested. A typical Cherenkov photon hit pattern of pions at $p = 1$~GeV/c, $\theta = 23.66^\circ$ and $\phi = 15^\circ$ is displayed in Fig.~\ref{FIG:DTOF_TOP_Resolution}. A clear correlation between the time of propagation (TOP) and the hit position is demonstrated by the simulation. There are two bands in the figure: the lower left band represents direct photons without reflection, and the upper right band represents indirect photons with one reflection off the lateral side. %Editor: Please ensure that the intended meaning has been maintained in this edit.
Obviously, good separation between the two bands is obtained, except from a few sensors close to the edge.
% (as shown in Fig. \ref{FIG:DTOF_TOP_Resolution}).

\begin{figure}[!htb]
	\centering
	\includegraphics[width=0.55\textwidth]{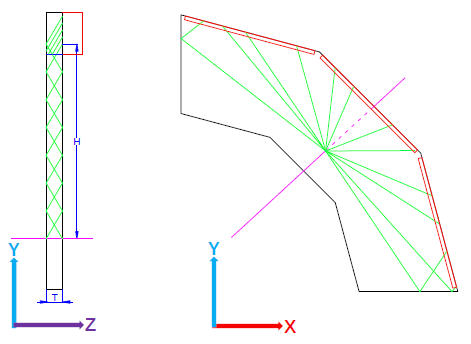}
	\includegraphics[width=0.35\textwidth]{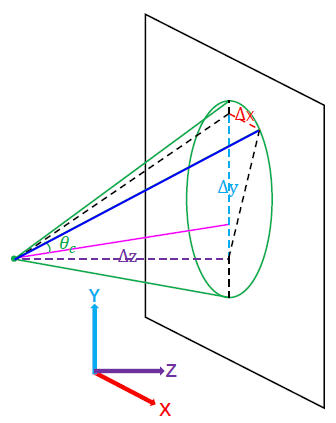}
\caption{The coordinate system used in DTOF reconstruction (left) and the direction of Cherenkov photons (deep blue line).}
	\label{FIG:DTOF_Reco_Coor}
\end{figure}

The DTOF reconstruction is performed in the coordinates shown in Fig.~\ref{FIG:DTOF_Reco_Coor} for one DTOF quadrant. According to the Cherenkov angle relation
\begin{equation}
\cos(\bar{\theta_{c}}) = \frac{1}{n_{p}\beta} = \frac{{\vec{v_{t}}} \cdot {\vec{v_{p}}}}{|\vec{v_{t}}| \cdot |\vec{v_{p}}|},
\end{equation}
where $\vec{v_{t}}=(a,b,c)$ is the incident particle velocity vector when the particle hits the radiator, $\vec{v_{p}}$ is the velocity vector of the emitted Cherenkov photons, $n_{p}$ is the refractive index of the radiator, and $\beta$ is the reduced speed of the particle.
The directional components of $\vec{v_{p}}$ can be expressed as $(\Delta{X},\Delta{Y},\Delta{Z})$, representing
the 3D spatial difference between the photon sensor pixel and the incident position of the particle on the radiator surface, as depicted in Fig. \ref{FIG:DTOF_Reco_Coor} (right). Although the 2D (X and Y) difference can be readily obtained, $\Delta{Z}$ must be deduced with a certain particle species hypothesis. If $V = \cos(\bar{\theta_{c}})$ is known, the equation for $\Delta{Z}$ can be expressed as
\begin{equation}
\label{eq::DTOF-DZ-Solver}
(c^{2}-V^{2})\Delta{Z}^{2} + 2c(a\Delta{X}+b\Delta{Y})\Delta{Z} + (a\Delta{X}+b\Delta{Y})^{2} - V^{2}(\Delta{X}^{2}+\Delta{Y}^{2}) = 0.
\end{equation}
By solving this equation, we find $\Delta{Z}=\frac{-B\pm\sqrt{B^{2}-4AC}}{2A}$, with $A=c^{2}-V^{2}$, $B=2c(a\Delta{X}+b\Delta{Y})$ and $C=(a\Delta{X}+b\Delta{Y})^{2} - V^{2}(\Delta{X}^{2}+\Delta{Y}^{2})$. To obtain a real solution, $\Delta = B^{2}-4AC \ge 0$ is required. Furthermore, after the physical cuts $V > 0$ (Cherenkov photons are forwardly emitted) and $\frac{\Delta{X}^{2}+\Delta{Y}^{2}}{\Delta{X}^{2}+\Delta{Y}^{2}+\Delta{Z}^{2}} \ge \frac{1}{n_{p}^{2}}$ (internal total reflection is ensured) are applied, the minimal solution of Eq.~\ref{eq::DTOF-DZ-Solver}, ($\Delta{Z} = \min{|\Delta{Z}_{1}|,|\Delta{Z}_{2}|}$) is taken as the optimum.

\begin{figure}[!htb]
	\centering
	\includegraphics[width=0.45\textwidth]{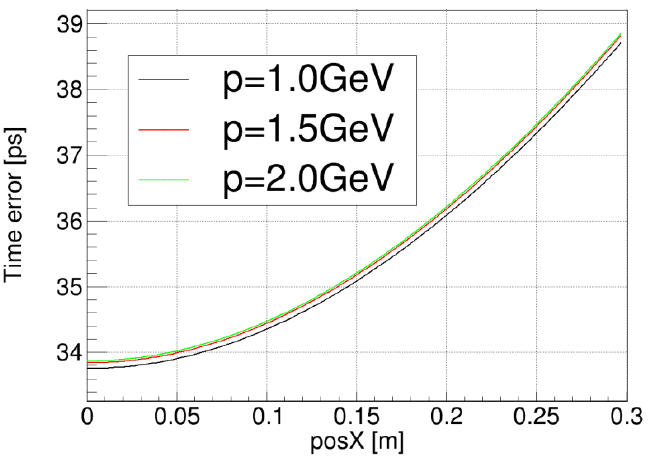}
	\includegraphics[width=0.45\textwidth]{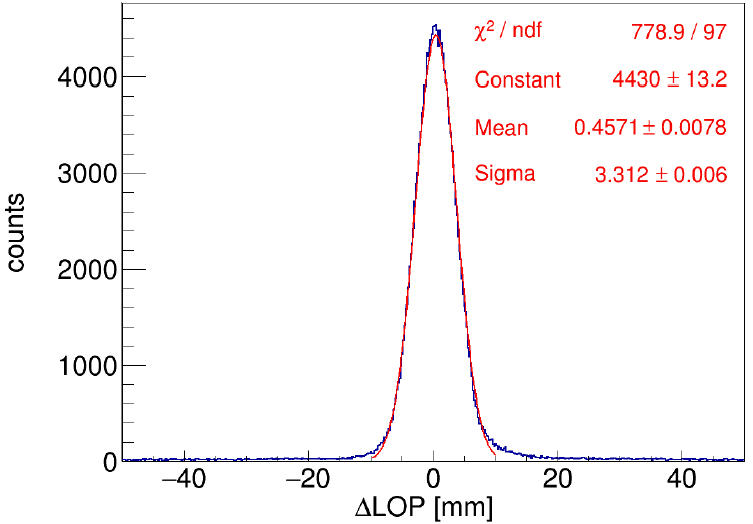}
\caption{The expected timing error and propagation length uncertainty of Cherenkov photons in a DTOF quadrant.}
	\label{FIG:DTOF_Reco_TLerr}
\end{figure}

The timing error of such an approach is estimated by adding up the possible factors, such as the dispersion effect, the finite photon sensor size and the propagation length of photons inside the radiator. The expected timing uncertainty for a pion of $p \sim 1 - 2$~GeV/c crossing the radiator perpendicularly, with $H = 0.5$~m (in the coordinate system defined in Fig.~\ref{FIG:DTOF_Reco_Coor}), is shown in Fig.~\ref{FIG:DTOF_Reco_TLerr} for sensors at different positions. The pitch of the photon sensor is $5.5$~mm. No multiple Coulomb scattering (MCS) effects are accounted for. It is obvious that the intrinsic detector timing uncertainty is no more than $40$~ps with this DTOF structure. Furthermore, we find that the reconstructed length of propagation (LOP) of light inside the radiator agrees with the MC truth to a precision of $\sim 3.3$~mm, as also shown in Fig.~\ref{FIG:DTOF_Reco_TLerr}. The reconstruction algorithm works well for most photon sensors independent of the incident position of the particles, except for a few sensors near the lateral side.

\begin{figure}[!htb]
	\centering
	\includegraphics[width=0.45\textwidth]{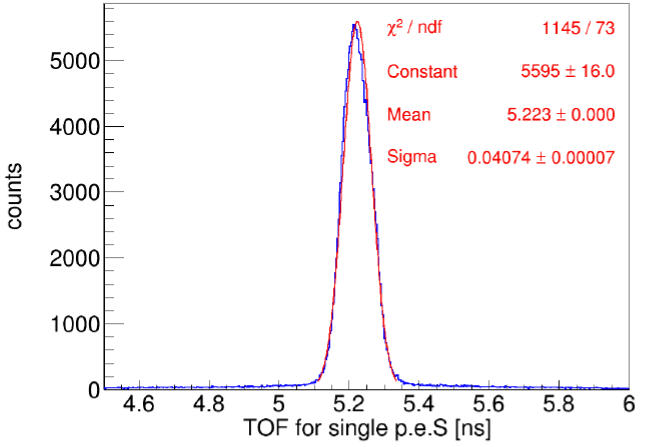}
	\includegraphics[width=0.45\textwidth]{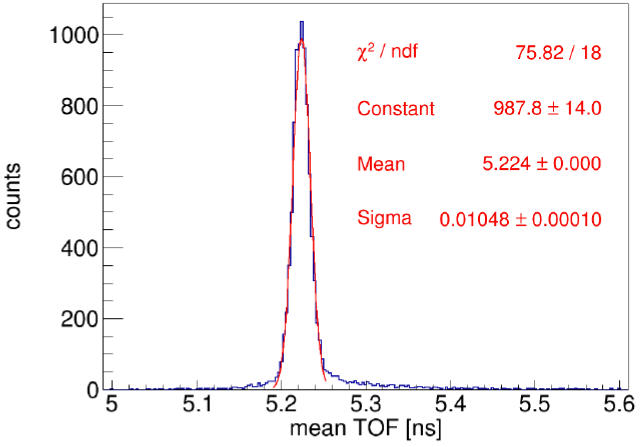}
\caption{The TOF resolution of the DTOF detector for a single photoelectron and the average of all photons.}
	\label{FIG:DTOF_Reco_TOFreso}
\end{figure}

By applying the formula below
\begin{equation}
\label{eq::DTOF-TOF}
TOF = T - TOP - T_{0} = T - \frac{LOP}{v_{g}} - T_{0} = = T - \frac{LOP\times\bar{n_{g}}}{c} - T_{0},
\end{equation}
where $v_{g}$ is the group velocity of Cherenkov light in the radiator, the TOF information is obtained. Figure~\ref{FIG:DTOF_Reco_TOFreso} shows the time resolution of the DTOF detector for a SPE and the average of all photons without taking into account the timing jitter of the MCP-PMTs and electronics. For an SPE, the intrinsic time resolution of the DTOF is $\sim 41$~ps. Averaging the timing information over $\sim$~18 detected photons, the timing jitter shrinks to $\sim$~10~ps. It is also noted that in the TOF distribution plot, a low (but visible) long tail appears on both sides of the main peak. The tail is mainly caused by secondary particles along the primary pion, mostly $\delta$-electrons.

\begin{figure}[!htb]
	\centering
	\includegraphics[width=0.45\textwidth]{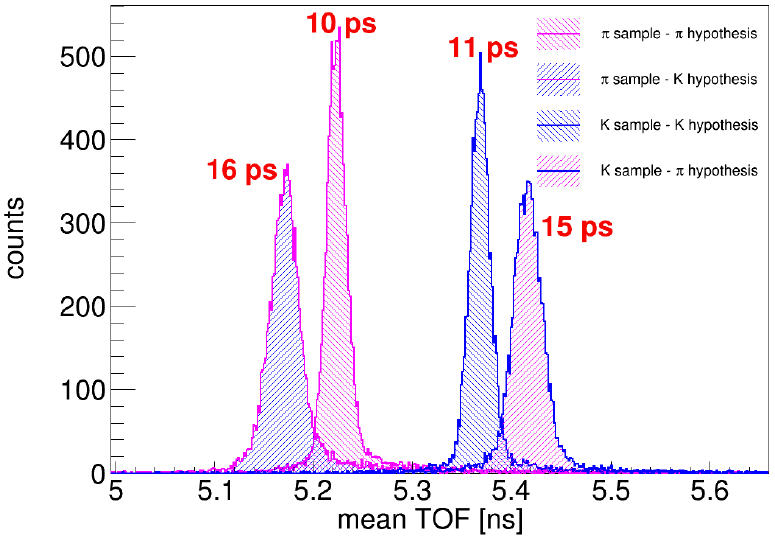}
	\includegraphics[width=0.45\textwidth]{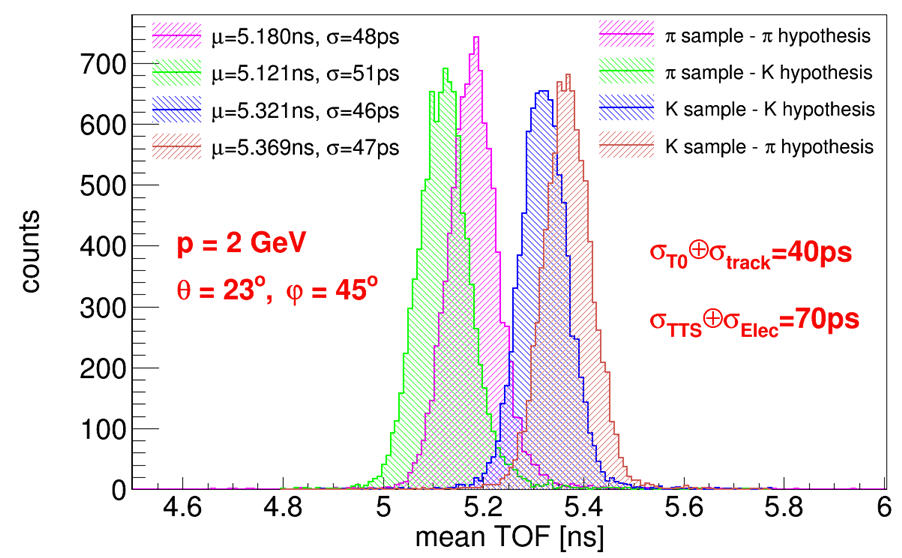}
\caption{The TOF PID capabilities of the DTOF detector for $\pi/K$ separation at $2$~GeV/c, without (left) and with (right) contributions from other timing uncertainties.}
	\label{FIG:DTOF_Reco_TOFPID}
\end{figure}

The TOF information is deduced and compared to the expectation of each particle hypothesis. Figure~\ref{FIG:DTOF_Reco_TOFPID} shows the reconstructed intrinsic TOF distributions of both pions and kaons at $2~$GeV/c. We can easily find if the particle hypothesis is correct the reconstructed TOF peak is at the correct position, with a resolution of $\sim$~10~ps. However, if the hypothesis is not correct, the reconstructed TOF peak is shifted with respect to the expectation. The shift makes the separation between the pion and kaon TOF peaks even larger, which may benefit the PID power. When convoluting all contributing factors, the overall reconstructed TOF time resolution is $45\sim50$~ps, as shown in Fig. \ref{FIG:DTOF_Reco_TOFPID}. Directly comparing the TOF information shows that a $3.0\sigma$ separation power for $\pi/K$ at $2$~GeV/c is achieved. Furthermore, the separation power becomes stronger if we compare the reconstructed TOFs of various hypotheses for the same set of particles, mainly due to the beneficial time shift under the incorrect hypothesis (as in Fig. \ref{FIG:DTOF_Reco_TOFPID}).

\subsubsection{Expected Performance}

\begin{figure}[!htb]
	\centering
	\includegraphics[width=0.45\textwidth]{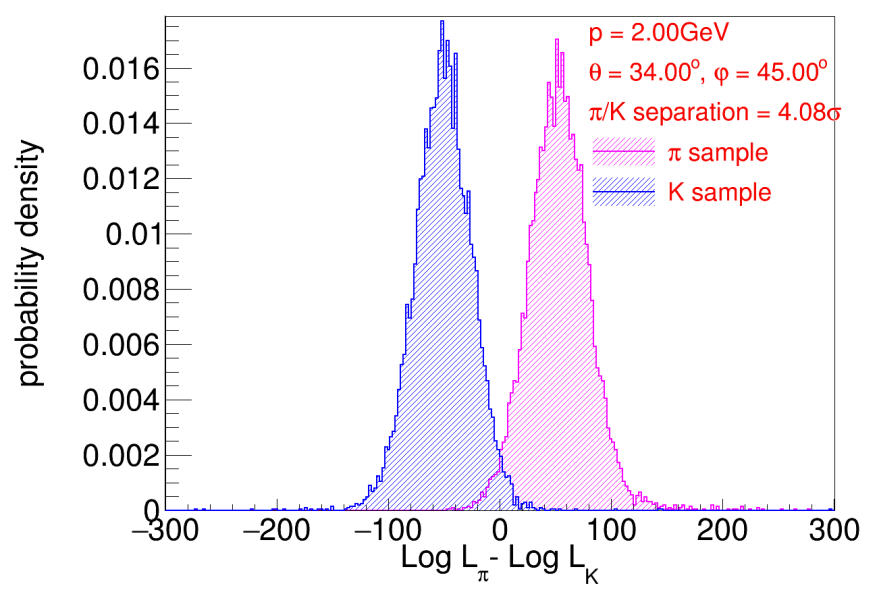}
	\includegraphics[width=0.45\textwidth]{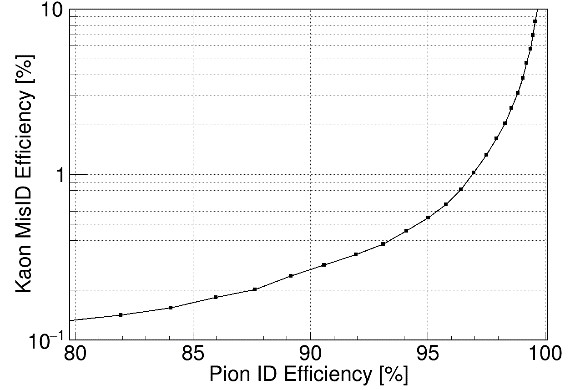}
\caption{The likelihood PID capabilities of the DTOF detector for $\pi/K$ separation at $2$~GeV/c emitted at different angles.}
	\label{FIG:DTOF_Reco_LHPID}
\end{figure}

\begin{figure}[!htb]
	\centering
	\includegraphics[width=0.95\textwidth]{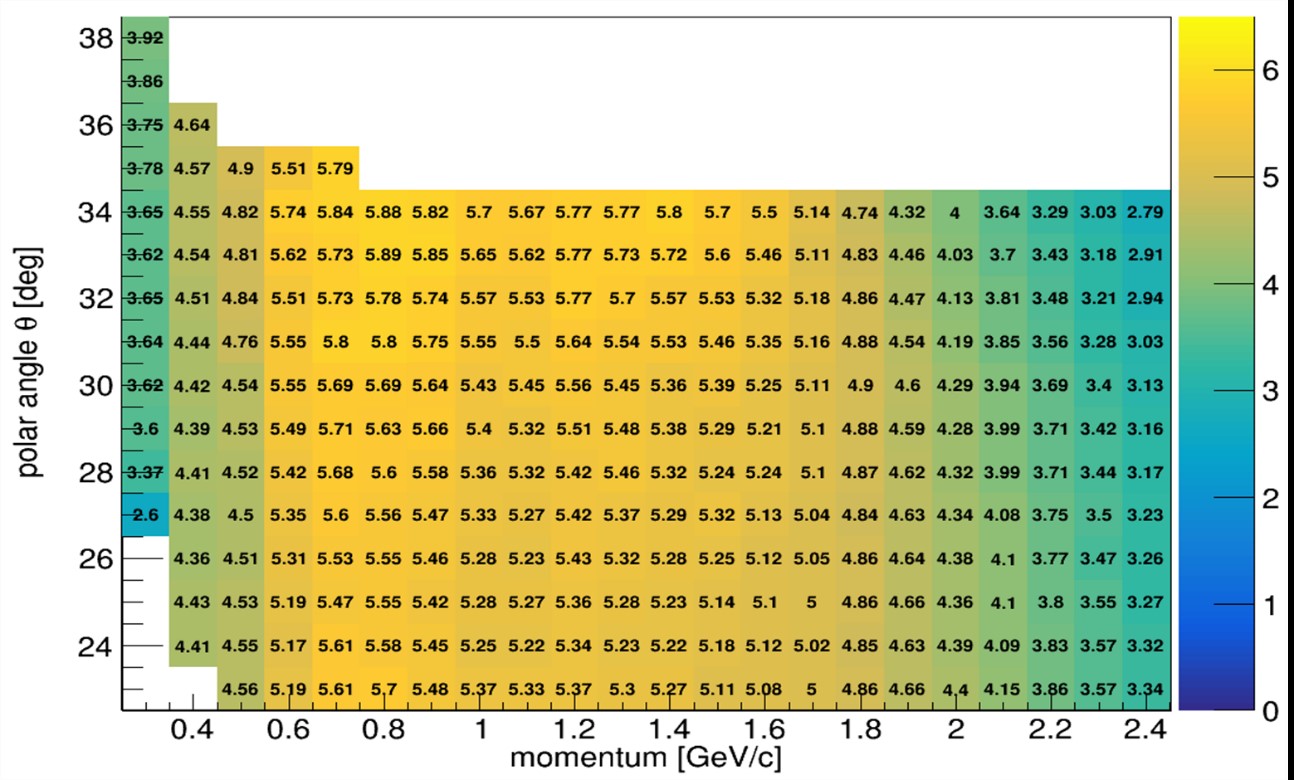}
\caption{The likelihood PID capabilities of the DTOF detector for $\pi/K$ separation in different directions and at different momenta.}
	\label{FIG:DTOF_Reco_LHPID_Scan}
\end{figure}

To evaluate the PID capabilities of the DTOF detector, we apply a likelihood method. The likelihood function is constructed by
\begin{equation}
\label{eq::DTOF-Likelihood-Function}
\mathcal{L}_{h} = \Pi^{i}_{i=1}f_{h}(TOF^{h}_{i}), ~~~~~
\Delta\mathcal{L} = \mathcal{L}_{\pi} - \mathcal{L}_{K},
\end{equation}
where $h$ denotes hadron species (in our case, $\pi$ and $K$) and $i$ accounts for each detected photon. The probability density function $f_{h}$ is taken as a Gaussian fit to the expected TOF distribution (as in Fig. \ref{FIG:DTOF_Reco_TOFPID} (right)), plus a constant background of 0.05. Fig.~\ref{FIG:DTOF_Reco_LHPID} (left) shows an example of the reconstructed $\Delta\mathcal{L}$ for $2$~GeV/c $\pi$ and $K$, where a $\pi/K$ separation power better than $4\sigma$ is demonstrated. Fig.~\ref{FIG:DTOF_Reco_LHPID} (right) shows the corresponding identification efficiency under this condition. When a kaon misidentification efficiency of $<2\%$ is required, a pion identification efficiency of $>98\%$ can be achieved, fulfilling the STCF PID requirement. Furthermore, with improved PID algorithm that includes also the spatial hit pattern information, the PID performance of DTOF can extend to even higher momentum range. Fig.~\ref{FIG:DTOF_Reco_LHPID_Scan} shows the likelihood PID capabilities of the DTOF detector for $\pi/K$ separation in different directions and different momenta. Despite the very different particle directions and momenta, a separation power of $\sim4\sigma$ or better over the full DTOF sensitive area is achieved.

\subsubsection{$T_0$ Determination}
\label{sec:dtof_t0}
\begin{figure}[!htb]
	\centering
	\includegraphics[width=0.95\textwidth]{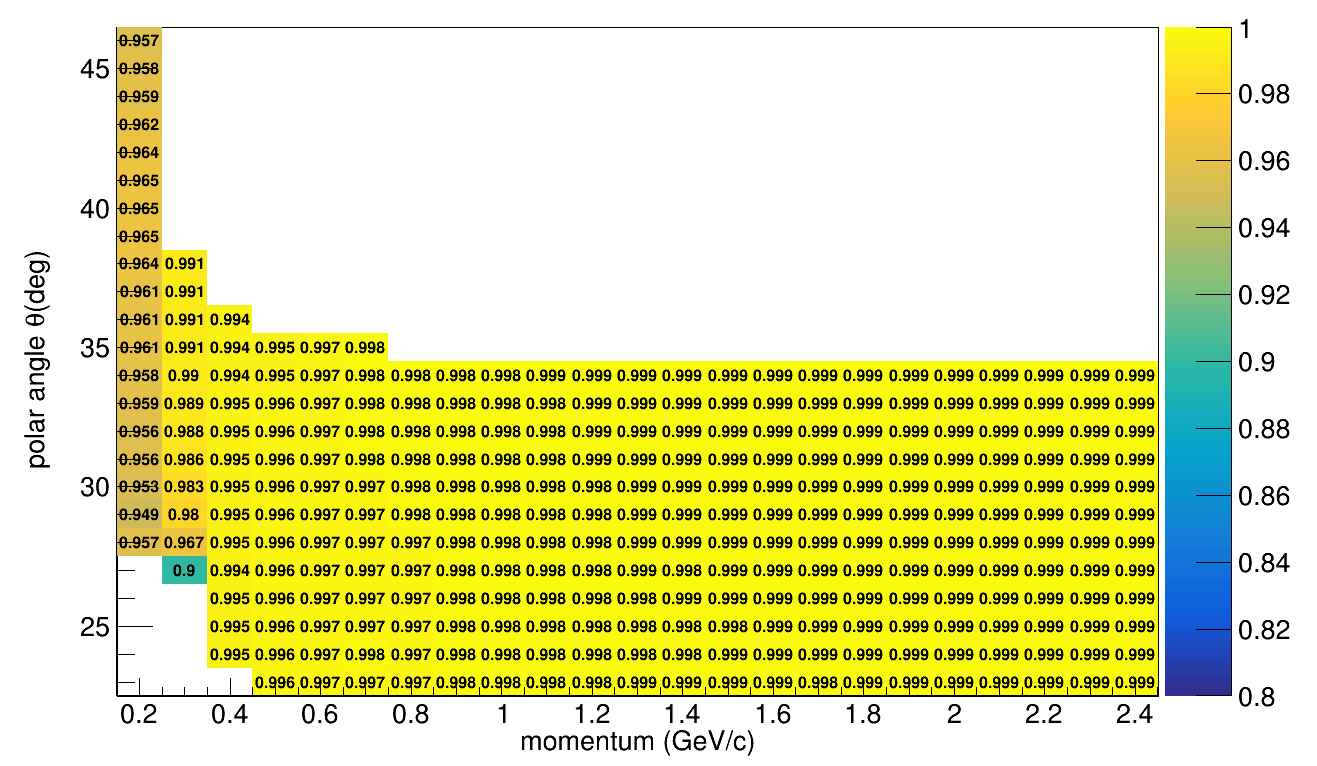}
\caption{The rate at which $T_{0}$ is determined correctly using the DTOF detector for $\pi$ samples in different directions and at different momenta.}
	\label{FIG:DTOF_T0_Reco}
\end{figure}

The DTOF detector's excellent timing performance can help determine the $e^{+}e^{-}$ collision time ($T_{0}$). According to Eq.~\ref{eq::DTOF-TOF}, the TOF of incident particles and the TOP of Cherenkov photons can be obtained from simulations or calculations with different particles and different $T_{0}$ hypotheses, provided that the track information is known. To evaluate different hypotheses, we define a likelihood function,

\begin{equation}
	\label{eq::T0-Likelihood-Function}
	\mathcal{L}_{h_{1}, h_{2}} = \Pi^{i}_{i=0}f_{h_{1}, h_{2}} (ch_{i}, T_{i}), 
\end{equation}

where $h_{1}$ denotes hadron species (in our case, $\pi$ and $K$) and $h_{2}$ denotes different $T_{0}$ (the bunches collide every $4 ns$, i.e., $T_{0} = 0$, $\pm 4 ns$, $\pm 8 ns$,...). The probability density function $f_{h_{1}, h_{2}} (ch_{i}, T_{i})$ is the photon arrival-time distribution of different channels. The ${L}_{h_{1}, h_{2}}$ with different particle and $T_{0}$ hypotheses is compared, and the $T_{0}$ candidate with the maximum likelihood is determined.
Fig.~\ref{FIG:DTOF_T0_Reco} shows the $T_{0}$ determination efficiency using the DTOF detector for $\pi$ samples in different directions and at different momenta. Despite the very different particle directions or momenta, $T_{0}$ can be determined correctly with an efficiency $>99\%$. Note that if the particle velocity is below the Cherenkov threshold, the corresponding $T_{0}$ cannot be correctly pinpointed. With decreasing momentum, the time resolution of the DTOF detector worsens, resulting in a decrease in the $T_{0}$ determination efficiency.

\subsection{DTOF Structure Optimization}

Different geometric parameters of the DTOF detector are tested to study their effects on PID. The $\pi$/K separation powers of different geometry configurations are compared with the reconstruction algorithm and likelihood method described above. The geometry configurations studied are listed in Table~\ref{TAB:DTOF_GEO_CONFIG}. We study the effects of three main factors: radiator shape/size, radiator thickness and the setting of mirrors.

%%%%%%%%%%%%%%%%%  TABLE  %%%%%%%%%%%%%%%%%%%%%%%%
\begin{table*}[hptb]
	\small
	\caption{Description of the different DTOF geometry configurations, where A stands for absorber and R for reflective mirror.}
	\label{TAB:DTOF_GEO_CONFIG}
	\vspace{0pt}
	\centering
	\begin{tabular}{llllllll}
	\hline
	\thead[l]{Configuration/Geometry ID} & \thead[l]{0} & \thead[l]{1} & \thead[l]{2} & \thead[l]{3} & \thead[l]{4} & \thead[l]{5} & \thead[l]{6} \\
	\hline
	Radiator shapes (sector number)	&4        &12	&24	&4	&4	    &4  &4 \\
	Radiator thickness (mm)	    &15	    &15   	&15      &10     &20    &15   	&15   \\
	Outer side surface	&A	&A    	&A    	&A      	&A     	&R   &$45^{\circ}$ R  \\
	Inner side surface	 &A	&A    	&A    	&A      	&A     	&A   &A  \\
	Lateral side surface	&R	&R    	&R    	&R      	&R    	&R   &R  \\
	\hline	\end{tabular}
\end{table*}
%%%%%%%%%%%%%%%%%%%%%%%%%%%%%%%%%%%%%%%%%%%%%%%%%%

\begin{figure}[!htb]
	\centering
	\includegraphics[width=0.75\textwidth]{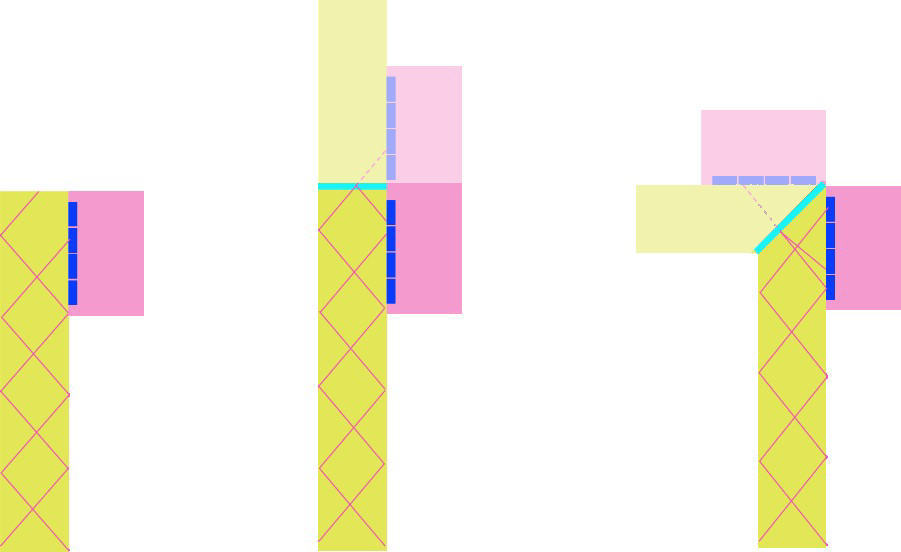}
\caption{Three different configurations on the outer surface of the radiator. An absorber (left) or mirror (middle) on the outer surface and a mirror on the $45^\circ$ chamber of the outer side surface (right).}
	\label{FIG:DTOF_OPTIMIZATION}
\end{figure}

In Table~\ref{TAB:DTOF_GEO_CONFIG}, three different radiator shapes (and sizes) correspond to Geometries 0, 1 and 2, where the DTOF disc of Geometry 0 is made up of 4 quadrant sectors, as in Sec.~\ref{sec:dtof_layout}, and for Geometries 1 and 2, the DTOF disc is made up of 12 and 24 trapezoidal sectors, respectively. Each sector includes readouts from 18 and 8 MCP-PMTs for Geometries 1 and 2, respectively. The effect of the radiator thickness is studied by comparing Geometries 0, 3 and 4. The radiator thicknesses are 15 mm, 10 mm and 20 mm, respectively. For all the geometric configurations listed in Table~\ref{TAB:DTOF_GEO_CONFIG}, the inner/lateral side surfaces of the radiator are covered by an absorber/reflective mirror, labeled A/R, respectively. For the outer side surface, a mirror can extend the acceptance of Cherenkov light, which increases the number of detected photons. As shown in Fig.~\ref{FIG:DTOF_OPTIMIZATION}, in Geometry 5, a mirror is attached to the outer side surface of the radiator, which is equivalent to putting mirror MCP-PMTs parallel to the real tubes, and in Geometry 6, a mirror is placed on the $45^{\circ}$ chamfer, which is equivalent to putting mirror MCP-PMTs perpendicular to the real tubes.

The key results regarding the DTOF performance for different geometry configurations are listed in Table ~\ref{TAB:DTOF_GEO_PERFORM} for $\pi/K$ mesons at $p = 2$~GeV/c, $\theta = 24^{\circ}$ and $\phi = 45^{\circ}$.

%%%%%%%%%%%%%%%%%  TABLE  %%%%%%%%%%%%%%%%%%%%%%%%
\begin{table*}[hptb]
	\small
	\caption{Performance of different geometries at $p = 2$~GeV/c, $\theta = 24^{\circ}$ and $\phi = 45^{\circ}$.}
	\label{TAB:DTOF_GEO_PERFORM}
	\vspace{0pt}
	\centering
	\begin{tabular}{llllllll}
		\hline
		\thead[l]{Configuration/Geometry ID} & \thead[l]{0} & \thead[l]{1} & \thead[l]{2} & \thead[l]{3} & \thead[l]{4} & \thead[l]{5} & \thead[l]{6} \\
		\hline
		$N_{pe}$ for pions	&21.8       &21.9	&17.0	&15.5	&25.7	    &33.2  &38.7 \\
		Accumulated charge density on     &10.8	    &10.5   	&9.6      &8.8     &11.8    &17.0   	&25.6   \\
		MCP-PMT anode ($C/{\rm cm}^{2}$)	    &	    &   	&      &     &    &   	&   \\
		$\pi$/K separation power	&$4.17 \sigma$	&$4.08 \sigma$	&$3.66 \sigma$	&$3.99 \sigma$	&$4.27 \sigma$	&$4.26 \sigma$   &$4.19 \sigma$  \\
		\hline	\end{tabular}
\end{table*}
%%%%%%%%%%%%%%%%%%%%%%%%%%%%%%%%%%%%%%%%%%%%%%%%%%

The effect of radiator shape/size is studied by comparing the DTOF performance with Geometries 0, 1 and 2. As the radiator becomes smaller, the reflection time of Cherenkov light off the lateral-side mirror increases, which causes more photon losses and, more importantly, ``confusion'' in LOP reconstruction. Geometry 0 has the best $\pi/K$ separation power, $4.17 \sigma$, while Geometry 2 has the worst, $3.66 \sigma$, indicating that a larger radiator is preferred.

Geometries 3 and 4 have different radiator thicknesses than Geometry 0, which affects the photon yield. Although more detected photons mean better time resolution and PID performance, we favor the 15 mm thick radiator, as it offers the best balance between reducing the impact of the material budget on EMC and providing performance redundancy, e.g., for reducing the influence of the detector aging effect in long-term operation.

To increase the photon yield, mirrors are attached to the outer side surface of the radiator in different ways in Geometries 5 and 6. The numbers of p.e. received in these two geometries are $\sim33$ and $\sim39$, which are much higher than that in Geometry 0. However, the mirror also increases the number of possible light paths, which causes ``confusion'' similar to the effect of multiple reflections off the lateral-side mirror and degrades the time resolution. Therefore, even with more detected photons, the $\pi/K$ separation powers of Geometries 5 and 6 are similar to that of Geometry 0. In addition, the accumulated charge densities of these two geometry configurations are much higher, which affects the lifetime of the MCP-PMTs. Thus, options with mirrors attached to the outer-side surface of the radiator are not favorable.

According to the above comparison, the optimum geometry configuration of the DTOF detector is Geometry 0.
It is worth noting that the large radiator of Geometry 0 may result in inefficiency when two tracks hit the same radiator in one event. The rate of inefficiency due to such pile-ups is studied in two kinds of MC samples, $J/\psi\to anything$ and $\psi(3770)\to D^{0}\bar{D}^{0}\to anything$. By requiring the polar angle
of the charged tracks to be within $(22^{\circ}, 36^{\circ})$, the multiplicity of these endcap tracks is examined, where
one track is dominant and three tracks are negligible. For the case of two tracks, if the difference in their
azimuth angles is within $(-45^{\circ}, 45^{\circ})$, it is treated as a pile-up event and causes inefficiency.
These studies show that the rates of inefficiency for these two MC samples are 0.36\% and 0.63\%, respectively.
It can be concluded that the large radiator of the DTOF design does not affect the reconstruction efficiency for most
physics programs and is acceptable at the STCF.
Therefore, Geometry 0 can be chosen as the baseline design.

\subsection{Background Simulation}

\begin{figure}[!htb]
	\centering
	\includegraphics[width=0.65\textwidth]{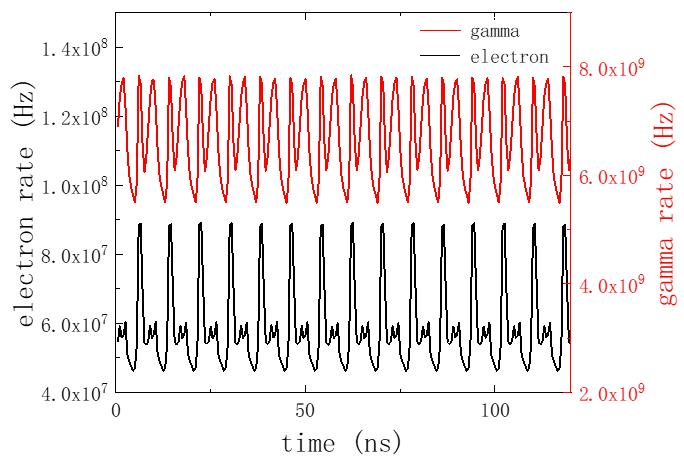}
\caption{Overall time distribution of background particles hitting the DTOF detector.}
	\label{FIG:DTOF_ALLBKG_T}
\end{figure}

The distributions of background particles obtained by the dedicated MDI and background study, as discussed in Sec.~\ref{sec:mdi_bkg}, are used as input in a DTOF {\sc Geant4} simulation and then to simulate the effect of background on DTOF performance. The background particles influencing the DTOF detector are mainly secondary gammas and electrons. Their hit rates are approximately $7\times10^{9}$~Hz and $6\times10^{7}$~Hz, respectively, including two main parts: the beam-induced background ($75\%$) and the physical background ($25\%$). The probability that the background particle generates single photon-electron signal is very low,  and the simulated background value is approximately 10 hits per quadrantal sectors in a 100~ns time window, based on the following simulations. This is consistent with data shown in Table~\ref{tab:TIDNIEL_mean} and Table~\ref{tab:TIDNIEL_max}. The beam-induced background is uniformly distributed in time, while the physical background is related to collisions during bunch crossing (once per 8 ns), exhibiting a characteristic time structure. With Monte Carlo sampling of the physical background time distribution combined with the uniform time distribution of the beam-induced background, the overall time distribution of background hits on the DTOF detector can be obtained, as shown in Fig.~\ref{FIG:DTOF_ALLBKG_T}.

$\pi$ and K particles are generated in the {\sc Geant4} simulation along with the background samples to study the effect of the background. In the {\sc Geant4} simulation, the time window of the signal acquisition is 100 ns and is placed within the interval [-40 ns, 60 ns] so that the real signal is in the middle of the time window. In this time window, the number of background particles is given by a Poisson distribution, and the time distribution is sampled according to Fig~\ref{FIG:DTOF_ALLBKG_T}. The background may greatly increase the number of photoelectrons detected by the DTOF detector for a single event, resulting in an increased possibility of multiple hits in a single channel. The correction of multiple hits is applied, which means that in the time window of [-40 ns, 60 ns], only the first arriving photoelectron signal is taken, and all other hits are dismissed. With the background hits taken into account and assuming an MCP-PMT gain of $10^{6}$, the average accumulated charge density on the MCP-PMT anode is 12~C/cm$^{2}$ over 10 years of STCF operation ($50\%$ run time). Under such a background level, the radiation effects of the quartz radiator and the MCP-PMTs should be small according to studies of other DIRC and DIRC-like detectors.

\begin{figure}[!htb]
	\centering
	\includegraphics[width=0.45\textwidth]{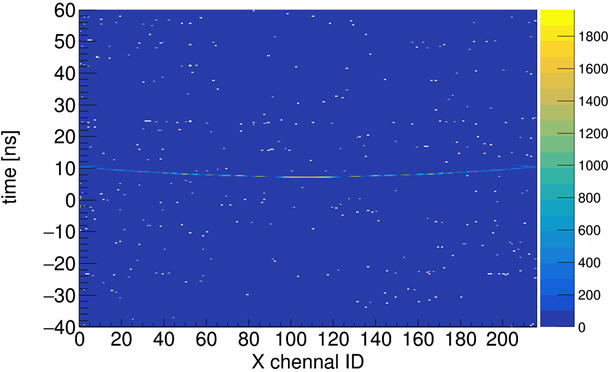}
	\includegraphics[width=0.45\textwidth]{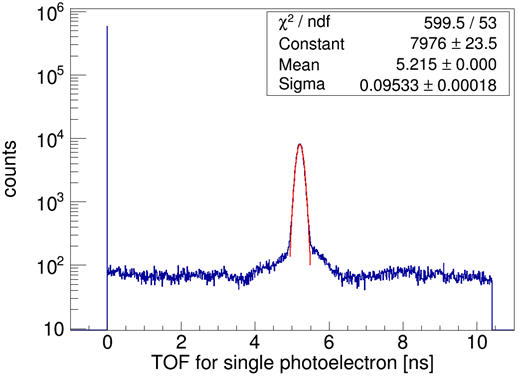}
\caption{2-D time-position map of DTOF hits (left) and reconstructed TOF distribution of a single photoelectron signal (right), with multiple-hit correction.}
	\label{FIG:DTOF_XTHIT_BKG}
\end{figure}

\begin{figure}[!htb]
\centering
\includegraphics[width=0.65\textwidth]{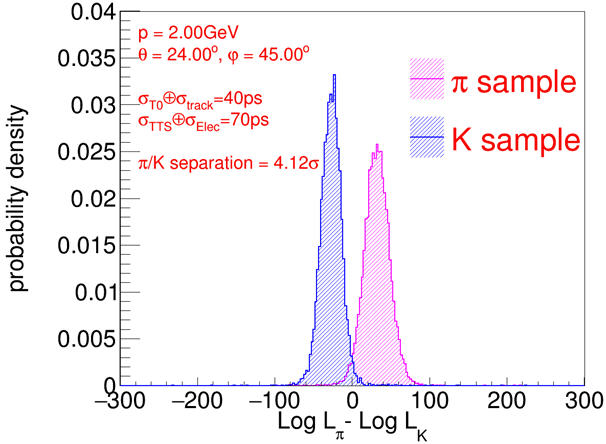}
\caption{$\pi/K$ identification capabilities (at $2$~GeV/c) of DTOF with multiple-hit correction.}
\label{FIG:DTOF_PID_BKG}
\end{figure}

Fig.~\ref{FIG:DTOF_XTHIT_BKG} shows the 2D time-position map of DTOF hits. Hits by the background particles are uniformly distributed throughout the phase space, while the real signal hits are concentrated as bands. After the time reconstruction, the TOF distribution of a single photoelectron signal can be obtained, as shown in Fig.~\ref{FIG:DTOF_XTHIT_BKG}. The reconstructed TOF of the real signal is a Gaussian distribution ($\sigma \sim100$ ps), while the TOFs of background particles are uniformly distributed. Some single-photon electrons with zero TOF do not meet the reconstruction conditions and are taken as background. Due to the uniform distribution of the reconstructed background signal, the influence of the background can be eliminated by using the maximum likelihood method. The $\pi$/K resolution is found to be $4.1 \sigma$, as shown in Fig.~\ref{FIG:DTOF_PID_BKG}.

Considering the Poisson fluctuation of the background count, the $\pi/K$ resolution remains at
4.12$\sigma$ even under an extreme condition of three standard deviations above the average background
level, {\it i.e.}, the background hit rate is increased to $2.1\times10^{8}$~Hz per single DTOF disc.
 Although the number of photoelectrons from the background increases dramatically, the final effect on $\pi/K$ separation is fairly small.

\subsection{Readout Electronics}
\label{FTOF_ELECTRONICS}

In MCP-PMT-based DTOF detectors, particle identification relies on the Cherenkov photon arrival time rather than the position information of the MCP-PMTs. Therefore, high timing resolution is an essential feature of DTOF detectors. Furthermore, DTOF detectors usually have a large channel number and require a track timing resolution below 30~ps, which is a great challenge for front-end electronics.

\begin{figure}[!htb]
	\centering
	\includegraphics[width=0.8\textwidth]
{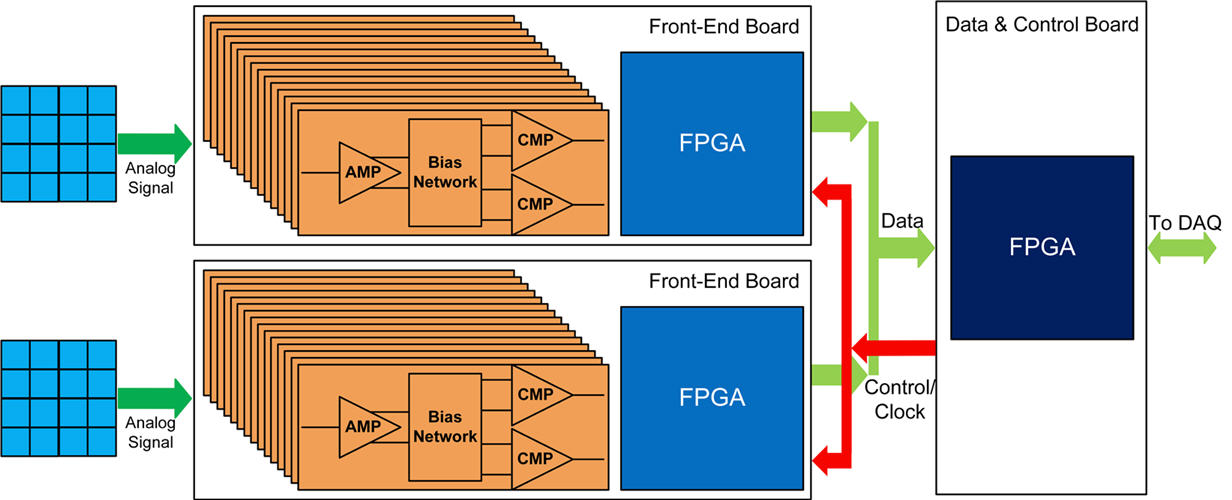}
\caption{The preliminary structure of the DTOF readout electronic system.}
	\label{FTOF_electronics_architecture}
\end{figure}

The preliminary structure of the DTOF readout electronics is shown in Fig.~\ref{FTOF_electronics_architecture}. The readout electronics consist of the front-end board and the data-control board. The front-end board utilizes a time over multithreshold scheme to extract timing information from analog signals. The signals from the MCP-PMTs are preamplified first. The gains of the amplifiers are set independently to compensate for the gain variations of the individual MCP-PMT channels. The amplified signals are fed into high-performance comparators, each of which has a different threshold. The outputs of these comparators are fed into the FPGA. The time-to-digital converter~(TDC) module is implemented in the FPGA. The TDC measures the arrival time of both edges of the comparator outputs with high accuracy. Then, the front-end board passes the resulting binary data stream to the data-control board. The data-control board not only collects data streams from front-end boards but also sends out a high-performance clock and control signal to the front-end boards. According to the structure described above, we can briefly estimate the amount of data that the readout electronic system feeds into the subsequent DAQ system. Assuming that the event rate of the DTOF detector is 80M hits/s, the data rate output to the subsequent DAQ system can be simply calculated to be 960 MB/s without considering the effects of the detector background noise and crosstalk.

For high time resolution, it is crucial that the front-end electronics have high bandwidth and a high SNR. The time resolution of the FPGA-based TDC was found to be 3.9~ps in a previous work of ours~\cite{fastTDC}. We therefore implement the TDC that can measure narrow pulse widths, of which the time resolution will be better than 5~ps. The TDC performance in the radiation environment will be evaluated. The results of the evaluation will affect our system structure design. A stable, low-jitter detector-wide clock distribution network also needs to be developed. Its long-term stability, short-term stability, and temperature stability will be carefully evaluated. The jitter of the clock distribution network, which is better than 15~ps, satisfies the design requirements. To further improve the measurement resolution of the leading-edge timing, the multithreshold TOT 
method may be used.

The DTOF readout electronic system consists of many front-end boards that need to be synchronized, so the jitter of the distribution clock network is a key parameter of the electronic system. It is necessary to develop a clock distribution network that meets the requirements of the DTOF detector based on existing technology. The conventional White Rabbit network
can provide subnanosecond accuracy and tens of picosecond precision for synchronization.
To achieve a high synchronization accuracy, a new optical fiber
link and synchronization scheme based on optical circulators and
serial transceivers embedded in FPGAs is proposed~\cite{clock}.
Through a prototype implementation and performance evaluation, the distributed clock
has shown a synchronization accuracy better than 15 ps, which
can meet the requirements of most current large-scale physics
experiments.

Finally, the readout electronics must sustain the radiation loads during the operational lifetime of the STCF. Among them, the radiation influence on the FPGA-based TDC needs to be regarded from two aspects, i.e., the damage to the FPGA-based TDC from a single event and the long-term changes in the FPGA-based TDC performance under the radiation environment.

\subsection{Summary and Outlook}

A conceptual DTOF detector design for an endcap PID detector providing effective $\pi/K/p$ identification at the STCF is presented. 
An optimum quadrantal radiator with a
thickness of 15~mm with an absorber attached to its outer surface is chosen as the baseline
design. The performance of the DTOF is investigated through MC simulation, and a
reconstruction algorithm for the DTOF is developed.
The simulation indicates an overall
reconstructed TOF time resolution of $\sim50$~ps with an average of $\sim20$ photons detected by the
MCP-PMT arrays when all the contributing factors are converted. 
It is worth noting that the
uncertainty of $T_{0}$ dominates the overall timing error; therefore, an optimal design of the STCF
bunch size is crucial. 
By applying the likelihood method, a $\pi/K$ separation power of DTOF of $\sim4\sigma$
or better at a momentum of 2~GeV/c is achieved over the entire DTOF sensitive area. This fulfills
the requirements for the PID detector at the STCF. Extensive R$\&$D works concerning DTOF
are underway to verify the 
designs and the key technical aspects.

\newpage

\clearpage
\newpage
\section{ElectroMagnetic Calorimeter (EMC)}
\label{sec:emc}

\subsection{Introduction}
\label{sec:emcintro}

The electromagnetic calorimeter (EMC) of the STCF detector is a cylindrical array of scintillating crystals that provides energy and position measurements for photons with high resolution and 4$\pi$ coverage. It can also identify particles including photons, electrons and hadrons. The primary detector requirements leading to the conceptual design of the EMC are listed as follows:

\begin{itemize}
\item Energy resolution of approximately 2.5\% for 1 GeV photons and good energy linearity from 25 MeV to 3.5 GeV.
\item Position resolution of about 5~mm for 1 GeV photons.
\item Fast response to cope with the expected high rate environment.
\item Time resolution of about 300 ps for 1 GeV energy deposits to suppress background and identify particles.
\item Good radiation resistance against the radiation dose anticipated of 10 years of operation.
\item Providing critical input to the trigger system.
\item Precise luminosity measurement. 
\end{itemize}

%%%%%%%%%%%%%%%%%%%%%%%%%%%%%%%%%%%%%%%%%%%%%%%%%%%%%%%%%%%%%%%%%%%%%%

\subsection{EMC Conceptual Design}

\subsubsection{Crystal and Photo-Detector Choices}

%\quad\\
With the advantage of high light-yield, CsI(Tl) crystal based calorimeters have been widely used in collider experiments in the $\tau$-charm energy region, for example, in BESIII~\cite{bes}, Belle~\cite{belle} and Belle II~\cite{belle2}. However, the decay time of this crystal is so long that it cannot meet the requirements of a high luminosity experiment. In the STCF, it is necessary to select a faster crystal for the EMC. Pure CsI (pCsI) crystals, with a short decay time and excellent radiation resistance, are promising candidates. The light yield of pCsI crystal is approximately 2\% of that from CsI(Tl)~\cite{1} and would further decrease by 20\% when the received irradiation dose reaches 100~krad (the total irradiation dose of the STCF EMC endcap in 10 years would be approximately 45 krad (see Table~\ref{tab:TIDNIEL_max})).

For the readout electronics of a crystal scintillator, semiconductor photodetectors are favored to operate in a strong magnetic field of 1~T. Mature commercial semiconductor photodetectors mainly include photodiodes~(PDs), avalanch photodiodes~(APDs) and silicon photomultiplifiers~(SiPMs). Considering the gain performance, large dynamic response range, linear output signal amplitude, sensitive area and light collection efficiency of APDs, the photodetectors are more suitable as light collection devices of pCsI crystals.

A pCsI crystal scintillator with an APD readout is chosen as a possiable candidate  satisfying the physics requirements of the STCF EMC.Two of the most important factors regarding the performance of the EMC are the energy resolution and position resolution. The relevant considerations for the conceptual design of the EMC are discussed below.

\begin{figure*}[htb]
 \centering
 \mbox{
  %\vskip -1.5cm
  \begin{overpic}[width=0.5\textwidth, height=0.33\textwidth]{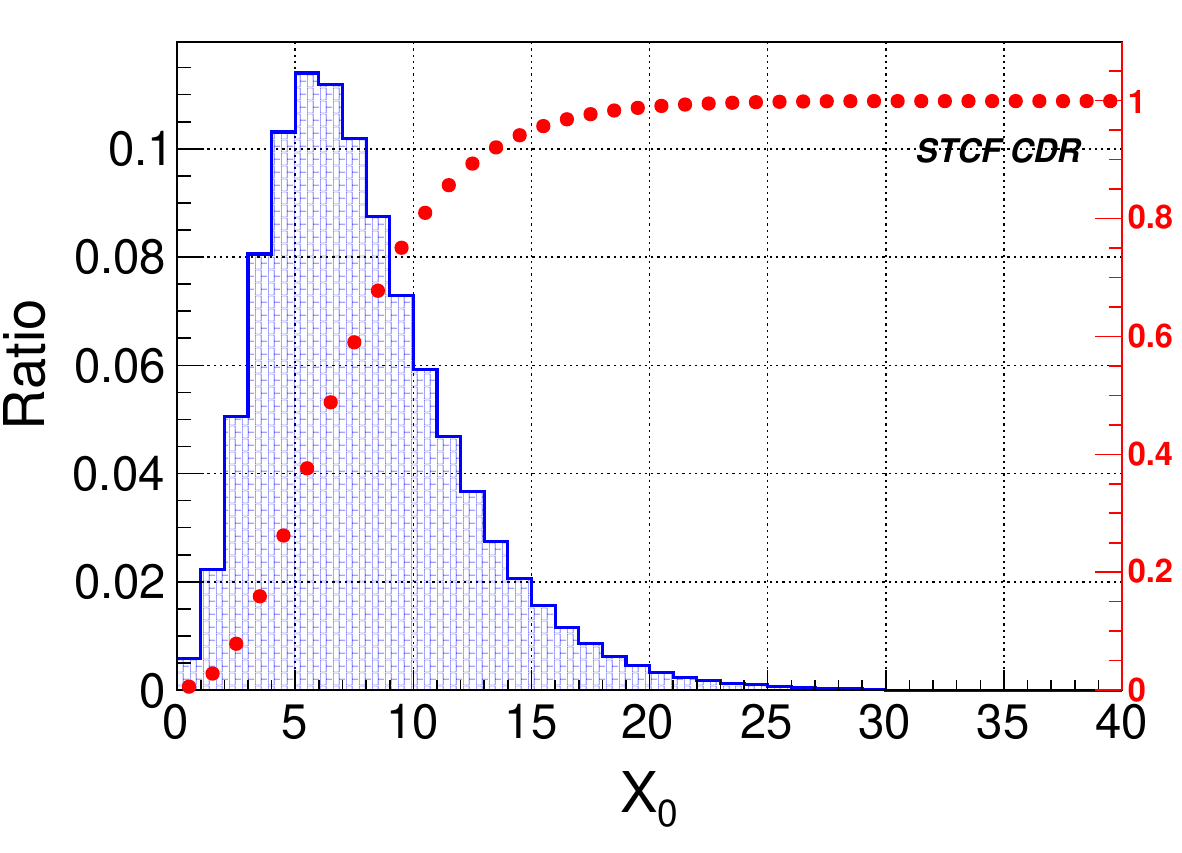}
  \end{overpic}
 }
\caption{The shower longitudinal distribution.}
\label{Fig4:ECAL-TotalRadiationLength}
\end{figure*}

\iffalse
\newcommand{\tabincell}[2]
{\begin{tabular}{@{}#1@{}}#2\end{tabular}}
\begin{tiny}
\begin{table}
\begin{center}
\caption{The Parameters of Crystals.}\label{Tab:List-1st}
\begin{tabular}{|c|c|c|c|c|c|c|}
\hline Crystal&pCsI&LYSO&PWO&YAP&GSO&BaF:Y\\
\hline Density ($g/cm^{3}$)&4.51&7.40&8.30&5.37&6.71&4.89\\
\hline Melting Point (${}^{o}C$)&621&2050&1123&1872&1950&1280\\
\hline Radiation Length (cm)&1.86&1.14&0.89&2.70&1.38&2.03\\
\hline Moliere Radius (cm)&3.57&2.07&2.00&4.50&2.23&3.10\\
\hline Reflective index&1.95&1.82&2.20&1.95&1.85&1.50\\
\hline Hygroscopicity&Slight&No&No&No&No&No\\
\hline Luminosity (nm)&310&402&\tabincell{c}{425\\420}&370&430&\tabincell{c}{300\\220}\\
\hline Decay Time (ns)&\tabincell{c}{30\\6}&40&\tabincell{c}{30\\10}&30&60&\tabincell{c}{600\\1.2}\\
\hline Light Yield(\%)&\tabincell{c}{3.6\\1.1}&85&\tabincell{c}{0.3\\0.1}&65&20&\tabincell{c}{1.7\\4.8}\\
\hline Dose Rate Dependent&No&No&Yes&TBA&TBA&No\\
\hline D(LY)/dT ($\%/{}^{o}C$)&-1.4&-0.2&-2.5&TBA&-0.4&TBA\\
\hline Experiment&\tabincell{c}{kTeV\\Mu2e}&-&\tabincell{c}{CMS\\ALICE\\PANDA}&-&-&-\\
\hline
\end{tabular}
\end{center}
\end{table}
\end{tiny}
\fi
%--------------------------------------------------------%
%\quad\\
The energy resolution of a calorimeter is affected by the shower leakage, which is mainly determined by the total radiation length ($X_{0}$) of the EMC. Figure~\ref{Fig4:ECAL-TotalRadiationLength} shows the longitudinal development of the shower of a 3.5\,GeV photon (close to the most energetic photon in the STCF) in a pure CsI crystal. Approximately 95\% of the shower energy is deposited within 15 $X_{0}$ (28\,cm). Then, with increasing total radiation length, the total energy deposition increases slowly, and the improvement in the energy resolution is very small. Considering the total crystal cost and performance of the EMC, we choose the crystal radiation length of 15 $X_{0}$.

%%%%%%%%%%%%%%%%%%%%%%%%%%%%%%%%%%%%%%%%%%%%%%%%%%%%%%%%%%%%%%%%%%%%%%%%%%%%%%
%\quad\\
The position resolution (angular resolution) is affected by the shower transverse development and the calorimeter segmentation and is an important parameter for the reconstruction of $\pi^{0}$ particles. According to the simulation, the minimum angle between two photons from the decay of a 1.5 GeV $\pi^{0}$ (about the most energetic $\pi^{0}$ in the STCF experiment) is about 10 degrees. To separate these two photons, the maximum coverage of each crystal should be $\sim$ 3 degrees.  In principle, finer dimensions improve the angular resolution. However, the packaging and supporting materials of the crystal increase, and the lateral leakage increases, which worsens the energy resolution. Optimization of the crystal size is performed using simulation and is discussed in Sec.~\ref{sec:emc_opt}.

\iffalse
\begin{tiny}
\begin{table}
\begin{center}
\caption{The photodetector parameters.}\label{Tab:List-2nd}
\begin{tabular}{|c|c|c|c|}
\hline Device&Typical Gain&Typical QE or PDE (\%)&Typical Size ($cm^2$)\\
\hline Photodiode&1&80&1\\
\hline Avalanche Photodiode&50&80&0.25\\
\hline SiPM&$10^{5}$&25&0.36\\
\hline
\end{tabular}
\end{center}
\end{table}
\end{tiny}
\fi
%%%%%%%%%%%%%%%%%%%%%%%%%%%%%%%%%%%%%%%%%%%%%%%%%%%%%%%%%%%%%%%%%%%%%%%%%%%%%%%%%%%
\subsubsection{Crystal Size Optimization}
\label{sec:emc_opt}

\begin{figure*}[htb]
 \centering
 \mbox{
  \begin{overpic}[width=0.5\textwidth, height=0.33\textwidth]{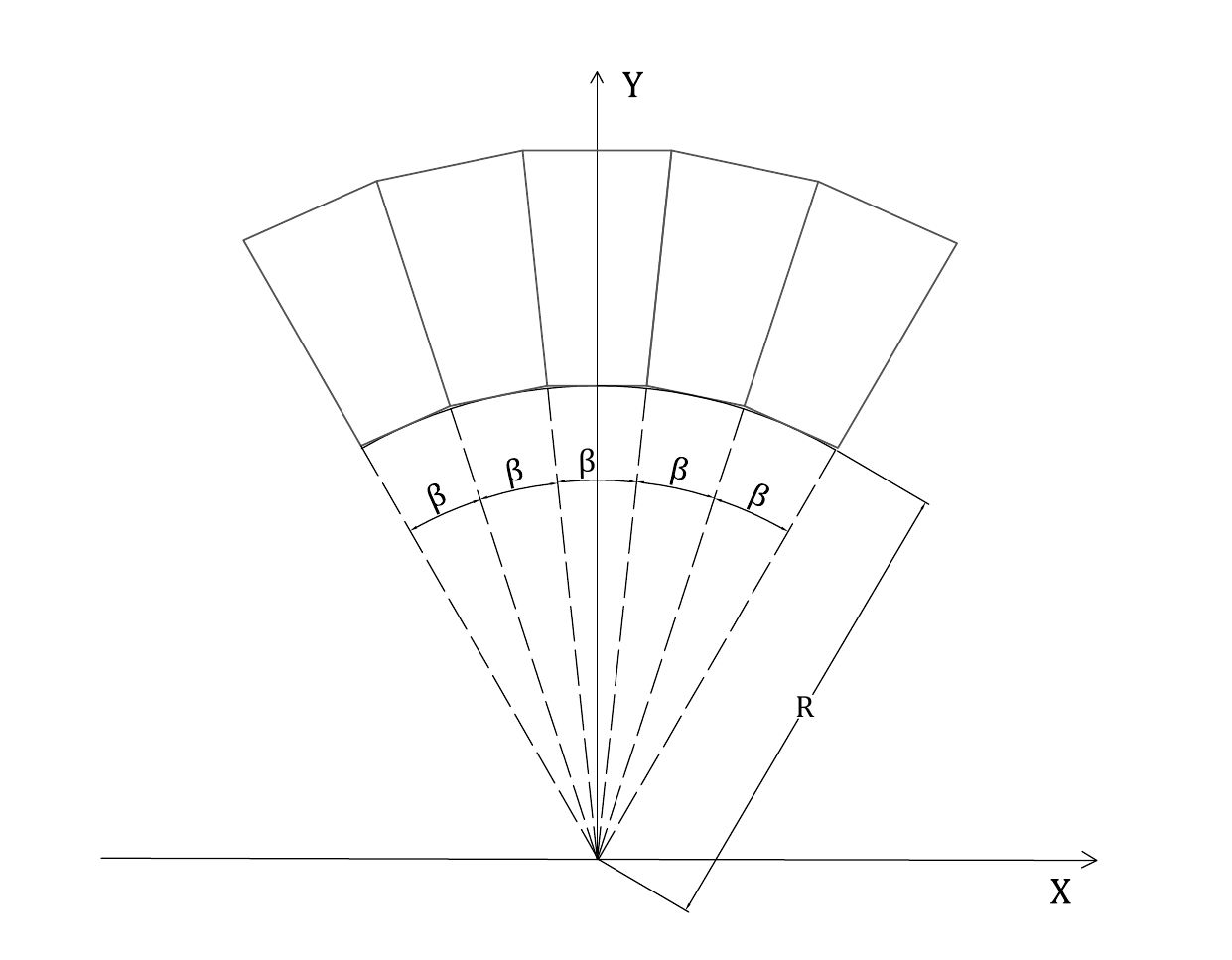}
  \end{overpic}
 }
\caption{The schematic arrangement of crystals.}
\label{Fig4:ECAL-CrystalCellSize}
\end{figure*}

Figure~\ref{Fig4:ECAL-CrystalCellSize} shows the schematic arrangement of crystals in the $X-Y$ plane of the barrel EMC. The inner radius R of the calorimeter is 105 cm, and the length of the crystal is 28 cm. The size of the cyrstal, denoted by the dimension of the front face where particles enter, is a key design parameter to be optimized.

\begin{figure*}[htbp]
 \centering
  \subfloat[][]{\includegraphics[width=0.4\textwidth]{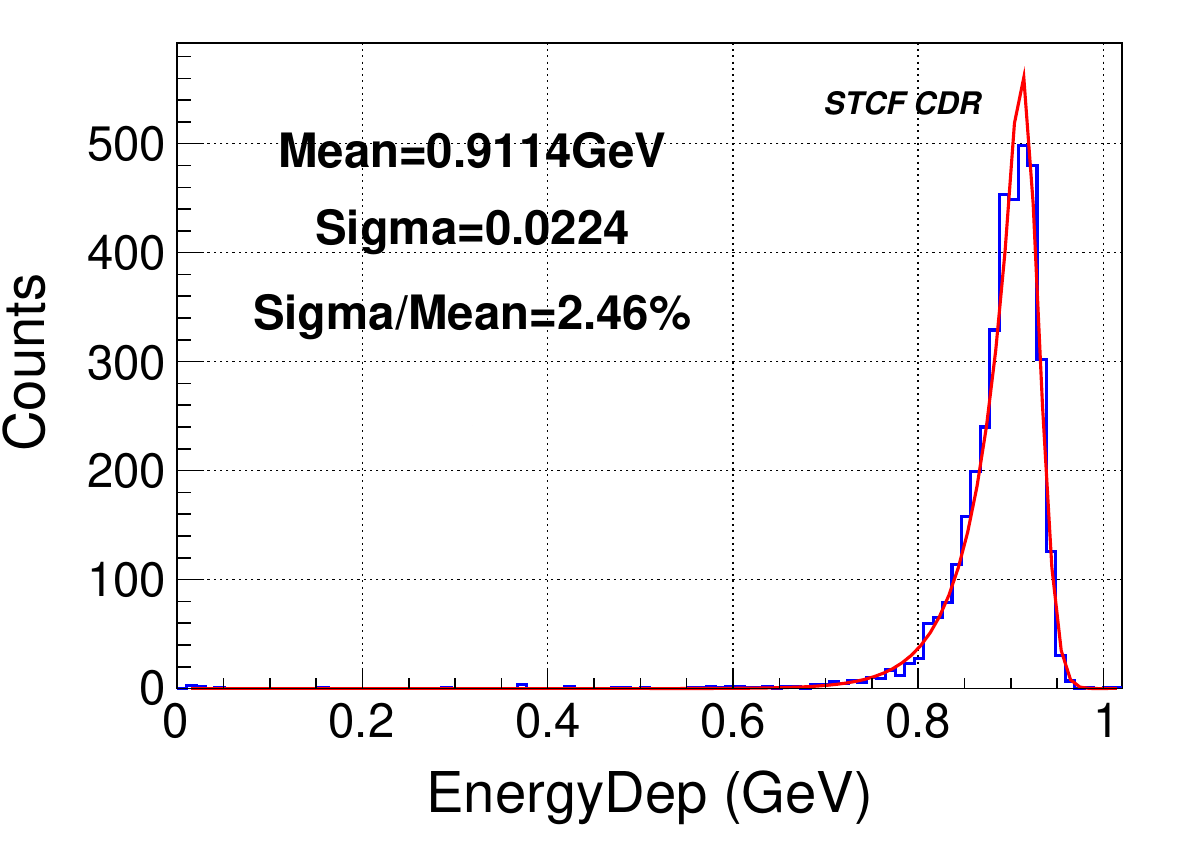}}
  \subfloat[][]{\includegraphics[width=0.4\textwidth]{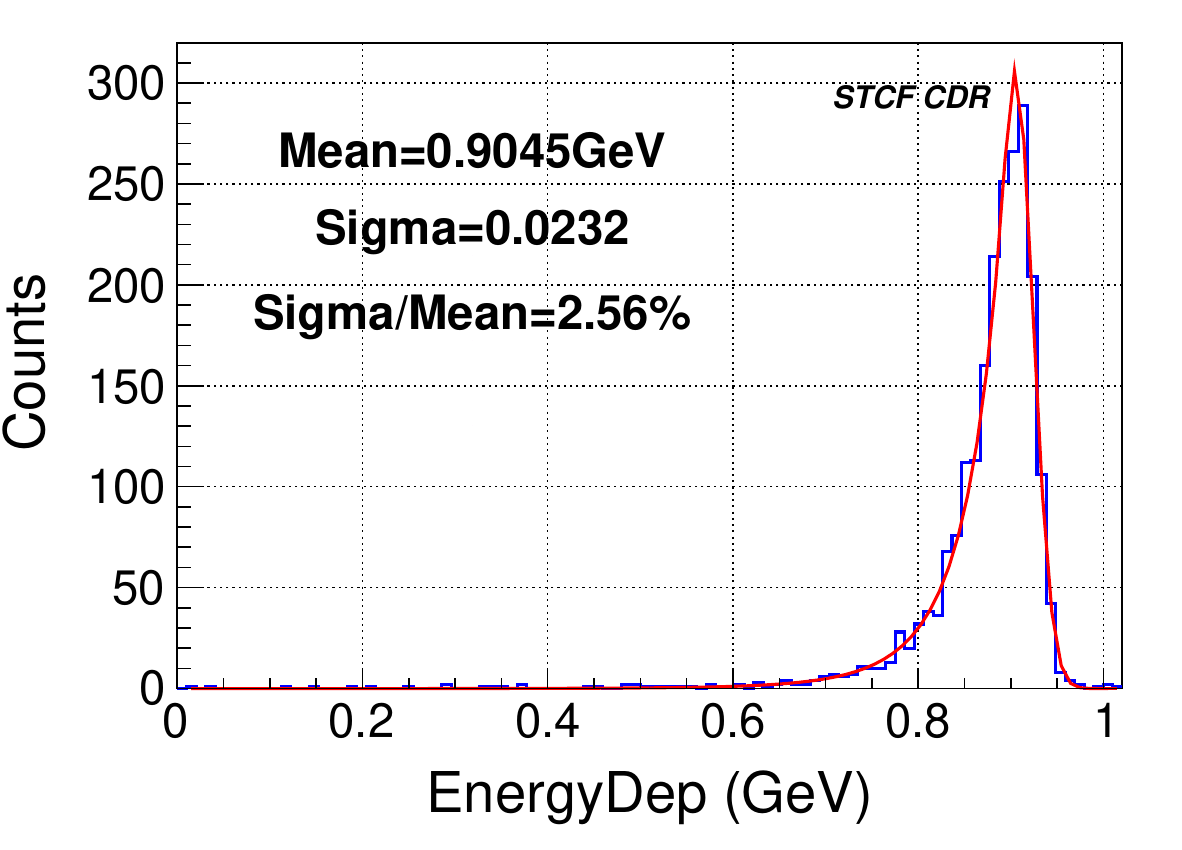}}
\caption{The energy resolution of the EMC. The pCsI front face size is 5~cm  (a) and 3~cm (b).}
\label{Fig4:ECAL AngRes-1stfactor}
\end{figure*}

\subparagraph{Energy Resolution}
The influence of the front face size of the crystal on the energy resolution of the EMC is also investigated. In principle, the angular resolution of the EMC can be improved by reducing the crystal front face size, but more packaging and supporting materials would be introduced, which would degrade the energy resolution of the EMC. Figure~\ref{Fig4:ECAL AngRes-1stfactor} shows the energy resolution of 1~GeV gamma ray reconstruction, comparing the energy deposition in two layouts: a 3 $\times$ 3 crystal array with a 5~cm crystal front face size and a $5\times5$ array with a 3~cm crystal front face size. The reconstructed energies are 0.911~GeV and 0.905~GeV, respectively, where a smaller crystal gives a lower reconstructed energy, but the difference is not significant.
To obtain the energy resolution, a Crystal Ball function is fit to the reconstructed energy distribution, and the FWHM divided by 2.35 is used as the energy resolution.
For the two layouts, the energy resolutions are found to be 2.46\% and 2.56\%, respectively, and the smaller crystal results in a larger energy uncertainty. The total effective areas of the EMC used in both cases are the same. However, more packaging materials with a smaller crystal size are needed. The packaging materials here are 300 $\mu$m Teflon film and 200 $\mu$m carbon fiber as crystal support materials.

%%%

\begin{figure*}[htbp]
 \centering
  \subfloat[][]{\includegraphics[width=0.4\textwidth]{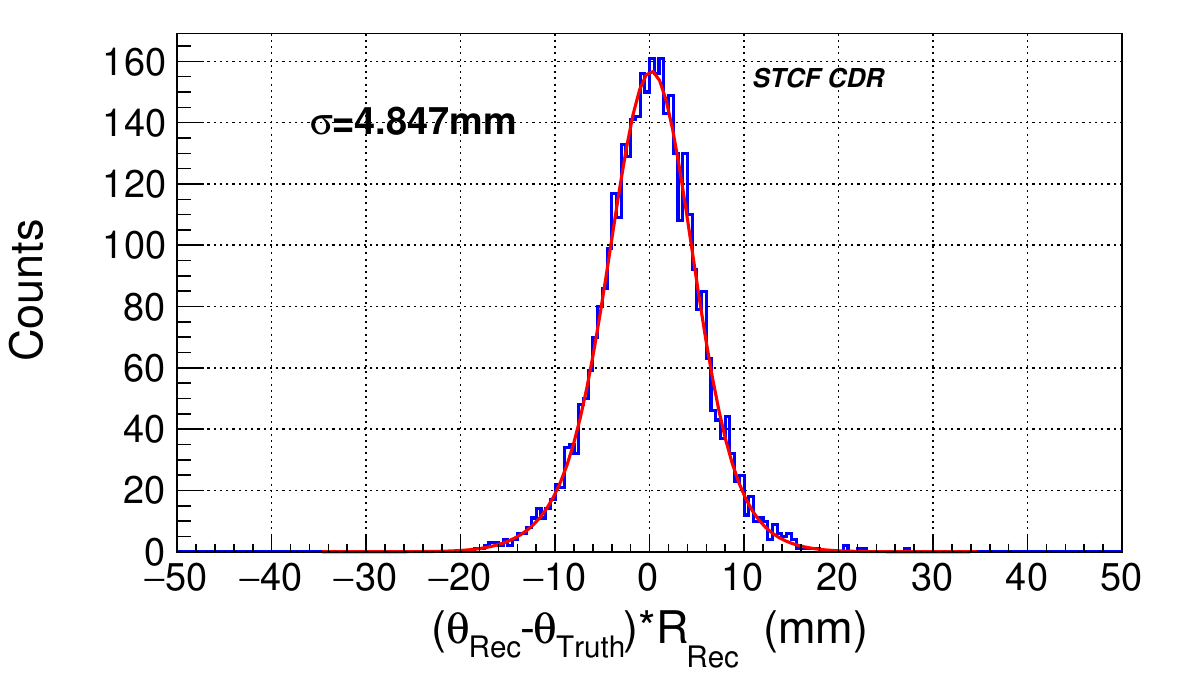}}
  \subfloat[][]{\includegraphics[width=0.4\textwidth]{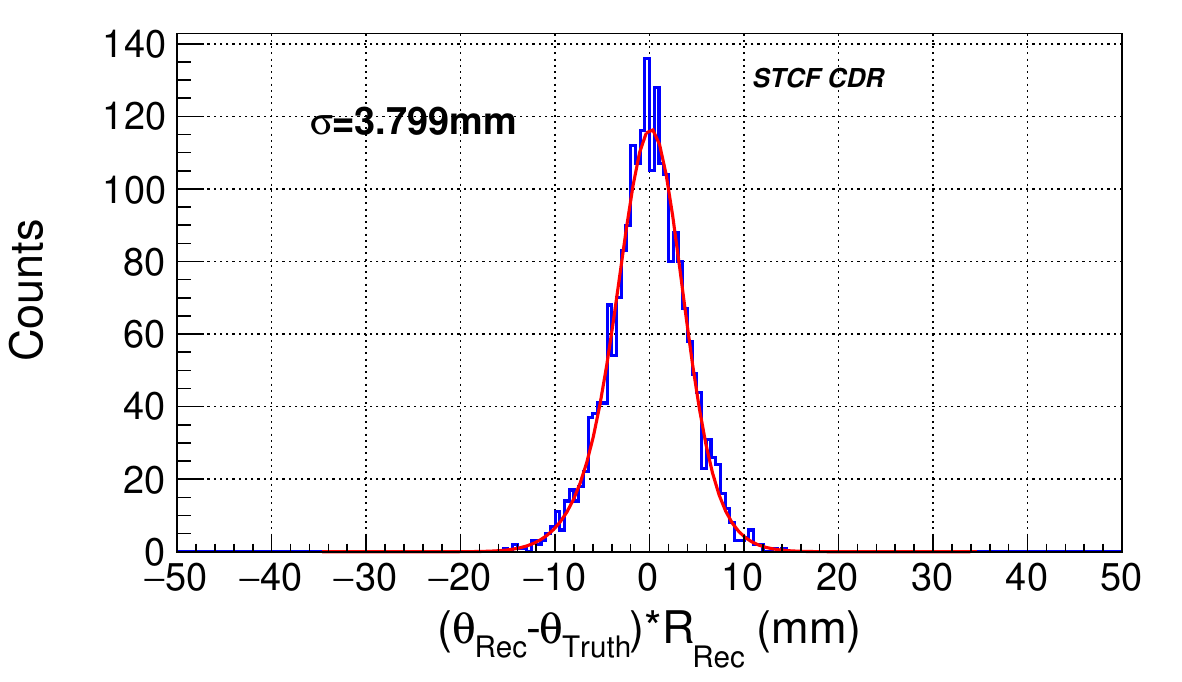}}
\caption{The EMC position resolution based on the logarithmic energy weighting method, with (a) 5 cm $\times$ 5 cm and (b) 3 cm $\times$ 3 cm crystal front face sizes.}
\label{Fig4:ECAL PosRes-2-2ndfactor}
\end{figure*}

\subparagraph{Position Resolution and $\pi^{0}$ Reconstruction Efficiency}
\quad
The position resolution of incident photons is directly dependent on the crystal front face size. The shower position measurement is the key to reconstrcuting the incident angle, where the barycenter method is commonly used, with the following formula:
\begin{equation}
\begin{array}{cll}
x_{c} = \sum\limits_{j}^{N} w_{j}(E_{j}) x_{j}/\sum\limits_{j}^{N} w_{j}(E_{j}),
\end{array}
\end{equation}
\noindent
where $x_{c}$ is the reconstruction position, $x_{j}$ is the spatial coordinate of the $j-th$ crystal, and $w_{j}$ ($E_{j}$) is the weight of the $j-th$ crystal involved in the position reconstruction. The weight is related to the energy deposition in that crystal.
The angle can be determined by reconstructing the incident point on the crystal front surface and the collision point. Figure~\ref{Fig4:ECAL PosRes-2-2ndfactor} compares the position resolution of 1~GeV $\gamma$ reconstructed by the logarithmic weight of energy in crystals with two different front face sizes of (a) 5~cm $\times$ 5~cm and (b) 3~cm $\times$ 3~cm. The position resolutions are 4.8~mm and 3.8~mm, respectively, clearly showing that the position or angular resolution can be improved by decreasing the crystal size.\\

\begin{figure*}[htbp]
 \centering
 \mbox{
  \begin{overpic}[width=0.5\textwidth, height=0.33\textwidth]{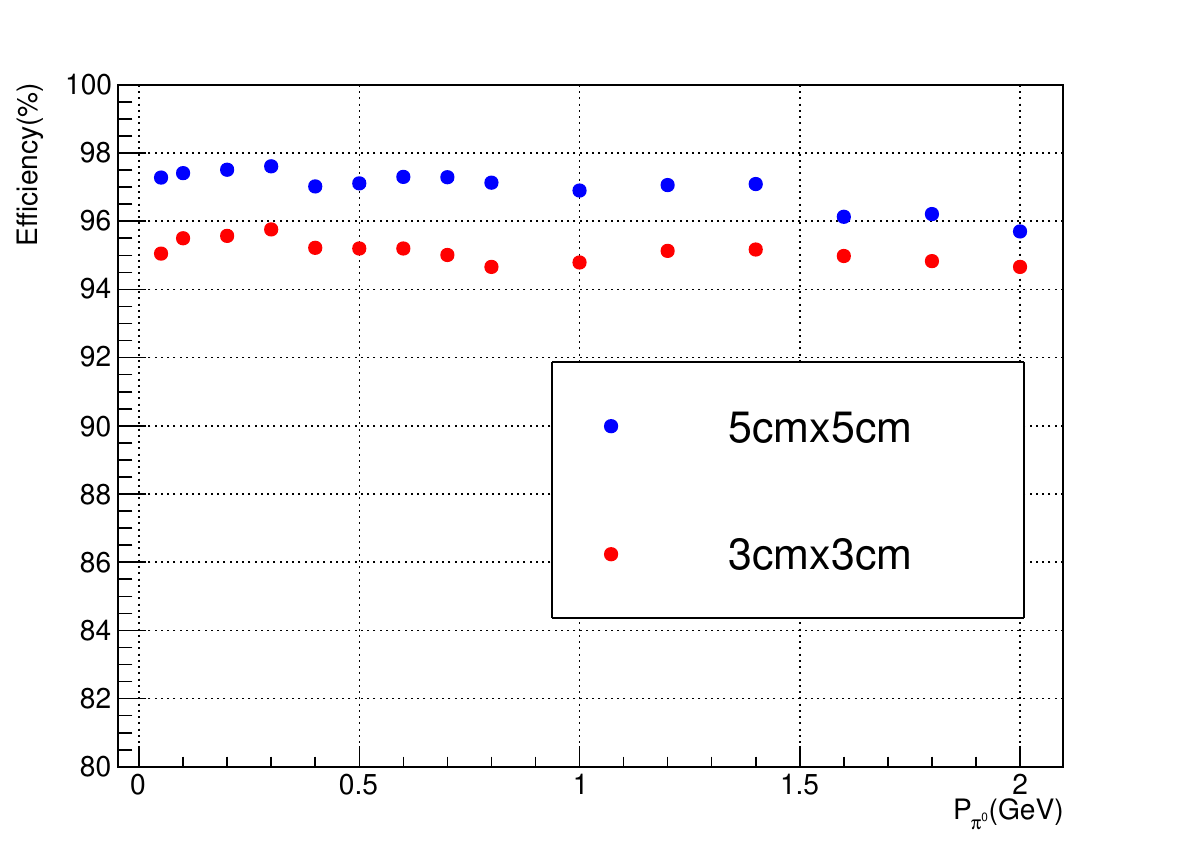}
  \end{overpic}
 }
\caption{The efficiency of the EMC for $\pi^{0}$.}
\label{Fig4:ECAL AngRes-2-5thfactor}
\end{figure*}

With decreasing crystal front face size, the proportion of packaging materials between crystals increases, which leads to degraded reconstruction efficiency for $\pi^{0}$. The reconstruction efficiencies of $\pi^{0}$ with different crystal front face sizes are compared, as shown in Fig.~\ref{Fig4:ECAL AngRes-2-5thfactor}. Here, it is required that the two photons generated by $\pi^{0}$ decay are both within the effective acceptance of EMC, and to reconstruct the $\pi^{0}$ mass.
It is clear that the reconstruction efficiency of $\pi^{0}$ is higher for larger crystal front face size. The average opening angle of the crystal with the front  face size of 5 cm relative to the collision point is about 2 degrees, which can distinguish the two photons generated from $\pi^{0}$ ($\pi^{0}$ of 1.5~GeV, where the minimum included angle of photons is about 10 degrees). Moreover, the energy resolution of the 5 cm crystal is better than that of the small-sized crystal, so the reconstruction efficiency of $\pi^{0}$ in the same mass width range is slightly higher.

\subparagraph{Fake Photon Discrimination}
In contrast with the normal photons from the collision point, most of the fake photons are generated at other positions, and their direction of incidence on the EMC deviates greatly from the axial direction of the crystal, which causes the hit number and secondary moment of the shower in the EMC to be different from that of real photons. Another important difference is that the time information is different from the real signal. This information can be used to identify fake photons. In addition, whether the direction reconstruction of photons can be realized by longitudinal sampling of the EMC is still an interesting topic, but considering the importance of energy resolution, the scheme of longitudinal sampling needs to be carefully designed.

In conclusion, considering the energy resolution, position resolution, reconstruction efficiency and total number of readout channels, based on the premise of meeting the STCF EMC requirements, the end face size of $\sim$5~cm is selected as the typical size of the crystal.

%%%%%%%%%%%%%%%%%%%%%%%%%%%%%%%%%%%%%%%%%%%%%%%%%%%%%%%%%%%%%%
%%%%%%%%%%%%%%%%%%%%%%%%%%%%%%%%%%%%%%%%%%%%%%%%%%%%%%%%%%%%%%%%%%%%%%%%%%%%%%%%%
\subsubsection{EMC Layout}
\begin{figure*}[htbp]
 \centering
 \mbox{
  %\vskip -1.5cm
  \begin{overpic}[width=0.8\textwidth]{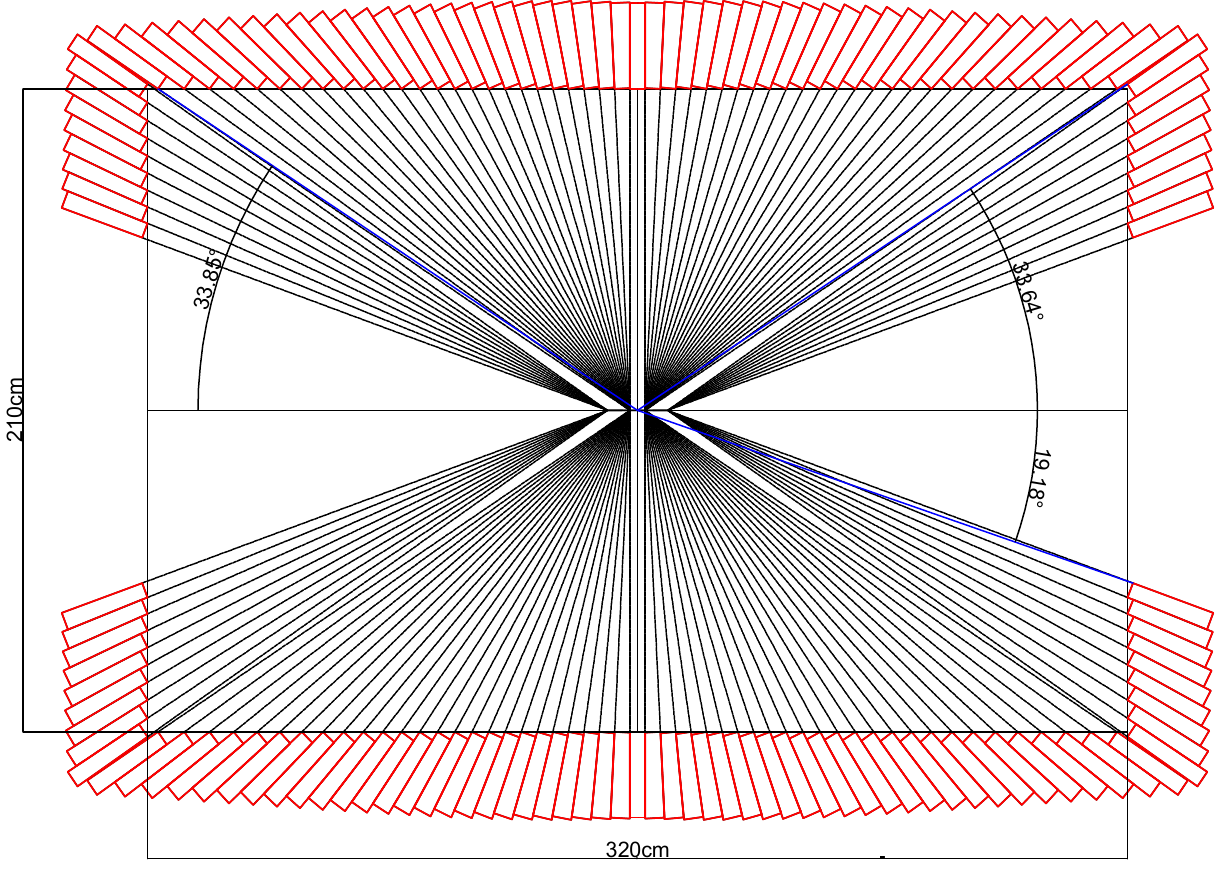}
  \end{overpic}
 }
\caption{The EMC layout design.}
\label{Fig4:ECAL Layout}
\end{figure*}

%{The Layout Design}
The calorimeter is composed of a barrel part and an endcap cover, and the crystal arrangement diagram is shown in Fig.~\ref{Fig4:ECAL Layout}. The barrel part covers polar angles from $33.85^{\circ}$ to $146.15^{\circ}$, with an inner radius of 105~cm and length of 320~cm along the beam direction. The endcap part is located at 160~cm $\le \left|z\right| \le$ 190~cm, with an inner radius of 56~cm and an outer radius of 105~cm. The endcap covers polar angles from 19.18$^{\circ}$ to 33.64$^{\circ}$ and 156.15$^{\circ}$ to 160.82$^{\circ}$. The entire EMC provides a 94.45\% solid angle coverage of 4 $\pi$.

There are 51 circles of crystals in the barrel along the beam (Z) direction, 132 crystals per circle, and 6732 crystals in total.
The crystal size and shape in the same circle are the same, and the crystal shapes in different circles are similar but different. The crystal is in an irregular trapezoid platform shape, in which the sizes of the front and back faces are about $5\times5$ and $6.5\times6.5$~cm$^2$ respectively, and the longitudinal length is 28~cm. The endcap is divided into two identical parts, located to the left and right side of the barrel part, each with 969 crystals, with a total of 1938 crystals. The crystal in the endcap is also in an irregular trapezoid platform shape. The sizes of the front and back faces are about $4.5\times4.5$~cm$^2$ and $5.2\times5.2$~cm$^2$ respectively, and the longitudinal length is also 28~cm.

\begin{figure*}[htbp]
 \centering
 \mbox{
  %\vskip -1.5cm
  \begin{overpic}[width=0.5\textwidth]{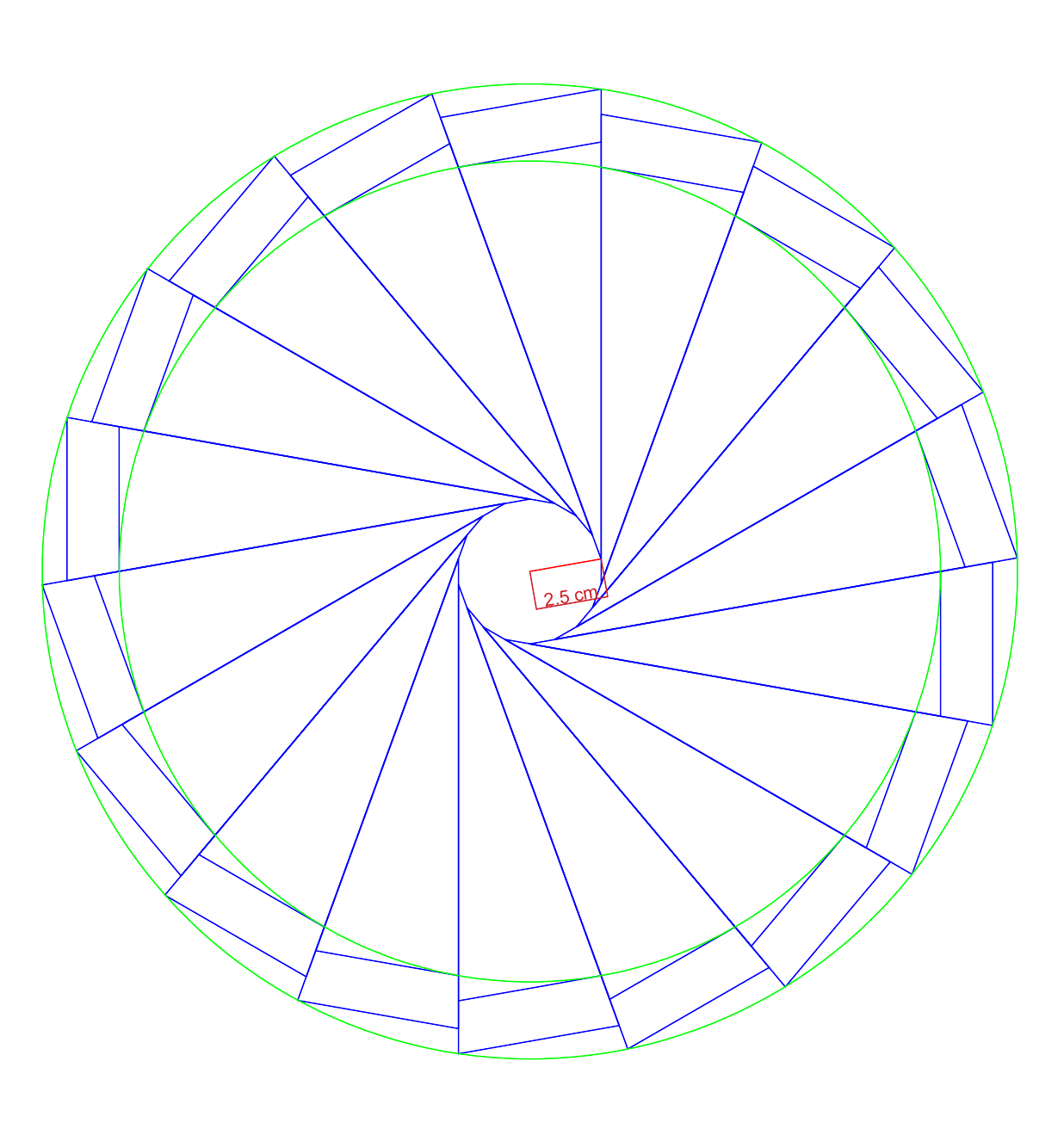}
  \end{overpic}
 }
\caption{The EMC defocus design.}
\label{Fig4:ECAL Defocus-phi}
\end{figure*}

To reduce the probability of secondary particles escaping from the gap between crystals, a defocus design is added to the geometric structure of the EMC to improve the detection efficiency. In the direction of $\theta$, except for the middle circle of crystals in the barrel, each circle of crystals points to $\pm2.5$~cm away from the collision point, and each circle of crystals in the endcap points to $\pm10$~cm away from the collision point, which can also be seen in Fig.~\ref{Fig4:ECAL Layout}. In the azimuth direction ($\phi$), the crystal in each circle deflects 1.36$^{\circ}$ in the direction of phi. In this way, the crystal points to the circumference with a radius of 2.5~cm centered on the beam line, as shown Fig.~\ref{Fig4:ECAL Defocus-phi}.

\subsection{Expected Performance of the EMC}
\label{sec:emc_perf}
\subsubsection{Energy Response}
\begin{figure*}[htbp]
 \centering
  \subfloat[][]{\includegraphics[width=0.4\textwidth]{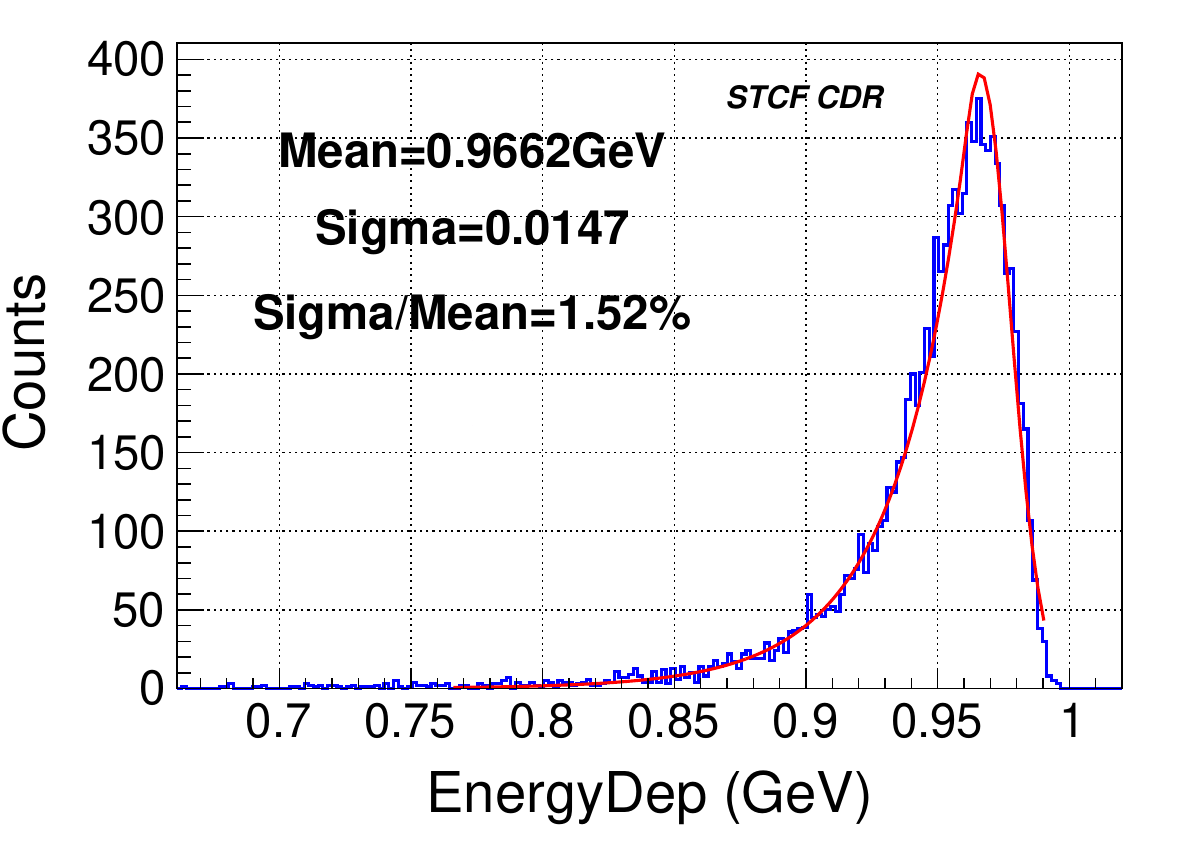}}
  \subfloat[][]{\includegraphics[width=0.4\textwidth]{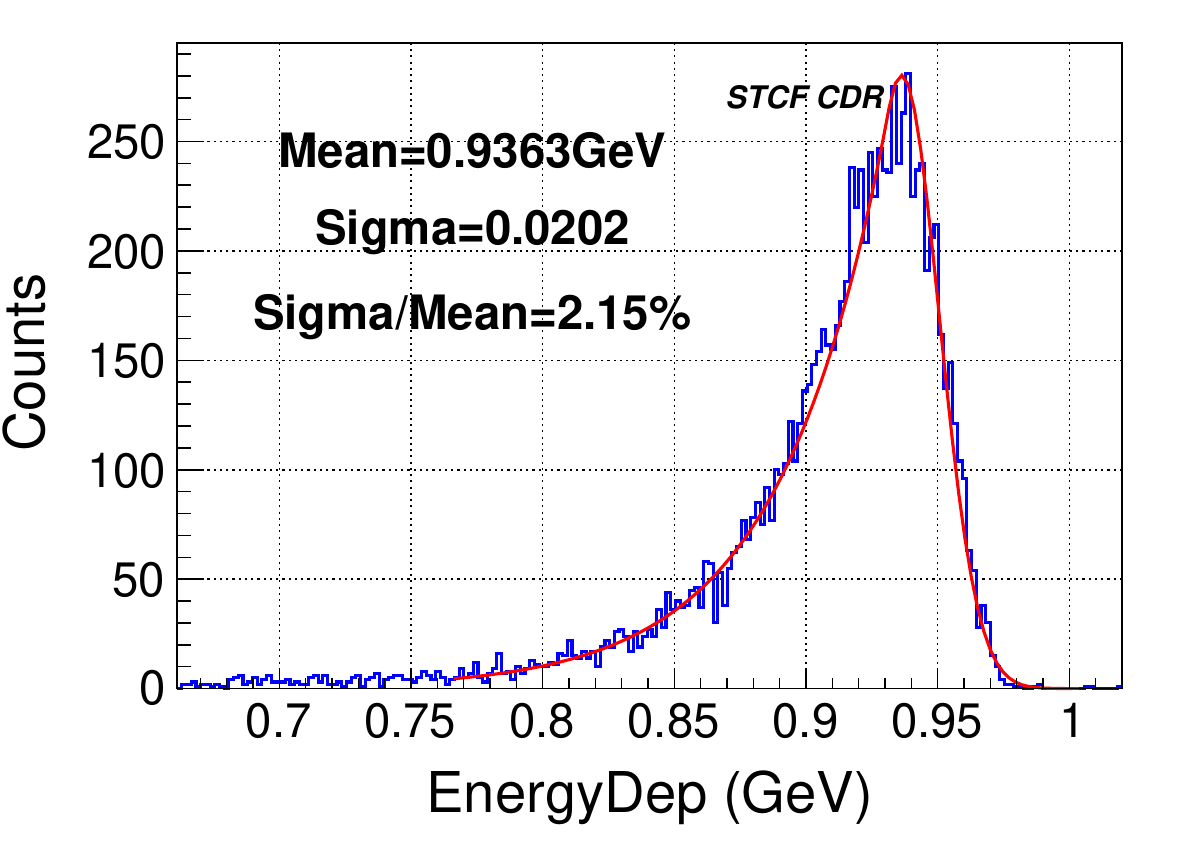}}
\caption{The expected energy resolution of the EMC, (a) the intrinsic performance without considering material effects and (b) with several main factors.}
\label{Fig4:ECAL EneRes-1stfactor}
\end{figure*}

Based on the conceptual design, the response of the EMC to photons with different energies is studied via {\sc Geant4} simulation using 1~GeV photons.
As shown in Fig.~\ref{Fig4:ECAL EneRes-1stfactor}(a), the intrinsic energy resolution is about 1.52\% with only considering the interaction between photons and the crystal. Here, the energy reconstruction is in the range of the $5\times5$ array and each crystal has a cross-section of $5\times 5$~cm$^2$ at the front face. The energy deposition fluctuation is mostly caused by backscattering and leakage of the shower tail. The energy resolution is simulated by taking into account several main factors. The influence of these factors on the energy resolution is given in Table~\ref{Tab:List-3rd}. When the light yield of the crystal is set to 100~pe/MeV (preliminary measurement shows that the light yield of the pCsI crystal can reach about 150 pe/MeV, see Sec.~\ref{sec:emc_cosmic}), the simulated energy resolution is 1.52\%. When a 200~$\mu$m thick carbon fiber material is introduced as the support unit, the energy resolution becomes 1.96\%. The light collection nonuniformity of large crystals can generally reach a few percent. For example, in the BESIII experiment, the average nonuniformity is about 3\%-4\%. After taking into account a 5\% nonuniformity, the simulated energy resolution increases to 2.06\%. Considering that the secondary particles generated in the shower process hit the APD and produce electron-hole pairs, which are superimposed on the signals, the energy resolution is 2.11\%. When electronic noise of 1 MeV is added, the final energy resolution is 2.15\%, as shown in Fig.~\ref{Fig4:ECAL EneRes-1stfactor}(b).

\begin{tiny}
\begin{table}[htb]
\begin{center}
\caption{The energy resolution considering different effects.}\label{Tab:List-3rd}
\begin{tabular}{|c|c|c|c|c|c|}
\hline Condition&Intri&Carbon Fiber (200 um)&Uni (5\%)&APD&Noise (1 MeV)\\
\hline EneRes @1GeV (\%)&1.52&1.96&2.06&2.11&2.15\\
\hline
\end{tabular}
\end{center}
\end{table}
\end{tiny}

\begin{figure*}[htbp]
 \centering
  %\vskip -1.5cm
  \subfloat[][]{\includegraphics[width=0.4\textwidth]{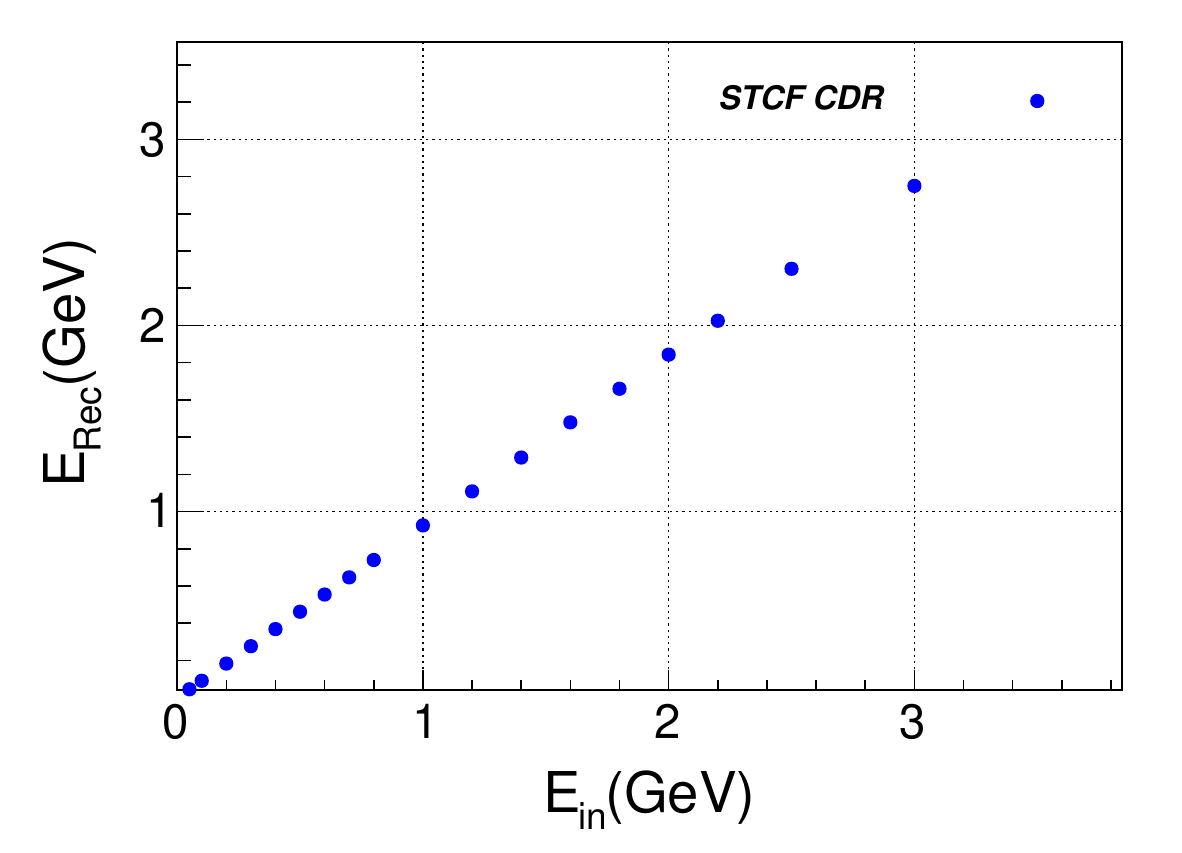}}
  \subfloat[][]{\includegraphics[width=0.4\textwidth]{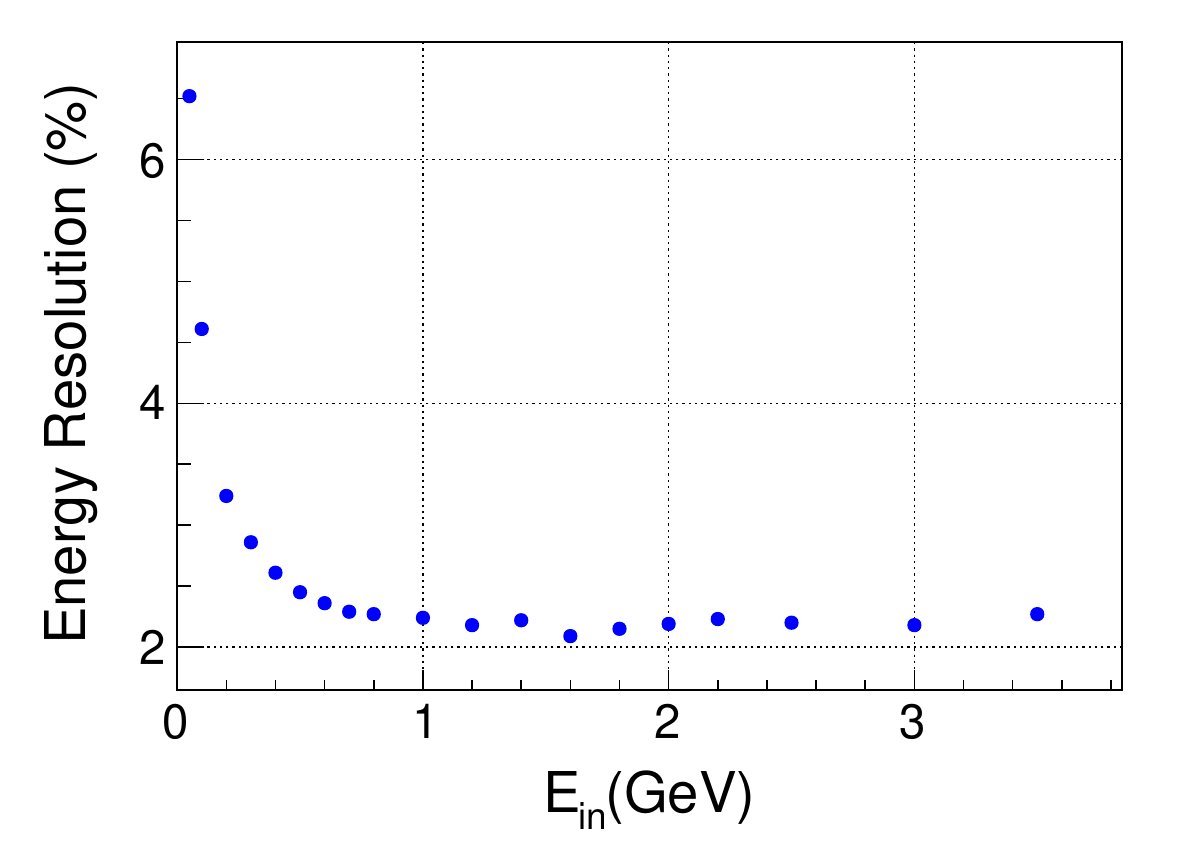}}
\caption{The expected performance of (a) the energy linearity, and (b) the energy resolution.}
\label{Fig4:EMC Energy Response Curve}
\end{figure*}

Figure~\ref{Fig4:EMC Energy Response Curve}(a) shows the photon energy response from 50 MeV to 3.5 GeV. The results show excellent linearity, with a nonlinearity of about 1\%. Figure~\ref{Fig4:EMC Energy Response Curve}(b) shows the energy resolution curve. The results show that from 50 MeV - 2 GeV, the energy resolution gradually improves with increasing energy, but it begins to deteriorate slightly at 2 GeV because of energy leakage in the EMC back end. At 1 GeV, the energy resolution is better than 2.5 \%, which meets the requirement of the EMC.

\subsubsection{Position Resolution}
\begin{figure*}[htbp]
 \centering
  %\vskip -1.5cm
  \subfloat[][]{\includegraphics[width=0.4\textwidth]{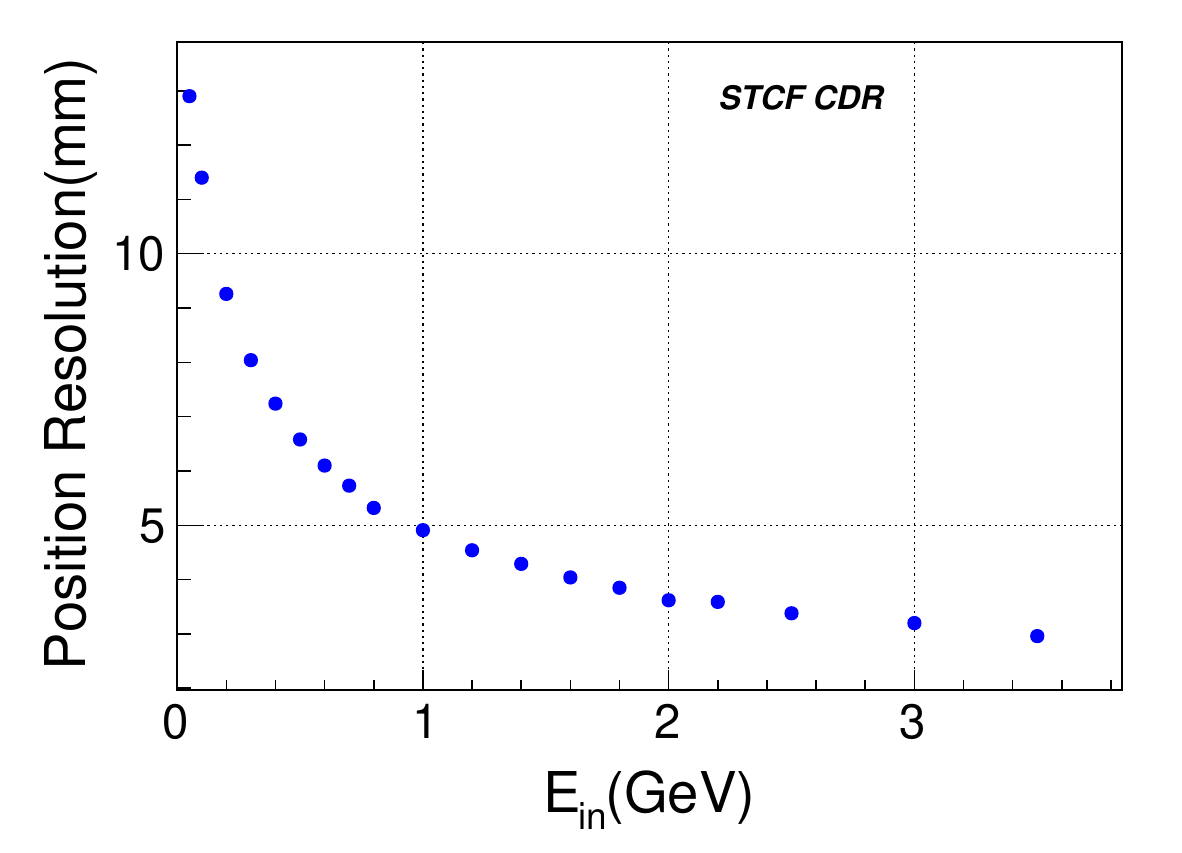}}
  \subfloat[][]{\includegraphics[width=0.4\textwidth]{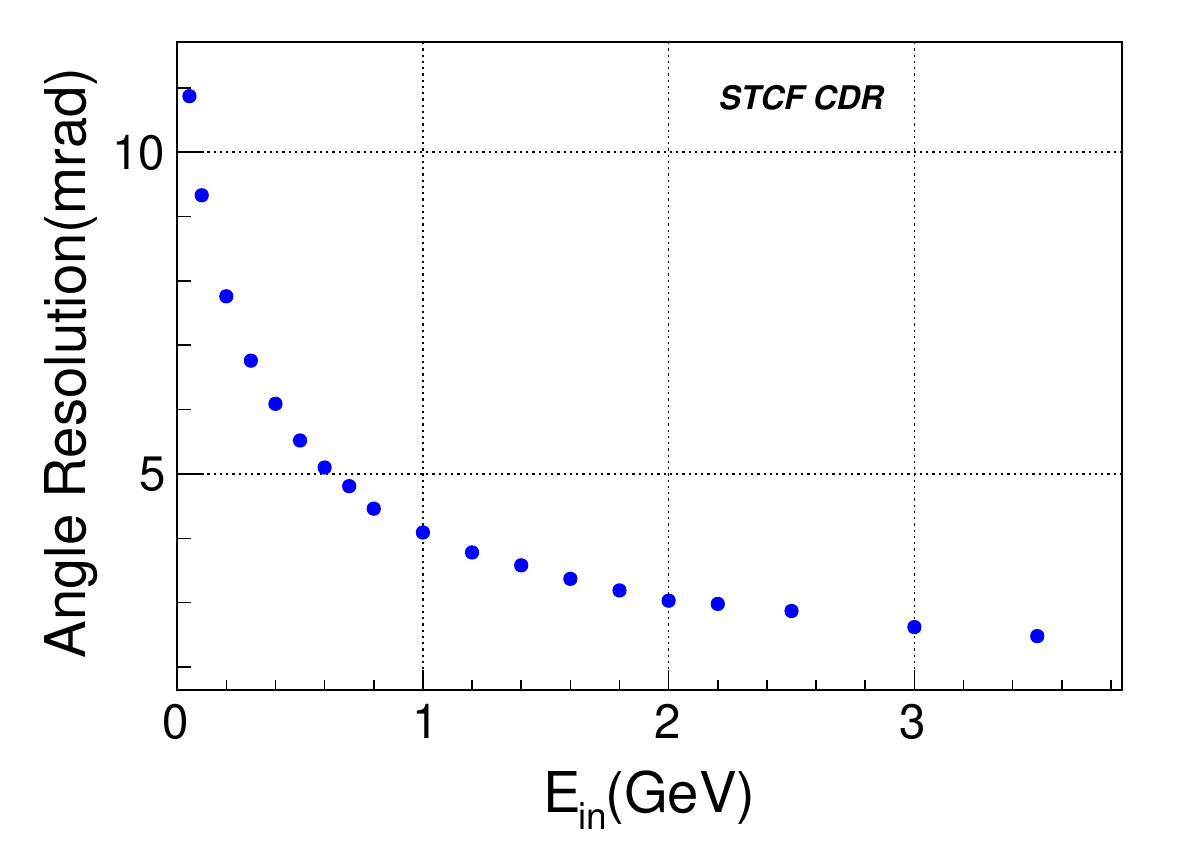}}
\caption{The expected (a) position resolution and (b) angular resolution for 1~GeV photons.}
\label{Fig4:EMC Position and Angular Reslution Curve}
\end{figure*}

Figure~\ref{Fig4:EMC Position and Angular Reslution Curve}(a) shows the performance of the EMC position resolution. The results show that the position resolution gradually improves with increasing energy. At 1~GeV, the expected positon resolution is about 5~mm, which meets the design requirements. With the position resolution, the angular resolution can be obtained, as shown in Fig.~\ref{Fig4:EMC Position and Angular Reslution Curve}(b). The angular resolution is 4~mrad at 1~GeV.

\subsubsection{Time Resolution}
\begin{figure*}[htbp]
 \centering
  %\vskip -1.5cm
  \subfloat[][]{\includegraphics[width=0.4\textwidth]{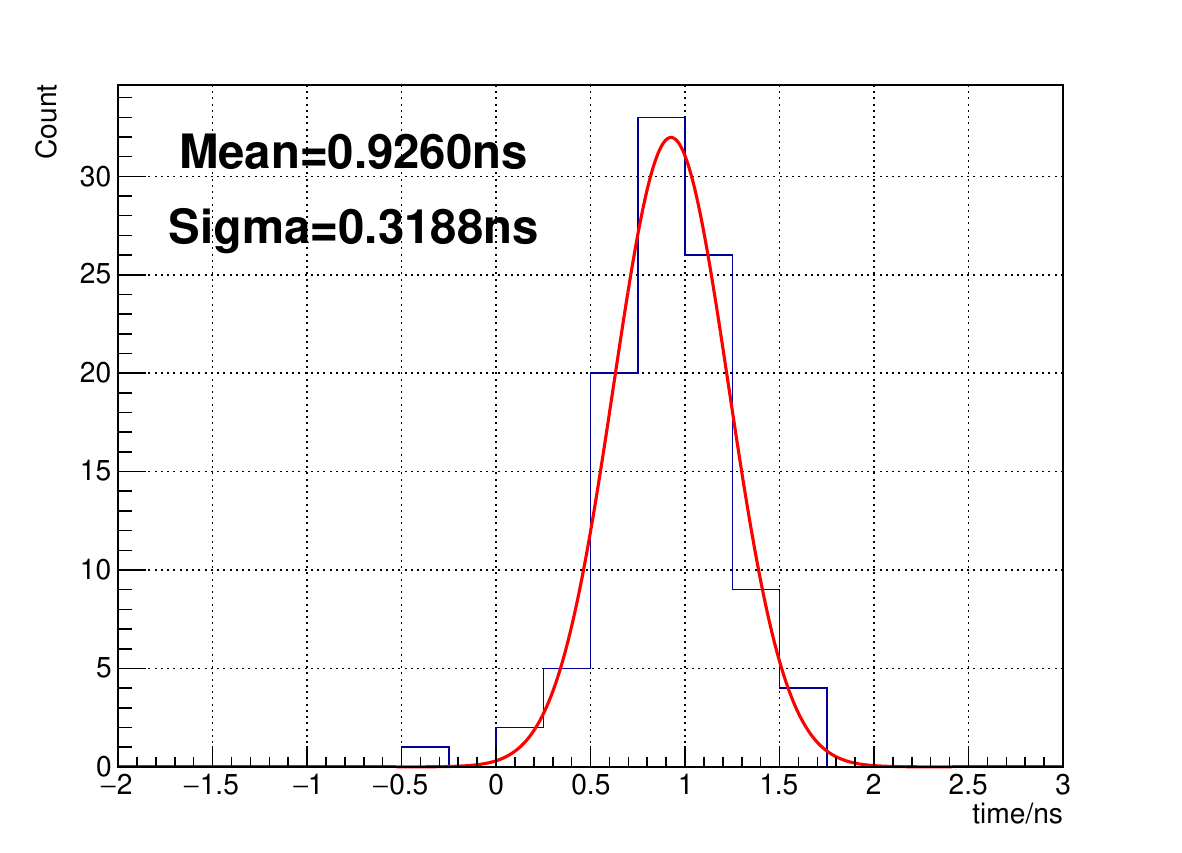}}
  \subfloat[][]{\includegraphics[width=0.4\textwidth]{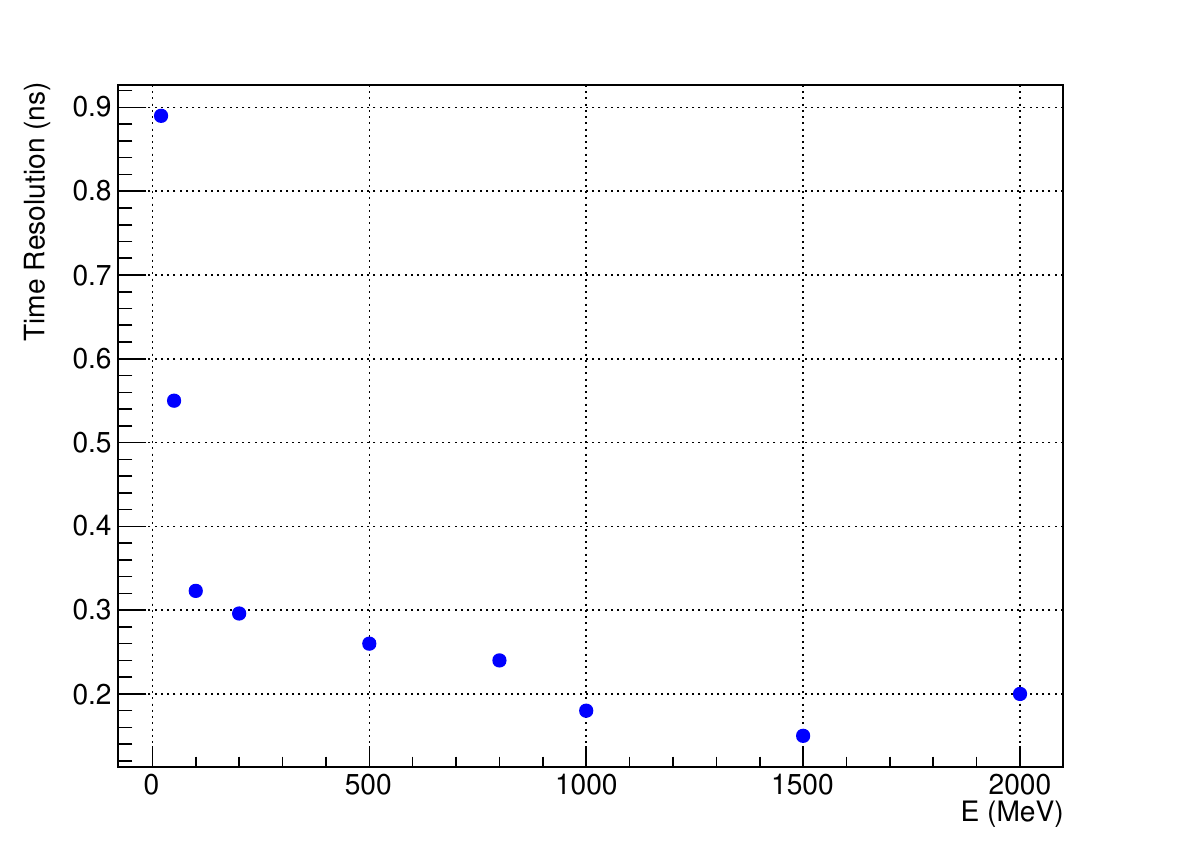}}
\caption{(a) The time resoluiton of 100 MeV gamma rays and (b) time resolution curve.}
\label{Fig4:EMC Time Reslution Curve}
\end{figure*}

The full response of the EMC including the optical processes, the response of APD and readout electronics is simulated with {\sc Geant4}. The output waveform from the simulation was processed using the template fitting method (see Sec.~\ref{sec:emc_waveformfitting}) to extract the arrival time of the output signal (corresponding to the hit time of incident particles up to a certain delay). Assuming that the light yield is 100 pe/MeV, the distribution of the hit time on the seed crystal in a shower for 100 MeV photons is shown in Fig.~\ref{Fig4:EMC Time Reslution Curve}(a). The width of this distribution representative of the EMC time resolution is 318.8~ps (no electronic noise is considered here). Figure~\ref{Fig4:EMC Time Reslution Curve}(b) shows the time resolution for different energy deposits. The time resolution improves as the deposited energy increases and can reach 200~ps at 1 GeV. Our preliminary study has demonstrated a light yield beyond 100 pe/MeV for the pCsI unit of the EMC. In view of the enhanced light yield, the time resolution of the EMC could be even better. For muons or charged hadrons that penetrate the EMC without shower, the  deposited energy is about 100 MeV, and hence the EMC time resolution for these particles would be about 300 ps. If either charged or neutral hadrons produced shower in the EMC, the time resolution for these particles may be subject to the uncertainty of the shower starting point.

\subsubsection{Impact of Upstream Materials}
\begin{figure*}[htbp]
 \centering
  %\vskip -1.5cm
  \subfloat[][]{\includegraphics[width=0.4\textwidth]{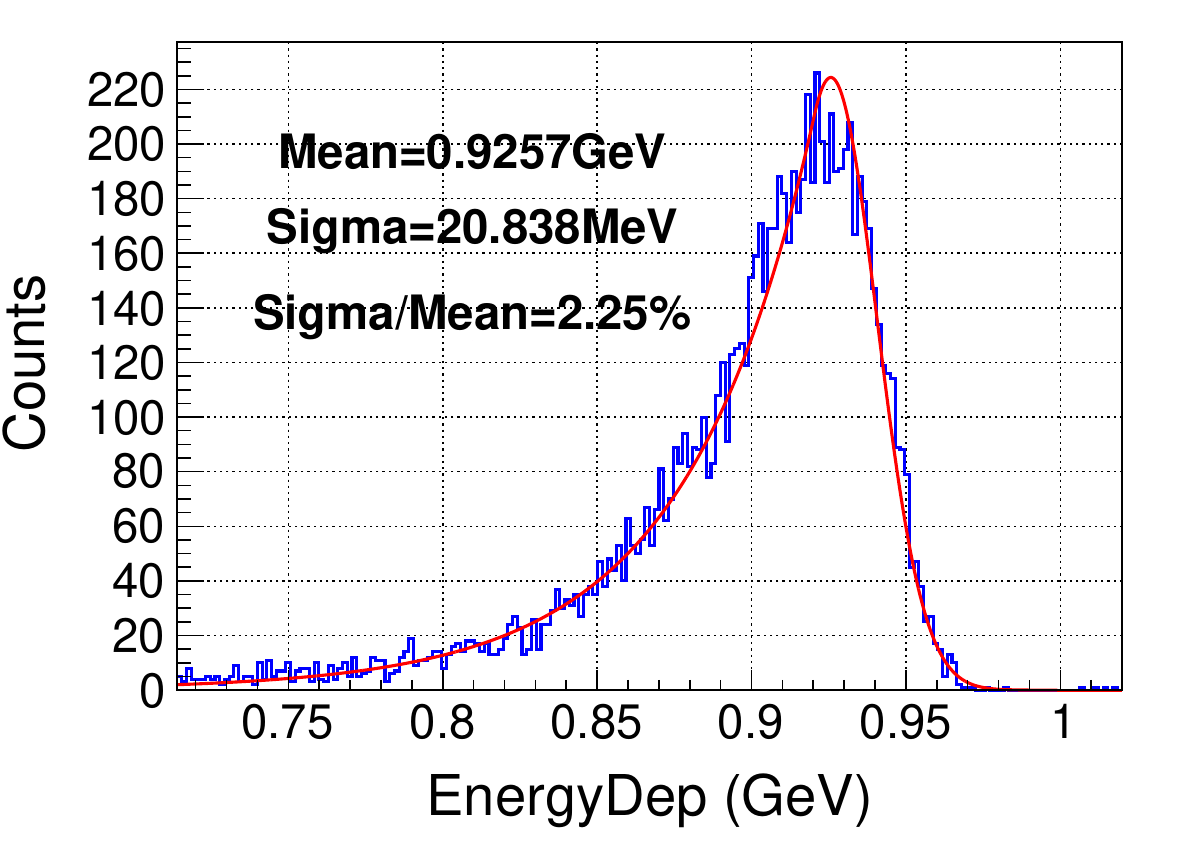}}
  \subfloat[][]{\includegraphics[width=0.4\textwidth]{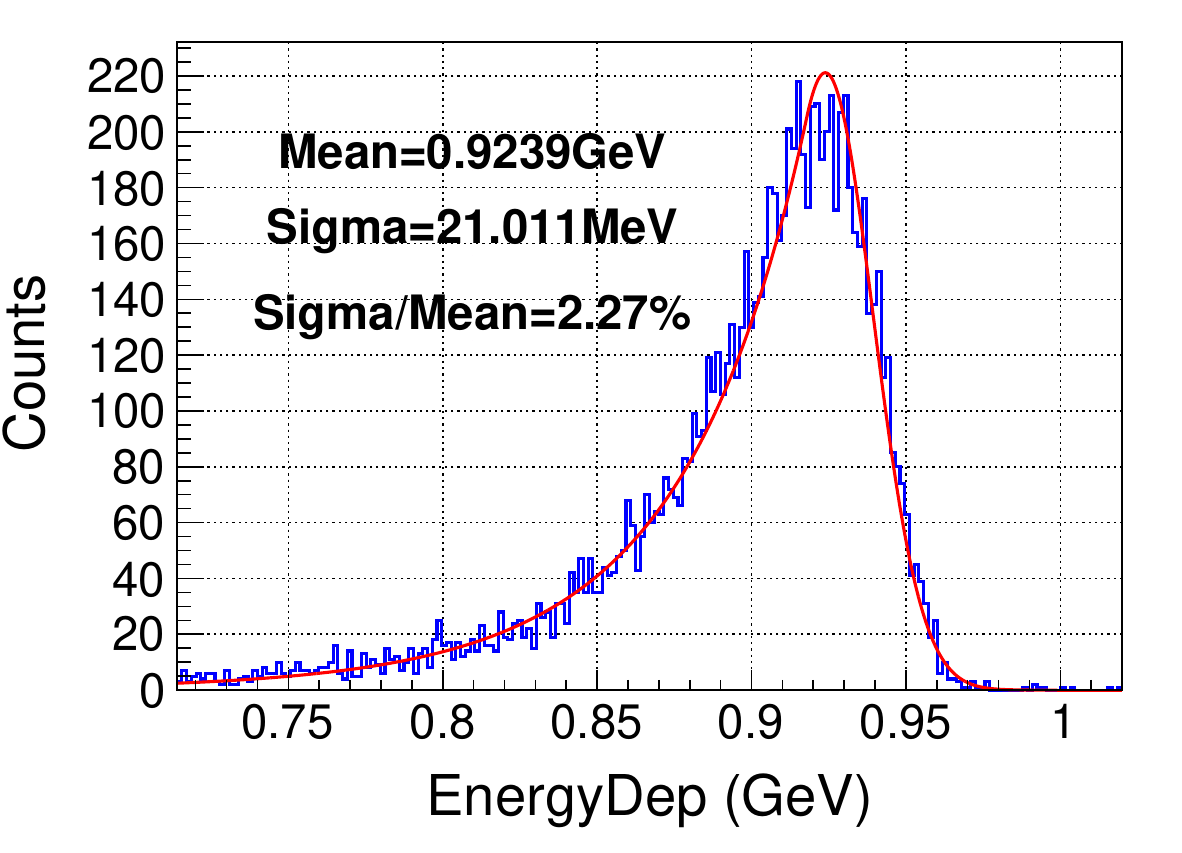}}

  \subfloat[][]{\includegraphics[width=0.4\textwidth]{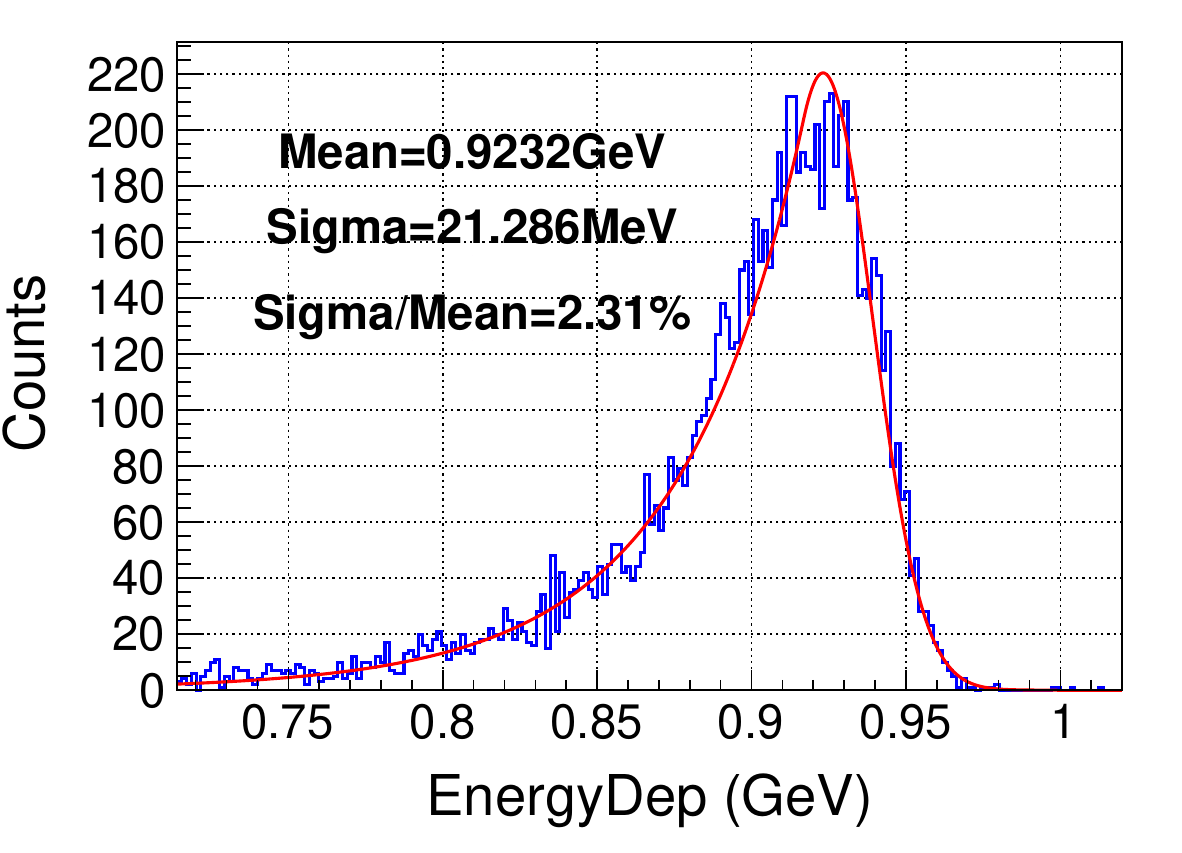}}
  \subfloat[][]{\includegraphics[width=0.4\textwidth]{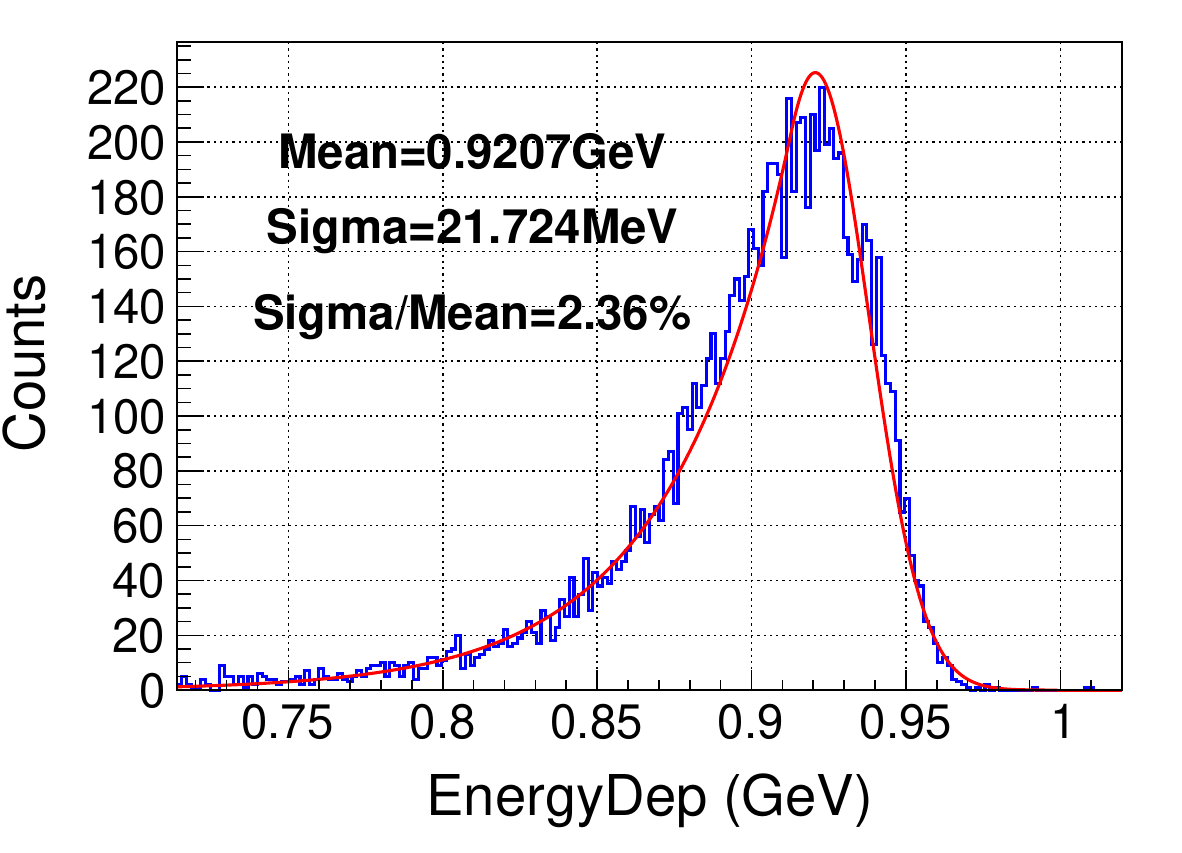}}
\caption{The energy resolution of EMC with different thicknesses of upstream materials. (a) 23\% $X_{0}$, (b) 27\% $X_{0}$, (c) 31\% $X_{0}$ and (d) 35\% $X_{0}$.}
\label{Fig4:EMC-ER-UpStreamMaterial}
\end{figure*}

\begin{figure*}[htbp]
 \centering
 \mbox{
  %\vskip -1.5cm
  \begin{overpic}[width=0.5\textwidth, height=0.33\textwidth]{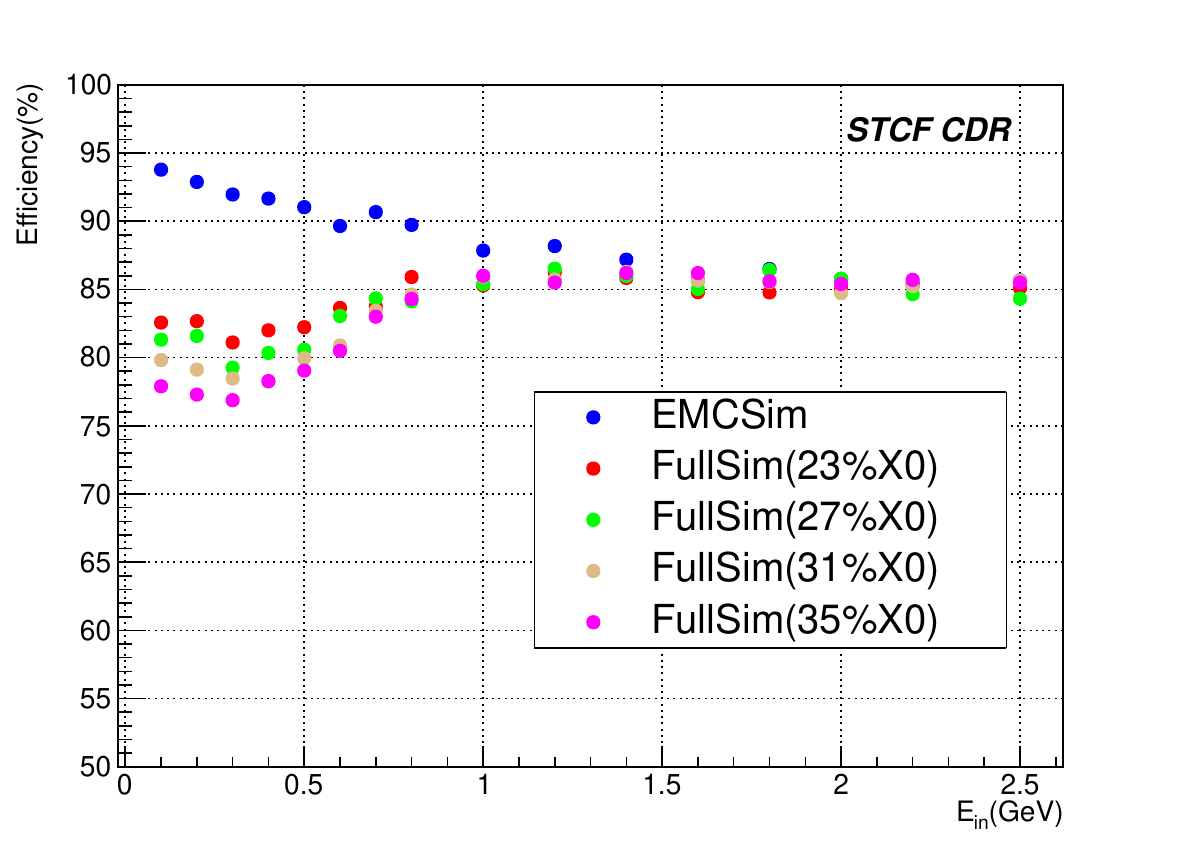}
  \end{overpic}
 }
\caption{The Reconstruction Efficiency curve of the EMC with upstream materials.}
\label{Fig4:EMC-Effi-UpStreamMaterial}
\end{figure*}

%{Influence of upstream materials}
Before arriving at the EMC, photons have a certain probability of interacting with the beam pipe or other subdetectors in front of the EMC. To study the influence of the upstream materials on the performance of the EMC, materials with different thicknesses are added in front of the EMC, and the resulting EMC performance is compared. Four cases of radiation length for materials in front of the EMC, 23\% $X_{0}$, 27\% $X_{0}$, 31\% $X_{0}$ and 35\% $X_{0}$ are considered. The equivalent mass of aluminum is placed at the RICH detector position, about 10 cm in front of the EMC. The simulated energy resolution is shown in Fig.~\ref{Fig4:EMC-ER-UpStreamMaterial}, and the energy resolution changes from 2.25\% to 2.36\%. The photon detection efficiency curve is shown in Fig.~\ref{Fig4:EMC-Effi-UpStreamMaterial}, and it can be seen that in the low energy region, below 1~GeV, with additional material, the detection efficiency decreases significantly, and the impact is found to be very small in the higher energy region.

\begin{figure*}[htbp]
 \centering
 \mbox{
  %\vskip -1.5cm
  \begin{overpic}[width=0.5\textwidth, height=0.33\textwidth]{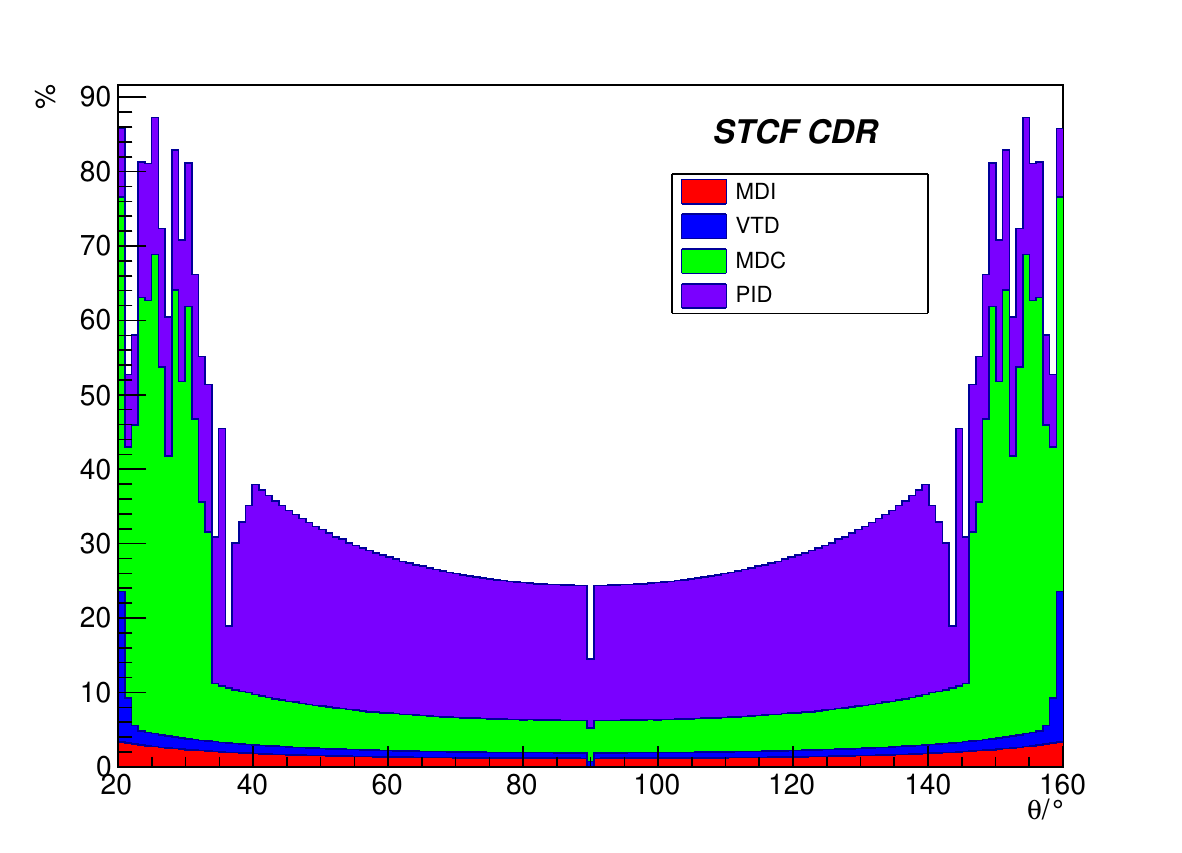}
  \end{overpic}
 }
\caption{The radiation length of the inner subdetectors in front of the EMC.}
\label{Fig4:STCF-Materials}
\end{figure*}

\begin{figure*}[htbp]
 \centering
  %\vskip -1.5cm
  \subfloat[][]{\includegraphics[width=0.4\textwidth]{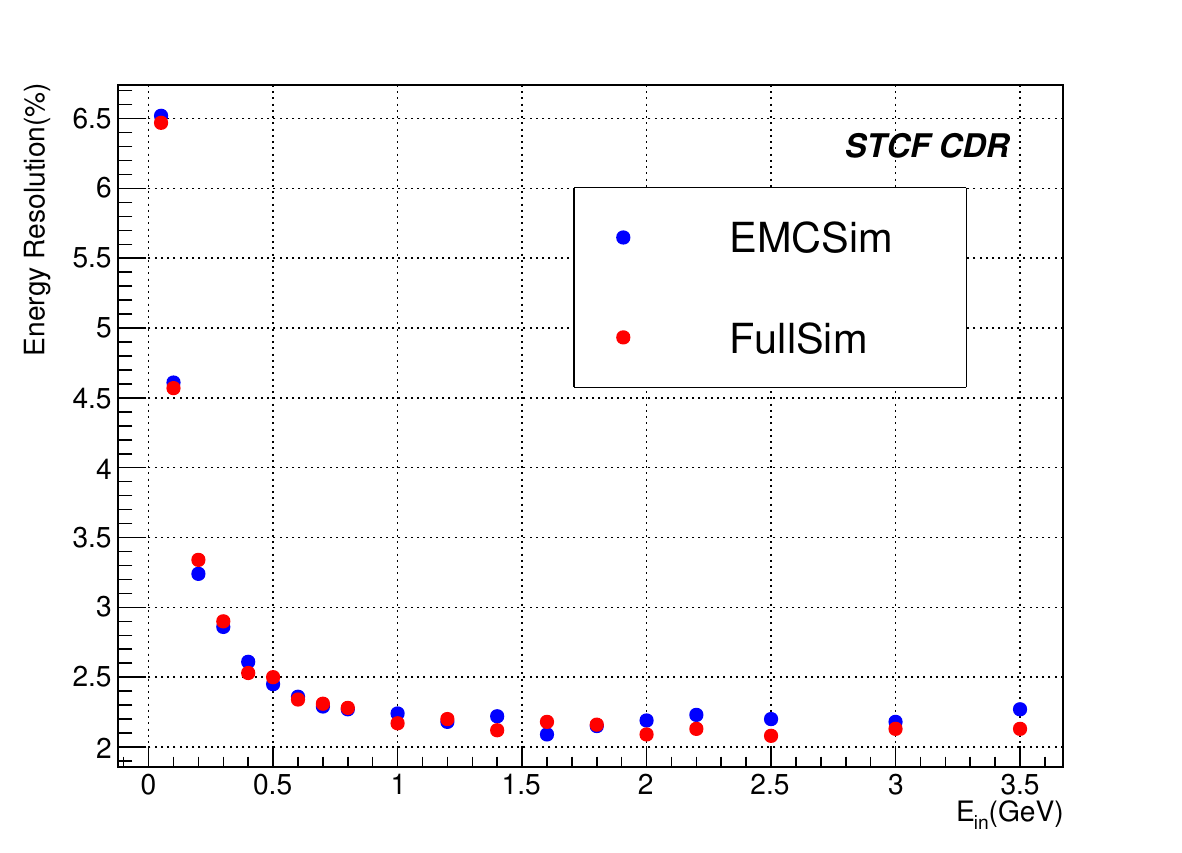}}
  \subfloat[][]{\includegraphics[width=0.4\textwidth]{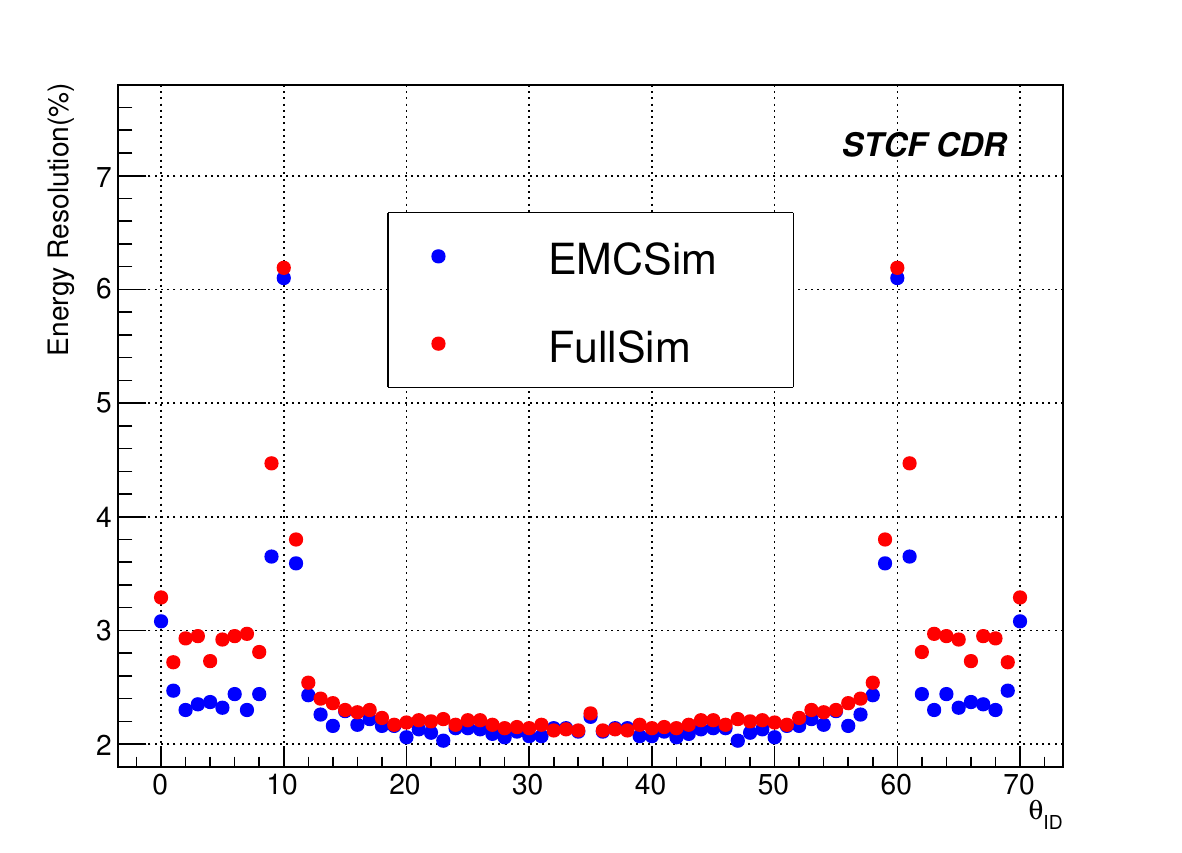}}
\caption{(a) Comparison of the energy resolution of the EMC (barrel) as a function of the energy with full detector simulation and EMC-only simulation. (b) The energy resolution as a function of the incident polar angle for 1~GeV photons.}
\label{Fig4:EMC-ERCurve-FullSim}
\end{figure*}

For the baseline design of the STCF detector system, the radiation length of each subdetector in front of the EMC is studied via simulation, as shown in Fig.~\ref{Fig4:STCF-Materials}. In the barrel, the total radiation length is about $0.3 X_0$, while in the endcap, the radiation length can reach $0.8 X_0$, which is mainly contributed by the MDC. To evaluate the effect of upstream materials, a Geant4 simulation study is carried out to preliminarily study the energy resolution of the EMC under the existing structural design. A full STCF detector simulation is performed, and the performance of the EMC is compared to that of an EMC-only simulation. The results are shown in Fig.~\ref{Fig4:EMC-ERCurve-FullSim}. Figure~\ref{Fig4:EMC-ERCurve-FullSim}(a) shows that upstream materials have little effect on the energy resolution of the EMC when considering only the barrel EMC. Figure~\ref{Fig4:EMC-ERCurve-FullSim}(b) shows the energy resolution at different polar angles of incident photon with an energy of 1~GeV. In the endcap region, due to material effects of inner subdetectors, the expected energy resolution decreases from about 2.5\% to about 3.0\%. The significant degradation of the energy resolution around $\theta=10^{\circ}$ and $\theta=60^{\circ}$ is due to the large energy leakage at the transition region of the barrel and the endcap.

\subsection{Pileup Mitigation}

\subsubsection{Challenges of High Background}
In Sec.~\ref{sec:emc_perf}, when studying the expected performance of the EMC, the background contribution is normally not considered in the simulation. The photon reconstruction is based on a simple clustering algorithm used in BESIII. The reconstruction algorithm searches the related crystals in a shower, adds their energies together, and calculates the hit position.

\begin{figure*}[htbp]
 \centering
  %\vskip -1.5cm
  \subfloat[][]{\includegraphics[width=0.4\textwidth]{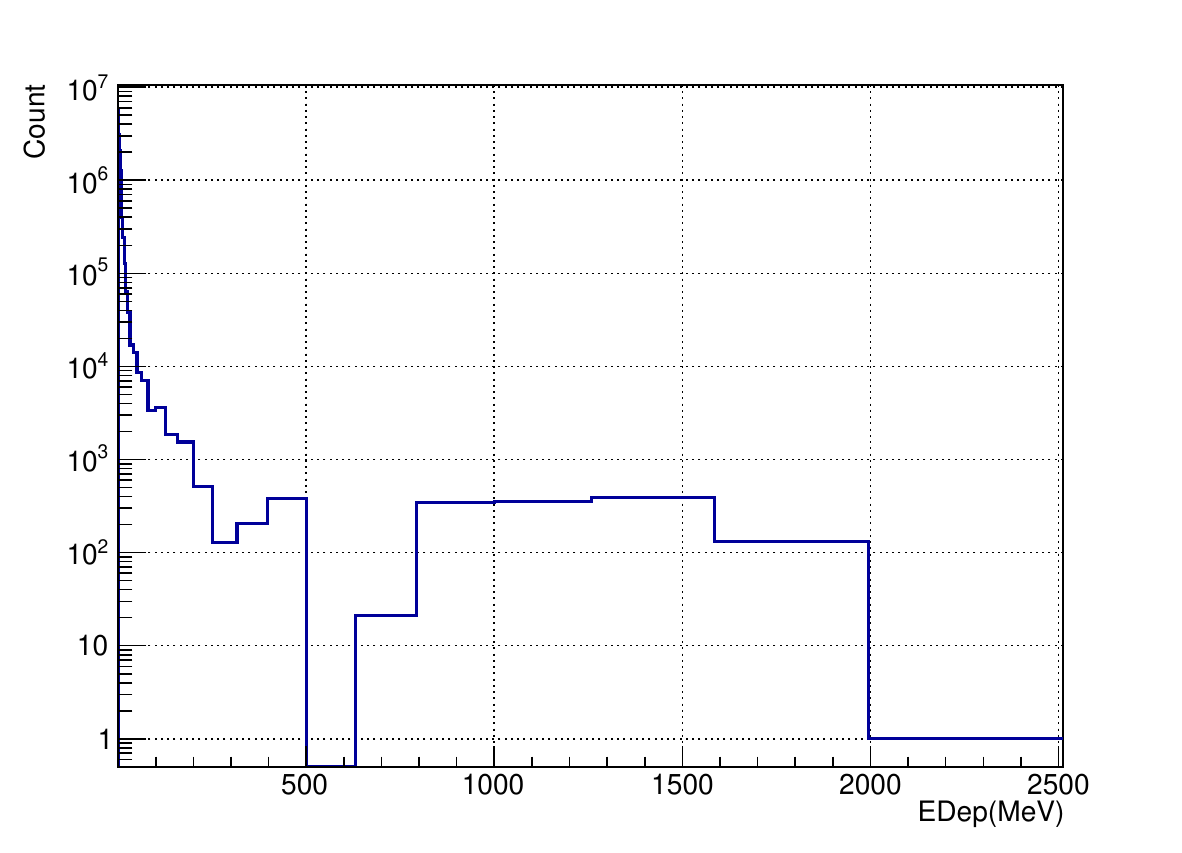}}
  \subfloat[][]{\includegraphics[width=0.4\textwidth]{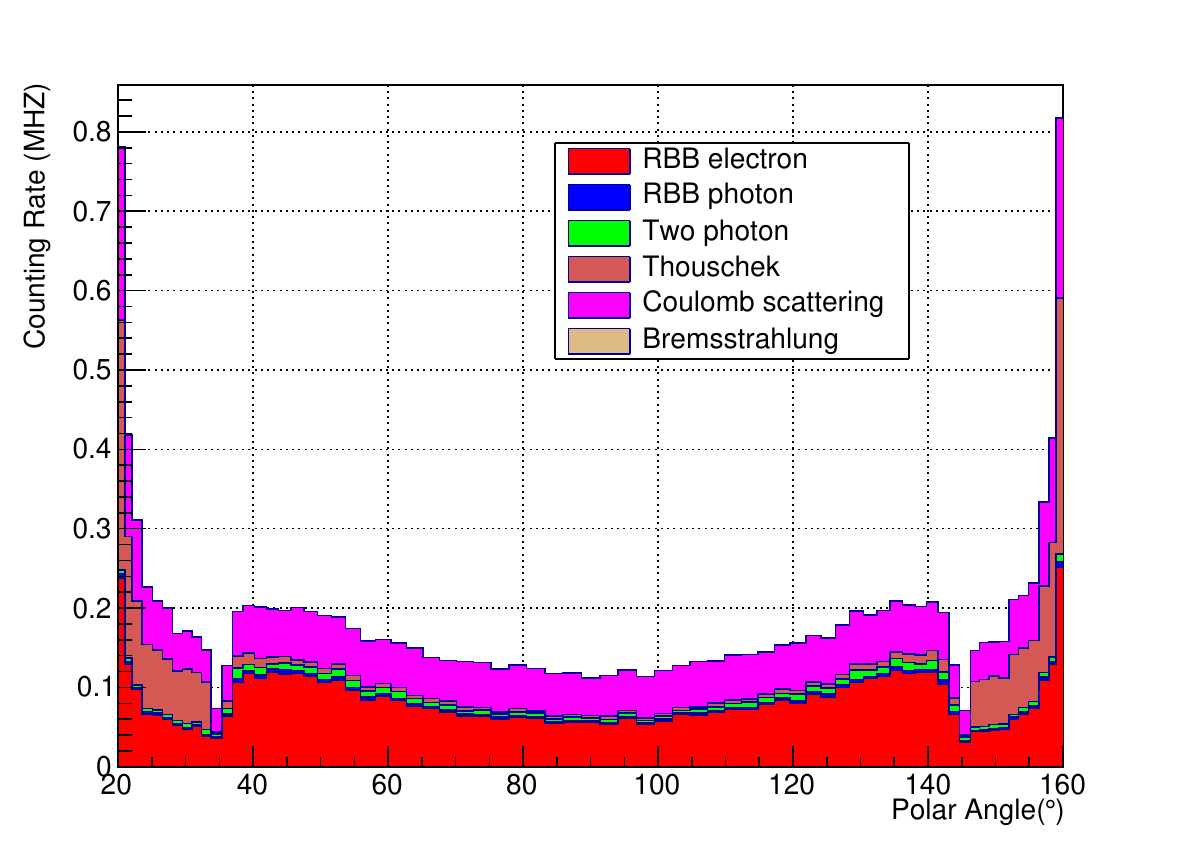}}
\caption{The background simulation in the EMC: (a) the deposited energy distribution and (b) the background counting rate.}
\label{Fig4:ECAL Background-Energy-CR}
\end{figure*}

Considering the high background at the STCF, the EMC background is studied by Monte Carlo simulation. The simulated background energy distribution of the EMC is shown in Fig.~\ref{Fig4:ECAL Background-Energy-CR}(a). Figure~\ref{Fig4:ECAL Background-Energy-CR}(b) shows the background counting rate at each position of the calorimeter (with a threshold value of 1~MeV). The background counting rate is close to about 1~MHz. This result is consistent with the simulation data shown in Table~\ref{tab:TIDNIEL_max}, which uses a 0.5~MeV threshold.

\begin{figure*}[htbp]
 \centering
  %\vskip -1.5cm
  \subfloat[][]{\includegraphics[width=0.4\textwidth]{Figures/Figs_04_00_DetectorSubSystems/Figs_04_04_ElectromagneticCalorimeter/ECAL-EneRes-Noise-1MeV.pdf}}
  \subfloat[][]{\includegraphics[width=0.4\textwidth]{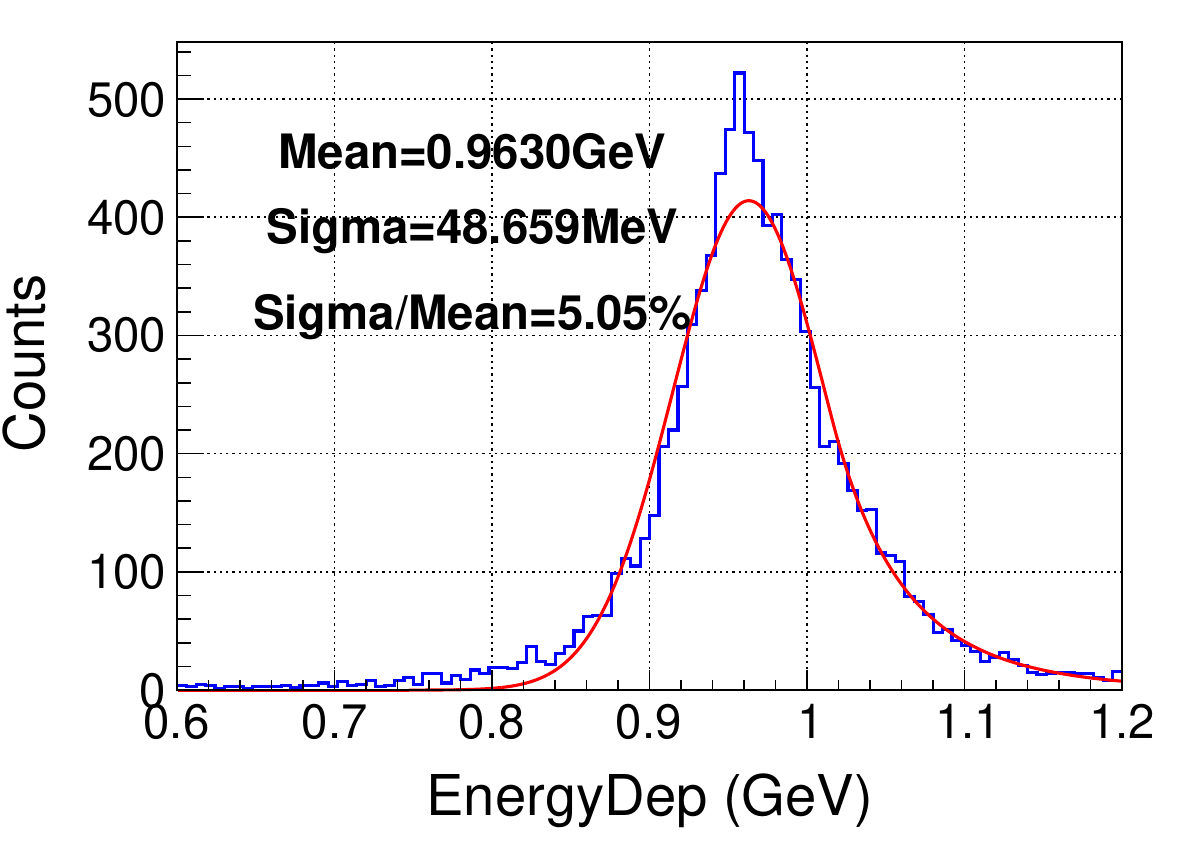}}
\caption{The EMC energy response for 1~GeV photons (a) without background and (b) with background included in the simulation.}
\label{Fig4:ECAL EnergyRes Before and After BG}
\end{figure*}

Such a high background counting rate has a great impact on the energy measurement of the calorimeter. As shown in Fig.~\ref{Fig4:ECAL EnergyRes Before and After BG}, the energy spectrum of the 1~GeV gamma-ray is reconstructed without and with considering the background. The results show that the energy resolution values are 2.15\% and 5.05\%, respectively. After the background is introduced, the energy resolution is degraded by more than a factor of two.

\subsubsection{Waveform Fitting}
\label{sec:emc_waveformfitting}
%{Waveform Sampling and Fitting Method}
To correct the uncertainty of the energy measurement caused by a high intensity background, one feasible scheme is to reconstruct the amplitude and time information by using a waveform fitting algorithm; this was used in the electromagnetic calorimeter of the CMS experiment at the LHC~\cite{CMS-CAL}. The CMS template fitting technique, named ``multifit'', was motivated by the reduction of the LHC bunch spacing from 50 to 25~ns and by the higher instantaneous luminosity of Run II, which led to a substantial increase in both the in-time and out-of-time pileup, where the latter refers to the overlapping signals from neighboring bunch crossings. It has been demonstrated by CMS that, with the multifit method, the contribution of out-of-time pileup to the signal reconstruction is found to be negligible, both in data and in simulated samples. The energy resolution and response are improved with respect to those of the Run I method where the amplitude reconstructed as a weighted sum of the ten digitized samples.

For the STCF EMC, considering the design parameters of the readout circuit (see Sec.~\ref{sec:emc_elec}), the total width of the signal after shaping is about 500~ns, and the leading edge is about 50~ns. To sample the leading edge and consider that the total amount of data is controllable, we use a sampling rate of 40 MHz as the baseline design here. In the multiwaveform fitting algorithm, the time window of the waveform fitting is set to 1000~ns (-250~ns - 750~ns, and the signal start time is set to 0~ns). The pulse template, $p_j$, for both the signal and the background is built by convoluting the pCsI fluorescence signal with the CSA impulse response function. The total template, $P$ is shown in Eq.~\ref{eq:multifit}; this template is an overlay of the signal template and the background template but shifted in time, in which the normalizations are free parameters ($A_j$). The total template is then used to fit the waveform readout by the electronics. Referring to the CMS multifit method, $N=40$ templates with a repetition period of 12.5~ns (CMS uses 25~ns, half of which is temporarily taken here) are used to fit our results. Starting at -250~ns, one template is placed every 12.5~ns, and a total of 40 templates are used for a multitemplate fitting.
\begin{equation}
P = \sum_{j=0}^{N} A_j p_j,\label{eq:multifit}
\end{equation}

\begin{figure*}[htbp]
 \centering
  %\vskip -1.5cm
  \subfloat[][]{\includegraphics[width=0.4\textwidth]{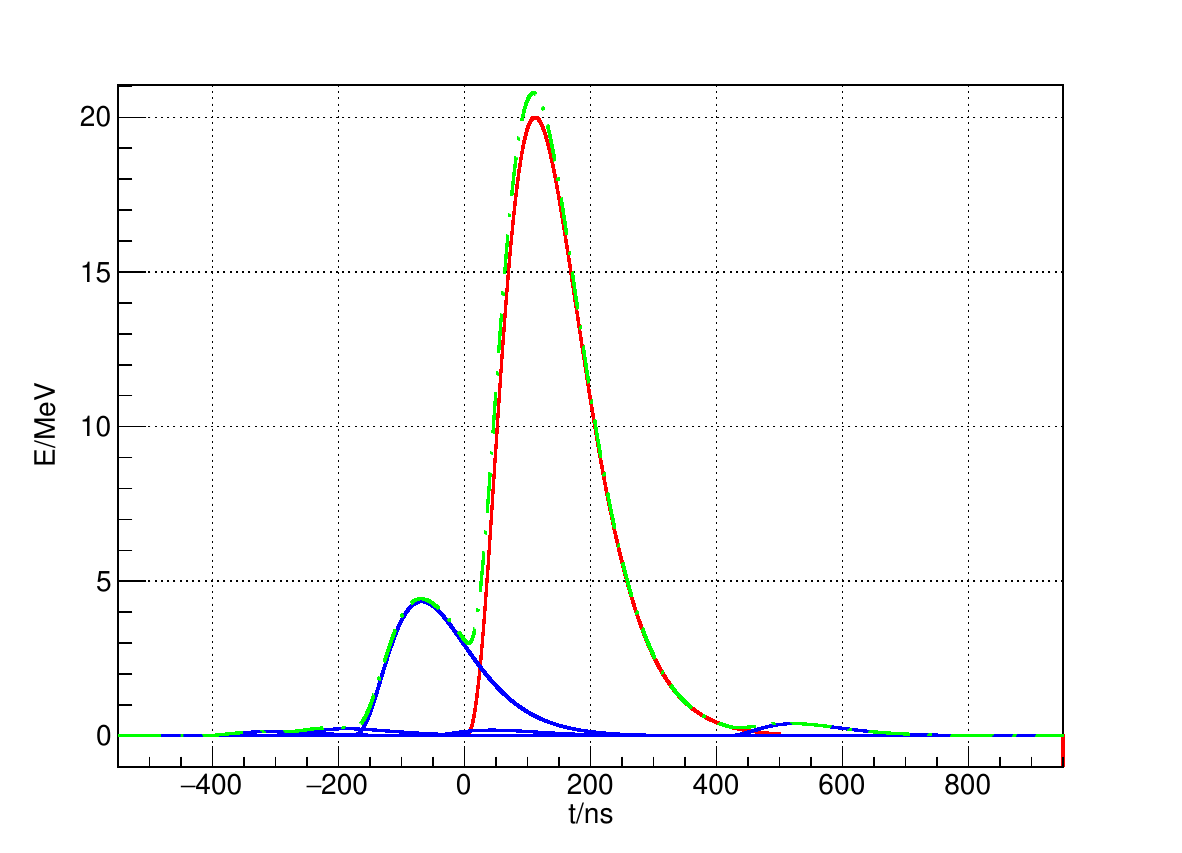}}
  \subfloat[][]{\includegraphics[width=0.4\textwidth]{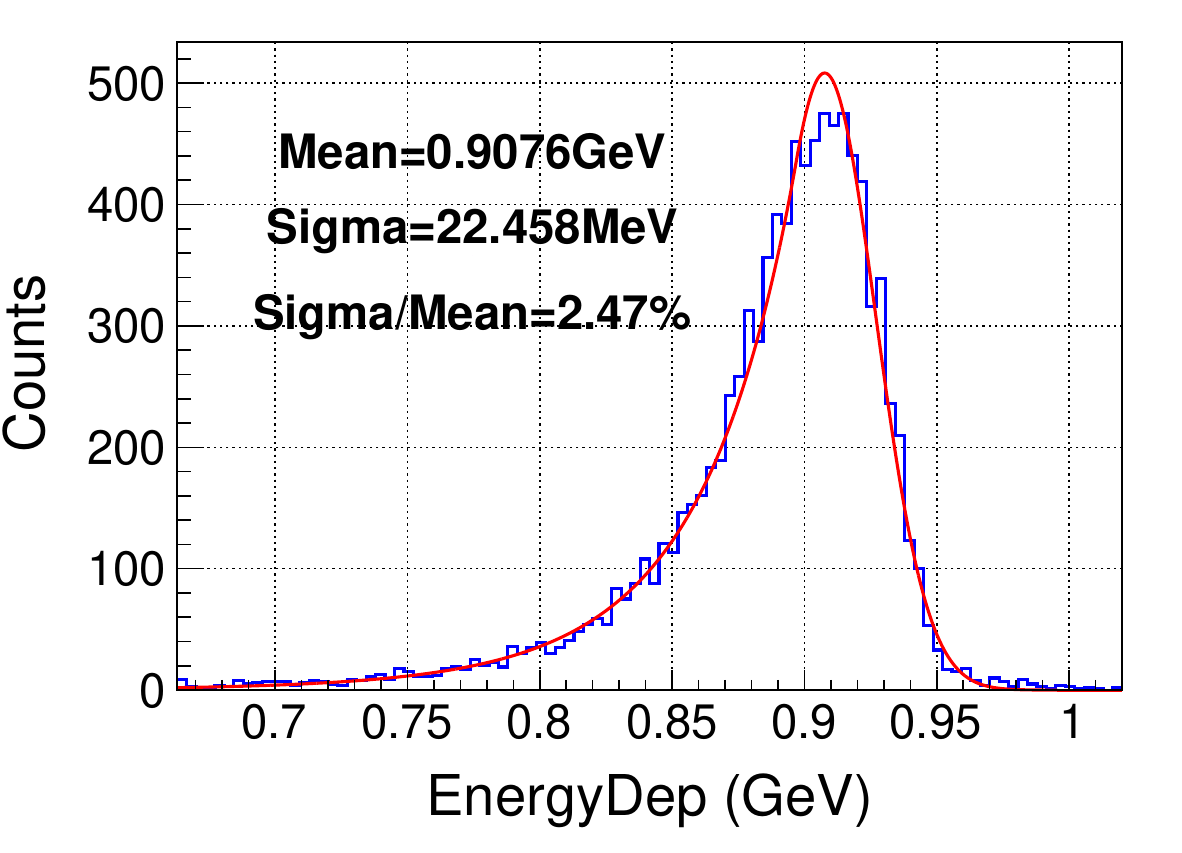}}
\caption{(a) An example output pulse of the EMC with multiwaveform fitting. The dotted green curve is a simulated waveform, which is a superposition of the signal and background spectra. The red curve represents the signal template, and the blue represents the fitting results of the background, (b) The energy resolution of 1~GeV $\gamma$ rays in the EMC using the multifit method.}
\label{Fig4:ECAL EnergyRes with Multifit}
\end{figure*}

A set of toy events are generated to test the performance of the multiwaveform fitting.
In Fig.~\ref{Fig4:ECAL EnergyRes with Multifit}(a), the dotted green curve is the superposition of the signal and background spectra of one simulated event. The signal peak is around 120 ns (the integration time is set as 40 ns), and the rest represent the backgrounds. The background event rate is assumed to be $\sim$~MHz, and the energy is obtained by sampling according to Fig.~\ref{Fig4:ECAL Background-Energy-CR}(a). Because of the background, the peak value of the spectrum is larger than the real value. With multiwaveform fitting, the original energy can be reconstructed more precisely. Additionally as shown in Fig.~\ref{Fig4:ECAL EnergyRes with Multifit}(a), the red curve represents the signal template, and blue represents the fitting results of the background templates. Figure~\ref{Fig4:ECAL EnergyRes with Multifit}(b) shows the energy resolution of 1~GeV gamma-ray events, obtained by multiwaveform fitting. The result is 2.47\%, which is 50\% better than that without waveform fitting. This result is close to the result without considering the background.

Figure~\ref{Fig4:ECAL EnergyRes with BG and Multi-Fitting}(a) shows the EMC energy resolution curve from the Geant4 simulation for photons reconstructed in the barrel only. The expected energy resolution is compared for three scenarios: the traditional clustering reconstruction algorithm without the influence of background, that with the influence of the background, and the multifit method with the background included. The result shows that the pileup background has a significant impact on the energy resolution, especially in the low energy region (\textless 100~MeV), and the energy resolution changes from $\sim$ 4.6\% to $\sim$ 22\% at the energy of 100~MeV when the background effect is included. With increasing energy, the impact of the background  gradually decreases, which is mainly due to the relatively low energy of the background. By using the waveform sampling and fitting method, the energy resolution is greatly improved, from $\sim$ 22\% to $\sim$ 5.2\%, which is close to the result without considering the background. The result for the endcap, which suffers from an even higher pileup background of about 10~MHz, is shown in Fig.~\ref{Fig4:ECAL EnergyRes with BG and Multi-Fitting}(b). Even for a 10~MHz background rate, the multiwaveform fitting method works well and can greatly improve the energy resolution.

\iffalse
\begin{figure*}[htbp]
 \centering
 \mbox{
  %\vskip -1.5cm
  \begin{overpic}[width=0.6\textwidth, height=0.45\textwidth]{Figures/Figs_04_00_DetectorSubSystems/Figs_04_04_ElectromagneticCalorimeter/ECAL-Background-Pulse.pdf}
  \end{overpic}
 }
\caption{An example output pulse of the EMC with multiwaveform fitting. The dotted green curve is a simulated waveform, which is a superposition of the signal and background spectra. The red curve represents the signal template, and the blue represents the fitting results of the background.}
\label{Fig4:ECAL Background-Pulse}
\end{figure*}

\begin{figure*}[htbp]
 \centering
 \mbox{
  %\vskip -1.5cm
  \begin{overpic}[width=0.6\textwidth, height=0.45\textwidth]{Figures/Figs_04_00_DetectorSubSystems/Figs_04_04_ElectromagneticCalorimeter/ECAL-ER-AftMFit.pdf}
  \end{overpic}
 }
\caption{The energy resolution of 1~GeV $\gamma$ rays in the EMC using the multifit method.}
\label{Fig4:ECAL EnergyRes After MultiFit with BG}
\end{figure*}
\fi

\begin{figure*}[htbp]
 \centering
  %\vskip -1.5cm
  \subfloat[][]{\includegraphics[width=0.4\textwidth]{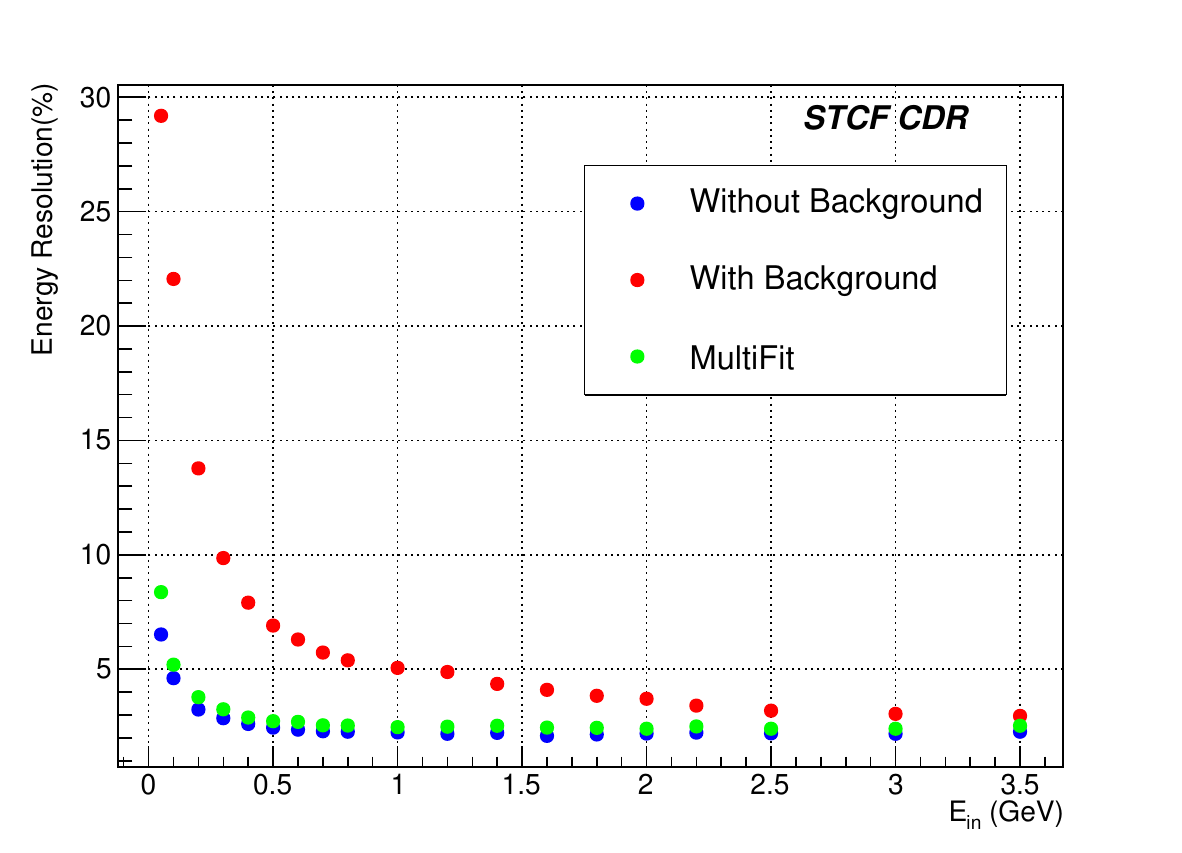}}
  \subfloat[][]{\includegraphics[width=0.4\textwidth]{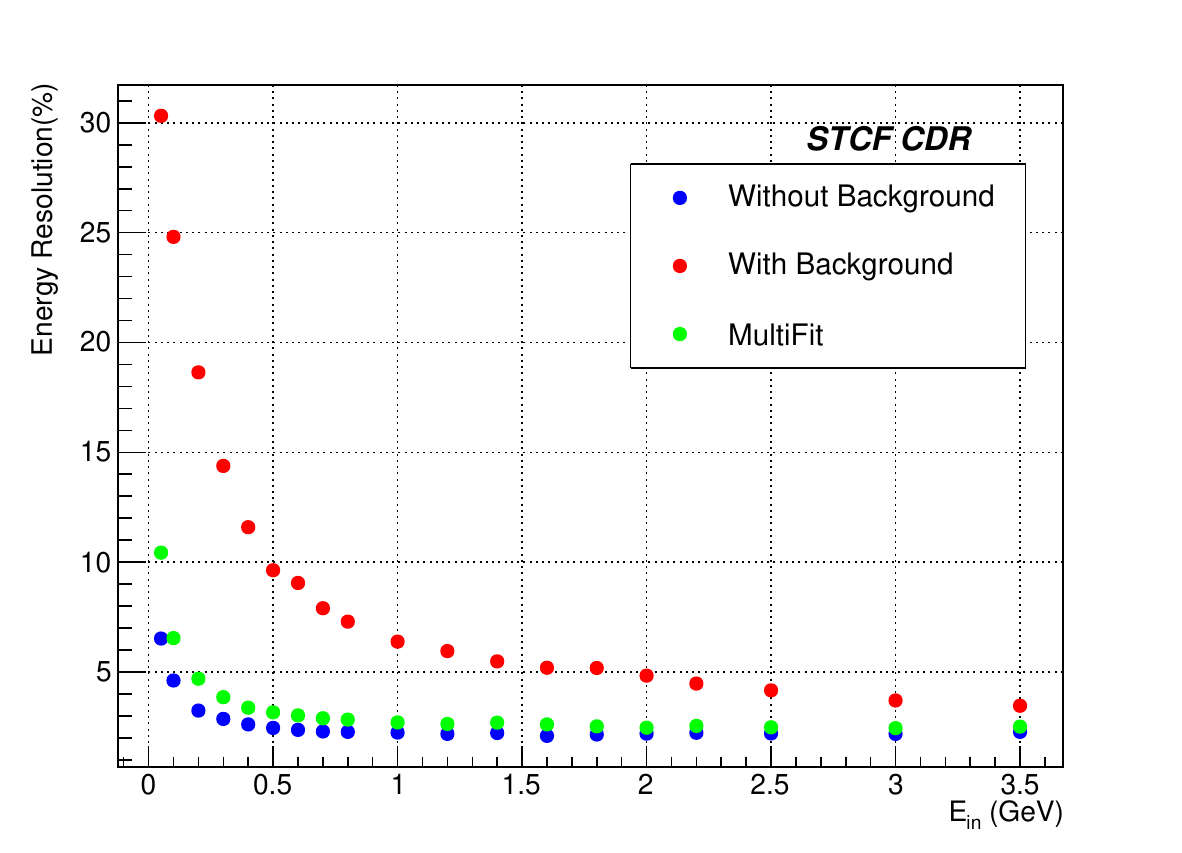}}
\caption{The expected EMC energy resolution with the multifit method for (a) the barrel region and (b) the endcap region.}
\label{Fig4:ECAL EnergyRes with BG and Multi-Fitting}
\end{figure*}

%%%%%%%%%%%%%%%%%%%%%%%%%%%%%%%%%%%%%%%%%%%%%%%%%%%%%%%%%%%%%%%%%%%%%%%%%%%%%%%%%
%%%%%%%%%%%%%%%%%%%%%%%%%%%%%%%%%%%%%%%%%%%%%%%%%%%%%%%%%%%%%%%%%%%%%%%%%%%%%%%%%
\FloatBarrier

%%%%%%%%%%%%%%%%%%%%%%%%%%%%%%%%%%%%%%%%%%%%%%%%%%%%%%%%%%%%%%%%%%%%%%%%%%%%%%%%%
%%%%%%%%%%%%%%%%%%%%%%%%%%%%%%%%%%%%%%%%%%%%%%%%%%%%%%%%%%%%%%%%%%%%%%
\subsection{Readout Electronics}
\label{sec:emc_elec}
\begin{figure*}[htb]
 \centering
 \mbox{
  %\vskip -1.5cm
  \begin{overpic}[width=0.5\textwidth, height=0.33\textwidth]{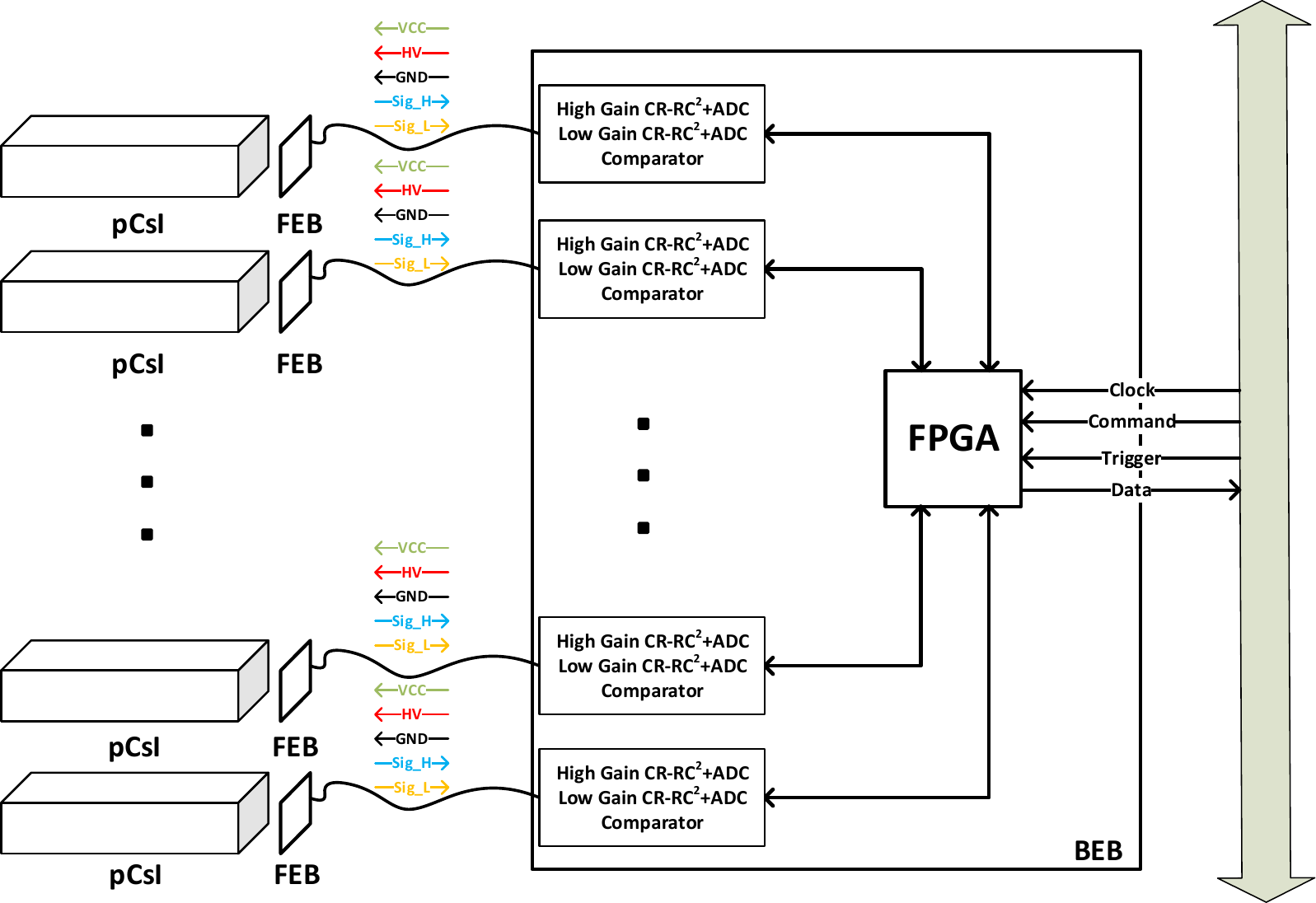}
  \end{overpic}
 }
\caption{The structure of CSA-based readout electronics.}
\label{Fig4:ECAL-electronics-1st}
\end{figure*}

The electronics system of the EMC provides the photon detector signal readout, analog-to-digital conversion and data acquisition. According to the physics requirements and detector characteristics, some demands placed to the electronics system, which are listed below:
\begin{itemize}
\item The energy deposition on each crystal ranges from 2.5~MeV to 2500~MeV. Considering that the light yield of pCsI is about 100 p.e./MeV and the gain of the photodetector is about 50, the dynamic range of each electronics channel should range from 2~fC to 2000~fC.
\item The high luminosity of the STCF results in a high event rate in the detector. It is estimated that the event rate in the barrel can reach as high as hundreds kHz, and it can be even higher in the endcap. Therefore, the deadtime of the readout system should be shorter than 1~$\mu s$.
\item In addition to energy measurement, time measurement is needed and the precision should be better than 200 ps at 1 GeV.
\end{itemize}
%%%%%%%%%%%%%%%%%%%%%%%%%%%%%%%%%%%%%%%%%%%%%%%%%%%%%%%%%%%%%%%%%%%%%%%%%%%%%%%
%\subsubsection{CSA-based Design}
According to the requirements discussed above, a charge-sensitive amplifier (CSA) based design is proposed. The structure of the CSA-based readout electronics is shown in Fig.~\ref{Fig4:ECAL-electronics-1st}, and it mainly consists of a front-end board (FEB) and a back-end board (BEB). The FEB is placed at the outer end of a pCsI crystal with 4 APDs on it to receive the fluorescence light. Multiple APDs are used to improve the light yield and the system robustness. On the other side of the FEB, 4 CSAs read out the signals of 4 APDs. Then, the outputs of the CSAs are added by two adders, providing in a dual-gain (high/low) outputs. One BEB can connect several FEBs via cables. The BEB provides power and high voltage to FEBs and obtains high-gain and low-gain signals from FEBs. Signals pass through the CR-RC$^2$ shaping circuits and are then digitized by ADCs on the BEB. In addition, there is one comparator corresponding to each channel that can compare the input signal with the threshold and generate a hit for the FPGA-TDC for time measurement. All data produced by the ADCs and the FPGA-TDCs are aggregated, packaged and transmitted by the FPGA. Considering the high event rate, we use the waveform sampling readout method to suppress the high background. The waveform sampling can retain the original waveform information, which is also necessary for the waveform fitting mentioned in the previous background study.

Due to the relatively large size ($\sim6.5\times6.5$~cm) of the crystal end, it is favorable to couple the APD with a large sensitive area to improve the light yield. Presently, there are only a few commercial models with large areas available. The HAMAMATSU company mainly provides two models: S8664-55 (effective area of $5\times5$~mm) and S8664-1010 (effective area of $10\times10$~mm).\\

%%%%%%%%%%%%%%%%%%%%%%%%%%%%%%%%%%%%%%%%%%%%%%%%%%%%%%%%%%%%%%%%%%%%%%%%%%%%%%%%%
%%%%%%%%%%%%%%%%%%%%%%%%%%%%%%%%%%%%%%%%%%%%%%%%%%%%%%%%%%%%%%%%%%%%%%%%%%%%%%%%%
\subsection{EMC R\&D}
\subsubsection{pCsI Crystal}
\begin{figure*}[htbp]
 \centering
 \mbox{
  %\vskip -1.5cm
  \begin{overpic}[width=0.5\textwidth, height=0.33\textwidth]{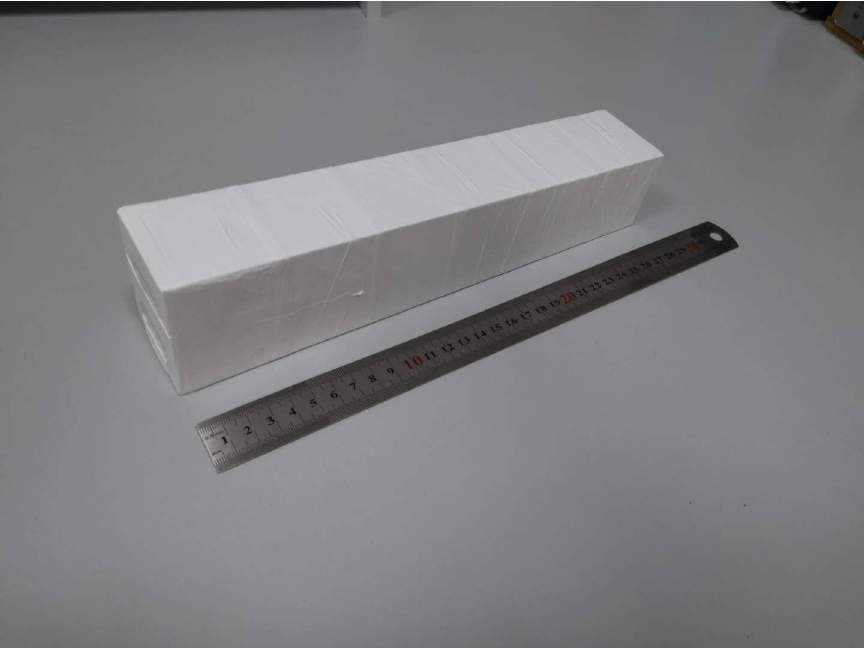}
  \end{overpic}
 }
\caption{The pCsI crystal for the EMC.}
\label{Fig4:ECAL CR-1st}
\end{figure*}

To understand the properties of the pCsI crystal, relevant tests are carried out in the laboratory. The crystals, as shown in Fig.~\ref{Fig4:ECAL CR-1st}, are produced at the Shanghai Institute of Ceramics, Chinese Academy of Sciences (SIC CAS). The wavelength of emission spectrum spreads from $\sim$ 260 nm to $\sim$ 700 nm, with its main component around 310 nm accounting for $\sim$ 3/4 of the total light intensity.

\iffalse
\begin{figure*}[htbp]
 \centering
  %\vskip -1.5cm
  \subfloat[][]{\includegraphics[width=0.4\textwidth]{Figures/Figs_04_00_DetectorSubSystems/Figs_04_04_ElectromagneticCalorimeter/ECAL-pCsIEmissionSpec.pdf}}
  \subfloat[][]{\includegraphics[width=0.4\textwidth]{Figures/Figs_04_00_DetectorSubSystems/Figs_04_04_ElectromagneticCalorimeter/ECAL-pCsITransmission.pdf}}
\caption{The fluorescence properties of the pCsI crystal, including (a) the emission spectrum and (b) the longitudinal light transmission.}
\label{Fig4:ECAL pCsI-1st}
\end{figure*}
\fi

\subsubsection{Reflective Material}
The efficiency of light collection is an important factor in achieving a high-precision energy resolution in crystal calorimeters. Since the fluorescence emitted by pCsI is mainly in the ultraviolet wavelength range (about 310 nm) and is very easily to be absorbed. Considering the excellent reflection coefficient of Teflon, which is basically independent of the wavelength, we chose Teflon with a thickness of 300 um to package the crystal.

\subsubsection{Cosmic Ray Test}
\label{sec:emc_cosmic}
\begin{figure*}[htbp]
 \centering
  %\vskip -1.5cm
  \subfloat[][]{\includegraphics[width=0.4\textwidth,angle=-90]{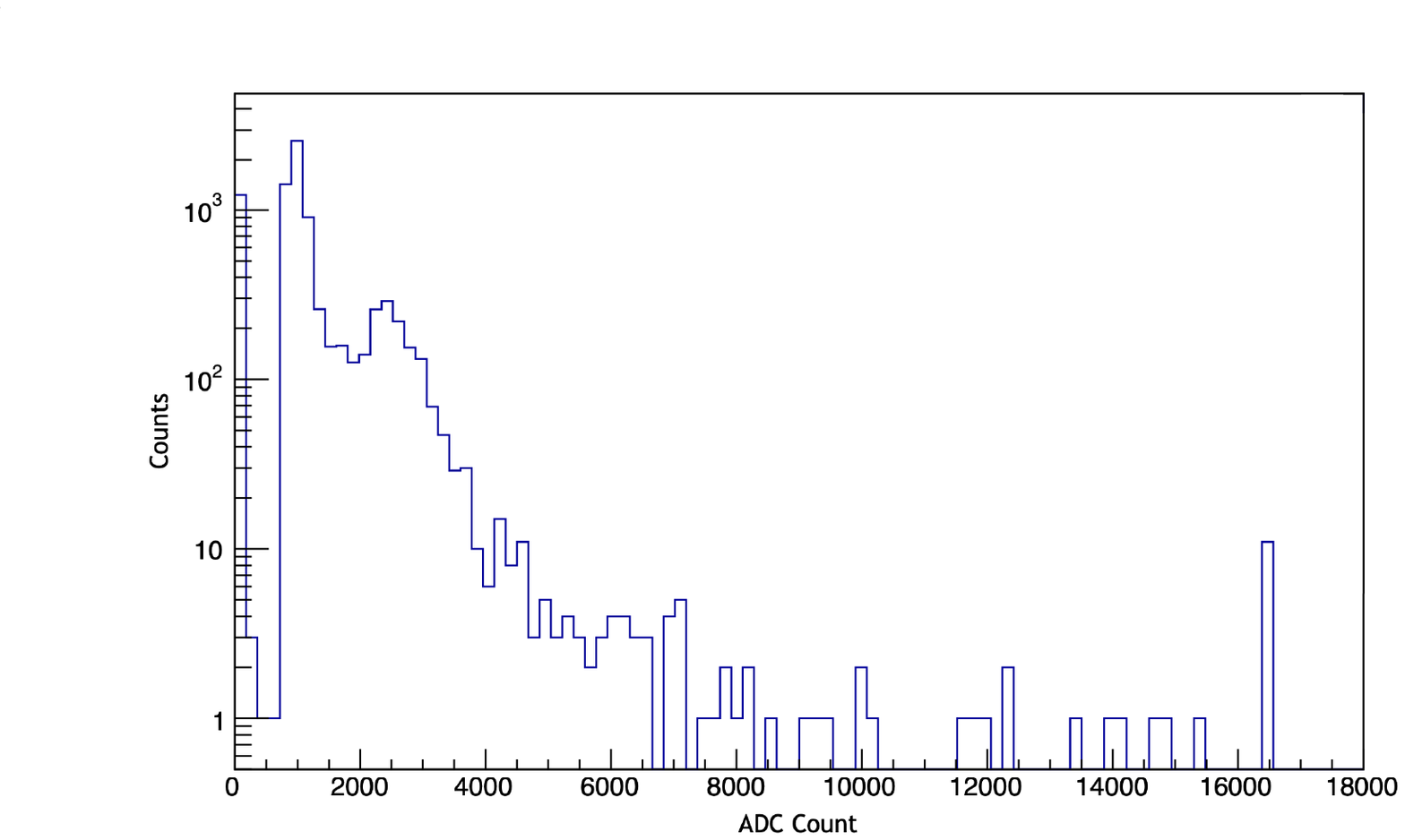}}
  \subfloat[][]{\includegraphics[width=0.4\textwidth,angle=-90]{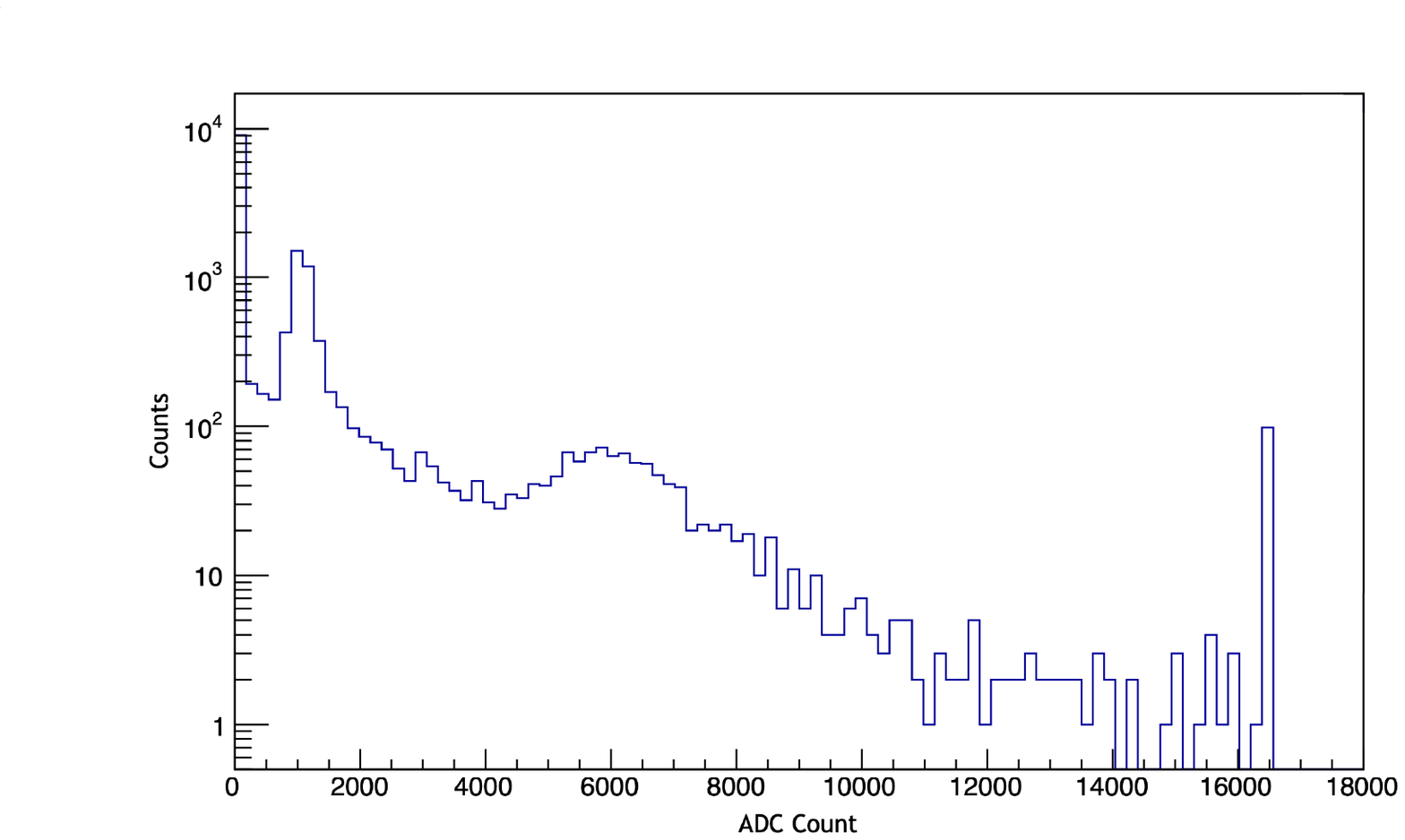}}
\caption{The energy deposition of MIPs. (a) S8664-55 result and (b) S8664-1010 result.}
\label{Fig4:ECAL CR-2nd}
\end{figure*}

The performance of the sensitive unit (pCsI + APD) is tested with the cosmic rays. A typical-sized pCsI crystal is coupled with four APDs. In the test, the crystal is wrapped with three layers of BC642 material. Two large-area APD models from HAMAMATSU are used, S8664-55 and S8664-1010, with sensitive areas of 5~mm~$\times$~5~mm and 10~mm~$\times$ 10~mm, respectively. The APD is coupled with silicone (EJ-550, 310 nm wavelength transmission is greater than 90\%) at the back end of the crystal. The deposition energy of Minimum Ionization Particles (MIPs, $\mu$) passing through the pCsI crystal is about 30 MeV. The results are shown in Fig.~\ref{Fig4:ECAL CR-2nd}. The light yield of pCsI is calculated to be 54 p.e./MeV (with S8664-55 APD) and 156~p.e./MeV (with S8664-1010 APD).\\

%%%%%%%%%%%%%%%%%%%%%%%%%%%%%%%%%%%%%%%%%%%%%%%%%%%%%%%%%%%%%%%%%%%%%%%%%%%%%%%%%

%%%%%%%%%%%%%%%%%%%%%%%%%%%%%%%%%%%%%%%%%%%%%%%%%%%%%%%%%%%%%%%%%%%%%%%%%%%%%%%%%
\subsubsection{Readout Electronics}

\begin{figure*}[htbp]
 \centering
 \mbox{
  %\vskip -1.5cm
  \begin{overpic}[width=0.8\textwidth, height=0.3\textwidth]{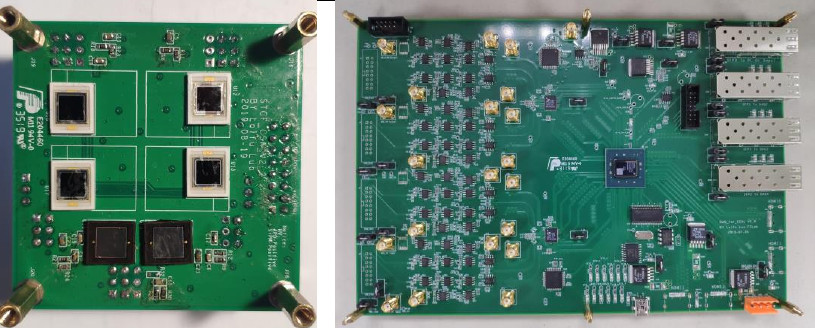}
  \end{overpic}
 }
\caption{Prototype electronics of the CSA-based method (FEE on the left and BEU on the right).}
\label{Fig4:ECAL-electronics-3rd}
\end{figure*}

%At present, the CSA has made a preliminary attempt.
A prototype readout out electronics system for the CSA-based method, described in Sec.~\ref{sec:emc_elec}, has been implemented, as shown in Fig.~\ref{Fig4:ECAL-electronics-3rd}. The dynamic range and noise performance is studied based on a CSA with a 3-JFET as the input stage with different APDs: Hamamatsu S8664-0505 and S8664-1010.

\begin{figure*}[htbp]
 \centering
 \mbox{
  %\vskip -1.5cm
  \begin{overpic}[width=0.5\textwidth, height=0.33\textwidth]{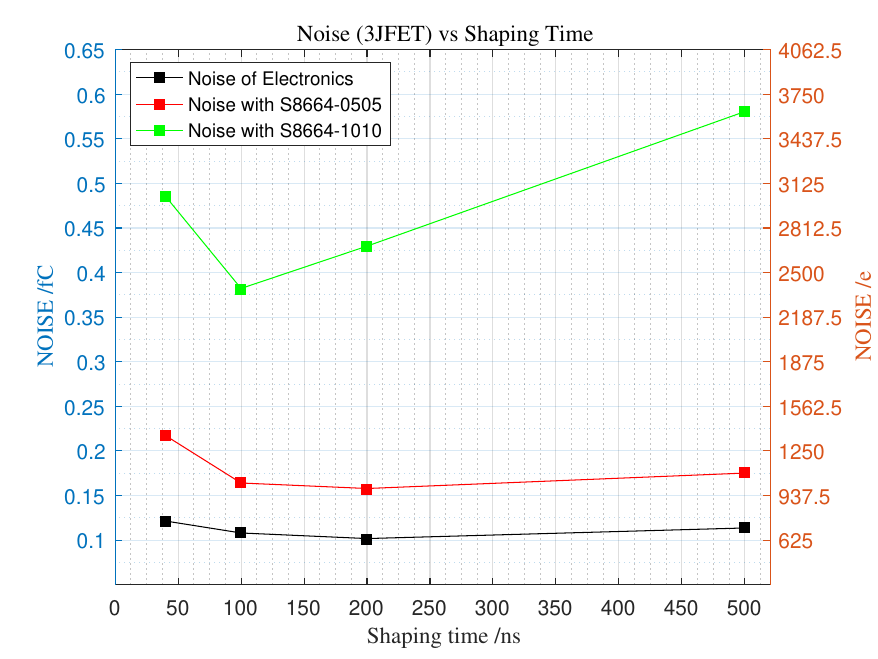}
  \end{overpic}
 }
\caption{Noise of the readout system at different shaping times.}
\label{Fig4:ECAL-electronics-4th}
\end{figure*}

%%%%%%%%%%%%%%%%%%%%%%%%%%%%%%%%%%%%%%%%%%%%%%%%%%%%%%%%%%%%%%%%%%%%%%%%%%%
%\subsubsection{The Noise and Dynamic Range of Electronics}
The electronic noise of the readout system with two types of APDs at different shaping times is measured, with an APD gain of 50. Type S8664-1010, which creates twice as much noise as S8664-0505, achieves similar performance considering the size of the area. The detailed experimental results are shown in {Fig.~\ref{Fig4:ECAL-electronics-4th}}. The noise of S8664-1010 is lower than 0.4~fC when the shaping time is 100~ns, which means the noise performance would be better than 0.8~fC when using 4 APDs and 4 CSAs. The equivalent noise energy is 1 MeV when the light yield reaches 100~p.e./MeV.\\
Given that the upper limit of the charge measurement of the high-gain channel is 120 fC, the dynamic range of the high-gain channel can cover the range of 3~MeV (2.4~fC, 3 times noise) - 150~MeV (120~fC). Considering that the gain ratio of the high- and low-gain channels is 20 and the low-gain noise is close to 2~fC, the dynamic range of the low-gain channel is 10~MeV (6~fC, 3 times noise) - 3000 MeV (2400 fC). The dynamic range that can be realized by this dual gain design is shown in Table~\ref{Tab:ECAL Dynamic Range}.

\begin{tiny}
\begin{table}[htb]
\begin{center}
\caption{The Dynamic Range of EMC}
\label{Tab:ECAL Dynamic Range}
\begin{tabular}{|c|c|c|}
\hline Channel&Low Limit (MeV)&High Limit (MeV)\\
\hline High Gain&3&150\\
\hline Low Gain&10&3000\\
\hline
\end{tabular}
%\end{scriptsize}
\end{center}
\end{table}
\end{tiny}

%%%%%%%%%%%%%%%%%%%%%%%%%%%%%%%%%%%%%%%%%%%%%%%%%%%%%%%%%%%%%%%%%%%%%%%%%%%
\iffalse
The time resolution of EMC is of great significance to eliminating the background and distinguishing photons from neutral hadrons. The time measurement of the readout electronics (including the APD) is carried out using an LED test, and the test results are shown in Fig.~\ref{Fig4:ECAL-electronics-TR-PF}. The results show that with increasing charge (the LED light intensity enhancement), the time resolution gradually improves, and the corresponding time resolution at 200 fc ($\sim$ 1 GeV energy deposition) is 150 ps.
\begin{figure*}[htbp]
 \centering
 \mbox{
  %\vskip -1.5cm
  \begin{overpic}[width=0.5\textwidth, height=0.33\textwidth]{Figures/Figs_04_00_DetectorSubSystems/Figs_04_04_ElectromagneticCalorimeter/ECAL-Electronics-TR-PF.pdf}
  \end{overpic}
 }
\caption{Electronics time resolution.}
\label{Fig4:ECAL-electronics-TR-PF}
\end{figure*}
\fi
In summary, the conceptual design of the electronics can meet the requirements of the STCF calorimeter in terms of noise, dynamic range and time resolution.

\subsection{Summary}
The baseline design of the STCF EMC adopts a pCsI crystal scintillator coupled with large area APDs, and a charge-sensitive readout scheme is chosen for the readout electronics. The preliminary Monte Carlo simulation and experimental test results show that the conceptual designs can meet the requirements of the STCF. Extensive R\&D works are underway to verify the designs and the key technical aspects.

\clearpage
\newpage
\section{Muon Detector~(MUD)}
\label{sec:muc}

\subsection{Introduction}
The muon detector~(MUD), as the outermost part of the STCF detector system,
is used to provide muon identification in the presence of a significant pion background.
It can also be used for neutral hadron identification to complement the EMC identification, for example, of neutrons and $K_L$.
A MUD usually
has a sandwich-like structure that consists of a hadron absorber and a detector array. Multiple layers of steel plates are used as both the magnetic flux return yoke and the hadron absorber. The muon detection array is inserted into the gap of these hadron absorbers.

\subsubsection{Performance Requirements}
%{Physics requirements}
For the STCF MUD, a high detection efficiency of muons and a good suppression power for muons/pions are the main requirements~\cite{mud1}.
The momenta of the final state $\mu$ and $\pi$ produced at the STCF are mostly below 2.0~GeV/c, as shown in Sec.~\ref{sec:phys_requirements}.
Because of the EMC and solenoid
material preceding the MUD, muons with momenta less than 0.4 GeV/c cannot be detected by the MUD.
In contrast, muons with low momenta can be identified well by the PID system, as introduced in Sec.~\ref{sec:rich}.
According to the physics requirements, in the momentum range of $p>0.7$~GeV/c, the ideal muon detection efficiency should be higher than 95\%; in the range of $0.5<p<0.7$~GeV/c, the muon detection efficiency should be higher than 70\%, with the muon/hadron suppression power being better than 30.

Additionally, the identification of neutral hadrons with sufficiently high detection efficiency is important for
STCF physics, especially for particles with momenta in the range of [0.2, 1.2]~GeV/c.
The probability of obtaining a hadronic shower in the EMC or MUD for neutrons or \KLs\ is quite high, generating a high multiplicity of photons, neutrons, protons and other hadrons. Thus, a detector array with good photon and neutron sensitivity is required.

For the STCF, with a luminosity of $1\times10^{35}$~cm$^{-2}$s$^{-1}$, the MUD receives a nonnegligible background contribution dominated by neutrons with an energy of [10, 100]~keV \cite{mud2, belle2}. This background may cause more hits in the MUD, which may affect the track finding and identification efficiency of muons. Thus, the MUD must tolerate the very high background rate of the STCF~\cite{mud4, mud5}. In addition, the low energy photon and neutron background may cause significant contamination in the identification of neutral hadrons~\cite{mud6}. As mentioned in Section~\ref{sec:expcon}, the simulated highest background level is approximately 3.99 Hz/cm$^{2}$ and 265 Hz/cm$^{2}$ for barrel and endcap MUD, respectively. As a consequence, optimization of the detector layout to high-rate capabilities and excellent background suppression power is important in the MUD design.

\subsubsection{Technology Choices}
In particle physics experiments, the resistive plate chamber (RPC) and the plastic scintillator are among the most widely used detector technologies for the detection of muon particles.

%\subparagraph{Resistive Plate Chamber}
%\quad\\
A RPC is a traditional option for large-area muon detection with centimeter-level spatial resolution~\cite{mud7}, and has been used in many experiments as the MUD, such as ATLAS~\cite{mud8}, BaBar~\cite{mud9}, Belle~\cite{mud10}, STAR~\cite{star}, BESIII \cite{mud11} and Daya Bay~\cite{DayaBay:2015kir}. RPCs are robust and low cost for large detection areas and have simple manufacturing and maintenance processes. In addition, their centimeter-level spatial resolution can satisfy the demand of particle detection with an area of hundreds of square meters.
A Bakelite-RPC, which operates in the avalanche mode, has
count rate capabilities over 1~kHz/cm$^{2}$, and it also offers the advantage of low
background sensitivity.
Thus, Bakelite-RPC is chosen as a candidate for the STCF MUD.

A plastic scintillator with a silicon photo-multiplier (SiPM) is also a choice for the MUD.
In the Belle II experiment, a combination of polystyrene scintillator strips, wavelength shifting fibers and SiPM was selected as the upgrade of the $K_L$ and muon detector to replace the low-count rate RPCs \cite{mud13}. Compared with the Bakelite-RPC, the plastic scintillator detector has a much higher count rate and is more sensitive to photons and neutrons, resulting in powerful neutral/hadron separation. However, the plastic scintillator detector suffers from higher background counts, leading to worse track-finding and particle identification. As a result,
 the design of the MUD should be optimized to achieve a balance between the rate
capabilities and the identification of low-momentum muons.
Consequently, a hybrid muon detector design that combines a Bakelite-RPC with a plastic scintillator is proposed for the STCF.
Details regarding the design, optimization, and particle identification performance are presented in the following.

\subsection{MUD Conceptual Design}
\subsubsection{Detector Layout}
%\quad\\
The baseline design of the MUD is a combination of a Bakelite-RPC and plastic scintillator detector: 3 layers of Bakelite-RPC are placed in the innermost part, and 7 layers of the plastic scintillator detector form the outer layers.

Considering that the dominant background sources are photons and neutrons, the usage of the Bakelite-RPC in inner layers can decrease the background level in the MUD and help to separate the charged particle tracks and neutral particle shower signals.
The plastic scintillator detector is more sensitive to photon and neutron particles, which are the main background contributors to neutral hadron detection and identification. The {\sc Geant4}~\cite{Geant4_ref} simulation results indicate that with the expected STCF background level described in Sec.~\ref{sec:expcon}, the muon detection efficiency in the momentum range of [0.4, 0.6] GeV/c decreases by 10-20\% if the MUD detectors are all plastic scintillators compared with that of the STCF MUD baseline design.

\begin{figure*}[htb]
    \centering
    {
        \includegraphics[width=0.7\textwidth]{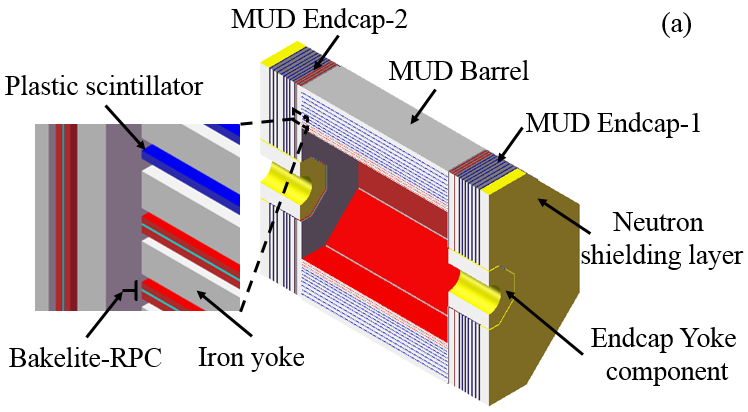}
    }
    \hspace{5mm}   
    {
        \includegraphics[width=0.7\textwidth]{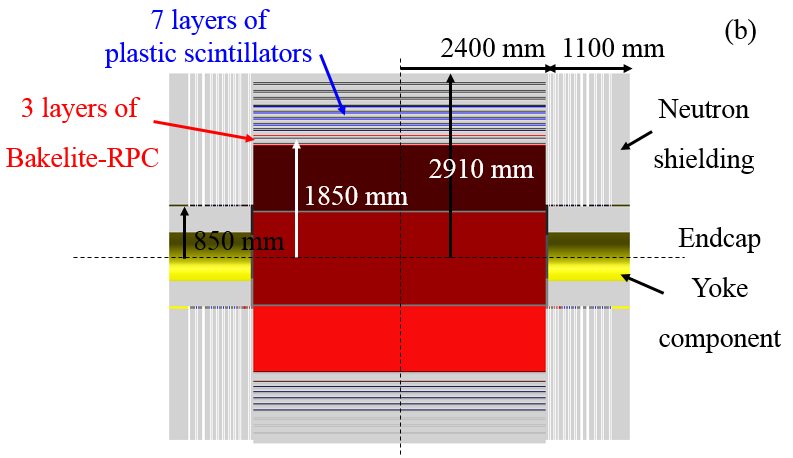}
    }
    \vspace{0cm}
\caption{Schematic of the MUD design. (a) Half-section view of the MUD, and partial enlarged view of the sandwich placement of the Bakelite-RPC, plastic scintillator, and iron yoke. (b) Cutaway view of the MUD and the setting of the main structural parameters.}
    \label{fig:4.5.02}
\end{figure*}

Fig.~\ref{fig:4.5.02} shows a schematic of the baseline design of the MUD. The barrel MUD covers the solid angle of 79.2\%$\times$4$\pi$ (37.63$^{\circ}$$<$$\theta$$<$142.37$^{\circ}$), and the endcap MUD covers the solid angle of 14.8\%$\times$4$\pi$ (20$^{\circ}$$<$$\theta$$<$37.63$^{\circ}$ and 142.37$^{\circ}$$<$$\theta$$<$160$^{\circ}$). Both the barrel and endcap MUD contain 10 layers of detectors, and the iron yoke layout can be seen at Sec.~\ref{sec:yoke}.

\begin{figure*}[htb]
    \centering
    {
        \includegraphics[width=0.85\textwidth]{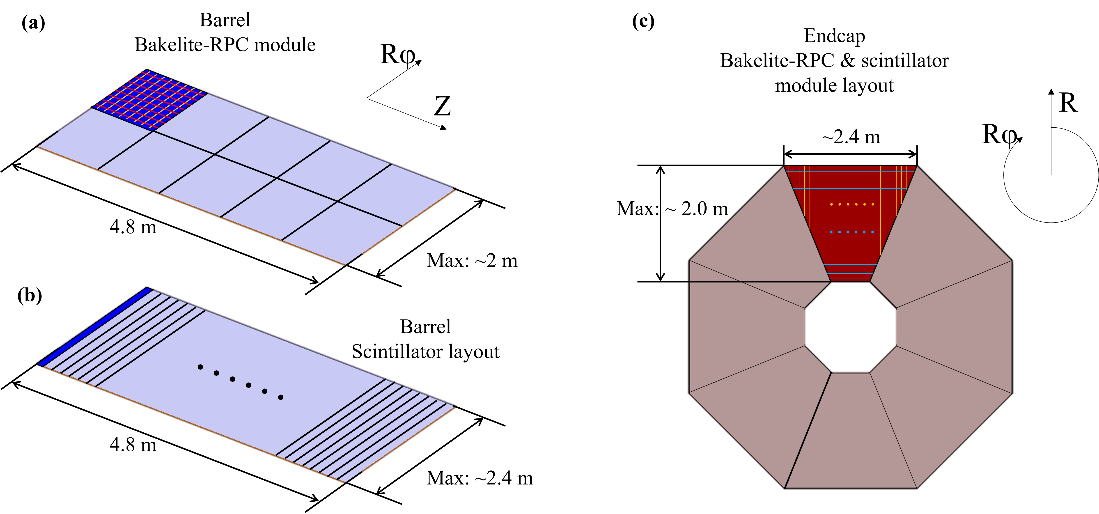}
    }
    \vspace{0cm}
\caption{Module layout of the MUD design. (a) Bakelite-RPC in barrel MUD. (b) Scintillator in barrel MUD. (c) Bakelite-RPC and scintillator in endcap MUD}
    \label{fig:4.5.01}
\end{figure*}

In MUD, the width of the RPC $X/Y$ readout strips and the width of the plastic scintillator are both 4~cm. The maximum length of the Bakelite-RPC module is approximately 1.1 m, and the maximum length of the plastic scintillator strip is 2.4~m~\cite{mud13}. As shown in Fig.~\ref{fig:4.5.01}, each layer of barrel MUD consists of 8 rectangle detector module. For Bakelite-RPC, the module is divided into 5 sub-modules along Z direction, and 2 sub-modules along R$\phi$ direction to control the maximum length of readout strips around 1.1 m. In each sub-module, the 2-D readout strips are perpendicularly arranged. For plastic scintillator, the axis of the scintillator strip is perpendicular to the Z direction. Each layer of the endcap MUD consists of 8 trapezoidal modules. For both Bakelite-RPC and plastic scintillator, the module has 60 strips in R$\phi$ direction, and 49 to 43 strips in R direction, due to the increase of the inner radius.

%%%%%%%%%%%%%%%%%%% Fig %%%%%%%%%%%%%%%%%%%%%%%%%%
%%%%%%%%%%%%%%%%%%%%%%%%%%%%%%%%%%%%%%%%%%%%%%%%%%

%%%%%%%%%%%%%%%%%%% Fig %%%%%%%%%%%%%%%%%%%%%%%%%%
%%%%%%%%%%%%%%%%%%%%%%%%%%%%%%%%%%%%%%%%%%%%%%%%%%

\subsubsection{Neutron Shield and Cylindrical Yoke Component}
%\quad\\
Previous studies by Belle II indicated that the detector system was affected by the neutron background generated by the GeV level electrons and positrons that escape from the beam background~\cite{mud5}.
To suppress the effect of the neutron background, the outer face of the endcap MUD is covered by a 15 cm neutron shielding layer, as shown in Fig.~\ref{fig:4.5.02}. The outer layer of the neutron shielding is 5 cm lead, and the inner layer is 10 cm boron-doped polyethylene (10\%wt of nat-boron). Lead can moderate fast neutrons to an energy of approximately 1 MeV, and boron-doped polyethylene can moderate neutrons to the thermal neutron level. Neutrons with energies less than 1~eV have a large probability of being absorbed by the boron atoms.
 A {\sc Geant4} simulation demonstrated that the 15 cm-thick composite neutron shielding could
decrease the MUD hits by approximately 90\% for neutrons with a kinetic energy less than 1~MeV.

A cylindrical yoke component is arranged between the beamline and endcap MUD detector array, which is designed to make the magnet field uniform in the STCF detector system.
The thickness of this cylindrical yoke is approximately $40\sim45$~cm. {\sc Geant4} simulations indicate that the yoke provides background suppression capabilities similar to those of composite shielding.

\FloatBarrier

\subsubsection{MUD Optimization}

\begin{figure*}[htb]
    \centering
    {
        \includegraphics[width=0.5\textwidth]{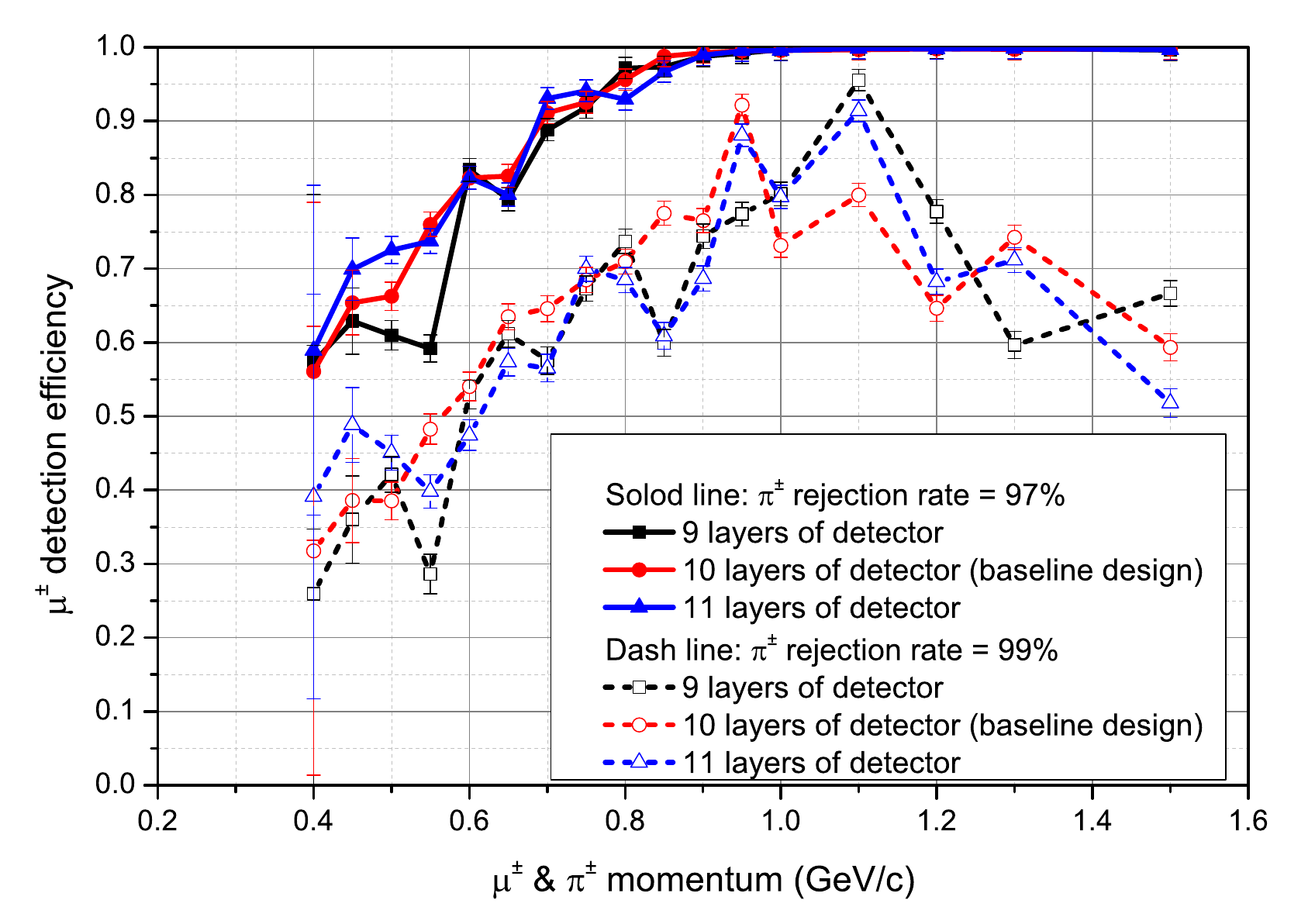}
    }
\caption{The muon detection efficiency curve from the Geant4 simulation. Efficiencies are compared with different detector layer settings.
    The results for two scenarios of muon/pion suppression power, 33 and 100, are shown.}
    \label{fig:4.5.03a}
\end{figure*}

The thickness of the iron yoke and the number of detector layers are optimized via a simulation study, and the detailed parameters are summarized in Table~\ref{tab:4.5.02}. Fig.~\ref{fig:4.5.03a} illustrates the muon detection efficiency curves for different detector layer settings and a yoke with a thickness of 51~cm. In the simulation, 9 to 11 layers are applied
and evaluated. The result indicates that 10 or 11 detector layers can produce a higher and smoother muon detection efficiency. Considering the detector complexity and manufacturing costs, the MUD baseline design incorporates 10 detector layers.
It should be noted that in the muon detection efficiency curves, $\mu/\pi$ suppression powers of both 33 and 100 are applied, while the former is sufficient for most physics processes containing muons at the STCF.

%%%%%%%%%%%%%%%%%  TABLE  %%%%%%%%%%%%%%%%%%%%%%%%
\begin{table*}[htb]
\small
    \caption{The structure parameters of the conceptual baseline design of the MUD. R$_{in}$ and R$_{out}$ are the inner and outer radius of the barrel MUD, respectively, including the 15 cm-thick iron plate shielding outside the detector system. R$_{e}$ is the inner radius of the endcap MUD. L$_{Barrel}$ and T$_{Endcap}$ are the length of the barrel and endcap MUD in the z-direction, respectively. The size of neutron shielding layer is not included.}
    \label{tab:4.5.02}
    \vspace{0pt}
    \centering
    \begin{tabular}{ll}
        \hline
        \thead[l]{Parameter} & \thead[l]{Baseline design}\\
        \hline
            R$_{in}$ [cm]	           &185 \\
            R$_{out}$ [cm]	          &291 \\
            R$_{e}$ [cm]	                &85 \\
            L$_{Barrel}$ [cm]	           &480 \\
            T$_{Endcap}$ [cm]	           &107 \\
            Segmentation in $\phi$	         &8 \\
            Number of detector layers	      &10 \\
            Iron yoke thickness [cm]    &4/4/4.5/4.5/6/6/6/8/8 cm \\
            ($\lambda$=16.77 cm)            &Total: 51 cm, 3.04$\lambda$ \\
            Solid angle	                &79.2\%$\times$4$\pi$ in barrel \\
                                &14.8\%$\times$4$\pi$ in endcap \\
                                &94\%$\times$4$\pi$ in total \\
            Total area [m$^2$]	         &Barrel $\sim$717 \\
                                &Endcap $\sim$520 \\
                                &Total $\sim$1237 \\
        \hline
    \end{tabular}
\end{table*}
%%%%%%%%%%%%%%%%%%%%%%%%%%%%%%%%%%%%%%%%%%%%%%%%%%

%%%%%%%%%%%%%%%%%%% Fig %%%%%%%%%%%%%%%%%%%%%%%%%%
%%%%%%%%%%%%%%%%%%%%%%%%%%%%%%%%%%%%%%%%%%%%%%%%%%

\begin{figure*}[htb]
    \centering
    {
        \includegraphics[width=0.5\textwidth]{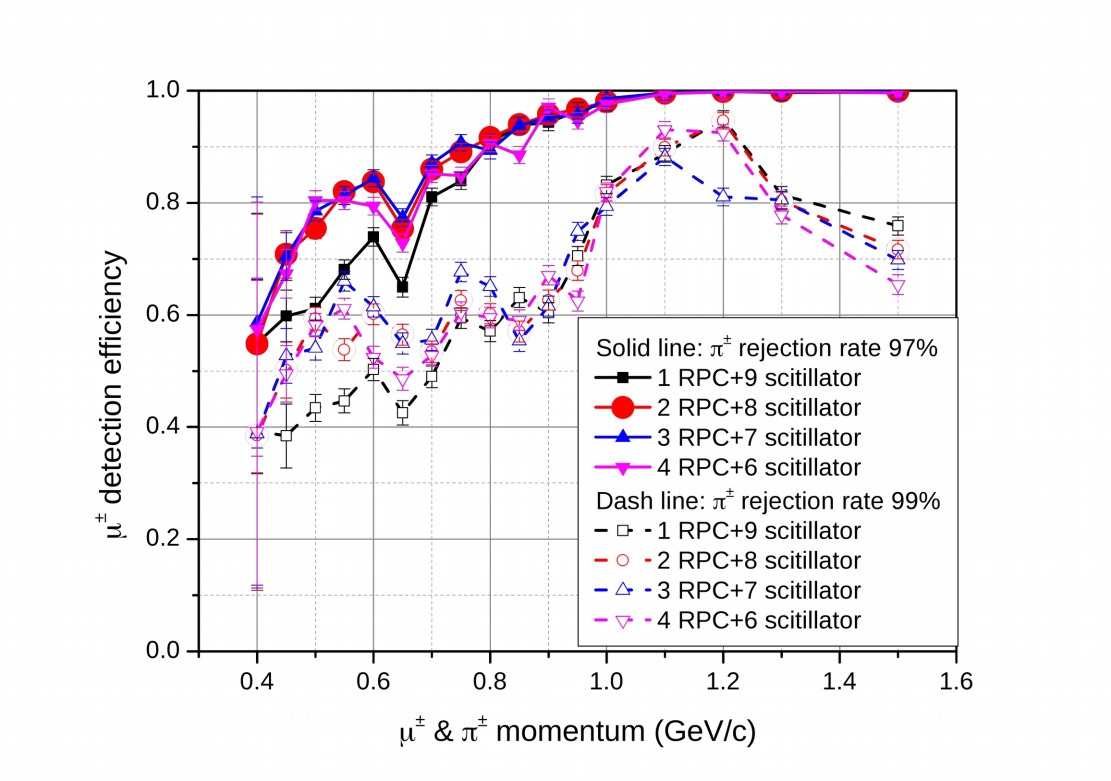}
    }
\caption{The muon detection efficiency curves with different combinations of Bakelite-RPC and plastic scintillator
in the MUD design along the direction of $\theta= 90^{\circ}$ and $\phi = 90^{\circ}$ ($\theta$: polar angle, $\phi$: azimuth angle) including the background.}
    \label{fig:combination}
\end{figure*}

The arrangement of the Bakelite-RPC and plastic scintillator in the ten layers affects the performance
of the MUD. On one hand, the plastic scintillator has higher robustness and higher detection
efficiency for high-momentum muon. However, the {\sc Geant4} simulation indicates that the
Bakelite-RPC exhibits better performance in detecting low-momentum muons in the high-background region
of the MUD.
Fig.~\ref{fig:combination} shows the {\sc Geant4}-simulated muon detection efficiency as a function of momentum for
different MUD layouts at the full STCF luminosity. The
MUD designs with two-, three-, or four-layer Bakelite-RPCs exhibit similar $\mu/\pi$ separation
power under the current luminosity and background level.
For the design with the two-layer Bakelite-RPC, a high count rate and significant interference may occur.
in the 3rd (plastic scintillator) layer due to the potential fluctuations in the background level or the
future upgrades of the STCF. As a result, the hybrid MUD design with the three-layer Bakelite-RPC
and a seven-layer plastic scintillator is considered the optimal choice.

\begin{figure*}[htb]
    \centering
    {
        \includegraphics[width=0.5\textwidth]{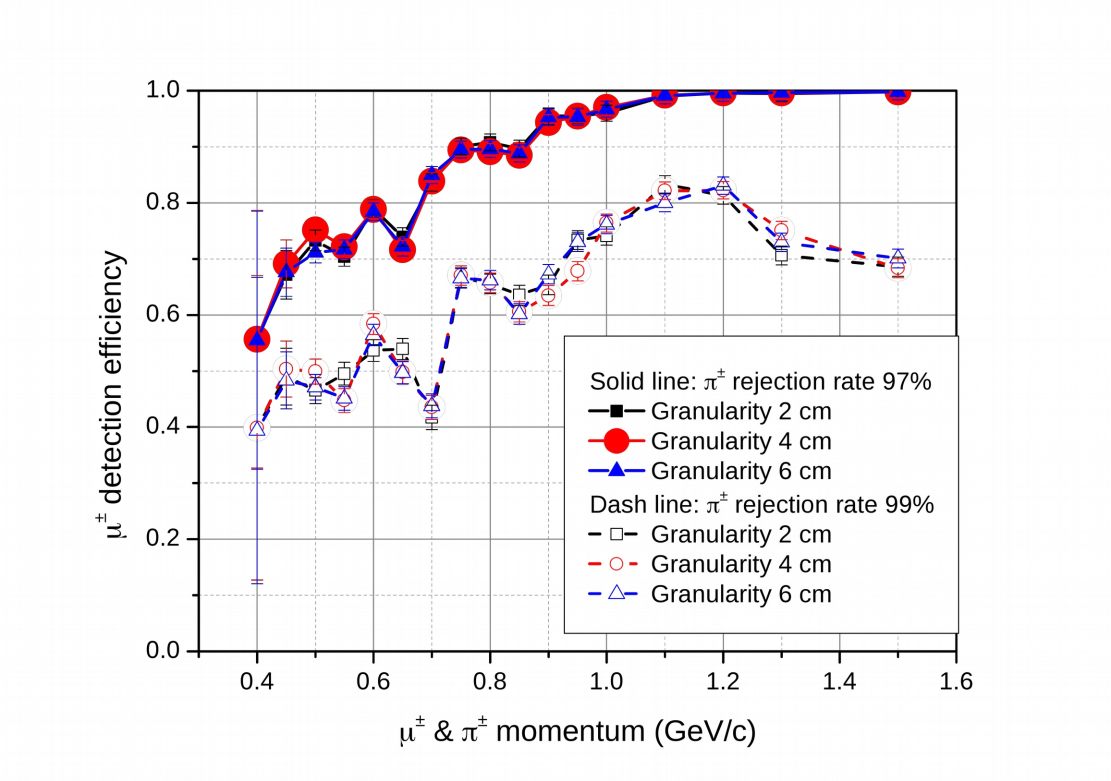}
    }
\caption{{\sc Geant4}-simulated muon detection efficiency with different granularities along the zenith direction.}
    \label{fig:granularity}
\end{figure*}

The granularity is determined by the readout strip pitch of the Bakelite-RPC and the size of the
plastic scintillator strips. Physical simulations of the measurement precision for reconstructed
muons indicate that a spatial resolution of 1–2 cm is required in the MUD, equivalent to a detector
granularity of 3.5 - 7 cm. Fig.~\ref{fig:granularity} presents simulated muon detection efficiency curves with
different granularities, indicating similar particle detection performances. As a smaller granularity
implies additional electronic channels, both the readout strip pitch of the Bakelite-RPC and the width of the plastic scintillator strips are chosen to be 4~cm.

\subsection{Expected Performance}
To evaluate the expected performance of the MUD, {\sc Geant4}-based full simulations are studied. The identification efficiency of muons and neutral hadrons is obtained using boosted decision tree~(BDT) algorithms. The details of BDT algorithms can be found in Ref.~\cite{Fang:2021fhm}. In the MUD, both the Bakelite-RPC and the plastic scintillator have very high sensitivity to charged muons and pions, and the required spatial resolution is within 2~cm, which can be ensured by the 4~cm detector granularity.

\subsubsection{Muon Identification Efficiency}
\begin{figure*}[htb]
    \centering
    \includegraphics[width=70mm]{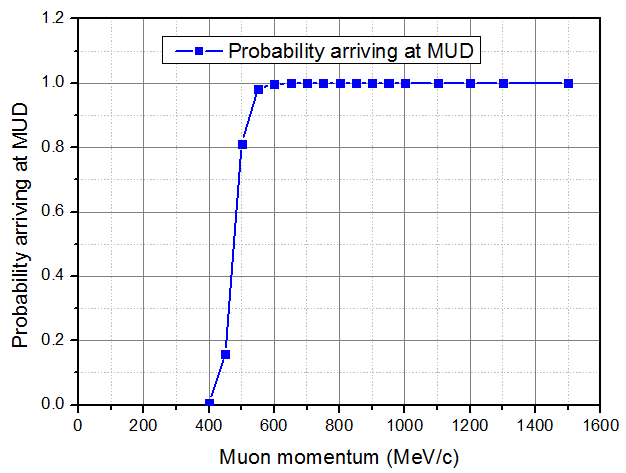}
    \vspace{0cm}
\caption{The probability that a muon arrives at the MUD as a function of the muon momentum in the zenith direction.}
    \label{fig:4.5.prob}
\end{figure*}

%\quad\\
Fig.~\ref{fig:4.5.prob} shows the probability of a muon arriving at the MUD as a function of the muon momentum. In the simulation, the bin width is 50~MeV/c. The result indicates that a muon with momentum between 400 and 450~MeV/c only has a probability of 0.68\% of arriving at the MUD. When the muon momentum reaches 550~MeV/c, almost all of muons can generate signals in the MUD.
%%%%%%%%%%%%%%%%%%% Fig %%%%%%%%%%%%%%%%%%%%%%%%%%
%%%%%%%%%%%%%%%%%%%%%%%%%%%%%%%%%%%%%%%%%%%%%%%%%%

\begin{figure*}[htb]
    \centering
    \includegraphics[width=70mm]{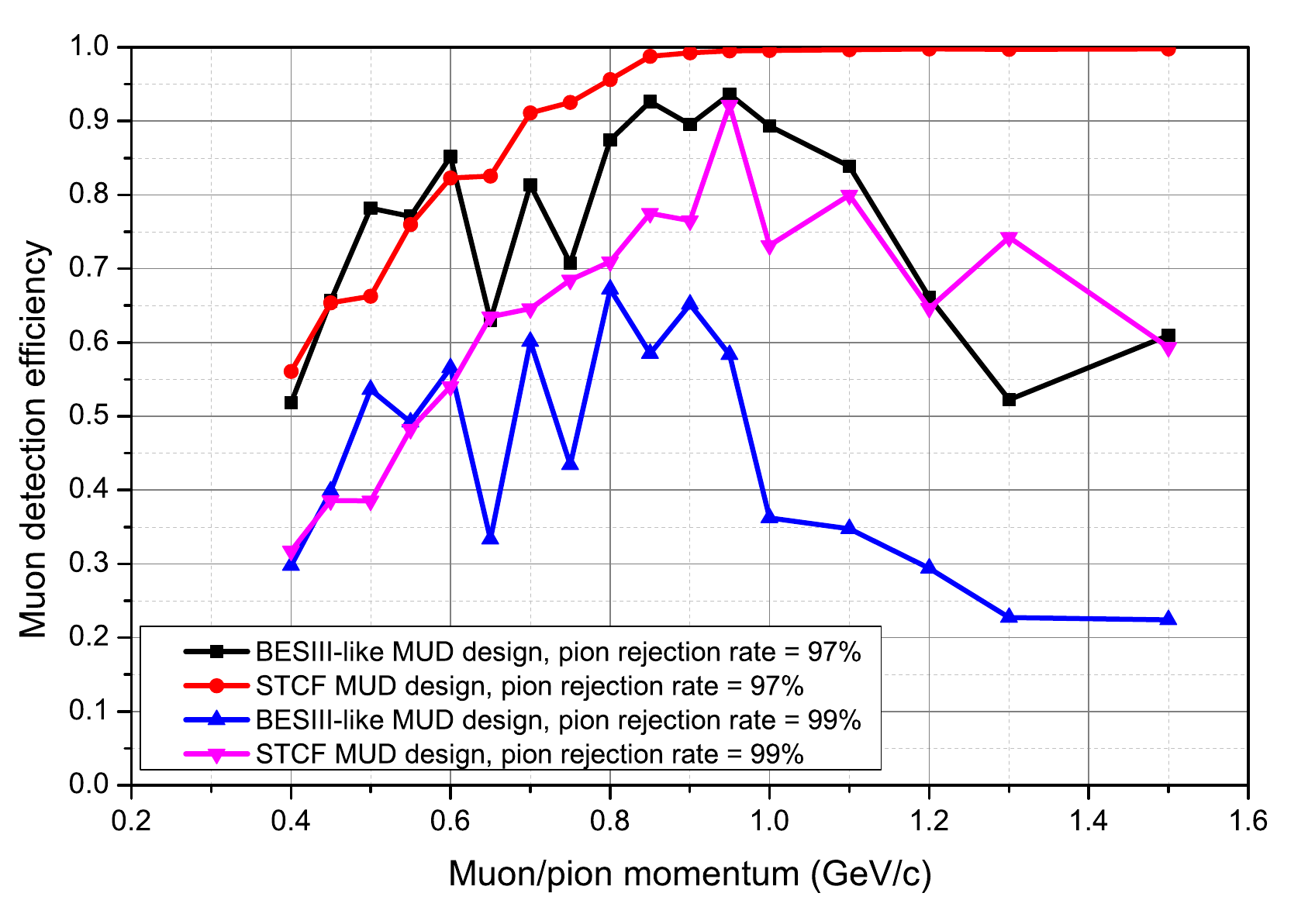}
    \vspace{0cm}
\caption{The muon detection efficiency curve from the {\sc Geant4} simulation, with a polar angle of 90 degree.
    Results for two scenarios of muon/pion suppression power, 33 and 100, are also shown.}
    \label{fig:4.5.03b}
\end{figure*}

\begin{figure*}[htb]
    \centering
    {
        \includegraphics[height=50mm]{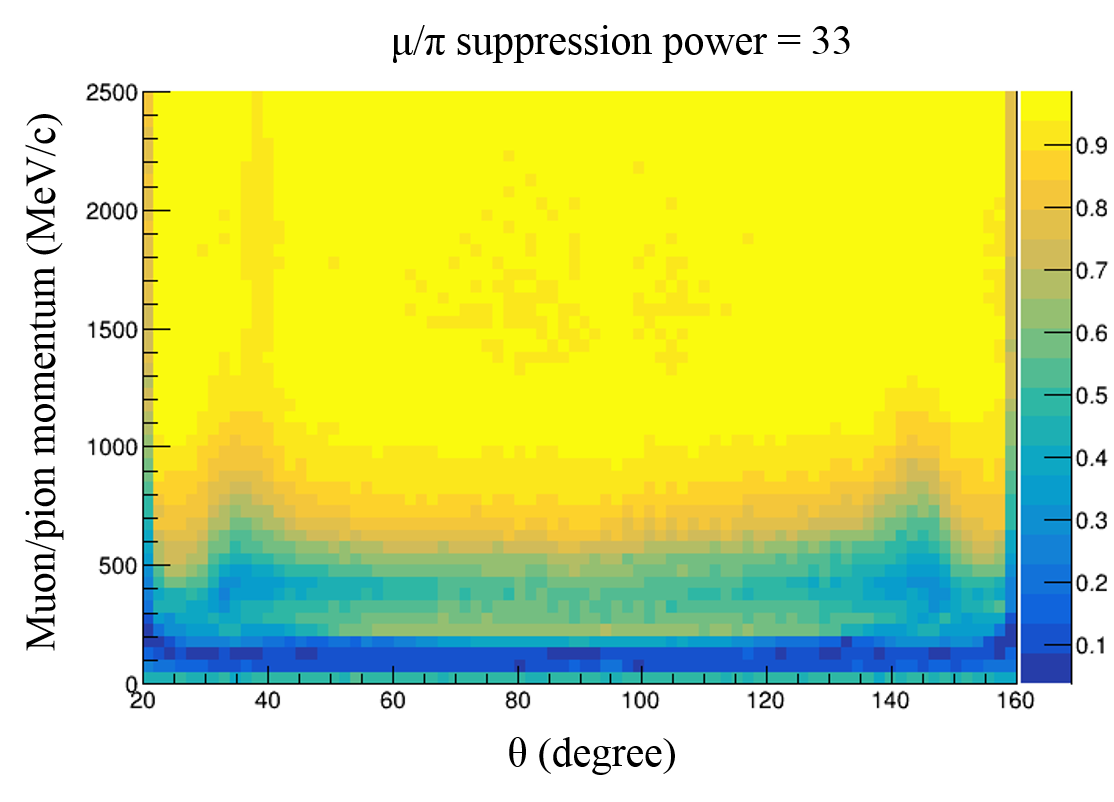}
    }
    \hspace{5mm}
    {
        \includegraphics[height=50mm]{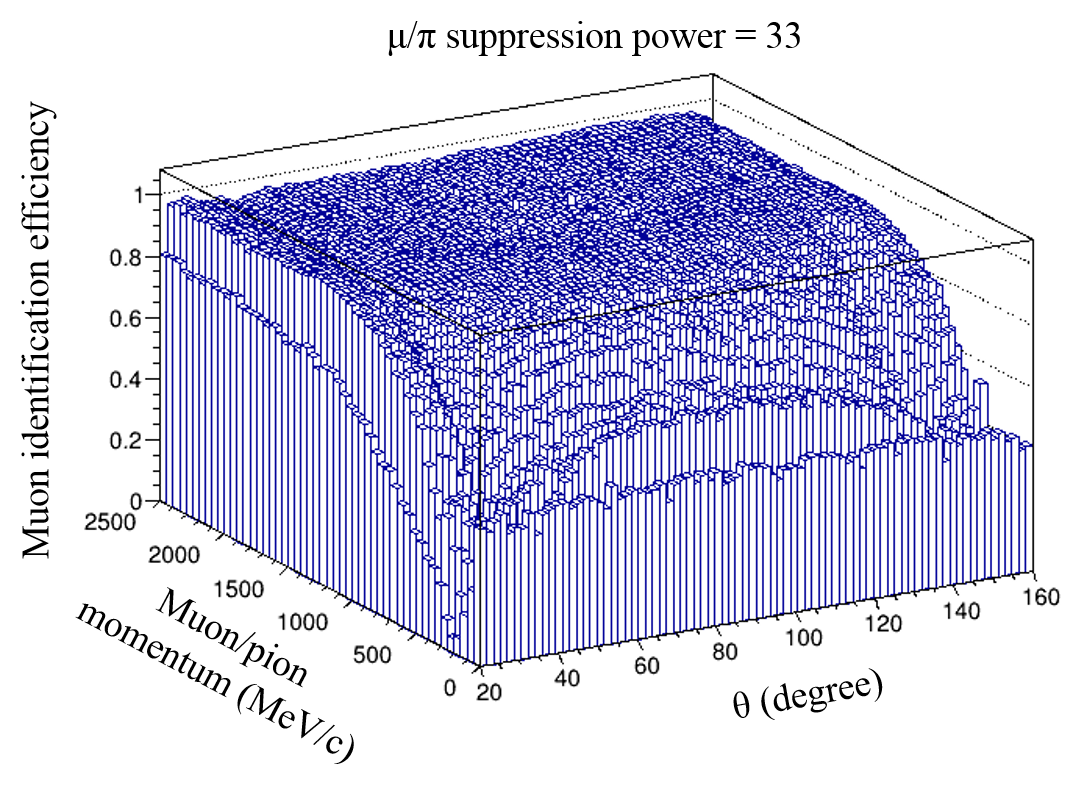}
    }
    \vspace{0cm}
\caption{2D map of the muon identification efficiency as a function of the momentum and the $\theta$ angle from {\sc Geant4} simulation.
    The $\mu/\pi$ suppression power is assumed to be 33.}
    \label{fig:4.5.04}
\end{figure*}

Fig.~\ref{fig:4.5.03b} shows the simulated muon detection efficiency with a polar angle of 90 deg and with $\mu/\pi$ suppression power of 33 and 100.
A significant increase in the muon detection efficiency can be observed in the momentum range of [0.65, 1.5]~GeV/c
owing to a thicker yoke and an optimized detector setting.
This result indicates that the muon detection efficiency curve of the baseline design is smoother than that of the BESIII-like MUD geometry. In the low momentum range [0.4-0.6]~GeV/c, the STCF MUD design exhibits a performance similar to that of the BESIII geometry, ensuring an acceptable muon detection efficiency. Fig.~\ref{fig:4.5.04} shows the muon identification efficiency of the baseline design, with the particle momentum in [0, 2.5]~GeV/c and polar angle in [20$^{\circ}$, 160$^{\circ}$]
considering a pion rejection rate of 97\% and background influence.

%%%%%%%%%%%%%%%%%%% Fig %%%%%%%%%%%%%%%%%%%%%%%%%%
%%%%%%%%%%%%%%%%%%%%%%%%%%%%%%%%%%%%%%%%%%%%%%%%%%

%%%%%%%%%%%%%%%%%%% Fig %%%%%%%%%%%%%%%%%%%%%%%%%%
%%%%%%%%%%%%%%%%%%%%%%%%%%%%%%%%%%%%%%%%%%%%%%%%%%

\subsubsection{Neutral Hadron Detection and Identification}
\begin{figure*}[htb]
    \centering
    {
        \includegraphics[height=50mm]{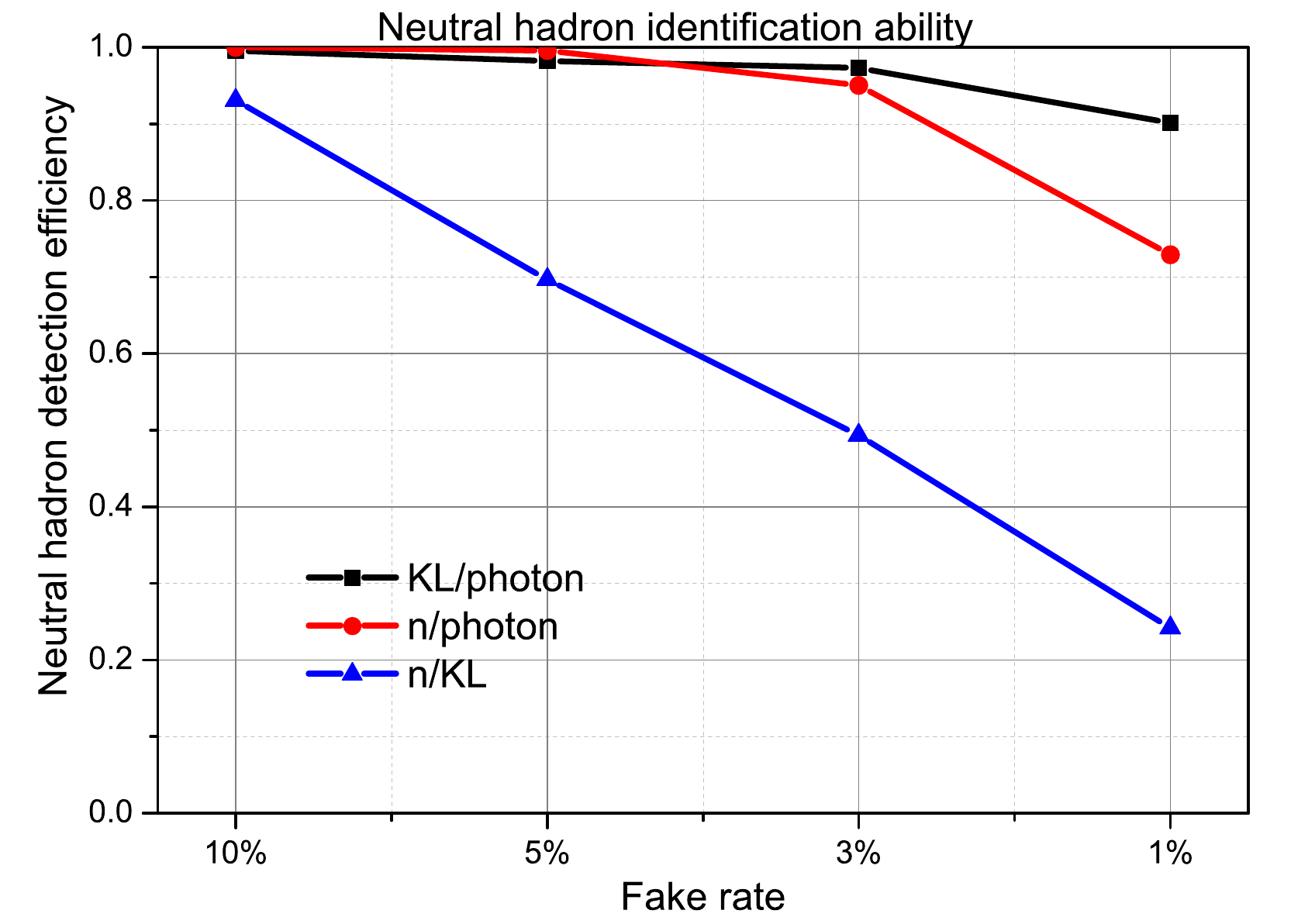}
    }
    \vspace{0cm}
\caption{The detection efficiency curves of various neutral hadrons in the MUD (notice: all of these neutral hadrons deposit less than 40 MeV energy in the EMC).}
    \label{fig:4.5.05}
\end{figure*}

%\quad\\
At the STCF, the MUD is set as an auxiliary neutral hadron detector to complement the EMC. In this baseline MUD design, the 3-layer Bakelite-RPC acts as a filter, preventing too many secondary gamma and neutron hits from being generated by background interference. The {\sc Geant4} simulation illustrates that approximately 40\% of the neutrons and $K_L$ deposit a very small amount of energy, less than 40 MeV, in the EMC. Thus, the MUD is responsible for detecting them. Table~\ref{tab:4.5.03} shows the Geant4 simulated cluster size of neutral hadrons in the 7 layers of plastic scintillator detectors, and Fig.~\ref{fig:4.5.05} presents the neutral hadron detection efficiency curves with different fake rates. When a neutral hadron cannot be detected by EMC directly, the MUD has a quite high efficiency in terms of detection and identification of the hadron, demonstrating the excellent identification power of the MUD for neutral hadrons.

%%%%%%%%%%%%%%%%%  TABLE  %%%%%%%%%%%%%%%%%%%%%%%%
\begin{table*}[htb]
\small
    \caption{The {\sc Geant4} simulated neutron and $K_L$ cluster parameters in plastic scintillator detector layers.}
    \label{tab:4.5.03}
    \vspace{0pt}
    \centering
    \begin{tabular}{lllll}
        \hline
        \thead[l]{ } & \thead[l]{Neutron\\Average cluster size\\ in scintillators}&\thead[l]{\\Probability of cluster\\ size $\ge$2 } & \thead[l]{$K_L$\\Average cluster size\\ in scintillators}&\thead[l]{\\Probability of cluster\\ size $\ge$2 }\\
        \hline
            200 MeV/c	&2.42	&5\%	&4.42	&32\% \\
            400 MeV/c	&4.07	&31\%	&6.48	&50\% \\
            600 MeV/c	&5.57	&49\%	&7.88	&68\% \\
            800 MeV/c	&7.23	&66\%	&9.20	&74\% \\
            1000 MeV/c	&8.31	&74\%	&8.96	&76\% \\
            1200 MeV/c	&9.03	&79\%	&11.18	&84\% \\
        \hline
    \end{tabular}
\end{table*}
%%%%%%%%%%%%%%%%%%%%%%%%%%%%%%%%%%%%%%%%%%%%%%%%%%

%%%%%%%%%%%%%%%%%%% Fig %%%%%%%%%%%%%%%%%%%%%%%%%%
%%%%%%%%%%%%%%%%%%%%%%%%%%%%%%%%%%%%%%%%%%%%%%%%%%

\subsubsection{Background Simulation}
%\quad\\
Table~\ref{tab:4.5.04} shows the background count rate in the barrel MUD, calculated based on the STCF full {\sc Geant4} simulations. The background estimation demonstrates that the rate capabilities of both the barrel and endcap MUD detectors can meet the requirements. For good neutral hadron detection performance, the usage of a plastic scintillator in the MUD is necessary. The simulation result indicates that with the compound detector design of the MUD, the first layer of the Bakelite-RPC and the first layer of the plastic scintillator have background count rates with the same order of magnitude, ensuring minimal background interference over the whole MUD detector volume. In this case, a higher accuracy of identification for both muons/pions and neutral hadrons can be obtained. The background level in the endcap MUD is almost 2 to 3 times that in the barrel MUD, indicating the necessity of a background shield on the outer surface of the endcap MUD.

%%%%%%%%%%%%%%%%%  TABLE  %%%%%%%%%%%%%%%%%%%%%%%%
\begin{table*}[htb]
\small
    \caption{The Geant4 simulated barrel MUD background count rate.}
    \label{tab:4.5.04}
    \vspace{0pt}
    \centering
    \begin{tabular}{lllllllllll}
        \hline
        \thead[l]{Detector type} &\multicolumn{3}{l}{Bakelite-RPC}&\multicolumn{7}{l}{Plastic scintillator}\\
        \hline
            Detector layer &1&2&3&4&5&6&7&8&9&10\\
            Simulated background & \multirow{3}*{9.2}&\multirow{3}*{3.54} &\multirow{3}*{1.42}&\multirow{3}*{4.25}&\multirow{3}*{6.50}&\multirow{3}*{2.80}&\multirow{3}*{1.77}&\multirow{3}*{0.76}&\multirow{3}*{0.39}&\multirow{3}*{0.36}\\
            count rate in the barrel &&&&&&&&&& \\
            (Hz/cm2) &&&&&&&&&& \\
        \hline
    \end{tabular}
\end{table*}
%%%%%%%%%%%%%%%%%%%%%%%%%%%%%%%%%%%%%%%%%%%%%%%%%%

As shown in Tables \ref{tab:TIDNIEL_max} and \ref{tab:TIDNIEL_eletronic}, the highest TID and NIEL damage values in the MUD are 0.37~Gy/y and $2.79\times10^{12}$ n/y, respectively. Good working conditions can be maintained in the Bakelite-RPC by flushing the working gas. Other studies indicate that the plastic scintillator can tolerate more than 1000~Gy~\cite{mud15,mud16}, indicating that the MUD detector system can work well under background radiation for decades.

\subsubsection{Readout Design}
%\quad\\
In the MUD conceptual design, a 3-layer Bakelite-RPC and 7-layer plastic scintillator are used. The plastic scintillator strips in the barrel are double-ended readouts, and those in the endcap are single-ended readouts. The numbers of readout channels in the MUD are estimated in Table~\ref{tab:4.5.05}, with a total number of readout channels of 16000 for the Bakelite-RPC and 25280 for the plastic scintillator, considering a 4 cm granularity in the detector.

In the MUD, the signal position information can be obtained by the number of channels, so it is only necessary to record the time information of signals whose amplitude exceeds the threshold for particle discrimination. For the Bakelite-RPC in avalanche mode, the front-end electronics are required to achieve an effective gain of 10-fold and a pulse time width of 50~ns. Considering that the count rate of the MUD is not high, an ordinary charge-sensitive amplifier and integral shaping circuit can satisfy the requirements. For the plastic scintillator, a SiPM is used as an electron multiplier. With 5~p.e. thresholds, the noise rate can be controlled within 1~kHz, and the measurement efficiency of MIP could exceed 99\%. This can meet our needs. Additionally, in the barrel-MUD, a double-ended readout is required to determine the hit position, which requires a time resolution of approximately 0.5~ns. Using the waveform sampling method, a 0.25~ns time resolution can be achieved, and we are still studying a lower-cost implementation for the barrel-MUD.

%%%%%%%%%%%%%%%%%  TABLE  %%%%%%%%%%%%%%%%%%%%%%%%
\begin{table*}[htb]
\small
    \caption{The estimated readout channel requirement in the MUD conceptual design.}
    \label{tab:4.5.05}
    \vspace{0pt}
    \centering
    \begin{tabular}{llllllllll}
        \hline
        \thead[l]{ } & \thead[l]{Detector } & \thead[l]{ }& \thead[l]{Barrel}& \thead[l]{MUD }& \thead[l]{ }& \thead[l]{ }& \thead[l]{Endcap}& \thead[l]{MUD }&\thead[l]{ } \\
         &layer &{Half-} &{Half-} &{Channel} &{Channel} &{Inner } &{Outer } &{Channel} &{Channel} \\
         & &{length} &{length} &{number} &{number} &{radius} &{radius} &{number} &{number} \\
         & &{in X} &{in Z} &{in X} &{in Z} &{(cm)} &{(cm)} &{in X} &{in Z} \\
         & &{(cm)} &{(cm)} &{} &{} &{} &{} &{} &{} \\
         \hline
            Bakelite-	&1	&76.6	&240	&1535	&1920	&94	&290	&960	&784 \\
            RPC	&2	&79.9	&240	&1600	&1920	&94	&290	&960	&784 \\
            	&3	&83.3	&240	&1670	&1920	&98	&290	&960	&768 \\
            Plastic 	&4	&86.8	&240	&0	&1920	&98&	290	&960	&768 \\
            scintillator	&5	&90.3	&240	&0	&1920	&102	&290	&960	&752 \\
            	&6	&94.4	&240	&0	&1920	&102	&290	&960	&752 \\
            	&7	&98.6	&240	&0	&1920	&106	&290	&960	&736 \\
            	&8	&102.7	&240	&0	&1920	&110	&290	&960	&720 \\
            	&9	&107.7	&240	&0	&1920	&114	&290	&960	&704 \\
            	&10	&112.7	&240	&0	&1920	&118	&290	&960	&688 \\
        \hline
    \end{tabular}
\end{table*}
%%%%%%%%%%%%%%%%%%%%%%%%%%%%%%%%%%%%%%%%%%%%%%%%%%

According to the background simulation results, the highest count rate per channel in the barrel MUD appears in the first layer of the Bakelite-RPC, with an average count rate of 4~Hz/cm$^{2}$. The corresponding Bakelite-RPC readout strip has a size of 1.1~m$~\times~$ 4~cm, leading to the highest count rate of 1.76~kHz. The highest count rate per channel in the endcap MUD is in the first layer of the plastic scintillator close to the MDI, which is approximately 102~Hz/cm$^{2}$. With the longest scintillator strip (2.4~m$~\times~$4~cm), the highest count rate per channel is calculated to be 98~kHz.

In the reconstruction of tracks and clusters in MUDs, the position information of hits can be reconstructed by the number of readout channels or strips; thus, only the hit-time information should be transported from the detector to the data storage system, corresponding to 4~bytes/signal (16 bits for time information, 12 bits for the channel number, 4 bits for quality). Considering the background level of the MUD and the 300 ns time window, the data stream size is approximately 105~MB/s for the MUD system.

\subsection{Conclusion and Outlook}
The conceptual baseline design of the MUD is a combination of Bakelite-based RPC and plastic scintillator detectors. This design can improve the neutral hadron identification efficiency with the proper $\mu/\pi$ separation capabilities. Detailed detector technologies and further optimization will be studied in the future, such as the optimization of the muon/pion identification algorithm, the detection and identification of neutral hadrons, and R\&D of MUD detectors.

\clearpage
\newpage
\section{Superconducting Solenoid~(Solenoid)}
\label{sec:solenoid}
The STCF detector magnet is a superconducting solenoid coil which is surrounded by iron yoke to provide an axial magnetic field of 1.0 Tesla over the tracking volume. Particle detectors within this volume will measure the trajectories of charged tracks emerging from the collisions. Particle momentum is determined from the measured curvature of these tracks in the field.

The main parameters of the STCF detector superconducting magnet are listed in Table~\ref{tab:4.6.01}. Its room temperature bore diameter is approximately 2.98 m. The conceptual design of the magnet, including the design of the magnetic field, solenoid coil, cryogenics, quench protection, power supply and the iron Yoke are briefly described in this section.

%%%%%%%%%%%%%%%%%  TABLE  %%%%%%%%%%%%%%%%%%%%%%%%
\begin{table*}[hptb]
\small
    \caption{Main parameters of the STCF detector superconducting magnet.}
    \label{tab:4.6.01}
    \vspace{0pt}
    \centering
    \begin{tabular}{ll}
        \hline
        \thead[l]{Cryostat} & \thead[l]{ }\\
        \hline
        Inner radius&	1.450 m \\
        Outer radius&	1.850 m \\
        Length&	4.760 m \\
        \hline
        Coil & \\
        \hline
        Mean radius	&1.565 m \\
        Length	&4.000 m \\
        Conductor dimension	&4.67$\times$15.0 mm$^{2}$ \\
        \hline
        Electrical parameters & \\
        \hline
        Central field	&1.0 T \\
        Nominal current	&3820 A \\
        Inductance	&1.68 H \\
        Stored energy	&12.3 MJ \\
        Cold mass	&4.6 ton \\
        Radiation thickness	&1.9 X$_{0}$ \\
        Cool down time from room temperature	&$\leq$7 days \\
        Quench recovery time	&$\leq$7 hours \\
        \hline
    \end{tabular}
\end{table*}
%%%%%%%%%%%%%%%%%%%%%%%%%%%%%%%%%%%%%%%%%%%%%%%%%%

\subsection{Magnetic Field Design}
\begin{figure*}[hp]
    \centering
{
        \includegraphics[width=0.7\linewidth]{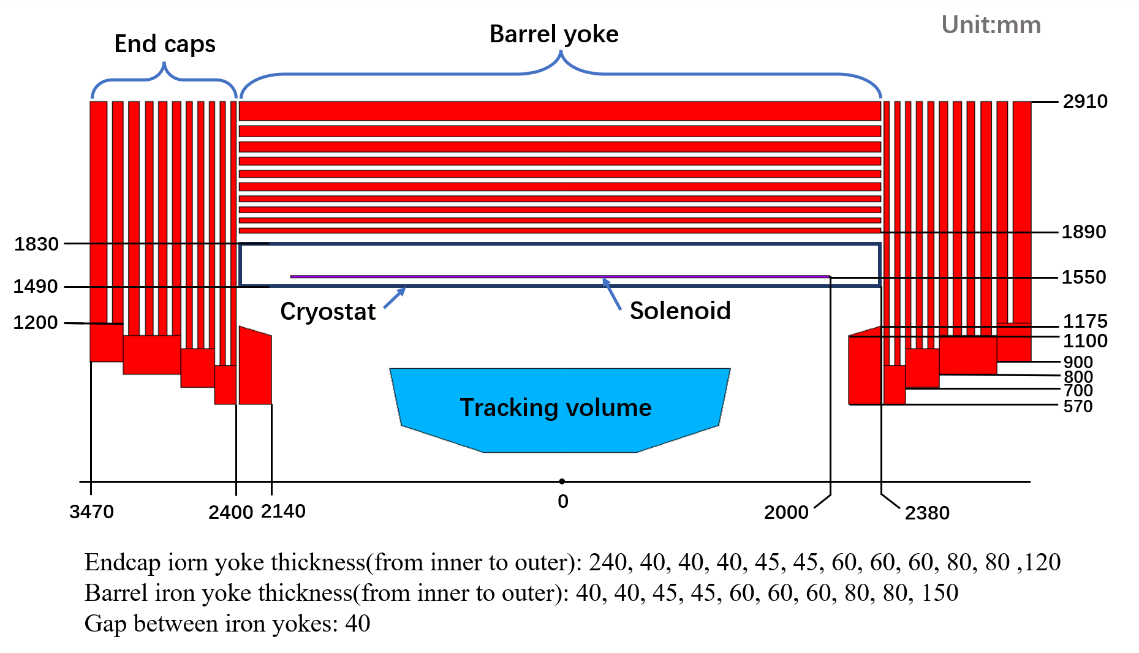}
}
\vspace{0cm}
\caption{2D geometrical layout of the STCF detector magnet, which consists of a superconducting coil and an iron yoke with a barrel yoke and two end-cap yokes.}
    \label{fig:4.6.01}
\end{figure*}

%\paragraph{Main parameters}
The STCF detector magnet will comprise the design concepts similar to TOPAZ magnet in Tristan and CMS magnet in LHC~\cite{1,2,3}. The magnet system consists of the superconducting coil and the iron yoke with a barrel yoke and two end-cap yokes. The superconducting coil is designed as a single layer solenoid wound on the inner surface of an aluminum cylinder. It will run in an operating current 3820 A which corresponding to 1 T magnetic field at the interaction point. The geometrical layout of magnet are shown in Fig.~\ref{fig:4.6.01}.

%%%%%%%%%%%%%%%%%%% Fig %%%%%%%%%%%%%%%%%%%%%%%%%%
%%%%%%%%%%%%%%%%%%%%%%%%%%%%%%%%%%%%%%%%%%%%%%%%%%

%\paragraph{Magnetic field design}
The magnetic field simulation has been calculated in 2D FEA model, with a fine structure of the barrel Yokes and end-cap Yokes with pole tips at each end side. The iron yoke act as the absorber plates of the MUD and provide the magnetic flux return. There are two main field parameters concerned, one is the uniformity in the tracking volume, and the other is the fringe field along the beam axis outside the detector.

\begin{figure*}[hp]
    \centering
{
        \includegraphics[width=0.6\linewidth]{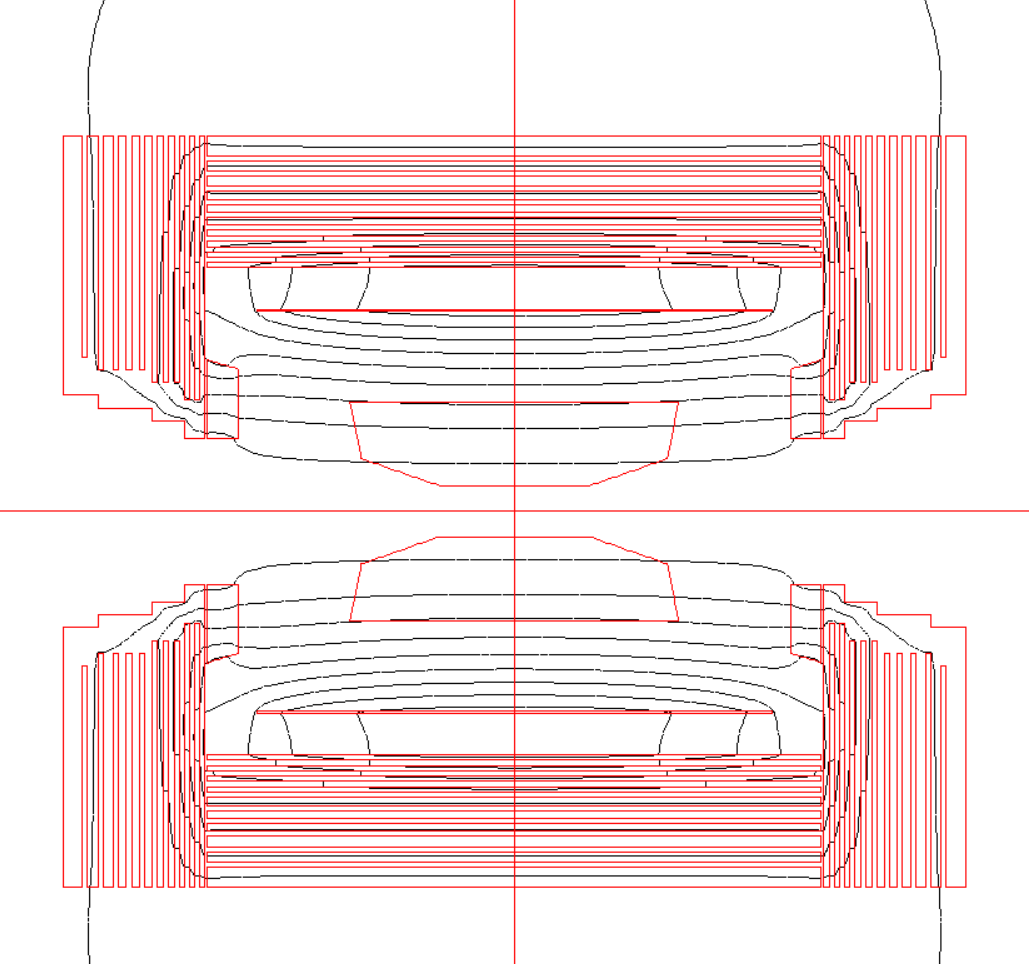}
}
\vspace{0cm}
\caption{Flux of the magnetic field.}
    \label{fig:4.6.02}
\end{figure*}

\begin{figure*}[hp]
    \centering
{
\subfloat[][]{\includegraphics[width=0.35\linewidth]{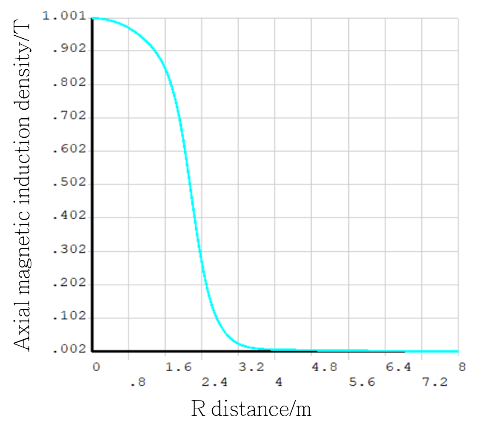}}
\vspace{0.5cm}
\subfloat[][]{\includegraphics[width=0.45\linewidth]{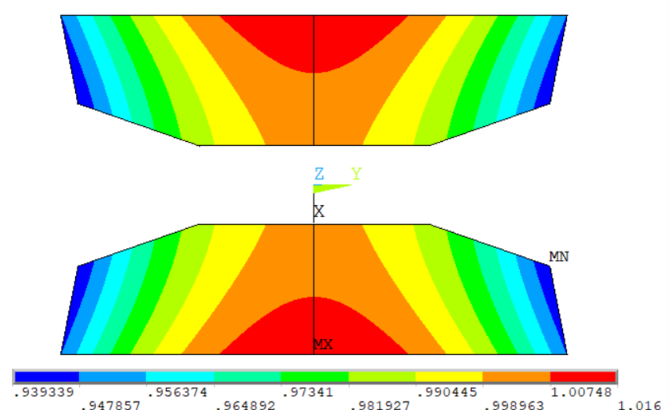}}
}
\vspace{0cm}
\caption{(a) Bz as a function of Z along the beam axis.
(b) The field uniformity in the tracking area.
}
    \label{fig:4.6.03}
\end{figure*}

Fig.~\ref{fig:4.6.02} shows the magnetic field flux of the magnet. The distribution of the magnetic field along the beam line is shown in Fig.~\ref{fig:4.6.03}(a), and the field uniformity is presented in Fig.~\ref{fig:4.6.03}(b). The uniformity in the tracking volume is approximately 2\%. The fringe field remains at 50 Gauss at a distance of 5.1~m from the IP. It decreases sharply to less than 50 Gauss with the addition of some iron material to the gap between the barrel yoke and end yoke.

%%%%%%%%%%%%%%%%%%% Fig %%%%%%%%%%%%%%%%%%%%%%%%%%
%%%%%%%%%%%%%%%%%%%%%%%%%%%%%%%%%%%%%%%%%%%%%%%%%%

%%%%%%%%%%%%%%%%%%% Fig %%%%%%%%%%%%%%%%%%%%%%%%%%
%%%%%%%%%%%%%%%%%%%%%%%%%%%%%%%%%%%%%%%%%%%%%%%%%%

%\FloatBarrier

\subsection{Solenoid Coil Design}
By referring to similar magnets in other HEP laboratories, we decide to adopt a single layer of coil, indirect cooling by liquid helium, a pure aluminum-based stabilizer and a NbTi/Cu superconductor in the STCF superconducting magnet. The overall dimensions of the magnet should be a length of 4.76~m, with an inner diameter of 2.9~m and an outer diameter of 3.7~m; the coil effective length is 4.0~m, and the coil mean diameter is 3.13~m.

Assuming the nominal current as \emph{I}=4000~A, \emph{B$_{0}$=$\mu$$_{0}$nI}, and \emph{B$_{0}$}=1~T, the number of turns in a 1~m long coil should be n $\approx$ 208, and the width of the cable should be 4.67~mm. In total, 832 turns should be needed.
The energy stored by the solenoid is:
\begin{eqnarray}
E = (\frac{1}{2}H \cdot B)V = \frac{1}{2}\cdot\frac{B^2}{\mu_0}\cdot S\cdot l = \frac{1}{2} \cdot \frac{B^2}{\mu_0}\cdot\frac{\pi D^2}{4}\cdot l = 12.3 MJ
\end{eqnarray}

The inductance can be derived as follows:
\begin{eqnarray}
\Phi= B \cdot S \cdot n = B \cdot \frac{\pi D^2}{4} \cdot n = 6274.5 WB
\end{eqnarray}
\begin{eqnarray}
\frac{\mathrm{d}\Phi}{\mathrm{d}t} = L\frac{\mathrm{d}I}{\mathrm{d}t}
\end{eqnarray}
\begin{eqnarray}
L=\frac{\Phi}{I}
\end{eqnarray}
\quad\\
which gives \emph{L}=1.68~H.

From the limitation of the maximum temperature rise of 70~K after quenching, the enthalpy difference of the cable material can be obtained, and the height of the superconductor should be 15~mm.

To address the hoop stress, an aluminum cylinder is necessary; material A5083, which is good for welding and has high mechanical strength, is used; this cylinder is the primary structural component of the cold mass, and coolant is supplied through tubes attached to the outer surface of the restraining hoop cylinder. This eliminates the large cryogen inventory and thick cryostat necessary for cooled coils. The conductor is cooled by thermal conduction through the thickness of the hoop restraint. Two coaxial aluminum cylinders with close endplates cooled by liquid nitrogen act as the radiation heat shield.

\begin{figure*}[hp]
    \centering
{
        \includegraphics[width=0.7\linewidth]{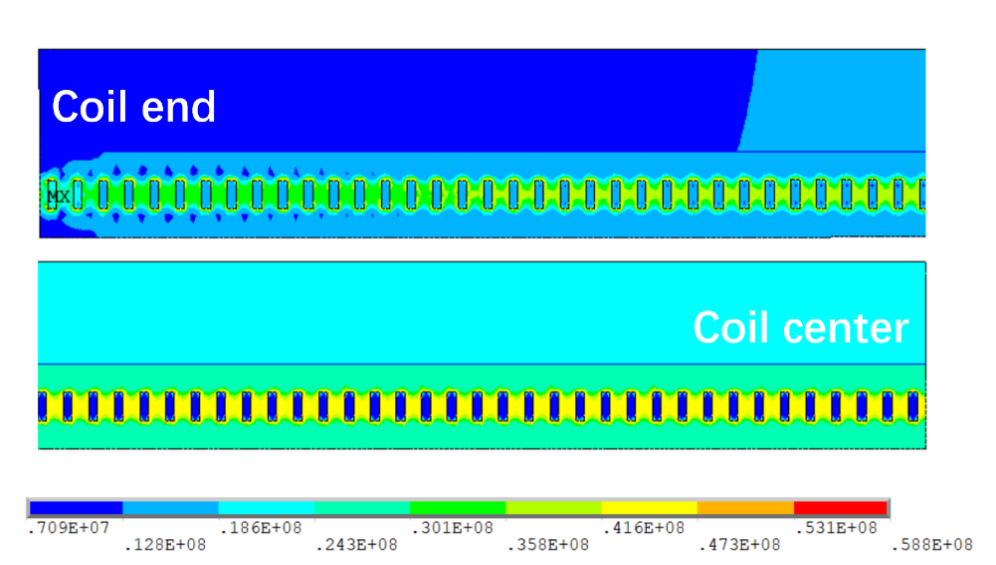}
}
\vspace{0cm}
\caption{The equivalent stress distribution in the coil.}
    \label{fig:4.6.04.1}
\end{figure*}

\begin{figure*}[hp]
    \centering
{
        \includegraphics[width=0.7\linewidth]{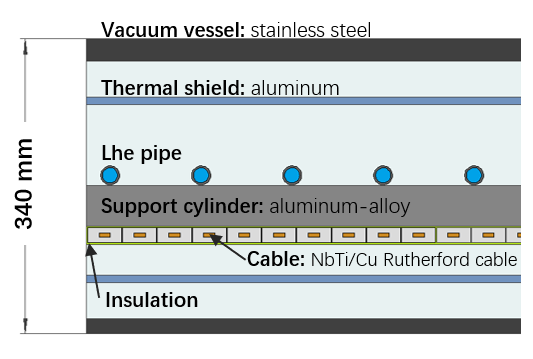}
}
\vspace{0cm}
\caption{Conceptual structure layout of the solenoid.}
    \label{fig:4.6.04.2}
\end{figure*}

From calculations, the thickness of this support cylinder should be 18 mm. After cooling-down and excite, the maximum equivalent stress on the support cylinder is 23.9 MPa. The equivalent stress distribution in the coil is shown in the Fig.~\ref{fig:4.6.04.1}, and the peak stress is 58.8 MPa. The total weight of the cold mass is 4.6 tons, including the cable, the support cylinder, the end flange and the cooling tube. Figure ~\ref{fig:4.6.04.2} shows the conceptual structure layout of the solenoid magnet.

%%%%%%%%%%%%%%%%%%% Fig %%%%%%%%%%%%%%%%%%%%%%%%%%
%%%%%%%%%%%%%%%%%%%%%%%%%%%%%%%%%%%%%%%%%%%%%%%%%%

%%%%%%%%%%%%%%%%%%% Fig %%%%%%%%%%%%%%%%%%%%%%%%%%
%%%%%%%%%%%%%%%%%%%%%%%%%%%%%%%%%%%%%%%%%%%%%%%%%%

\begin{figure*}[hp]
    \centering
{
        \includegraphics[width=0.5\linewidth]{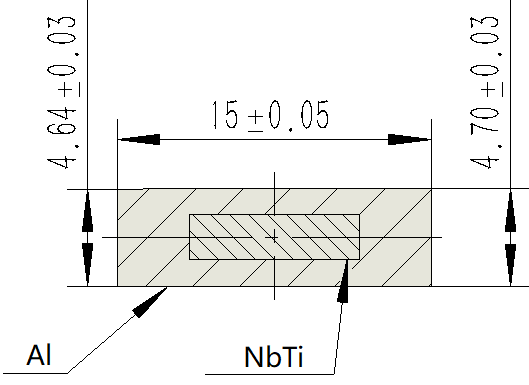}
}
\vspace{0cm}
\caption{Cross-sectional view of the superconductor for the STCF detector magnet.}
    \label{fig:4.6.06}
\end{figure*}

The superconducting conductor is composed of NbTi/Cu~(1:1) Rutherford cable and a high purity ($>$99.99\%, RRR$>$500) pure aluminum stabilizer. The Rutherford cable contains 16 NbTi strands. The Main parameters of the superconducting conductor are listed in Table~\ref{tab:4.6.02}. The cross section of the superconductor is shown in the Fig.~\ref{fig:4.6.06}, the superconducting wire is located in the center of the aluminum stabilizer. With this configuration, the coil temperature kept below 70 K after quenching at operating current.

%%%%%%%%%%%%%%%%%  TABLE  %%%%%%%%%%%%%%%%%%%%%%%%
\begin{table*}[hptb]
\small
    \caption{Main parameters of the superconducting conductor.}
    \label{tab:4.6.02}
    \vspace{0pt}
    \centering
    \begin{tabular}{ll}
        \hline
        \thead[l]{Rated current} & \thead[l]{3820 A}\\
        \hline
        Critical current at 4.2 k \& 2 T&	$\geq$15000 A \\
        Conductor length&	9.15 km \\
        Cable dimension&	4.67 mm $\times$ 15 mm \\
        \hline
        Rutherford cable parameters & \\
        \hline
        Number of strands	&16 \\
        Cable transposition pitch	&100$\pm$5 mm \\
        Cu:NbTi	&~1:1 \\
        NbTi filament diameter	&30$\pm$5 $\mu$m \\
        Number of filaments	&$\geq$600 \\
        N value@2T	&$\geq$35 \\
        \hline
        Aluminum stabilizer parameters & \\
        \hline
        RRR@0T,4.2K	&$\geq$500 \\
        Yield strength@4.2k	&$\geq$60MPa \\
        Impurity content	&$\textgreater$1000ppm \\
        Cross-section ratio of aluminum	&$\textgreater$80\% \\
        \hline
    \end{tabular}
\end{table*}
%%%%%%%%%%%%%%%%%%%%%%%%%%%%%%%%%%%%%%%%%%%%%%%%%%

%%%%%%%%%%%%%%%%%%% Fig %%%%%%%%%%%%%%%%%%%%%%%%%%
%%%%%%%%%%%%%%%%%%%%%%%%%%%%%%%%%%%%%%%%%%%%%%%%%%

The coil windings are wound by the inner winding technique with the aluminum-alloy cylinder support, which acts as an external supporting mandrel and removes part of the heat energy induced by quenching. To maintain the operating temperature of the detector magnet, the cooling tubes for the circular flow of liquid helium are welded on the outer surface of the aluminum-alloy cylinder.

The titanium tie rods will be used for the coil suspension system and provide axial and radial fixation, to ensure the precise and rigid suspension of the cold mass inside the vacuum vessel. The loads to be supported are the self weight of the cold mass and the magnetic forces due to the decentering and misalignment of the coil with respect to the return yoke. The design must also take into count the contraction of the coil during cooling and its deformation under magnetic forces, with enough strengths and smallest heat conducting. 8 vertical rods are used to counteract the electromagnetic and cold shrinkage forces, and the diameter of the rods is designed to be 25 mm.

\subsection{Iron Yoke Design}
\begin{figure*}[hp]
    \centering
{
        \includegraphics[width=0.7\linewidth]{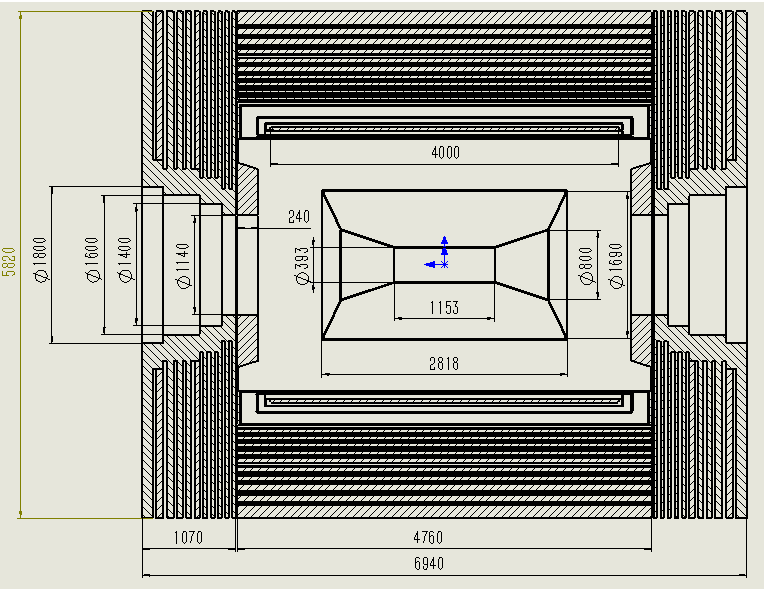}
}
\vspace{0cm}
\caption{Configuration of the iron yoke with pole tips for the STCF detector.}
    \label{fig:4.6.07}
\end{figure*}

The iron yoke has three functions. First, it provides a magnetic flux return path to shield the leaked field. Second, it is used as the absorber material for the muon detector, which is sandwiched between the multilayer iron plates. Finally, it is the mechanical support structure of the overall detector. Therefore, a high permeability material with high mechanical strength is required for the yoke material to account for the mechanical and magnetic field performance. Low-carbon steel is chosen for the yoke based on its magnetic properties. However, this material has relatively low strength. Thus, a balance between good magnetic properties and moderate strength must be achieved. The yoke is divided into two main components, one barrel yoke and two end-cap yokes. A detailed design would include data cables, cooling pipes, and gas pipes passing through the yoke from the inner detectors. The total weight of the iron yoke assembly is approximately 800 tons.
Fig.~\ref{fig:4.6.07} shows the iron yoke configuration with the main dimensions.

%%%%%%%%%%%%%%%%%%% Fig %%%%%%%%%%%%%%%%%%%%%%%%%%
%%%%%%%%%%%%%%%%%%%%%%%%%%%%%%%%%%%%%%%%%%%%%%%%%%

The barrel yoke is an octagonal-shaped structure with a length of 4,760~mm. The outer and inner heights of the octagon are 5,820~mm and 3,760~mm, respectively. The barrel yoke supports the magnet cryostat. The inner vacuum vessels of the cryostat host inner detectors. The barrel yoke is subdivided into 10 layers, with 40~mm gaps between the layers for the muon detector.
The end-cap yoke also consists of 11 layers of steel plates. Used as the end door of the detector, it is designed to slide apart to provide an opening for access to the inner detectors. 
The end-cap muon chambers are mounted in the 4~cm gap between the vertical end-cap yoke disks and are normally inaccessible.
The details can be found in Sec.~\ref{sec:yoke}.

\subsection{Quench Protection and Power Supply}
The protection of the solenoid is based on the classical concept of the extraction of stored energy by an external resistor.
Approximately 12~MJ of energy is stored in the magnet.
Selected voltage signals from the STCF detector magnet coil and power leads are monitored by a special quench detection device.
If a quench occurs, the power supply is switched off, and a dump resistor is switched into the electrical circuit; the large stored energy is extracted mainly by the dump resistor and partially by the coil itself. The value of the dump resistor limits the maximum voltage across the solenoid terminals.
A 0.1~Ohms dump resistor is mounted in parallel to the breakers, which are doubled for safety. As the operating current is approximately 3820 A, the maximum voltage is 382 V with respect to ground.
The power supply is designed to operate with 10~V and 4000~A, with a current stability of less than 10~ppm within 8 hours, a current ripple of less than 10~ppm and an accuracy of 20~ppm. The repeatability should reach 10~ppm.

\subsection{Magnet Cryogenics}
Solenoid cryogenics cooling is based on the thermosiphon cooling method, where the superconducting coil is indirectly cooled with saturated liquid helium. The thermosiphon circuit consists of a helium phase separator located in an elevated position and cooling tubes. A horizontal cryostat is designed, including a vacuum tank, an inner thermal shield, and an outer thermal shield. The stainless steel vacuum vessel is a cylinder with a length of 4.76~m. The thickness of the inner wall, outer wall, and endplates is 6 mm, 16 mm, and 32 mm, respectively. Figure 11.5.1 presents the vacuum vessel stress and deformation near the endplate, indicating a 41.80 MPa stress and 0.23 mm strain in the middle of the inner vessel.

%%%%%%%%%%%%%%%%%%% Fig %%%%%%%%%%%%%%%%%%%%%%%%%%
\begin{figure*}[htbp]
    \centering
{
        \includegraphics[width=0.6\linewidth]{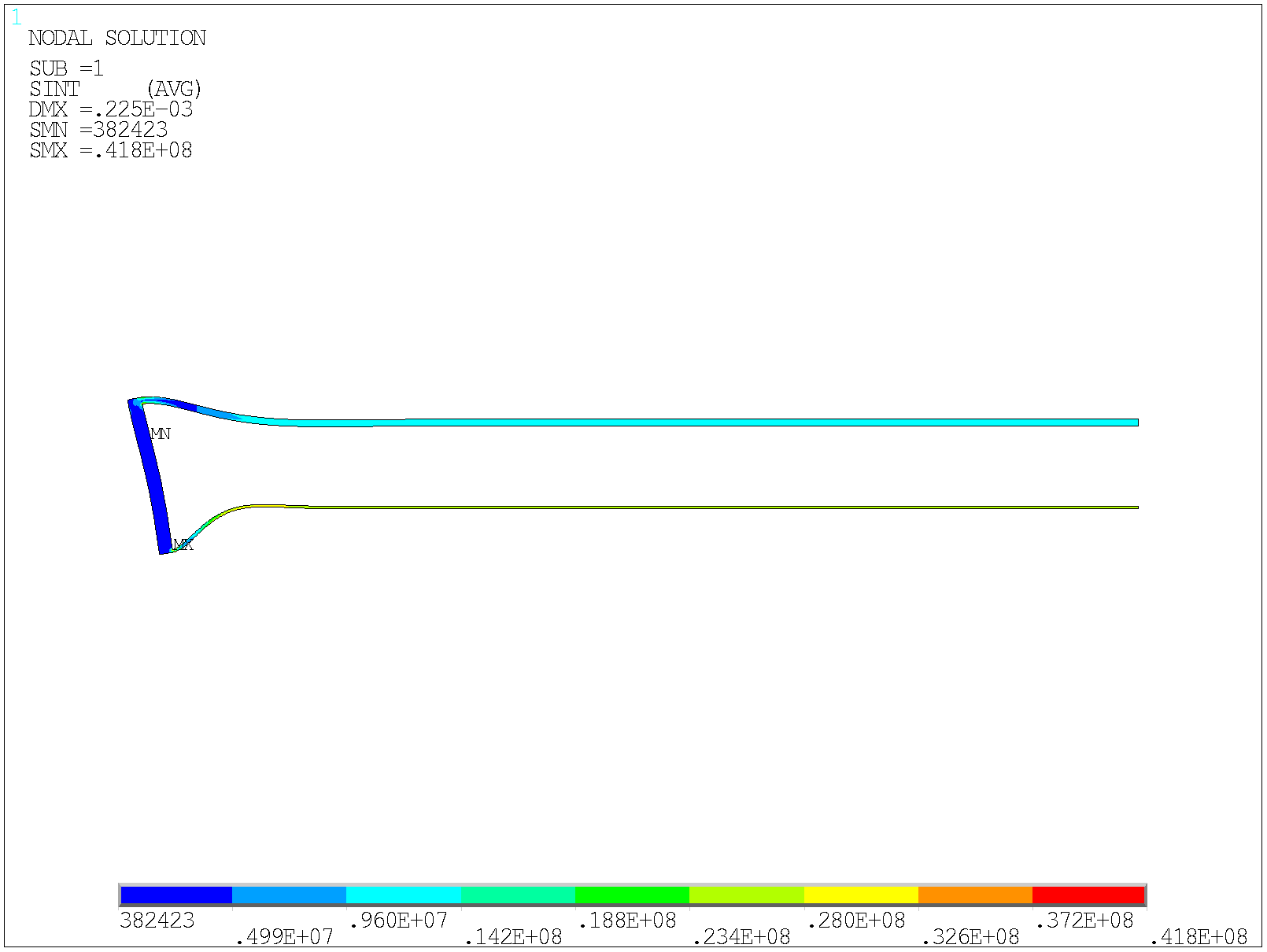}
}
\vspace{0cm}
\caption{Stress and deformation of vacuum vessel(2D 1/2 Model).}
    \label{fig:4.6.09}
\end{figure*}
%%%%%%%%%%%%%%%%%%%%%%%%%%%%%%%%%%%%%%%%%%%%%%%%%%

The service tower is designed on the top section of the barrel iron yoke. Table~\ref{tab:11.5.1} shows the heat load estimation of the magnet.

%%%%%%%%%%%%%%%%%%%%%%%%%%%%%%%%%%%%%%%%%%%%%%%%%%
\begin{table*}[hptb]
\small
    \caption{Heat load estimation.}
    \label{tab:11.5.1}
    \vspace{0pt}
    \centering
    \begin{tabular}{lll}
        \hline
Heat Load Components & 77~K & 4.5~K \\
        \hline
Caused by the support rods in cryostat	& 27~W & 	1.0~W \\
Caused by the radiation in cryostat	& 74~W &	3.2~W \\
Caused by the current leads		& ——	& 7.9~W + 0.4~g/s \\
Caused by the radiation in chimney \& SP	&	10~W &	0.4~W \\
Caused by the support rods in chimney \& SP	&	4~W	& 0.1~W \\
Caused by the bayonet and valves in SP	& 46~W	&	13~W \\
Caused by the measuring wires		& 5~W	& 0.8~W \\
        \hline
Total	& 166~W	26.4~W + 0.4~g/s \\
        \hline
Adopted heat load~($\times 1.5$) & 249~W	& 39.6~W + 0.6~g/s \\
        \hline
    \end{tabular}
\end{table*}
%%%%%%%%%%%%%%%%%%%%%%%%%%%%%%%%%%%%%%%%%%%%%%%%%%

The STCF detector cryogenic system consists of a helium refrigerator, liquid and gas transfer lines, liquid and gas storage, and a nitrogen system. The major components include compressors, oil removal systems, cold boxes and control systems. All the helium supplied by this system, except for normal leakage and necessary venting, is circulated and reliquefied. Liquid nitrogen usage includes the cooling of the transfer line thermal shield, the cooling of the thermal shield and thermal intercepts in the superconducting solenoid magnet, and the cooling of the high-pressure helium feed stream in the refrigerator.

\clearpage
\newpage
\section{Iron Yoke and Mechanical Structure}
\label{sec:yoke}
\subsection{Mechanical Structure of the Iron Yoke}
The iron yoke is the base structure of the MUD and the base support of the subdetector components attached to it. It has two important functions: one is absorbing all the high-energy particles except muons in the MUD, and the other is conducting a magnetic field from the superconducting solenoid that serves as leak shielding. The iron yoke material and structure must be magnetically permeable and have high mechanical strength. Low carbon steel is selected as the baseline iron yoke material and is measured piece by piece with strict requirements for permeability before machining.

\begin{figure*}[htb]
	\centering
\subfloat[][]{\includegraphics[height=0.45\linewidth]{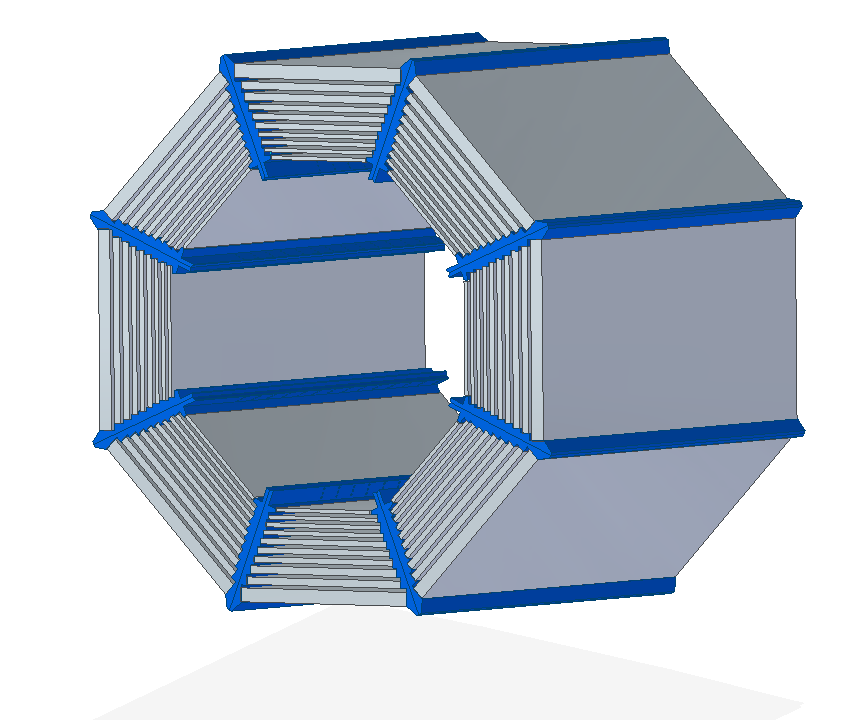}}
\subfloat[][]{\includegraphics[height=0.45\linewidth]{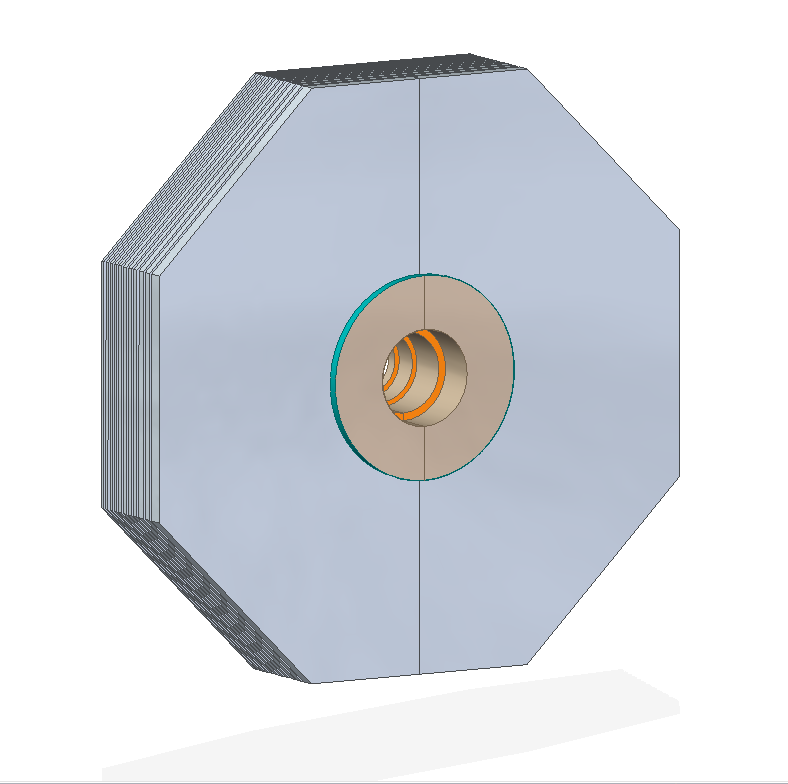}}
\vspace{0cm}
\caption{A schematic view of the iron yoke in the (a) barrel and (b) endcap regions.}
    \label{fig:4.7.04}
\end{figure*}

%\quad\\
The iron yoke consists of a barrel part and an endcap part, both of which have a multilayer structure. The barrel yoke, as shown in Fig.~\ref{fig:4.7.04}(a), has ten layers with thicknesses of 40, 40, 45, 45, 60, 60, 60, 80, 80, and 150~mm from the innermost layer to the outermost layer. Every layer has an octagonal column configuration and is formed by eight pieces of separate iron plates. The net inner height of the 1st layer is 3.78~m and that of the 10th layer is 5.52~m.

%\quad\\
The endcap yoke, as shown in Fig.~\ref{fig:4.7.04}(b), consists of eleven layers of iron plates in an octagonal shape. Every layer is formed by two half-octagonal iron plates. The thicknesses of the endcap yoke layers are 40, 40, 40, 45, 45, 60, 60, 60, 80, 80, and 150~mm along the beampipe direction. Polar iron is situated in the center area of the endcap yoke and is separated into two halves: one half is fixed to the endcap iron, and the other half can be pushed and pulled during installation and maintenance. An adequate gap between iron plates is maintained for the installation of cables and pipes.

%%%%%%%%%%%%%%%%%%% Fig %%%%%%%%%%%%%%%%%%%%%%%%%%
%%%%%%%%%%%%%%%%%%%%%%%%%%%%%%%%%%%%%%%%%%%%%%%%%%

%\quad\\
The total weight of the iron yoke is up to approximately 800~tons. The iron yoke is subject to a large electrodynamic force when the superconducting solenoid is in operation. The structure of the iron yoke should be strong enough to resist this strong force. Welding is avoided in the iron plate connections to mitigate deformation effects. Bolting is convenient for position adjustment. The dead zone of the MUD should be minimized when considering connections.

%\quad\\
The general structure specifications of the iron yoke are as follows:
\begin{itemize}
\item The center of the iron yoke geometry should be below 3.9~m.
\item The final assembly of the iron yoke will be completed in the detector hall, and the yoke will be transported to and positioned at the positron-electron collision point.
\item The interior areas of the iron yoke should be easily accessible, and the vacuum conditions should be maintained during maintenance.
\item The support structure of the solenoid should guarantee that the magnetic field is reproducible after detector maintenance. Electric cables should be routed along magnetic lines to avoid additional stray magnetic fields.
\item Enough space should be left for electric cables and media coming in and out of the iron yoke.
\end{itemize}

\subsection{Mechanical Movement of the Iron Yoke}
The iron yoke assembly stands on a steel base support that includes the main frame, a guiding frame, an enforcement frame and a lifting jack. The mechanical support allows the detector to move in both horizontal and vertical direction. In the final positioning step, the detector is transported horizontally to the collision point on a guiding rail and raised in the vertical direction on the hydraulically driven jack system. The position precision is guaranteed by the straightness of the guiding rail and the accuracy of the motor controller system. The base support structure should be strong enough to withstand the load of the whole detector, which may be as heavy as 800~tons. At the end of the guiding rail, a disc spring is mounted as a buffer to prevent the detector from going out of range.

The endcap iron yoke can be opened and closed to allow people to enter the detector to perform maintenance. To make this easy, half of the polar ring iron in the center is pulled out before the endcap iron opens, and the four endcap iron yokes move independently and are controlled by servo motors. On the top of the detector, extra guiding rails are mounted to increase the position precision and stability of the movement.

\subsection{Detector Hall and Infrastructure}
The detector hall is the main place where the whole setup will be assembled and the detector will be positioned in its final working location. The requirements of the detector hall are as follows:
\begin{itemize}
\item The detector hall dimensions should be be no less than 20~m along the beam direction, 7~m in height, and 30~m in the transverse beam direction.
\item The collision point should be approximately 3.9~m above the ground floor.
\item The load capacity of the ground floor should be 400~kN/m${^2}$.
\item The sum of the floor vibration amplitude should be less than 100~$\mu$m in the range of $1\sim60$~Hz.
\item The ceiling crane should have a 50~ton lifting load capacity.
\item The main gate should be no less than 5.5~m in width and 6.0~m in height. The dimensions of the detector component and subsystem that need to be brought into the hall through the main gate are restricted by the gate size.
\end{itemize}

%%%%%%%%%%%%%%%%%%% Fig %%%%%%%%%%%%%%%%%%%%%%%%%%
\begin{figure*}[htb]
	\centering
    \includegraphics[height=0.6\linewidth]{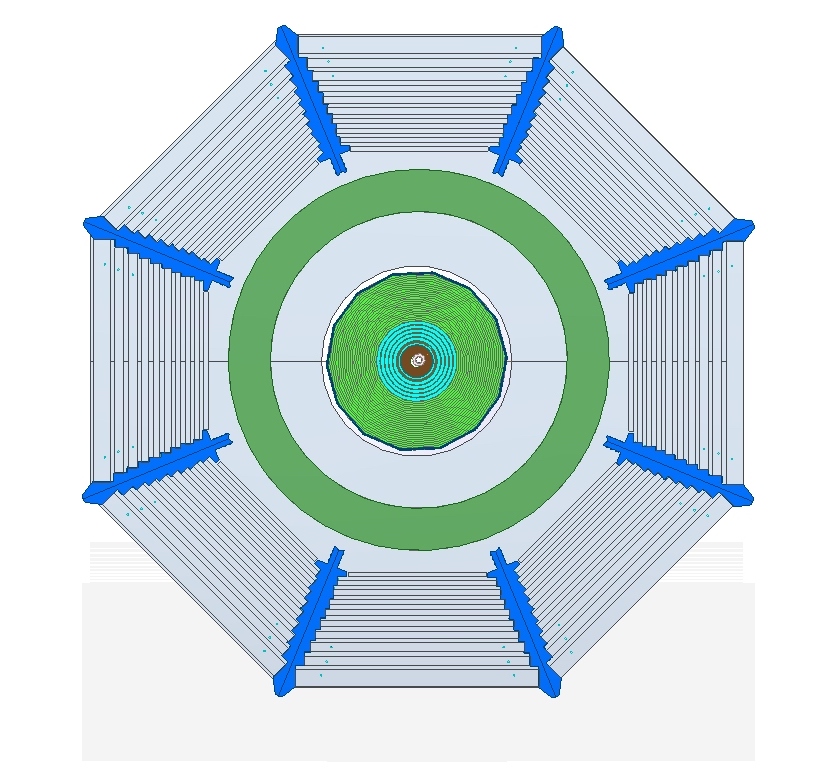} \\
    \includegraphics[height=0.6\linewidth]{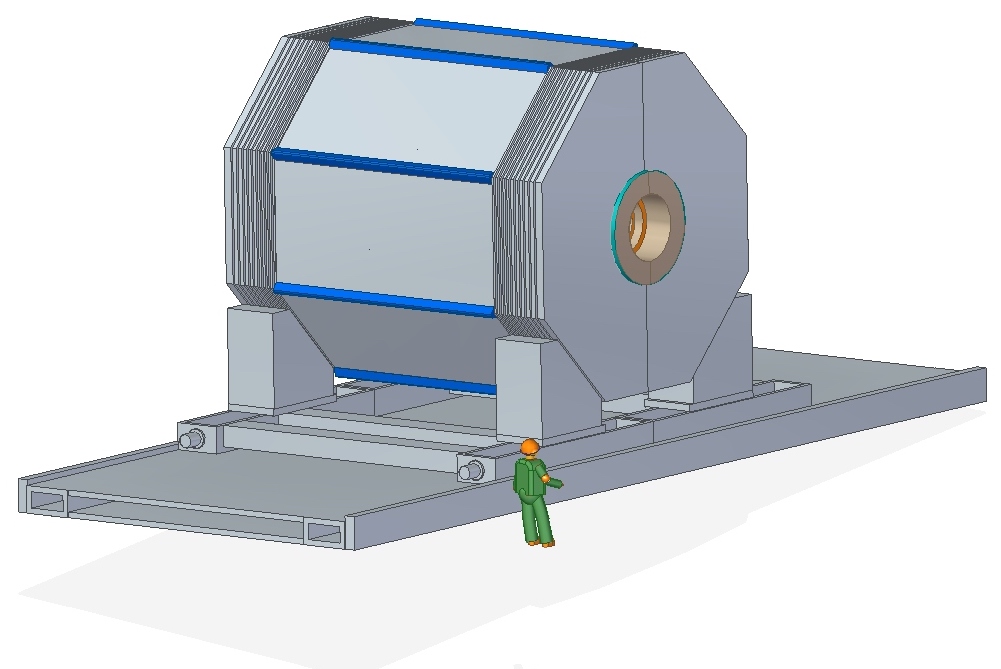}
\vspace{0cm}
\caption{A cross-sectional view of the detector (upper) and an overall view of the detector (lower).}
    \label{fig:4.7.03}
\end{figure*}
%%%%%%%%%%%%%%%%%%%%%%%%%%%%%%%%%%%%%%%%%%%%%%%%%%

\clearpage
\newpage
\section{Trigger, Clock and Data Acquisition~(TDAQ)}
\label{sec:tdaq}

The CME region of the STCF spans from 2.0 to 7.0~GeV, and the corresponding event rate
varies with respect to different CMEs.
The event rates of different CMEs can be divided into two categories: on-$J/\psi$ peak
and off-$J/\psi$ peak. The total cross-sections and event rates at the
on-$J/\psi$ peak, $\sqrt{s}=3.097$~GeV, is approximately one magnitude larger than that at the off-$J/\psi$ peak due to the
large production cross-section of $J/\psi$.
Table~\ref{eventrate} summarizes the cross-sections and event rates
at $\sqrt{s}=3.097$~GeV and 3.773~GeV, respectively.
The production cross-section of $J/\psi$ is related to the energy spread of the beams. According to the
designed machine parameters of the STCF, which are described in Section~\ref{sec:expcon}, the beam energy spread is
$4.0\times10^{-4}$, corresponding to a beam energy spread of $\Delta E_{\rm beam}=$0.6~MeV
and a total energy spread of $\Delta  E=\sqrt{2}\cdot\Delta E_{\rm beam}=0.848$~MeV at $\sqrt{s}=3.097$~GeV.
The event rate is calculated from the product of the cross-section and luminosity, which is supposed to be 75\% of the peaking luminosity at the optimized c.m energy $\sqrt{s}=4.0$~GeV, that is $0.75\times10^{35}$\,cm$^{-2}$s$^{-1}$ at $\sqrt{s}=3.097$~GeV. The total rate of physics events is expected to be approximately $\stcftrigrate$ at the on-$J/\psi$ peak and to be approximately 60~kHz at the off-$J/\psi$ peak.

\begin{table}[htbp]
\caption{Summary of the cross-sections and event rates from physics processes at $\sqrt{s}=3.097$~GeV and
$\sqrt{s}=3.773$~GeV.
}
\label{eventrate}
\footnotesize
\begin{center}
\begin{spacing}{1.3}
\begin{tabular}{ccc}
\hline
\hline
\vspace{0.2cm}
Physics Process~~~~~              &	  ~~~~~~~~Cross-section~(nb)~~~~~~~~          &  ~~~~~~~Rate~(Hz)~~~~~~~~ \\
\hline
\hline
\multicolumn{3}{c}{$\sqrt{s}=3.097$~GeV, ~~$\mathcal{L}=0.75\times10^{35}$~cm$^{-2}$s$^{-1}$,~~$\Delta E=0.848$~MeV }\\
\hline
$J/\psi$            &    4500                   &  337500 \\
\hspace{1.25cm} $\to e^{+}e^{-}$  &  \hspace{0.5cm}  270  &  \hspace{0.5cm}  20000\\
\hspace{1.25cm} $\to \mu^{+}\mu^{-}$ &\hspace{0.5cm}  270  & \hspace{0.5cm}  20000 \\
Bhabha $(\theta\in(20^{\circ}, 160^{\circ})$  &  734    & 55000 \\
$\gamma\gamma$~$(\theta\in(20^{\circ}, 160^{\circ})$ & 36 & 2700 \\
$\mu^{+}\mu^{-}$    &  11.4       & 900  \\
Hadronic from continuum   &  25.6  &  2000  \\
$2\gamma$ process~$(\theta\in(20^{\circ}, 160^{\circ}), E>0.1$~GeV  &  $\sim$23.3 &  $1740$\\
\hline
Total                      &      $\sim$5300  &  $\sim$400000 \\
\hline
\hline
\multicolumn{3}{c}{$\sqrt{s}=3.773$~GeV, ~~$\mathcal{L}=1.0\times10^{35}$cm$^{-2}$s$^{-1}$}\\
\hline
$\psi(3770)$            &    9                   &  900 \\
Bhabha $(\theta\in(20^{\circ}, 160^{\circ})$  &  517    & 51700 \\
$\gamma\gamma$~$(\theta\in(20^{\circ}, 160^{\circ})$ & 24.5 & 2450 \\
$\mu^{+}\mu^{-}$    &  7.9       & 790  \\
Hadronic from continuum   &  18  &  1800  \\
$2\gamma$ process~$(\theta\in(20^{\circ}, 160^{\circ}), E>0.1$~GeV  &  $\sim$25 &  $2500$\\
\hline
Total                      &      $\sim$601  &  $\sim$60100 \\
\hline
\hline
\end{tabular}
\end{spacing}
\end{center}
\end{table}

\subsection{Trigger}

\subsubsection{Trigger System}
\begin{figure*}[htb]
    \centering
    {
        \includegraphics[width=100mm]{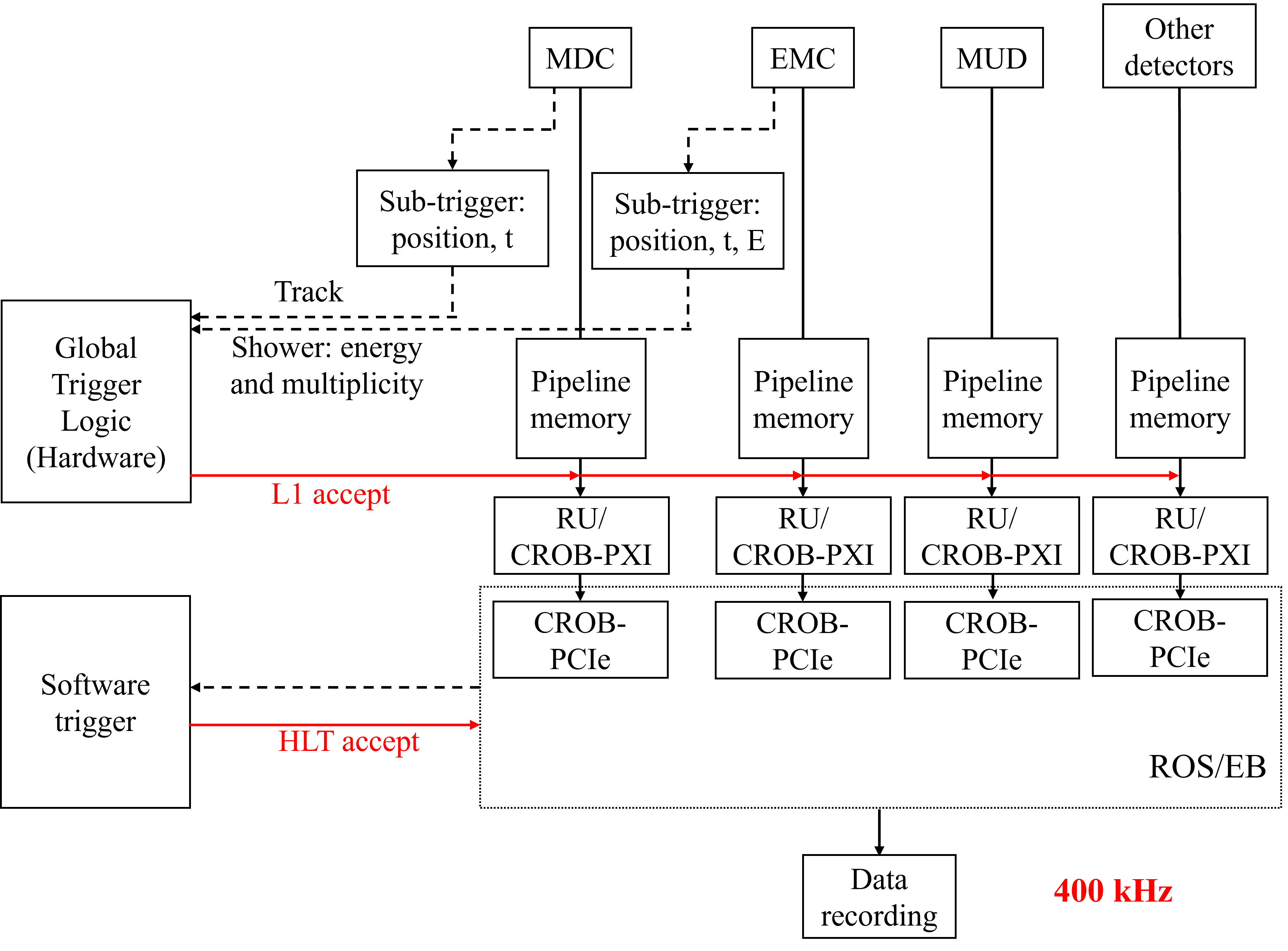}
    }
    \vspace{0cm}
\caption{The schematic of STCF trigger system.}
    \label{fig:trigger_system}
\end{figure*}

\quad\\
The expected peak event rate at the STCF of 400~kHz  places challenging requirements on the trigger system.
The baseline design of the STCF trigger system is based on a two-level trigger concept: a hardware-based ~(Level 1, L1) trigger and a software-based high-level trigger~(HLT), as shown in Fig.~\ref{fig:trigger_system}.
The L1 trigger mainly utilizes the information from the MDC and EMC subdetectors.
The MDC provides charged track information (momentum, position, charge, multiplicity, and so on).
The EMC gives energy deposit information and alternative cluster information (position, energy, size, multiplicity, and so on).
The trigger signal generated by L1 will be sent back to each sub-detector system.
The HLT uses infomation from all sub-detector systems. Typically, the MUD provides the fast reconstructed cluster information (layer number, hit number, and position). EMC also calculates the cluster shape information and gets used in HLT. In this level, the reconstructed tracks and clusters will be marked with a preliminary tab and get identified as various trigger types. The HLT will be implemented as part of the DAQ system.

%%%%%%%%%%%%%%%%%%% Fig %%%%%%%%%%%%%%%%%%%%%%%%%%
%%%%%%%%%%%%%%%%%%%%%%%%%%%%%%%%%%%%%%%%%%%%%%%%%%

It is worth noting that the \stcftrigrate\ event rate only occurs at the $J/\psi$ peak. At other CMEs, the event rate would be much lower. For instance, according to Table~\ref{eventrate}, the expected event rate is approximately 60~kHz at $\sqrt{s}=3.773$~GeV ($\psi(3770)$). At the event rate of \stcftrigrate\ expected on the $J/\psi$ peak, the probability of more than one physics event occuring within a time window of 200 (500) ns is approximately 8 (18)\%. This poses major challenges to the STCF detector in terms of separating the physics events close in time. This calls for fast detector response to minimize the event overlapping probability. The MDC is the part of the STCF detector that is most susceptible to the high event rate given its rather long drift time and hence a long integration time window (up to ~1 $\mu$s). However, timing capability of other components of the STCF detector (for example, the CMOS-ITK and EMC) can be exploited to resolve the overlapping events recorded by the MDC as described in Section~\ref{sec:mdc_pileup_effect}. It should be noted that the current conceptual design of the STCF detector has not been fully optimized to effectively distinguish close-by events at a \stcftrigrate\ event rate. Further design studies of both detectors and readout electronics are needed to fully meet the challenge posed by the very high event rate.

\subsubsection{Trigger Electronics}
\begin{figure*}[htb]
	\centering
    \includegraphics[width=0.9\linewidth]{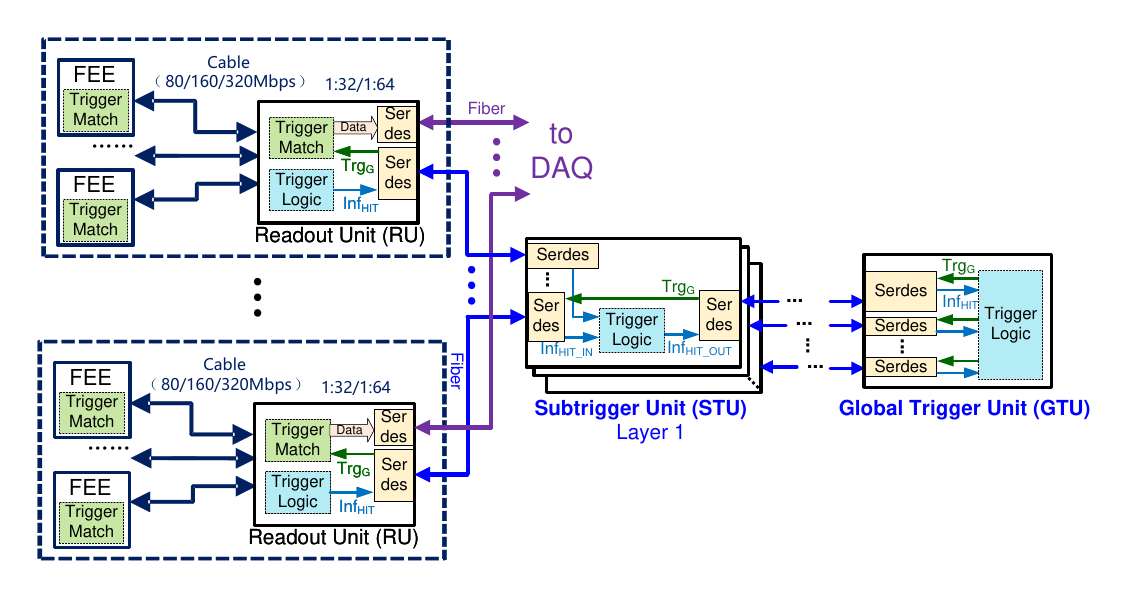}
\vspace{0cm}
\caption{Block diagram of the trigger electronics for the STCF.}
    \label{fig:trig_elec}
\end{figure*}

As shown in Fig.~\ref{fig:trig_elec}, the trigger electronics are organized into several hierarchies. The first step of trigger processing is finished within the electronics of each subdetector (i.e., in ``trigger logic'' in each readout unit of Fig.~\ref{fig:trig_elec}), which contributes to the generation of the ``global trigger'' signal. The results (marked as ``InfHIT'' in Fig.~\ref{fig:trig_elec}) are then sent to the subtrigger unit~(STU) for further processing, and the final information is gathered by the global trigger unit~(GTU) for the decision of whether to generate a global trigger (marked as ``TrgG" in Fig.~\ref{fig:trig_elec}). Actually, there may be several layers of STUs, according to the final system scale.
Optical fibers will be used for communication between the RU, STU, and GTU, as this is suitable for signal transmission over long distances and can isolate the ground of different modules over a large area. Based on the wavelength division multiplexing~(WDM) technique, bidirectional signal transmission can be performed over one optical fiber.
As for the downlink distribution of TrgG, TrgG is fanned out through the same optical fiber path for the uplink from the RU to the STU and finally to the GTU, just backward. Once the FEE receive TrgG, the ``trigger match" function is executed to locate the valid data, which are then read out to the DAQ. This ``trigger match" process would be conducted in the RU or in FEE, according to the electronics designs of different subdetectors. The uplink delay from the subdetector electronics to GTU and the downlink delay backwards contribute to the final trigger latency. Besides, a proper trigger match window should be selected to perform the ``trigger match” process. According to the trigger latency and match window size, the times tamp region of valid data can be calculated, and then the trigger match process is accomplished by searching data within the caches of the subdetector electronics.

\subsection{Clock System for the Readout Electronics}
\begin{figure*}[htb]
	\centering
    \includegraphics[width=0.9\linewidth]{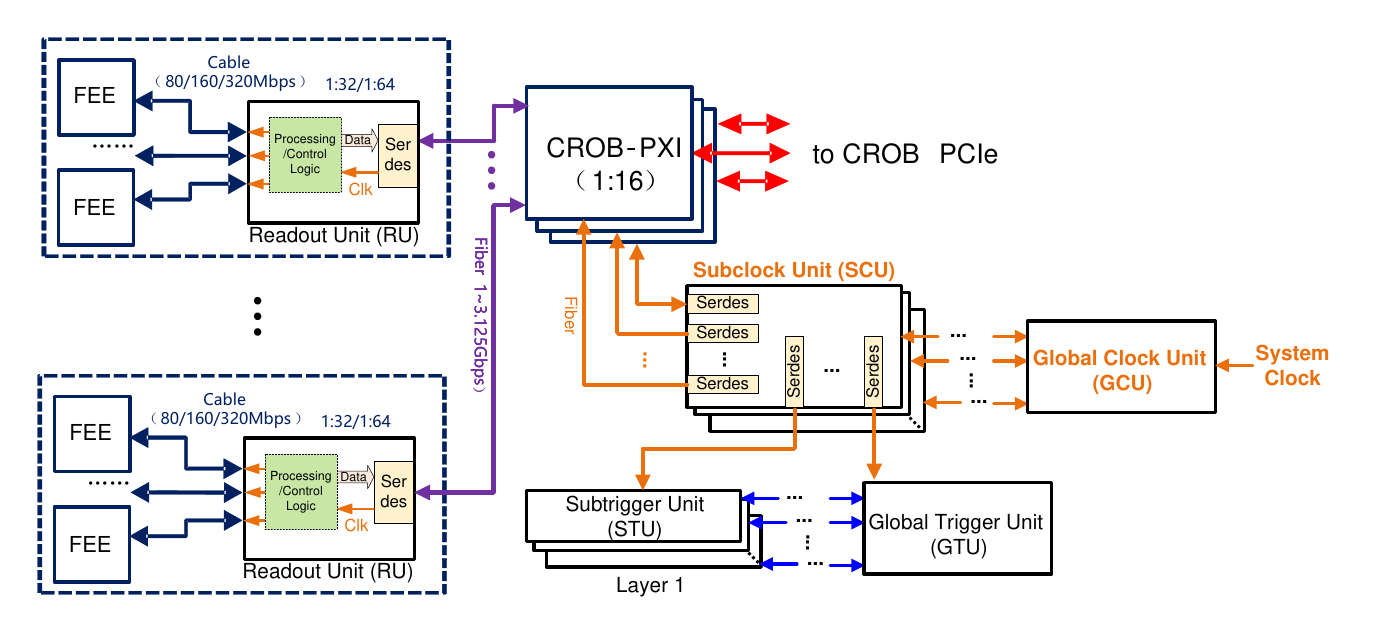}
\vspace{0cm}
\caption{Block diagram of the clock system for the readout electronics in the STCF.}
    \label{fig:trig_clk}
\end{figure*}

\quad\\
The structure of the clock system for the readout electronics in the STCF is shown in Fig.~\ref{fig:trig_clk}.
The readout electronics should be synchronized for time measurement or event building. This is achieved in two steps. First, all the FEE should be synchronized to a system clock which functions as the global reference for the time stamping or precise time measurement in all the subdetector electronics, and this system clock will be imported from the clock system of the accelerator in STCF; second, the time stamp should be started (or cleared) at the same time point.
For the first step, as shown in Fig.~\ref{fig:trig_clk}, the global clock unit (GCU) receives the system clock and then fans it out to multiple subclock units (SCUs). These SCUs further send the synchronized clocks to the Common ReadOut Board based on PXI (CROB-PXI) modules in the DAQ and then finally transmit them to the RUs \& FEE of different subdetector electronics. Considering the large scale of electronics, optical fibers will also be used to guarantee high-quality clock signal transmission over long distances.

For the second step ({\it i.e.}, the synchronous starting of the time stamp), once the DAQ receives a command from the operator, it transmits this command to the GTU, as shown in Fig.~\ref{fig:trig_elec}. The GTU translates this command and generates a ``Start" signal, and then distributes this signal through the same path for the global trigger signal in normal working mode. When the FEE receive this ``Start" signal, the counters within all the FEEs are cleared to zero, and then start counting driven by the aforementioned system clock. The outputs of the counters are used as the time stamps, which are well synchronized among all the FEEs. Besides, a periodic checking process will also be implemented: the GTU sends out global time stamp information which is compared with the local time stamps within the FEEs. Errors detected in this checking process will be reported in time, and the local time stamp will be corrected simultaneously. The time stamp information is added to the data packages from the FEEs, which will be further used in event building.

\FloatBarrier

\subsection{Data Acquisition System}

\subsubsection{Data Size Requirements}
The data size requirements are based on requests from subdetectors, as summarized in Table~\ref{tab:stcf_data_size}.

\begin{table*}[htb]
    \caption{Estimated data size of each detector system.
     The readout time window is adjusted for each detector system to collect the stored events after receiving the trigger.
     A 50~ns jitter for the event starting time is assumed.
     The estimated event size includes both the physics events and the background contribution within the corresponding readout time.}
    \label{tab:stcf_data_size}
    \centering
    \begin{tabular}{lrrrr}
        \hline
        Component  & Num. of channels & Readout time window & Event size (B) & Total (B/s) \\
        \hline
        ITK (Silicon)   & 50M & 500~ns &  14300 & 5.72G \\
        ITK ($\mu$RWELL) & 10552 & 500~ns & 17232 & 6.89G  \\
        MDC & 11520 & 1~$\mu$s & 20400 & 8.16G \\
        PID (RICH) & 518400 & 500~ns & 15600 & 6.24G \\
        PID (DTOF) & 6912 & 500~ns  & 7380 & 2.95G\\
        EMC &  8670 & 500~ns & 15000 & 6.00G\\
        MUD & 41280 & 500~ns & 262 & 105M\\
        \hline
        Total(Silicon) & 50.6M & -- & 72.9k & 29.2G\\
        Total($\mu$RWELL) & 594k & -- & 75.9k & 30.4G\\
        \hline
    \end{tabular}
\end{table*}

\subsubsection{Design Goal}
The data acquisition (DAQ) system performs data processing and system control in the STCF experiment. The main functionalities of the STCF DAQ system are as follows:
\begin{itemize}
\item Reads out the detector data after the Level-1~(L1) trigger matching from the front-end electronics (FEE) or the merged L1 triggered data from the FEE readout unit~(RU).
\item Implements several steps of the data processing, including event building, event data compression or information extraction, and high-level trigger~(HLT) computing, before the physics events of interest are ultimately saved to the storage system.
\item Organizes and manages the FEE modules, configures the FEE working parameters, and controls the working flow of the system.
\item Monitors the running status online, including the FEE working status, the status of the link, etc.
\item Decimates and analyzes the event data to assess the working status of the system when needed.
\end{itemize}
Of the points mentioned above, the reading out and processing of FEE data is the most significant functionality of a DAQ system.

In the STCF, as shown in Table~\ref{tab:stcf_data_size}, the total trigger rate is expected to be approximately \stcftrigrate, which produces a total data rate of approximately
\stcftotdata\ when the average event size is \stcfeventsize.
Furthermore, the design of the DAQ architecture should consider other factors, such as the budget, the development time, the difficulty of development, the scalability, and the maintenance cost in hardware, software and human resources.

\subsubsection{Conceptual Hardware Architecture of the DAQ System}
\quad\\
The conceptual DAQ system is composed of two layers. One is the FPGA-based processing layer (FPGA layer, FL), where the FEE data reading, preprocessing and merging are performed in real time with low time latency by the FPGA. The other part is the CPU/GPU-based processing layer (software layer, SL), where the software part of the DAQ, such as event building, data compressing/information extracting, high-level trigger~(HLT) computing, is running,
The FPGA layer is described in Fig.~\ref{fig:daq_fl}:

\begin{figure*}[htb]
	\centering
    \includegraphics[width=0.9\linewidth]{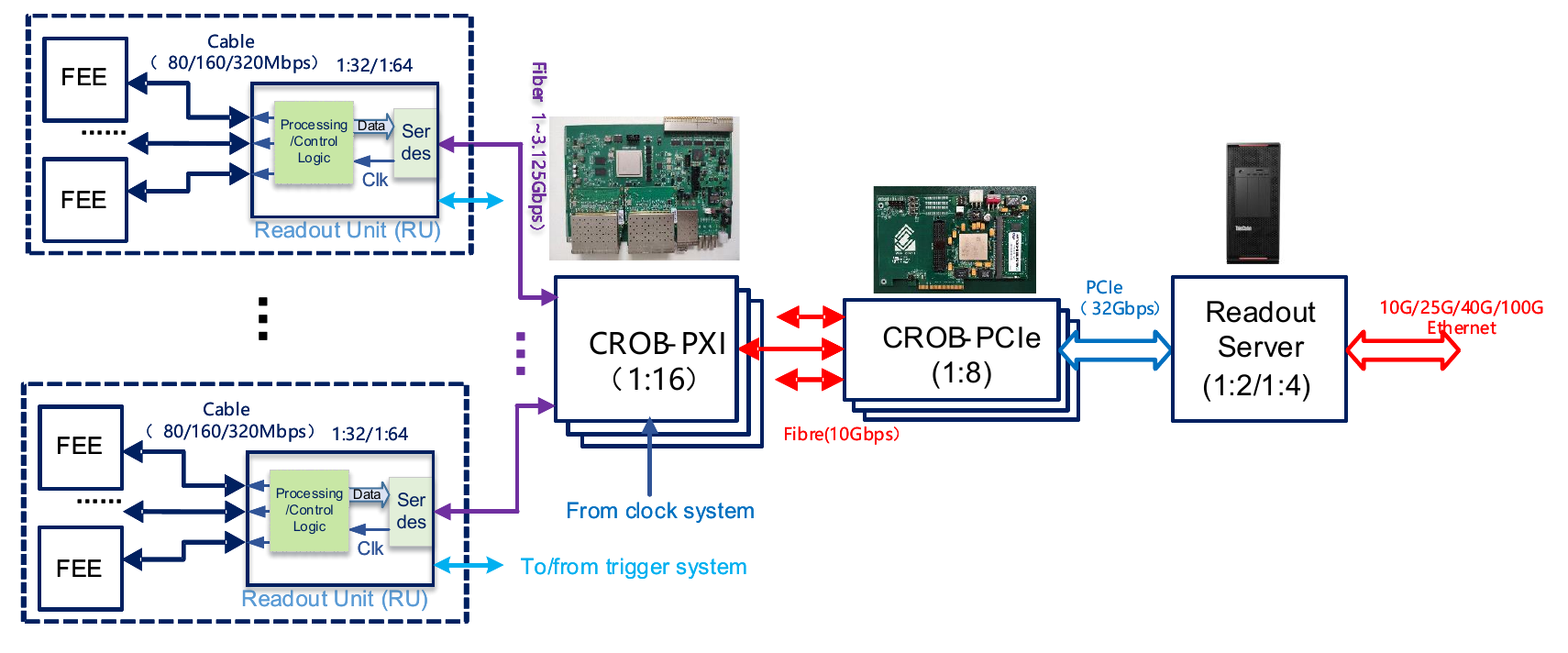}
\vspace{0cm}
\caption{Conceptual hardware architecture (FL).}
    \label{fig:daq_fl}
\end{figure*}

As shown in Fig.~\ref{fig:daq_fl}, a data merging stage named readout units (RUs) %Please note that this abbreviation has already been defined as ``readout units". --By J.Yang: All the ``Read Unit" in the figures should be replaced by the ``Readout Unit"
is used to collect the data from the FEE, extract the trigger information, send the trigger information to the trigger system (TS), receive the trigger signals (from the trigger system) and the commands (from the DAQ system), and forward them to the FEE. The RUs are connected to the FEE through cables, such as twisted pairs or coaxial cables, at a typical data rate of 80/160/320~Mbps, through which a typical merging ratio of 1:32 or 1:64 can be provided.

After the RUs, two other data merging stages (CROB-PXI and CROB-PCIe) are used to merge the data from the RUs, transmit the commands, and distribute the clock and trigger signal (optional) to the RUs and FEE. By using GTX/GTH transceivers in the FPGA, each fiber link in CROB-PXI and CROB-PCIe can transfer data at a rate up to 10~Gb/s, while each PCIe Gen2$\times$8 interface can provide a theoretical transmission capability of 32~Gb/s. Additionally, the FPGAs on CROB-PXI and CROB-PCIe can perform real-time data preprocessing and data merging in the pipeline, so the entire FL can provide sufficient transfer bandwidth, processing ability, and scalability for the FEE of different detectors.

\begin{figure*}[htb]
	\centering
    \includegraphics[width=0.9\linewidth]{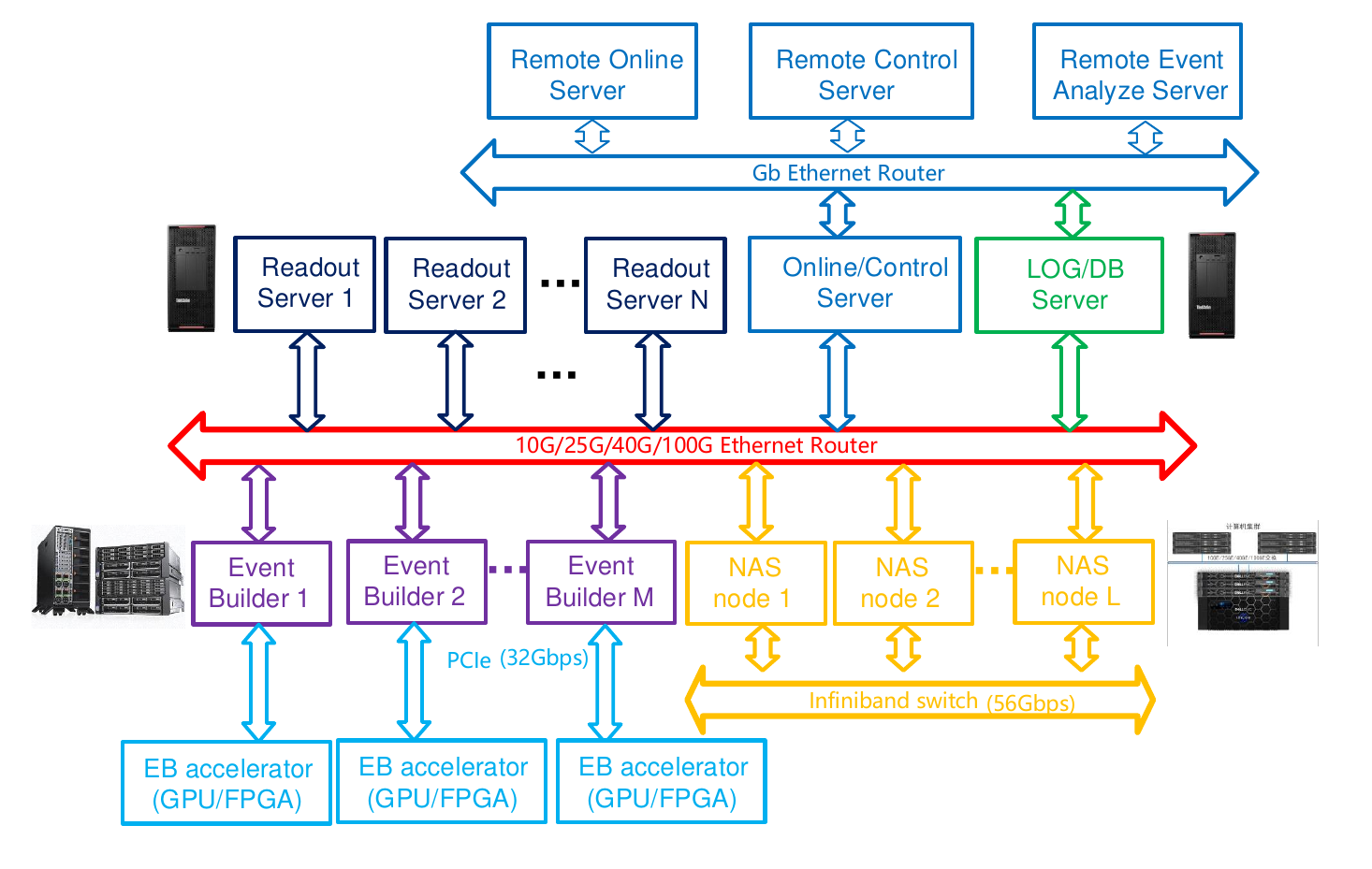}
\vspace{0cm}
\caption{Conceptual hardware architecture (SL).}
    \label{fig:daq_sl}
\end{figure*}

The SL is mainly composed of workstations or servers, and the conceptual hardware architecture of the SL is shown in Fig.~\ref{fig:daq_sl}.

There are 3 main types of server nodes in the SL: the readout servers (ROS), the event builders (EBs), and the NAS nodes. All the nodes are interconnected with 10G/25G/40G Ethernet. The ROSs are equipped with CROB-PCIe boards, acting as the entrance of the uplink streams (such as the data stream, the status stream) and the output port of the downlink streams (such as the command stream). The EBs perform high-performance computing (HPC) in terms of data compression, information extraction, and event building by integrating all kinds of computing resources, including CPUs, GPUs, and FPGA hardware accelerators. The NAS provides the necessary storage capacity and bandwidth for the STCF experiment.

In addition to the server nodes mentioned above, there are some other servers in the SL. The LOG/DB server (LOG) provides the system database for recording the system log information, including the system topology, running status, working parameters, and warning and error messages. The online/control server (OLC) is the command hub of the whole system and can receive, process, and forward the commands sent by the remote control server (RCS). It is also an online information controller that can merge the status stream, extract the online information for each detector system, and decimate the event data. All of this online information is processed in the remote online server (ROL) and the remote event analysis server (REA).

\subsubsection{Conceptual Architecture of the Firmware and Software}
\label{sec:daq_fl_sl}
\begin{figure*}[htb]
	\centering
    \includegraphics[width=0.7\linewidth]{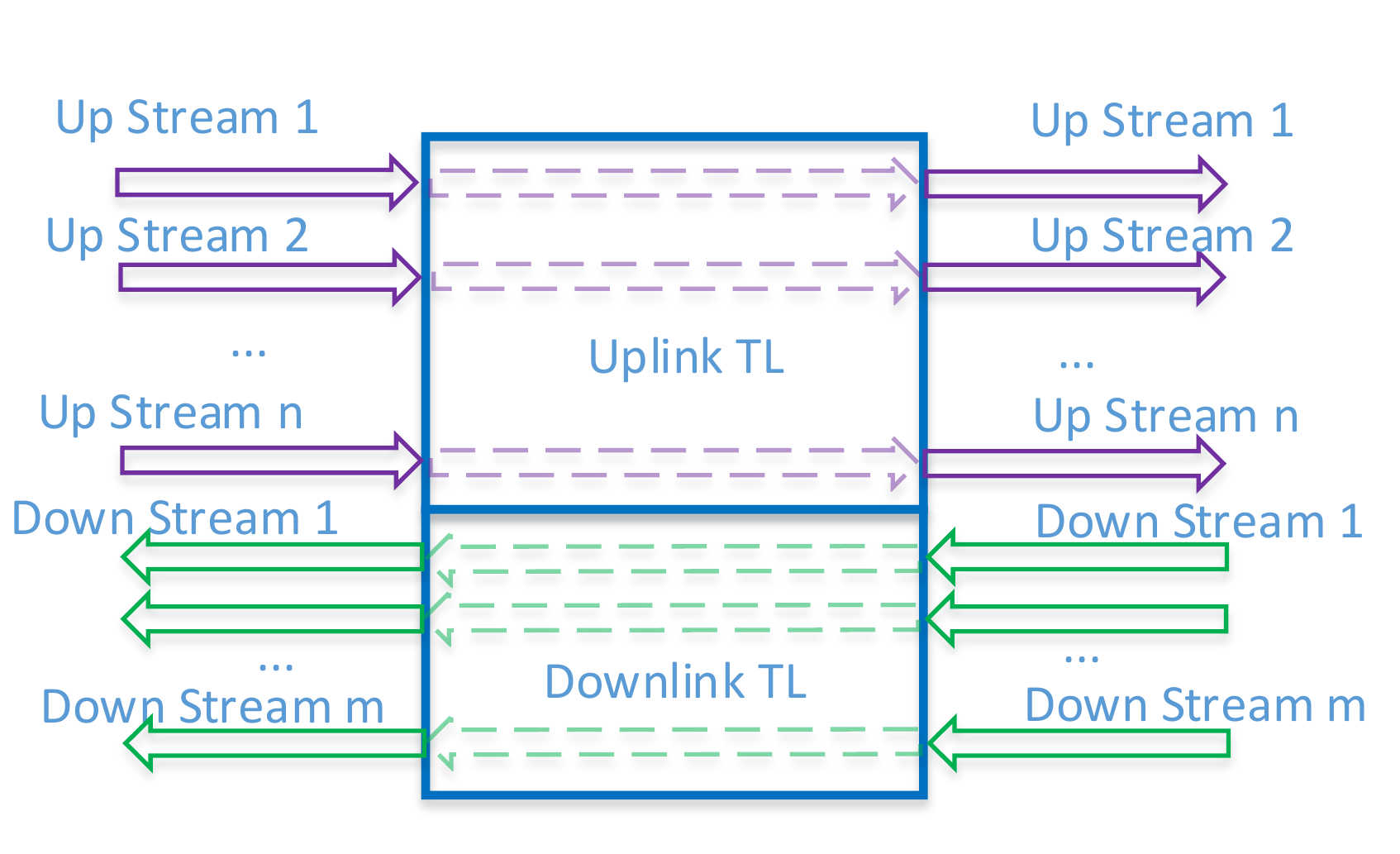}
\vspace{0cm}
\caption{Multiple point-to-point (MPP) model.}
    \label{fig:daq_mpp}
\end{figure*}

\quad\\
The firmware and the software of the STCF conceptual DAQ system are based on a generic stream-processing DAQ framework named the D-Matrix. The design philosophies of the D-Matrix are as follows:
\begin{itemize}
\item The workflows of a DAQ system are abstracted into several independent streams, such as the data stream, the command stream, the command feedback stream, the status stream, the online information stream, the log stream, and the emergency message stream, etc.
\item Each stream has its own processing map constructed by the cascading of some standard or custom stream processing nodes named ``logic nodes" (LNs). LNs can be FPGA firmware modules or software processes, work in the application layer of the OSI model, and can be distributed to ``physical nodes (PNs)", which means the hardware entities where the LNs run, such as FPGAs or servers. The distribution of an LN to a PN is based on the factors including the LN type, the resources and the processing ability that PN can provide and the real connection map among PNs.
With a feasible function abstract, most of the LNs can be reused in different streams with different working parameters. With the standard interface and the standard data structure definition, the LNs with matched parameters can be connected freely.
\item For the transport layer interfaces between two PNs, the ``multiple point-to-point" (MPP) model is used, which means that the interfaces based on different mediums and different protocols are encapsulated to a unified model, as shown in Fig.~\ref{fig:daq_mpp}. Under this model, each stream crossing the PNs has a channel ``independent" from other streams, and the interface between PNs seems to be ``transparent" for each stream. In this way, the connection of the LNs for each stream is independent of the transport layer, the link layer and the physical layer of the hardware.
\end{itemize}

Take the data stream as an example.

\begin{figure*}[htb]
	\centering
    \includegraphics[width=0.9\linewidth]{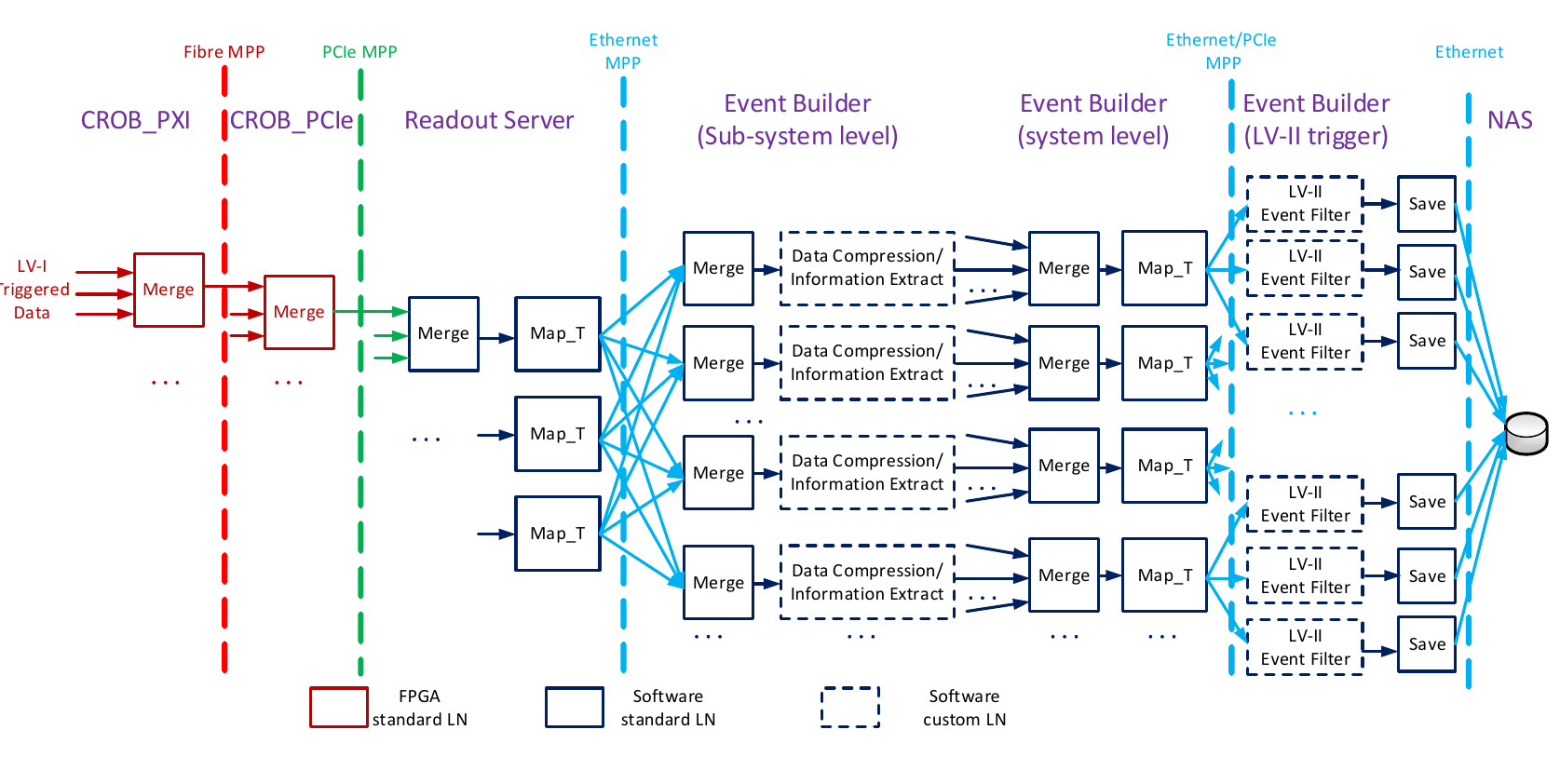}
	\vspace{0cm}
	\caption{Processing Map for Data stream}
    \label{fig:daq_proc_map_data}
\end{figure*}

One of the core functionalities of a DAQ system is the event building. In D-Matrix, this functionality is implemented by the cascading of the “Merge” nodes, which can merge the data fragments for some continuous small spatio-temporal ranges into an intact data frame. The process of the event building can be divided into multiply steps, shown as Figure~\ref{fig:daq_proc_map_data}. In the CROB-PXI and CROB-PCIe, a firmware “Merge” node is used severally to aggregate the data with same trigger number from multiple RUs. In the Readout Server, the data from multiply CROB-PCIe boards can be packaged in a software “Merge” node and then be distributed to diverse EBs according to the trigger number by a “Map-T” node (mapping the stream to multiple output according to the “Time index” of the frame). All the data sections with same trigger number will be sent to the same EB, where a sub-system level event building (building an event section with all the data from one detector system) and a system level event building (generating the final event file collected all the data with the same trigger number) are accomplished. Among these event-building steps, two custom LNs may be inserted to perform the possible data compressing or information extracting after the sub-system level event building, or to do the Level-II trigger computing through a “Filter” way when needed. At the same time, optional “Map-T” nodes may be inserted to promote the depth of parallelism. Benefit from the flexible stream-processing architecture, the software LNs can also be replaced by the GPU version or the FPGA version to accelerate the event building computing.

Following the D-Matrix framework, the processing function of the STCF DAQ can be adjusted flexibly by reusing the standard LNs and inserting the custom LNs, which greatly reduces the difficulty of development and maintenance.

\subsection{Event Start Time ($T_0$) Determination}
\label{sec:daq_t0}

The time of an $e^{+}e^{-}$ collision corresponding to an event that fired a given trigger, defined as the event start time $T_0$, will be determined offline.
$T_0$ is essential for event reconstruction and particle identification.
For example, the measurement of time of flight~(TOF) of a charged particle by the DTOF detector and that of the drift time by the MDC both require a precise determination of $T_0$.
The precise  $T_0$ is actually provided by the accelerator beam arrival monitor with a time resolution around $100$~fs. The bunch spacing of the STCF accelerator is 4~ns in the preliminary conceptual design of the accelerator, while the the trigger clock is designed to have a period of 24~ns.. Thus there will be 6 or 7 collisions within one trigger clock, and the task of $T_{0}$ determination is to resolve the bunching crossings within one trigger clock, which requires at least 800~ps time resolution.

In the current conceptual design of the STCF detector, no dedicated $T_0$ detectors are considered.
However, a precise determination of $T_{0}$ would still be viable by exploiting the timing capability (including offline timing capability) of the subdetectors.
The MDC is expected to provide a time resolution of approximately 1.0~ns for a charged track traversing most of the wire layers, which gives a $T_{0}$ precision better than 1 ns when multiple tracks are present in an event~\cite{MaXiang_2008}. 
The DTOF detector has been demonstrated to be able to distinguish bunch crossings correctly with an efficiency of $> 99$\% (Section~\ref{sec:dtof_t0}).
However, the DTOF detector has a momentum threshold for Cherenkov production that prevents it from determining $T_{0}$ with low-momentum charged particles.  
The EMC detector is capable of providing a 300 ps time resolution for each through-going charged track  as well as a photon with energy above 100 MeV (as discussed in Section~\ref{sec:emc_perf}). Thus, the EMC has the potential to cover the whole $T_{0}$ determination task alone. 
More investigation and studies are needed to explore the full potential of the STCF detector in its current design in terms of determining $T_{0}$.

\FloatBarrier

\clearpage
\newpage

\newpage
\clearpage

\section{Offline Software }

\subsection{Introduction }

The offline software is extensively used to design and optimize a detector, 
to generate large amounts of MC data, to help in developing reconstruction algorithms and analysis procedures, to understand and demonstrate 
that a conceptual detector satisfies the goals of the STCF experiment at the experiment investigation stage as well as to support physics analysis at the experiment operation stage.

According to the design of the STCF,  the luminosity will be above $\stcflum$, 100 times higher than that of BESIII, and the total trigger rate as well as the data rate  are expected to be approximately $\stcftrigrate$ and  $\stcftotdata$ respectively, which are also much larger than those of the BESIII. Therefore, an Offline Software System of Super Tau-Charm Facility~(OSCAR) is designed and developed to provide the unified computing environment and platform to facilitate design of the STCF detector, conduct detector performance study as well as physics potential study. This chapter is going to talk about strategies and possible technologies to be used for development of STCF offline software system.

\subsection{Architecture Design }

\begin{figure}[htbp]
\begin{center}
\includegraphics[width=0.6\textwidth]{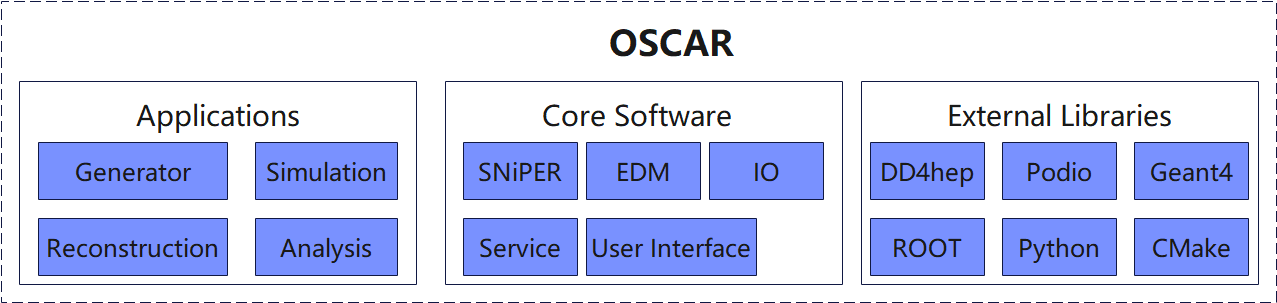}
\caption{Overview of the STCF offline software system}
\label{fig:oscar}
\end{center}
\end{figure}

Fig.~\ref{fig:oscar} shows the architecture of the whole offline software system, which consists of three components: External Libraries, Core Software and Applications. The  External Libraries include frequently used third-party software and tools, such as DD4hep, PODIO, ROOT, {\sc Geant4} and so on. The Core Software provides the common functionalities for all data processing and MC production, such as event data model, event data management, data processing control, data input and output, stable interfaces between different components, job configuration, user interfaces and the tools for the compiling, building and deployment of the whole software system. The Applications include the components specific to the STCF experiment, including extensions to SNiPER, Generator, Simulation, Calibration, Reconstruction and Analysis.

The offline software system is designed to meet the various requirements of data processing.
Based on our experience, Linux OS and GNU compilers are the first choice for the STCF software platform.
However, other popular OSs (such as Ubuntu, MacOS, ...) development environments should be considered for the sake of compatibility.
CMT and CMAKE, which can calculate package dependencies and generate makefiles,
are used for the software configuration and management.
An automated installation tool can ease the deployment of the STCF software and is also helpful for daily compilation and testing.
Users can concentrate on the implementation of application software, free from the interference of different development environments.

Currently, mixed programming with multiple languages is practical,
therefore, we choose different technologies for different parts of our software.
The main features of the application are implemented via C++ to guarantee efficiency.
User interfaces are provided in Python for additional flexibility.
Boost Python, as a widely used library, is a good choice for integrating C++ and Python.
If it is included properly at the beginning of the system design,
most users will be able to enjoy the benefits of mixed programming without knowing the details of Boost Python.

\subsection{Core Software}

\subsubsection{SNiPER Framework}

SNiPER ~\cite{ref_sniper} is the foundation of the STCF software system and determines the performance, flexibility and extensibility of the system. It provides common functionalities for offline data processing and the standard interfaces for developing different applications via \emph{Task, Algorithm, Service and Tool}. An algorithm provides a specific procedure for data processing. A sequence of algorithms forms the whole data processing chain. A service provides certain useful features, such as access to the detector geometry or database, which are necessary during data processing. Both algorithms and services are plugged in and executed dynamically, and they can be flexibly selected and combined to perform certain data processing tasks. All applications in terms of algorithms request event data from a Data Store and push new event data back to the Data Store, finally all event data in the Data Store will be automatically written into files via the File Output System. Meanwhile, The File Input System is responsible for reading the event data from the output files and placing them into the Data Store for downstream processing. SNiPER also provides many frequently used functions, such as the logging mechanism, particle property lookup, system resource loading, database access and histogram booking, etc.

 \subsubsection{Event Data Model}
 
 The event data model (EDM) serves as the central part of the whole offline software system; it defines the event information at the different data processing stages and builds the inter-event and intra-event correlations, which are very useful for optimization of reconstruction algorithms, event selection criteria and physics analysis. Two options of EDM are investigated, one is based on ROOT TObject and implemented with an XML Object Description (XOD) toolkit~\cite{XOD}, the other is based on PODIO~\cite{podio}, which is a new merging tool developed by the future collider experiments and supposed to have good supports on the concurrency of EDM. The EDMs for detector simulation and reconstruction have been implemented in both XOD and PODIO, their performances are under investigation by running detector simulation and reconstruction algorithms.

\subsubsection{Event Data Management }
Event data management system is designed and implemented with the Data Store which provides a common place for data sharing between applications via standard interfaces to get event objects from the Data Store, and to push new event objects into the Data Store.The event objects in the Data Store can be written to the ROOT files and read back from the ROOT files later. The event data management system based on PODIO consists of three components: DataHandle, PodioDataSvc and PodioSvc. PodioDataSvc is used to manage event data objects, while PodioSvc is used to read/write event data from/into ROOT files. User code is not involved with any data management details, except that DataHandle is used to register/retrieve event data to/from the Data Store.

\subsubsection{Parallel Computing}
Based on the estimation above, the STCF is expected to deliver more than 1~$\invab$ of data per year. Assuming a lifetime of 10 years for the STCF, a total of 10~${\invab}$ of data is expected. Therefore, 
it is vital to adopt parallel computing techniques to make use of all possible computing resources and accelerate the offline data processing and physics analysis. At the moment, multi-threaded programming has been supported and implemented in SNiPER by integration of multi-task with the TBB technology,  which can extract the capacity of multi-core CPUs to speed up a single job significantly. Meanwhile, a server-client system will be deployed to run detector simulation and produce massive MC data on both local computing clusters and the distributed computing infrastructure.

\subsection{Detector Simulation}

\subsubsection{Physics Generator}

The goals of the STCF are to study  $\tau$-charm physics, to precisely test the standard model and to search for new physics.
MC simulations are used to study physics potentials, to determine detection efficiencies and to study backgrounds. To meet these challenges and to reach unprecedented precision, a comprehensive MC study is necessary. Event generators with high quality and precision are essential for performing a reliable MC simulation and removing systematic uncertainty as much as possible.
For high-precision measurements, we expect MC generators to simulate the processes under study as realistically as possible.
Hence, the generators with only kinematic information (e.g., a pure phase space) do not meet this requirement.
Recently, high-precision generators for QED processes have been developed based on the Yennie-Frautchi-Suura exponentiation technique to describe the process $e^+e^-\to f\bar f$ ($f$: fermion).
The official precision tags of these generators are approximately 1\% or less, e.g., KKMC and BHLUMI.
Generators with dynamical information for hadron decays have also been developed,
such as EvtGen for BaBar and CLEO collaborations to study B physics.
All these generators have been integrated within OSCAR and provide us with a large room to make a choice among them to simulate $\tau$-charm physics processes.
Furthermore, one generator framework is developed to provide the uniform format and standardize interface for all generators. The application developers or physicists can easily and quickly call generators in a coherent manner without knowing much details of generators.

\subsubsection{Detector Geometry }
\label{section:geo}

Detector simulation requires a precision description of the detector, including materials, geometry and structure, and this description is used not only for detector simulation but also for reconstruction and visualization. The STCF detector is composed of the five sub-detectors and two auxiliary devices. There are more than  200,000 volumes in total constructing the whole spectrometer, so it is a very complicated task to design the STCF detector and requires collaboration between different working groups, thus, a geometry management system (GMS)~\cite{GMS_dete_geo_mana} is designed to provide a consistent detector-geometry description for different offline applications. A new Detector Description Toolkit, {\sc DD4Hep}~\cite{GMS_dd4hep}, is adopted to describe STCF detector geometry for GMS, with all geometric parameters stored in the compact files with eXtensible Markup Language (XML)~\cite{GMS_xml}, which is more human readable and can be edited by any text editor. 

\begin{figure}[htbp]
\begin{center}
\includegraphics[scale=0.2]
{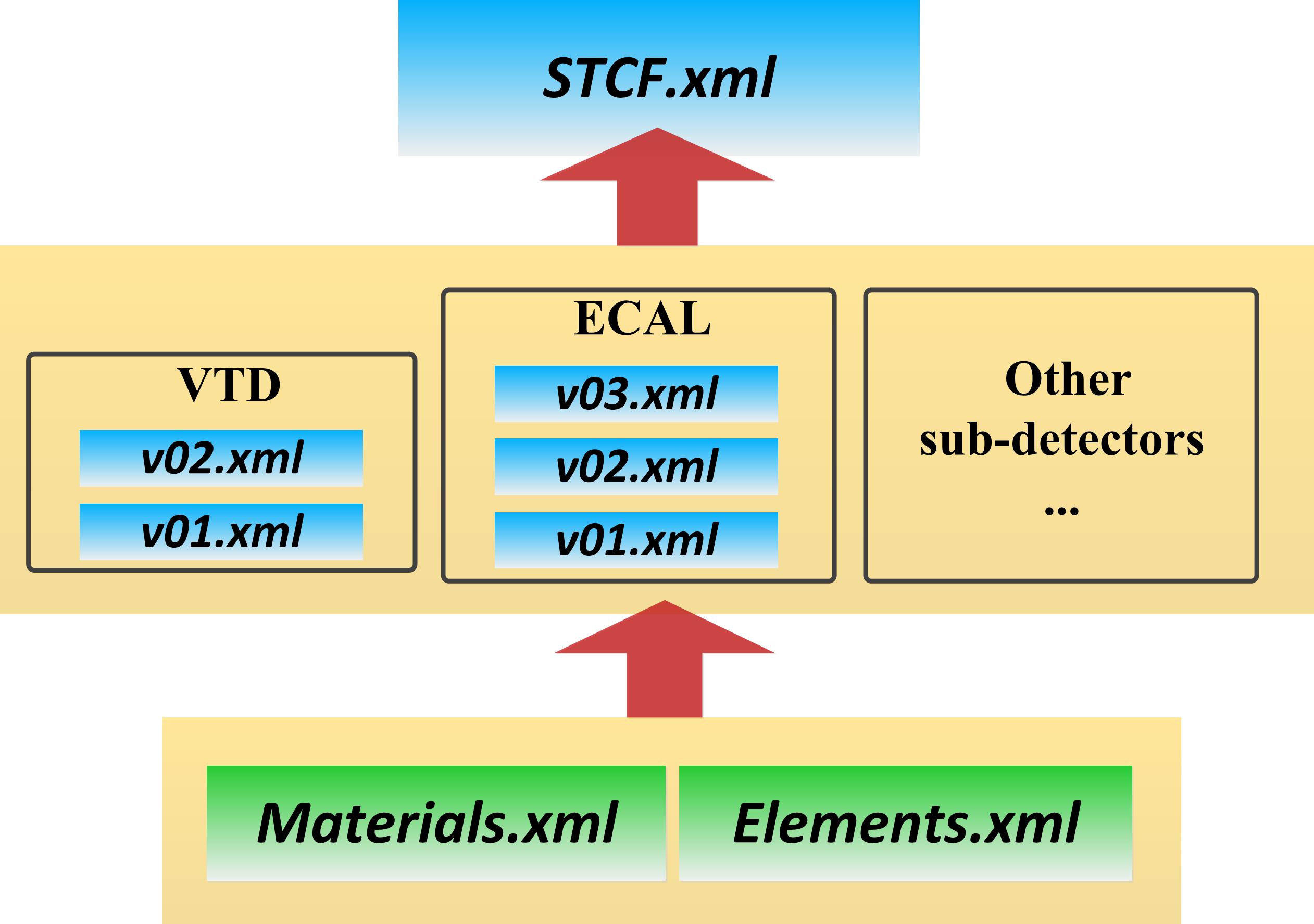}
\caption{Structure of the geometry parameters repository. The library of elements and materials is shared
by all sub-detectors. Different sub-detector designs are separately managed in different XML files; if a
sub-detector has more than one designs, the parameters for different designs are stored in different XML files
with version numbers. A mother XML file can include several daughter XML files.}
\label{fig:detectorrepository}
\end{center}
\end{figure}

\begin{figure}[htbp]
\begin{center}
\includegraphics[scale=0.15]{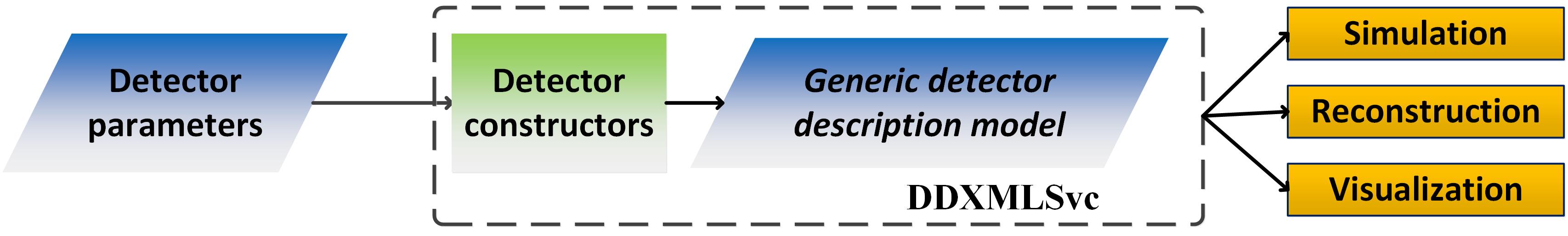}
\caption{Workflow of the DDXMLSvc. The geometry parameters in the repository
are parsed by detector constructors and converted to formats for simulation, reconstruction and visualization.}
\label{fig:detctorconstruction}
\end{center}
\end{figure}

\begin{figure}[htbp]
	\centering
	{
		\begin{minipage}{0.4\linewidth}
		\includegraphics[scale=0.25]{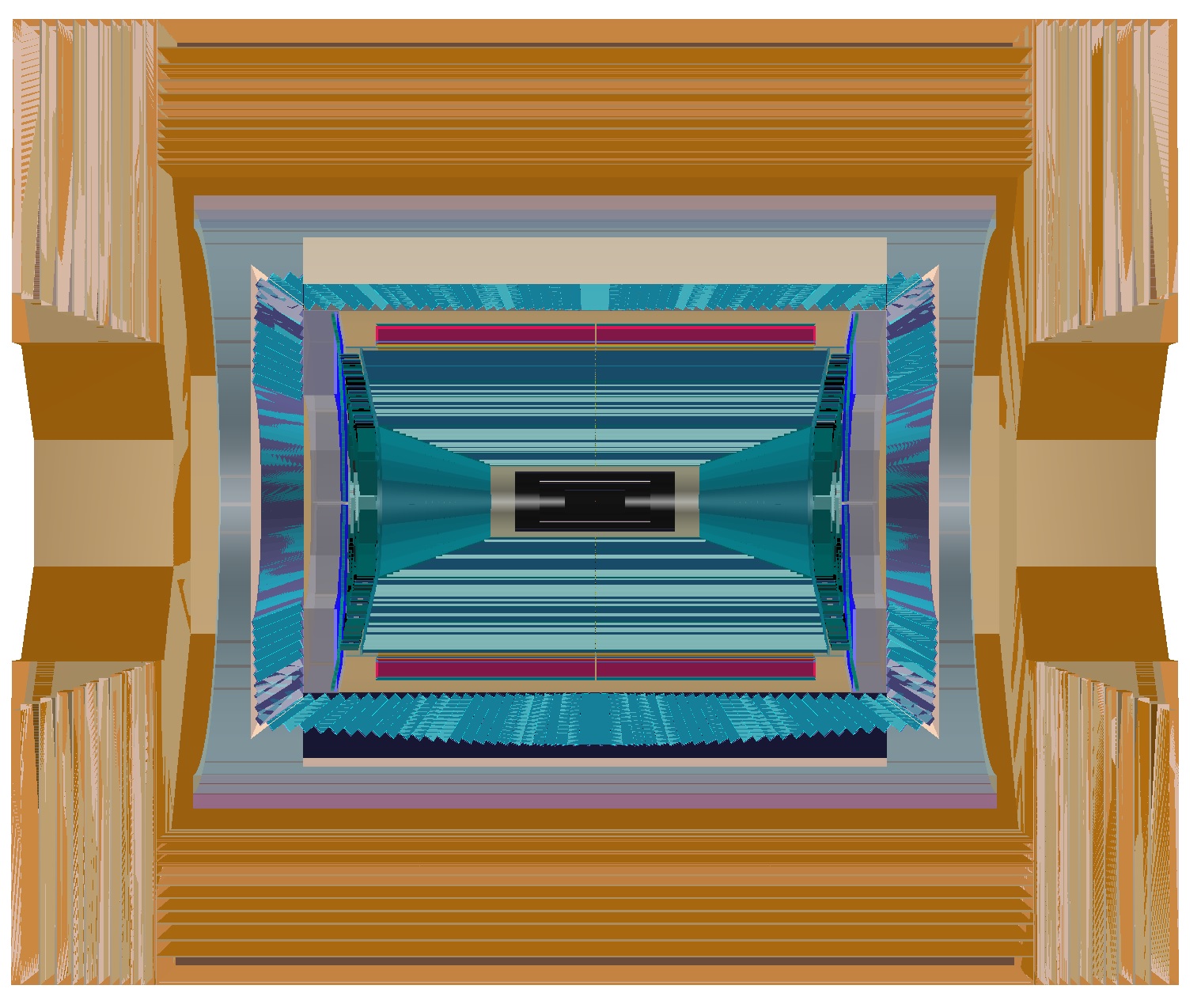}
		\end{minipage}
	}
	{
		\begin{minipage}{0.4\linewidth}
		\includegraphics[scale=0.25]{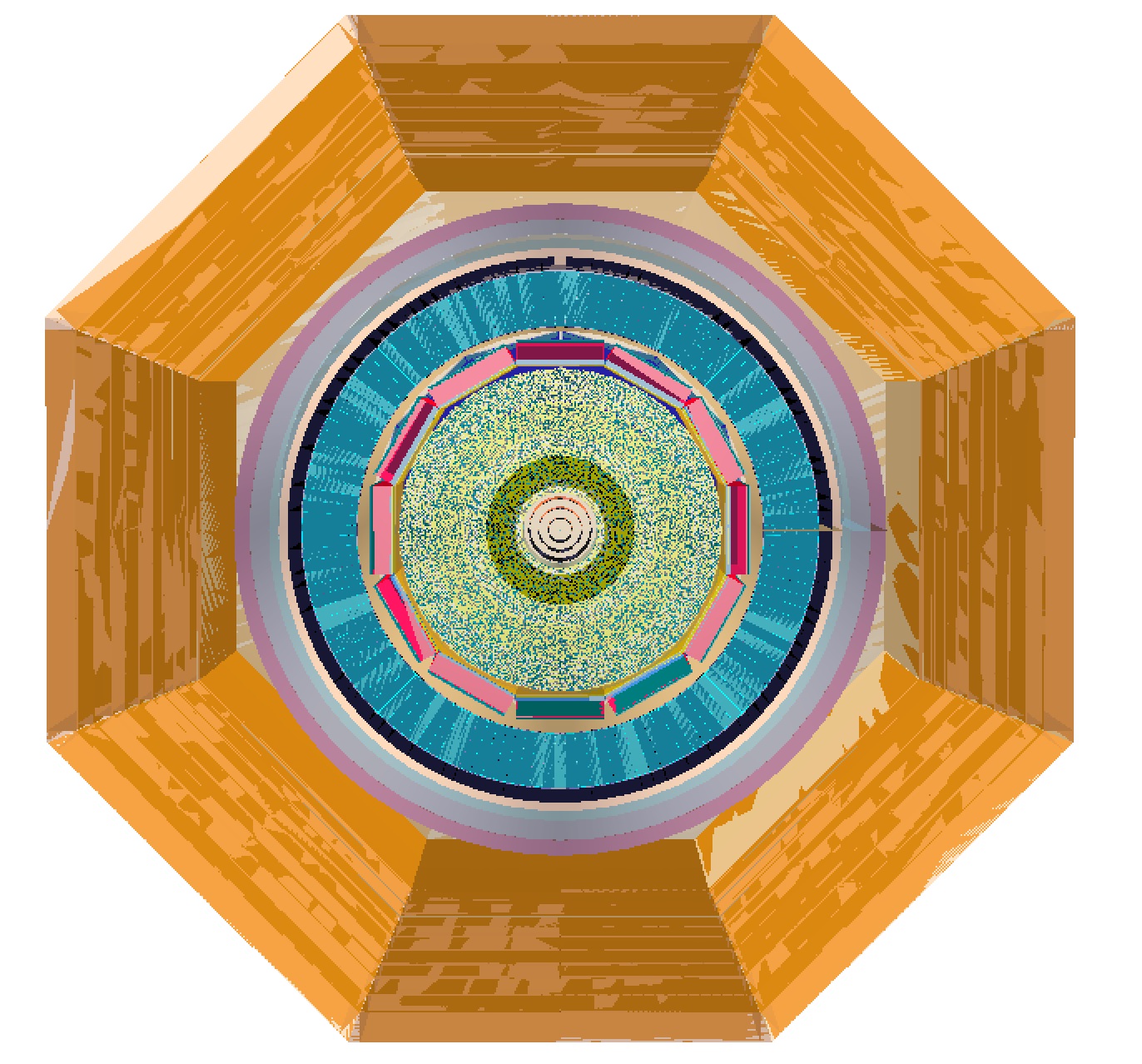}
		\end{minipage}
	}

	\caption{Full detector described by the GMS and printed using the default 3D visualization plugin of DD4Hep. 2D sections are viewed from different directions: (left) Z-R view and (right) X-Y view}
	\label{fig:fulldetector}
\end{figure}

To hold and manage geometric parameters, a customized repository with hierarchical structure is designed in the GMS, which is shown in  Fig.~\ref{fig:detectorrepository}. The feature of XML allows a flexible configuration for the assembly of different sub-detectors. As shown in Fig.~\ref{fig:detctorconstruction}, OSCAR provides DDXMLSvc to transform XML descriptions to the offline applications, including simulation, reconstruction and visualization within two steps. First, specialized C++ code fragments, called detector constructors, read and parse the geometry parameters from the XML-based repository and construct the generic detector description model based on the ROOT geometry modeller (TGeo) in memory. Second, the DDXMLSvc converts TGeo to  geometry format of simulation, reconstruction or visualization. Based on the GMS, the whole STCF detector has been build, as shown in Fig.~\ref{fig:fulldetector}.

\subsubsection{Standalone Detector Simulation}

First, dedicated standalone simulation packages are implemented based on  {\sc Geant4}~\cite{Geant4_ref} for the purpose of R\&D for each sub-detector. This provides guidance for choosing the best sub-detector option that can fulfill the desired physics goals. Based on these packages, we conduct very detailed studies on the performance of these sub-detectors, including the energy resolution, momentum resolution, tracking efficiency and PID.

\subsubsection{Full Detector Simulation }

\begin{figure}[htbp]
\begin{center}
\includegraphics[width=0.5\textwidth]{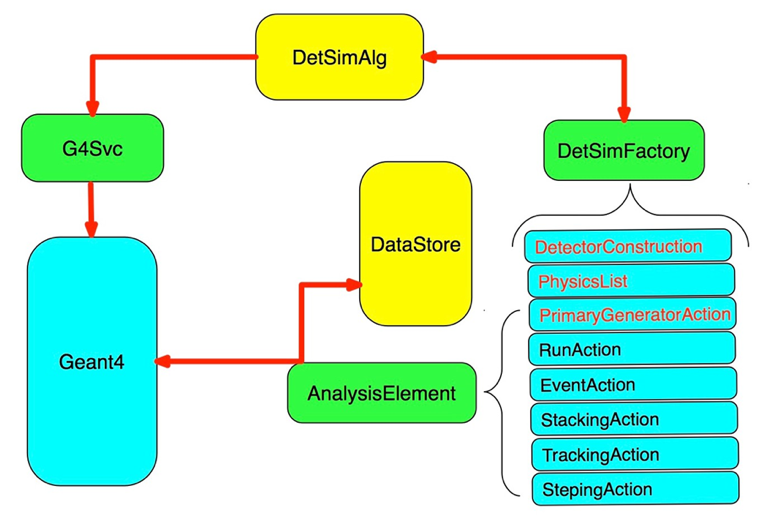}
\caption{Overview of the full detector simulation framework for STCF}
\label{fig:sim}
\end{center}
\end{figure}

To study the performance of the whole detector, the full detector simulation framework is developed to serve as a bridge between {\sc Geant4} and OSCAR.
It consists of the integration of {\sc Geant4} with SNiPER, a configurable user interface, geometry management and modularized user actions.
With this extra layer as a middleware, detector simulation becomes a seamless component of OSCAR. The Fig.~\ref{fig:sim} shows the architecture of the full detector simulation framework, the algorithm named {\sc DetSimAlg} is designed to invoke {\sc Geant4} event loop in OSCAR, and it invokes the service named {\sc G4SvcRunManager}, which is derived from {\sc G4RunManager}, to initialize simulation and start tracking.
To decouple run manager and user code, such as detector construction and physics list, the factory class named {\sc DetSimFactory} is designed and implemented to construct user actions and pass them to {\sc G4SvcRunManager}.
Now we have setup the full detector simulation chain which includes detector geometry description, physics processes, hit recording and user interfaces. The detector geometry is described in Section~\ref{section:geo}. At the moment, a uniform magnetic field of 1 Tesla is currently defined in the detector geometry  xml file, and a realistic magnetic field map will be implemented later.
The physics processes include standard electromagnetic, ionization, multiple scattering, bremsstrahlung and optical photon processes.
The hit information and particle history information, including the initial primary particles, energy deposition, direction and position of the hit as well as the relationship between track and hits, for each sub-detector has been defined in the simulation EDM, and all these information can be saved into the ROOT files and used for reconstruction. 

\subsection{Reconstruction}

The reconstruction of track, photon, muon and particle identification plays very important roles for achieving STCF physics goals with unprecedented precision. At the moment, the reconstruction methods and algorithms are under-study with the simulated hit information.The reconstruction chain starts with the track reconstruction, followed by the particle flow interpretation of tracks and calorimeter hits and finally the reconstruction of compound physics objects such as converted photons.

\subsubsection{Track Reconstruction}
Track reconstruction consists of two main procedures, track finding and track fitting.
The purpose of track finding is to find hits produced by the same particle.
The 2D track candidates in r-$\phi$ plane are searched first,
then $z$ direction information is used to infer 3D track.
The commonly used methods are conformal transformation and Hough transformation.
Other methods, such as hypothesis test method and machine learning technique, are also under investigation.
The  track fitting is performed by the Deterministic Annealing Filter (DAF) method, which is an extension of Kalman Filter and can reduce the influence of bad points on reconstruction
by setting weights of measurement points and updating iteratively.

\subsubsection{PID Reconstruction  }
PID systems (including DTOF and RICH detectors) are used to identify charged particles based on the information of  the Cherenkov light  which is generated when tracks pass through the DTOF and RICH detectors. For DTOF, the reconstruction algorithms mainly focus on the reconstruction of the Cherenkov angle, and the flight time of the charged particle can also be used to separate different particles. For RICH, the Cherenkov photons are collected by anode pads of the detector and the log-likelihood values in all pads for a certain particle hypothesis, including $\pi, K, p, e, \mu$, are calculated, respectively. The reconstruction algorithm of RICH calculates the sum of the log-likelihood difference between two particle hypotheses and the one with larger log-likelihood difference is chosen as the optimal candidate.

\subsubsection{EMC Reconstruction}

EMC is used to detect the energy deposition of neutral particles (mainly photons) and charged particles, and also to measure the time information. The reconstruction process is divided into several steps, including cluster searching, seed finding, and cluster splitting into showers. While reconstructing the energy of the shower, the position of photons can be reconstructed by the barycenter method with logarithmic weight. On this basis, the timing algorithm included waveform simulation, leading edge timing or waveform fitting is introduced into the reconstruction framework.

\subsubsection{MUD Reconstruction}
MUD is used to detect and separate muons from other charged particles as well as identify neutral particles. The reconstruction algorithm includes three steps: firstly, the information from MDC, EMC and MUD is combined together to judge whether the gathering hits are generated by a charged track or a neutral particle; secondly, the direction from MDC extrapolation or EMC is used to do a pre-screening of the hits, then these hits are fitted and some hits with large deviations are removed; finally, after reconstruction of a track, BDT is adopted to identify which particle does the track belong to.

\subsection{Validation System}

Software validation at different levels is vital in the long lifecycle of OSCAR.To make sure its quality and performance, an automated software validation system is developed based on Python. It includes a unit test system and a data production system, covering the software validation from code quality monitoring to physics validation. The validation system supports defining and profiling unit test cases, results validation based statistical methods, as well as automated data production for high level physical validation.

\subsection{Summary}

OSCAR is developed based on SNiPER framework and several state-of-art tools, including DD4hep, PODIO and TBB, and have been adopted to well meet the requirements and challenges from the large amount of data of the STCF. Currently, the main functions of the core software have been implemented, including the event data model, event data management, IO system and interfaces between different components, and the full detector simulation chain has been set up to optimize the detector options and study the detector performance. Development of event reconstruction methods and algorithms is under going , including reconstruction of the charged tracks, electromagnetic showers and particle identifications for further physics analysis.

\newpage
\clearpage
\newpage
\chapter{Physics Performance}
\label{chap_phyper}

\section{Fast Simulation}

To facilitate physics potential studies, a fast simulation package~\cite{Fast_simu_ref2} has been developed and used to produce physics signals and background samples to investigate the physics potential capabilities of the STCF.
The basic idea of this fast simulation package is to model the detector responses, including the efficiencies,
resolutions, particle identification and other responses needed in data analysis, instead of simulating all details
and physical interactions with {\sc Geant4}. To more accurately simulate the detector's response, these model
shapes are extracted from the full simulation of the BESIII detector and parameterized based on empirical
formulas or extracted from histograms, which are defined with different momentum and polar angle (with
with respect to the beam direction) regions.

In this package, the modeling of responses in each subdetector is individually implemented for the charged
particles $e$, $\mu$, $\pi$, $K$ and $p$, as well as the neutral particles $\gamma$, $n$, $\bar{n}$ and $K_{L}^{0}$. The detector efficiency is simulated by a sampling according to its curves as a function of a two-dimensional variable,
{\it i.e.}, momentum versus $\cos\theta$, where $\theta$ is the polar angle of objects in the laboratory frame. For the
observables with measurement uncertainty, such as energy, momentum, space and time, the expected
value is the overlay of the detection resolution on top of the MC truth value, where the corresponding
detection resolution is extracted by a sampling according to the distribution. The reliability and stability
of the fast simulation package have been validated in terms of many aspects. At the object level, input and output
checks for all the observations of different types of particles with various input parameters have been
performed. In addition, full physics analyses for some interesting physical processes are performed in
in the fast simulation package by setting the same parameters as those of the BESIII detector and comparing the
results from BESIII physical programs, {\it e.g.} event selection efficiency and distribution of some physical
variables of interest. Good consistency is found in the above validation. In addition, an interface
is provided for users to flexibly adjust the responses to easily estimate the detector's performance. With
this fast simulation package, many physics analysis processes are ongoing, as introduced in the next
Section.

\section{Selected Physics Performance}
\begin{figure}[htbp]
\begin{center}
\begin{overpic}[width=15cm, height=5.cm, angle=0]{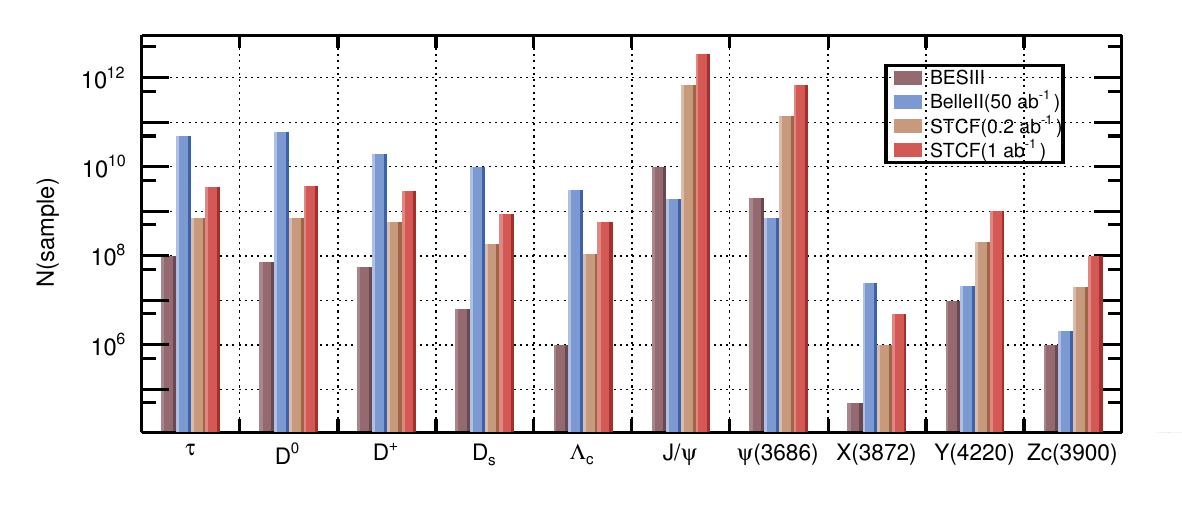}
\end{overpic}
\end{center}
\caption{Number of expected samples at STCF under 0.2~ab$^{-1}$ and 1~ab$^{-1}$ integrated luminosity, compared with current BESIII statistics and Belle II 50~ab$^{-1}$ expected. }
\label{fig1}
\end{figure}

As discussed in Chapter~\ref{CDR_phys}, the physics highlights of the STCF can be grouped into three categories:
QCD and hadronic physics, flavor physics and CP violation and the search for new physics. 
The statistics is a crucial factor for learning the properties of the exotic
particles. Figure~\ref{fig1} shows the number of expected samples
at STCF under 0.2~ab$^{-1}$ and 1~ab$^{-1}$ integrated luminosity. 
STCF can collect world leading statistics for the charmonium and
charmonium-like samples. Moreover, the clean environment of STCF will
provide an ideal platform for the tau physics and study of charmed hadrons.

\begin{figure}[htbp!]
\begin{center}
\begin{overpic}[width=15cm, height=6.5cm, angle=0]{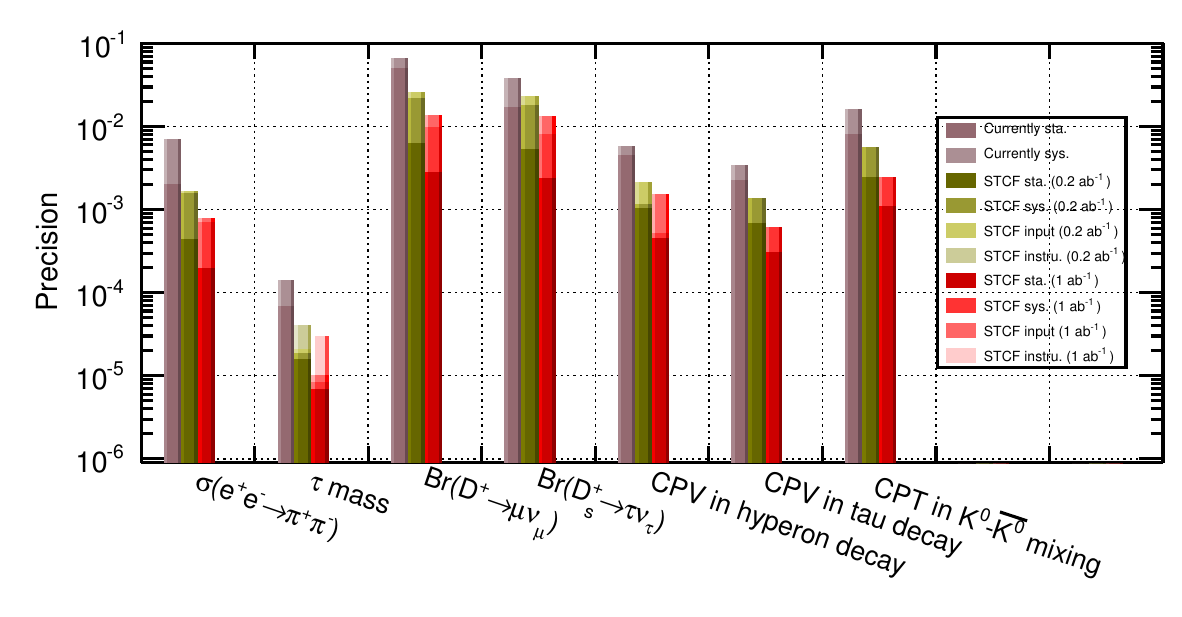}
\end{overpic}
\end{center}
\caption{Precision of various measurements to test SM, such as muon g-2, tau mass, CKM matrix and $CPV$, from current precision and STCF expected with 0.2~ab$^{-1}$ and 1~ab$^{-1}$ integrated luminosity. The uncertainties of STCF expected consider the sources from statistics (sta.), reducible systematic (sys.) such as tracking, PID, and other selection criteria, irreducible systematic from theoretic input and instrument effects such as beam energy and beam spread.  }
\label{fig2}
\end{figure}

\begin{figure}[htbp!]
\begin{center}
\begin{overpic}[width=17cm, height=6.5cm, angle=0]{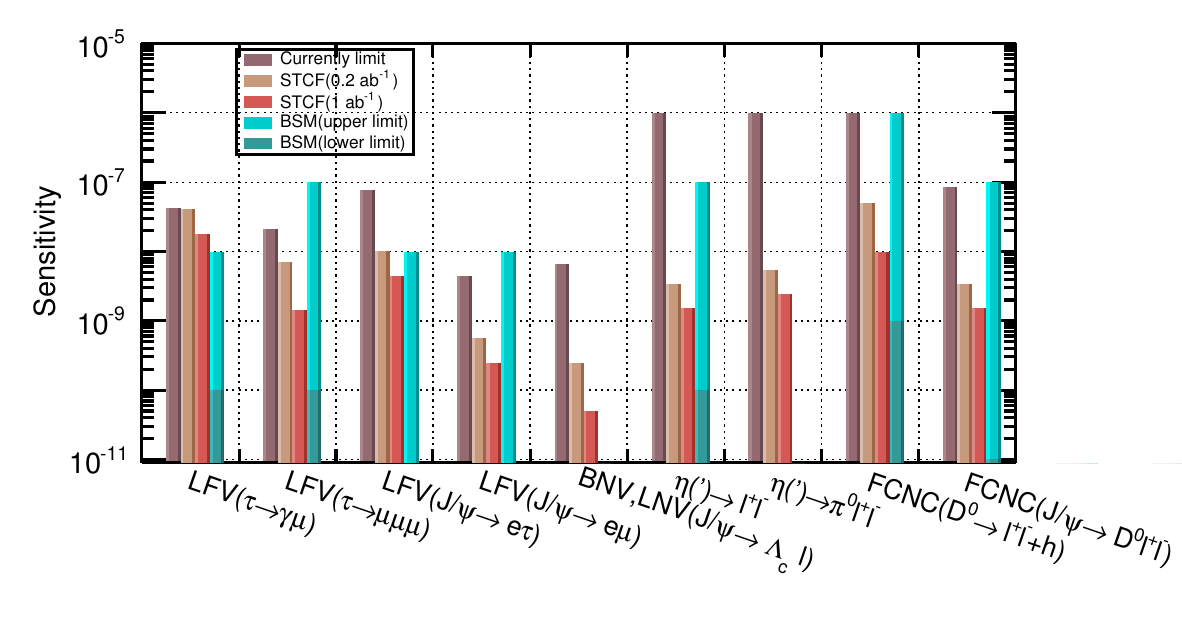}
\end{overpic}
\end{center}
\caption{Sensitivity of processes that are forbidden or rare in SM prediction, from current results and STCF expect with 0.2~ab$^{-1}$ and 1~ab$^{-1}$, and compared with predictions from theoretical models beyond the SM.}
\label{fig3}
\end{figure}

With the expected luminosity collected at STCF, the key parameters
from EW test and new physics probe are shown in Fig.~\ref{fig2} and Fig~\ref{fig3}. The statistical sensitivity for flavor and CP violation test can be significantly improved compared with current world best result. However, the systematic uncertainty will be an 
essential limitation of the precision by then. The discussion 
of systematic uncertainty is presented in Sec.~\ref{sec:sysunc}.
For the new physics probe, an increased statistics will no doubt help to test various models beyond SM. The sensitivity of various rare of forbidden decay can be improved with a magnitude factor of 2 to 3, and is lying in the range of beyond SM model predictions.

Table~\ref{phyper} lists the statistical sensitivities for several
benchmark processes and compared with current operating BESIII and
Belle II experiments.

\begin{table}[htbp!]
\caption{Summary of the statistical sensitivities for some benchmark physics processes, not inclusive yet.}
\label{phyper}
\footnotesize
\begin{center}
\begin{spacing}{1.3}
\begin{tabular}{cccc}
\hline
\hline
\vspace{0.2cm}
Observable                       &  BESIII (2020)~~~~~     & Belle II (50~ab$^{-1}$)~~~~~              &  STCF (1~ab$^{-1}$)~~~~~   \\           
\hline
{\it Charmonium(like) spectroscopy:}\\
Luminosity between 4-5~GeV        &     20~fb$^{-1}$          &      0.23~ab$^{-1}$                       &    1~ab$^{-1}$  \\            
\hline
{\it Collins fragmentation functions:}\\
Asymmetry in $e^{+}e^{-}\to KK+X$  &      0.3~\cite{BESIII:2015fyw}                 &        -                                  &  $<0.002$~\cite{collins_stcf} \\       
\hline 
{\it CP violations:} \\
$A_{cp}$ in hyperon              & 0.014~\cite{BESIII:2018cnd}          &  -                                  &      $0.00023$    \\
$A_{cp}$ in $\tau$                  &   -               &  $\mathcal{O}(10^{-3})/\sqrt{70}$~\cite{Chen:2020uxi}      &      0.0009~\cite{Sang:2020ksa}    \\
\hline
{\it Leptonic decays of $D(_{s})$:}  \\
$V_{cd}$                          &    0.03~\cite{BESIII:2019vhn}            &     -          & 0.0015     \\
$f_{D}$ 		                  &    0.03            &     -          & 0.0015  \\ 
$\frac{\mathcal{B}(D\to\tau\nu)}{\mathcal{B}(D\to\mu\nu)}$  &   0.2  &     -   &  0.005 \\
$V_{cs}$                          &    0.02~\cite{BESIII:2018hhz}            &   0.005        & 0.0015  \\
$f_{D_{s}}$                       &     0.02           &    0.005       & 0.0015  \\
$\frac{\mathcal{B}(D_{s}\to\tau\nu)}{\mathcal{B}(D_{s}\to\mu\nu)}$  &   $0.04$  &     $0.009$   &  0.0038 \\
\hline
{\it D mixing parameter:}  \\
$x$                          &     -         & 0.03    &   0.05~\cite{Cheng:2022tog}  \\
$y$                          &     -         & 0.02    &   0.05 \\
\hline
{\it $\tau$ properties: }\\
$m_{\tau}$ (MeV/c$^{-2}$)                 &   0.12~\cite{BESIII:2014srs}   &  -    &   0.012  \\
$d_{\tau}$ (e~cm)             & -          &  $2.02\times10^{-19}$   &   $5.14\times10^{-19}$ \\
\hline
{\it cLFV decays of $\tau$(U.L at 90\% C.L.):}  \\
$\tau\to lll$        &              -                 &  $1\times10^{-9}$                &   $1.4\times10^{-9}$  \\  
$\tau\to \gamma\mu$  &              -                 &  $5\times10^{-9}$                 &   $1.8\times10^{-8}$  \\
$J/\psi\to e\tau$     &                $7.5\times10^{-8}$      &    -                                     &  $7.1\times10^{-10}$  \\
\hline 
\hline
\end{tabular}
\end{spacing}
\end{center}
\end{table}

\subsection{Prospects of the Collins Fragmentation Function}
The measurement of the Collins fragmentation function~(FF) represents an important test for understanding strong interaction dynamics and thus is of fundamental interest in understanding QCD.
There have been several semi-inclusive deep inelastic scattering (SIDIS) measurements of Collins asymmetry from HERMES~\cite{Airapetian:2004tw, Airapetian:2010ds}, COMPASS~\cite{Adolph:2012sn} and JLab~\cite{Qian:2011py}. Direct information on the Collins FF can be observed from $\ee$ annihilation experiments such as Belle~\cite{Abe:2005zx, Seidl:2008xc} and BaBar~\cite{TheBABAR:2013yha}, which give consistent nonzero asymmetries. However, the $\ee$ Collins asymmetry obtained from Belle and BaBar corresponds to a $Q^2(\approx 100\GeV^2)$ higher than the typical energy scale of the existing SIDIS data. The STCF studies $\ee$ at a moderate energy scale ($Q^2$ in 4$\sim$49~GeV$^2$).
The results can be directly connected to future SIDIS experiments, such as EIC and EicC. In addition, it is crucial to explore the $Q^2$ evolution of the Collins FF and to improve our knowledge of strong interaction dynamics.

\begin{figure}[htbp]
\begin{center}
\begin{overpic}[width=7.5cm, height=6.cm, angle=0]{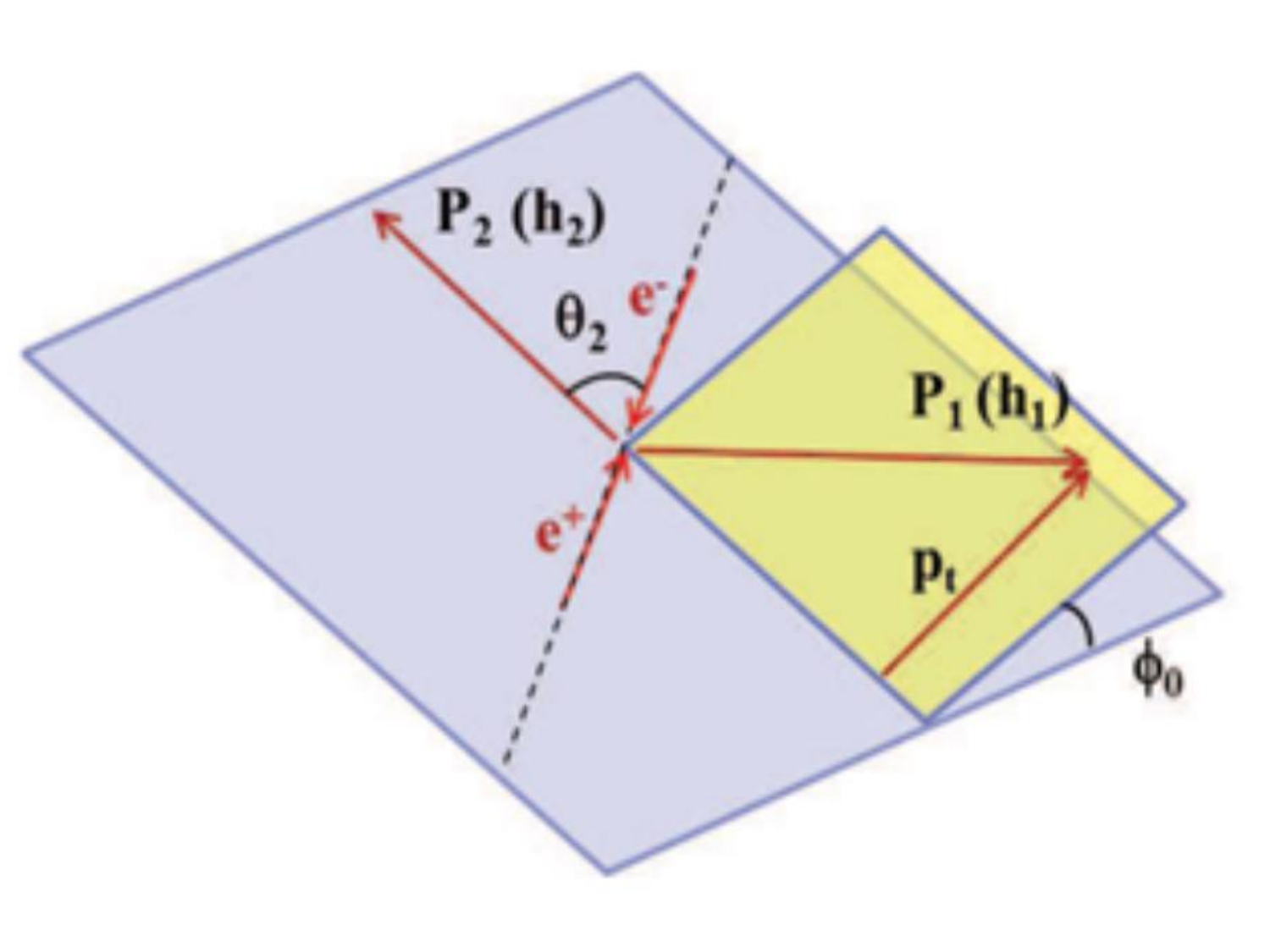}
\put(20,70){\small{(a)}}
\end{overpic}
\begin{overpic}[width=7.5cm, height=6.cm, angle=0]{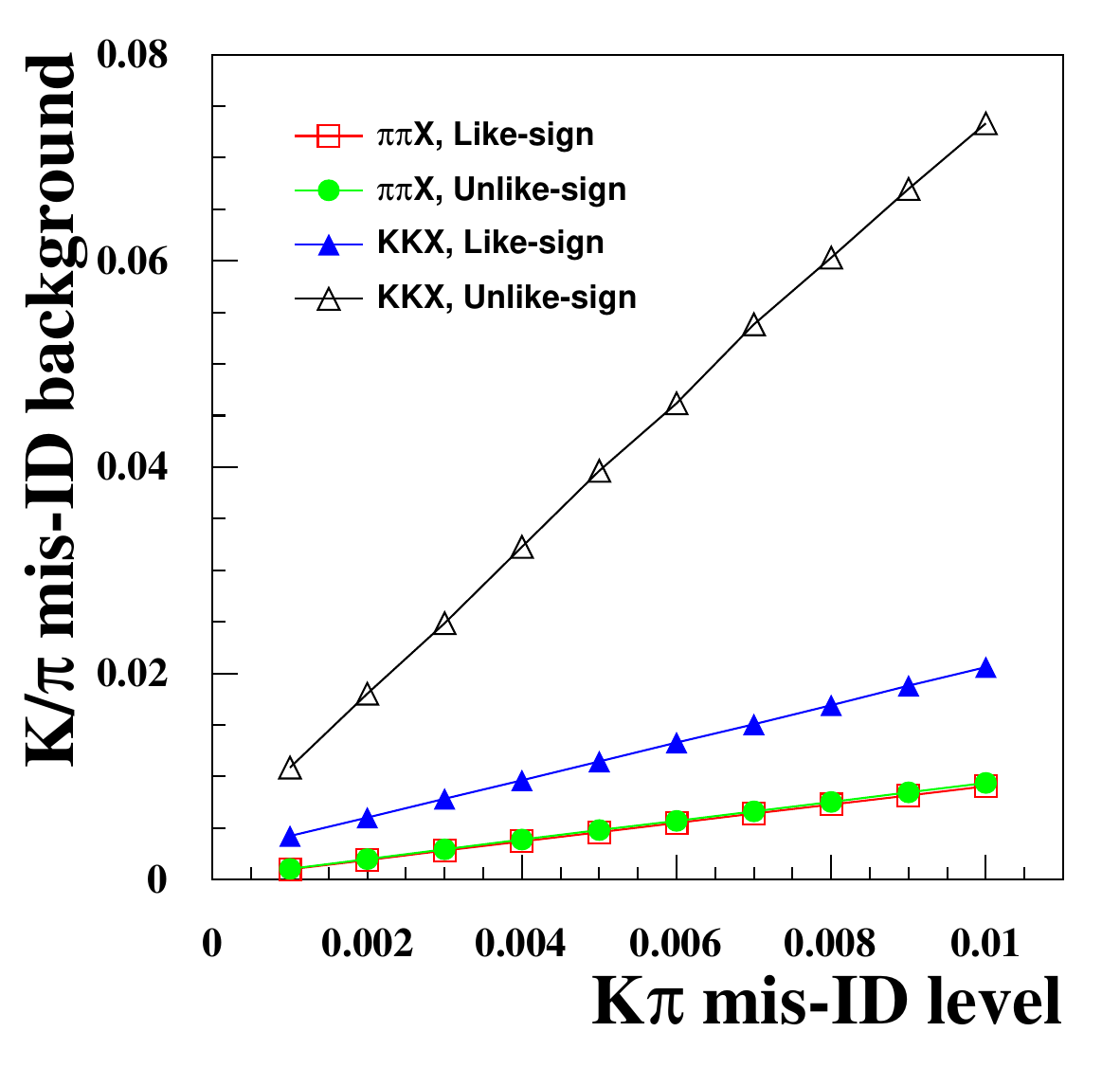}
\put(20,70){\small{(b)}}
\end{overpic}
\end{center}
\vspace{-0.5cm}
\caption{(a) Definition of $\phi_{0}$ as the angle between the plane spanned by the beam axis and the momentum of the second hadron ($P_2$) and the plane spanned by the transverse momentum $p_t$ of the first hadron relative to the second hadron.
(b) Different $K\pi$ mis-IDs versus the background level $f_{K\pi}$. }
\label{kpimisidbackground}
\end{figure}

To study the Collins FF, we introduce the $2\phi_0$ normalized ratio,
$R=\frac{N(2\phi_0)}{\langle N_0\rangle}$, where the azimuthal angle $\phi_{0}$ is defined as in Fig.~\ref{kpimisidbackground}(a),
$N(2\phi_0)$ is the dipion yield in each $2\phi_0$ subdivision, and $\langle N_0\rangle$ is the averaged bin content. The normalized ratios are built for unlike signs ($\pi^\pm \pi^\mp $) and like signs ($\pi^\pm \pi^\pm $), defined as $R^U$, $R^L$, where different combinations of favored FFs and disfavored FFs are involved. A favored fragmentation process refers to the fragmentation of a quark into a hadron containing the valence quark of the same flavor, {\it e.g.}, $u(\bar d)\to \pi^+$, while the corresponding $u(\bar d)\to \pi^-$ is disfavored. A double ratio $R^U/R^L$ is used to extract the azimuthal asymmetries since $R$ is strongly affected by detector acceptance. In the double ratio, charge-independent instrumental effects cancel out, while the charge-dependent Collins asymmetries are kept. $R^U/R^L$ follows the expression $\frac{R^U}{R^L} = 1 + A_{UL}\cos(2\phi_0)$, while $A_{UL}$ denotes the asymmetry for the UL ratio.

The measured asymmetry of $KK+X$ is significantly diluted by the $K\pi$ background. We use $A^{KK}_{\rm true}$ and $A^{K\pi}_{\rm true}$ to express the true Collins asymmetries for two processes $KK+X$ and $K\pi+X$, respectively, with $f_{K\pi}$ and $f_{\rm flat}$ for the level of the $K\pi$ background and the other background contributing zero asymmetry. Then, we can obtain the Collins measurement by unfolding
\begin{equation}
  A^{KK}_{\rm meas} = (1-f_{\rm flat}-f_{K\pi}) A^{KK}_{\rm true} + f_{K\pi}A^{K\pi}_{\rm true}.
\end{equation}
The $f_{K\pi}$ levels for $K\pi$ mis-ID levels from $0.001$ to $0.01$ are shown in Fig.~\ref{kpimisidbackground}(b). $f_{K\pi}$ can reach up to 8\% for $K\pi$ mis-ID at the 1\% level. Since $A^{KK}_{\rm meas}$ and $A^{\pi\pi}_{\rm meas}$ are at the (0.1, 0.2) level, the systematic background coming from $f_{K\pi}$ is approximately 0.02.

With the $K/\pi$ misidentification rate required to be 1\% up to 2~GeV, the statistical uncertainty for $\pi\pi+X$ is $(2\sim7)\times10^{-4}$, and the statistical uncertainty for $KK+X$ is $(7\sim20)\times10^{-4}$ for $1~\rm {ab}^{-1}$ luminosity at $\sqrt{s}=7$~GeV. The momentum dependent precision 
for the Collins effect is described in Ref.~\cite{Wang:2021pus}, and it is found the precision of Collins effect is better than 0.2\% if the azimuthal asymmetries in the inclusive production of $\pi^{+}\pi^{-}$ is over 0.02 and that of $K^{+}K^{-}$ is over 0.07 in the kinematics bin $z=E/E_{\rm beam}\in [0.5, 0.9]$. 

\subsection{Prospects of $CP$ Violations in the Lepton/Baryon Sector}

Within the SM, there is no direct $CP$ violation in hadronic $\tau$ decays at the tree level in weak
interaction; however, the well-measured $CP$ asymmetry in $K_{L}\to\pi^{\mp}l^{\pm}\nu$ produces a difference between $\Gamma(\tau^{+}\to K_{L}\pi^{+}\bar{\nu})$
and $\Gamma(\tau^{-}\to K_{L}\pi^{-}{\nu})$ due to the $K^0-\bar K^0$ oscillation.
The same asymmetry also appears between $\Gamma(\tau^{+}\to K_{S}\pi^{+}\bar{\nu})$ and $\Gamma(\tau^{-}\to K_{S}\pi^{-}{\nu})$ and is calculated to be $(0.36\pm0.01)$\%~\cite{taucp5,taucp6}.

The BaBar experiment has found evidence for $CP$ violation in $\tau\to K_{S}\pi^{-}\nu[\geq 0\pi^{0}]$ with an asymmetry rate of $(-0.36\pm0.23\pm0.11)$\%~\cite{CPbabar},
which is $2.8$ standard deviations from the theoretical prediction.
Apart from the above BaBar measurement, a collaboration of CLEO~\cite{CPcleo} and Belle~\cite{CPbelle} focuses on the $CP$ violation that could arise from a charged scalar boson exchange~\cite{theocp}; this type of $CP$ violation can be detected as a difference in the $\tau$ decay angular distributions. The results are found to be compatible with zero with a precision of $\mathcal{O}(10^{-3})$ in each mass bin~\cite{CPbelle}.
In all these experimental measurements, the statistical uncertainty is at the level of $\mathcal{O}(10^{-3})$, and the current experimental sensitivity cannot yield a conclusion regarding the $CP$ from $\tau$ decay. Therefore, a higher-precision result is strongly required for clarifying the new physics features.

\begin{figure}[htbp]
\begin{center}
\begin{overpic}[width=7.5cm, height=5.cm, angle=0]{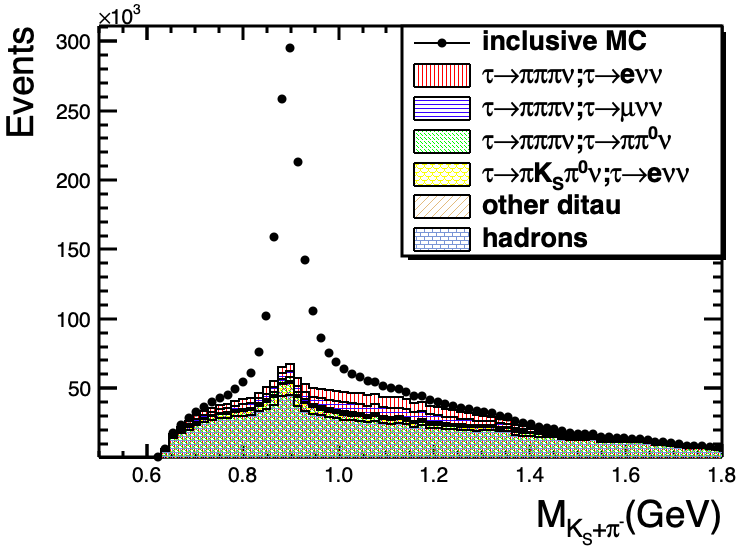}
\put(26,58){\small{(a)}}
\end{overpic}
\begin{overpic}[width=7.5cm, height=5.cm, angle=0]{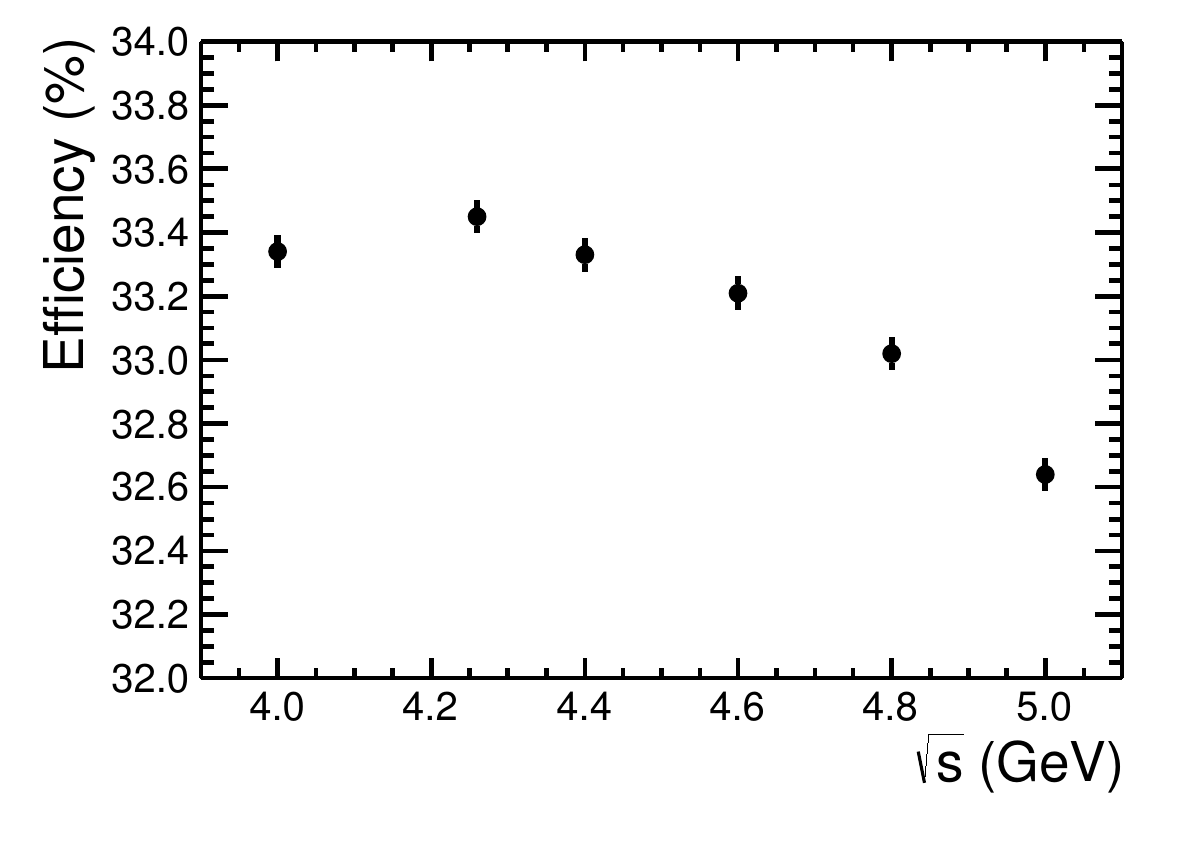}
\put(26,58){\small{(b)}}
\end{overpic}
\end{center}
\vspace{-0.5cm}
\caption{(a) Invariant mass of $K_{S}\pi^{-}$ with combined $e$-tag and $\mu$-tag from $\tau^{+}$ decay. (b) Selection efficiencies for the signal process at different CMEs from $\sqrt{s}=4.0$~GeV to 5.0~GeV.}
\label{tautokspinu}
\end{figure}

MC samples normalized to 1~ab$^{-1}$ luminosity at $\sqrt{s}=4.26$~GeV are simulated.
The $e^{+}e^{-}\to \tau^{+}\tau^{-}$ process is generated with {\sc KKMC}~\cite{KKMC}, which implements {\sc Tauola} to describe the production of the $\tau$ pair.
Passage of the particles through the detector is simulated by the fast simulation software.
We select signal events with one $\tau^{+}$ decay to leptons, $\tau^{+}\to l^{+}\nu_{l}\bar{\nu}_{\tau},(l=e,\mu)$,
denoted as the tag side.
The other is $\tau^{-}\to K_{S}\pi^{-}\nu_{\tau}$ with $K_{S}\to\pi^{+}\pi^{-}$, denoted as the signal side.
The charge conjugate decays are implied.
The distribution of $M_{K_{s}\pi^{-}}$ after selection is shown in Fig.~\ref{tautokspinu}(a), and the selection
efficiency for the signal is 32.8\% after detector optimization including improving the low-momentum tracking
efficiency by 10\% and requiring the $\pi/\mu$ misidentification to be 3\%.

The efficiency corrected numbers for $\tau^{-}\to K_{S}\pi^{-}\nu_{\tau}$ and $\tau^{+}\to K_{S}\pi^{+}\bar{\nu}_{\tau}$ are well consistent with the input value.
The statistical sensitivity of $CP$ asymmetry with decay rate is calculated to be $9.7\times10^{-4}$.
In addition, the selection efficiency for this process with CME from $\sqrt{s}=4.0$~GeV to 5.0~GeV, where the cross-section for $e^{+}e^{-}\to\tau^{+}\tau^{-}$ is the
maximum, is studied, as shown in Fig.~\ref{tautokspinu}(b).
It is found that the efficiency varies from 32.6\% to 33.5\% (without a likelihood requirement), which is very stable in this energy region.
Therefore, the statistics of the signal process can be increased linearly with more data collected.
Since this process is not free of a background, the sensitivity of the $CP$ asymmetry is proportional to $1/\sqrt{L}$.
With 10~ab$^{-1}$ integrated luminosity collected at the STCF, the sensitivity of the $CP$ asymmetry is estimated
to be $3.1\times10^{-4}$, comparable to the uncertainty for theoretical prediction.\\

According to the CKM mechanism, the $CP$ violation of $\Lambda$ was predicted to be on the order of $\mathcal{O}(10^{-4}) \sim \mathcal{O}(10^{-5})$ ~\cite{SM27}. Recently, BESIII has measured this $CP$ violation by studying the spin correlation
in $J/\psi\to\Lambda\bar{\Lambda}$ with 1.3~billion $J/\psi$~\cite{BESIII:2018cnd}. The sensitivity of the $CP$ violation is $\mathcal{O}(10^{-2})$,
which is far from the sensitivity of theoretical prediction. At the STCF, more than 1 trillion $J/\psi$ are expected in
one year, and the $CP$ violation of hyperons can be studied with high precision.

For the process $J/\psi \to \Lambda \bar{\Lambda} \to p \pi^{-} \bar{n} \pi^{0}$, the selection efficiency is 13.1\%,
which is increased by a factor of 22.1\% after a series of detector response optimizations. With the fitting of the joint angular distribution with the spin-correlation functions, the precision of the $\alpha$ asymmetry is estimated to be $8\times10^{-4}$ with a 1 trillion $J/\psi$ MC sample.
The $CP$ violation is constructed with the difference of $\alpha$ from the decay of $\Lambda$ and $\bar{\Lambda}$.
From this study, we can also estimate the sensitivity of $CP$ violation for the process $J/\psi\to\Lambda\bar{\Lambda}\to p\bar{p}\pi^{+}\pi^{-}$; by improving the tracking efficiency for low-momentum tracks by 10\%, the sensitivity of $CP$ study for this process is approximately $3\times10^{-4}$, which is comparable to the sensitivity from theoretical prediction.

It should be noted that, 
if the electron beam is polarized, the polarization translates nearly 100\% into a well understood polarization of the two final-state tau leptons. 
Thus, the STCF operation with a polarized electron beam just above the $tau$-pair threshold would enable a high sensitivity search for CP-violating asymmetries in many $\tau$ decay
modes, such as $\tau\to\pi\pi^{0}\nu$~\cite{Tsai:1994rc}. Besides, the polarization of electron beams can also improve the sensitivity of CP in hyperon decays with a factor of three~\cite{Salone:2022lpt}.

\subsection{Prospects of Leptonic Decay $D_{s}^{+}\to l^{+}\nu_{l}$}

In the SM, pure leptonic decay $D_{s}^{+} \to l^{+} \nu~(l=e,~\mu,~ \tau)$ is
described by the annihilation of the initial quark-antiquark pair into a virtual $W^{+}$ that materializes as a $l^+ \nu$ pair. Since there is no strong
interaction present in the leptonic final state $l^+ \nu$, this decay provides a clean way to probe the complex strong interaction hiding the quark and antiquark within the initial-state meson, where the strong interaction effects can be parameterized by the decay constants $\fdsp$ and $|V_{cs}|$.

For a given value of $|V_{cs}|$, we can obtain $f_{D_s^+}$, and vice versa. Improved measurements of $f_{D_s^+}$ are important for testing and calibrating lattice QCD calculations, and $|V_{cs}|$ is important for testing the CKM matrix unitarity.
In addition, the ratio of the decay rates of $D_s^+ \to \tau^+ \nu_{\tau}$ and $D_s^+ \to \mu^+ \nu_{\mu}$ can be
used to test lepton flavor universality (LFU).
The current experimental results of $D_{s}\to l^{+}\nu$ are limited by statistics; therefore, it is important
to study the $D_{s}$ leptonic decay with high statistics.

\begin{figure}[htbp]
\begin{center}
\begin{overpic}[width=7.5cm, height=5.cm, angle=0]{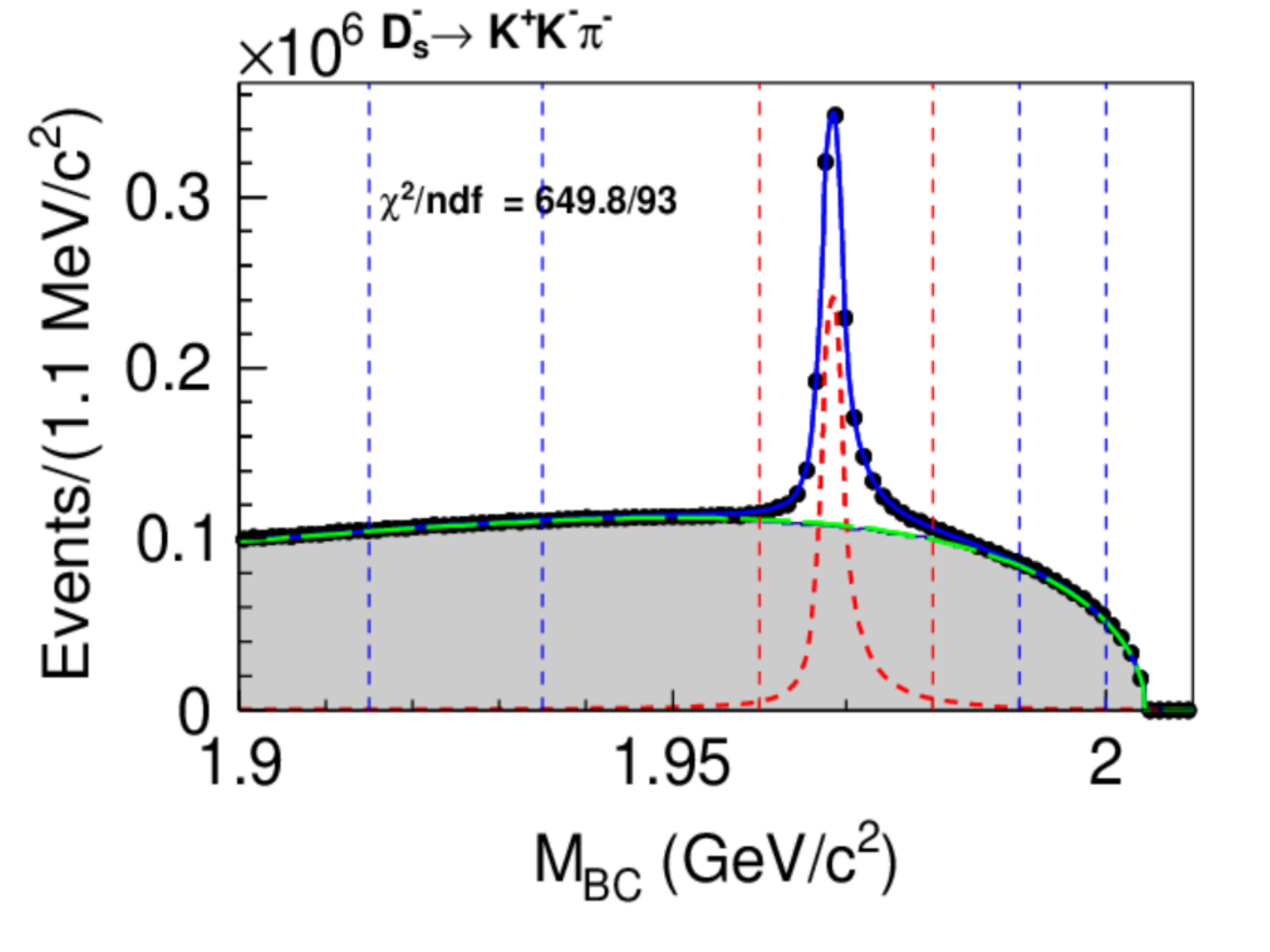}
\put(23,50){\small{(a)}}
\end{overpic}
\begin{overpic}[width=7.5cm, height=5.cm, angle=0]{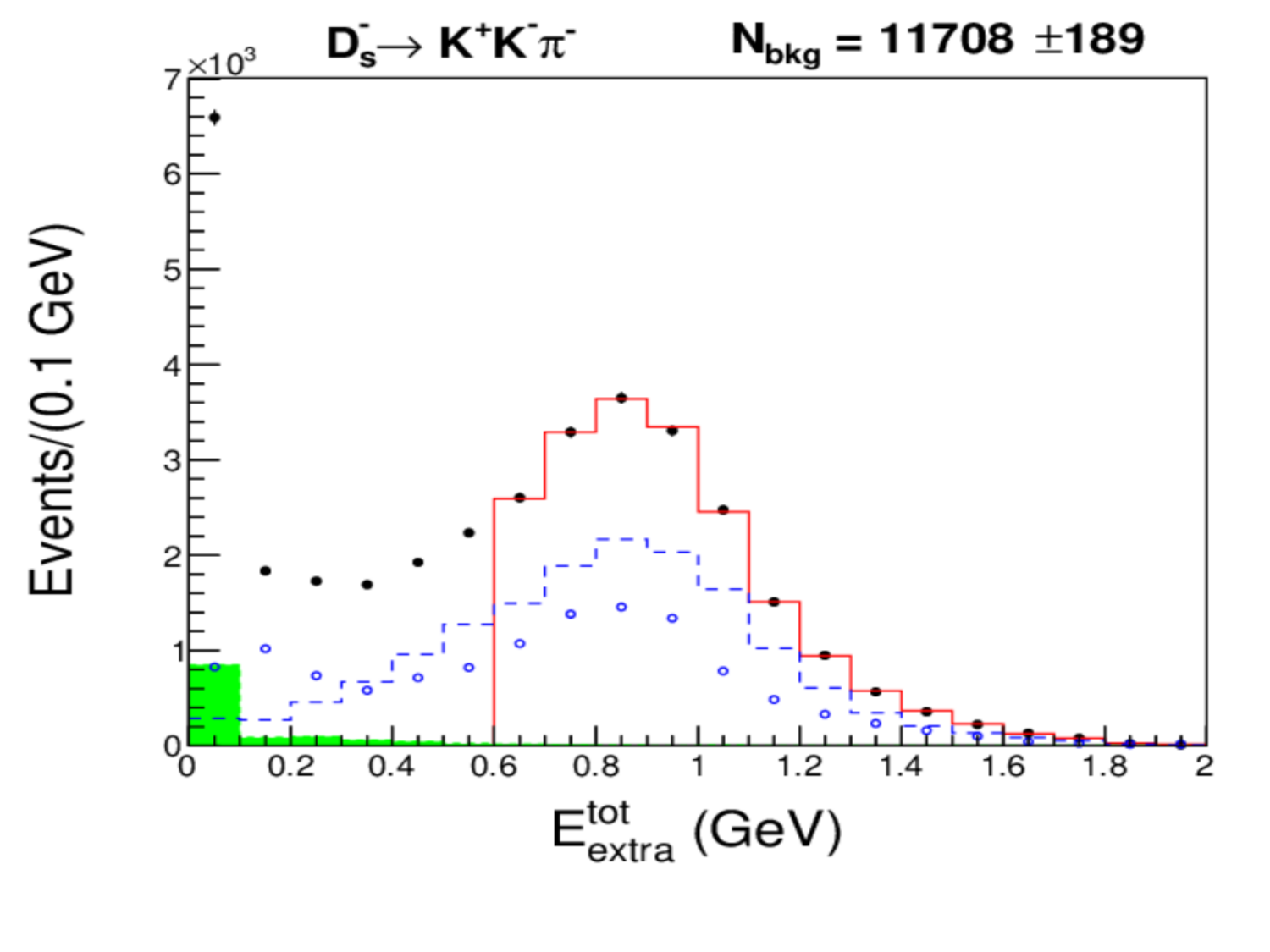}
\put(23,50){\small{(b)}}
\end{overpic}
\end{center}
\vspace{-0.5cm}
\caption{(a) Distribution of $M_{bc}$ from signal tag $D_{s}^{-}\to K^{+}K^{-}\pi^{-}$.
(b) Distribution of extra energy for $D_{s}\to\tau\nu$ with $\tau\to e\nu\nu$, where the open dots come from class I background,
the shaded green comes from class II background, and the dashed blue comes from class III background. The plots are depicted with
0.1~ab$^{-1}$ simulated cocktail MC. }
\label{Ds}
\end{figure}

We study $D_{s}^{+}\to l^{+}\nu~(l=\mu,\tau)$ at $\sqrt{s}=4.009$~GeV, where the $D_s^{+} D_s^{-}$ pairs are produced with no other accompanying particles, which provides extremely clean and pure $D_s^{\pm}$ signals. The cross-section
for $e^{+}e^{-}\to D_s^{+} D_s^{-}$ is approximately 0.3~nb~\cite{crosssection_DsDs_cleo}.
The double tag~(DT) technique is used to perform absolute measurement of the branching fraction.
The single tag~(ST) candidate events are selected by reconstructing a $D_s^-$ in the following 11 tag modes: $\dstoksk$, $\kkpi$, $\kkpipiz$, $\kskpipi$, $\kskpipim$, $\pipipi$, $\pieta$, $\pipizeta$, $\pietapgam$, $\pietaprho$, and $\kpipi$.
Figure~\ref{Ds}(a) shows the distribution of $M_{BC}$ for the $D_{s}$ candidates from the $K^{+}K^{-}\pi^{-}$ mode.

To reconstruct $D_{s}^{+}\to\tau^{+}\nu$, the decay $\tau^{+}\to e^{+}\nu\bar{\nu}$ is studied. Only one good charged track is required; this track is identified as a positron and has a charge opposite that of the ST $D_s^-$ meson. The variable $\etot$ is used to demonstrate the $D_s^+$ candidate, which is defined as the total energy of good showers in the EMC, excluding those used in the tag side and photons from the positron.
The distribution of $\etot$ is shown in Fig.~\ref{Ds}(b).
The signal events form a peak at 0 on the $\etot$ spectrum. The background sources can be categorized into three classes. The class I background is from the events with the wrong tag $D_s^-$, which is represented by the events in the $M_{\rm BC}$ sideband regions; the class II background arises from the peak of $D_s^+ \to K_L^0 e^+ \nu_e$ decay with little or no energy from $K_L^0$ deposited in the EMC; the class III background is dominant and includes other $D_s^+$ semileptonic decays. The class II and class II backgrounds are depicted with MC simulations. The signal region of $\etot$ is defined to be $\etot<0.4$ GeV to maximize the signal significance. To mitigate the signal shape dependency on the tag mode, we use the cut-count method, {\it i.e.}, fit $\etot$ in the high region of $\etot>0.6$ GeV, and then subtract the extrapolated background from the number of observed events in the signal of $\etot<0.4$ GeV to obtain the DT yield and DT efficiencies.
To reconstruct $D_{s}\to\mu\nu$, one charged track is selected and identified as $\mu$, and the missing mass of neutrinos is used to distinguish signals from background.

\begin{figure}[htbp]
\begin{center}
\includegraphics[width=14.5cm,height=10.cm,angle=0]{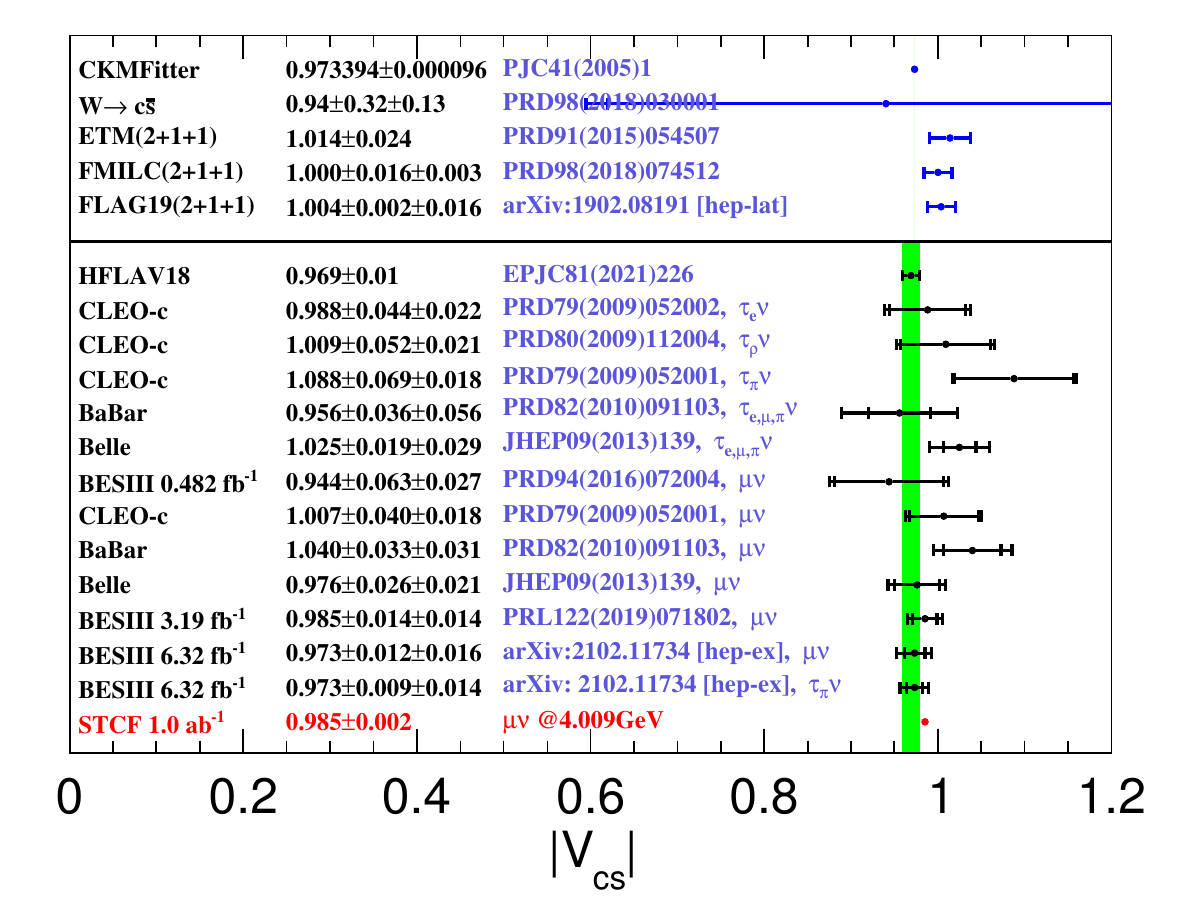}
\end{center}
\vspace{-0.5cm}
\caption{Comparison of $|V_{cs}|$.}
\label{expectvcs1}
\end{figure}

\begin{figure}[htbp]
\begin{center}
\includegraphics[width=14.5cm,height=10.cm,angle=0]{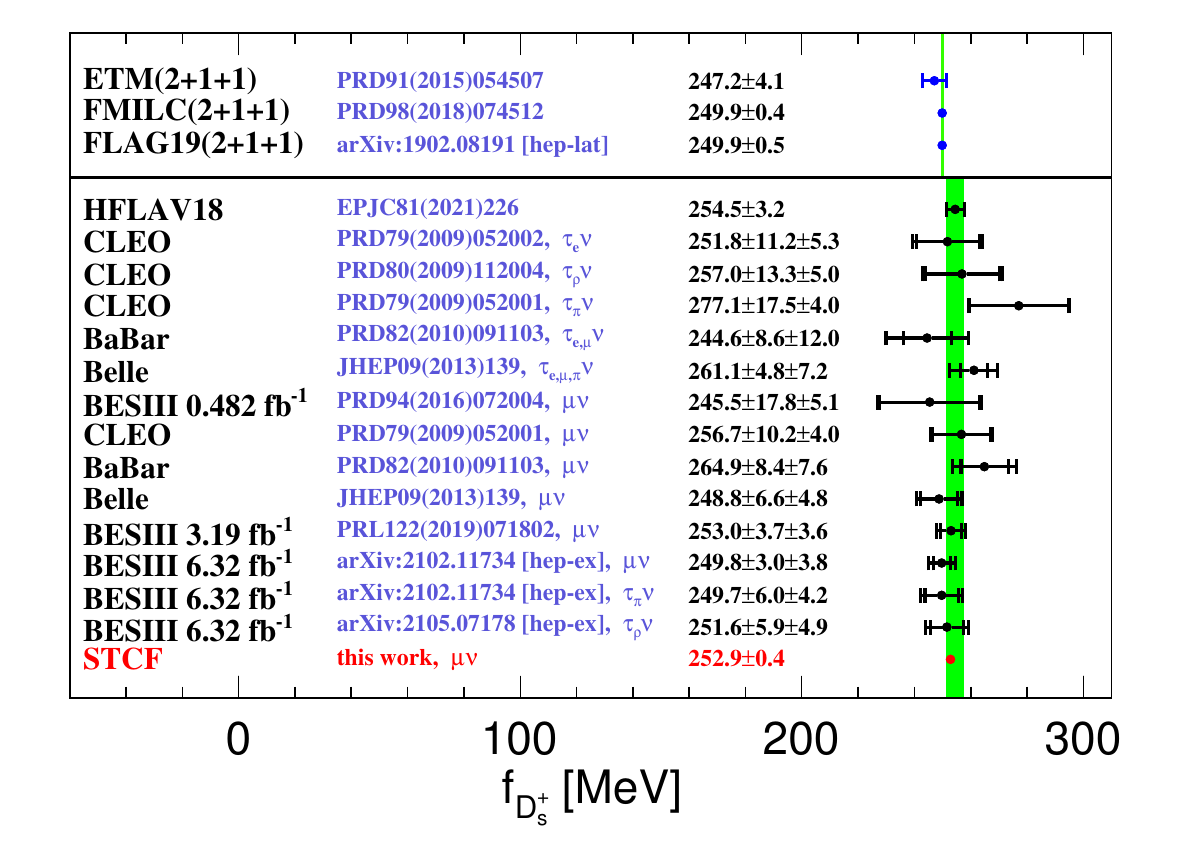}
\end{center}
\vspace{-0.5cm}
\caption{Comparison of $f_{D_{s}}$.}
\label{expectvcs2}
\end{figure}

With the 0.1~ab$^{-1}$ cocktail MC at 4.009~GeV, the statistical uncertainty for $D_{s}\to\mu\nu$ is estimated to be 0.89\%,
and the statistical uncertainty for $D_{s}\to\tau\nu$ is 1.3\%. This is comparable with the expected precision
at Belle II, with 50~ab$^{-1}$ luminosity.
Moreover, at the STCF, with 1~ab$^{-1}$ luminosity collected in one year, the uncertainty can
be reduced to 0.28\% and 0.41\% for $D_{s}\to\mu \nu$ and $D_{s}\to\tau\mu$ with $\tau\to e\nu\nu$, respectively.
The obtained $\fdsp$ and $|V_{cs}|$ have an uncertainty of less than 0.2\%, which is comparable to the uncertainty of the theoretical calculation.
Figure~\ref{expectvcs1} and Fig.~\ref{expectvcs2} shows the expected $|V_{cs}|$ and $f_{D_{s}}$ obtained with 1~ab$^{-1}$ luminosity collected at 4.009~GeV at the future STCF.

\subsection{Prospect of cLFV in $\tau$ Decay}

The charged lepton flavor violation (cLFV) process is forbidden in the SM and is highly suppressed even when taking neutrino oscillation into account since the rate of the cLFV process is suppressed by $(m_{\nu}/m_{W})^{4}$, where $m_\nu$ and $m_W$ are the masses of neutrinos and $W$ bosons, respectively. Any observation of cLFV in experiment would be an unambiguous signature of new physics. On the other hand, lepton flavor conservation, differing from other conservation laws in SM, is not associated with an underlying conserved current symmetry. Therefore, many BSM models naturally introduce lepton flavor violation processes, and some of them predict branch fractions that are almost as high as the current experimental sensitivity~\cite{tauto3lepton2,tauto3lepton3,tauto3lepton4, tauto3lepton5, tauto3lepton6, tauto3lepton7}.
Of all the possible cLFV processes, $\tau \to \gamma \mu$ and $\tau\to lll$ are two golden channels as
the most likely decay modes in a wide variety of theoretical predictions.
Currently, the most stringent upper limits of the branch ratios of the two channels are given by B-factories \SI{4.4e-8} at \SI{90}{\percent} C.L. \cite{tautogammuBABAR} and \SI{4.5e-8} at \SI{90}{\percent} C.L. \cite{tautogammuBelle} for $\tau\to\gamma\mu$ and  $(4-8)\times10^{-8}$ at 90\% C.L. for $\tau\to lll$. The next generation of electron-position colliders, such as Belle II and the STCF, aim to push the sensitivity down another several orders of magnitude into the prediction range of BSM models.

\begin{figure}[htbp]
\begin{center}
\begin{overpic}[width=7.5 cm, height=6.0 cm, angle=0]{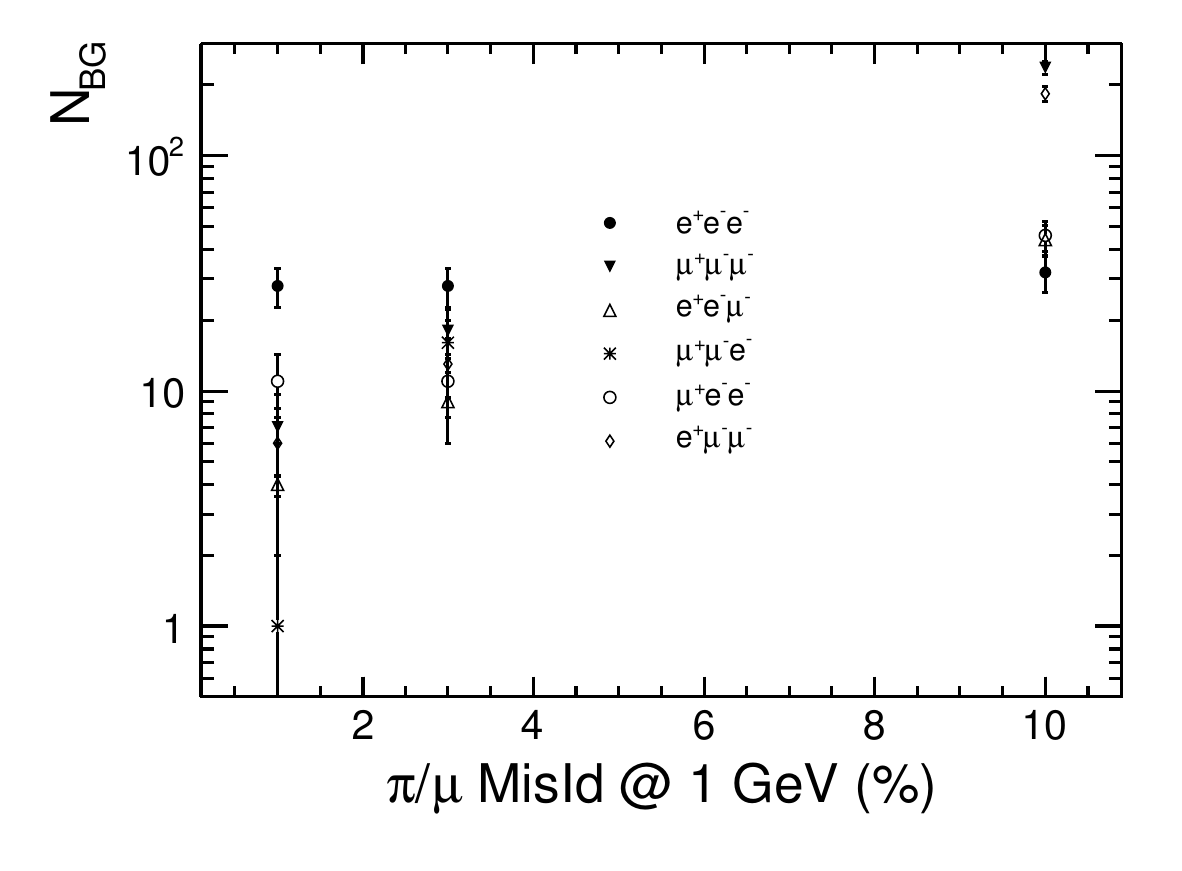}
\end{overpic}
\begin{overpic}[width=7.5 cm, height=6.0 cm, angle=0]{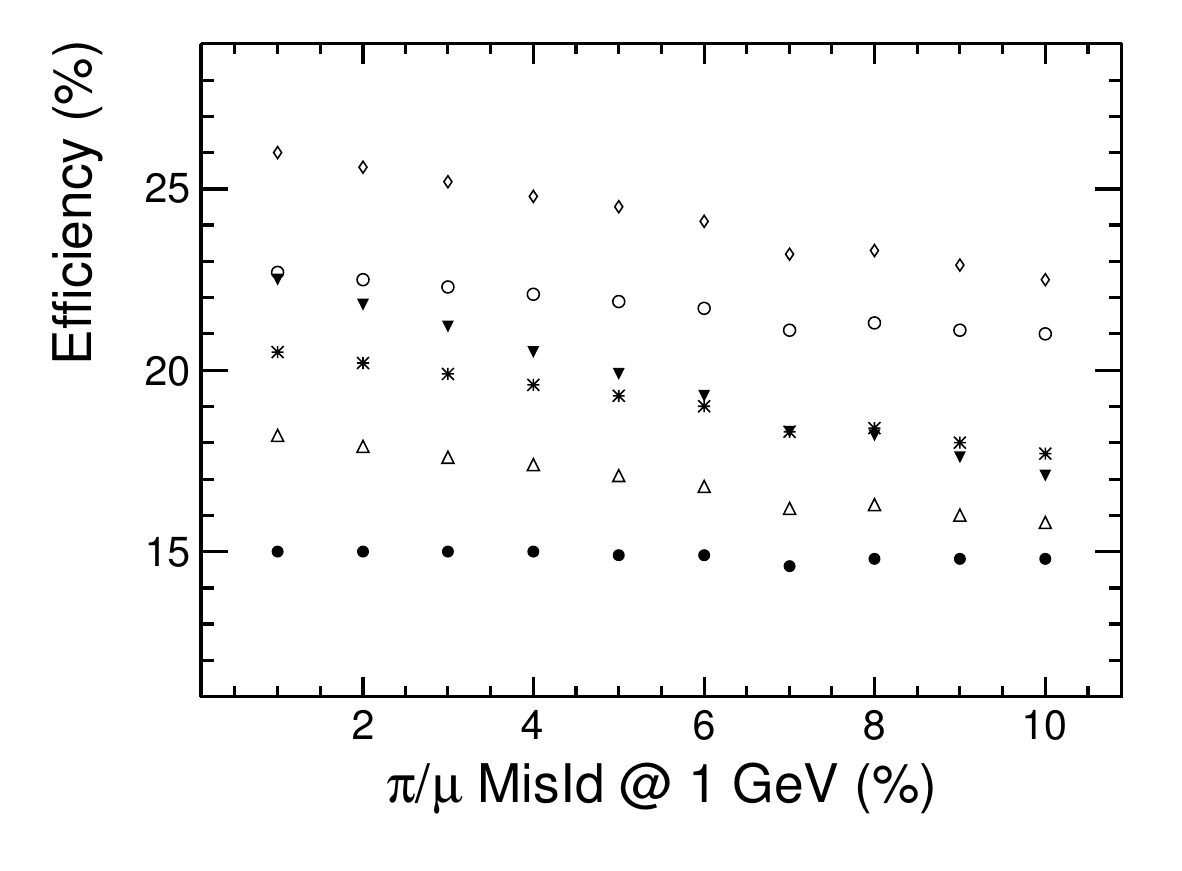}
\end{overpic}
\caption{(a) The number of surviving background events $N_{BG}$ and (b) the selection efficiency with different $\pi/\mu$ mis-ID rates for $\tau\to lll$. }
\label{tauto3leptonresults}
\end{center}
\end{figure}

We use the DT technique for cFLV decays of $\tau$, in which the tag of $\tau^{+}$ is the large branching fraction
processes, including one charged track
($\tau^{+}\to\pi^{+}/e^{+}/\mu^{+}+n\pi^{0}+n\nu)$. For $\tau^{-}\to
l^{+}l^{-}l^{-}$, 82.62\% of the total $\tau^{+}$ branching fraction is tagged, and $54\%$ of the total $\tau^{+}$ branching fraction is tagged for $\tau^{-}\to\gamma\mu$ (excluding the channels $\tau\to\mu\nu\mu$ and $\tau\to\pi \pi^{0}\pi^{0}\nu$). The tag side can be reconstructed with missing energy from neutrinos, while the signal side can be fully reconstructed.
For the process $\tau^{-}\to l^{+}l^{-}l^{-}$, there are six modes ($e^{-}e^{+}e^{-}$, $\mu^{+}e^{-}e^{-}$, $\mu^{-}e^{+}e^{-}$, $e^{+}\mu^{-}\mu^{-}$, $e^{-}\mu^{+}\mu^{-}$, $\mu^{-}\mu^{+}\mu^{-}$).
The number of surviving events $N_{BG}$ using 1~ab$^{-1}$ cocktail MC at $\sqrt{s}=4.26~$GeV is obtained, with scaling of the $\pi/\mu$ mis-ID rate to 10\%, 3\%,
and 1\%, as illustrated in Fig.~\ref{tauto3leptonresults}. The MC selection efficiencies are also obtained by
varying the $\pi/\mu$ mis-ID rate from 10\% to 1\%. From the plot, we find that the detection efficiency
of $\tau^{-}\to\mu^{-}\mu^{+}\mu^{-}$ is most sensitive to the $\pi/\mu$ mis-ID rate and is increased from
15.4\% to 22.5\%, with the corresponding background rate decreasing from 237 to 8.
With $3.5\times10^{9}$ $\tau^{+}\tau^{-}$ pairs collected at the STCF per year, under the assumption that the $\pi/\mu$ mis-ID rate is 1\%, the upper limit is predicted to be $1.4\times10^{-10}$.

For $\tau\to\gamma\mu$, there are two dominant backgrounds, from $e^{+}e^{-}\to\mu^{+}\mu^{-}$
and $e^{+}e^{+}\to\tau^{+}\tau^{-}$ with $\tau^{+}\to\pi \pi^{0}\nu$ and $\tau^{-}\to\mu\nu\nu$.
Stringent selection criteria should be applied to remove the background.
With the $\pi/\mu$ misidentification required to be 1\% and
the position resolution of photons required to be 4~mm, we obtain a background-free sample
with a selection efficiency of approximately 8\%.
The upper limit for
$\tau\to\gamma\mu$ is calculated to be around $1.8\times10^{-8}$ with 1~ab$^{-1}$
at $\sqrt{s}=4.26$~GeV~\cite{Xiang:2023mkc}.

\section{Discussion of Systematic Uncertainties}
\label{sec:sysunc}
To shed light on the physics programs at the STCF, especially those that require
precise measurements, it is essential to have precise knowledge about the systematic
uncertainty sources.
However, a full systematic uncertainty study requires both experimental data and MC simulation,
which are not possible for this conceptual design report.
Here, we only provide a general discussion of the leading systematic uncertainty sources in physics analyses.
A more precise estimation of systematic uncertainty is expected when the design and construction of the detector is completed.

\subsection{Luminosity Measurement}
The precise measurement of the production cross-section at the STCF
requires a high-precision measurement of the luminosity.
Usually, the luminosity measurement utilizes the Bhabha scattering process
$e^{+}e^{-}\to (\gamma)e^{+}e^{-}$ due to its clear signature and large production
cross-section. A cross check of the luminosity can be performed using the di-photon process $e^{+}e^{-}\to\gamma\gamma$.
At a tau-charm factory, large-angle Bhabha scattering events are selected.
The integrated luminosity is calculated with
$\mathcal{L}=N_{\rm obs}/(\sigma\times\varepsilon)$, where $N_{\rm obs}$ is the
number of observed signal events, $\sigma$ is the cross-section of the process
$e^{+}e^{-}\to (\gamma)e^{+}e^{-}$ or $e^{+}e^{-}\to\gamma\gamma$, and
$\varepsilon$ is the detection efficiency.

Sources of systematic uncertainties of luminosity measurements necessitate requirements
regarding event selection, MC statistics, trigger efficiency and the MC generator.
Regarding the requirements for event selection, {\it e.g.}, the tracking efficiency,
the EMC energy and track acceptance requirements can
be studied by an alternative selection criterion using only information from the EMC,
varying these parameters, respectively. The MC
statistics can be negligible with the generation of enough MC samples. The trigger efficiency
will be discussed after the design of the detector is complete.
For the MC generator, the observed cross-sections $\sigma$ for the two processes are currently provided by the {\sc Babayaga@NLO} generator~\cite{Balossini:2006wc,Balossini:2008xr} with a precision of 0.1\%.
An even higher-precision QED calculation of the MC generator is expected with more accurate luminosity measurements required in the future~\cite{Balossini:2006wc}.

\subsection{Tracking/PID Uncertainty}

Tracking and PID play a key role in physics analyses, and comprehensive study of the data/MC difference for the tracking and PID efficiencies is important.
With the high luminosity of the STCF, large samples of charged and neutral final states
can be selected from the QED processes, $J/\psi$ decay, and charm meson decay.
The tracking and PID efficiencies can be studied in different kinematic regions with
pure samples. These process-independent efficiencies can be used to
correct the data/MC differences and estimate the corresponding systematic uncertainties.

A large sample of electrons and muons can be selected from radiative QED processes
$e^{+}e^{-}\to\gamma e^{+}e^{-}$ and $e^{+}e^{-}\to\gamma \mu^{+}\mu^{-}$.
Charged pions and kaons can be selected from $J/\psi$ decay, {\it e.g.}, $J/\psi\to\pi^{+}\pi^{-}\pi^{0}$ and $J/\psi\to K_{s}K^{\mp}\pi^{\pm}$,
or charm meson decay, $D^{0}\to K^{-}\pi^{+}$.
Protons can be selected from $J/\psi\to p\bar{p}\pi^{+}\pi^{-}$, while this process can also
be used to study the tracking and PID efficiencies of pions.
The detection efficiency of photons can be selected from processes $e^{+}e^{-}\to\gamma\mu^{+}\mu^{-}$, $J/\psi\to \pi^{+}\pi^{-}\pi^{0}$, $\psi(2S)\to\pi^{0}\pi^{0} J/\psi$, and $D^{0}\to K^{-}\pi^{+}\pi^{0}$.
The detection efficiency of neutrons can be studied from $J/\psi\to p\bar{n}\pi^{-}+c.c$.
The selection efficiency for the intermediate states $K_{s}$ and $\Lambda$ can be studied
with $J/\psi\to K_{s}\pi^{\pm}K^{\mp}$ and $J/\psi\to\Lambda\bar{\Lambda}$, respectively.

\subsection{Radiative Correction}

The radiative correction includes both initial-state radiative~(ISR) and final-state radiative~(FSR) corrections, and the current precision for the radiative corrections is model dependent.
For models used in {\sc Phokhara}, the precision is 0.5\%~\cite{Rodrigo:2001kf}.
A comprehensive discussion of radiation corrections is presented in Ref.~\cite{WorkingGrouponRadiativeCorrections:2010bjp}.
With increasing experimental accuracy, better modeling with radiative corrections is necessary.

\subsection{Vacuum Polarization}
In the calculation of the Born cross-section or the bare cross-section, the vacuum polarization~(VP) effect should be taken into account, which affects the EM fine structure constant $\alpha(s)$ as follows:
\begin{equation}
\sigma^{\rm bare} = \frac{\sigma^{\rm dressed}}{|1-\Pi(s)|^{2}} = \sigma^{\rm dressed}\cdot \left(\frac{\alpha(0)}{\alpha(s)}\right)^{2}
\end{equation}
The leptonic VP contributions to $\Pi(s)$ are calculated to four-loop accuracy~\cite{Sturm:2013uka},
the dominant uncertainty comes from the hadronic VP contribution that relies on the hadronic cross-sections,
which is currently within 0.2\% ~\cite{Jegerlehner:2008rs}. With more precise cross-sections obtained in experiments, especially at low CMEs, the accuracy of the VP effect will be further improved.

\subsection{Others}
In addition to the above systematic uncertainty sources widely involved in physics analyses,
there are other uncertainty sources depending on the physics processes involved.
In $\tau$ mass measurements, minimizing the uncertainties in the beam energy scale and energy spread is
so essential that a dedicated detector for these measurements is needed.
In the leptonic decay of charm mesons, radiative processes, for example, $D_{s}^{+}\to\gamma l^{+} \nu_{l}$,
are considered to have 100\% uncertainty on the branching fraction, which yields 1\% uncertainty
in the measurement of $D_{s}^{+}\to l^{+} \nu_{l}$. With the large sample collected at the STCF,
radiative processes can be measured with high precision, and its effect on the leptonic
decay will be limited to within 0.1\%.
Other uncertainty sources include those from background estimation, fitting procedures,
trigger efficiency, etc. Only when each uncertainty source is carefully examined and understood can the experimental accuracy match the high statistical precision achieved at the STCF.

\subsection{Conclusion}
Among the various physics highlights discussed at the beginning of this
section, the precision frontier needs careful systematic uncertainty 
studies. These uncertainties can be categorized into three aspects,
the first is the reducible uncertainties, that comes from tracking, PID, and other selection criteria, which will be scaled down 
according to the statistics of control sample. 
The second aspects from theoretical input as discussed above, such as initial radiative correction, vacuum polarization correction, luminosity measurement etc. The third comes from instrumental effects,  such as beam energy
measurement, energy spread etc. The latter two are irreducible 
uncertainties that should be studied in detail during R\&D. 
\newpage
\clearpage
\chapter{Future Plans}
\label{chap_plan}
\label{sec:future_RD}
\label{sec:cost_timeline}

As presented in this CDR, substantial efforts on detector R\&D and physics performance are underway.
It has been demonstrated that the proposed conceptual detectors can meet the physics requirements and can feasibly be built on the timescale of the STCF project.
Moving forward, more in-depth R\&D and studies are required for the preparation of the Technical Design Report~(TDR).
These include sharpening the physics case, finalizing detector technological choices, further optimizing to improve performance and
reduce cost, designing mechanical, electrical and thermal systems, and developing installation and integration schemes.

\section{Detector cost and project timeline}
\label{sec:costs}
\label{sec:timeline}

The cost of the STCF detector is estimated to be 550~million RMB based on its conceptual design. The leading cost driver is the electromagnetic calorimeter with pure CsI crystals. 
%The current estimates of the cost of the STCF project are given in %Table~\ref{tab:stcf_cost}.

The current timeline for the STCF project is shown in Fig.~5.1.  With the completion of the conceptual design reports of the accelerator and the detector, the next phase of the project will be a 5-year period for accelerator and detector R\&D toward the completion of the Technical Design Reports of the accelerator and the detector.  The project will then move on to the construction phase spanning approximately 7 years that will eventually see both the accelerator and the detector fully constructed. An approximately 15 year data-taking period is envisaged to achieve the primary physics targets of the STCF project, and there is a possibility to upgrade the STCF with a polarized electron beam afterwards.

%The current timeline for the STCF project is shown in Fig.~\ref{fig:stcf_timeline}.
%The first stage of the project is the completion of this CDR.
%The next stage is the 5-year period for detector R\&D toward the completion of the Technical Design Report (TDR).
%The construction of both the accelerator and the detectors will commence immediately after the approval of the TDR and is expected to take approximately 7 years.
%An approximately 15 year data-taking period is envisaged to achieve the physics targets of the STCF, and there is also a possibility to further upgrade the STCF project with polarized electron beam during the data-taking period.

\begin{figure}[htbp]
\begin{center}
\includegraphics[width=0.8\textwidth]{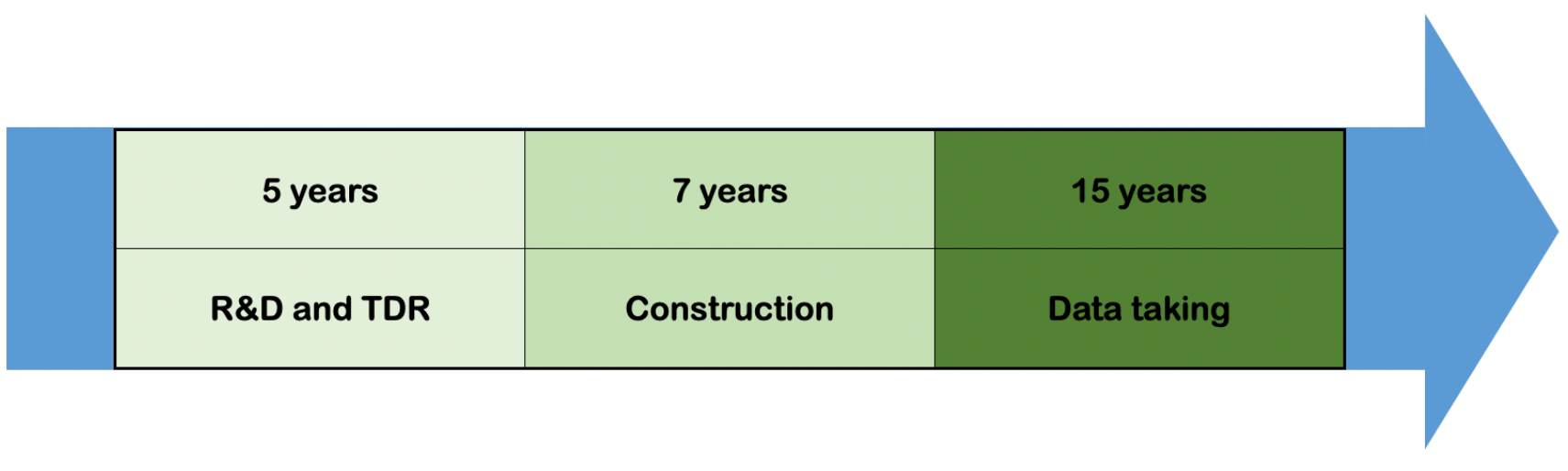}
\caption{Expected timeline for the STCF project.}
\label{fig:stcf_timeline}
\end{center}
\end{figure}

\section{R\&D Prospects}

\subsection{ITK}
\label{sec:rd_itk}
The inner tracker is designed to have a low material budget and fast readout and to provide high detection efficiencies for charged particles,
especially those with very low momentum, below 100 MeV/c.
It must also satisfy the stringent requirements imposed by the high luminosity and trigger rate conditions.
Two conceptual designs of the ITK have been proposed: a cylindrical $\mu$RWELL-based detector and a monolithic active pixel sensor-based detector.
Several critical R\&D items have been identified for the TDR:
\begin{itemize}
  \item $\mu$RWELL-based ITK:
    \begin{itemize}
      \item Manufacturing methods for large area diamond-like carbon (DLC) resistive electrodes;
      \item Electrode structure design and manufacturing methods for a large-area WELL detector and a large-area canal type (GROOVE) detector;
      \item Mechanical structure and installation method for a cylindrical detector;
      \item Design of front-end ASICs to integrate analog signal manipulation, ADC, charge and time calculations, and a high-speed data transfer interface;
      \item Design of the readout electronics, including the front-end electronics with customized ASICs, readout units, and subclock and subtrigger modules;
      \item Design, construction and characterization of a small cylindrical $\mu$RWELL detector prototype.
    \end{itemize}
  \item MAPS-based ITK:
    \begin{itemize}
      \item Track reconstruction performance study and further optimization of the layout;
      \item Design of the MAPS pixel layout with optimal sensor parameters, such as low capacitance, high charge collection efficiency and short charge collection time;
      \item Design of a low-power low-noise in-pixel circuit;
      \item Design of the readout strategy for the hit signals and the architecture of on-chip readout circuitry;
      \item Design of the readout system, including readout units, power units and common readout units, for data receiving, collecting, transferring, and configuring of the MAPS;
      \item Design of the stave module, including support mechanics and a cooling system;
      \item Design, construction and characterization of a MAPS stave prototype module.
    \end{itemize}
\end{itemize}

\subsection{MDC}
\label{sec:rd_mdc}
The main drift chamber (MDC) is the central part of the tracking system of the STCF detector. The key factors affecting the tracking performance in the STCF experiment are multiple scattering and energy loss of charged particles traversing the MDC. Therefore, the driving force in the design of the MDC is reducing its material budget as much as possible. This also represents the core of the R\&D work required for the MDC detector. The MDC provides ionization measurements for charged particle identification as well as precise position measurements for charged particle tracking. Thus, precise time and charge measurements are both required for the MDC readout electronics. In addition, the high counting rates expected at the MDC inner layers and the extremely high rate of physics events of interest expected when running at the $J/\psi$ peak impose stringent requirements on the MDC readout electronics as well as the MDC detector. A vigorous R\&D program needs to be developed and carried out for the MDC to meet all the technical challenges. Such an R\&D program should cover the following aspects where some critical R\&D items have been identified.
\begin{itemize}
\item \textbf{Detector design optimization}\\
The material budget of the detector should be maximally reduced to enhance the tracking performance for low-momentum particles by optimizing the detector design in terms of various aspects, including the working gas, wire material and size, configuration of wire layers, chamber structure and material. Drift cells that are much smaller than regular small cells and yet maintain a very low material budget for the whole detector should be designed. The adoption of such drift cells would shorten the maximum drift time in a drift cell and thus make the response of the detector faster, allowing it to better cope with the high count rate and physics event rate. The structure  of the drift chamber should be designed, including a detailed deformation analysis that fully takes into account wire tensions and their possible creep effects.

\item \textbf{R\&D of low-mass wires} \\
Electrode wires are one of the primary contributors to the material budget of a drift chamber. It is therefore of great importance to develop low-mass wires for use in a drift chamber. Invention of such low-mass wires could represent a breakthrough in drift chamber technology. The use of low-mass wires would also require much less tension to be applied to the wires than necessary when using regular metal wires. This would effectively reduce the total load on the endplates of the drift chamber due to the wire tension and hence leave much more room for the design and engineering of a light chamber structure. Ideas regarding using light polymeric fibers or carbon monofilaments coated with low-mass metals as wires for a drift chamber have been proposed and are being explored by some Italian groups. This could serve as one direction to pursue in the R\&D of low-mass wires for the STCF drift chamber.

\item \textbf{R\&D of high-density wiring}\\
Very small drift cells imply a very large number of closely spaced wires in a drift chamber. This poses a great challenge to wiring the drift chamber, which includes threading wires through the chamber and fixing them at the endplates. In this case, manual wiring is unlikely to be adequate, and regular feedthroughs can no longer be used to hold wires. Thus, an automatic wiring method along with the corresponding key devices need to be developed to enable efficient and accurate wiring operations. A novel method to fix wires without feedthroughs also needs to be developed.

\item \textbf{R\&D of readout electronics}\\
The MDC readout electronics consist of a transimpedance amplifier (TIA) followed by a shaping circuit and an analog-to-digital converter (ADC). The digitized signals are further filtered and processed to reduce the pile-up and enhance the SNR by a data processing circuit. The charge and time information of the signals are extracted with the processed data. The hardware components of the MDC readout electronics include front-end electronics (FEE), readout units (RUs), and subclock and subtrigger modules. Dedicated R\&D work on the readout electronics is needed to meet the requirements for precise time and charge measurements under high rate conditions. Two technical approaches are planned for R\&D, one based on discrete devices and the other on ASIC chips. The primary weight would be placed on the design and development of an ASIC chip suitable for the STCF drift chamber. The proposed ASIC chip incorporates a TIA, a discriminator and a shaping circuit. The transistor parameters and feedback resistance of the TIA should be carefully optimized in terms of circuit input impedance, bandwidth, and dynamic range.

\item \textbf{Development and characterization of a full-size drift chamber prototype}\\
A full-length drift chamber prototype designed for the STCF needs to be developed using all the key techniques and components developed for the drift chambers. The design of the prototype should closely follow the optimized design of the STCF drift chamber. The prototype would also be fully instrumented with the readout electronics developed for the STCF drift chamber. The prototype along with its readout electronics will be fully tested and characterized to validate the design of the STCF MDC and the key techniques and components of the drift chamber.
\end{itemize}

\subsection{RICH}
\label{sec:rd_rich}
The STCF PID barrel detector, a RICH detector, is designed to provide a $\pi/K$ misidentification rate lower than $2\%$ in the momentum range up to $p=2.0\gevc$, with the corresponding identification efficiency being larger than 97\% under a high luminosity environment. Comprehensive research is needed on the following items.

\begin{itemize}
\item \textbf{Radiator study}

The radiator is essential for the RICH detector. For the STCF momentum range, liquid C$_6$F$_{14}$ is chosen as the radiator.
Although this type of radiator has been successfully used in the ALICE experiment, several aspects need further study to fulfill the STCF requirements, for example, the refractive index varies with pressure and temperature and needs to be calibrated, and the UV transmission rate and the radiation hardness need to be investigated.
Additionally, purification systems and online monitoring systems are essential to guarantee long-term stable operation and need to be further researched.
In addition, research on new radiators, such as photonic crystals and silicon aerogels, is being carried out.

\item \textbf{Large area gaseous UV-photon detector} \\
The hybrid micropattern gaseous detector THGEM with a CsI-coated photocathode and Micromegas is the baseline design for the large-area UV-photon detector.
To achieve a high detection efficiency for Cherenkov radiation, the quantum efficiency of the CsI cathode should be relatively high and uniform.
Thus, it is necessary to study the decrease in quantum efficiency for different working gases and electric extraction fields and different pre- and post-treatments for the CsI coating.
The amplification of the gaseous detector must be large enough for single photon-electron detection, and the ion back flow must be low.
The detector also must be expandable with minimum dead space.

\item \textbf{Readout electronics} \\
Considering the large channel number, on the order of 10$^4$, and the requirement of high-precision signal measurements, it is preferable to implement an application-specific integrated circuit (ASIC) that can integrate front-end analog circuits and analog-to-digital conversion circuits within the chips to reduce the system complexity.

\item \textbf{RICH prototype} \\
To verify the performance and PID capabilities of the RICH detector, a large RICH prototype will be developed to demonstrate the feasibility of the detector.
The prototype will be expandable toward future STCF experiments with a connected liquid radiator purification system.
A beam test for a small prototype with $160\times160\,$mm$^2$ sensitive area has been performed at DESY with $5$ GeV/$c$ electron beam. Full-size module will be built and the beam test will be performed. The result will be compared with Monte Carlo simulations.
The FEE and DAQ electronics will also be commissioned with the prototype.

\end{itemize}

\subsection{DTOF}
\label{sec:rd_dtof}
According to the operation environment and physics requirements of the STCF imposed on the endcap PID subsystem, a DTOF detector, we will carry out extensive $R\&D$ on the key technologies of the DTOF detector and readout electronics. In the following, the main research items are listed.
\begin{itemize}
  \item \textbf{The DTOF detector}\\
In terms of achieving high-precision time measurement, key technologies of the DTOF detector include the optical design of the radiator and photosensor, fine processing of a large area radiator (fused silica) and fast-response photoelectric detection technology.
The design of the radiator structure, especially its optical performance, is very important for a DIRC detector. Another technical challenge is the high-precision processing and surface control of large-area radiators, which will be studied in cooperation with the Beijing Special Glass Research Institute.
Research on fast-response photoelectric detection technology includes the design and development of MCP-PMT readout circuits and the suppression of signal oscillation and crosstalk.

  \item \textbf{Readout electronics}\\
The output signals of the MCP-PMTs in the DTOF detector demonstrate a very short rise time and a very narrow width, which poses a significant challenge for high-precision readout electronics. To fully exploit the timing potential of the MCP-PMTs, the following $R\&D$ topics are important: multithreshold high-precision timing technology based on an FPGA TDC, including studies on the ultrafast signal processing and timing circuits; high-precision and highly integrated TDC technology development based on FPGAs; application specific integrated circuit (ASIC) development for the DTOF detector; multichannel high-precision clock synchronization and distribution technology; and the design and realization of high-bandwidth data acquisition system.

  \item \textbf{DTOF prototyping} \\
Verifications of the key technologies and the system-level detector performance are important to demonstrate the feasibility of the DTOF detector. Through the operation of a large-size DTOF prototype and comparison of experimental test data to simulation results, the performance of the technology and the important features in the design, processing, installation and testing of the DTOF detector will be explored and understood. Further optimization of the DTOF engineering design can be achieved to fulfill the required PID capabilities of the STCF experiment.

  \item \textbf{Radiation resistance and aging}\\
In STCF operation, the spectrometer system will encounter an unprecedented radiation dose, which poses a great challenge in terms of the radiation resistance of the detector and the front-end electronics (FEE). Particularly important are the radiation hardness of the ASIC chip and the aging properties of the photosensors, ({\it i.e.}, the MCP-PMTs). The R\&D of the ASIC chip needs to include radiation reinforcement technology.
\end{itemize}

\subsection{EMC}
\label{sec:rd_emc}
The main function of the electromagnetic calorimeter (EMC) of the STCF is to realize accurate measurements of photon energy, position and arrival time under the conditions of high background count rates. For 1~GeV photons, the energy resolution should be better than 2.5\%, and the position resolution should be better than 5~mm. In addition, the time resolution should reach 300~ps @ 1~GeV to distinguish neutral particles (neutrons, photons). The pure cesium iodide crystal (pCsI) has the advantages of radiation hardness and fast time response. It is a very promising option for the EMC of the STCF spectrometer. In the following, the main R\&D items are listed.
\begin{itemize}
\item \textbf{Light Yield Study}\\
For the measurement of low-energy photons, the light yield of a crystal directly determines its energy measurement accuracy. It is necessary to study the crystal light yield according to the fluorescence characteristics of the pCsI crystal (wavelength band, decay time, etc.). On the basis of previous research, further study on wavelength shifting materials is likely to improve the light yield of the pCsI crystal.

  \item \textbf{Electronics Study}\\
The EMC readout electronics mainly include two modules: a preamplifier module and a digital processing module. The preamplifier module includes a photoelectric conversion device, an APD, and a front-end amplification circuit.
According to the requirements for high precision and large dynamic range detection of the EMC, a design yielding low noise and large dynamic range needs to be obtained for the preamplifier module. Additionally, the preamplifier module needs an antistacking design to solve the problem of preamplifier circuit saturation caused by a high background and high signal counting rate.

  \item \textbf{Pileup Study}\\
To cope with the high background, in addition to reducing the recovery time of the readout circuit as much as possible in the hardware design, the analysis method still needs to be studied.
It is planned to use a multiwaveform fitting method in the readout of the EMC, and it is also essential to verify the feasibility of this method with simulations and experiments.

  \item \textbf{Time Measurement Study}\\
It is planned to use waveform sampling readout and online waveform fitting to realize high-precision time measurement based on the premise of controllable data size. The detailed algorithm and technical implementation need to be studied.

\item \textbf{Prototype Study}\\
To verify the performance of the EMC, a small prototype needs to be developed, such as a 3 $\times$ 3 or 5 $\times$ 5 crystal array. This includes batch testing of crystals, photoelectronic devices and electronics, and finally, the test process and test standards must be established.
\end{itemize}

\subsection{MUD}
\label{sec:rd_mud}
According to the intensive background at the STCF, the conceptual design of the muon detector (MUD) adopts an innovative approach combining an inner RPC detector and an outer plastic scintillator (PS) detector to avoid interference from background while retaining the required muon-ID abilities. To realize such a design, R\&D on various subjects is necessary, including PS+SiPM technology, RPC technology and readout electronics. A large prototype of the hybrid MUD is needed to verify the design and basic performance of the detector. The studies will involve the following:
\begin{itemize}
\item R\&D of a high-rate (up to approximately 100~kHz/channel) and large-area (size $>0.5$~m$^2$ and length $>1$~m for a single module) RPC detector;

\item R\&D of the large-size (size $>0.5$~m$^2$ and length $>1$~m for a single module) MUD module based on a plastic scintillator + wavelength shifting fiber (WLS) + SiPM technology;

\item R\&D of an electronic system suitable for both RPC and SiPM signal processing, which is required to exhibit a compact structure, low power consumption, high efficiency, radiation resistance and precision timing ($<100$~ps);

\item Construction and testing of an MUD prototype with 6 or more layers, equipped with a complete readout electronics system, to study the feasibility of the MUD.

\item Study on the MUD performance in a high rate and high radiation environment to explore the dependence of muon-ID abilities with a high background level.
\end{itemize}

\subsection{Solenoid}
\label{sec:rd_magnet}
The proposed 1~T solenoid with a 3~m diameter bore for the STCF detector solenoid magnet can be realized by adopting a self-supporting aluminum stabilized low temperature NbTi superconductor. However, a low-mass superconductor and thin-wall structure are important to make the solenoid more transparent so that particles can more easily cross the solenoid.
R\&D activities will focus on key technologies such as special superconductors, large superconducting coil manufacturing processes and liquid helium thermosiphon cooling.

\begin{itemize}
\item In the first stage, to develop a special low-temperature superconductor, an innovative coextruding technique will be used. A Rutherford cable that consists of strands of NbTi wires will be inserted into a pure aluminum stabilizer, forming an aluminum stabilized cable. This cable will then be inserted into high mechanical strength aluminum alloy reinforcement. The critical current $I_c$ must exceed 6~kA@4.2 K@4T.

\item In the second stage, automatic winding equipment with the ability to wind a superconducting coil with a 3 m aperture will be developed.

\item In the third stage, a method of liquid helium thermosiphon cooling for large superconducting solenoids will be developed. To study the phase transition process of helium in the circuit, the changes in the temperature distribution and the density distribution over time, a superconducting prototype of a suitable-scale thermosiphon circuit will be established for simulation and verification before formal solenoid construction.

\end{itemize}

\subsection{TDAQ}
\label{sec:rd_tdaq}
The data acquisition (DAQ) system performs the data processing and the system control of the STCF experiment, such as the L1 trigger data readout, data compression, information extraction, event building, high-level trigger (HLT) computing, parameter configuration, status monitoring, and online data decimating. In the STCF, the DAQ system is expected to process the data from $\sim$50.6 M FEE channels at an L1 trigger rate of approximately 400~kHz, with a total data rate after the L1 trigger of $\sim30$~GB/s.
Aiming at such design goals, the following studies on the STCF DAQ system design are being planned:
\begin{itemize}
\item Design of high-performance electronic modules used in the DAQ system to provide more interfaces, higher transmission speed, more resources, and lower average power consumption and cost for each channel;
\item High-speed data transmission and processing techniques supported by high-speed interfaces, high-speed networks, high-performance computing systems, high-performance firmware and software and high-speed storage techniques;
\item Design of the DAQ architecture to perform real-time trigger generation, trigger matching, and event building at a high trigger rate level of $10^5\sim 10^6$~Hz, supporting both the trigger mode and trigger-less mode;
\item The trigger algorithm and its real-time hardware or software implementation techniques in trigger-less mode;
\item Techniques that can improve the robustness, reliability, maintainability and scalability of the system;
\item Design of system operation and management of both the DAQ system and other related systems, such as the trigger system (under trigger mode), slow control system, fast control system, and online system.
\end{itemize}

\subsection{Software}
\label{sec:rd_software}
Offline software is designed and developed for Monte Carlo simulation and offline data processing. It mainly consists of a software framework, detector simulation, calibration and reconstruction as well as physics analysis tools. In the prestudy phase of the STCF experiment, offline software is deployed to optimize the detector options and study the detector performance as well as the physics potentials. After the experiment is running, it will be used to conduct complicated offline data processing on the data collected with the detector and to convert them into physics results.

The STCF, with a peak luminosity of $\stcflum$ or higher, producing a data sample approximately 100 times larger than current $\tau$-charm factories, presents a very large challenge for offline software and computing in terms of both rate and complexity; therefore, a specific offline software needs to be redesigned and developed with the state-of-art technologies to meet the STCF requirements. The main tasks are listed below.

\begin{itemize}

\item \textbf{Development of a high-performance software framework} \\
One of the great challenges of the STCF offline software is the management and processing of a higher volume of data (approximately several petabytes per day) than the present $\tau$-charm factories; the software framework, providing common functions for offline data processing and integrating all the applications into the unified software platform, plays a very important role. Therefore, one of the most crucial tasks is to design and develop a new framework that supports heterogeneous computing, including algorithms, data models and workflows running in heterogeneous environments, and provides interfaces to new toolkits, such as machine learning toolkits, IO systems, and simulation engines.

\item \textbf{Development of fast and accurate detector simulation suites} \\
Detector simulations serve many purposes at each point in the lifecycle of the STCF facility. The new toolkit DD4hep is adopted for the detector description, including the ITK, MDC, RICH, DTOF, EMC and MUD, as well as the MDI and support systems. A detector geometry management system is needed to manage different versions of subdetector options, to support fast iterations of detector performance studies and to provide consistent geometry information for different applications, such as detector simulation, reconstruction and visualization. Further study on the accuracy of detector description, the physics interaction of the different types of particles with the detector medium, and a realistic electronics response is the part of the detector simulation most crucial to achieving a high degree of compliance between the simulation and the data. Additionally, it is necessary to explore emerging technologies, such as parallel computing, heterogeneous computing and machine learning toolkits, to speed up the detector simulation and improve its performance.

\item \textbf{Development of the calibration methods and algorithms}\\
The main task is to study the calibration methods for the key measurements from each subdetector, such as the relationship between drift distance and drift time, event start time, energy loss of the MDC, refractive index of the PID radiator, energy and position of the shower from the EMC, and noise level of the MUD, to develop the corresponding calibration algorithms and establish the complete calibration system to perform accurate conversions between electronic readouts and physical quantities, minimizing the influence of external factors of the experiment and the operating status of the detector itself on the physical measurements.

\item \textbf{Development of event reconstruction methods and algorithms}\\
Event reconstruction is a very complicated and challenging task in offline data processing, including reconstruction of the charged tracks, electromagnetic showers and particle identifications to produce the momentum, energy and type of the particles for further physics analysis. For the charge tracks, we develop a track finding method with conformal transformation and Hough transformation, a track fitting method based on the deterministic annealing filter (DAF) and track extrapolation based on Geant4. The likelihood-based PID methods are also studied for the RICH and DTOF detectors. The procedure for the EMC shower is also built up from clustering, seed finding, cluster splitting and the correction of the energy. For the MUD, one algorithm is developed based on the BDT method. Further study of these methods is crucial to achieving the design specifications of the detector hardware.

\item \textbf{Development of physics simulation software}\\
In the prestudy stage of the STCF, a parameterized (fast) simulation toolkit is necessary for detector optimization and determining the physics potential capabilities of the STCF. The fast simulation takes as inputs the response of physical objects in each subsystem of the detector, including the resolution, efficiency and related variables for the kinematic fit and the secondary vertex reconstruction algorithm. Therefore, the physics signal significance can be used as a metric to evaluate the detector options and the physics reach studies. The further optimization of the current fast simulation tool according to the new requirements of the physics study and detector design is one of the key tasks of offline software.

\end{itemize}

\subsection{Physics}
\label{sec:rd_physics}
The physical potentials at the STCF are presented in Chapter~\ref{CDR_phys}.
During R\&D, it is necessary to provide a feasibility study of these physics
programs under the software framework developed at the STCF, which can be achieved
by establishing and optimizing the reconstruction algorithm, developing the physics
tools for partial-wave analysis or Argand diagram analysis, analyzing the systematic
uncertainty sources and finally extracting the key physical parameters.
The physics simulation will be expanded in terms the following three aspects.
\begin{itemize}
\item The hadron spectrum and hadronic structure, including studies on properties
of $XYZ$ particles; spectrum analysis of light hadrons; charmonium states and
charm hadrons; precise tests of SM parameters such as
muon anomaly magnetic moments, R values and $\tau$-mass; and
hadronic structures from electromagnetic form factors and fragmentation functions. At higher CME $\sqrt{s}>5$~GeV, a feasibility study of searching for penta-quark, doubly charm baryon, and di-charmonium production will be carried out under
the guidance of theory. Useful physical tools will be developed during the R\&D for
a highly efficient physics analysis.

\item Flavor physics and $CP$ violations.
The STCF will be a flavor factory with very large amounts of charm hadrons produced.
In R\&D, a sensitivity study of the fundamental parameters will be performed
including the CKM matrix elements $|V_{cs}|$, $|V_{cd}|$,
$\gamma/\phi_{3}$ of the
CKM triangle, $D^{0}-\bar{D}^{0}$ mixing, decay of charm hadrons, etc.
In addition, the sensitivity of $CP$ violations at the STCF from
various aspects in the hyperon, $\tau$-lepton and charm hadron sectors
will be studied under unpolarized and polarized electron beams.
In these studies, the sources of systematic uncertainty should be
carefully examined to match the unprecedented statistical
uncertainty accuracy.

\item Probing of new physics beyond the SM. During R\&D, the mixing strengths of new particles, such as dark photons and millicharged particles, will be studied at the STCF
to test various models beyond the SM.
Sensitivity studies of the processes that violate quantum number conservation
such as lepton flavor violations, lepton number violations or baryon number violations,
and flavor-changing-neutral-current processes
will be carried out. Moreover, the rare decays from $J/\psi$, charm meson decays will be
studied with large data samples. The background needs to be carefully analyzed
in these studies to achieve a high level of sensitivity.
This is expected to extend the reach of the current experimental efforts in both the energy
and intensity frontiers and
to make several quantitative estimations
of the sensitivities in the probing of these new physics to test several scenarios beyond the SM.

\end{itemize}

\newpage
\clearpage
%\end{comment}     

\chapter*{Acknowledgement}
\addcontentsline{toc}{chapter}{Acknowledgement}
%The authors thank Andrzej Kupsc, Sergey Barsuk, Olivier Callot and Wolfgang K{\"u}hn for their contribution on the CDR draft.
%The authors thank the international review committee XXX for their great effort in reading the CDR draft and providing valuable suggestions. 
%The STCF working group thanks all 
%the colleagues in the world-wide community for many profitable discussions
%and expresses gratitude to the Hefei Comprehensive National Science Center for their strong support.  This work is supported by: international 
%partnership program of the Chinese Academy of Sciences Grant No. 211134KYSB20200057.

We would like to thank  Sergey Barsuk( IN2P3-CNRS \& Université Paris 11, France), Oliver Callot(IN2P3/CNRS \& Université Paris 11, France), Wolfgang K{\"u}hn (Justus-Liebig-Universitaet Giessen, II. Physikalisches Institut, Germany), Andrzej Kupsc(National Centre for Nuclear Research, Poland;  Uppsala University,  Sweden), Cheng Li(University of Science and Technology of China, China), Jin Li (Institute of High Energy Physics,  China ), Weiguo Li (Institute of High Energy Physics, China),  Changzheng Yuan (Institute of High Energy Physics, China; University of Chinese Academy of Sciences, China) for their contribution to this report. We thank the University of Science and Technology of China, the Hefei Comprehensive National Science Center, State Key Laboratory of Particle Detection and Electronics, and National Synchrotron Radiation Laboratory for their strong support. The research work leading to this report was supported by the National Key R\&D Program of China under Contract No.~2022YFA1602200, the International Partnership Program of the Chineses Academy of Sciences under Grant No.~211134KYSB20200057 and  the STCF key technology research and development project.
\bibliographystyle{apsrev4-1}
\bibliography{ref_introduction_det}

\end{document}